\documentclass [12pt,notitlepage]{report}
\usepackage{amsmath,amssymb,cite}
\usepackage[]{hyperref}
\makeatletter
\@addtoreset{chapter}{part}
\makeatother  
\usepackage{appendix}
\usepackage{tabularx}
\usepackage[english]{babel}
\usepackage{graphicx}
\usepackage{bmpsize}
\usepackage{indentfirst}
\usepackage{epsfig}
\usepackage{slashed}
\usepackage{fancyhdr}
\usepackage{pdfpages}
\usepackage[section]{placeins}
\numberwithin{equation}{chapter}

\fancypagestyle{plain}{\fancyhead{}
}

\begin{document}

\title{
Computational Physics: An Introduction to\\
Monte Carlo Simulations of Matrix Field Theory
}

\author{Badis Ydri\\
Department of Physics, Faculty of Sciences, BM Annaba University,\\
 Annaba, Algeria.
}

\maketitle
\date
\abstract{This book is divided into two parts. In the first part we give an elementary introduction to computational physics consisting of $21$ simulations which originated from a formal course of lectures and laboratory simulations delivered since  $2010$ to physics students at Annaba University. The second part is much more advanced  and deals with the problem of how to set up working Monte Carlo simulations of matrix field theories which involve finite dimensional matrix regularizations of noncommutative and fuzzy field theories, fuzzy spaces and matrix geometry.   The study of matrix field theory in its own right  has also become very important to the proper understanding of all noncommutative, fuzzy and matrix  phenomena.  The second  part, which consists of $9$ simulations, was delivered informally to doctoral students who are working on various problems in matrix field theory. Sample codes as well as sample key solutions are also provided for convenience and completness.  An appendix containing an executive arabic summary of the first part is added at the end of the book.}

\tableofcontents

\chapter*{Introductory Remarks}
\addcontentsline{toc}{chapter}{Introductory Remarks}  
\section*{Introducing Computational Physics}
\addcontentsline{toc}{section}{Introducing Computational Physics}  

Computational physics is a subfield of computational science and scientific computing in which we combine elements from physics (especially theoretical), elements from mathematics (in particular applied mathematics such as numerical analysis) and elements from computer science (programming) for the purpose of solving a physics problem. In physics there are traditionally two approaches which are followed: $1)$ The experimental approach and  $2)$ The theoretical approach. Nowadays, we may consider ``The computational approach'' as a third approach in physics. It can even be argued that the computational approach is independent from the first two approaches and it is not simply a bridge between the two.

The most important use of computers in physics is {\it simulation}. Simulations are suited for nonlinear problems which can not generally solved by analytical methods. The starting point of a simulation is an idealized model of a physical system of interest. We want to check whether or not the behaviour of this model is consistent with observation. We specify an algorithm for the implementation of the model on a computer. The execution of this implementation is a simulation. Simulations are therefore virtual experiments. The comparison between computer simulations and laboratory experiments goes therefore as follows:
\begin{center}
\begin{tabular}{||p{4.5cm}|p{4.5cm}||}
\hline
Laboratory experiment & Simulation \\ \hline
 sample  & model \\ \hline
physical apparatus  & computer program (the code) \\ \hline
calibration  & testing of code \\ \hline
measurement  & computation \\ \hline
data analysis  & data analysis \\ \hline
\end{tabular}
\end{center}
A crucial tool in computational physics is programming languages. In simulations as used by the majority of research physicists codes are written in a high-level compiled language such as Fortran and C/C++. In such simulations we may also use calls to routine libraries such as Lapack. The use of mathematical software packages such as Maple, Mathematica and Matlab is only suited for relatively small calculations. These packages are interpreted languages and thus the code they produce run generally far too slowly compared to compiled languages. In this book we will mainly follow the path of developping and writing all our codes in a high-level compiled language and not call any libraries. As our programming language we will use Fortran $77$ under the Linux operating system. We adopt exclusively the Ubuntu distribution of Linux. We will use the Fortran compilers  f$77$ and gfortran. As an editor we will use mostly Emacs and sometimes Gedit and Nano while for graphics we will use mostly Gnuplot.

\section*{References}
\addcontentsline{toc}{section}{References}  

The main  references which we have followed in developing the first part of this book include the following items:
\begin{enumerate}
\item N.J.Giordano, H. Nakanishi, Computational Physics (2nd edition), Pearson/Prentice Hall, (2006).
\item H.Gould, J.Tobochnick, W.Christian, An Introduction To Computer Simulation Methods: Applications to Physical Systems (3rd Edition), Addison-Wesley (2006). 
\item R.H.Landau, M.J.Paez, C.C. Bordeianu,  Computational Physics: Problem Solving with Computers (2nd edition), John Wiley and Sons (2007).
\item R.Fitzpatrick, Introduction to Computational Physics,\\ \url{http://farside.ph.utexas.edu/teaching/329/329.html}.
\item Konstantinos Anagnostopoulos, Computational Physics: A Practical Introduction to Computational Physics and Scientific Computing, Lulu.com (2014).

\item J. M. Thijssen, Computational Physics, Cambridge University Press (1999). 
\item M. Hjorth-Jensen,Computational Physics, CreateSpace Publishing (2015).
\item Paul L.DeVries, A First Course in Computational Physics (2nd edition), Jones and Bartlett Publishers (2010).
\end{enumerate}

\section*{Codes and Solutions}
\addcontentsline{toc}{section}{Codes and Solutions}  
The Fortran codes relevant to the problems considered in the first part of the book as well as some key sample solutions can be found at the URL:\\
\url{http://homepages.dias.ie/ydri/codes_solutions/}

\section*{Matrix Field Theory}
\addcontentsline{toc}{section}{Matrix Field Theory}  
The second part of this book, which is   effectively the main part, deals with the important problem of how to set up working Monte Carlo simulations of  matrix field theories in a, hopefully, pedagogical way. The subject of matrix field theory involves non-perturbative matrix regularizations, or simply matrix representations, of noncommutative field theory and noncommutative geometry, fuzzy physics and fuzzy spaces, fuzzy field theory, matrix geometry and gravity and random matrix theory.  The subject of matrix field theory may even include matrix regularizations of supersymmetry, string theory and M-theory. These matrix regularizations employ necessarily finite dimensional matrix algebras so that the problems are amenable and are accessible to Monte Carlo methods. 

The matrix regulator should be contrasted with the, well established, lattice regulator with advantages and disadvantages which are discussed in their places in the literature. However, we note that only $5$ simulations among the $7$ simulations considered in this part of the book use the matrix regulator whereas the other $2$, closely related simulations, use the usual lattice regulator. This part contains also a special chapter on the Remez and conjugate gradient algorithms which are required for the simulation of dynamical fermions. The study of matrix field theory in its own right, and not thought of as regulator, has also become very important to the proper understanding of all noncommutative, fuzzy and matrix  phenomena.  Naturally, therefore, the mathematical, physical and numerical aspects, required for the proper study of matrix field theory, which are found in this part of the book are quite advanced by comparison with what is found in the first part of the book.  

The set of references for each topic consists mainly of research articles and is included at the end of each chapter. Sample numerical calculations are also included as a section or several sections in each chapter. Some of these solutions are quite detailed whereas others are brief. The relevant Fortran codes for this part of the book are collected in the last chapter for convenience and completeness. These codes are, of course, provided as is and no warranty should be assumed.

\section*{Appendices}
\addcontentsline{toc}{section}{Appendices}  
We attach two appendices at the end of this book relevant to the first part of this book. In the first appendix we discuss the floating point representation of numbers, machine precision and roundoff and systematic errors. In the second appendix we give an executive summary of the simulations of part I translated into arabic.
\section*{Acknowledgments}
\addcontentsline{toc}{section}{Acknowledgments}  
Firstly, I would like to thank both the ex-head as well as the current-head of the physics department, professor M.Benchihab and professor A.Chibani, for their critical help in formally launching the computational physics course at  BM Annaba University during the academic year 2009-2010 and thus making the whole experience possible. This three-semester course, based on the first part of this book, has become since a fixture of the physics curriculum at both the Licence (Bachelor) and Master levels. Secondly, I should also thank  doctor A.Bouchareb and doctor R.Chemam who had helped in a crucial way with the actual teaching of the course, especially the laboratory simulations,  since the beginning. Lastly, I would like to thank my doctoral students and doctor A.Bouchareb for their patience and contributions during the development of the second part of this book in the weekly informal meeting we have organized for this purpose.

\part{Introduction to Computational Physics}

\chapter{Euler Algorithm}

\section{Euler Algorithm}
It is a well appreciated fact that first order differential equations are commonplace in all branches of physics. They appear virtually everywhere and some of the most fundamental problems of nature obey simple first order differential equations or second order differential equations. It is so often possible to recast second order differential equations as first order differential equations with a doubled number of unknown. From the numerical standpoint the problem of solving first order differential equations is a conceptually simple one as we will now explain.

We consider the general first order ordinary differential equation
\begin{eqnarray}
y^{'}=\frac{dy}{dx}=f(x,y).
\end{eqnarray}
We impose the general initial-value boundary condition is
\begin{eqnarray}
y(x_0)=y_0.
\end{eqnarray}
We solve for the function $y=y(x)$ in the unit $x-$interval starting from $x_0$. We make the $x-$interval discretization
\begin{eqnarray}
x_n=x_0+n{\Delta}x~,~n=0,1,...
\end{eqnarray} 
The Euler algorithm is one of the oldest known numerical recipe. It  consists in replacing the function $y(x)$ in the interval $ [x_n,x_{n+1}]$ by the straight line connecting the points $(x_n,y_n)$ and $(x_{n+1},y_{n+1})$. This comes from the definition of the derivative at the point $x=x_n$ given by
\begin{eqnarray}
\frac{y_{n+1}-y_n}{x_{n+1}-x_n}=f(x_n,y_n).
\end{eqnarray}
This means that we replace the above first order differential equation by the finite difference equation
\begin{eqnarray}
y_{n+1}\simeq y_n+{\Delta}x f(x_n,y_n).
\end{eqnarray}
This is only an approximation. The truncation error is given by the next term in the Taylor's expansion of the function $y(x)$ which is given by
\begin{eqnarray}
y_{n+1}\simeq y_n+{\Delta}x f(x_n,y_n)+\frac{1}{2}\Delta x^2 \frac{df(x,y)}{dx}|_{x=x_n}+....
\end{eqnarray}
The error then reads
\begin{eqnarray}
\frac{1}{2}({\Delta}x)^{2} \frac{df(x,y)}{dx}|_{x=x_n}.
\end{eqnarray}
The error per step is therefore proportional to $({\Delta}x)^{2}$. In a unit interval we will perform $N=1/{\Delta}x$ steps. The total systematic error is therefore proportional to
\begin{eqnarray}
N({\Delta}x)^{2}=\frac{1}{N}. 
\end{eqnarray}
\section{First Example and Sample Code}
\subsection{Radioactive Decay}
It is an experimental fact that radioactive decay obeys a very simple first order differential equation. In a spontaneous radioactive decay a particle with no external influence will decay into other particles. A typical example is the nuclear isotope uranium $235$. The exact moment of decay of any one particle is random. This means that the number $-d{\cal N}(t)={\cal N}(t)-{\cal N}(t+dt)$ of nuclei which will decay during a time inetrval  $dt$ must be proportional to $dt$ and to the number ${\cal N}(t)$ of particles present at time $t$, i.e.
\begin{eqnarray}
-d{\cal N}(t)\propto {\cal N}(t) dt.
\end{eqnarray}
In other words the probability of decay per unit time given by $(-d{\cal N}(t)/{\cal N}(t))/dt$ is a constant which we denote $1/\tau$. The minus sign is due to the fact that  $d{\cal N}(t)$ is negative since the number of particles decreases with time. We write
\begin{eqnarray}
\frac{d{\cal N}(t)}{dt}=-\frac{{\cal N}(t)}{\tau}.\label{decaylaw0}
\end{eqnarray}
The solution of this first order differential equation is given by a simple exponential function, viz
\begin{eqnarray}
{\cal N}(t)={\cal N}_0 \exp(-t/\tau).\label{decaylaw}
\end{eqnarray}
The number ${\cal N}_0$ is the number of particles at time $t=0$. The time $\tau$ is called the mean lifetime. It is the average time for decay. For the uranium $235$ the mean lifetime is around $10^{9}$ years.

The goal now is to obtain an approximate numerical solution to  the problem of radioactivity using the Euler algorithm. In this particular case we can  compare to an exact solution given by the exponential decay law (\ref{decaylaw}).  We start evidently from the Taylor's expansion

\begin{eqnarray}
{\cal N}(t+\Delta t)={\cal N}(t)+\Delta t \frac{d{\cal N}}{dt} +\frac{1}{2}({\Delta}t)^{2}\frac{d^{2}{\cal N}}{dt^{2}}+...
\end{eqnarray}
We get in the limit $\Delta t \longrightarrow 0$
\begin{eqnarray}
\frac{d{\cal N}}{dt}={\rm Lim}_{\Delta t\longrightarrow 0}\frac{{\cal N}(t+\Delta t)-{\cal N}(t)}{\Delta t}.
\end{eqnarray}
We take $\Delta t$ small but non zero. In this case we obtain the approximation
\begin{eqnarray}
\frac{d{\cal N}}{dt}{\simeq}\frac{{\cal N}(t+\Delta t)-{\cal N}(t)}{\Delta t}.
\end{eqnarray}
Equivalently
\begin{eqnarray}
{\cal N}(t+\Delta t)\simeq {\cal N}(t)+\Delta t \frac{d{\cal N}}{dt}. 
\end{eqnarray}
By using (\ref{decaylaw0}) we get
\begin{eqnarray}
{\cal N}(t+\Delta t)\simeq {\cal N}(t)-\Delta t \frac{{\cal N}(t)}{\tau}.\label{decaylaw1} 
\end{eqnarray}
We will start from the number of particles at time $t=0$ given by ${\cal N}(0)={\cal N}_0$ which is known. We substitute $t=0$ in (\ref{decaylaw1}) to obtain ${\cal N}(\Delta t)={\cal N}(1)$ as a function of ${\cal N}(0)$. Next the value ${\cal N}(1)$ can be used in equation (\ref{decaylaw1}) to get ${\cal N}(2\Delta t)={\cal N}(2)$, etc. We are thus led to the  time discretization
\begin{eqnarray}
t\equiv t(i)=i\Delta t~,~i=0,...,N.\label{time}
\end{eqnarray}
In other words
\begin{eqnarray}
{\cal N}(t)={\cal N}(i).
\end{eqnarray}
The integer $N$ determine the total time interval $T=N\Delta t$. The numerical solution (\ref{decaylaw1}) can be rewritten as 
\begin{eqnarray}
{\cal N}(i+1)= {\cal N}(i)-\Delta t \frac{{\cal N}(i)}{\tau}~,~i=0,...,N.\label{decaylaw2} 
\end{eqnarray}
This is Euler algorithm for radioactive decay. For convenience we shift the integer $i$ so that the above equation takes the form
\begin{eqnarray}
{\cal N}(i)= {\cal N}(i-1)-\Delta t \frac{{\cal N}(i-1)}{\tau}~,~i=1,...,N+1.
\end{eqnarray}
We  introduce $\hat{\cal N}(i)={\cal N}(i-1)$, i.e $\hat{\cal N}(1)={\cal N}(0)={\cal N}_0$. We get
\begin{eqnarray}
\hat{\cal N}(i+1)= \hat{\cal N}(i)-\Delta t \frac{\hat{\cal N}(i)}{\tau}~,~i=1,...,N+1.\label{rad1}
\end{eqnarray}
The corresponding times are
\begin{eqnarray}
\hat{t}(i+1)=i\Delta t~,~i=1,...,N+1.\label{rad2}
\end{eqnarray}
 The initial number of particles at time $\hat{t}(1)=0$ is $\hat{\cal N}(1)={\cal N}_0$. This approximate solution should be compared with the exact solution (\ref{decaylaw}).
\subsection{A Sample Fortran Code}
The goal in this section is to provide a sample Fortran code which implements the above  algorithm (\ref{rad1}).  The reasons behind choosing Fortran were explained in the introduction. Any Fortran program, like any other programing language, must start with some {\it program statement} and conclude with an {\it end statement}. The program statement allows us to give a name to the program. The end statement may be preceded by a {\it return statement}. This looks like
\begin{verbatim}
program radioactivity 

c Here is the code

return
end
\end{verbatim} 
We have chosen the name ``radioactivity'' for our program. The ``c'' in the second line indicates that the sentence ``here is the code'' is only a comment and not a part of the code.  

After the program statement come the {\it declaration statements}. We state the {\it variables} and their {\it types} which are used in the program. In Fortran we have the {\it integer type} for integer variables and the {\it double precision type} for real variables. In the  case of (\ref{rad1}) the variables $\hat{\cal N}(i)$, $\hat{t}(i)$, $\tau$, $\Delta t$, ${\cal N}_0$ are real numbers while the variables $i$ and $N$ are integer numbers. 

An {\it array} $A$ of dimension $K$ is an ordered list of $K$ variables of a given type called the elements of the array and denoted $A(1)$, $A(2)$,...,$A(K)$. In our above example $\hat{\cal N}(i)$ and $\hat{t}(i)$ are real arrays of dimension $N+1$. We declare that $\hat{\cal N}(i)$ and $\hat{t}(i)$ are real for all $i=1,...,N+1$ by writing $\hat{\cal N}(1:N+1)$ and $\hat{t}(1:N+1)$.

Since an array is declared at the begining of the program it must have a fixed size. In other words the upper limit must be a constant and not a variable. In Fortran a constant is declared with a {\it parameter statement}. In our above  case the upper limit is $N+1$ and hence $N$ must be declared in parameter statement.

In the Fortran code we choose to use the notation $A=\hat{\cal N}$, $A0=\hat{\cal N}_0$, ${\rm time}=\hat{t}$, $\Delta=\Delta t$ and ${\rm tau}=\tau$. By putting all declarations together we get the following preliminary lines of code
\begin{verbatim}
program radioactivity 
integer i,N
parameter (N=100)
doubleprecision  A(1:N+1),A0,time(1:N+1),Delta,tau

c Here is the code

return
end
\end{verbatim}

The {\it input} of the computation in our case are obviously given by the parameters ${\cal N}_0$, $\tau$, $\Delta t$ and $N$.

For the radioactivity problem the main part of the code consists of equations (\ref{rad1}) and (\ref{rad2}). We start with the known quantities $\hat{\cal N}(1)= {\cal N}_0$ at $\hat{t}(1)=0$ and generate via the successive use of (\ref{rad1}) and (\ref{rad2}) $\hat{\cal N}(i)$ and $\hat{t}(i)$ for all $i> 1$. This will be coded using a {\it do loop}. It begins with a {\it do statement} and ends with an {\it enddo statement}. We may also indicate a step size. 

The {\it output} of the computation can be saved to a  file using a {\it write statement} inside the do loop. In our case the output is the number of particles $\hat{\cal N}(i)$ and the time $\hat{t}(i)$. The write statement reads explicitly  
\begin{eqnarray}
{\rm write}(10,*)~\hat{t}(i),\hat{\cal N}(i). \nonumber
\end{eqnarray}
The data will then be saved to a file called {\it fort.10}. 

By including the initialization, the do loop and the write statement we obtain the complete code
\begin{verbatim}
program radioactivity 
integer i,N
parameter (N=100)
doubleprecision  A(1:N+1),A0,time(1:N+1),Delta,tau
parameter (A0=1000,Delta=0.01d0,tau=1.0d0)

A(1)=A0
time(1)=0
do i=1,N+1,1
A(i+1)=A(i)-Delta*A(i)/tau
time(i+1)=i*Delta
write(10,*) time(i+1),A(i+1)
enddo

return
end
\end{verbatim}

\section{More Examples}
\subsection{Air Resistance}
We consider an athlete riding a bicycle moving on a flat terrain. The goal is to determine the velocity. Newton's second law is given by
\begin{eqnarray}
m\frac{dv}{dt}=F.
\end{eqnarray}
$F$ is the force exerted by the athlete on the bicycle. It is clearly very difficult to write down a precise expression for $F$. Formulating the problem in terms of the power generated by the athlete will avoid the use of an explicit formula for $F$. Multiplying the above equation by $v$ we obtain
\begin{eqnarray}
\frac{dE}{dt}=P.
\end{eqnarray}
$E$ is the kinetic energy and $P$ is the power, viz
\begin{eqnarray}
E=\frac{1}{2}mv^2~,~P=Fv.
\end{eqnarray}
Experimentaly we find that the output of well trained athletes is around $P=400$ watts over periods of $1h$. The above equation can also be rewritten as
 \begin{eqnarray}
\frac{dv^{2}}{dt}=\frac{2P}{m}.
\end{eqnarray}
For $P$ constant we get the solution
\begin{eqnarray}
v^{2}=\frac{2P}{m}t+v_0^{2}.
\end{eqnarray}
We remark the unphysical effect that $v\longrightarrow \infty$ as $t\longrightarrow \infty$. This is due to the absence of the effect of friction and in particular air resistance.

The most important form of friction is air resistance. The force due to air resistance (the drag force) is
\begin{eqnarray}
F_{\rm drag}=-B_1v-B_2v^{2}.
\end{eqnarray}
At small velocities the first term dominates whereas at large velocities it is the second term that dominates. For very small velocities the dependence on $v$ given by $F_{\rm drag}=-B_1v$ is known as Stockes' law. For reasonable velocities the drag force is dominated by the second term, i.e. it is given for most objects by 
\begin{eqnarray}
F_{\rm drag}=-B_2v^{2}.
\end{eqnarray}
The coefficient $B_2$ can be calculated as follows. As the bicycle-rider combination moves with velocity $v$ it pushes in a time $dt$ a mass of air given by $dm_{\rm air}=\rho A vdt$ where $\rho$ is the air density and $A$ is the frontal cross section. The corresponding kinetic energy is 
\begin{eqnarray}
dE_{\rm air}=dm_{\rm air}v^{2}/2. 
\end{eqnarray}
This is equal to the work done by the drag force, i.e. 
\begin{eqnarray}
-F_{\rm drag}v dt=dE_{\rm air}. 
\end{eqnarray}
From this we get
 \begin{eqnarray}
B_2=C\rho A.
\end{eqnarray}
The drag coefficient is $C=\frac{1}{2}$. The drag force becomes
\begin{eqnarray}
F_{\rm drag}=-C\rho A v^{2}.
\end{eqnarray}
Taking into account the force due to air resistance we find that Newton's law becomes 
\begin{eqnarray}
m\frac{dv}{dt}=F+F_{\rm drag}.
\end{eqnarray}
Equivalently
\begin{eqnarray}
\frac{dv}{dt}=\frac{P}{mv}-\frac{C\rho Av^{2}}{m}.
\end{eqnarray}
It is not obvious that this equation can be solved exactly in any easy way. The Euler algorithm gives the approximate solution
\begin{eqnarray}
v(i+1)&=&v(i)+\Delta t\frac{dv}{dt}(i).
\end{eqnarray}
In other words
\begin{eqnarray}
v(i+1)&=&v(i)+\Delta t\bigg(\frac{P}{mv(i)}-\frac{C\rho Av^2(i)}{m}\bigg)~,~i=0,...,N.
\end{eqnarray}
This can also be put in the form (with $\hat{v}(i)=v(i-1)$)
\begin{eqnarray}
\hat{v}(i+1)&=&\hat{v}(i)+\Delta t\bigg(\frac{P}{m\hat{v}(i)}-\frac{C\rho A\hat{v}^2(i)}{m}\bigg)~,~i=1,...,N+1.
\end{eqnarray}
The corresponding times are
\begin{eqnarray}
t\equiv \hat{t}(i+1)=i\Delta t~,~i=1,...,N+1.
\end{eqnarray}
The initial velocity $\hat{v}(1)$ at time $t(1)=0$ is known.

\subsection{Projectile Motion}

There are two forces acting on the projectile. The weight force and the drag force. The drag force is opposite to the velocity. In this case Newton's law is given by

\begin{eqnarray}
m\frac{d\vec{v}}{dt}&=&\vec{F}+\vec{F}_{\rm drag}\nonumber\\
&=&m\vec{g}-B_2v^{2}\frac{\vec{v}}{v}\nonumber\\
&=&m\vec{g}-B_2v\vec{v}.
\end{eqnarray}
The goal is to determine the position of the projectile and hence one must solve the two equations
\begin{eqnarray}
\frac{d\vec{x}}{dt}=\vec{v}.
\end{eqnarray}
\begin{eqnarray}
m\frac{d\vec{v}}{dt}=m\vec{g}-B_2v\vec{v}.
\end{eqnarray}
In components (the horizontal axis is $x$ and the vertical axis is $y$) we have $4$ equations of motion given by

\begin{eqnarray}
\frac{dx}{dt}={v}_x.\label{proj1}
\end{eqnarray}
\begin{eqnarray}
m\frac{d{v}_x}{dt}&=&-B_2v v_x.\label{proj2}
\end{eqnarray}
\begin{eqnarray}
\frac{dy}{dt}={v}_y.\label{proj3}
\end{eqnarray}
\begin{eqnarray}
m\frac{d{v}_y}{dt}&=&-mg-B_2v v_y.\label{proj4}
\end{eqnarray}
We recall the constraint
\begin{eqnarray}
v=\sqrt{v_x^{2}+v_y^{2}}.\label{proj5}
\end{eqnarray}
The numerical approach we will employ in order to solve the $4$ equations of motion (\ref{proj1})-(\ref{proj4}) together with (\ref{proj5}) consists in using  Euler algorithm. This yields the approximate solution given by the equations
\begin{eqnarray}
x(i+1)=x(i)+\Delta t {v}_x(i).
\end{eqnarray}
\begin{eqnarray}
v_x(i+1)&=&v_x(i)-\Delta t \frac{B_2v(i) v_x(i)}{m}.
\end{eqnarray}
\begin{eqnarray}
y(i+1)=y(i)+\Delta t v_y(i).
\end{eqnarray}
\begin{eqnarray}
v_y(i+1)&=&v_y(i)-\Delta t g-\Delta t \frac{B_2v(i) v_y(i)}{m}.
\end{eqnarray}
The constraint is
\begin{eqnarray}
v(i)=\sqrt{v_x(i)^{2}+v_y(i)^{2}}.
\end{eqnarray}
In the above equations the index $i$ is such that $i=0,...,N$. The initial position and velocity are given, i.e. $x(0)$, $y(0)$, $v_x(0)$ and $v_y(0)$ are known.

\section{Periodic Motions and Euler-Cromer and Verlet Algorithms}

As discussed above at each iteration using the Euler algorithm there is a systematic error proportional to $1/N$. Obviously this error will accumulate and may become so large that it will alter the solution drastically at later times. In the particular case of periodic motions, where the true nature of the motion can only become clear after few elapsed periods, the large accumulated error can lead to diverging results. In this section we will discuss simple variants of the Euler algorithm which perform much better than the plain Euler algorithm for periodic motions.

\subsection{Harmonic Oscillator}
We consider a simple pendulum: a particle of mass $m$ suspended by a massless string from a rigid support. There are two forces acting on the particle. The weight and the tension of the string. Newton's second law reads
\begin{eqnarray}
m\frac{d^{2}\vec{s}}{dt}&=&m\vec{g}+\vec{T}.
\end{eqnarray}
The parallel (with respect to the string) projection reads
\begin{eqnarray}
0&=&-m{g}\cos\theta+{T}.
\end{eqnarray}
The perpendicular projection reads
\begin{eqnarray}
m\frac{d^{2}{s}}{dt^2}&=&-m{g}\sin\theta.
\end{eqnarray}
The $\theta$ is the angle that the string makes with the vertical. Clearly $s=l\theta$. The force $mg\sin\theta$ is a restoring force which means that it is always directed toward the equilibrium position (here $\theta=0$) opposite to the displacement and hence the minus sign in the above equation. We get by using $s=l\theta$ the equation
\begin{eqnarray}
\frac{d^{2}{\theta}}{dt^2}&=&-\frac{g}{l}\sin\theta.
\end{eqnarray}
For small $\theta$ we have $\sin\theta\simeq \theta$. We obtain
\begin{eqnarray}
\frac{d^{2}{\theta}}{dt^2}&=&-\frac{g}{l}\theta.\label{sp}
\end{eqnarray}
The solution is a sinusoidal function of time with frequency $\Omega=\sqrt{g/l}$. It is given by
\begin{eqnarray}
{\theta}(t)&=&{\theta}_0\sin(\Omega t+\phi).
\end{eqnarray}
The constants ${\theta}_0$ and $\phi$ depend on the initial displacement and velocity of the pendulum. The frequency is independent of the mass $m$ and the amplitude of the motion and depends only on the length $l$ of the string. 

\subsection{Euler Algorithm}
The numerical solution is based on Euler algorithm. It is found as follows. First we replace the equation of motion (\ref{sp}) by the following two equations
\begin{eqnarray}
\frac{d{\theta}}{dt}=\omega.
\end{eqnarray}
\begin{eqnarray}
\frac{d{\omega}}{dt}=-\frac{g}{l}\theta.
\end{eqnarray}
We use the definition of a derivative of a function, viz
\begin{eqnarray}
\frac{df}{dt}=\frac{f(t+{\Delta}t)-f(t)}{\Delta t}~,~{\Delta}t\longrightarrow 0.
\end{eqnarray}
We get for small but non zero $\Delta t$ the approximations
\begin{eqnarray}
&&{\theta}(t+\Delta t)\simeq{\theta}(t)+{\omega}(t){\Delta}t\nonumber\\
&&{\omega}(t+\Delta t)\simeq {\omega} (t)-\frac{g}{l}{\theta}(t){\Delta}t.
\end{eqnarray}
 We consider the time discretization
\begin{eqnarray}
t\equiv t(i)=i\Delta t~,~i=0,...,N.
\end{eqnarray}
In other words
\begin{eqnarray}
\theta(t)=\theta(i)~,~\omega(t)=\omega(i).
\end{eqnarray}
The integer $N$ determine the total time interval $T=N\Delta t$. The above numerical solution can be rewritten as 

\begin{eqnarray}
&&{\omega}(i+1)={\omega} (i)-\frac{g}{l}{\theta}(i){\Delta}t\nonumber\\
&&{\theta}(i+1)={\theta}(i)+{\omega}(i){\Delta}t.
\end{eqnarray}
We shift the integer $i$ such that it takes values in the range $[1,N+1]$. We obtain
\begin{eqnarray}
&&{\omega}(i)={\omega} (i-1)-\frac{g}{l}{\theta}(i-1){\Delta}t\nonumber\\
&&{\theta}(i)={\theta}(i-1)+{\omega}(i-1){\Delta}t.
\end{eqnarray}
We introduce $\hat{\omega}(i)=\omega(i-1)$ and $\hat{\theta}(i)=\theta(i-1)$. We get with $i=1,...,N+1$ the equations
\begin{eqnarray}
&&\hat{\omega}(i+1)=\hat{\omega} (i)-\frac{g}{l}\hat{\theta}(i){\Delta}t\nonumber\\
&&\hat{\theta}(i+1)=\hat{\theta}(i)+\hat{\omega}(i){\Delta}t.\label{om}
\end{eqnarray}
By using the values of $\theta$ and $\omega$ at time  $i$ we calculate the corresponding values at time $i+1$. The initial angle and angular velocity $\hat\theta(1)=\theta(0)$ and $\hat\omega(1)=\omega(0)$ are known. This process will be repeated until the functions $\theta$ and $\omega$ are determined for all times.

\subsection{Euler-Cromer Algorithm}
As it turns out the above Euler algorithm does not conserve energy. In fact Euler's method is not good for all oscillatory systems. A simple modification of Euler's algorithm due to Cromer will solve this problem of energy non conservation. This goes as follows. We use the values of the angle $\hat{\theta} (i)$ and the angular velocity $\hat{\omega} (i)$ at time step $i$ to calculate the angular velocity $\hat{\omega} (i+1)$ at time step $i+1$. This step is the same as before. However we use $\hat{\theta} (i)$ and $\hat{\omega} (i+1)$ (and not $\hat{\omega} (i)$) to calculate $\hat{\theta} (i+1)$ at time step $i+1$. This procedure as shown by Cromer's will conserve energy in oscillatory problems. In other words equations (\ref{om}) become

\begin{eqnarray}
&&\hat{\omega}(i+1)=\hat{\omega} (i)-\frac{g}{l}\hat{\theta}(i){\Delta}t\nonumber\\
&&\hat{\theta}(i+1)=\hat{\theta}(i)+\hat{\omega}(i+1){\Delta}t.
\end{eqnarray}
The error can be computed as follows. From these two equations we get
\begin{eqnarray}
\hat{\theta}(i+1)&=&\hat{\theta}(i)+\hat{\omega}(i){\Delta}t-\frac{g}{l}\hat{\theta}(i)\Delta t^2\nonumber\\
&=&\hat{\theta}(i)+\hat{\omega}(i){\Delta}t+\frac{d^2\hat\theta}{dt}|_{i}\Delta t^2.
\end{eqnarray}
In other words the error per step is still of the order of $\Delta t^2$. However the Euler-Cromer algorithm does better than Euler algorithm with periodic motion. Indeed at each step $i$ the energy conservation condition reads
\begin{eqnarray}
E_{i+1}=E_i+\frac{g}{2l}(\omega_i^2-\frac{g}{l}\theta_i^2)\Delta t^2.
\end{eqnarray}
The energy of the simple pendulum is of course by
\begin{eqnarray}
E_i=\frac{1}{2}\omega_i^2+\frac{g}{2l}\theta_i^2.
\end{eqnarray}
The error at each step is still proportional to $\Delta t^2$ as in the Euler algorithm. However the coefficient is precisely equal to the difference between the values of the kinetic energy and the potential energy at the step $i$. Thus the accumulated error which is obtained by summing over all steps vanishes since the average kinetic energy is equal to the average potential energy. In the Euler algorithm the coefficient is actually equal to the sum of the kinetic and potential energies and as consequence no cancellation can occur. 
\subsection{Verlet Algorithm}
Another method which is much more accurate and thus very suited to periodic motions is due to Verlet. Let us consider the forward and backward Taylor expansions
\begin{eqnarray}
  \theta(t_i+\Delta t)=\theta(t_i)+\Delta t \frac{d\theta}{dt}|_{t_i}+\frac{1}{2}(\Delta t)^2\frac{d^2\theta}{dt^2}|_{t_i}+\frac{1}{6}(\Delta t)^3\frac{d^3\theta}{dt^3}|_{t_i}+...
\end{eqnarray}
\begin{eqnarray}
  \theta(t_i-\Delta t)=\theta(t_i)-\Delta t \frac{d\theta}{dt}|_{t_i}+\frac{1}{2}(\Delta t)^2\frac{d^2\theta}{dt^2}|_{t_i}-\frac{1}{6}(\Delta t)^3\frac{d^3\theta}{dt^3}|_{t_i}+...
\end{eqnarray}
Adding these expressions we get
\begin{eqnarray}
  \theta(t_i+\Delta t)=2\theta(t_i)-\theta(t_i-\Delta t)+(\Delta t)^2\frac{d^2\theta}{dt^2}|_{t_i}+O(\Delta ^4).
\end{eqnarray}
We write this as
\begin{eqnarray}
  \theta_{i+1}=2\theta_i-\theta_{i-1}-\frac{g}{l}(\Delta t)^2\theta_i.
\end{eqnarray}
This is the Verlet algorithm for the harmonic oscillator. First we remark that the error is proportional to $\Delta t^4$ which is less than the errors in the Euler, Euler-Cromer (and even less than the error in the second-order Runge-Kutta)  methods so this method is much more accurate. Secondly in this method we do not need to calculate the angular velocity $\omega=d\theta/d t$. Thirdly this method is not self-starting. In other words given the initial conditions $\theta_1$ and $\omega_1$ we  need also to know $\theta_2$ for the algorithm to start. We can for example determine $\theta_2$ using the Euler method, viz $\theta_2=\theta_1+\Delta t~\omega_1$.
\section{Exercises}
\paragraph{Exercise $1$:}

We give the differential equations
\begin{eqnarray}
\frac{dx}{dt}=v.
\end{eqnarray}
\begin{eqnarray}
\frac{dv}{dt}=a-bv.
\end{eqnarray}
\begin{itemize}
\item{}
Write down the exact solutions.
\item{}
Write down the numerical solutions of these differential equations using Euler and Verlet methods  and determine the corresponding errors. 
\end{itemize}
\paragraph{Exercise $2$:}
The equation of motion of the solar system in polar coordinates is
\begin{eqnarray}
\frac{d^2r}{dt^2}=\frac{l^2}{r^3}-\frac{GM}{r^2}.
\end{eqnarray}
Solve this equation using Euler, Euler-Cromer and Verlet methods.
\paragraph{Exercise $3$:}
The equation of motion of a free falling object is
\begin{eqnarray}
\frac{d^2z}{dt^2}=-g.
\end{eqnarray}
\begin{itemize}
\item{}
Write down the exact solution.
\item{}
Give a solution of this problem in terms of Euler method  and determine the error. 
\item{}
We choose the initial conditions $z=0$, $v=0$ at $t=0$. Determine the position and the velocity between $t=0$ and $t=1$ for $N=4$. Compare with the exact solution and compute the error in each step. Express the result in terms of $l=g\Delta t^2$.
\item{}
Give a solution of this problem in terms of Euler-Cromer and Verlet methods  and determine the corresponding errors. 
\end{itemize}
\paragraph{Exercise $4$:}
The equation governing population growth is
\begin{eqnarray}
\frac{dN}{dt}=aN-bN^2.
\end{eqnarray}
The linear term represents the rate of birth while the quadratic term represents the rate of death. Give a solution of this problem in terms of the Euler and Verlet methods and determine the corresponding errors.

\section{Simulation $1$: Euler Algorithm- Air Resistance}

The equation of motion of a cyclist exerting a force on his bicycle corresponding to a constant power $P$ and moving against the force of air resistance is given by
\[\frac{dv}{dt}=\frac{P}{mv}-\frac{C\rho A v^2}{m} .\]
The numerical approximation of this first order differential equation which we will consider in this problem is based on Euler algorithm. 

\begin{itemize}
\item[$(1)$]Calculate the speed $v$ as a function of time in the case of zero air resistance and then
 in the case of non-vanishing air resistance. What do you observe.  We will take $P=200$ and $C=0.5$. We also give the values

\[m=70{\rm kg}~,~A=0.33m^2~,~\rho=1.2{\rm kg}/m^3~,~\Delta t=0.1s~,~T=200s.\]
The initial speed is

\[\hat{v}(1)=4m/s~,~\hat{t}(1)=0.\]

\item[$(2)$]What do you observe if we change the drag coefficient and/or the power. What do you observe
if we decrease the time step.

\end{itemize} 

\section{Simulation $2$: Euler Algorithm- Projectile Motion}

The numerical approximation based on the Euler algorithm of the equations of motion of a projectile moving under the effect of the forces of gravity and  air resistance  is given by the equations

\[v_x(i+1)=v_x(i)-\Delta t \frac{B_2v(i) v_x(i)}{m}.\]
\[v_y(i+1)=v_y(i)-\Delta t g-\Delta t \frac{B_2v(i) v_y(i)}{m}.\]
\[v(i+1)=\sqrt{v_x^2(i+1)+v_y^2(i+1)}.\]
\[x(i+1)=x(i)+\Delta t ~{v}_x(i).\]
\[y(i+1)=y(i)+\Delta t ~v_y(i).\]

\begin{itemize}
 \item[$(1)$]Write a Fortran code which implements the above Euler algorithm.

\item[$(2)$]We take the values

\[\frac{B_2}{m}=0.00004m^{-1}~,~g=9.8m/s^2.\]
\[v(1)=700m/s~,~\theta=30~{\rm degree}.\]
\[v_x(1)=v(1)\cos\theta~,~v_y(1)=v(1)\sin\theta.\]
\[N=10^5~,~\Delta t=0.01s.\]
Calculate the trajectory of the projectile with and without air resistance. What do you observe.
\item[$(3)$]We can determine numerically the range of the projectile by means of the conditional instruction if. This can be done by adding inside the do loop the following condition
\[{\rm if}~(y(i+1).{\rm le}.0)~{\rm exit}\]
Determine the range of the projectile with and without air resistance.
\item[$(4)$]In the case where air resistance is absent we know that the range is maximal when the initial angle is $45$ degrees. Verify this fact numerically by considering several angles. More precisely add a do loop over the initial angle in order to be able to study the range as a function of the initial angle.
\item[$(5)$]In the case where air resistance is non zero calculate the angle for which the range is maximal.
\end{itemize}

\section{Simulation $3$: Euler, Euler-Cromer and Verlet Algorithms}

We will  consider the numerical solutions of the equation of motion of a simple harmonic oscillator given by the  Euler, Euler-Cromer and Verlet algorithms which take the form
\[\omega_{i+1}=\omega_i-\frac{g}{l}\theta_i~\Delta t~,~\theta_{i+1}=\theta_i+\omega_i~\Delta t~,~{\rm Euler}.\]
\[\omega_{i+1}=\omega_i-\frac{g}{l}\theta_i~\Delta t~,~\theta_{i+1}=\theta_i+\omega_{i+1}~\Delta t~,~{\rm Euler}-{\rm Cromer}.\]
\[\theta_{i+1}=2\theta_i-\theta_{i-1}-\frac{g}{l}\theta_i(\Delta t)^2~,~{\rm Verlet}.\]
\begin{itemize}
 \item[$(1)$] Write a Fortran code which implements the Euler, Euler-Cromer and Verlet algorithms for the harmonic oscillator problem.
\item[$(2)$]Calculate the angle, the angular velocity and the energy of the harmonic oscillator as functions of time. The energy of the harmonic oscillator is given by
\[E=\frac{1}{2}\omega^2+\frac{1}{2}\frac{g}{l}\theta^2.\]
We take the values
\[g=9.8m/s^2~,l=1m~.\]
We take the number of iterations $N$ and the time step $\Delta t$ to be 
\[N=10000~,~\Delta t=0.05s.\]
The initial angle and the angular velocity are given by
\[\theta_1=0.1~{\rm radian}~,~\omega_1=0.\]
By using the conditional instruction if we can limit the total time of motion to be equal to say $5$ periods as follows 
\[{\rm if}~(t(i+1).{\rm ge}.5*{\rm period})~{\rm exit}.\]

\item[$(3)$]Compare between the value of the energy calculated with the Euler method and the value of the energy calculated with the Euler-Cromer method. What do you observe and what do you conclude.
\item[$(4)$]Repeat the computation using the Verlet algorithm. Remark that this method can not self-start from the initial values $\theta_1$ and $\omega_1$ only.
We must also provide the angle $\theta_2$ which can be calculated using for example Euler, viz
\[\theta_2=\theta_1+\omega_1~\Delta t.\] 
We also remark that the Verlet algorithm does not require the calculation of the angular velocity. However in order to calculate the energy we need to evaluate the angular velocity which can be obtained from the expression
\[\omega_i=\frac{\theta_{i+1}-\theta_{i-1}}{2\Delta t}.\]
\end{itemize}

\chapter{Classical Numerical Integration
}
\section{Rectangular Approximation}
We consider a generic one dimensional integral of the form

\begin{eqnarray}
F=\int_a^b f(x)dx.
\end{eqnarray}
In general this can not be done analytically. However this integral is straightforward to do  numerically. The starting point is Riemann definition of the integral $F$ as the area under the curve of the function $f(x)$ from $x=a$ to $x=b$. This is obtained as follows. We discretize the $x-$interval so that we end up with $N$ equal small intervals of lenght $\Delta x$, viz
\begin{eqnarray}
x_n=x_0+n{\Delta}x~,~\Delta x=\frac{b-a}{N}
\end{eqnarray}
Clearly $x_0=a$ and $x_N=b$.  Riemann definition is then given by the following limit
\begin{eqnarray}
F={\rm lim}_{\big(\Delta x\longrightarrow 0~,~N\longrightarrow \infty ~,~b-a={\rm fixed}\big)}\bigg(\Delta x\sum_{n=0}^{N-1} f(x_n)\bigg).
\end{eqnarray}
The first approximation which can be made is to drop the limit. We get the so-called rectangular approximation given by
\begin{eqnarray}
F_N=\Delta x\sum_{n=0}^{N-1} f(x_n).\label{rectangular00}
\end{eqnarray}
General integration algorithms approximate the integral $F$ by 
\begin{eqnarray}
F_N=\sum_{n=0}^{N} f(x_n)w_n.
\end{eqnarray}
In other words we evaluate the function $f(x)$ at $N+1$ points in the interval  $[a,b]$ then we sum the values $f(x_n)$ with some corresponding weights $w_n$. For example in the rectangular approximation (\ref{rectangular00}) the values $f(x_n)$ are summed with equal weights $w_n=\Delta x$, $n=0,N-1$ and $w_N=0$. It is also clear that the estimation $F_N$ of the integral $F$ becomes exact only in the large $N$ limit. 

\section{Trapezoidal Approximation}

The trapezoid rule states that we can approximate the integral by a sum of trapezoids. In the subinterval $[x_n,x_{n+1}]$ we replace the function $f(x)$ by a straight line connecting the two points $(x_n,f(x_n))$ and $(x_{n+1},f(x_{n+1}))$. The trapezoid has as vertical sides the two straight lines $x=x_n$ and $x=x_{n+1}$. The base is the interval $\Delta x=x_{n+1}-x_n$. It is not difficult to convince ourselves that the area of this trapezoid is

\begin{eqnarray}
\frac{(f(x_{n+1})-f(x_n))\Delta x}{2}+f(x_n)\Delta x=\frac{(f(x_{n+1})+f(x_n))\Delta x}{2}.
\end{eqnarray}
The integral $F$ computed using the trapezoid approximation is therefore given by summing the contributions from all the $N$ subinterval, viz
\begin{eqnarray}
T_N=\sum_{n=0}^{N-1}\frac{(f(x_{n+1})+f(x_n))\Delta x}{2}=\bigg(\frac{1}{2}f(x_0)+\sum_{n=1}^{N-1}f(x_n)+\frac{1}{2}f(x_N)\bigg)\Delta x.
\end{eqnarray}
We remark that the weights here are given by $w_0=\Delta x/2$, $w_n=\Delta x$, $n=1,...,N-1$ and $w_N=\Delta x/2$.
\section{Parabolic Approximation or Simpson's Rule}
In this case we approximate the function in the subinterval $[x_n,x_{n+1}]$ by a parabola given by
\begin{eqnarray}
f(x)=\alpha x^2 +\beta x +\gamma.
\end{eqnarray}
The area of the corresponding box is thus given by
\begin{eqnarray}
\int_{x_n}^{x_{n+1}}dx(\alpha x^2 +\beta x +\gamma)=\bigg(\frac{\alpha x^3}{3} +\frac{\beta x^2}{2} +\gamma x\bigg)_{x_n}^{x_{n+1}}.\label{simps}
\end{eqnarray}
Let us go back and consider the integral
\begin{eqnarray}
\int_{-1}^{1}dx(\alpha x^2 +\beta x +\gamma)=\frac{2\alpha }{3} +2\gamma .
\end{eqnarray}
We remark that
\begin{eqnarray}
f(-1)=\alpha-\beta+\gamma~,~f(0)=\gamma~,~f(1)=\alpha+\beta+\gamma.
\end{eqnarray}
Equivalently
\begin{eqnarray}
\alpha=\frac{f(1)+f(-1)}{2}-f(0)~,~\beta=\frac{f(1)-f(-1)}{2}~,~\gamma=f(0).
\end{eqnarray}
Thus
\begin{eqnarray}
\int_{-1}^{1}dx(\alpha x^2 +\beta x +\gamma)=\frac{f(-1)}{3}+\frac{4f(0)}{3}+\frac{f(1)}{3} .
\end{eqnarray}
In other words we can  express the integral of the function $f(x)=\alpha x^2 +\beta x +\gamma$ over the interval $[-1,1]$ in terms of the values of this function $f(x)$ at $x=-1,0,1$. Similarly we can express the  integral of $f(x)$ over the adjacent subintervals   $[x_{n-1},x_{n}]$ and  $[x_{n},x_{n+1}]$ in terms of the values of $f(x)$ at $x=x_{n+1},x_n,x_{n-1}$, viz

\begin{eqnarray}
\int_{x_{n-1}}^{x_{n+1}}dx~f(x)&=&\int_{x_{n-1}}^{x_{n+1}}dx(\alpha x^2 +\beta x +\gamma)\nonumber\\
&=&\Delta x\bigg(\frac{f(x_{n-1})}{3}+\frac{4f(x_n)}{3}+\frac{f(x_{n+1})}{3}\bigg) .
\end{eqnarray}
By adding the contributions from each pair of adjacent subintervals we get the full integral
\begin{eqnarray}
S_N&=&\Delta x\sum_{p=0}^{\frac{N-2}{2}}\bigg(\frac{f(x_{2p})}{3}+\frac{4f(x_{2p+1})}{3}+\frac{f(x_{2p+2})}{3}\bigg).
\end{eqnarray}
Clearly we must have $N$ (the number of subintervals) even. We compute
\begin{eqnarray}
S_N&=&\frac{\Delta x}{3}\bigg(f(x_0)+4f(x_1)+2f(x_2)+4f(x_3)+2f(x_4)+...+2f(x_{N-2})+4f(x_{N-1})+f(x_N)\bigg).\nonumber\\
\end{eqnarray}
 It is trivial to read from this expression the weights in this approximation. 

Let us now recall the trapezoidal approximation given by

\begin{eqnarray}
T_N=\bigg(f(x_0)+2\sum_{n=1}^{N-1}f(x_n)+f(x_N)\bigg)\frac{\Delta x}{2}.
\end{eqnarray}
Let us also recall that $N\Delta x=b-a$ is the length of the total interval which is always kept fixed. Thus by doubling the number of subintervals we halve the width, viz
\begin{eqnarray}
4T_{2N}&=&\bigg(2f(\hat{x}_0)+4\sum_{n=1}^{2N-1}f(\hat{x}_n)+2f(\hat{x}_{2N})\bigg)\frac{\Delta x}{2}\nonumber\\
&=&\bigg(2f(\hat{x}_0)+4\sum_{n=1}^{N-1}f(\hat{x}_{2n})+4\sum_{n=0}^{N-1}f(\hat{x}_{2n+1})+2f(\hat{x}_{2N})\bigg)\frac{\Delta x}{2}\nonumber\\
&=&\bigg(2f({x}_0)+4\sum_{n=1}^{N-1}f({x}_{n})+4\sum_{n=0}^{N-1}f(\hat{x}_{2n+1})+2f({x}_{N})\bigg)\frac{\Delta x}{2}.
\end{eqnarray}
In above we have used the identification $\hat{x}_{2n}=x_n$, $n=0,1,...,N-1,N$. Thus
\begin{eqnarray}
4T_{2N}-T_N
&=&\bigg(f(x_0)+2\sum_{n=1}^{N-1}f(x_{n})+4\sum_{n=0}^{N-1}f(\hat{x}_{2n+1})+f(x_{N})\bigg)\Delta{\hat{x}}\nonumber\\
&=&3S_N.
\end{eqnarray}
\section{Errors}
The error estimates for numerical integration are computed as follows. We start with the Taylor expansion
\begin{eqnarray}
f(x)=f(x_n)+(x-x_n)f^{(1)}(x_n)+\frac{1}{2!}(x-x_n)^2f^{(2)}(x_n)+...
\end{eqnarray}
Thus
\begin{eqnarray}
\int_{x_n}^{x_{n+1}}dx~f(x)=f(x_n)\Delta x+\frac{1}{2!}f^{(1)}(x_n)(\Delta x)^2+\frac{1}{3!}f^{(2)}(x_n)(\Delta x)^3+...
\end{eqnarray}
The error in the interval $[x_n,x_{n+1}]$ in the rectangular approximation is
\begin{eqnarray}
\int_{x_n}^{x_{n+1}}dx~f(x)-f(x_n)\Delta x=\frac{1}{2!}f^{(1)}(x_n)(\Delta x)^2+\frac{1}{3!}f^{(2)}(x_n)(\Delta x)^3+...
\end{eqnarray}
This is of order $1/N^2$. But we have $N$ subintervals. Thus the total error is of order $1/N$.

The error in the interval $[x_n,x_{n+1}]$ in the trapezoidal approximation is
\begin{eqnarray}
\int_{x_n}^{x_{n+1}}dx~f(x)-\frac{1}{2}(f(x_n)+f(x_{n+1}))\Delta x&=&\int_{x_n}^{x_{n+1}}dx~f(x)\nonumber\\
&-&\frac{1}{2}(2f(x_n)+\Delta x f^{(1)}(x_n)+\frac{1}{2!}(\Delta x)^2f^{(2)}(x_n)+...)\Delta x\nonumber\\
&=&(\frac{1}{3!}-\frac{1}{2}\frac{1}{2!})f^{(2)}(x_n)(\Delta x)^3+...
\end{eqnarray}
This is of order $1/N^3$ and thus the total error is of order $1/N^2$.

In order to compute the error in the interval $[x_{n-1},x_{n+1}]$ in the parabolic approximation we compute
\begin{eqnarray}
\int_{x_{n-1}}^{x_{n}}dx~f(x)+\int_{x_{n}}^{x_{n+1}}dx~f(x)&=&2f(x_n)\Delta x+\frac{2}{3!}(\Delta x)^3f^{(2)}(x_n)+\frac{2}{5!}(\Delta x)^5f^{(4)}(x_n)+...\nonumber\\
\end{eqnarray}
Also we compute
\begin{eqnarray}
\frac{\Delta x}{3}(f(x_{n+1})+f(x_{n-1})+4f(x_n))&=&2f(x_n)\Delta x+\frac{2}{3!}(\Delta x)^3f^{(2)}(x_n)+\frac{2}{3.4!}(\Delta x)^5f^{(4)}(x_n)+...\nonumber\\
\end{eqnarray}
Hence the error in the interval $[x_{n-1},x_{n+1}]$ in the parabolic approximation is
\begin{eqnarray}
\int_{x_{n-1}}^{x_{n+1}}dx~f(x)-\frac{\Delta x}{3}(f(x_{n+1})+f(x_{n-1})+4f(x_n))&=&(\frac{2}{5!}-\frac{2}{3.4!})(\Delta x)^5f^{(4)}(x_n)+...\nonumber\\
\end{eqnarray}
This is of order $1/N^5$. The total error is therefore of order  $1/N^4$.

\section{Simulation $4$: Numerical Integrals}
\begin{itemize}
 \item[$(1)$] We take the integral
\[I=\int_0^1 f(x) dx~;~f(x)=2x+3x^2+4x^3.\]
Calculate the value of this integral using the rectangular approximation. Compare with the exact result.

Hint: You can code the function using either "subroutine" or "function".
\item[$(2)$]Calculate the numerical error as a function of $N$. Compare with the theory.
\item[$(3)$]Repeat the computation  using the trapezoid method and the Simpson's rule.

\item[$(4)$]Take now the integrals
\[~I=\int_0^{\frac{\pi}{2}}\cos x dx~,~~I=\int_1^e \frac{1}{x} dx~,~I=\int_{-1}^{+1}\lim_{\epsilon\longrightarrow 0}\bigg( \frac{1}{\pi}\frac{\epsilon}{x^2+\epsilon^2}\bigg) dx.\]
\end{itemize}

\chapter{Newton-Raphson Algorithms and Interpolation
}

\section{Bisection Algorithm}
Let $f$ be some function. We are interested in the solutions (roots) of the equation

\begin{eqnarray}
f(x)=0.
\end{eqnarray}
The bisection algorithm works as follows. We start with two values of $x$ say $x_+$ and $x_-$ such that
\begin{eqnarray}
f(x_-) < 0~,~f(x_+)> 0.
\end{eqnarray}
In other words the function changes sign in the interval between $x_-$ and $x_+$  and thus there must exist a root between $x_{-}$ and $x_+$. If the function changes from  positive to negative as we increase $x$ we conclude that $x_+\leq x_-$. We bisect the interval $[x_+,x_-]$  at
\begin{eqnarray}
x=\frac{x_++x_-}{2}.
\end{eqnarray}
If $f(x)f(x_+)>0$ then $x_+$ will be changed to the point $x$ otherwise $x_-$ will be changed to the point $x$. We continue this process until the change in $x$ becomes insignificant or until the error becomes smaller than some tolerance. The relative error is defined by
\begin{eqnarray}
{\rm error}=\frac{x_+-x_-}{x}.
\end{eqnarray}
Clearly the absolute error $e=x_i-x_f$ is halved at each iteration and thus the rate of convergence of the bisection rule is linear. This is slow.
\section{Newton-Raphson Algorithm}
We start with a guess $x_0$. The new guess $x$ is written as $x_0$ plus some unknown correction $\Delta x$, viz
\begin{eqnarray}
x=x_0+\Delta x.
\end{eqnarray}
Next we expand the function $f(x)$ around $x_0$, namely
\begin{eqnarray}
f(x)=f(x_0)+\Delta x \frac{df}{dx}|_{x=x_0}.
\end{eqnarray}
The correction $\Delta x$ is determined by finding the intersection point of this linear approximation of $f(x)$ with the $x$ axis. Thus
\begin{eqnarray}
f(x_0)+\Delta x \frac{df}{dx}|_{x=x_0}=0\Longrightarrow \Delta x=-\frac{f(x_0)}{({df}/{dx})|_{x=x_0}}.
\end{eqnarray}
The derivative of the function $f$ is required in this calculation. In complicated problems it is much simpler to evaluate the derivative numerically than analytically. In these cases the derivative may be given by the forward-difference approximation (with some $\delta x$ not necessarily equal to $\Delta x$)
\begin{eqnarray}
\frac{df}{dx}|_{x=x_0}=\frac{f(x_0+\delta x)-f(x_0)}{\delta x}.
\end{eqnarray}
In summary this method works by drawing the tangent to the function $f(x)$ at the old guess $x_0$ and then use the intercept with the $x$ axis as the new hopefully better guess $x$. The process is repeated until the change in $x$ becomes insignificant.

Next we compute the rate of convergence of the Newton-Raphson algorithm. Starting from $x_i$ the next guess is $x_{i+1}$ given by
\begin{eqnarray}
x_{i+1}=x_i-\frac{f(x_i)}{f^{'}(x)}.
\end{eqnarray}
The absolute error at step $i$ is ${\epsilon}_i=x-x_i$ while the absolute error at step $i+1$ is ${\epsilon}_{i+1}=x-x_{i+1}$ where $x$ is the actual root. Then
\begin{eqnarray}
{\epsilon}_{i+1}={\epsilon}_i+\frac{f(x_i)}{f^{'}(x)}.
\end{eqnarray}
By using Taylor expansion we have
\begin{eqnarray}
f(x)=0=f(x_i)+(x-x_i)f^{'}(x_i)+\frac{(x-x_i)^2}{2!}f^{''}(x_i)+...
\end{eqnarray}
In other words
\begin{eqnarray}
f(x_i)=-{\epsilon}_if^{'}(x_i)-\frac{{\epsilon}_i^2}{2!}f^{''}(x_i)+...
\end{eqnarray}
Therefore the error is given by
\begin{eqnarray}
{\epsilon}_{i+1}=-\frac{{\epsilon}_i^2}{2}\frac{f^{''}(x_i)}{f^{'}(x_i)}.
\end{eqnarray}
This is quadratic convergence. This is faster than the bisection rule.

\section{Hybrid Method}
We can combine the certainty of the bisection rule in finding a root with the fast convergence of the  Newton-Raphson algorithm into a hybrid algorithm as follows. First we must know that the root is bounded in some interval $[a,c]$. We can use for example a graphical method. Next we start from some initial guess $b$. We take a  Newton-Raphson step
\begin{eqnarray}
b^{'}=b-\frac{f(b)}{f^{'}(b)}.
\end{eqnarray}
We check whether or not this step is bounded in the interval $[a,c]$. In other words we must check that
\begin{eqnarray}
a{\leq}b-\frac{f(b)}{f^{'}(b)}{\leq}c~{\Leftrightarrow}~(b-c)f^{'}(b)-f(b){\leq}0{\leq}(b-a)f^{'}(b)-f(b).
\end{eqnarray}
Therefore if
\begin{eqnarray}
\bigg((b-c)f^{'}(b)-f(b)\bigg)\bigg((b-a)f^{'}(b)-f(b)\bigg)<0
\end{eqnarray}
Then the Newton-Raphson step is accepted else we take instead a bisection step.

\section{Lagrange Interpolation}
Let us first recall that taylor expansion allows us to approximate a function at a point $x$ if the function and its derivatives are known in some neighbouring point $x_0$. The lagrange interpolation tries to approximate a function at a point $x$ if only the values of the function in several other points are known. Thus this method does not require the knowledge of the derivatives of the function.  We start from taylor expansion 
\begin{eqnarray}
f(y)=f(x)+(y-x)f^{'}(x)+\frac{1}{2!}(y-x)^2f^{''}(x)+..
\end{eqnarray}
Let us assume that the function is known at three points $x_1$, $x_2$ and $x_3$. In this case we can approximate the function $f(x)$ by some function $p(x)$ and write
\begin{eqnarray}
f(y)=p(x)+(y-x)p^{'}(x)+\frac{1}{2!}(y-x)^2p^{''}(x).
\end{eqnarray}
We have
\begin{eqnarray}
&&f(x_1)=p(x)+(x_1-x)p^{'}(x)+\frac{1}{2!}(x_1-x)^2p^{''}(x)\nonumber\\
&&f(x_2)=p(x)+(x_2-x)p^{'}(x)+\frac{1}{2!}(x_2-x)^2p^{''}(x)\nonumber\\
&&f(x_3)=p(x)+(x_3-x)p^{'}(x)+\frac{1}{2!}(x_3-x)^2p^{''}(x).
\end{eqnarray}
We can immediately find
\begin{eqnarray}
p(x)=\frac{1}{1+a_2+a_3}f(x_1)+\frac{a_2}{1+a_2+a_3}f(x_2)+\frac{a_3}{1+a_2+a_3}f(x_3).
\end{eqnarray}
The coefficients $a_2$ and $a_3$ solve the equations
\begin{eqnarray}
&&a_2(x_2-x)^2+a_3(x_3-x)^2=-(x_1-x)^2\nonumber\\
&&a_2(x_2-x)+a_3(x_3-x)=-(x_1-x).
\end{eqnarray}
We find
\begin{eqnarray}
a_2=\frac{(x_1-x)(x_3-x_1)}{(x_2-x)(x_2-x_3)}~,~a_3=-\frac{(x_1-x)(x_2-x_1)}{(x_3-x)(x_2-x_3)}.
\end{eqnarray}
Thus
\begin{eqnarray}
1+a_2+a_3=\frac{(x_3-x_1)(x_2-x_1)}{(x_2-x)(x_3-x)}.
\end{eqnarray}
Therefore we get
\begin{eqnarray}
p(x)=\frac{(x-x_2)(x-x_3)}{(x_1-x_2)(x_1-x_3)}f(x_1)+\frac{(x-x_1)(x-x_3)}{(x_2-x_1)(x_2-x_3)}f(x_2)+\frac{(x-x_1)(x-x_2)}{(x_3-x_1)(x_3-x_2)}f(x_3).\nonumber\\
\end{eqnarray}
This is a quadratic polynomial.

Let $x$ be some independent variable with tabulated values $x_i$, $i=1,2,...,n.$. The dependent variable is a function $f(x)$ with tabulated values $f_i=f(x_i)$. Let us then assume that we can approximate $f(x)$ by a polynomial of degree $n-1$ , viz
\begin{eqnarray}
p(x)= a_0+a_1x+a_2x^2+...+a_{n-1}x^{n-1}.
\end{eqnarray} 
A polynomial which goes through the $n$ points $(x_i,f_i=f(x_i))$ was given by Lagrange. This is given by
\begin{eqnarray}
p(x)= f_1{\lambda}_1(x)+f_2{\lambda}_2(x)+...+f_n{\lambda}_n(x).
\end{eqnarray} 
\begin{eqnarray}
{\lambda}_i(x)={\prod}_{j({\neq}i)=1}^{n}\frac{x-x_j}{x_i-x_j}.
\end{eqnarray} 
We remark
\begin{eqnarray}
{\lambda}_i(x_j)={\delta}_{ij}.
\end{eqnarray} 
\begin{eqnarray}
\sum_{i=1}^{n}{\lambda}_i(x)=1.
\end{eqnarray} 
The Lagrange polynomial can be used to fit the entire table with $n$ equal the number of points in the table. But it is preferable  to use   the Lagrange polynomial to to fit only a small region of the table with a small value of $n$. In other words use several polynomials to cover the whole table and the fit considered here is local and not global.

\section{Cubic Spline Interpolation}
We consider $n$ points $(x_1,f(x_1))$,$(x_2,f(x_2))$,...,$(x_n,f(x_n))$ in the plane. In every interval $x_j{\leq}x{\leq}x_{j+1}$ we approximate the function $f(x)$ with a cubic polynomial of the form
\begin{eqnarray}
p(x)=a_j(x-x_j)^3+b_j(x-x_j)^2+c_j(x-x_j)+d_j.
\end{eqnarray} 
We assume that 
\begin{eqnarray}
p_j=p(x_j)=f(x_j).
\end{eqnarray} 
In other words the $p_j$ for all  $j=1,2,...,n-1$ are known. From the above equation we conclude that
\begin{eqnarray}
d_j=p_j.
\end{eqnarray}
We compute
\begin{eqnarray}
p^{'}(x)=3a_j(x-x_j)^2+2b_j(x-x_j)+c_j.
\end{eqnarray} 
\begin{eqnarray}
p^{''}(x)=6a_j(x-x_j)+2b_j.
\end{eqnarray}
Thus we get by substituting $x=x_{j}$ into $p^{''}(x)$ the result
\begin{eqnarray}
b_j=\frac{p^{''}_j}{2}.
\end{eqnarray}
By substituting $x=x_{j+1}$ into $p^{''}(x)$ we get the result
\begin{eqnarray}
a_j=\frac{p^{''}_{j+1}-p^{''}_j}{6h_j}.
\end{eqnarray}
By substituting $x=x_{j+1}$ into $p(x)$ we get
\begin{eqnarray}
p_{j+1}=a_jh_j^3+b_jh_j^2+c_jh_j+p_j.
\end{eqnarray} 
By using the values of $a_j$ and $b_j$ we obtain
\begin{eqnarray}
c_{j}=\frac{p_{j+1}-p_j}{h_j}-\frac{h_j}{6}(p^{''}_{j+1}+2p^{''}_j).
\end{eqnarray} 
Hence
\begin{eqnarray}
p(x)=\frac{p^{''}_{j+1}-p^{''}_j}{6h_j}(x-x_j)^3+\frac{p^{''}_j}{2}(x-x_j)^2+\bigg(\frac{p_{j+1}-p_j}{h_j}-\frac{h_j}{6}(p^{''}_{j+1}+2p^{''}_j)\bigg)(x-x_j)+p_j.\nonumber\\
\end{eqnarray} 
In other words the polynomials are determined from $p_j$ and $p^{''}_j$. The $p_j$ are known given by $p_j=f(x_j)$. It remains to determine $p^{''}_j$. We take the derivative of the above equation
\begin{eqnarray}
p^{'}(x)=\frac{p^{''}_{j+1}-p^{''}_j}{2h_j}(x-x_j)^2+p^{''}_j(x-x_j)+\bigg(\frac{p_{j+1}-p_j}{h_j}-\frac{h_j}{6}(p^{''}_{j+1}+2p^{''}_j)\bigg).
\end{eqnarray} 
This is the derivative in the interval $[x_j,x_{j+1}]$. We compute
\begin{eqnarray}
p^{'}(x_j)=\bigg(\frac{p_{j+1}-p_j}{h_j}-\frac{h_j}{6}(p^{''}_{j+1}+2p^{''}_j)\bigg).
\end{eqnarray}
The derivative in the interval $[x_{j-1},x_j]$ is
\begin{eqnarray}
p^{'}(x)=\frac{p^{''}_{j}-p^{''}_{j-1}}{2h_{j-1}}(x-x_{j-1})^2+p^{''}_{j-1}(x-x_{j-1})+\bigg(\frac{p_{j}-p_{j-1}}{h_{j-1}}-\frac{h_{j-1}}{6}(p^{''}_{j}+2p^{''}_{j-1})\bigg).
\end{eqnarray} 
 We compute
\begin{eqnarray}
p^{'}(x_j)=\frac{p^{''}_{j}-p^{''}_{j-1}}{2}h_{j-1}+p^{''}_{j-1}h_{j-1}+\bigg(\frac{p_{j}-p_{j-1}}{h_{j-1}}-\frac{h_{j-1}}{6}(p^{''}_{j}+2p^{''}_{j-1})\bigg).
\end{eqnarray} 
By matching the two expressions for $p^{'}(x_j)$ we get
\begin{eqnarray}
h_{j-1}p^{''}_{j-1}+2(h_j+h_{j-1})p^{''}_j+h_jp^{''}_{j+1}=6\bigg(\frac{p_{j+1}-p_j}{h_j}-\frac{p_j-p_{j-1}}{h_{j-1}}\bigg).\label{tri1}
\end{eqnarray} 
These are $n-2$ equations since $j=2,...,n-1$ for $n$ unknown $p^{''}_j$. We need two more equations. These are obtained by computing the first derivative $p^{'}(x)$ at $x=x_1$ and $x=x_n$. We obtain the two equations
\begin{eqnarray}
h_1(p^{''}_2+2p^{''}_1)=\frac{6(p_2-p_1)}{h_1}-6p^{'}_1.\label{tri2}
\end{eqnarray} 
\begin{eqnarray}
h_{n-1}(p^{''}_{n-1}+2p^{''}_n)=-\frac{6(p_n-p_{n-1})}{h_{n-1}}+6p^{'}_n.\label{tri3}
\end{eqnarray} 
The $n$ equations (\ref{tri1}), (\ref{tri2}) and (\ref{tri3}) correspond to a tridiagonal linear system. In general $p^{'}_1$ and $p^{'}_n$ are not known. In this case we may use natural spline in which the second derivative vanishes at the end points and hence
 \begin{eqnarray}
\frac{p_2-p_1}{h_1}-p^{'}_1=\frac{p_n-p_{n-1}}{h_{n-1}}-p^{'}_n=0.
\end{eqnarray} 
\section{The Method of Least Squares}
We assume that we have $N$ data points $(x(i),y(i))$. We want to fit this data to some curve say a straight line $y_{\rm fit}=mx+b$. To this end we define the function
\begin{equation}
{\Delta}=\sum_{i=1}^{N}(y(i)-y_{\rm fit}(i))^2=\sum_{i=1}^N(y(i)-mx(i)-b)^2.
\end{equation}
The goal is to minimize this function with respect to $b$ and $m$. We have
\begin{equation}
\frac{\partial{\Delta}}{\partial m}=0~,\frac{\partial{\Delta}}{\partial b}=0.
\end{equation}
We get the solution
\begin{equation}
b=\frac{\sum_ix(i)\sum_jx(j)y(j)-\sum_ix(i)^2\sum_jy(j)}{(\sum_ix(i))^2-N\sum_ix_i^2}.
\end{equation}
\begin{equation}
m=\frac{\sum_ix(i)\sum_jy(j)-N\sum_ix(i)y(i)}{(\sum_ix(i))^2-N\sum_ix_i^2}.
\end{equation}

\section{Simulation $5$: Newton-Raphson Algorithm}
A particle of mass $m$ moves inside a potential well of height $V$ and length $2a$ centered around $0$.
We are interested in the states of the system which have energies less than $V$, i.e. bound states. The states of the system can be even or odd. The energies associated with the even wave functions are solutions of the transcendental equation 

\[\alpha\tan\alpha a=\beta.\]
\[\alpha=\sqrt{\frac{2mE}{\hbar^2}}~,~\beta=\sqrt{\frac{2m(V-E)}{\hbar^2}}.\]
In the case of the infinite potential well we find the solutions
\[E_n=\frac{(n+\frac{1}{2})^2\pi^2\hbar^2}{2ma^2}~,~n=0,1....\]
We choose (dropping units)
\[\hbar=1~,~a=1~,~2m=1.\]
In order to find numerically the energies $E_n$  we will use the Newton-Raphson algorithm which allows us to find the roots of the equation $f(x)=0$ as follows. From an initial guess $x_0$, the first approximation $x_1$ to the solution is determined from the intersection of the tangent to the function $f(x)$ at $x_0$ with the $x-$axis. This is given by

\[x_1=x_0-\frac{f(x_0)}{f^{'}(x_0)}.\]
Next by using $x_1$ we repeat the same step in order to find the second approximation $x_2$ to the solution. In general the approximation $x_{i+1}$ to the desired solution in terms of the approximation $x_i$ is given by the equation
\[x_{i+1}=x_i-\frac{f(x_i)}{f^{'}(x_i)}.\]

\begin{itemize}
 \item[$(1)$] For $V=10$, determine the solutions using the graphical method. Consider the two functions
\[f(\alpha)=\tan\alpha a~,~g(\alpha)=\frac{\beta}{\alpha}=\sqrt{\frac{V}{\alpha^2}-1}.\]
\item[$(2)$]Find using the method of Newton-Raphson the two solutions with a tolerance equal  $10^{-8}$.
For the first solution we take the initial guess $\alpha=\pi/a$ and for the second solution we take the initial guess $\alpha=2\pi/a$.

\item[$(3)$]Repeat for $V=20$.
\item[$(4)$]Find the $4$ solutions for $V=100$. Use the graphical method to determine the initial step each time.
\item[$(5)$]Repeat the above questions using the bisection method.
\end{itemize}



\chapter{The Solar System-The Runge-Kutta Methods}

\section{The Solar System}
\subsection{Newton's Second Law}We consider the motion of the Earth around the Sun. Let $r$ be the distance and $M_s$ and $M_e$ be the masses of the Sun and the Earth respectively. We neglect the effect of the other planets and the motion of the Sun (i.e. we assume that $M_s>>M_e$). The goal is to calculate the position of the Earth as a function of time. We start from Newton's second law of motion

\begin{eqnarray}
M_e\frac{d^{2}\vec{r}}{dt^{2}}&=&-\frac{GM_eM_s}{r^{3}}\vec{r}\nonumber\\
&=&-\frac{GM_eM_s}{r^{3}}(x\vec{i}+y\vec{j}).
\end{eqnarray}
We get the two equations
\begin{eqnarray}
\frac{d^{2}x}{dt^{2}}=-\frac{GM_s}{r^{3}}x.
\end{eqnarray}
\begin{eqnarray}
\frac{d^{2}y}{dt^{2}}=-\frac{GM_s}{r^{3}}y.
\end{eqnarray}
We replace these two second-order differential equations by the four first-order differential equations
\begin{eqnarray}
\frac{dx}{dt}=v_x.\label{e1}
\end{eqnarray}
\begin{eqnarray}
\frac{dv_x}{dt}=-\frac{GM_s}{r^{3}}x.\label{e2}
\end{eqnarray}
\begin{eqnarray}
\frac{dy}{dt}=v_y.\label{e3}
\end{eqnarray}
\begin{eqnarray}
\frac{dv_y}{dt}=-\frac{GM_s}{r^{3}}y.\label{e4}
\end{eqnarray}
We recall
\begin{eqnarray}
r=\sqrt{x^2+y^2}.\label{e5}
\end{eqnarray}
\subsection{Astronomical Units and Initial Conditions} The distance will be measured in astronomical units (AU) whereas time will be measured in years. One astronomical unit of lenght ($1$ {\rm AU}) is equal to the average distance between the earth and the sun, viz $1 {\rm AU}=1.5\times 10^{11}m$. The astronomical unit of mass can be found as follows. Assuming a circular orbit we have
\begin{eqnarray}
\frac{M_ev^{2}}{r}=\frac{GM_sM_e}{r^{2}}.
\end{eqnarray}
Equivalently
\begin{eqnarray}
GM_s=v^{2}{r}.
\end{eqnarray}
The radius is $r=1{\rm AU}$. The velocity of the earth is $v=2\pi r/{\rm yr}=2\pi {\rm AU}/{\rm yr}$. Hence
\begin{eqnarray}
GM_s=4{\pi}^{2}{\rm AU}^{3}/{\rm yr}^{2}.
\end{eqnarray}
For the numerical simulations it is important to determine the correct initial conditions. The orbit of Mercury is known to be an ellipse with eccentricity $e=0.206$ and radius (semimajor axis) $a=0.39~{\rm AU}$ with the Sun at one of the foci. The distance between the Sun and the center is $ea$. The first initial condition is $x_0=r_1$, $y_0=0$ where $r_1$ is the maximum distance from Mercury to the Sun,i.e. $r_1=(1+e)a=0.47~{\rm AU}$. The second initial condition is the velocity $(0,v_1)$ which can be computed using conservation of energy and angular momentum. For example by comparing with the point $(0,b)$  on the orbit where $b$ is the semiminor axis, i.e $b=a\sqrt{1-e^{2}}$ the velocity  $(v_2,0)$ there can be obtained in terms of $(0,v_1)$ from conservation of angular momentum as follows
\begin{eqnarray}
r_1v_1=bv_2\Leftrightarrow v_2=\frac{r_1v_1}{b}.
\end{eqnarray}
Next conservation of energy yields
\begin{eqnarray}
-\frac{GM_sM_m}{r_1}+\frac{1}{2}M_mv_1^{2}=-\frac{GM_sM_m}{r_2}+\frac{1}{2}M_mv_2^{2}.
\end{eqnarray}
In above $r_2=\sqrt{e^{2}a^{2}+b^{2}}$ is the distance between the Sun and Mercury when at the point $(0,b)$. By substituting the value of $v_2$ we get an equation for $v_1$. This is given by
\begin{eqnarray}
v_1=\sqrt{\frac{GM_s}{a}\frac{1-e}{1+e}}=8.2~{\rm AU}/{\rm yr}.
\end{eqnarray}
\subsection{Kepler's Laws} Kepler's laws are given by the following three statements:
\begin{itemize}
\item{}The planets move in elliptical orbits around the sun. The sun resides at one focus.
\item{}The line joining the sun with any planet sweeps out equal areas in equal times.
\item{}Given an orbit with a period $T$ and a semimajor axis $a$ the ratio ${T}^{2}/a^{3}$ is a constant.
\end{itemize}
The derivation of these three laws proceeds as follows. We work in polar coordinates. Newton's second law reads

\begin{eqnarray}
M_e\ddot{\vec{r}}=-\frac{GM_sM_e}{r^2}\hat{r}.
\end{eqnarray}
We use $\dot{\hat{r}}=\dot{\theta}\hat{\theta}$ and $\dot{\hat{\theta}}=-\dot{\theta}\hat{r}$ to derive $\dot{\vec{r}}=\dot{r}\hat{r}+r\dot{\theta}\hat{\theta}$ and $\ddot{\vec{r}}=(\ddot{r}-r\dot{\theta}^2)\hat{r}+(r\ddot{\theta}+2\dot{r}\dot{\theta})\hat{\theta}$. Newton's second law decomposes into the two equations
\begin{eqnarray}
r\ddot{\theta}+2\dot{r}\dot{\theta}=0.\label{eqtke1}
\end{eqnarray}
\begin{eqnarray}
\ddot{r}-r\dot{\theta}^2=-\frac{GM_s}{r^2}.\label{eqtke2}
\end{eqnarray}
Let us recall that the angular momentum by unit mass is defined by $\vec{l}=\vec{r}\times\dot{\vec{r}}=r^2\dot{\theta}\hat{r}\times \hat{\theta}$. Thus $l=r^2\dot{\theta}$. Equation (\ref{eqtke1}) is precisely the requirement that angular momentum is conserved. Indeed we compute
 \begin{eqnarray}
\frac{dl}{dt}=r(r\ddot{\theta}+2\dot{r}\dot{\theta})=0.
\end{eqnarray}
Now we remark that the area swept by the vector $\vec{r}$ in a time interval $dt$ is $dA=(r\times rd\theta)/2$ where $d\theta$ is the angle traveled by $\vec{r}$ during $dt$. Clearly
 \begin{eqnarray}
\frac{dA}{dt}=\frac{1}{2}l .\label{eqtke3}
\end{eqnarray}
In other words the planet sweeps equal areas in equal times since $l$ is conserved. This is Kepler's second law.

The second equation (\ref{eqtke2}) becomes now
\begin{eqnarray}
\ddot{r}=\frac{l^2}{r^3}-\frac{GM_s}{r^2}
\end{eqnarray}
By multiplying this equation with $\dot{r}$ we obtain
\begin{eqnarray}
\frac{d}{dt}E=0~,~E=\frac{1}{2}\dot{r}^2+\frac{l^2}{2r^2}-\frac{GM_s}{r}.
\end{eqnarray}
This is precisely the statement of conservation of energy. $E$ is the energy per unit mass. Solving for $dt$ in terms of $dr$ we obtain
 \begin{eqnarray}
dt=\frac{dr}{\sqrt{2\bigg(E-\frac{l^2}{2r^2}+\frac{GM_s}{r}\bigg)}}
\end{eqnarray}
However $dt=(r^2d\theta)/l$. Thus
 \begin{eqnarray}
d\theta=\frac{ldr}{r^2\sqrt{2\bigg(E-\frac{l^2}{2r^2}+\frac{GM_s}{r}\bigg)}}
\end{eqnarray}
By integrating this equation we obtain (with $u=1/r$)
 \begin{eqnarray}
\theta&=&\int \frac{ldr}{r^2\sqrt{2\bigg(E-\frac{l^2}{2r^2}+\frac{GM_s}{r}\bigg)}}\nonumber\\
&=&-\int \frac{du}{\sqrt{\frac{2E}{l^2}+\frac{2GM_s}{l^2}u-u^2}}.
\end{eqnarray}
This integral can be done explicitly. We get
 \begin{eqnarray}
\theta&=&-\arccos\bigg(\frac{u-C}{eC}\bigg)+\theta^{'}~,~e=\sqrt{1+\frac{2l^2E}{G^2M_s^2}}~,~C=\frac{GM_s}{l^2}.
\end{eqnarray}
By inverting this equation we get an equation of ellipse with eccentricity $e$ since $E<0$, viz
 \begin{eqnarray}
\frac{1}{r}=C(1+e\cos(\theta-\theta^{'})).
\end{eqnarray}
This is Kepler's first law. The angle at which $r$ is maximum is $\theta-\theta^{'}=\pi$. This distance is precisely $(1+e)a$ where $a$ is the semi-major axis of the ellipse since $ea$ is the distance between the Sun which is at one of the two foci and the center of the ellipse.  Hence we obtain the relation
\begin{eqnarray}
(1-e^2)a=\frac{1}{C}=\frac{l^2}{GM_s}.\label{eqtke4}
\end{eqnarray}
From equation (\ref{eqtke3}) we can derive Kepler's third law. By integrating both sides of the equation over a single period $T$ and then taking the square we get
 \begin{eqnarray}
A^2=\frac{1}{4}l^2 T^2.
\end{eqnarray}
$A$ is the area of the ellipse, i.e. $A=\pi a b$ where the semi-minor axis $b$ is related the semi-major axis $a$ by $b=a\sqrt{1-e^2}$. Hence
 \begin{eqnarray}
\pi^2 a^4(1-e^2)=\frac{1}{4}l^2 T^2.
\end{eqnarray}
By using equation  (\ref{eqtke4}) we get the desired formula
\begin{eqnarray}
\frac{T^2}{a^3}=\frac{4\pi^2}{GM_s}.
\end{eqnarray}

\subsection{The inverse-Square Law and Stability of Orbits}
Any object with mass generates a gravitational field and thus gravitational field lines will emanate from the object and radiate outward to infinity. The number of field lines $N$ is proportional to the mass. The density of field lines crossing a sphere of radius $r$ surrounding this object is given by $N/4\pi r^{2}$. This is the origin of the inverse-square law. Therefore  any other object placed in this gravitational field will experience a gravitational force proportional to the number of field lines which intersect  it. If the distance between this second object and the  source is increased the force on it will become weaker because the number of field lines which intersect it will decrease as we are further away from the source.

 \section{Euler-Cromer Algorithm}
 The time discretization is
\begin{eqnarray}
t\equiv t(i)=i\Delta t~,~i=0,...,N.
\end{eqnarray}
The total time interval is $T=N\Delta t$. We define $x(t)=x(i)$, $v_x(t)=v_x(i)$, $y(t)=y(i)$, $v_y(t)=v_y(i)$. Equations (\ref{e1}), (\ref{e2}), (\ref{e3}),(\ref{e4}) and (\ref{e5}) become (with $i=0,...,N$)
\begin{eqnarray}
v_x(i+1)=v_x(i)-\frac{GM_s}{(r(i))^{3}}x(i)\Delta t.
\end{eqnarray}
\begin{eqnarray}
x(i+1)=x(i)+v_x(i)\Delta t.
\end{eqnarray}
\begin{eqnarray}
v_y(i+1)=v_y(i)-\frac{GM_s}{(r(i))^{3}}y(i)\Delta t.
\end{eqnarray}
\begin{eqnarray}
y(i+1)=y(i)+v_y(i)\Delta t.
\end{eqnarray}
\begin{eqnarray}
{r}(i)=\sqrt{x(i)^2+y(i)^2}.
\end{eqnarray}
This is Euler algorithm. It can also be rewritten with $\hat{x}(i)=x(i-1)$, $\hat{y}(i)=y(i-1)$, $\hat{v}_x(i)=v_x(i-1)$, $\hat{v}_y(i)=v_y(i-1)$, $\hat{r}(i)=r(i-1)$ and $i=1,...,N+1$ as
\begin{eqnarray}
\hat{v}_x(i+1)=\hat{v}_x(i)-\frac{GM_s}{(\hat{r}(i))^{3}}\hat{x}(i)\Delta t.
\end{eqnarray}
\begin{eqnarray}
\hat{x}(i+1)=\hat{x}(i)+\hat{v}_x(i)\Delta t.
\end{eqnarray}
\begin{eqnarray}
\hat{v}_y(i+1)=\hat{v}_y(i)-\frac{GM_s}{(\hat{r}(i))^{3}}\hat{y}(i)\Delta t.
\end{eqnarray}
\begin{eqnarray}
\hat{y}(i+1)=\hat{y}(i)+\hat{v}_y(i)\Delta t.
\end{eqnarray}
\begin{eqnarray}
\hat{r}(i)=\sqrt{\hat{x}(i)^2+\hat{y}(i)^2}.
\end{eqnarray}
In order to maintain energy conservation we employ Euler-Cromer algorithm. We calculate as in the Euler's algorithm the velocity at time step $i+1$ by using the position and velocity at time step $i$. However we compute the position at time step $i+1$ by using the position at time step $i$ and the velocity at time step $i+1$, viz
\begin{eqnarray}
\hat{v}_x(i+1)=\hat{v}_x(i)-\frac{GM_s}{(\hat{r}(i))^{3}}\hat{x}(i)\Delta t.
\end{eqnarray}
\begin{eqnarray}
\hat{x}(i+1)=\hat{x}(i)+\hat{v}_x(i+1)\Delta t.
\end{eqnarray}
\begin{eqnarray}
\hat{v}_y(i+1)=\hat{v}_y(i)-\frac{GM_s}{(\hat{r}(i))^{3}}\hat{y}(i)\Delta t.
\end{eqnarray}
\begin{eqnarray}
\hat{y}(i+1)=\hat{y}(i)+\hat{v}_y(i+1)\Delta t.
\end{eqnarray}

\section{The Runge-Kutta Algorithm}
\subsection{The Method}
The problem is still trying to solve the first order differential equation
\begin{eqnarray}
\frac{dy}{dx}=f(x,y).
\end{eqnarray}
In the Euler's method we approximate the function $y=y(x)$ in each interval $[x_n,x_{n+1}]$ by the straight line
\begin{eqnarray}
y_{n+1}=y_n+\Delta x f(x_n,y_n).
\end{eqnarray}
The slope $f(x_n,y_n)$ of this line is exactly given by the slope of the function $y=y(x)$ at the begining of the inetrval $[x_n,x_{n+1}]$. 

Given the value $y_n$ at $x_n$ we evaluate the value $y_{n+1}$ at  $x_{n+1}$ using the method of Runge-Kutta  as follows. First the middle of the interval $[x_n,x_{n+1}]$ which is at the value $x_n+\frac{1}{2}\Delta x$ corresponds to the $y$-value $y_{n+1}$ calculated using the Euler's method, viz $y_{n+1}=y_n+\frac{1}{2}k_1$ where
\begin{eqnarray}
k_1=\Delta xf(x_n,y_n).
\end{eqnarray} 
Second the slope at this middle point $(x_n+\frac{1}{2}\Delta x,y_n+\frac{1}{2}k_1)$ which is given by 
\begin{eqnarray}
\frac{k_2}{{\Delta}x}=f(x_n+\frac{1}{2}\Delta x,y_n+\frac{1}{2}k_1) 
\end{eqnarray}
is the value of the slope which will be used to estimate the correct value of $y_{n+1}$ at $x_{n+1}$ using again Euler's method, namely
\begin{eqnarray}
y_{n+1}=y_n+k_2.
\end{eqnarray}
In summary the Runge-Kutta algorithm is given by
\begin{eqnarray}
&&k_1=\Delta xf(x_n,y_n)\nonumber\\
&&k_2=\Delta x f(x_n+\frac{1}{2}\Delta x,y_n+\frac{1}{2}k_1)\nonumber\\
&&y_{n+1}=y_n+k_2.
\end{eqnarray} 
The error in this method is proportional to ${\Delta}x^{3}$. This can be shown as follows. We have
\begin{eqnarray}
y(x+\Delta x)&=&y(x)+\Delta x \frac{dy}{dx}+\frac{1}{2}(\Delta x)^2\frac{d^2y}{dx^2}+...\nonumber\\
&=&y(x)+\Delta x f(x,y)+\frac{1}{2}(\Delta x)^2\frac{d}{dx}f(x,y)+...\nonumber\\
&=&y(x)+\Delta x \bigg(f(x,y)+\frac{1}{2}\Delta x \frac{\partial f}{\partial x}+\frac{1}{2}\Delta x f(x,y)\frac{\partial f}{\partial y}\bigg)+...\nonumber\\
&=&y(x)+\Delta x f(x+\frac{1}{2}\Delta x,y+\frac{1}{2}\Delta x f(x,y))+O(\Delta x^3)\nonumber\\
&=&y(x)+\Delta x f(x+\frac{1}{2}\Delta x,y+\frac{1}{2}k_1)+O(\Delta x^3)\nonumber\\
&=&y(x)+k_2+O(\Delta x^3).
\end{eqnarray}
Let us finally note that the above Runge-Kutta method is strictly speaking the second-order Runge-Kutta method. The first-order Runge-Kutta method is the Euler algorithm. The higher-order Runge-Kutta methods will not be discussed here.
\subsection{Example $1$: The Harmonic Oscillator}
Let us apply this method to the problem of the harmonic oscillator. We have the differential equations

\begin{eqnarray}
&&\frac{d\theta}{dt}=\omega\nonumber\\
&&\frac{d\omega}{dt}=-\frac{g}{l}\theta.
\end{eqnarray}
Euler's equations read
\begin{eqnarray}
&&{\theta}_{n+1}={\theta}_n+\Delta t{\omega}_n\nonumber\\
&&{\omega}_{n+1}={\omega}_n-\frac{g}{l}{\theta}_n\Delta t.
\end{eqnarray}
First we consider the function $\theta=\theta(t)$. The middle point is $(t_n+\frac{1}{2}\Delta t,{\theta}_n+\frac{1}{2}k_1)$ where $k_1=\Delta t {\omega}_n$. For the function $\omega=\omega(t)$ the middle point is $(t_n+\frac{1}{2}\Delta t,{\omega}_n+\frac{1}{2}k_3)$ where $k_3=-\frac{g}{l}\Delta t {\theta}_n$. Therefore we have
\begin{eqnarray}
&&k_1=\Delta t {\omega}_n\nonumber\\
&&k_3=-\frac{g}{l}\Delta t {\theta}_n.
\end{eqnarray}
The slope of the function $\theta(t)$ at its middle point is
\begin{eqnarray}
&&\frac{k_2}{\Delta t}={\omega}_n+\frac{1}{2}k_3.
\end{eqnarray}
The slope of the function $\omega(t)$ at its middle point is
\begin{eqnarray}
&&\frac{k_4}{\Delta t}=-\frac{g}{l}({\theta}_n+\frac{1}{2}k_1).
\end{eqnarray}
The Runge-Kutta solution is then given by

\begin{eqnarray}
&&{\theta}_{n+1}={\theta}_n+k_2\nonumber\\
&&{\omega}_{n+1}={\omega}_n+k_4.
\end{eqnarray}
 \subsection{Example $2$: The Solar System}
Let us consider the  equations
\begin{eqnarray}
\frac{dx}{dt}=v_x.
\end{eqnarray}
\begin{eqnarray}
\frac{dv_x}{dt}=-\frac{GM_s}{r^{3}}x.
\end{eqnarray}

\begin{eqnarray}
\frac{dy}{dt}=v_y.
\end{eqnarray}
\begin{eqnarray}
\frac{dv_y}{dt}=-\frac{GM_s}{r^{3}}y.
\end{eqnarray}
First we consider the function $x=x(t)$. The middle point is $(t_n+\frac{1}{2}\Delta t,{x}_n+\frac{1}{2}k_1)$ where $k_1=\Delta t~ {v}_{xn}$. For the function $v_x=v_x(t)$ the middle point is $(t_n+\frac{1}{2}\Delta t,{v}_{xn}+\frac{1}{2}k_3)$ where $k_3=-\frac{GM_s}{r_n}\Delta t~ {x}_n$. Therefore we have
\begin{eqnarray}
&&k_1=\Delta t~ {v}_{xn}\nonumber\\
&&k_3=-\frac{GM_s}{r_n^3}\Delta t~ {x}_n.
\end{eqnarray}
The slope of the function $x(t)$ at the middle point is
\begin{eqnarray}
&&\frac{k_2}{\Delta t}={v}_{xn}+\frac{1}{2}k_3.
\end{eqnarray}
The slope of the function $v_x(t)$ at the middle point is
\begin{eqnarray}
&&\frac{k_4}{\Delta t}=-\frac{GM_s}{R_n^{3}}({x}_n+\frac{1}{2}k_1).
\end{eqnarray}
Next we consider the function $y=y(t)$. The middle point is $(t_n+\frac{1}{2}\Delta t,{y}_n+\frac{1}{2}k_1^{'})$ where $k_1^{'}=\Delta t~ {v}_{yn}$. For the function $v_y=v_y(t)$ the middle point is $(t_n+\frac{1}{2}\Delta t,{v}_{yn}+\frac{1}{2}k_3^{'})$ where $k_3^{'}=-\frac{GM_s}{r_n}\Delta t~ {y}_n$. Therefore we have
\begin{eqnarray}
&&k_1^{'}=\Delta t~ {v}_{yn}\nonumber\\
&&k_3^{'}=-\frac{GM_s}{r_n^3}\Delta t~ {y}_n.
\end{eqnarray}
The slope of the function $y(t)$ at the middle point is
\begin{eqnarray}
&&\frac{k_2^{'}}{\Delta t}={v}_{yn}+\frac{1}{2}k_3^{'}.
\end{eqnarray}
The slope of the function $v_y(t)$ at the middle point is
\begin{eqnarray}
&&\frac{k_4^{'}}{\Delta t}=-\frac{GM_s}{R_n^{3}}({y}_n+\frac{1}{2}k_1^{'}).
\end{eqnarray}
In the above equations
\begin{eqnarray}
R_n=\sqrt{({x}_n+\frac{1}{2}k_1)^{2}+({y}_n+\frac{1}{2}k_1^{'})^{2}}.
\end{eqnarray}
The Runge-Kutta solutions are then given by

\begin{eqnarray}
&&{x}_{n+1}={x}_n+k_2\nonumber\\
&&{v}_{x(n+1)}={v}_{xn}+k_4\nonumber\\
&&{y}_{n+1}={y}_n+k_2^{'}\nonumber\\
&&{v}_{y(n+1)}={v}_{yn}+k_4^{'}.
\end{eqnarray}

\section{Precession of the Perihelion of Mercury} 

The orbit of Mercury is elliptic. The orientation of the axes of the ellipse rotate with time. This is the precession of the perihelion (the point of the orbit nearest to the Sun) of Mercury. Mercury's perihelion makes one revolution every  $23000$ years. This is approximately $566$ arcseconds per century. The gravitational forces of the other planets (in particular Jupiter) lead to a precession of $523$ arcseconds per century. The remaining $43$ arcseconds per century are accounted for by general relativity.

For objects too close together (like the Sun and Mercury) the force of gravity predicted by general relativity deviates from the inverse-square law. This force is given by
\begin{eqnarray}
F=\frac{GM_sM_m}{r^{2}}(1+\frac{\alpha}{r^{2}})~,~\alpha=1.1\times 10^{-8} {\rm AU}^{2}.
\end{eqnarray} 
We discuss here some of the numerical results obtained with the Runge-Kutta method for different values of $\alpha$. We take the time step and the number of iterations to be $N=20000$ and $dt=0.0001$. The angle of the line joining the Sun and  Mercury with the horizontal axis when  mercury is at the perihelion is found to change linearly with time.  We get the following rates of precession 
\begin{eqnarray}
&&\alpha=0.0008~,~\frac{d\theta}{dt}=8.414\pm 0.019\nonumber\\
&&\alpha=0.001~,~\frac{d\theta}{dt}=10.585\pm 0.018\nonumber\\
&&\alpha=0.002~,~\frac{d\theta}{dt}=21.658\pm 0.019\nonumber\\
&&\alpha=0.004~,~\frac{d\theta}{dt}=45.369\pm 0.017.
\end{eqnarray}
Thus
\begin{eqnarray}
\frac{d\theta}{dt}=a \alpha~,~\alpha=11209.2\pm 147.2 ~{\rm degrees}/({\rm yr}.\alpha).
\end{eqnarray}
By extrapolating to the value provided by general relativity, viz $\alpha=1.1\times 10^{-8}$ we get
\begin{eqnarray}
\frac{d\theta}{dt}=44.4\pm 0.6 ~{\rm arcsec}/{\rm century}.
\end{eqnarray}
\section{Exercises}
\paragraph{Exercise $1$:}
Using the Runge-Kutta method solve the following differential equations 
\begin{eqnarray}
\frac{d^2r}{dt^2}=\frac{l^2}{r^3}-\frac{GM}{r^2}.
\end{eqnarray}
\begin{eqnarray}
\frac{d^2z}{dt^2}=-g.
\end{eqnarray}
\begin{eqnarray}
\frac{dN}{dt}=aN-bN^2.
\end{eqnarray}

\paragraph{Exercise $2$:} 
The Lorenz model is a chaotic system given by three coupled first order differential equations
\begin{eqnarray}
&&\frac{dx}{dt}=\sigma(y-x)\nonumber\\
&&\frac{dy}{dt}=-xz+rx-y\nonumber\\
&&\frac{dz}{dt}=xy-bz.
\end{eqnarray}
This system is a simplified version of the system of Navier-Stokes equations of fluid mechanics which are relevant for the Rayleigh-B\'enard problem. Write down the numercial solution of these equations according to the Runge-Kutta method.

\section{Simulation $6$: Runge-Kutta Algorithm- The Solar System }
\paragraph{Part I}
We consider a solar system consisting of a single planet moving around the Sun. We suppose that the Sun is very heavy compared
to the planet that we can safely assume that it is not moving at the center of the system. Newton's second law gives the following
 equations of motion

\[v_x=\frac{dx}{dt}~,~\frac{dv_x}{dt}=-\frac{GM_s}{r^3}x~,~v_y=\frac{dy}{dt}~,~\frac{dv_y}{dt}=-\frac{GM_s}{r^3}y.\]
We will use here the astronomical units defined by $GM_s=4\pi^2 {\rm AU}^3/{\rm yr}^2$.

\begin{itemize}
 \item[$(1)$] 
Write a Fortran code in which we implement the Runge-Kutta algorithm for the problem of solving the equations of motion of the the solar system.

\item[$(2)$]
Compute the trajectory, the velocity and the energy as functions of time. What do you
observe for the energy.   

\item[$(3)$]
According to Kepler's first law the orbit of any planet is an ellipse with the Sun at one of the two foci.
In the following we will only consider planets which are known to have circular orbits to a great accuracy. 
These planets are Venus, Earth, Mars, Jupiter and Saturn. The radii in astronomical units are given by

\[a_{\rm venus}=0.72~,~a_{\rm earth}=1~,~a_{\rm mars}=1.52~,~a_{\rm jupiter}=5.2~,~a_{\rm saturn}=9.54.\]
Verify that Kepler's first law indeed holds for these planets.

In order to answer questions $2$ and $3$ above we take the initial conditions

\[x(1)=a~,~y(1)=0~,~v_x(1)=0~,~v_y(1)=v.\]
The value chosen for the initial velocity is very important to get a correct orbit and must be determined for example
by assuming that the orbit is indeed circular and as a consequence the centrifugal force is balanced by the force of gravitational
attraction. We get $v=\sqrt{{GM_s}/{a}}$. 

We take the step and the number of iterations $\Delta t=0.01~{\rm yr}~,~N=10^3-10^4$.
\end{itemize}
\paragraph{Part II}
\begin{itemize}
\item[$(1)$]
According to Kepler's third law the square of the period of a planet is directly proportional to the cube of the semi-major axis of its 
orbit. For circular orbits the proportionality factor is equal $1$ exactly. Verify this fact for the planets mentioned above.
We can measure the period of a planet by monitoring when the planet returns to its farthest point from the sun.

\item[$(2)$]
By changing the initial velocity appropriately we can obtain an elliptical orbit. Check this thing.

\item[$(3)$]
The fundamental laws governing the motion of the solar system are Newton's law of universal attraction and Newton's second 
law of motion. Newton's law of universal attraction states that the force between the Sun
and a planet is inversely proportioanl to the square of the distance between them and it is directed from the planet to the Sun.
We will assume in the following that this force is inversely proportional to a different power of the distance.
Modify the code accordingly and calculate the new orbits for powers between $1$ and $3$. What do you observe and
what do you conclude.
\end{itemize}

\section{Simulation $7$: Precession of the perihelion of Mercury}
According to Kepler's first law the orbits of all planets are ellipses with the Sun at one of the two foci. This law can be obtained from applying Newton's second law to the system consisting of the Sun and a single planet.
The effect of the other planets on the motion  will lead to a change of orientation of the orbital ellipse within the orbital plane of the planet. Thus the point of closest approach (the perihelion) will  precess, i.e. rotate around the sun. All planets suffer from this effect but because they are all farther from the sun  and all have longer periods than Mercury the amount of precession observed for them is smaller than that of Mercury.

However it was established earlier on that the precession of the perihelion of Mercury due to Newtonian effects deviates from the observed precession by the amount $43~{\rm arcsecond}/{\rm century}$. As it turns out this can only be explained within general relativity. The large mass of the Sun causes space and time around it to be curved which is felt the most by Mercury because of its proximity. This spacetime curvature can be approximated by the force law

\[F=\frac{GM_sM_m}{r^2}(1+\frac{\alpha}{r^2})~,~\alpha=1.1.10^{-8} AU^2.\]

\begin{itemize}
 \item[$(1)$] Include the above force in the code. The initial position and velocity of Mercury are 

\[x_0=(1+e)a~,~y_0=0.\] 
\[v_{x0}=0~,~v_{y0}=\sqrt{\frac{GM_s}{a}\frac{1-e}{1+e}}.\]
Thus initially Mercury is at its farthest point from the Sun since $a$ is the semi-major axis of Mercury ($a=0.39$ AU) and $e$ is its eccentricity ($e=0.206$) and hence $ea$ is the distance between the Sun and the center of the ellipse.  The semi-minor axis is defined by $b=a\sqrt{1-e^2}$.   The initial velocity was calculated from applying the principles
of conservation of angular momentum and conservation of energy between the above initial point and the point $(0,b)$.

\item[$(2)$]The amount of precession of the perihelion of Mercury is very small because $\alpha$ is very small. In fact it can not be measured directly in any numerical simulation with a limited amount of time. Therefore we will choose a larger value of $\alpha$ for example $\alpha=0.0008$ AU$^2$. We also work with $N=20000~,~dt=0.0001$. Compute the orbit for these values. Compute the angle $\theta$ made between the vector position of Mercury and the horizontal axis as a function of time. Compute also the distance between Mercury and the sun and its derivative with respect to time given by

\[\frac{dr}{dt}=\frac{xv_x+yv_y}{r}.\]
This derivative will vanish each time Mercury reaches its farthest point from the sun or its closest  point from the sun (the perihelion). Plot the angle $\theta_p$ made between the vector position of Mercury at its farthest point and the horizontal axis as a function of time. What do you observe. Determine the slope $d\theta_p/dt$ which is precisely the amount of precession of the perihelion of Mercury for the above value of $\alpha$.

\item[$(3)$] Repeat the above question for other values of $\alpha$ say $\alpha=0.001,0.002,0.004$. Each time compute $d\theta_p/dt$. Plot $d\theta_p/dt$ as a function of $\alpha$. Determine the slope. Deduce the amount of precession of the perihelion of Mercury for the value of $\alpha=1.1.10^{-8}$AU$^2$.

\end{itemize}

\chapter{Chaotic Pendulum}

\section{Equation of Motion}
We start from  a simple pendulum. The equation of motion is given by
\begin{eqnarray}
ml\frac{d^{2}{\theta}}{d t^2}&=&-mg\sin\theta.
\end{eqnarray}
We consider the effect of air resistance on the motion of the mass $m$. We will assume that the force of air resistance is given by Stokes' law. We get
\begin{eqnarray}
ml\frac{d^{2}{\theta}}{d t^2}&=&-mg\sin\theta-m l q \frac{d\theta}{d t}.
\end{eqnarray}
The air friction will drain all energy from the pendulum. In order to maintain the motion against the damping effect of air resistance we will add a driving force. We will choose a periodic force with amplitude $ml F_D$ and frequency $\omega_D$. This arise for example if we apply a periodic electric field with  amplitude $E_D$ and frequency $\omega_D$ on the mass $m$ which is assumed to have  an electric charge $q$, i.e $ml F_D=q E_D$. It can also arise from the periodic oscillations of the pendulum's pivot point. By adding the driving force we get then the equation of motion
\begin{eqnarray}
ml\frac{d^{2}{\theta}}{d t^2}&=&-mg\sin\theta-ml q\frac{d\theta}{d t}+ml F_D\cos\omega_D t.
\end{eqnarray}
The natural frequency of the oscillations is given by the frequency of the simple pendulum, viz
\begin{eqnarray}
\omega_0=\sqrt{\frac{g}{l}}.
\end{eqnarray}
We will always take $\omega_0=1$, i.e. $l=g$. The equation of motion becomes
\begin{eqnarray}
\frac{d^{2}{\theta}}{d t^2}+\frac{1}{Q}\frac{d\theta}{d t}+\sin\theta&=&F_D\cos\omega_D t.
\end{eqnarray}
The coefficient $Q=1/q$ is known as the quality factor. It measures how many oscillations the pendulum without driving force will make before its energy is drained.  We will write the above second order differential equation as two first order differential equations, namely 
\begin{eqnarray}
&&\frac{d\theta}{d t}=\Omega\nonumber\\
&&\frac{d{\Omega}}{d t}=-\frac{1}{Q}\Omega-\sin\theta+F_D\cos\omega_D t.
\end{eqnarray}
This system of differential equations does not admit a simple analytic solution. The linear approximation corresponds to small amplitude oscillations, viz
\begin{eqnarray}
\sin\theta\simeq \theta.
\end{eqnarray}
The differential equations become linear given by
\begin{eqnarray}
&&\frac{d\theta}{d t}=\Omega\nonumber\\
&&\frac{d{\Omega}}{d t}=-\frac{1}{Q}\Omega-\theta+F_D\cos\omega_D t.
\end{eqnarray}
Or equivalently
\begin{eqnarray}
&&\frac{d{\theta}^2}{d t^2}=-\frac{1}{Q}\frac{d\theta}{d t}-\theta+F_D\cos\omega_D t.
\end{eqnarray}
For $F_D=0$ the solution is given by
\begin{eqnarray}
\theta_{t0}=\bigg(\theta(0)\cos\omega_*t +\frac{1}{\omega_*}\big(\Omega(0)+\frac{\theta(0)}{2Q}\big)~\sin\omega_*t \bigg)~e^{-\frac{t}{2Q}}~,~\omega_*=\sqrt{1-\frac{1}{4Q^2}}.
\end{eqnarray}
For $F_D\neq 0$ a particular solution is given by
\begin{eqnarray}
\theta_{\infty}=F_D(a\cos\omega_D t+b \sin\omega_D t).
\end{eqnarray}
We find
\begin{eqnarray}
a=\frac{1}{(1-\omega_D^2)^2+\frac{\omega_D^2}{Q^2}}(1-\omega_D^2)~,b=\frac{1}{(1-\omega_D^2)^2+\frac{\omega_D^2}{Q^2}}\frac{\omega_D}{Q}.
\end{eqnarray}
For $F_D\neq 0$ the general solution is given by
\begin{eqnarray}
\theta&=&\theta_{\infty}+\theta_t.
\end{eqnarray}
\begin{eqnarray}
\theta_t=\bigg[\bigg(\theta(0)-\frac{F_D(1-\omega_D^2)}{(1-\omega_D^2)^2+\frac{\omega_D^2}{Q^2}}\bigg)\cos\omega_*t +\frac{1}{\omega_*}\bigg(\Omega(0)+\frac{\theta(0)}{2Q}-\frac{1}{2Q}\frac{F_D(1-3\omega_D^2)}{(1-\omega_D^2)^2+\frac{\omega_D^2}{Q^2}}\bigg)\sin\omega_*t\bigg] ~e^{-\frac{t}{2Q}}.\nonumber\\
\end{eqnarray}
The last two terms depend on the initial conditions and will vanish exponentially at very large times $t\longrightarrow\infty$, i.e. they are transients. The asymptotic motion is given by $\theta_{\infty}$. Thus  for $t\longrightarrow\infty$ we get
\begin{eqnarray}
\theta&=&\theta_{\infty}=F_D(a\cos\omega_D t+b \sin\omega_D t).
\end{eqnarray}
Also for $t\longrightarrow\infty$ we get
\begin{eqnarray}
\Omega&=&\frac{d\theta}{d t}=F_D\omega_D(-a \sin\omega_D t+b \cos\omega_D t).
\end{eqnarray}
We compute in the limit of large times $t\longrightarrow\infty$
\begin{eqnarray}
\theta^2+\frac{\Omega^2}{\omega_D^2}=\tilde{F}_D^2=F_D^2(a^2+b^2)=\frac{F_D^2}{(1-\omega_D^2)^2+\frac{\omega_D^2}{Q^2}}.
\end{eqnarray}
In other words the orbit of the system in  phase space is an ellipse. The motion is periodic with period equal to the period of the driving force. This ellipse is also called a periodic attractor because regardless of the initial conditions the trajectory of the system will tend at large times to this ellipse. 

Let us also remark that the maximum angular displacement is $\tilde{F}_D$. The function $\tilde{F}_D=\tilde{F}_D(\omega_D)$ exhibits resonant behavior as the driving frequency approaches the natural frequency which is equivalent to the limit $\omega_D\longrightarrow 1$. In this limit $\tilde{F}_D=Q F_D$. The width of the resonant window is  proportional to $1/Q$ so for $Q\longrightarrow \infty$ we observe that $\tilde{F}_D\longrightarrow \infty$ when $\omega_D\longrightarrow 1$ while for $Q\longrightarrow 0$ we observe that $\tilde{F}_D\longrightarrow 0$  when $\omega_D\longrightarrow 1$.  

In general the time-asymptotic response of any linear system to a periodic drive is periodic with the same period as the driving force. Furthermore when the driving frequency approaches one of the natural frequencies the response will exhibits resonant behavior.  

The basic ingredient in deriving the above results is the linearity of the dynamical system. As we will see shortly periodic motion is not the only possible time-asymptotic response of a dynamical system to a periodic driving force.

\section{Numerical Algorithms}
The equations of motion are
\begin{eqnarray}
&&\frac{d\theta}{d t}=\Omega\nonumber\\
&&\frac{d{\Omega}}{d t}=-\frac{1}{Q}\Omega-\sin\theta+F(t).
\end{eqnarray}
The external force is periodic and it will be given by one of the following expressions
\begin{eqnarray}
F(t)=F_D\cos\omega_D t.
\end{eqnarray}
\begin{eqnarray}
F(t)=F_D\sin\omega_D t.
\end{eqnarray}

\subsection{Euler-Cromer Algorithm}
Numerically we can employ the Euler-Cromer algorithm in order to solve this system of differential equations. The solution goes as follows. First we choose the initial conditions. For example 
\begin{eqnarray}
&&\Omega(1)=0\nonumber\\
&&\theta(1)=0\nonumber\\
&&t(1)=0.
\end{eqnarray}
For $i=1,...,N+1$ we use
\begin{eqnarray}
&&\Omega(i+1)=\Omega(i)+\Delta t\bigg(-\frac{1}{Q}\Omega(i)-\sin\theta(i)+F(i)\bigg)\nonumber\\
&&\theta(i+1)=\theta(i)+\Delta t ~\Omega(i+1)\nonumber\\
&&t(i+1)=\Delta t ~i.
\end{eqnarray}
\begin{eqnarray}
F(i)\equiv F(t(i))=F_D\cos\omega_D \Delta t (i-1).
\end{eqnarray}
\begin{eqnarray}
F(i)\equiv F(t(i))=F_D\sin\omega_D \Delta t (i-1).
\end{eqnarray}
\subsection{Runge-Kutta Algorithm}

In order to achieve better precision we employ the Runge-Kutta  algorithm. For $i=1,...,N+1$ we use
\begin{eqnarray}
&&k_1=\Delta t~ \Omega(i)\nonumber\\
&&k_3=\Delta t\bigg[-\frac{1}{Q}\Omega(i)-\sin\theta(i)+F(i)\bigg]\nonumber\\
&&k_2=\Delta t\bigg(\Omega(i)+\frac{1}{2}k_3\bigg)\nonumber\\
&&k_4=\Delta t\bigg[-\frac{1}{Q}\bigg(\Omega(i)+\frac{1}{2}k_3\bigg)-\sin\bigg(\theta(i)+\frac{1}{2}k_1\bigg)+F(i+\frac{1}{2})\bigg]\nonumber\\
\end{eqnarray}
\begin{eqnarray}
&&\theta(i+1)=\theta(i)+k_2\nonumber\\
&&\Omega(i+1)=\Omega(i)+k_4\nonumber\\
&&t(i+1)=\Delta t ~i.
\end{eqnarray}
\begin{eqnarray}
F(i)\equiv F(t(i))=F_D\cos\omega_D \Delta t (i-1).
\end{eqnarray}
\begin{eqnarray}
F(i)\equiv F(t(i))=F_D\sin\omega_D \Delta t (i-1).
\end{eqnarray}
\begin{eqnarray}
F(i+\frac{1}{2})\equiv F(t(i)+\frac{1}{2}\Delta t)=F_D\cos\omega_D \Delta t (i-\frac{1}{2}).
\end{eqnarray}
\begin{eqnarray}
F(i+\frac{1}{2})\equiv F(t(i)+\frac{1}{2}\Delta t)=F_D\sin\omega_D \Delta t (i-\frac{1}{2}).
\end{eqnarray}

\section{Elements of Chaos}

\subsection{Butterfly Effect: Sensitivity to Initial Conditions} 

The solution in the linear regime (small amplitude) reads
\begin{eqnarray}
\theta=\theta_{\infty}+\theta_t.
\end{eqnarray}
The transient is of the form
\begin{eqnarray}
\theta_t=f(\theta(0),\Omega(0))e^{-t/2Q}.
\end{eqnarray}
This goes to zero at large times $t$. The time-asymptotic is thus given by
\begin{eqnarray}
\theta_{\infty}=F_D(a \cos\omega_D t+ b \sin  \omega_D t).
\end{eqnarray}
The motion in the phase space is periodic with period equal to the period of the driving force. The orbit in phase space is precisley an ellipse of the form
\begin{eqnarray}
\theta_{\infty}^2+\frac{\Omega_{\infty}^2}{\omega_D^2}={F}_D^2(a^2+b^2).
\end{eqnarray}
Let us consider a perturbation of the initial conditions. We can imagine that we have two pendulums $A$ and $B$ with slightly different initial conditions. Then the difference between the two trajectories is
\begin{eqnarray}
\delta \theta=\delta f(\theta(0),\Omega(0))e^{-t/2Q}.
\end{eqnarray}
This goes to zero at large times. If we plot $\ln\delta\theta$ as a function of time we find a straight line with a negative slope. The time-asymptotic motion is not sensitive to initial conditions. It converges at large times to $\theta_{\infty}$ no matter what the initial conditions are. The curve $\theta_{\infty}=\theta_{\infty}(\Omega_{\infty})$ is called a (periodic) attractor. This is because any perturbed trajectory will decay exponentially in time to the attractor.

In order to see chaotic behavior we can for example increase $Q$ keeping everything else fixed. We observe that the slope of the line $\ln\delta \theta=\lambda t$ starts to decrease until at some value of $Q$ it becomes positive. At this value the variation between the two pendulums increases exponentially  with time. This is the chaotic regime. The value $\lambda=0$ is the value where chaos happens. The coefficient $\lambda$ is called Lyapunov exponent. 

The chaotic pendulum is a deterministic system (since it obeys ordinary differential equations) but it is not predictable in the sense that given two identical pendulums their motions will diverge from each other in the chaotic regime if there is the slightest error in determining their initial conditions. This high sensitivity to initial conditions is known as the butterfly effect and could be taken as the definition of chaos itself.

However we should stress here that the motion of the chaotic pendulum is not random. This can be seen by inspecting Poincare sections.
\subsection{Poincare Section and Attractors}

The periodic motion of the linear system with period equal to the  period of the driving force is called a period-$1$ motion. In this motion the trajectory repeats itself exactly every one single period of the external driving force. This is the only possible motion in the low amplitude limit.

Generally a period-${\cal N}$ motion corresponds to an orbit of the dynamical system which repeats itself every ${\cal N}$ periods of the external driving force. These orbits exist in the non-linear regime of the pendulum. 

The Poincare section is defined as follows. We plot in the $\theta$-$\Omega$ phase space only one point per period of the external driving force. We plot for example $(\theta,\Omega)$ for 
\begin{eqnarray}
\omega_D t=\phi +2\pi n.
\end{eqnarray}
The angle $\phi$ is called the Poincare phase and $n$ is an integer. For period-$1$ motion the Poincare section consists of one single point. For period-${\cal N}$ motion the Poincare section consists of ${\cal N}$ points.

Thus in the linear regime if we plot $(\theta,\Omega)$ for $\omega_D t=2\pi n$ we get a single point since the motion is periodic with period equal to that of the driving force. The single point we get as a Poincare section is  also an attractor since all pendulums with almost the same initial conditions will converge onto it.

In the chaotic regime the Poincare section is an attractor known as strange attractor. It is a complicated curve which could have fractal structure and  all pendulums with almost the same initial conditions will converge onto it. 

\subsection{Period-Doubling Bifurcations}
In the case of the chaotic pendulum we encounter between the linear regime and the emergence of chaos the so-called period  doubling phenomena. In the linear regime the Poincare section is a point $P$ which corresponds to a period-$1$ motion with period equal $T_D=2\pi/\omega_D$. The $\theta$ or $\Omega$ coordinate of this point $P$ will trace a line as we increase $Q$ while keeping everything fixed.  We will eventually reach a value $Q_1$ of $Q$ where this line bifurcates into two lines. By close inspection we see that at $Q_1$ the motion becomes period-$2$ motion, i.e. the period becomes equal to $2T_D$. 

In a motion where the period is $T_D$ (below $Q_1$) we get the same value of $\theta$ each time $t=mT_D$ and since we are plotting $\theta$ each time $t=2n\pi/\omega_D=nT_D$ we will get a single point in the Poincare section. In a motion where the period is $2T_D$ (at $Q_2$) we get the same value of $\theta$ each time $t=2mT_D$, i.e. the value of $\theta$ at times $t=mT_D$ is different and hence we get two points in the Poincare section.  

As we increase $Q$ the motion becomes periodic with period equal $4T_D$, then with period equal $8T_D$ and so on. The motion with period $2^{\cal N}T_D$ is called period-${\cal N}$  motion. The corresponding Poincare section consists of ${\cal N}$ distinct points.

The diagram of $\theta$ as a function of $Q$ is called a bifurcation diagram. It has a  fractal structure. Let us point out here that normally in ordinary  oscillations we get harmonics with periods equal to the period of the driving force divided by $2^{\cal N}$. In this case we obtained in some sense subharmonics with periods equal to the period of the driving force times $2^{\cal N}$. This is very characteristic of chaos. In fact chaotic behavior corresponds to the limit ${\cal N}\longrightarrow \infty$. In other words chaos is period-$\infty$ (bounded) motion which could be taken as another definition of chaos.
\subsection{Feigenbaum Ratio}
Let $Q_{\cal N}$ be the critical value of $Q$ above which the ${\cal N}$th bifurcation is triggered. In other words $Q_{\cal N}$ is the value where the transition to period-${\cal N}$ motion happens. We define the Feigenbaum ratio by
\begin{eqnarray}
F_{\cal N}=\frac{Q_{{\cal N}-1}-Q_{{\cal N}-2}}{Q_{\cal N}-Q_{{\cal N}-1}}.
\end{eqnarray}
It is shown that $F_{\cal N}\longrightarrow F=4.669 $ as ${\cal N}\longrightarrow\infty$. This  is a universal ratio  called the Feigenbaum ratio and it characterizes many chaotic systems which suffer a transition to chaos via an infinite series of period-doubling bifurcations. The above equation can be then rewritten as
\begin{eqnarray}
Q_{\cal N}=Q_1+(Q_2-Q_1)\sum_{j=0}^{{\cal N}-2}\frac{1}{F^j}
\end{eqnarray}
Let us define the accumulation point by $Q_{\infty}$ then 
\begin{eqnarray}
Q_{\infty}=Q_1+(Q_2-Q_1)\frac{F}{F-1}
\end{eqnarray}
This is where chaos occur. In the bifurcation diagram the chaotic region is a solid black region.
\subsection{Spontaneous Symmetry Breaking}
The bifurcation process is associated with a deep phenomenon known as spontaneous symmetry breaking. The first period-doubling bifurcation corresponds to the breaking of the symmetry $t\longrightarrow t+T_D$. The linear regime respects this symmetry. However period-$2$ motion and in general period-${\cal N}$ motions with ${\cal N}>2$ do not respect this symmetry.

There is another kind of spontaneous symmetry breaking which occurs in the chaotic pendulum and which is associated with a bifurcation diagram. This happens in the   region of period-$1$ motion and it is the breaking of spatial symmetry or parity $\theta\longrightarrow -\theta$. Indeed there exists solutions of the equations of motion that are either left-favoring or right-favoring. In other words the pendulums in such solutions spend much of its time in the regions to the left of the pendulum's vertical ($\theta <0$) or to the right of the pendulum's vertical ($\theta >0$). This breaking of left-right symmetry can be achieved by a gradual increase of $Q$. We will then reach either the left-favoring solution or the right-favoring solution starting from a left-right symmetric solution depending on the initial conditions. The symmetry $\theta\longrightarrow -\theta$ is also spontaneously broken in period-${\cal N}$ motions.

\section{Simulation $8$: The Butterfly Effect}
We consider a pendulum of a mass $m$ and a length $l$ moving under the influence of the force of gravity, the force of air resistance and  a driving periodic force. 
 Newton's second law of motion reads
\[\frac{d^2\theta}{dt^2}=-\frac{g}{l}\sin\theta -q\frac{d\theta}{dt}+F_D\sin 2\pi\nu_D t.\]
We will always take the angular frequency $\sqrt{g/l}$ associated with simple oscillations of the pendulum equal $1$, i.e. $l=g$. The numerical solution we will consider here is based on the Euler-Cromer algorithm. 

The most important property of a large class of solutions of this differential equation is hyper sensitivity to  initial conditions known also as the butterfly effect which is the defining characteristic of chaos. For this reason the driven non-linear pendulum is also known as the chaotic pendulum. 

The chaotic pendulum can have two distinct behaviors.
In the linear regime the motion (neglecting the initial transients) is periodic with a period equal to the period of the external driving force.  In the chaotic regime the motion never repeats and any error even infinitesimal in determining the initial conditions will lead to a completely different orbit in the phase space.

\begin{itemize}
 \item[$(1)$] Write a code which implements the Euler-Cromer algorithm for the chaotic pendulum. The angle $\theta$ must always be taken between $-\pi$ and $\pi$ which can be maintained as follows
\[{\rm if}(\theta_i.{\rm lt}.\mp\pi)~ \theta_i=\theta_i\pm 2\pi.\]
\item[$(2)$]
We take the values and initial conditions
\[dt=0.04s~,~2\pi\nu_D=\frac{2}{3}s^{-1}~,~q=\frac{1}{2}s^{-1}~,~N=1000-2000.\]
\[\theta_1=0.2~{\rm radian}~,~\Omega_1=0~{\rm radian}/s.\]
\[F_D=0~{\rm radian}/s^2~,~F_D=0.1~{\rm radian}/s^2~,~F_D=1.2~{\rm radian}/s^2.\]
Plot $\theta$ as a function of time. What do you observe for the first value of $F_D$. What is the period of oscillation for small and large times for the second value of $F_D$
. Is the motion periodic for the third value of $F_D$.

\end{itemize}

\section{Simulation $9$: Poincar\'e Sections}

In the chaotic regime the motion of the pendulum although deterministic is not predictable. This however does not mean that the motion of the pendulum is random which can clearly be seen from the Poincare sections. 

A Poincare section is a curve in the phase space obtained by plotting one point of the orbit per period of the external drive. Explicitly we plot points $(\theta,\Omega)$ which corresponds to times $t=n/\nu_D$ where $n$ is an integer.  In the linear regime of the pendulum  the Poincare section  consists of a single point. Poincare section in the chaotic regime is a curve  which does not depend on the initial conditions thus confirming that the motion is not random and which may have a fractal structure. As a consequence this curve is called a strange attractor.

\begin{itemize}

\item[$(1)$]
We consider two identical chaotic pendulums $A$ and $B$ with slightly different initial conditions. For example we take

\[\theta_1^{A}=0.2~{\rm radian}~,~ \theta_1^{B}=0.201~{\rm radian}.\]
The difference between the two motions can be measured by 
\[\Delta\theta_i=\theta_i^A-\theta_i^B.\]
Compute $\ln\Delta\theta$ as a function of time for
\[ F_D=0.1~{\rm radian}/s^2~,~F_D=1.2~{\rm radian}/s^2.\]
What do you observe. Is the two motions identical. What happens for large times. Is the motion of the pendulum predictable. For the second value of $F_D$ use
\[N=10000~,~dt=0.01s.\] 

\item[$(2)$]

Compute the angular velocity $\Omega$ as a function of $\theta$ for
\[F_D=0.5~{\rm radian}/s^2~,~F_D=1.2~{\rm radian}/s^2.\]
What is the orbit in the phase space for small times and what does it represent. What is the orbit for large times. Compare between the two pendulums $A$ and $B$. Does the orbit for large times depend on the initial conditions.

\item[$(3)$] 
A Poincare section is obtained numerically by plotting the points $(\theta,\Omega)$ of the orbit at the times at which the function $\sin \pi\nu_D t$ vanishes. These are the times at which this function changes sign. This is implemented as follows

\[{\rm if}(\sin\pi\nu_D t_i \sin\pi\nu_D t_{i+1}.{\rm lt}.0) {\rm then}\]
\[{\rm write}(*,*)t_i,\theta_i,\Omega_i.\]

Verify that Poincare section in the linear regime is given by a single point in the phase space. Take and use $F_D=0.5~{\rm radian}/s^2~,~N=10^{4}-10^7~,~dt=0.001s$. Verify that Poincare section in the chaotic regime is also an attractor. Take and use $F_D=1.2~{\rm radian}/s^2~,~N=10^{5}~,~dt=0.04s$. Compare between Poincare sections of the pendulums $A$ and $B$. What do you observe and what do you conclude. 

\end{itemize}

\section{Simulation $10$: Period Doubling}
Among the most important chaotic properties of the driven non-linear pendulum is the phenomena of period doubling. The periodic orbit with period equal to the period of the external driving force are called period-$1$ motion. There exist however other periodic orbits with periods equal twice, four times and in general $2^{\cal N}$ times  the period of the external driving force. The orbit with period equal $2^{\cal N}$ times the period of the external driving force is called period-${\cal N}$ motion. The period doubling observed in the driven non-linear pendulum is a new phenomena which belongs to the world of chaos. In the standard phenomena of mixing the response of  a non-linear system to a single frequency external driving force will contain components with periods equal to the period of the driving force divided by $2^{\cal N}$. In other words we get "harmonics" as opposed to the  "subharmonics" we observe in the chaotic pendulum.

For period-${\cal N}$ motion we expect that there are ${\cal N}$ different values of the angle $\theta$ for every value of $F_D$. The function $\theta=\theta(F_D)$ is called a bifurcation diagram. Formally the transition to chaos occurs at ${\cal N}\longrightarrow\infty$. In other words chaos is defined as period-infinity motion.

\begin{itemize}
\item[$(1)$]
We take the values and initial conditions
\[l=g~,~2\pi\nu_D=\frac{2}{3}s^{-1}~,~q=\frac{1}{2}s^{-1}~,~N=3000-100000~,~dt=0.01s.\]
\[\theta_1=0.2~{\rm radian}~,~\Omega_1=0~{\rm radian}/s.\]
Determine the period of the motion for

\[F_D=1.35~{\rm radian}/s^2~,~F_D=1.44~{\rm radian}/s^2~,~F_D=1.465~{\rm radian}/s^2.\]
What happens to the period when we increase $F_D$. Does the two second values of $F_D$ lie in the linear or chaotic regime of the chaotic pendulum.

\item[$(2)$]
Compute the angle $\theta$ as a function of  $F_D$ for the times $t$ which satisfy the condition $2\pi\nu_D t=2n\pi$. We take $F_D$ in the interval
\[F_D=(1.34+0.005k)~{\rm radian}/s^2~,~k=1,...,30.\]
Determine the interval of the external driving force in which the orbits are period-$1$, period-$2$ and period-$4$ motions.

In this problem it is very important to remove the initial transients before we start measuring the bifurcation diagram. This can be done as follows. We calculate the motion for $2N$ steps but then only consider the last $N$ steps in the computation of the Poincare section for every value of $F_D$.
\end{itemize}

\section{Simulation $11$: Bifurcation Diagrams}
\paragraph{Part I}
The chaotic pendulum is given by the equation
\[\frac{d^2\theta}{dt^2}=-\sin\theta -\frac{1}{Q}\frac{d\theta}{dt}+F_D\cos 2\pi\nu_D t.\]
In this simulation we take the values $F_D=1.5~{\rm radian}/s^2$ and $2\pi\nu_D=\frac{2}{3}s^{-1}$. In order to achieve a better numerical precision we use the second-order Runge-Kutta algorithm. 

In the linear regime the orbits are periodic with period equal to the period $T_D$ of the external driving force and are symmetric under $\theta\longrightarrow -\theta$. There exists other solutions which are periodic with period equal $T_D$ but are not symmetric under $\theta\longrightarrow -\theta$. In these solutions the pendulum spends  the majority of its time in the region to the left of its vertical ($\theta<0$) or in the region to the right of its vertical ($\theta>0$). 

These symmetry breaking solutions can be described by a bifurcation diagram  $\Omega=\Omega(Q)$. For every value of the quality factor $Q$ we calculate the Poincare section. We observe that the Poincare section will bifurcate at some value $Q_*$ of $Q$. Below this value we get one line whereas above this value we get two lines corresponding to the two symmetry breaking solutions in which the pendulum spends the majority of its time in the regions ($\theta>0$) and ($\theta<0$). 
\begin{itemize}
\item[$(1)$]
Rewrite the code for the chaotic pendulum using Runge-Kutta algorithm.
\item[$(2)$]
We take two different sets of initial conditions
\[\theta=0.0~{\rm radian}~,~\Omega=0.0~{\rm radian}/s.\]
\[\theta=0.0~{\rm radian}~,~\Omega=-3.0~{\rm radian}/s~.\]
Study the nature of the orbit for the values $Q=0.5$s, $Q=1.24$s and $Q=1.3$s. What do you observe. 

\item[$(3)$]
Plot the bifurcation diagram $\Omega=\Omega(Q)$ for values of $Q$ in the interval $[1.2,1.3]$. What is the value $Q_*$ at which the symmetry  $\theta\longrightarrow -\theta$ is spontaneously broken.
\end{itemize}
\paragraph{Part II}
As we have seen in the previous simulation period doubling can also be described by a bifurcation diagram. This phenomena is also an example of a spontaneous symmetry breaking. In this case the symmetry is $t\longrightarrow t+T_D$. Clearly only orbits with period $T_D$ are symmetric under this transformation. 

Let  $Q_{\cal N}$ be the value of $Q$ at which the ${\cal N}$th bifurcation occurs. In other words this is the value at which the orbit goes from being a period-$({\cal N}-1)$ motion to a period-${\cal N}$ motion. The Feigenbaum ratio is defined by
 
\[F_{\cal N}=\frac{Q_{{\cal N}-1}-Q_{{\cal N}-2}}{Q_{{\cal N}}-Q_{{\cal N}-1}}.\]
As we approach the chaotic regime, i.e. as ${\cal N}\longrightarrow \infty$ the ratio $F_{\cal N}$ converges rapidly to the constant value  $F=4.669$. This is a general result which holds for many chaotic systems. Any dynamical system which can exhibit a transition to chaos via an infinite series of period-doubling bifurcations is characterized by a Feigenbaum ratio which approaches $4.669$ as ${\cal N}\longrightarrow \infty$.
 
\begin{itemize}
\item[$(1)$]
Calculate the orbit and Poincare section for $Q=1.36$s. What is the period of the motion. Is the orbit symmetric under $t\longrightarrow t+T_D$. Is the orbit symmetric under $\theta\longrightarrow -\theta$.
\item[$(2)$]
Plot the bifurcation diagram $\Omega=\Omega(Q)$ for two different sets of initial conditions for values of $Q$ in the interval $[1.3,1.36]$. What is the value $Q$ at which the period gets doubled. What is the value of $Q$ at which the symmetry $t\longrightarrow t+T_D$ is spontaneously broken.
\item[$(3)$]
In this question we use the initial conditions
\[\theta=0.0~{\rm radian}~,~\Omega=0.0~{\rm radian}/s.\]
Calculate the orbit and Poincare section and plot the bifurcation diagram $\Omega=\Omega(Q)$ for values of $Q$ in the interval $[1.34,1.38]$. 
Determine from the bifurcation diagram the values $Q_{\cal N}$ for
${\cal N}=1,2,3,4,5$. Calculate the Feigenbaum ratio. Calculate the accumulation point $Q_{\infty}$ at which the transition to chaos occurs.
\end{itemize}


\chapter{Molecular Dynamics}
\section{Introduction}
In the molecular dynamics approach we attempt to understand the behavior of a classical many-particle system by simulating the trajectory of each particle in the system. In practice this can be applied to systems containing $10^9$ particles at most. The molecular dynamics approach is complementary to the more powerful Monte Carlo method. The Monte Carlo method deals with systems that are in thermal equilibrium with a heat bath. The molecular dynamics approach on the other hand is useful in studying how fast in real time a system moves from one microscopic state to another. 

We consider a box containing a collection of atoms or molecules. We will use Newton's second law to calculate the positions and velocities of all the molecules as functions of time. Some of the questions we can answer with the molecular dynamics approach are:
\begin{itemize}
\item{}The melting transition.
\item{}The rate of equilibration.
\item{}The rate of diffusion.
\end{itemize}
As state above molecular dynamics allows us to understand classical systems. A classical treatment can be justified as follows. We consider the case of liquid argon as an example. The energy required to excite an argon atom is of the order of $10$eV while the typical kinetic energy of the center of mass of an argon atom is $0.1$eV. Thus a collision between two argon atoms will not change  the electron configuration of either atoms.  Hence for all practical purposes we can ignore the internal structure of argon atoms. Furthermore the wavelength of an argon atom which is of the order of $10^{-7}$A is much smaller than the spacing  between argon atoms typically of the order of $1$A which again justifies a classical treatment. 
\section{The Lennard-Jones Potential}
We consider a box containing $N$ argon atoms. For simplicity we will assume that our argon atoms move in two dimensions. The equations of motion of the $i$th atom which is located at the position $(x_i,y_i)$ with velocity $(v_{i,x},v_{i,y})$ read

\begin{eqnarray}
\frac{dv_{i,x}}{dt}=a_{x,i}~,~\frac{dx_i}{dt}=v_{i,x}.
\end{eqnarray}
\begin{eqnarray}
\frac{dv_{i,y}}{dt}=a_{y,i}~,~\frac{dy_i}{dt}=v_{i,y}.
\end{eqnarray}
Each argon atom experience a force from all other argon atoms. In order to calculate this force we need to determine the interaction potential. We assume that the interaction potential between any pair of argon atoms depend only on the distance between them. Let $r_{ij}$ and $u(r_{ij})$ be the distance and the interaction potential between atoms $i$ and $j$. The total potential is then given by
\begin{eqnarray}
U=\sum_{i=1}^{N-1}\sum_{j=i+1}^Nu(r_{ij}).
\end{eqnarray}
The precise form of $u$ can be calculated from first principles, i.e. from quantum mechanics. However this calculation is very complicated and in most circumstances a phenomenological form of $u$ will be sufficient. 

For large separations $r_{ij}$ the potential $u(r_{ij})$ must be weakly attractive given by the Van der Walls force which arises from electrostatic interaction between the electric dipole moments of the two argon atoms. In other words  $u(r_{ij})$ for large $r_{ij}$ is attractive due to the mutual polarization of the two atoms. The Van der Walls potential can be computed from quantum mechanics where it is shown that it varies as $1/r_{ij}^6$. For small separations $r_{ij}$ the potential $u(r_{ij})$ must become strongly repulsive due to the overlap of the electron clouds of the two argon atoms. This repulsion known also as core repulsion is a consequence of Pauli exclusion principle. It is a common practice to choose the repulsive part of the potential $u$ to be proportional to $1/r_{ij}^{12}$. The total potential takes the form
\begin{eqnarray}
u(r)=4\epsilon\bigg[\bigg(\frac{\sigma}{r}\bigg)^{12}-\bigg(\frac{\sigma}{r}\bigg)^6\bigg].
\end{eqnarray}
This is the Lennard-Jones potential. The parameter $\sigma$ is of dimension length while $\epsilon$ is of dimension energy. We observe that at $r=\sigma$ the potential is $0$ identically while for $r>2.5\sigma$ the potential approaches zero rapidly. The minimum of the potential occurs at $r=2^{1/6}\sigma$. The depth of the potential at the minimum is $\epsilon$.

The force of atom $k$ on atom $i$ is

\begin{eqnarray}
\vec{f}_{k,i}=-\vec{\nabla}_{k,i}u(r_{k,i})=\frac{24\epsilon}{r_{ki}}\bigg[2\bigg(\frac{\sigma}{r_{ki}}\bigg)^{12}-\bigg(\frac{\sigma}{r_{ki}}\bigg)^6\bigg]\hat{r}_{ki}.
\end{eqnarray}
The acceleration of the $i$th atom is given by
\begin{eqnarray}
a_{x,i}=\frac{1}{m}\sum_{k\neq i}{f}_{k,i}\cos\theta_{k,i}&=&\frac{1}{m}\sum_{k\neq i}{f}_{k,i}\frac{x_i-x_k}{r_{ki}}\nonumber\\
&=&\frac{24\epsilon}{m}\sum_{k\neq i}\frac{x_i-x_k}{r_{ki}^2}\bigg[2\bigg(\frac{\sigma}{r_{ki}}\bigg)^{12}-\bigg(\frac{\sigma}{r_{ki}}\bigg)^6\bigg].
\end{eqnarray}
\begin{eqnarray}
a_{y,i}=\frac{1}{m}\sum_{k\neq i}{f}_{k,i}\sin\theta_{k,i}&=&\frac{1}{m}\sum_{k\neq i}{f}_{k,i}\frac{y_i-y_k}{r_{ki}}\nonumber\\
&=&\frac{24\epsilon}{m}\sum_{k\neq i}\frac{y_i-y_k}{r_{ki}^2}\bigg[2\bigg(\frac{\sigma}{r_{ki}}\bigg)^{12}-\bigg(\frac{\sigma}{r_{ki}}\bigg)^6\bigg].
\end{eqnarray}

\section{Units, Boundary Conditions and Verlet Algorithm}
\paragraph{Reduced Units} We choose $\sigma$ and $\epsilon$ as the units of distance and energy respectively. We also choose the unit of mass to be the mass $m$ of a single argon atom. Everything else is measured in terms of $\sigma$, $\epsilon$ and $m$. For example velocity is measured in units of $(\epsilon/m)^{1/2}$ and time in units of $\sigma (\epsilon/m)^{1/2}$. The reduced units are given by
\begin{eqnarray}
\sigma=\epsilon=m=1.
\end{eqnarray}
For argon atoms we have the values
\begin{eqnarray}
\sigma=3.4\times 10^{-10}m~,~\epsilon=1.65\times 10^{-21}J=120k_BJ~,~m=6.69\times 10^{-26}kg.
\end{eqnarray}
Thus
\begin{eqnarray}
\sigma\sqrt{\frac{m}{\epsilon}}=2.17\times 10^{-12}s.
\end{eqnarray}
Hence a molecular dynamics simulation which runs for $2000$ steps with a reduced time step $\Delta t=0.01$ corresponds to a total reduced time $2000\times 0.01=20$ which is equivalent to a real time $20\sigma (\epsilon/m)^{1/2}=4.34\times 10^{-11}s$.
\paragraph{Periodic Boundary Conditions}
The total number of atoms in a real physical system is huge of the order of $10^{23}$. If the system is placed in a box the fraction of atoms of the system near the walls of the box is negligible compared to the total number of atoms. In typical simulations the total number of atoms is only of the order of $10^3-10^5$ and in this case the fraction of atoms near the walls is considerable and their effect can not be neglected.   

In order to reduce edge effects we use periodic boundary conditions. In other words the box is effectively a torus and there are no edges. Let $L_x$ and $L_y$ be the lengths of the box in the $x$ and $y$ directions respectively. If an atom crosses the walls of the box in a particular direction we add or subtract the length of the box in that direction as follows
\begin{eqnarray}
&&{\rm if}~(x>L_x)~{\rm then}~x=x-L_x\nonumber\\
&&{\rm if}~(x<0)~{\rm then}~x=x+L_x.
\end{eqnarray}
\begin{eqnarray}
&&{\rm if}~(y>L_y)~{\rm then}~y=y-L_y\nonumber\\
&&{\rm if}~(y<0)~{\rm then}~y=y+L_y.
\end{eqnarray}
The maximum separation in the $x$ direction between any two particles is only $L_x/2$ whereas the  maximum separation in the $y$ direction between any two particles is only $L_y/2$. This can be implemented as follows
\begin{eqnarray}
&&{\rm if}~(x_{ij}>+L_x/2)~{\rm then}~x_{ij}=x_{ij}-L_x\nonumber\\
&&{\rm if}~(x_{ij}<-L_x/2)~{\rm then}~x_{ij}=x_{ij}+L_x.
\end{eqnarray}
\begin{eqnarray}
&&{\rm if}~(y_{ij}>+L_y/2)~{\rm then}~y_{ij}=y_{ij}-L_y\nonumber\\
&&{\rm if}~(y_{ij}<-L_y/2)~{\rm then}~y_{ij}=y_{ij}+L_y.
\end{eqnarray}
\paragraph{Verlet Algorithm}
The numerical algorithm we will use is Verlet algorithm. Let us consider the forward and backward Taylor expansions of a function $f$ given by
\begin{eqnarray}
  f(t_n+\Delta t)=f(t_n)+\Delta t \frac{df}{dt}|_{t_n}+\frac{1}{2}(\Delta t)^2\frac{d^2f}{dt^2}|_{t_n}+\frac{1}{6}(\Delta t)^3\frac{d^3f}{dt^3}|_{t_n}+...
\end{eqnarray}
\begin{eqnarray}
  f(t_n-\Delta t)=f(t_n)-\Delta t \frac{df}{dt}|_{t_n}+\frac{1}{2}(\Delta t)^2\frac{d^2f}{dt^2}|_{t_n}-\frac{1}{6}(\Delta t)^3\frac{d^3f}{dt^3}|_{t_n}+...
\end{eqnarray}
Adding these expressions we get
\begin{eqnarray}
  f(t_n+\Delta t)=2f(t_n)-f(t_n-\Delta t)+(\Delta t)^2\frac{d^2f}{dt^2}|_{t_n}+O(\Delta t^4).
\end{eqnarray}
We remark that the error is proportional to $\Delta t^4$ which is less than the errors in the Euler, Euler-Cromer and second-order Runge-Kutta methods so this method is more accurate. We have therefore for the $i$th atom

\begin{eqnarray}
  x_{i,n+1}=2x_{i,n}-x_{i,n-1}+(\Delta t)^2a_{x,i,n}.
\end{eqnarray}
\begin{eqnarray}
  y_{i,n+1}=2y_{i,n}-y_{i,n-1}+(\Delta t)^2a_{y,i,n}.
\end{eqnarray}
The force and the acceleration are given by
\begin{eqnarray}
{f}_{k,i,n}=\frac{24\epsilon}{r_{ki,n}}\bigg[2\bigg(\frac{\sigma}{r_{ki,n}}\bigg)^{12}-\bigg(\frac{\sigma}{r_{ki,n}}\bigg)^6\bigg].
\end{eqnarray}
\begin{eqnarray}
a_{x,i,n}&=&\frac{1}{m}\sum_{k\neq i}{f}_{k,i,n}\frac{x_{i,n}-x_{k,n}}{r_{ki,n}}.
\end{eqnarray}
\begin{eqnarray}
a_{y,i,n}&=&\frac{1}{m}\sum_{k\neq i}{f}_{k,i,n}\frac{y_{i,n}-y_{k,n}}{r_{ki,n}}.
\end{eqnarray}
The separation $r_{ki,n}$ between the two atoms $k$ and $i$ is given by
\begin{eqnarray}
r_{ki,n}=\sqrt{(x_{i,n}-x_{k,n})^2+(y_{i,n}-y_{k,n})}.
\end{eqnarray}
In the Verlet method it is not necessary to calculate the components $dx_{i,n}/dt$ and $dy_{i,n}/dt$ of the velocity. However since the velocity will be needed for other purposes we will also compute it using the equations
\begin{eqnarray}
  v_{x,i,n}=\frac{x_{i,n+1}-x_{i,n-1}}{2\Delta t}.
\end{eqnarray}
\begin{eqnarray}
  v_{y,i,n}=\frac{y_{i,n+1}-y_{i,n-1}}{2\Delta t}.
\end{eqnarray}
Let us remark that the Verlet method is not self starting. In other words given the initial conditions $x_{i,1}$, $y_{i,1}$, $v_{x,i,1}$ and $v_{y,i,1}$  we  need also to know $x_{i,2}$, $y_{i,2}$, $v_{x,i,2}$ and $v_{y,i,2}$ for the algorithm to start which can be  determined using the Euler method.

\section{Some Physical Applications}
\subsection{Dilute Gas and Maxwell Distribution}
A gas in thermal equilibrium is characterized by a temperature $T$. Molecular dynamics allows us to study how a dilute gas approaches equilibrium. The temperature of the gas can be computed using the molecular dynamics simulations as follows.
 According to the equipartition theorem the average thermal energy of each quadratic degree of freedom in a gas in thermal equilibrium is equal $k_BT/2$. In other words
\begin{eqnarray}
 \frac{1}{2}k_BT=\frac{1}{d}<\frac{1}{2}m\vec{v}^2>.
\end{eqnarray}
The average $<>$ can be understood in two different but equivalent ways. We can follow the motion of a single atom and take the time average of its kinetic energy. The same result can be obtained by taking the average of the kinetic energy over the different atoms. In this latter case we write
\begin{eqnarray}
 \frac{1}{2}k_BT=\frac{1}{dN}\sum_{i=1}^N\frac{1}{2}m\vec{v}_i^2.
\end{eqnarray}
Another way of measuring the temperature $T$ of a dilute gas is through a study of the distribution of atom velocities. A classical gas in thermal equilibrium obeys Maxwell distribution. The speed and velocity distributions in two dimensions are given respectively by
\begin{eqnarray}
P(v)=C\frac{v}{k_BT}e^{-\frac{mv^2}{2k_BT}}.
\end{eqnarray}
\begin{eqnarray}
P(v_x)=C_x\frac{1}{\sqrt{k_BT}}e^{-\frac{mv_x^2}{2k_BT}}~,~P(v_y)=C_y\frac{1}{\sqrt{k_BT}}e^{-\frac{mv_y^2}{2k_BT}}.
\end{eqnarray}
Recall that the probability per unit $v$ of finding an atom with speed $v$ is equal $P(v)$ whereas the probability per unit $v_{x,y}$ of finding an atom with velocity $v_{x,y}$ is equal $P(v_{x,y})$. The constants $C$ and $C_{x,y}$ are determined from the normalization conditions. There are peaks in the distributions $P(v)$ and $P(v_{x,y})$.  Clearly the temperature is related to the location of the peak which occurs in $P(v)$. This is given by

\begin{eqnarray}
k_BT=mv_{\rm peak}^2.
\end{eqnarray}

\subsection{The Melting Transition}
This is a very important subject which we will discuss at great length in the second lab problem of this chapter.


\section{Simulation $12$: Maxwell Distribution}
We consider the motion in two dimensions of $N$ argon atoms in an $L\times L$ box. The interaction potential $u$ between any two atoms in the gas separated by a distance $r$ is given by the Lennard-Jones potential.
The numerical algorithm we will use is Verlet algorithm. 
 

In this problem we will always take $L$ odd and $N$  a perfect square. The lattice spacing is defined by

\[a=\frac{L}{\sqrt{N}}.\]
Clearly there are $N$ cells of area $a\times a$. We choose $L$ and $N$ such that $a>2\sigma$. For simplicity we will use reduced units $\sigma=\epsilon=m=1$. In order to reduce edge effects we use periodic boundary conditions. In other words the box is effectively a torus and there are no edges. 
Thus the maximum separation in the $x$ direction between any two particles is only $L/2$ and similarly the  maximum separation in the $y$ direction between any two particles is only $L/2$. 

The initial positions of the atoms are fixed as follows. The atom $k=\sqrt{N}(i-1)+j$ will be placed at the center of the cell with corners $(i,j)$, $(i+1,j)$, $(i,j+1)$ and $(i+1,j+1)$. Next we perturb  in a random way these  initial positions by adding random numbers in the interval $[-a/4,+a/4]$ to the $x$ and $y$ coordinates of the atoms. The initial velocities can be chosen in random directions with a speed equal $v_0$ for all atoms.

\begin{itemize}
\item[$(1)$]Write a molecular dynamics code along the above lines. Take $L=15$, $N=25$, $\Delta t=0.02$, ${\rm Time}=500$ and $v_0=1$. As a first test verify that the total energy is conserved. Plot the trajectories of the atoms. What do you observe.
\item[$(2)$]As a second test we propose to measure the temperature by observing how the gas approaches equilibrium. Use the equipartition theorem 
\[k_BT=\frac{m}{2N}\sum_{i=1}^N(v_{i,x}^2+v_{i,y}^2).\]
Plot $T$ as a function of time. Take ${\rm Time}=1000-1500$. What is the temperature of the gas at equilibrium.
\item[$(3)$]Compute the speed distribution of the argon atoms by constructing an appropriate histogram as follows. We take the value ${\rm Time}=2000$. We consider the speeds of all particles at all times. There are ${\rm Time}\times N$  values of the speed in this sample. Construct the histogram for this sample by $1)$ finding the maximum and minimum, $2)$ dividing the interval into bins, $3)$ determining the number of times a given value of the speed falls in a bin and $(4)$ properly normalizing the distribution. Compare with the Mawell distribution

\[P_{\rm Maxwell}(v)=C\frac{v^2}{k_BT}~e^{-\frac{mv^2}{2k_BT}}.\]
Deduce the temperature from the peak of the distribution given by $k_BT=mv_{\rm peak}^2$.  Compare with the value of the temperature obtained from the equipartition theorem. What happens if we increase the initial speed.   
\end{itemize}
\section{Simulation $13$: Melting Transition}
We would like to study the melting transition. First we need to establish the correct conditions for a solid phase. Clearly the temperature must be sufficiently low and the density must be sufficiently high. To make the temperature as low as possible we will start with all particles at rest. In order to obatin maximum attraction between atoms we choose a low density of approximately one particle per unit reduced area. In particular we choose $N=16$ and $L=4$. 

\begin{itemize}
\item[$(1)$]Show that with these conditions you obtain a crystalline solid with a triangular lattice.
\item[$(2)$]In order to observe melting we must heat up the system. This can be achieved by increasing the kinetic energy of the atoms by hand. A convenient way of doing this is to rescale the current and previous positions of the atoms periodically (say every $1000$ steps) as follows
\begin{eqnarray}
&&{\rm hh}={\rm int}(n/1000)\nonumber\\
&&{\rm if}~({\rm hh}*1000.{\rm eq}.n)~{\rm then}\nonumber\\
&&x(i,n)=x(i,n+1)-R(x(i,n+1)-x(i,n))\nonumber\\
&&y(i,n)=y(i,n+1)-R(y(i,n+1)-y(i,n))\nonumber\\
&&{\rm endif}.\nonumber
\end{eqnarray}
This procedure will rescale the velocity by the amount $R$. We choose $R=1.5$. Verify that we will indeed reach the melting transition by means of this method. What happens to the energy and the temperature.
\end{itemize}

\chapter{Pseudo Random Numbers and Random Walks}
\section{Random Numbers}

A sequence of numbers $r_1$, $r_2$,... is called random if there are no correlations between the numbers. The sequence is called uniform if all numbers have an equal probability to occur. More precisely let the probability that a number $r_i$ in the sequence  occurs between $r$ and $r+dr$ be $P(r)dr$ where $P(r)$ is the probability distribution. A uniform distribution corresponds $P(r)={\rm constant}$. 

Most random number generators on computers generate uniform distributions between $0$ and $1$. These are sequences of pseudo random numbers since given $r_i$ and its preceding elements we can compute $r_{i+1}$. Therefore these sequences are not really random and correlations among the numbers of the sequence exist. True random numbers  can be found in tables of random numbers determined during say radioactive decay or other naturally occurring  random physical phenomena. 

\subsection{Linear Congruent or Power Residue Method}
In this method we generate a set of $k$ random numbers $r_1$,$r_2$,...,$r_k$ in the interval $[0,M-1]$ as follows. Given a random number $r_{i-1}$ we generate the next random number $r_i$ by the rule
\begin{eqnarray}
r_i=(ar_{i-1}+c)~{\rm mod}~M={\rm remainder}\bigg(\frac{ar_{i-1}+c}{M}\bigg).\label{random}
\end{eqnarray}
The notation $y=z~{\rm mod}~M$ means that we subtract $M$ from $z$ until $0{\leq}y{\leq}M-1$. The first random number $r_1$ is supplied by the user and it is called the seed. Also supplied are the multiplier $a$, the increment $c$ and the modulus $M$. The remainder is a built-in function in most computer languages. The largest possible integer number generated by the above rule is $M-1$. Thus the maximum possible period is $M$, i.e $k{\leq}M$. In general the period $k$ depends on $a$, $c$ and $M$. To get a uniform sequence in the interval $[0,1]$ we divide by $M-1$.

Let us take the following example $a=4$,$c=1$ and $M=9$ with seed $r_1=3$. We get a sequence of length $9$ given by
\begin{eqnarray}
3,4,8,6,7,2,0,1,5.
\end{eqnarray}
After the last number $5$ we get $3$ and therefore the sequence will repeat. In this case the period is $M=9$.

It is clear that we need to choose the parameters $a$, $c$ and $M$ and the seed $r_1$ with care so that we get the longest sequence of pseudo random numbers. The maximum possible period depends on the size of the computer word. A $32-$bit machine may use $M=2^{31}=2\times 10^{9}$. The numbers generated by (\ref{random}) are random only in the sense that they are evenly  distributed over their range. Equation (\ref{random}) is related to the logistic map which is known to exhibit chaotic behaviour. Although chaos is deterministic it looks random. In the same way although equation (\ref{random}) is deterministic the numbers generated by it look random. This is the reason why they are called pseudo random numbers.
\subsection{Statistical Tests of Randomness}

\paragraph{\it Period}: The first obvious test is to verify that the random number generator has a sufficiently long period for a given problem. We can use the random number generator to plot the position of a random walker. Clearly the plot will repeat itself when the period is reached. 

\paragraph{\it Uniformity}:  The $k$th moment of the random number distribution is 
\begin{eqnarray}
<x_i^{k}>=\frac{1}{N}\sum_{i=1}^{N}x_i^{k}.
\end{eqnarray}
Let $P(x)$ be the probability distribution of the random numbers. Then
\begin{eqnarray}
<x_i^{k}>=\int_0^1 dx ~x^{k}P(x)+O(\frac{1}{\sqrt{N}}).
\end{eqnarray}
For a uniform distribution $P(x)=1$ we must have
\begin{eqnarray}
<x_i^{k}>=\frac{1}{k+1}+O(\frac{1}{\sqrt{N}}).
\end{eqnarray}
In the words 
\begin{eqnarray}
\sqrt{N}\bigg(\frac{1}{N}\sum_{i=1}^{N}x_i^{k}-\frac{1}{k+1}\bigg)=O(1).
\end{eqnarray}
This is a test of uniformity as well as of randomness. To be more precise if $<x_i^k>$ is equal to $1/(k+1)$ then we can infer that the distribution is uniform  whereas if the deviation varies as $1/\sqrt{N}$ then we can infer that the distribution is random.

A direct test of uniformity is to divide the unit interval into $K$ equal subintevals (bins) and place each random number in one of these bins. For a uniform distribution we must obtain $N/K$ numbers in each bin where $N$ is the number of generated random numbers.

\paragraph{\it Chi-Square Statistic}: In the above test there will be statistical fluctuations about the ideal value $N/K$ for each bin. The question is whether or not these fluctuations are consistent with the laws of statistics.  The answer is based on the so-called chi-square statistic defined by
\begin{eqnarray}
{\chi}_{\rm m}^{2}=\sum_{i=1}^{K}\frac{(N_i-n_{\rm ideal})^{2}}{n_{\rm ideal}}.
\end{eqnarray}
In the above definition $N_i$ is the number of random numbers which fall into bin $i$ and $n_{\rm ideal}$ is the expected number of random numbers in each bin. 

The probability of finding any particular value ${\chi}^{2}$ which is less than ${\chi}_{\rm m}^{2}$ is found to be proportional to the incomplete gamma function $\gamma({\nu}/{2},{{\chi}_{\rm m}^{2}}/{2})$ where $\nu$ is the number of degrees of freedom given by $\nu=K-1$. We have 
\begin{eqnarray}
P(\chi^2\leq {\chi}_{\rm m}^{2})=\frac{\gamma({\nu}/{2},{{\chi}_{\rm m}^2}/{2})}{\Gamma({\nu}/{2})}\equiv P({\nu}/{2},{{\chi}_{\rm m}^2}/{2}).
\end{eqnarray}
The most likely value of  $\chi_m^2$, for some fixed number of degrees of freedom $\nu$, corresponds to the value  $P({\nu}/{2},{{\chi}_{\rm m}^2}/{2})=0.5$. In other words in half of the measurements (bin tests), for some fixed number of degrees of freedom $\nu$, the chi-square statistic predicts that we must find a value of  ${\chi}_{\rm m}^{2}$ smaller than the maximum. 

 
\paragraph{\it Randomness}: Let $r_1$, $r_2$,...,$r_N$ be a sequence of random numbers. A very effective test of randomness is to make a scatterplot of $(x_i=r_{2i},y_i=r_{2i+1})$ for many $i$. There must be no regularity in the plot otherwise the sequence is not random.

\paragraph{\it Short-Term Correlations}: Let us define the autocorrelation function
 \begin{eqnarray}
C(j)&=&\frac{<x_ix_{i+j}>-<x_i><x_{i+j}>}{<x_ix_{i}>-<x_i>^{2}}\nonumber\\
&=&\frac{<x_ix_{i+j}>-<x_i>^2}{<x_ix_{i}>-<x_i>^{2}}~,~j=1,2,...
\end{eqnarray}
In the above equation we have used the fact that $<x_{i+j}>=<x_i>$ for a large sample, i.e. the choice of the origin of the sequence is irrelevant in that case and 
 \begin{eqnarray}
<x_ix_{i+j}>=\frac{1}{N-j}\sum_{i=1}^{N-j}x_ix_{i+j}.
\end{eqnarray}
Again if $x_i$ and $x_{i+j}$ are independent random numbers which are distributed with the joint probability distribution $P(x_i,x_{i+j})$ then
\begin{eqnarray}
<x_ix_{i+j}>\simeq \int_0^1dx \int_0^1 dy xy P(x,y).
\end{eqnarray}
We have clearly assumed that $N$ is large. For a uniform distribution, viz $P(x,y)=1$ we get
\begin{eqnarray}
<x_ix_{i+j}>\simeq \frac{1}{4}.
\end{eqnarray}
For a random distrubution the deviation from this result is of order $1/\sqrt{N}$. Hence in the case that the random numbers are not correlated we have
\begin{eqnarray}
C(j)=0.
\end{eqnarray}
\section{Random Systems}
Both quantum and statistical physics deal with systems that are random or stochastic. These are non deterministic systems as opposed to classical systems. The dynamics of a deterministic system is given by a  unique solution to the equations of motion which describes the physics of the system at all times.

We take the case of the diffusion of fluid molecules. For example the motion of  dust particles in the atmosphere, the motion of perfume molecules in the air or the motion of milk molecules in a coffee. These  are all cases of a  Brownian motion. 

In the case of a drop of milk in a coffee the white mass of the drop of milk will slowly spread until the coffee  takes on a uniform brown color. At the molecular level each milk molecule collides with molecules in the coffee. Clearly it will change direction so frequently that its motion will appear random.  This trajectory can be described by a random walk. This is a system in which each milk molecule moves one step at a time in any direction with equal probability. 

The trajectory of a dust, perfume or milk molecule is not really random since it can in principle be computed by solving Newton's equations of motion for all  molecules which then allows us to know the evolution of the system in time. Although this is possible in principle it will not be feasible in practice. The  random walk is thus effectively an approximation. However the large number of molecules and collisions in the system makes the random walk a very good approximation. 

\subsection{Random Walks}
Let us consider a one dimensional random walk. It can take steps of lenght unity along a line. It begins at $s_0=0$ and the first step is chosen randomly to be either to the left or to right with equal probabilities. In other words there is a $50$ per cent chance that the walker moves to the point $s_1=+1$ and a $50$ per cent chance that it moves to the point $s_1=-1$.  Next the walker will again move either to the right or to the left  from the point $s_1$ to the point $s_2$ with equal probabilities. This process will be repeated $N$ times and we get the position of the walker $x_N$ as a function of the step number $N$. In the motion of a molecule in a solution the time between steps is a constant and hence the step number $N$ is proportional to time. Therefore $x_N$ is the position of the walker as a function of time. 

In general a one-dimensional random walker can move to the right with probability $p$ and to the left with probability $q=1-p$ with steps of equal lenght $a$. The direction of each step is independent of the previous one. The displacement or position of the walker after $N$ steps is
  
\begin{eqnarray}
x_N=\sum_{i=1}^Ns_i.
\end{eqnarray}
The walker for $p=q=1/2$ can be generated by flipping a coin $N$ times. The position is increased by $a$ for heads and decreased by $a$ for tails.

Averaging over many walks each consisting of $N$ steps we get
\begin{eqnarray}
<x_N>=\sum_{i=1} ^N<s_i>=N<s>.
\end{eqnarray}
In above we have used the fact that the average over every step is the same given by
\begin{eqnarray}
<s_i>=<s>=p(a)+q(-a)=(p-q)a.
\end{eqnarray}
For $p=q=1/2$ we get $<x_N>=0$. A better measure of the walk is given by

\begin{eqnarray}
x_N^2=\bigg(\sum_{i=1}^Ns_i\bigg)^2.
\end{eqnarray}
The mean square net displacement ${\Delta}x^2$ is defined by
\begin{eqnarray}
\Delta x^2=<(x_N-<x_N>)^2>=<x_N^2>-<x_N>^2.
\end{eqnarray}
We compute 
\begin{eqnarray}
\Delta x^2&=&\sum_{i=1}^N\sum_{j=1}^N<(s_i-<s>)(s_j-<s>)>\nonumber\\
&=&\sum_{i\neq j=1}^N<(s_i-<s>)(s_j-<s>)>+\sum_{i=1}^N<(s_i-<s>)^2>.
\end{eqnarray}
In the first term since $i\neq j$ we have $<(s_i-<s>)(s_j-<s>)>=<(s_i-<s>)><(s_j-<s>)>$. But $<(s_i-<s>)>=0$. Thus
\begin{eqnarray}
\Delta x^2&=&\sum_{i=1}^N<(s_i-<s>)^2>\nonumber\\
&=&N(<s_i^2>-<s>^2>)\nonumber\\
&=&N(a^2-(p-q)^2a^2)\nonumber\\
&=&4Npqa^2.
\end{eqnarray}
For $p=q=1/2$ and $a=1$ we get
\begin{eqnarray}
<x_N^2>&=&N.
\end{eqnarray}
The main point is that since $N$ is proportional to time we have $<x_N^2>\propto t$. This is an example of a diffusive behaviour.

\subsection{Diffusion Equation}
The random walk is successful in simulating many physical systems because it is related to the solutions of the diffusion equation. To see this we start from the probability $P(i,N)$ that the random walker is at site $s_i$ after $N$ steps. This is given by
\begin{eqnarray}
P(i,N)=\frac{1}{2}\bigg(P(i+1,N-1)+P(i-1,N-1)\bigg).
\end{eqnarray}
Let $\tau$ be the time between steps and $a$ the lattice spacing. Then $t=N\tau$ and $x=ia$. Also we define $P(x,t)=P(i,N)/a$. We get
\begin{eqnarray}
P(x,t)=\frac{1}{2}\bigg(P(x+a,t-\tau)+P(x-a,t-\tau)\bigg).
\end{eqnarray}
Let us rewrite this equation as
\begin{eqnarray}
\frac{1}{\tau}\bigg(P(x,t)-P(x,t-\tau)\bigg)=\frac{a^2}{2\tau}\bigg[P(x+a,t-\tau)-2P(x,t-\tau)+P(x-a,t-\tau)\bigg]\frac{1}{a^2}.\nonumber\\
\end{eqnarray}
In the limit $a\longrightarrow 0$, $\tau\longrightarrow 0$ with the ratio $D={a^2}/{2\tau}$ kept fixed we obtain the equation
\begin{eqnarray}
\frac{{\partial}P(x,t)}{{\partial}t}=D\frac{{\partial}^2P(x,t)}{{\partial}x^2}.\label{7.68}
\end{eqnarray}
This is the diffusion equation. Generalization to $3-$dimensions is
\begin{eqnarray}
\frac{{\partial}P(x,y,z,t)}{{\partial}t}=D{\nabla}^2P(x,y,z,t).
\end{eqnarray}
A particular solution of (\ref{7.68}) is given by
\begin{eqnarray}
P(x,t)=\frac{1}{\sigma}e^{-\frac{x^2}{2{\sigma}^2}}~,~\sigma=\sqrt{2Dt}.
\end{eqnarray}
In other words the spatial distribution of the diffusing molecules is always a gaussian with half-width $\sigma$ increasing with time as $\sqrt{t}$. 

The average of any function $f$ of $x$ is given by
\begin{eqnarray}
<f(x,t)>=\int f(x)P(x,t)dx.
\end{eqnarray}
Let us multiply both sides of (\ref{7.68}) by $f(x)$ and then integrate over $x$, viz
\begin{eqnarray}
\int f(x)\frac{{\partial}P(x,t)}{{\partial}t} dx=D\int f(x) \frac{{\partial}^2P(x,t)}{{\partial}x^2} dx.
\end{eqnarray}
Clearly
\begin{eqnarray}
\int f(x)\frac{{\partial}P(x,t)}{{\partial}t} dx=\int \frac{\partial}{\partial t}\big(f(x)P(x,t)\big) dx=\frac{d}{dt}\int f(x)P(x,t)dx=\frac{d}{dt}<f(x)>.
\end{eqnarray}
Thus
\begin{eqnarray}
\frac{d}{dt}<f(x)>&=&D\int f(x) \frac{{\partial}^2P(x,t)}{{\partial}x^2} dx\nonumber\\
&=&D\bigg(f(x)\frac{{\partial}P(x,t)}{{\partial}x}\bigg)|_{x=-\infty}^{x=+\infty}-D\int \frac{\partial f(x)}{\partial x} \frac{{\partial}P(x,t)}{{\partial}x} dx.
\end{eqnarray}
We have $P(x=\pm \infty,t)=0$ and also all spatial derivatives are zero at $x=\pm \infty$. We then get

\begin{eqnarray}
\frac{d}{dt}<f(x)>
&=&-D\int \frac{\partial f(x)}{\partial x} \frac{{\partial}P(x,t)}{{\partial}x} dx.
\end{eqnarray}
Let us choose $f(x)=x$. Then
\begin{eqnarray}
\frac{d}{dt}<x>
&=&-D\int \frac{{\partial}P(x,t)}{{\partial}x} dx=0.
\end{eqnarray}
In other words $<x>={\rm constant}$ and since $x=0$ at $t=0$ we must have ${\rm constant}=0$. Thus
\begin{eqnarray}
<x>=0.
\end{eqnarray}
Let us next choose $f(x)=x^2$. Then
\begin{eqnarray}
\frac{d}{dt}<x^2>
&=&-2D\int x \frac{{\partial}P(x,t)}{{\partial}x} dx\nonumber\\
&=&2D.
\end{eqnarray}
Hence
\begin{eqnarray}
<x^2>&=&2Dt.
\end{eqnarray}
This is the diffusive behaviour we have observed in the random walk problem.
\section{The Random Number Generators  {\rm RAN} $0,1,2$}
Linear congruential generators are of the form
\begin{eqnarray}
r_i=(ar_{i-1}+c)~{\rm mod}~M.
\end{eqnarray}
For $c>0$ the linear congruential generators are called mixed. They are denoted by ${\rm LCG}(a,c,M)$. The random numbers generated with ${\rm LCG}(a,c,M)$ are in the range $[0,M-1]$.  

For $c=0$ the linear congruential generators are called multiplicative. They are denoted by ${\rm MLCG}(a,M)$.  The random numbers generated with ${\rm MLCG}(a,M)$ are in the range $[1,M-1]$.  

In the case that $a$ is a primitive root modulo $M$ and $M$ is a prime the period of the generator is $M-1$. A number $a$ is a primitive root modulo $M$ means that for any integer $n$ such that ${\rm gcd}(n,M)=1$ there exists a $k$ such that $a^k=n~{\rm mod}~M$.

An example of ${\rm MLCG}$ is ${\rm RAN}0$ due to Park and Miller which is used extensively on IBM computers. In this case
\begin{eqnarray}
a=16807=7^5~,~M=2^{31}-1.
\end{eqnarray}
The period of this generator is not very long given by
\begin{eqnarray}
{\rm period}=2^{31}-2\simeq 2.15\times 10^9.
\end{eqnarray}
This generator can not be implemented directly in a high level language because of integer overflow. Indeed the product of $a$ and $M-1$ exceeds the maximum value for a $32-$bit integer. Assemply language implementation using $64-$bit product register is straightforward but not portable.

A better solution is given by Schrage's algorithm. This algorithm allows the multiplication of two $32-$bit integers without using any intermediate numbers which are larger than $32$ bits. To see how this works explicitly we factor $M$ as
\begin{eqnarray}
M=aq+r.
\end{eqnarray}
\begin{eqnarray}
r=M~{\rm mod}~a~,~q=[\frac{M}{r}].
\end{eqnarray}
In the above equation $[~]$ denotes integer part. Remark that
\begin{eqnarray}
r=M~{\rm mod}~a=M-[\frac{M}{a}]a.
\end{eqnarray}
Thus by definition $r<a$. We will also demand that $r<q$ and hence
\begin{eqnarray}
\frac{r}{qa}<<1.
\end{eqnarray}
We have also
\begin{eqnarray}
X_{i+1}=aX_i~{\rm mod}~M&=&aX_i-[\frac{aX_i}{M}]M\nonumber\\
&=&aX_i-[\frac{aX_i}{aq+r}]M.
\end{eqnarray}
We compute
\begin{eqnarray}
\frac{aX_i}{aq+r}=\frac{X_i}{q+\frac{r}{a}}&=&\frac{X_i}{q}\frac{1}{1+\frac{r}{qa}}\nonumber\\
&=&\frac{X_i}{q}(1-\frac{r}{qa})\nonumber\\
&=&\frac{X_i}{q}-\frac{X_i}{aq}\frac{r}{q}.
\end{eqnarray}
Clearly
\begin{eqnarray}
\frac{X_i}{aq}=\frac{X_i}{M-r}\simeq\frac{X_i}{M}<1.
\end{eqnarray}
Hence 
\begin{eqnarray}
[\frac{aX_i}{M}]
&=&[\frac{X_i}{q}],
\end{eqnarray}
if neglecting $\epsilon=(rX_i)/(aq^2)$ does not affect the integer part of $aX_i/M$ and 
\begin{eqnarray}
[\frac{aX_i}{M}]
&=&[\frac{X_i}{q}]-1,
\end{eqnarray}
if neglecting $\epsilon$ does affect the integer part of $aX_i/M$. Therefore we get
\begin{eqnarray}
X_{i+1}
&=&aX_i-[\frac{aX_i}{M}](aq+r)\nonumber\\
&=&a(X_i-[\frac{aX_i}{M}]q)-[\frac{aX_i}{M}]r\\
&=&a(X_i-[\frac{X_i}{q}]q)-[\frac{X_i}{q}]r\\
&=&a(X_i~{\rm mod}~q)-[\frac{X_i}{q}]r,
\end{eqnarray}
if 
\begin{eqnarray}
a(X_i~{\rm mod}~q)-[\frac{X_i}{q}]r\geq 0.
\end{eqnarray}
Also
\begin{eqnarray}
X_{i+1}
&=&aX_i-[\frac{aX_i}{M}](aq+r)\nonumber\\
&=&a(X_i-[\frac{aX_i}{M}]q)-[\frac{aX_i}{M}]r\\
&=&a(X_i-[\frac{X_i}{q}]q+q)-[\frac{X_i}{q}]r+r\\
&=&a(X_i~{\rm mod}~q)-[\frac{X_i}{q}]r+M,
\end{eqnarray}
if 
\begin{eqnarray}
a(X_i~{\rm mod}~q)-[\frac{X_i}{q}]r< 0.
\end{eqnarray}
The generator ${\rm RAN}$$0$ contains serial correlations. For example $D-$dimensional vectors $(x_1,...,x_D)$, $(x_{D+1},...,x_{2D})$,...which are obtained by successive calls of ${\rm RAN}$$0$ will lie on a small number of parallel $(D-1)-$dimensional hyperplanes. Roughly there will be $M^{1/D}$ such hyperplanes. In particular successive points $(x_i,x_{i+1})$ when binned into a $2-$dimensional plane for $i=1,...,N$ will result in a distribution which fails the $\chi^2$ test for $N\geq 10^7$ which is much less than the period $M-1$.

The ${\rm RAN}$$1$ is devised so that the correlations found in ${\rm RAN}$$0$ is removed using the Bays-Durham algorithm. The Bays-Durham algorithm shuffles the sequence to remove low-order
serial correlations. In other words it changes the order of the numbers so that the sequence 
is not dependent on order and a given number is not correlated with previous numbers. More precisely the $j$th random number is output not on the $j$th call but on a randomized later call which is on average the $j+32$th call on .

The ${\rm RAN}$$2$ is an improvement over ${\rm RAN}$$1$ and ${\rm RAN}$$0$  due to L'Ecuyer. It uses two sequences with different periods so as to obtain a new sequence with a larger period equal to the least common multiple of the two periods. In this algorithm we add the two sequences modulo the modulus $M$ of one of them. In order to avoid overflow we subtract rather than add and if the result is negative we add $M-1$ so as to wrap around into the inetrval $[0,M-1]$. L'Ecuyer uses the two sequences
\begin{eqnarray}
M_1=2147483563~,~a_1=40014~,~q_1=53668~,~r_1=12211.
\end{eqnarray}
\begin{eqnarray}
M_2=2147483399~,~a_2=40692~,~q_2=52774~,~r_2=3791.
\end{eqnarray}
The period is $2.3\times 10^{18}$. Let us also point out that  ${\rm RAN}$$2$ uses Bays-Durham algorithm in order to implement an additional shuffle.

We conclude this section by  discussing another generator based on the linear congruential method which is the famous random number generator ${\rm RAND}$ given by
\begin{eqnarray}
{\rm RAND}={\rm LCG}(69069,1,2^{32}).
\end{eqnarray}
The period of this generator is $2^{32}$ and lattice structure is present for higher dimensions $D\geq 6$. 

\section{Simulation $14$: Random Numbers}
\paragraph{Part I}
We consider a linear congruential pseudo-random number generator given by
\begin{eqnarray}
r_{i+1}={\rm remainder}\bigg(\frac{ar_i+c}{M}\bigg).\nonumber
\end{eqnarray}
We take the values
\begin{eqnarray}
&&a=899,c=0,M=32768,r_1=12~~"{\rm good}"\nonumber\\
&&a=57,c=1,M=256,r_1=10~,~"{\rm bad}".\nonumber
\end{eqnarray}
The function ``remainder'' is implemented in Fortran by
\begin{eqnarray}
{\rm remainder}~\frac{a}{b}={\rm mod}(a,b).\nonumber
\end{eqnarray}
\begin{itemize}
\item[$(1)$]Compute the sequence of the random numbers $r_i$ obtained using the above parameters. Plot $r_i$ as a function of $i$. Construct a scatterplot $(x_i=r_{2i},y_i=r_{2i+1})$.
\item[$(2)$]Compute the average of the random numbers. What do you observe.
\item[$(3)$]Let $N$ be the number of generated random numbers. Compute the correlation functions defined by
\begin{eqnarray}
{\rm sum}_1(k)=\frac{1}{N-k}\sum_{i=1}^{N-k}x_ix_{i+k}.\nonumber
\end{eqnarray}
\begin{eqnarray}
{\rm sum}_2=\frac{{\rm sum}_1(k)-<x_i>^2}{{\rm sum}_1(0)-<x_i>^2}.\nonumber
\end{eqnarray}
What is the behavior of these functions as a function of $k$.
\item[$(4)$]Compute the period of the above generators.
\end{itemize}
\paragraph{Part II}
We take $N$ random numbers in the interval $[0,1]$ which we divide into $K$ bins of length $\delta=1/K$. Let $N_i$ be the number of random numbers which fall in the $i$th bin. For a uniform sequence of random numbers  the number of random numbers in each bin is $n_{\rm ideal}={N}/{K}$.
\begin{itemize}
\item[$(1)$]Verify this result for the generator ``rand'' found in the standard Fortran library with seed  given by ${\rm seed}=32768$. We take $K=10$ and $N=1000$. Plot $N_i$ as a function of the position $x_i$ of the $i$th bin.
\item[$(2)$]The number of degrees of freedom is $\nu=K-1$. The most probable value of the chi-square statistics $\chi^2$ is $\nu$. Verify this result for a total number of bin tests equal $L=1000$ and $K=11$. Each time calculate the number of times $L_i$ in the $L=1000$ bin tests we get a specific value of $\chi^2$. Plot $L_i$ as a function of $\chi^2$. What do you observe. 
\end{itemize}

\section{Simulation $15$: Random Walks}
\paragraph{Part I}
We consider the motion of a random walker in one dimension. The walker can move with a step $s_i=a$ to the right with a probability $p$ or with a step $s_i=-a$ to the left with a probability $q=1-p$. After $N$ steps the position of the walker is $x_N=\sum_{i}s_i$. We take
\begin{eqnarray}
p=q=\frac{1}{2}~,~a=1.\nonumber
\end{eqnarray}
In order to simulate the motion of a random walker we need a generator of random numbers. In this problem we work with the generator ``rand'' found in the standard Fortran library. We call this generator as follows
 \begin{eqnarray}
&&{\rm call}~{\rm srand}({\rm seed})\nonumber\\
&&{\rm rand}()\nonumber
\end{eqnarray}
The motion of the random walker is implemented with the code
\begin{eqnarray}
&&{\rm if}~({\rm rand}()<p)~{\rm then}\nonumber\\
&&x_N=x_N+a\nonumber\\
&&{\rm else}\nonumber\\
&&x_N=x_N-a\nonumber\\
&&{\rm endif}.\nonumber
\end{eqnarray}
\begin{itemize}
\item[$(1)$]Compute the positions $x_i$ of three different random walkers as  functions of the step number $i$. We take $i=1,100$. Plot the three trajectories.
\item[$(2)$]We consider now the motion of $K=500$ random walkers. Compute the averages
\begin{eqnarray}
&&<x_N>=\frac{1}{K}\sum_{i=1}^Kx_N^{(i)}~,~<x_N^2>=\frac{1}{K}\sum_{i=1}^K(x_N^{(i)})^2.\nonumber
\end{eqnarray}
In the above equations $x_N^{(i)}$ is the position of the $i$th random walker after $N$ steps. Study the behavior of these averages as a function of $N$. Compare with the theoretical predictions.
\end{itemize}
\paragraph{Part II (optional)}
We consider next a random walker in two dimensions on an infinite lattice of points. From any point $(i,j)$ on the lattice the walker can reach one of the $4$ possible nearest neighbor sites $(i+1,j)$, $(i-1,j)$, $(i,j+1)$ and $(i,j-1)$ with probabilities $p_x$, $q_x$, $p_y$ and $q_y$ respectively such that $p_x+q_x+p_y+q_y=1$. For simplicity we will assume that $p_x=q_x=p_y=q_y=0.25$. 
\begin{itemize}
\item[$(1)$]Compute the averages $<\vec{r}_N>$ and $<\vec{r}^2_N>$ as function of the number of steps $N$ for a collection of $L=500$ two dimensional random walkers. We consider the values $N=10,...,1000$. 
\end{itemize}

\chapter{Monte Carlo Integration}
\section{Numerical Integration}
\subsection{Rectangular Approximation Revisted}
As usual let us start with something simple. The approximation of one-dimensional integrals by means of the rectangular approximation. This is a topic we have already discussed before.  

Let us then begin by recalling how the rectangular approximation of one dimensional integrals works. We consider the integral

\begin{eqnarray}
F=\int_a^b f(x)dx.
\end{eqnarray}
We discretize the $x-$interval so that we end up with $N$ equal small intervals of lenght $\Delta x$, viz
\begin{eqnarray}
x_n=x_0+n{\Delta}x~,~\Delta x=\frac{b-a}{N}
\end{eqnarray}
Clearly $x_0=a$ and $x_N=b$.  Riemann definition of the integral is given by the following limit
\begin{eqnarray}
F={\rm lim}~\Delta x\sum_{n=0}^{N-1} f(x_n)~,~\Delta x\longrightarrow 0~,~N\longrightarrow \infty ~,~b-a={\rm fixed}.
\end{eqnarray}
The first approximation which can be made is to simply drop the limit. We get the so-called rectangular approximation given by
\begin{eqnarray}
F_N=\Delta x\sum_{n=0}^{N-1} f(x_n).\label{rectangular}
\end{eqnarray}
The error can be computed as follows. We start with the Taylor expansion
\begin{eqnarray}
f(x)=f(x_n)+(x-x_n)f^{(1)}(x_n)+\frac{1}{2!}(x-x_n)^2f^{(2)}(x_n)+...
\end{eqnarray}
Thus
\begin{eqnarray}
\int_{x_n}^{x_{n+1}}dx~f(x)=f(x_n)\Delta x+\frac{1}{2!}f^{(1)}(x_n)(\Delta x)^2+\frac{1}{3!}f^{(2)}(x_n)(\Delta x)^3+...
\end{eqnarray}

The error in the interval $[x_n,x_{n+1}]$  is
\begin{eqnarray}
\int_{x_n}^{x_{n+1}}dx~f(x)-f(x_n)\Delta x=\frac{1}{2!}f^{(1)}(x_n)(\Delta x)^2+\frac{1}{3!}f^{(2)}(x_n)(\Delta x)^3+...
\end{eqnarray}
This is of order $1/N^2$. But we have $N$ subintervals. Thus the total error is of order $1/N$.

\subsection{Midpoint Approximation of Multidimensional Integrals}
Let us start with the two dimensional integral
\begin{eqnarray}
F=\int_R dx~dy~ f(x,y).
\end{eqnarray}
$R$ is the domain of integration. In order to give the midpoint approximation of this integral we imagine a rectangle of sides $x_b-x_a$ and $y_b-y_a$ which encloses the region $R$ and we divide it into squares of lenght $h$. The points in the $x/y$ direction are
\begin{eqnarray}
x_i=x_a+(i-\frac{1}{2})h~,~i=1,...,n_x.
\end{eqnarray}
\begin{eqnarray}
y_i=y_a+(i-\frac{1}{2})h~,~i=1,...,n_y.
\end{eqnarray}
The number of points in the  $x/y$ direction are
\begin{eqnarray}
n_x=\frac{x_b-x_a}{h}~,~n_y=\frac{y_b-y_a}{h}.
\end{eqnarray}
The number of cells is therefore
\begin{eqnarray}
n=n_xn_y=\frac{(x_b-x_a)(y_b-y_a)}{h^2}.
\end{eqnarray}
The integral is then approximated by
\begin{eqnarray}
F=h^2\sum_{i=1}^{n_x}\sum_{j=1}^{n_y}f(x_i,y_j)H(x_i,y_j).
\end{eqnarray}
The Heaviside function is defined by
\begin{eqnarray}
H(x_i,y_j)=1~{\rm if}~(x_i,y_j)\in R~{\rm otherwise}~H(x_i,y_j)=0.
\end{eqnarray}
The generalization to many dimensions is straightforward. We get
\begin{eqnarray}
F=h^d\sum_{i_1=1}^{n_1}...\sum_{i_d=1}^{n_d}f(x_1^{i_1},...,x_d^{i_d})H(x_1^{i_1},...,x_d^{i_d}).
\end{eqnarray}
The meaning of the different symbols is obvious.

The midpoint approximation is an improvement over the rectangular approximation. To see this let us consider a one dimensional integral
\begin{eqnarray}
F=\int_R dx~ f(x).
\end{eqnarray}
The midpoint approximation reads in this case as follows
\begin{eqnarray}
F=h\sum_{i=1}^{n_x}f(x_i)H(x_i)=h\sum_{i=1}^{n_x}f(x_i).
\end{eqnarray}
Let us say that we have $n_x$ intervals $[x_i,x_{i+1}]$ with $x_0=a$ and $x_i=x_a+(i-0.5)h$, $i=1,...,n_x-1$. The term $hf(x_{i+1})$ is associated with the interval  $[x_i,x_{i+1}]$. It is clear that we can write this approximation as
 \begin{eqnarray}
F=h\sum_{i=0}^{n_x-1}f(\frac{x_i+x_{i+1}}{2})~,~x_i=x_a+ih.
\end{eqnarray}
The error in the interval $[x_i,x_{i+1}]$ is given by
\begin{eqnarray}
\int_{x_i}^{x_{i+1}}f(x)~dx -f(\frac{x_i+x_{i+1}}{2})\Delta x=\frac{1}{24}f^{''}(x_i)(\Delta x)^3+...
\end{eqnarray}
The total error is thereore $1/n_x^2$ as opposed to the $1/n_x$ of the rectangular approximation. 

Let us do this in two dimensions. We write the error as
\begin{eqnarray}
\int_{x_i}^{x_{i+1}}\int_{y_j}^{y_{j+1}}f(x,y)~dx~dy -f(\frac{x_i+x_{i+1}}{2},\frac{y_j+y_{j+1}}{2})\Delta x\Delta y
\end{eqnarray}
As usual we use Taylor series in the form

\begin{eqnarray}
f(x,y)&=&f(x_i,y_j)+f^{'}_x(x_i,y_j)(x-x_i)+f^{'}_y(x_i,y_j)(y-y_j)+\frac{1}{2}f^{''}_x(x_i,y_j)(x-x_i)^2\nonumber\\
&+&\frac{1}{2}f^{''}_y(x_i,y_j)(y-y_j)^2+f^{''}_{xy}(x_i,y_j)(x-x_i)(y-y_j)+...
\end{eqnarray}
We find
\begin{eqnarray}
\int_{x_i}^{x_{i+1}}\int_{y_j}^{y_{j+1}}f(x,y)~dx~dy -f(\frac{x_i+x_{i+1}}{2},\frac{y_j+y_{j+1}}{2})\Delta x\Delta y&=&\frac{1}{24}f^{''}_x(x_i,y_j)(\Delta x)^3\Delta y+\frac{1}{24}f^{''}_y(x_i,y_j)\Delta x (\Delta y)^3\nonumber\\
&+&...
\end{eqnarray}
Since $\Delta x=\Delta y=h$. The individual error is proportional to $h^4$. The total error is $nh^4$ where $n=n_xn_y$ . Since $n$ is proportional to $1/h^2$, the total error in dimension two is proportional to $h^2$ or equivalently to $1/n$. As we have already seen the same method led to an error proportional to $1/n^2$ in dimension one. Thus as we increase the number of dimensions the error becomes worse. If in one dimension the error behaves as $1/n^a$ then in dimension $d$ it will behave as $1/n^{\frac{a}{d}}$. In other words classical numerical integration methods become impractical at sufficiently higher dimensions (which is the case of quantum mechanics and statistical mechanics).

\subsection{Spheres and Balls in $d$ Dimensions}
The volume of a ball of radius $R$ in $d$ dimensions is given by
\begin{eqnarray}
V_d&=&\int_{x_1^2+...+x_d^2\leq R^2}dx_1...dx_d\nonumber\\
&=&\int_{x_1^2+...+x_d^2\leq R^2}r^{d-1}~dr~d{\Omega}_{d-1}\nonumber\\
&=&\frac{R^d}{d}\int d{\Omega}_{d-1}\nonumber\\
&=&\frac{R^d}{d}\frac{2{\pi}^{\frac{d}{2}}}{\Gamma(\frac{d}{2})}.
\end{eqnarray}
The surface of a sphere of radius $R$ in $d$ dimensions is similarly given by
\begin{eqnarray}
S_{d-1}&=&\int_{x_1^2+...+x_d^2= R^2}dx_1...dx_d\nonumber\\
&=&R^{d-1}\frac{2{\pi}^{\frac{d}{2}}}{\Gamma(\frac{d}{2})}.
\end{eqnarray}
Here are some properties of the gamma function 
\begin{eqnarray}
\Gamma(1)=1~,~\Gamma(\frac{1}{2})=\sqrt{\pi}~,~\Gamma(n+1)=n\Gamma(n).
\end{eqnarray}
In order to compute  numerically the volume of the ball in any dimension $d$ we need a recursion formula which relates the volume of the ball in $d$ dimensions to the volume of the ball in $d-1$ dimensions. The derivation goes as follows

\begin{eqnarray}
V_d&=&\int_{-R}^{+R} dx_d~\int_{x_1^2+...+x_{d-1}^2\leq R^2-x_d^2}dx_1...dx_{d-1}\nonumber\\
&=&\int_{-R}^{+R} dx_d~\int_{0}^{\sqrt{R^2-x_d^2}}~r^{d-2}~dr~\int d{\Omega}_{d-2}\nonumber\\
&=&\frac{V_{d-1}}{R^{d-1}}\int_{-R}^{+R}dx_d~ (R^2-x_d^2)^{\frac{d-1}{2}}.
\end{eqnarray}
At each dimension $d$ we are thus required to compute only the remaining integral over $x_d$ using, for instance, the midpoint approximation while the volume $V_{d-1}$ is determined in the previous recursion step. The starting point of the recursion process, for example the volume in $d=2$, can be determined also using the midpoint approximation. As we will see in the lab problems this numerical calculation is very demanding with significant errors compared with the Monte Carlo method.

\section{Monte Carlo Integration: Simple Sampling}
Let us start with the one dimensional integral
\begin{eqnarray}
F=\int_a^b dx~ f(x).
\end{eqnarray}
A Monte Carlo method is any procedure which uses (pseudo) random numbers to compute or estimate the above integral. In the following we will describe two very simple Monte Carlo methods based on simple sampling which give an approximate value for this integral. As we progress we will be able to give more sophisticated Monte Carlo methods. First we start with the sampling (hit or miss) method then we go on to the  sample mean  method.
\subsection{Sampling (Hit or Miss) Method}
This method consists of the following three main steps:
\begin{itemize}
\item{} We imagine a rectangle of width $b-a$ and height $h$ such that $h$ is greater than the maximum value of $f(x)$, i.e the function is within the boundaries of the rectangle. 

\item{} To estimate the value $F$ of the integral we choose $n$ pairs of uniform random numbers $(x_i,y_i)$ where  $a\leq x_i\leq b$ and $0\leq y_i\leq h$.  

\item{}Then we evaluate the function $f$ at the points $x_{i}$. Let $n_{\rm in}$ be the number of random points  $(x_i,y_i)$ such that $y_i\leq f(x_i)$. The value $F$ of the integral is given by
\begin{eqnarray}
F=A\frac{n_{\rm in}}{n}~,~A=h(b-a).
\end{eqnarray} 
\end{itemize} 
\subsection{Sample Mean  Method}
 We start from the mean-value theorem of calculus, viz
\begin{eqnarray}
F=\int_a^b dx~ f(x)=(b-a)<f>.
\end{eqnarray}
$<f>$ is the average value of the function $f(x)$ in the range $a\leq x\leq b$. The sample mean method estimates the average $<f>$ as follows:
\begin{itemize}
\item{} We choose $n$ random points $x_i$ from the interval $[a,b]$ which are distributed uniformly.
\item{} We compute the values of the function $f(x)$ at these point.
\item{} We take their average. In other words
\begin{eqnarray}
F=(b-a)\frac{1}{n}\sum_{i=1}^nf(x_i).
\end{eqnarray}
\end{itemize}
This is formally the same as the rectangular approximation. The only difference is that here the points $x_i$ are chosen randomly from the interval $[a,b]$ whereas the points in the rectangular approximation are chosen with equal spacing. For lower dimensional integrals the rectangular approximation is more accurate whereas for higher dimensional integrals the sample mean method becomes more accurate.

\subsection{Sample Mean  Method in  Higher Dimensions}
We start with the two dimensional integral
\begin{eqnarray}
F=\int_R dx~dy~ f(x,y).
\end{eqnarray}
Again we consider a rectangle of sides $y_b-y_a$ and $x_b-x_a$ which encloses the region $R$. The Monte carlo sample mean method yields the approximation

\begin{eqnarray}
F=A\frac{1}{n}\sum_{i=1}^n f(x_i,y_i)H(x_i,y_i).
\end{eqnarray}
The points $x_i$ are random and uniformly distributed in the interval $[x_a,x_b]$ whereas the points $y_i$ are random and uniformly distributed in the interval $[y_a,y_b]$. $A$ is the areas of the rectangle, i.e $A=(x_b-x_a)(y_b-y_a)$. The Heaviside function is defined by
\begin{eqnarray}
H(x_i,y_i)=1~{\rm if}~(x_i,y_i)\in R~{\rm otherwise}~H(x_i,y_i)=0.
\end{eqnarray}
Generalization to higher dimensions is obvious. For example in three dimensions we would have
\begin{eqnarray}
F=\int_R dx~dy~dz~ f(x,y,z)\longrightarrow F=V\frac{1}{n}\sum_{i=1}^n f(x_i,y_i,z_i)H(x_i,y_i,z_i).
\end{eqnarray}
$V$ is the volume of the parallelepiped which encloses the three dimensional region $R$.
\section{The Central Limit Theorem}
Let $p(x)$ be a probability distribution function. We generate (or measure)  $n$ values $x_i$ of a certain variable $x$ according to the probability distribution function $p(x)$. The average $y_1=<x_i>$ is given by

\begin{eqnarray}
y_1=<x_i>=\frac{1}{n}\sum_{i=1}^nx_ip(x_i).
\end{eqnarray}
We repeat this measurement $N$ times thus obtaining $N$ averages $y_1$, $y_2$,...,$y_N$. The mean $z$ of the averages $y_i$ is
\begin{eqnarray}
z=\frac{1}{N}\sum_{i=1}^N y_i.
\end{eqnarray}
The question we want to answer is: what is the probability distribution function of $z$. 

Clearly the probability of obtaining a particular value $z$ is the product of the probabilities of obtaining the individual averages $y_i$ (which are assumed to be independent) with the constraint that the average of $y_i$ is $z$. 

Let $\tilde{p}(y)$ be the probability distribution function of the average $y$ and let $P(z)$ be the probability distribution of the average $z$ of the averages. We can then write $P(z)$ as 
\begin{eqnarray}
P(z)=\int dy_1...\int dy_N~\tilde{p}(y_1)...\tilde{p}(y_N)\delta(z-\frac{y_1+...+y_N}{N}).
\end{eqnarray}
The delta function expresses the constraint that $z$ is the average of $y_i$. The delta function can be written as
\begin{eqnarray}
\delta(z-\frac{y_1+...+y_N}{N})=\frac{1}{2\pi}\int dq e^{iq(z-\frac{y_1+...+y_N}{N})}.
\end{eqnarray} 
Let $\mu$ be the actual average of $y_i$, i.e.
\begin{eqnarray}
\mu=<y_i>=\int dy \tilde{p}(y) y.
\end{eqnarray} 
We write
\begin{eqnarray}
P(z)&=&\frac{1}{2\pi}\int dq e^{iq(z-\mu)}\int dy_1~\tilde{p}(y_1)e^{\frac{iq}{N}(\mu-y_1)}...\int dy_N~\tilde{p}(y_N)e^{\frac{iq}{N}(\mu-y_N)}\nonumber\\
&=&\frac{1}{2\pi}\int dq e^{iq(z-\mu)}\bigg[\int dy~\tilde{p}(y)e^{\frac{iq}{N}(\mu-y)}\bigg]^N.
\end{eqnarray}
But
\begin{eqnarray}
\int dy~\tilde{p}(y)e^{\frac{iq}{N}(\mu-y)}&=&\int dy~\tilde{p}(y)\bigg[1+\frac{iq}{N}(\mu-y)-\frac{q^2(\mu-y)^2}{2N^2}+...\bigg]\nonumber\\
&=&1-\frac{q^2\sigma^2}{2N^2}+...
\end{eqnarray}
We have used 
\begin{eqnarray}
\int dy~\tilde{p}(y)(\mu-y)^2=<y^2>-<y>^2=\sigma^2.
\end{eqnarray}
Hence
\begin{eqnarray}
P(z)
&=&\frac{1}{2\pi}\int dq e^{iq(z-\mu)}e^{-\frac{q^2\sigma^2}{2N}}\nonumber\\
&=&\frac{1}{2\pi}e^{-\frac{N}{2\sigma^2}(z-\mu)^2}\int dq e^{-\frac{\sigma^2}{2N}(q-\frac{iN}{\sigma}(z-\mu))^2}\nonumber\\
&=&\frac{1}{\sqrt{2\pi}}\frac{e^{-\frac{(z-\mu)^2}{2\sigma_N^2}}}{\sigma_N}.
\end{eqnarray}
\begin{eqnarray}
\sigma_N=\frac{\sigma}{\sqrt{N}}.
\end{eqnarray}
This is the normal distribution. Clearly the result does not depend on the original probability distribution functions $p(x)$ and $\tilde{p}(y)$.

The average $z$ of $N$ random numbers $y_i$ corresponding to a probability distribution function $\tilde{p}(y)$ is distributed according to the normal probability distribution function with average equal to the average value of $\tilde{p}(y)$ and variance equal to the variance of $\tilde{p}(y)$ divided by $\sqrt{N}$.
\section{Monte Carlo Errors and Standard Deviation}
In any Monte Carlo approximation method the error goes as $1/\sqrt{N}$ where $N$ is the number of samples. This behaviour is independent of the integrand and is independent of the number of dimensions. In contrast if the error in a classical numerical approximation method goes as $1/N^a$ in one dimension (where $N$ is now the number of intervals) then the error in the same approximation method will go as $1/N^{\frac{a}{d}}$ in $d$ dimensions. Thus as we increase the number of dimensions the error becomes worse.  In other words classical numerical integration methods become impractical at sufficiently higher dimensions.
 This is the fundamental appeal of Monte Carlo methods in physics (quantum mechanics and statistical mechanics) where we usually and so often encounter integrals of infinite dimensionality.

Let us again consider for simplicity the one dimensional integral as an example. We take
\begin{eqnarray}
F=\int_a^b dx~ f(x).
\end{eqnarray}
The Monte Carlo sample mean method gives the approximation
\begin{eqnarray}
F_N=(b-a)<f>~,~<f>=\frac{1}{N}\sum_{i=1}^Nf(x_i).
\end{eqnarray}
The error is by definition given by
\begin{eqnarray}
\Delta=F-F_N.
\end{eqnarray}
However in general we do not know the exact result $F$. The best we can do is to calculate the probability that the approximate result $F_N$ is within a certain range centered around the exact result $F$. 

The starting point is the central limit theorem. This states that the average $z$ of $N$ random numbers $y_{\alpha}$ corresponding to a probability distribution function $\tilde{p}(y)$ is distributed according to the normal probability distribution function.
Here the variable $y$ is  (we assume for simplicity that $b-a=1$)
\begin{eqnarray}
y=\frac{1}{N}\sum_{i=1}^Nf(x_{i}).
\end{eqnarray}
We make $M$ measurements  $y_{\alpha}$  of $y$. We write
\begin{eqnarray}
y_{\alpha}=\frac{1}{N}\sum_{i=1}^Nf(x_{i,\alpha}).
\end{eqnarray}
The mean $z$ of the averages is given by 
\begin{eqnarray}
z=\frac{1}{M}\sum_{\alpha=1}^My_{\alpha}.
\end{eqnarray}
According to the central limit theorem the mean $z$ is distributed according to the normal probability distribution function with average equal to the average value $<y>$ of $y_{\alpha}$ and variance equal to the variance of $y_{\alpha}$ divided by $\sqrt{M}$, viz

\begin{eqnarray}
    \sqrt{\frac{M}{2\pi\tilde{\sigma}_M^2}} \exp\!\left(-M\frac{(z-<y>)^2}{2\tilde{\sigma}_M^2} \right). 
\end{eqnarray}
 The $\tilde{\sigma}_M$ is the standard deviation of the mean given by the square root of the variance
\begin{eqnarray}
\tilde{\sigma}_M^2=\frac{1}{M-1}\sum_{\alpha=1}^M(y_{\alpha}-<y>)^2.
\end{eqnarray}
The use of $M-1$ instead of $M$ is known as Bessel's correction. The reason for this correction is 
  the fact that the computation of the mean $<y>$ reduces the number of independent data points $y_{\alpha}$ by one. For very large $M$ we can replace $\tilde{\sigma}_M$ with  ${\sigma}_M$ defined by
\begin{eqnarray}
\tilde{\sigma}_M^2\sim {\sigma}_M^2=\frac{1}{M}\sum_{\alpha =1}^M(y_{\alpha}-<y>)^2=<y^2>-<y>^2.
\end{eqnarray}
The standard deviation of the sample (one single measurement with $N$ data points) is given by the square root of the variance
\begin{eqnarray}
\tilde{\sigma}^2= \frac{1}{N-1}\sum_{i=1}^N(f(x_i)-<f>)^2.
\end{eqnarray}
Again since $N$ is large we can replace $\tilde{\sigma}$ with ${\sigma}$ defined by
\begin{eqnarray}
{\sigma}^2= \frac{1}{N}\sum_{i=1}^N(f(x_i)-<f>)^2=<f^2>-<f>^2.\label{34}
\end{eqnarray}
\begin{eqnarray}
<f>=\frac{1}{N}\sum_{i=1}^Nf(x_i)~,~<f^2>=\frac{1}{N}\sum_{i=1}^Nf(x_i)^2.\label{35}
\end{eqnarray}
The standard deviation of the mean $\tilde{\sigma}_M\sim {\sigma}_M$ is given in terms of the standard deviation of the sample   $\tilde{\sigma}\sim {\sigma}$ by the equation
\begin{eqnarray}
{\sigma}_M=\frac{{\sigma}}{\sqrt{N}}.
\end{eqnarray}
The proof goes as follows. We generalize equations (\ref{34}) and (\ref{35}) to the case of $M$ measurements each with $N$ samples. The total number of samples is $MN$. We have
\begin{eqnarray}
{\sigma}^2= \frac{1}{NM}\sum_{\alpha=1}^M\sum_{i=1}^N(f(x_{i,\alpha})-<f>)^2=<f^2>-<f>^2.
\end{eqnarray}
\begin{eqnarray}
<f>=\frac{1}{NM}\sum_{\alpha=1}^M\sum_{i=1}^Nf(x_{i,\alpha})~,~<f^2>=\frac{1}{NM}\sum_{\alpha=1}^M\sum_{i=1}^Nf(x_{i,\alpha})^2.
\end{eqnarray}
The standard deviation of the mean $\tilde{\sigma}_M\sim {\sigma}_M$ is given by
\begin{eqnarray}
{\sigma}_M^2&=&\frac{1}{M}\sum_{\alpha =1}^M(y_{\alpha}-<y>)^2\nonumber\\
&=&\frac{1}{M}\sum_{\alpha =1}^M\bigg(\frac{1}{N}\sum_{i=1}^Nf(x_{i,\alpha})-<f>\bigg)^2\nonumber\\
&=&\frac{1}{N^2M}\sum_{\alpha =1}^M\sum_{i=1}^N\sum_{j=1}^N\bigg(f(x_{i,\alpha})-<f>\bigg)\bigg(f(x_{i,\alpha})-<f>\bigg).
\end{eqnarray}
In above we have used the fact that $<y>=<f>$. For every set $\alpha$ the sum over $i$ and $j$ splits into two pieces. The first is the sum over the diagonal elements with $i=j$ and the second is the sum over the off diagonal elements with $i\neq j$. Clearly $f(x_{i,\alpha})-<f>$ and $f(x_{j,\alpha})-<f>$ are on the average equally positive and negative and hence for large numbers $M$ and $N$ the off diagonal terms will cancel and we end up with

 \begin{eqnarray}
{\sigma}_M^2
&=&\frac{1}{N^2M}\sum_{\alpha =1}^M\sum_{i=1}^N\bigg(f(x_{i,\alpha})-<f>\bigg)^2\nonumber\\
&=&\frac{{\sigma}^2}{N}.
\end{eqnarray}
The standard deviation of the mean ${\sigma}_M$ can therefore be interpreted as the probable error in the original $N$ measurements since if we make $M$ sets of measurements each with $N$ samples the standard deviation of the mean  ${\sigma}_M$ will estimate how much an average over $N$ measurements will deviate from the exact mean. 

This means in particular that the original measurement $F_N$ of the integral $F$ has a $68$ per cent chance of being within one standard deviation  ${\sigma}_M$ of the true mean and a $95$ per cent chance of being within $2{\sigma}_M$ and a $99.7$ per cent chance of being within $3{\sigma}_M$. In general the proportion of data values within $\kappa {\sigma}_M$ standard deviations of the true mean is defined by the error function

\begin{eqnarray}
   \int_{<y>-\kappa {\sigma}_M}^{<y>+\kappa {\sigma}_M} \frac{1}{\sqrt{2\pi{\sigma}_M^2}} \exp\!\left(-\frac{(z-<y>)^2}{2{\sigma}_M^2} \right)~dz=\frac{2}{\sqrt{\pi}}\int_{0}^{\frac{\kappa}{\sqrt{2}}} \exp\!\left(-x^2 \right)~dx={\rm erf}(\frac{\kappa}{\sqrt{2}}). \nonumber\\
\end{eqnarray}

\section{Nonuniform Probability Distributions}
\subsection{The Inverse Transform Method}
We consider two discrete events $1$ and $2$ which occur with probabilities $p_1$ and $p_2$ respectively such that $p_1+p_2=1$. The question is how  can we choose the two events with the correct probabilities using only a uniform probability distribution. The answer is as follows. Let $r$ be a uniform random number between $0$ and $1$. We choose the event $1$ if $r<p_1$ else we choose the event $2$.

Let us now consider three discrete events $1$, $2$ and $3$ with probabilities $p_1$, $p_2$ and $p_3$ respectively such that $p_1+p_2+p_3=1$. Again we choose a random number $r$ between $0$ and $1$. If $r<p_1$ then we choose event $1$, if $p_1<r<p_1+p_2$ we choose event $2$ else we choose event $3$.

We consider now $n$ discrete events with probabilities $p_i$ such that $\sum_{i=1}^np_i=1$. Again we choose a random number $r$ between $0$ and $1$. We choose the event $i$ if the random number $r$ satisfies the inequality
\begin{eqnarray}
\sum_{j=1}^{i-1}p_j\leq r\leq\sum_{j=1}^ip_j.\label{inequality}
\end{eqnarray}
In the continuum limit we replace the probability $p_i$ with $p(x)dx$ which is the probability that the event $x$ is found between $x$ and $x+dx$. The condition  $\sum_{i=1}^np_i=1$ becomes
\begin{eqnarray}
\int_{-\infty}^{+\infty}p(x)~dx=1.
\end{eqnarray}
The inequality (\ref{inequality}) becomes the identity
\begin{eqnarray}
P(x)\equiv \int_{-\infty}^xp(x^{'})~dx^{'}=r
\end{eqnarray}
Thus $r$ is equal to the cumulative probability distribution $P(x)$, i.e the probability of choosing a value less than or equal to $x$. This equation leads to the inverse transform method which allows us to generate a nonuniform probability distribution $p(x)$ from a uniform probability distribution $r$. Clearly we must be able to $1)$ perform the integral analytically to  find $P(x)$ then $2)$ invert the relation $P(x)=r$ for $x$. 

As a first example we consider the Poisson distribution
\begin{eqnarray}
p(x)=\frac{1}{\lambda}e^{-\frac{x}{\lambda}}~,~0\leq x\leq \infty.
\end{eqnarray}
We find
\begin{eqnarray}
P(x)=1-e^{-\frac{x}{\lambda}}=r.
\end{eqnarray}
Hence
\begin{eqnarray}
x=-\lambda\ln(1-r).
\end{eqnarray}
Thus given the uniform random numbers $r$ we can compute directly using the above formula the random numbers $x$ which are distributed according to the Poisson distribution $p(x)=\frac{1}{\lambda}e^{-\frac{x}{\lambda}}$. 

The next example is the Gaussian distribution in two dimensions
\begin{eqnarray}
p(x,y)=\frac{1}{2\pi {\sigma}^2}e^{-\frac{x^2+y^2}{2{\sigma}^2}}.
\end{eqnarray}
We can immediately compute that
\begin{eqnarray}
\frac{1}{2\pi {\sigma}^2}\int_{-\infty}^{+\infty}dx ~\int_{-\infty}^{+\infty}dy~e^{-\frac{x^2+y^2}{2{\sigma}^2}}=\int_0^1dw\int_0^1dv.
\end{eqnarray}
\begin{eqnarray}
x=r\cos\phi ~,~y=r\sin\phi.
\end{eqnarray}
\begin{eqnarray}
r^2=-2{\sigma}^2\ln v~,~\phi=2\pi w.
\end{eqnarray}
The random numbers $v$ and $w$ are clearly uniformly distributed between $0$ and $1$. The random numbers $x$ (or $y$) are distributed according to the Gaussian distribution in one dimension. This method is known as the Box-Muller method.
\subsection{The Acceptance-Rejection Method}
This was proposed by Von Neumann. The goal is to generate a sequence of random numbers distributed according to  some normalized probability density $y=p(x)$. This method consists of the following steps:
\begin{itemize}
\item{} We start by generating a  uniform random number $r_x$ in the range of interest $x_{\rm min}\leq r_x\leq x_{\rm max}$ where $[x_{\rm min},x_{\rm max}]$ is the interval in which $y=p(x)$ does not vanish. 
\item{} We evaluate $p(r_x)$. 
\item{} Then we generate another uniform random number $r_y$ in the range $[0,y_{\rm max}]$ where $y_{\rm max}$ is the maximum value of the distribution $y=p(x)$. 
\item{}If  $r_y<p(r_x)$ then we accept the random number $r_x$ else we reject it. 
\item{} We repeat this process a sufficient number of times. 
\end{itemize}
It is not difficult to convince ourselves that the accepted random numbers $r_x$ will be distributed according to $y=p(x)$.

\section{Simulation $16$: Midpoint and Monte Carlo Approximations}
\paragraph{Part I}

The volume of a ball of radius $R$ in $d$ dimensions is given by
\begin{eqnarray}
V_d&=&\int_{x_1^2+...+x_d^2\leq R^2}dx_1...dx_d\nonumber\\
&=&2\int dx_1...dx_{d-1}\sqrt{R^2-x_1^2-...-x_{d-1}^2}\nonumber\\
&=&\frac{R^d}{d}\frac{2{\pi}^{\frac{d}{2}}}{\Gamma(\frac{d}{2})}.\nonumber
\end{eqnarray}
\begin{itemize}

\item[$(1)$]Write a program that computes the three dimensional integral using the midpoint approximation. We take the stepsize $h=2R/N$, the radius $R=1$ and the number of steps in each direction to be  $N=N_x=N_y=2^p$ where $p=1,15$.

\item[$(2)$] Show that the error goes as $1/N$. Plot the logarithm of the absolute value of the absolute error versus the logarithm of $N$.

\item[$(3)$] Try out the two dimensional integral. Work in the positive quadrant and again take the stepsize $h=R/N$ where $R=1$ and $N=2^p$, $p=1,15$. We know that generically the theoretical error goes at least as $1/N^{2}$. What do you actually find? Why do you find a discrepancy?\\ 
Hint: the second derivative of the integrand is singular at $x=R$ which changes the dependence from $1/N^2$ to $1/N^{1.5}$.

\end{itemize}

\paragraph{Part II}
In order to compute  numerically the volume of the ball in any dimension $d$ we use the recursion formula

\begin{eqnarray}
V_d&=&\frac{V_{d-1}}{R^{d-1}}\int_{-R}^{+R}dx_d~ (R^2-x_d^2)^{\frac{d-1}{2}}.\nonumber
\end{eqnarray}
\begin{itemize}
\item[$(1)$] Find the volumes in $d=4,5,6,7,8,9,10,11$ dimensions. Compare with the exact result given above.
\end{itemize}
\paragraph{Part III}
\begin{itemize}

\item[$(1)$] Use the Monte Carlo sampling (hit or miss) method to find the integrals in $d=2,3,4$ and $d=10$ dimensions. Is the Monte Carlo method easier  to apply than the midpoint approximation?

\item[$(2)$] Use the Monte Carlo sample mean value method to find the integrals in $d=2,3,4$ and $d=10$ dimensions. For every $d$ we perform $M$ measurements each with $N$ samples. We consider $M=1,10,100,150$ and $N=2^p$, $p=10,19$. Verify that the exact error in this case goes like $1/\sqrt{N}$.\\ 
Hint: Compare the exact error which is known in this case with the standard deviation of the mean ${\sigma}_M$ and with ${\sigma}/\sqrt{N}$ where $\sigma$ is the standard deviation of the sample, i.e. of a single measurement. These three quantities must be identical. 
\end{itemize}

\paragraph{Part IV }
\begin{itemize}
\item[$(1)$]The value of $\pi$ can be given by the integral
\begin{eqnarray}
\pi=\int_{x^2+y^2\leq R^2}dx~dy.\nonumber
\end{eqnarray}
Use the Monte Carlo sampling (hit or miss) method to give an approximate value of $\pi$.

\item[$(2)$]The above integral can also be put in the form
\begin{eqnarray}
\pi=2\int_{-1}^{+1}dx~\sqrt{1-x^2}.\nonumber
\end{eqnarray}
Use the Monte Carlo sample mean value method to give another approximate value of $\pi$.
\end{itemize}

\section{Simulation $17$: Nonuniform Probability Distributions}
\paragraph{Part I}

The Gaussian distribution is given by
\begin{eqnarray}
P(x)=\frac{1}{\sqrt{2\pi{\sigma}^2}}~\exp{-\frac{(x-\mu)^2}{2{\sigma}}}.\nonumber
\end{eqnarray}
The parameter $\mu$ is the mean and $\sigma$ is the variance, i.e the square root of the standard deviation. We choose $\mu=0$ and $\sigma=1$.

\begin{itemize}

\item[$(1)$] Write a program that computes  a sequence of random numbers $x$ distributed according to $P(x)$ using the inverse transform method (Box-Muller algorithm) given by the equations
\begin{eqnarray}
x=r\cos\phi.\nonumber
\end{eqnarray}
\begin{eqnarray}
r^2=-2{\sigma}^2\ln v~,~\phi=2\pi w.\nonumber
\end{eqnarray}
The $v$ and $w$ are uniform random numbers in the interval $[0,1]$.

\item[$(2)$] Draw a histogram of the random numbers obtained in the previous question. The steps are as follows:
\begin{itemize}
\item[a-] Determine the range of the points $x$.
\item[b-] We divide the interval into $u$ bins. The lenght of each bin is $h={\rm interval}/u$. We take for example $u=100$.
\item[c-] We determine  the location of every point $x$ among the bins. We increase the counter of the corresponding bin by a unit. 
\item[d-] We plot the fraction of points as a function of $x$. The fraction of point is equal to the number of random numbers in a given bin divided by $hN$ where $N$ is the total number of random numbers. We take $N=10000$.
\end{itemize}

\item[$(3)$]  Draw the data on a logarithmic scale, i.e plot $\log({\rm fraction})$ versus $x^2$. Find the fit and compare with theory.

\end{itemize}
\paragraph{Part II }
\begin{itemize}
\item[$(1)$] Apply the acceptance-rejection method to the above problem.
\item[$(2)$] Apply  the Fernandez-Criado algorithm to the above problem. The procedure is as follows
\begin{itemize}
\item[a-] Start with $N$  points $x_i$ such that $x_i=\sigma$.
\item[b-] Choose at random a pair $(x_i,x_j)$ from the sequence and make the following change 
\begin{eqnarray}
&&x_i\longrightarrow \frac{x_i+x_j}{\sqrt{2}}\nonumber\\
&&x_j\longrightarrow -x_i+\sqrt{2}x_j.\nonumber
\end{eqnarray}
\item[c-] Repeat step $2$ until we reach equilibrium. For example try it $M$ times where $M=10,100,...$.

\end{itemize}
 \end{itemize}

\chapter{The Metropolis Algorithm and The Ising Model}

\section{The Canonical Ensemble}
We consider physical systems which are in thermal contact with  an environment.  The environment is usually much larger than the physical system of interest and as a consequence energy exchange between the two of them will not change the temperature  of the environement. The environement is called  heat bath or heat reservoir. When the system reaches equilibrium with the heat bath its temperature will be given by the temperature of the heat bath. 

A system in equilibrium with a heat bath is described statistically by the canonical ensemble in which the temperature is fixed. In contrast an isolated system is described statistically by the microcanonical ensemble in which the energy is fixed. Most systems in nature are not isolated but are in thermal contact with the environment. It is a  fundamental result of statistical mechanics  that the probability of finding  a system in equilibrium with a heat bath at temperature $T$ in a microstate $s$ with energy $E_s$ is given by the Boltzmann distribution
\begin{eqnarray}
P_s=\frac{1}{Z}e^{-\beta E_s}~,~\beta=\frac{1}{k_BT}.
\end{eqnarray}
The normalization connstant $Z$ is the partition function. It is defined by
\begin{eqnarray}
Z=\sum_s e^{-\beta E_s}.\label{partition}
\end{eqnarray}
The sum is over all the microstates of the system with a fixed $N$ and $V$. The Helmholtz free energy $F$ of a system is given by
\begin{eqnarray}
F=-k_BT\ln Z.
\end{eqnarray}
In equilibrium the free energy is minimum. All other thermodynamical quantities can be given by various derivatives of $F$. For example the internal energy $U$ of the system which is the expectation value of the energy can be expressed in terms of $F$ as follows
\begin{eqnarray}
U=<E>=\sum_s E_sP_s=\frac{1}{Z}\sum_s E_se^{-\beta E_s}=-\frac{\partial}{{\partial}\beta}\ln Z=\frac{\partial}{{\partial}\beta}(\beta F).
\end{eqnarray}
The specific heat is given by
\begin{eqnarray}
C_v=\frac{\partial}{{\partial}T}U.
\end{eqnarray}
In the definition of the partition function (\ref{partition}) we have implicitly assumed that we are dealing with a physical system with configurations (microstates) which have discrete energies. This is certainly true for many quantum systems. However for many other systems especially classical ones the energies are not discrete. For example the partition function of a gas of  $N$ distinguishable classical particles is given by

\begin{eqnarray}
Z=\int \prod_{i=1}^N\frac{d^3p_id^3q_i}{h^3}~e^{-\beta H(\vec{p}_i,\vec{q}_i)}.
\end{eqnarray}
For quantum dynamical field systems (in Euclidean spacetimes) which are of fundamental importance to elementary particles and their interactions the partition function is given by the so-called path integral which is essentially of the same form as the previous equation with the replacement of the Hamiltonian $H(\vec{p}_i,\vec{q}_i)$ by the action $S[\Phi]$ where $\Phi$ stands for the field variables and the replacement of the measure $\prod_{i=1}^N({d^3p_id^3q_i})/{h^3}$ by the relevant (infinite dimensional) measure ${\cal D}\Phi$ on the space of field configurations. We obtain therefore
\begin{eqnarray}
Z=\int {\cal D}\Phi ~e^{-\beta S[\Phi]}.
\end{eqnarray}
Similarly to what happens in statistical mechanics where all observables can be derived from the partition function the observables of a quantum field theory can all be derived from the path integral. The fundamental problem therefore is how to calculate the partition function or the path integral for a given physical system. Normally an analytic solution will be ideal. However  finding such a solution is seldom possible and as a consequence only the numerical approach remains available to us. The partition function and the path integral are essentially given by multidimensional integrals and thus one should seek numerical approaches to the problem of integration.
\section{Importance Sampling}
In any Monte Carlo integration the numerical error is   proportional to the standard deviation of the integrand and is inversely proportional to the number of samples. Thus in order to reduce the error we should either reduce the variance or increase the number of samples. The first option is preferable since it does not require any extra computer time. Importance sampling allows us to reduce the standard deviation of the integrand and hence the error by sampling  more often the important regions of the integral where the integrand is largest. Importance sampling uses also in a crucial way nonuniform probability distributions.

Let us again consider the one dimensional integral 
\begin{eqnarray}
F=\int_a^b dx~ f(x).
\end{eqnarray}
We introduce the probability distribution $p(x)$ such that
\begin{eqnarray}
1=\int_a^b dx~ p(x).
\end{eqnarray}
We write the integral as
\begin{eqnarray}
F=\int_a^b dx~p(x)~ \frac{f(x)}{p(x)}.
\end{eqnarray}
We evaluate this integral by sampling according to the probability distribution $p(x)$. In other words we find a set of $N$ random numbers $x_i$ which are distributed according to $p(x)$ and then approximate the integral by the sum
\begin{eqnarray}
F_N=\frac{1}{N}\sum_{i=1}^N \frac{f(x_i)}{p(x_i)}.
\end{eqnarray}
The probability distribution $p(x)$ is chosen such that the function $f(x)/p(x)$ is slowly varying   which reduces the corresponding standard deviation. 
 \section{The Ising Model}

We consider a $d-$dimensional periodic lattice with $n$ points in every direction so that there are $N=n^d$ points in total in this lattice. In every point (lattice site) we put a spin variable $s_i$ $(i=1,...,N)$ which can take either the value $+1$ or $-1$. A configuration of this system of $N$ spins is therefore specified by a set of numbers $\{s_i\}$. In the Ising model the energy of this system of $N$ spins in the configuration $\{s_i\}$ is given by 
\begin{eqnarray}
E_I\{s_i\}=-\sum_{<ij>}{\epsilon}_{ij}s_is_j-H\sum_{i=1}^Ns_i.
\end{eqnarray}
The parameter $H$ is the external magnetic field. The symbol $<ij>$ stands for nearest neighbor spins. The sum over $<ij>$ extends over ${\gamma N}/{2}$ terms where $\gamma$ is the number of nearest neighbors. In $2,3,4$ dimensions $\gamma=4,6,8$. The parameter ${\epsilon}_{ij}$ is the interaction energy between the spins $i$ and $j$. For isotropic interactions ${\epsilon}_{ij}=\epsilon$. For $\epsilon>0$ we obtain ferromagnetism while for $\epsilon <0$ we obtain antiferromagnetism. We consider only $\epsilon >0$. The energy  becomes with these simplifications given by
\begin{eqnarray}
E_I\{s_i\}=-{\epsilon}\sum_{<ij>}s_is_j-H\sum_{i=1}^Ns_i.
\end{eqnarray}
The partition function is given by
\begin{eqnarray}
Z=\sum_{s_1}\sum_{s_2}...\sum_{s_N}~e^{-\beta E_I\{s_i\}}.
\end{eqnarray}
There are $2^N$ terms in the sum and $\beta={1}/{k_BT}$.

In $d=2$ we have  $N=n^2$ spins in the square lattice. The configuration $\{s_i\}$ can be viewed as an $n\times n$ matrix. We impose periodic boundary condition as follows. We consider $(n+1)\times (n+1)$ matrix where the $(n+1)$th row is identified with the first row and  the $(n+1)$th column is identified with the first column. The square lattice is therefore a torus.

\section{The Metropolis Algorithm}
The internal energy $U=<E>$ can be put into the form
\begin{eqnarray}
<E>=\frac{\sum_s E_se^{-\beta E_s}}{\sum_s e^{-\beta E_s}}.
\end{eqnarray}
Generally given any physical quantity $A$ its expectation value $<A>$ can be computed using a similar expression, viz
\begin{eqnarray}
<A>=\frac{\sum_s A_se^{-\beta E_s}}{\sum_s e^{-\beta E_s}}.
\end{eqnarray}
The number $A_s$ is the value of $A$ in the microstate $s$. In general the number of microstates $N$ is very large. In any Monte Carlo simulation we can only generate a very small number $n$ of the total number $N$ of the microstates. In other words $<E>$ and $<A>$ will be approximated with
\begin{eqnarray}
<E>{\simeq}<E>_n=\frac{\sum_{s=1}^n E_se^{-\beta E_s}}{\sum_{s=1}^n e^{-\beta E_s}}.
\end{eqnarray}
\begin{eqnarray}
<A>{\simeq}<A>_n=\frac{\sum_{s=1}^n A_se^{-\beta E_s}}{\sum_{s=1}^n e^{-\beta E_s}}.
\end{eqnarray}
The calculation of $<E>_n$ and $<A>_n$ proceeds therefore by $1)$ choosing at random a microstate $s$, $2)$ computing $E_s$, $A_s$ and $e^{-\beta E_s}$ then $3)$ evaluating the contribution of this microstate to the expectation values $<E>_n$ and $<A>_n$. This general Monte Carlo procedure is however highly inefficient since the microstate $s$ is very improbable and therefore its contribution to the expectation values is negligible. We need to use importance sampling. To this end we introduce a probability distribution ${p}_s$ and rewrite the expectation value $<A>$ as
\begin{eqnarray}
<A>=\frac{\sum_s \frac{A_s}{p_s}e^{-\beta E_s}p_s}{\sum_s\frac{1}{p_s} e^{-\beta E_s}p_s}.
\end{eqnarray}
Now we generate the microstates $s$ with probabilities $p_s$  and approximate $<A>$ with $<A>_n$ given by
\begin{eqnarray}
<A>_n=\frac{\sum_{s=1}^n \frac{A_s}{p_s}e^{-\beta E_s}}{\sum_{s=1}^n \frac{1}{p_s} e^{-\beta E_s}}.
\end{eqnarray}
This is importantce sampling. The Metropolis algorithm is importance sampling with $p_s$ given by the Boltzmann distribution, i.e.
\begin{eqnarray}
p_s=\frac{e^{-\beta E_s}}{\sum_{s=1}^n e^{-\beta E_s}}.
\end{eqnarray}
We get then the arithmetic average
\begin{eqnarray}
<A>_n=\frac{1}{n}\sum_{s=1}^n A_s.
\end{eqnarray}
The Metropolis algorithm in the case of spin systems such as the Ising model can be summarized as follows:
\begin{itemize}
\item[$(1)$] Choose an initial microstate.
\item[$(2)$] Choose a spin at random and flip it.
\item[$(3)$] Compute $\Delta E=E_{\rm trial}-E_{\rm old}$. This is the change in the energy of the system due to the trial flip.
\item[$(4)$] Check if $\Delta E\leq 0$. In this case the trial microstate is accepted.
\item[$(5)$] Check if $\Delta E>0$. In this case compute the ratio of probabilities $w=e^{-\beta \Delta E}$.
\item[$(6)$] Choose a uniform random number $r$ in the inetrval $[0,1]$.
\item[$(7)$] Verify if $r\leq w$. In this case  the trial microstate is accepted, otherwise it is rejected.
\item[$(8)$] Repeat steps $2)$ through $7)$ until all spins of the system are tested. This sweep counts as one unit of Monte Carlo time.
\item[$(9)$] Repeat setps $2)$ through $8)$ a sufficient number of times until thermalization, i.e. equilibrium is reached.
\item[$(10)$] Compute the physical quantities of interest in $n$ thermalized microstates. This can be done periodically in order to reduce correlation between the data points.
\item[$(11)$] Compute averages.
\end{itemize}
The proof that this algorithm leads indeed to a sequence of states which are distributed according to the Boltzmann distribution goes as follows.

It is clear that the steps $2)$ through $7)$ corresponds to a transition probability between the microstates $\{s_i\}$ and $\{s_j\}$ given by
\begin{eqnarray}
W(i\longrightarrow j)={\rm min}(1,e^{-\beta\Delta E})~,~\Delta E=E_j-E_i.
\end{eqnarray}
Since only  the ratio of probabilities $w=e^{-\beta \Delta E}$ is needed it is not necessary to normalize the Boltzmann probability distribution. It is clear that this probability function satisfies the detailed balance condition

\begin{eqnarray}
W(i\longrightarrow j)~e^{-\beta E_i}=W(j\longrightarrow i)~e^{-\beta E_j}.
\end{eqnarray}
Any other probability function $W$ which satisfies this condition will generate a sequence of states which are distributed according to the Boltzmann distribution. This can be shown by summing over the index $j$ in the above equation and using  $\sum_j W(i\longrightarrow j) =1$. We get
\begin{eqnarray}
e^{-\beta E_i}=\sum_j W(j\longrightarrow i)~e^{-\beta E_j}.
\end{eqnarray}
The Boltzmann distribution is an eigenvector of $W$. In other words  $W$ leaves the equilibrium ensemble in equilibrium. As it turns out this equation is also  a sufficient condition for any ensemble to approach equilibrium.

\section{The Heat-Bath Algorithm}
The heat-bath algorithm is generally a less efficient algorithm than the Metropolis algorithm. The acceptance probability is given by
\begin{eqnarray}
W(i\longrightarrow j)={\rm min}(1,\frac{1}{1+e^{\beta\Delta E}})~,~\Delta E=E_j-E_i.
\end{eqnarray}
This acceptance probability satisfies also detailed balance for the Boltzmann probability distribution. In other words the detailed balance condition which is sufficient but not necessary for an ensemble to reach equilibrium does not have a unique solution.

\section{The Mean Field Approximation}
\subsection{Phase Diagram and Critical Temperature}
We consider $N=L^2$ spins on a square lattice where $L$ is the number of lattice  sites in each direction. Each spin can take only two possible values $s_i=+1$ (spin up) and $s_i=-1$ (spin down). Each spin interacts only with its $4$ neigbhors and also with a magnetic field $H$. The Ising model in $2$ dimensions is given by the energy
\begin{eqnarray}
E\{s\}=-J\sum_{<ij>}s_is_j-H\sum_is_i.
\end{eqnarray}
 The system is assumed to be in equilibrium with a heat bath with temperature $T$. Thermal equilibrium of the Ising model is  described by the canonical ensemble. The probability of finding the Ising model in a configuration $\{s_1,...,s_{2^N}\}$ is given by Boltzmann distribution
\begin{eqnarray}
P\{s\}=\frac{e^{-\beta E\{s\}}}{Z}.
\end{eqnarray}
The partition function is given by
\begin{eqnarray}
Z=\sum_{\{s\}}e^{-\beta E\{s\}}=\sum_{s_1}...\sum_{s_{2^N}}e^{-\beta E\{s\}}.
\end{eqnarray}
The magnetization $M$ in a configuration $\{s_1,...,s_{2^N}\}$ is the order parameter of the system. It is defined by
\begin{eqnarray}
M=\sum_i s_i.
\end{eqnarray}
The average of $M$ is given by
\begin{eqnarray}
<M>=\sum_i <s_i>=N<s>.
\end{eqnarray}
In above $<s_i>=<s>$ since all spins are equivalent. We have
\begin{eqnarray}
<M>=\frac{1}{\beta}\frac{\partial\log Z}{\partial H}=-\frac{\partial F}{\partial H}.
\end{eqnarray}
In order to compute $<M>$ we need to compute $Z$. In this section we use the mean  field approximation. First we rewrite the energy $E\{s\}$ in the form
\begin{eqnarray}
E\{s\}&=&(-J\sum_{<ij>}s_j)s_i-H\sum_is_i\nonumber\\
&=&\sum_i H_{\rm eff}^is_i-H\sum_is_i.
\end{eqnarray}
The effective magnetic field $H_{\rm eff}^i$ is given  by
\begin{eqnarray}
H_{\rm eff}^i=-J\sum_{j(i)}s_{j(i)}.
\end{eqnarray}
The index $j(i)$ runs over the four nearest neighbors of the spin $i$. In the mean field approximation we replace the spins $s_{j(i)}$ by their thermal average $<s>$. We obtain
\begin{eqnarray}
H_{\rm eff}^i=- J \gamma <s>~,~\gamma=4.
\end{eqnarray}
In other words
\begin{eqnarray}
E\{s\}
&=&-(H+ J\gamma <s>)\sum_is_i=H_{\rm eff}\sum_i s_i
\end{eqnarray}
The partition function becomes
\begin{eqnarray}
Z&=&\bigg(\sum_{s_1}e^{-\beta H_{\rm eff} s_i}\bigg)^N\nonumber\\
&=&\bigg(e^{-\beta H_{\rm eff}} +e^{\beta H_{\rm eff}}\bigg)^N\\
&=&\bigg(2\cosh \beta H_{\rm eff}\bigg)^N.
\end{eqnarray}
The free energy and magnetization are then given by
\begin{eqnarray}
F=-kT\ln Z&=&-kT N\ln\bigg(2\cosh \beta H_{\rm eff}\bigg).
\end{eqnarray}
\begin{eqnarray}
<M>=N<s>=N\tanh \beta H_{\rm eff}.
\end{eqnarray}
Thus for zero magnetic field we get from the second equation the constraint
\begin{eqnarray}
<s>=\tanh \gamma \beta J <s>.
\end{eqnarray}
Clearly $<s>=0$ is always a solution. This is the high temperature paramagnetic phase. For small temperature we have also a solution $<s>\neq 0$. This is the ferromagnetic phase. There must exist a critical temperature $T_c$ which separates the two phases. We expect $<s>$ to approach $<s>=0$ as $T$ goes to $T_c$ from below. In other words near $T_c$ we can treat $<s>$ as small and as a consequence we can use the expansion $\tanh x=x-\frac{1}{3}x^3$. We obtain
\begin{eqnarray}
<s>=\gamma \beta J<s>- \frac{1}{3}\big(\gamma \beta J<s>\big)^3.
\end{eqnarray}
Equivalently
\begin{eqnarray}
<s>\bigg(<s>^2-\frac{3}{T}\frac{1}{(\gamma \beta J)^3}\big(\frac{\gamma J}{k_B}-T\big)\bigg)=0.
\end{eqnarray}
We get the two solutions
\begin{eqnarray}
&&<s>=0~,~{\rm paramagnetic}~{\rm phase}\nonumber\\
&&<s>=\pm \sqrt{\frac{3}{T}\frac{1}{(\gamma \beta J)^3}}(T_c-T)^{\beta}~,~{\rm ferromagnetic}~{\rm phase}.
\end{eqnarray}
The critical temperature $T_c$ and the critical exponent $\beta$ are given by
\begin{eqnarray}
T_c=\frac{\gamma J}{k_B}~,~\beta=\frac{1}{2}.
\end{eqnarray}
The ferromagnetic solution can only exist for $T<T_c$. 
\subsection{Critical Exponents}
The free energy for zero magnetic field is

\begin{eqnarray}
F=-kT N\ln\bigg(2\cosh \gamma \beta J <s>\bigg).
\end{eqnarray}
We see that for $T<T_c$ the ferromagnetic solution has a lower free energy than the paramagnetic solution $<s>=0$. The phase $T<T_c$ is indeed ferromagnetic. The transition at $T=T_c$ is second order. The free energy is continuous at $T=T_c$, i.e. there is no latent heat while the specific heat is logarithmically divergent. The mean field theory yields the correct value $0$ for the critical exponent $\alpha$ although it does not reproduce the logarithmic divergence. The susceptibility diverges at $T=T_c$ with critical exponent $\gamma=1$. These latter statements can be seen as follows. 

The specific heat is given by

\begin{eqnarray}
C_v&=&-\frac{\partial }{\partial T}\bigg(k_BT^2\frac{\partial}{\partial T}(\beta F)\bigg)\nonumber\\
&=&-2k_BT\frac{\partial}{\partial T}(\beta F)-k_BT^2\frac{{\partial}^2}{\partial T^2}(\beta F).
\end{eqnarray}
Next we use the expression $\beta F=-N\ln (e^x+e^{-x})$ where $x=\gamma\beta J<s>$. We find
\begin{eqnarray}
\frac{C_v}{N}
&=&2k_BT\tanh x\frac{\partial x}{\partial T}+k_BT^2\tanh^2 x\frac{{\partial}^2x}{\partial T^2}+k_BT^2\frac{1}{\cosh^2 x}(\frac{\partial x}{\partial T})^2.
\end{eqnarray}
We compute
\begin{eqnarray}
x=\pm \sqrt{\frac{3k_B}{\gamma J}}(T_c-T)^{\frac{1}{2}}~,~\frac{\partial x}{\partial T}=\mp \frac{1}{2}\sqrt{\frac{3k_B}{\gamma J}}(T_c-T)^{-\frac{1}{2}}~,~\frac{{\partial}^2 x}{\partial T^2}=\mp \frac{1}{4}\sqrt{\frac{3k_B}{\gamma J}}(T_c-T)^{-\frac{3}{2}}.\nonumber\\
\end{eqnarray}
It is not difficult to show that the divergent terms cancel and as a consequence
\begin{eqnarray}
\frac{C_v}{N}\sim (T_c-T)^{-\alpha}~,~\alpha=0.
\end{eqnarray}
The  susceptibility is given by
\begin{eqnarray}
\chi=\frac{\partial}{\partial H}<M>.
\end{eqnarray}
To compute the behavior of $\chi$ near $T=T_c$ we consider the equation
\begin{eqnarray}
<s>=\tanh (\gamma \beta J <s>+\beta H).
\end{eqnarray}
For small magnetic field we can still assume that $\gamma \beta J <s>+\beta H$ is small near $T=T_c$ and as a consequence we can expand the above equation as
\begin{eqnarray}
<s>=(\gamma \beta J <s>+\beta H)-\frac{1}{3}(\gamma \beta J <s>+\beta H)^3.
\end{eqnarray}
Taking the derivative with respect to $H$ of both sides of this equation we obtain 
 \begin{eqnarray}
\hat{\chi}=(\gamma \beta J \hat{\chi}+\beta )-(\gamma \beta J \hat{\chi}+\beta )(\gamma \beta J <s>+\beta H)^2.
\end{eqnarray}
\begin{eqnarray}
\hat{\chi}=\frac{\partial}{\partial H}<s>.
\end{eqnarray}
Setting the magnetic field to zero we get
\begin{eqnarray}
\hat{\chi}=(\gamma \beta J \hat{\chi}+\beta )-(\gamma \beta J \hat{\chi}+\beta )(\gamma \beta J <s>)^2.
\end{eqnarray}
In other words
\begin{eqnarray}
\bigg(1-\gamma \beta J +\gamma \beta J (\gamma \beta J <s>)^2\bigg)\hat{\chi}=\beta -\beta (\gamma \beta J <s>)^2.
\end{eqnarray}
\begin{eqnarray}
2\frac{T_c-T}{T}\hat{\chi}=\frac{1}{k_BT}(1 -(\gamma \beta J <s>)^2).
\end{eqnarray}
Hence
\begin{eqnarray}
\hat{\chi}=\frac{1}{2k_B}(T_c-T)^{-\gamma}~,~\gamma=1.
\end{eqnarray}

\section{Simulation of The Ising Model and Numerical Results}
\subsection{The Fortran Code}
We choose to write our code in Fortran. The reason is simplicity and straightforwardness. A person who is not well versed in programming languages, who has a strong background in physics and maths, and who wants to get up and running quickly with the coding so that she starts doing physics (almost) immediately the choice of Fortran for her is ideal and thus it is only natural. The potential superior features which may be found in $C$ are peripheral to our purposes here.

 The spin found in the intersection point of the $i$th row and $j$th column  of the lattice will be represented with the matrix element $\phi(i,j)$. The energy will then read (with $N=n^2$ and $n\equiv L$)
\begin{eqnarray}
E=-\sum_{i,j=1}^n\bigg[\frac{J}{2}\phi(i,j)\bigg(\phi(i+1,j)+\phi(i-1,j)+\phi(i,j+1)+\phi(i,j-1)\bigg)+H\phi(i,j)\bigg].\nonumber\\
\end{eqnarray}
We impose periodic boundary condition in order to reduce edge and boundary effects. This can be done as follows. We consider $(n+1)\times (n+1)$ matrix where the $(n+1)$th row is identified with the first row and  the $(n+1)$th column is identified with the first column. The square lattice is therefore a torus.  The toroidal boundary condition will read explicitly as follows
\begin{eqnarray}
\phi(0,j)=\phi(n,j)~,~\phi(n+1,j)=\phi(1,j)~,~\phi(i,0)=\phi(i,n)~,~\phi(i,n+1)=\phi(i,1).\nonumber
\end{eqnarray}
The variation of the energy due to the flipping of the spin $\phi(i,j)$ is an essential ingredient in the Metropolis algorithm. This variation is explicitly given by

\begin{eqnarray}
\Delta E=2J\phi(i,j)\big(\phi(i+1,j)+\phi(i-1,j)+\phi(i,j+1)+\phi(i,j-1)\big)+2H \phi(i,j).
\end{eqnarray}

The Fortran code contains the following pieces:

\begin{itemize}
\item{}A subroutine which generates pseudo random numbers. We prefer to work with well established suboutines such as the RAN $2$ or the RANLUX.
\item{}A subroutine which implements the Metropolis algorithm for the Ising model. This main part will read (with some change of notation such as $J={\rm exch}$)
\begin{verbatim}
do i=1,L
    ip(i)=i+1
    im(i)=i-1
enddo
    ip(L)=1
    im(1)=L

do  i=1,L
    do   j=1,L    
      deltaE=2.0d0*exch*phi(i,j)*(phi(ip(i),j)+phi(im(i),j)+phi(i,ip(j))+phi(i,im(j)))
      deltaE=deltaE + 2.0d0*H*phi(i,j)
            if (deltaE.ge.0.0d0)then
                  probability=dexp(-beta*deltaE)   
               call ranlux(rvec,len)
               r=rvec(1)
               if (r.le.probability)then
                  phi(i,j)=-phi(i,j)
               endif        
            else
                  phi(i,j)=-phi(i,j)
           endif    
    enddo
enddo
\end{verbatim}
\item{} We compute  the energy $<E>$ and the magnetization $<M>$ of the Ising model in a separate subroutine. 
\item{}We compute the errors using for example the Jackknife method in a separate subroutine.
\item{}We fix the parameters of the model such as $L$, $J$, $\beta=1/T$ and $H$.
\item{}We choose an initial configuration. We consider both cold and  hot starts which are given respectively by
\begin{eqnarray}
\phi(i,j)=+ 1.
\end{eqnarray}
\begin{eqnarray}
\phi(i,j)={\rm random}~{\rm signs}.
\end{eqnarray}
\item{}We run the Metropolis algorithm for a given thermalization time and study the history of the energy and the magnetization for different values of the temperature. 

\item{}We add a Monte Carlo evolution with a reasonably large number of steps and compute the averages of $E$ and $M$.

\item{}We compute the specific heat and the susceptibility of the system. 

\end{itemize}
\subsection{Some Numerical Results}

\paragraph{Energy:} The energy is continuous through the transition point and as a consequence there is no latent heat. This indicates a second order behavior.

\paragraph{Specific Heat:}The critical exponent associated with the specific heat is given by $\alpha=0$.
 However the specific heat diverges logarithmically at $T=T_c$. This translates into the fact that the peak grows with $n$ logarithmically, namely
 \begin{eqnarray}
\frac{C_v}{n^2}\sim \log n.
\end{eqnarray}


\paragraph{Magnetization:} The magnetization near but below the critical temperature in the two-dimensional Ising model scales as
\begin{eqnarray}
\frac{<M>}{n^2}\sim (T_c-T)^{-\beta}~,~\beta={1}/{8}.
\end{eqnarray}
\paragraph{Susceptibility:}The susceptibility near the critical temperature  in the two-dimensional Ising model scales as
\begin{eqnarray}
\frac{\chi}{n^2}\sim |T-T_c|^{-\gamma}~,~\gamma={7}/{4}.
\end{eqnarray}
\paragraph{Critical Temperature:} From the behavior of the above observable we can measure the critical temperature, which marks the point where the second order ferromagnetic phase transition occurs, to be given approximately by 
\begin{eqnarray}
k_B T_c=\frac{2J}{\ln(\sqrt{2}+1)}.
\end{eqnarray}
\paragraph{Critical Exponents and $2-$Point Correlation Function:}

The $2-$point correlation function of the two-dimensional Ising model  is  defined by the expression
\begin{eqnarray}
f(x)&=&<s_0s_x>\nonumber\\
&=&<\frac{1}{4n^2}\sum_{i,j}\phi(i,j)\bigg(\phi(i+x,j)+\phi(i-x,j)+\phi(i,j+x)+\phi(i,j-x)\bigg)>.\nonumber\\
\end{eqnarray}
We can verify numerically the following statements:
\begin{itemize}
\item{} At  $T=T_c$ the behaviour of $f(x)$ is given by
\begin{eqnarray}
f(x)\simeq \frac{1}{x^{\eta}}~,~\eta={1}/{4}.
\end{eqnarray}
\item{} At $T$ less than $T_c$ the behavior of $f(x)$ is given by
\begin{eqnarray}
f(x)= <M>^2.
\end{eqnarray}
\item{} At $T$ larger than $T_c$ the behaviour of $f(x)$ is given by
\begin{eqnarray}
f(x)\simeq a~\frac{1}{x^{\eta}} e^{-\frac{x}{\xi}}.
\end{eqnarray}
\item{} Near $T_c$ the correlation lenght diverges as
\begin{eqnarray}
\xi\simeq \frac{1}{|T-T_c|^{\nu}}~,~\nu=1.
\end{eqnarray}
Note that near-neighbor lattice sites  which are a distance $x$ away in a given direction from a given index $i$ are given by\\
\begin{verbatim}
do x=1,nn
   if (i+x .le. n) then
      ipn(i,x)=i+x
   else
      ipn(i,x)=(i+x)-n
   endif 
   if ((i-x).ge.1)then
      imn(i,x)=i-x
   else
      imn(i,x)=i-x+n
   endif
enddo
\end{verbatim}
For simplicity we consider only odd lattices, viz $n=2nn+1$. Clearly because of the toroidal boundary conditions the possible values of the distance $x$ are $x=1,2,...,nn$.
\end{itemize}
\paragraph{First Order Transition and Hysteresis:}
We can also consider the effect of a magnetic field $H$ on the physics of the Ising model. We observe a first order phase transition at $H=0$ or $H$ near $0$ and a phenomena of hysteresis.
We observe the following:
\begin{itemize}
\item{}For $T<T_c$ we can observe a first order  phase transition. Indeed we observe a discontinuity in the energy and the magnetization which happens at a non-zero value of $H$ due to hysteresis.   The jumps in the energy and the magnetization are typical signal for a first order phase transition.

\item{}For $T>T_c$ the magnetization becomes a smooth function of $H$ near $H=0$ which means that above $T_c$ there is no distinction between the ferromagnetic states with $M\geq 0$ and $M\leq 0$.
 
\item{}We recompute the magnetization as a function of $H$ for a range of $H$ back and fourth. We observe the following:  
\begin{itemize}
\item{} A hysteresis loop.
\item{}The hysteresis window shrinks with increasing temperature or accumulating more Monte Carlo time. 
\item{}The hysteresis effect is independent of the size of the lattice. 
\end{itemize}
The phenomena of hysteresis indicates that the behaviour of the system depends on its initial state and history. Equivalently  we say that the system is trapped in a metastable state.
\end{itemize}


\section{Simulation $18$: The Metropolis Algorithm and The Ising Model}

\paragraph{Part I}
We consider $N=L^2$ spins on a square lattice where $L$ is the number of lattice  sites in each direction. Each spin can take only two possible values $s_i=+1$ (spin up) and $s_i=-1$ (spin down). Each spin interacts only with its $4$ neigbhors and also with a magnetic field $H$. The Ising model in $2$ dimensions is given by the energy
\begin{eqnarray}
E=-J\sum_{<ij>}s_is_j-H\sum_is_i.\nonumber
\end{eqnarray}
We will impose toroidal boundary condition. 
 The system is assumed to be in equilibrium with a heat bath with temperature $T$. Thermal fluctuations of the system will be simulated using the Metropolis algorithm.

\begin{itemize}

\item[$(1)$] Write a subroutine that computes  the energy $E$ and the magnetization $M$ of the Ising model in a configuration $\phi$. The   magnetization is the order parameter of the system. It is defined by
\begin{eqnarray}
M=\sum_i s_i.
\end{eqnarray}

\item[$(2)$]  Write a subroutine that implements the Metropolis algorithm for this system. You will need for this the variation of the energy  due to flipping the spin $\phi(i,j)$.



\item[$(3)$] We choose $L=10$, $H=0$, $J=1$, $\beta=1/T$. We consider both a cold start and a hot start. 

Run the Metropolis algorithm for a thermalization time ${\rm TTH}=2^{6}$ and study the history of the energy and the magnetization for different values of the temperature. The energy  and magnetization should approach the values $E=0$ and $M=0$ when $T\longrightarrow \infty$ and the values $E=-2JN$ and $M=+1$ when $T\longrightarrow 0$. 

\item[$(4)$] Add a Monte Carlo evolution with ${\rm TTM}=2^{10}$ and compute the averages of $E$ and $M$.

\item[$(5)$] Compute the specific heat and the susceptibility of the system. These are defined by
\begin{eqnarray}
C_v=\frac{\partial}{\partial \beta}<E>=\frac{\beta}{T}(<E^2>-<E>^2)~,~
\chi=\frac{\partial}{\partial H}<M>=\beta(<M^2>-<M>^2).\nonumber
\end{eqnarray}
\item[$(6)$] Determine the critical point. Compare with the theoretical exact result
\begin{eqnarray}
k_BT_c=\frac{2J}{\ln(\sqrt{2}+1)}.\nonumber
\end{eqnarray}
\end{itemize}
\paragraph{Part II} 
Add to the code a separate subroutine which implements the Jackknife method for any set of data points. Compute the errors in the energy, magnetization, specific heat and susceptibility of the Ising model using the Jackknife method.

\section{Simulation $19$: The Ferromagnetic Second Order Phase Transition}
\paragraph{Part I}
The critical exponent associated with the specific heat is given by $\alpha=0$, viz
\begin{eqnarray}
\frac{C_v}{L^2}\sim (T_c-T)^{-\alpha}~,~\alpha=0.\nonumber
\end{eqnarray}
However the specific heat diverges logarithmically at $T=T_c$. This translates into the fact that the peak grows with $L$ logarithmically, namely
 \begin{eqnarray}
\frac{C_v}{L^2}\sim \log L.\nonumber
\end{eqnarray}
Verify this behaviour numerically. To this end we take lattices between $L=10-30$ with ${\rm TTH}=2^{10}$, ${\rm TMC}=2^{13}$. The temperature is taken in the range
\begin{eqnarray}
T=T_c-10^{-2}\times {\rm step}~,~{\rm step}=-50,50.\nonumber
\end{eqnarray}
Plot the maximum of $C_v/L^2$ versus $\ln L$.

\paragraph{Part II}

 The magnetization near but below the critical temperature in $2$D Ising model scales as
\begin{eqnarray}
\frac{<M>}{L^2}\sim (T_c-T)^{-\beta}~,~\beta=\frac{1}{8}.\nonumber
\end{eqnarray}
We propose to study the magnetization near $T_c$ in order to determine the value of $\beta$ numerically. Towards this end we plot $|<M>|$ versus $T_c-T$ where $T$ is taken in the the range
\begin{eqnarray}
T=T_c-10^{-4}\times {\rm step}~,~{\rm step}=0,5000.\nonumber
\end{eqnarray}
We take large lattices say $L=30-50$ with ${\rm TTH}={\rm TMC}=2^{10}$. 

\paragraph{ Part III}The susceptibility near the critical temperature in $2$D Ising model scales as
\begin{eqnarray}
\frac{\chi}{L^2}\sim |T-T_c|^{-\gamma}~,~\gamma=\frac{7}{4}.\nonumber
\end{eqnarray}
Determine $\gamma$ numerically. Use ${\rm TTH}=2^{10}$, ${\rm TMC}=2^{13}$, $L=50$ with the two ranges
\begin{eqnarray}
T=T_c-5\times 10^{-4}\times {\rm step}~,~{\rm step}=0,100.\nonumber
\end{eqnarray}
\begin{eqnarray}
T=T_c-0.05-4.5\times 10^{-3}{\rm step}~,~{\rm step}=0,100.\nonumber
\end{eqnarray}

\section{Simulation $20$: The $2-$Point Correlator}
In this exercise we will continue our study of the ferromagnetic second order phase transition. In particular we will calculate the $2-$point correlator  defined by the expression
\begin{eqnarray}
f(n)=<s_0s_n>=<\frac{1}{4L^2}\sum_{i,j}\phi(i,j)\bigg(\phi(i+n,j)+\phi(i-n,j)+\phi(i,j+n)+\phi(i,j-n)\bigg)>.\nonumber
\end{eqnarray}
\begin{itemize}
\item[$(1)$] Verify that at $T=T_c$ the behaviour of $f(n)$ is given by
\begin{eqnarray}
f(n)\simeq \frac{1}{n^{\eta}}~,~\eta=\frac{1}{4}.\nonumber
\end{eqnarray}
\item[$(2)$] Verify that at $T$ less than $T_c$ the behaviour of $f(n)$ is given by
\begin{eqnarray}
f(n)= <M>^2.\nonumber
\end{eqnarray}
\item[$(3)$] Verify that at $T$ larger than $T_c$ the behaviour of $f(n)$ is given by
\begin{eqnarray}
f(n)\simeq a~\frac{1}{n^{\eta}} e^{-\frac{n}{\xi}}.\nonumber
\end{eqnarray}
In all the above questions we take odd lattices say $L=2LL+1$ with $LL=20-50$. We also consider the parameters  ${\rm TTH}=2^{10}$, ${\rm TTC}=2^{13}$.
\item[$(4)$] Near $T_c$ the correlation lenght diverges as
\begin{eqnarray}
\xi\simeq \frac{1}{|T-T_c|^{\nu}}~,~\nu=1.\nonumber
\end{eqnarray}

 In the above question we take $LL=20$. We also consider the parameters  ${\rm TTH}=2^{10}$, ${\rm TTC}=2^{15}$ and the temperatures 
\begin{eqnarray}
T=T_c+0.1\times {\rm step}~,~{\rm step}=0,10.\nonumber
\end{eqnarray}

\end{itemize}

\section{Simulation $21$: Hysteresis and The First Order Phase Transition}

In this exercise we consider the effect of the magnetic field on the physics of the Ising model. We will observe a first order phase transition at $H=0$ or $H$ near $0$ and a phenomena of hysteresis .
\begin{itemize}
\item[$(1)$] We  will compute the magnetization and the energy as functions of $H$ for a range of temperatures $T$. The initialization will be done once for all $H$. The thermalization will be performed once for the first value of the magnetic field $H$ say $H=-5$. After we compute the magnetization for $H=-5$, we start slowly (adiabatically) changing the magnetic field with small steps so we do not loose the thermalization of the Ising system of spins. We try out the range $H=-5,5$ with step equal $0.25$. 
\begin{itemize}
\item[a-]For $T<T_c$ say $T=0.5$ and $1.5$ determine the first order transition point from the discontinuity in the energy and the magnetization.  The transition should happen at a non-zero value of $H$ due to hysteresis. The jump in the energy is associated with a non-zero latent heat. The jumps in the energy and the magnetization are the typical signal for a first order phase transition. 

\item[b-]For $T>T_c$ say $T=3$ and $5$ the magnetization becomes a smooth function of $H$ near $H=0$ which means that above $T_c$ there is no distinction between the ferromagnetic states with $M\geq 0$ and $M\leq 0$.
\end{itemize}
 
\item[$(2)$] We recompute the magnetization as a function of $H$ for a range of $H$ from $-5$ to $5$ and back. You should observe a 
hysteresis loop.  
\begin{itemize}
\item[a-]Verify that the hysteresis window shrinks with increasing temperature or accumulating more Monte Carlo time. 
\item[b-]Verify what happens if we increase the size of the lattice. 
\end{itemize}
The phenomena of hysteresis indicates that the behaviour of the system depends on its initial state and history or equivalently  the system is trapped in metastable states.
\end{itemize}

\part{Monte Carlo Simulations of Matrix Field Theory}

\chapter{Metropolis Algorithm for Yang-Mills Matrix Models}
\section{Dimensional Reduction}
\subsection{Yang-Mills Action}
In a four dimensional Minkowski spacetime with metric  $g_{\mu\nu}=(+1,-1,-1,-1)$, the Yang-Mills action with a topological theta term  is given by 

\begin{eqnarray}
S&=&-\frac{1}{2g^2}\int d^4x {\rm Tr}F_{\mu\nu}F^{\mu\nu} -\frac{\theta}{16\pi^2}\int d^4x {\rm Tr} F_{\mu\nu}\tilde{F}^{\mu\nu}.
\end{eqnarray}
We recall the definitions
\begin{eqnarray}
D_{\alpha}=\partial_{\alpha}-i[A_{\alpha},...].
\end{eqnarray}
\begin{eqnarray}
F_{\mu\nu}=\partial_{\mu}A_{\nu}-\partial_{\nu}A_{\mu}-i[A_{\mu},A_{\nu}].
\end{eqnarray}
\begin{eqnarray}
\tilde{F}^{\mu\nu}=\frac{1}{2}\epsilon^{\mu\nu\alpha\beta}F_{\alpha\beta}.
\end{eqnarray}
The path integral of interest is
\begin{eqnarray}
Z=\int D A_{\mu}~\exp(iS).
\end{eqnarray}
This is invariant under the finite gauge transformations $A_{\mu}\longrightarrow g^{-1}A_{\mu}g+ig^{-1}\partial_{\mu}g$ with $g=e^{i\Lambda}$ in some group $G$ (we will consider mostly $SU(N)$).

We Wick rotate to Euclidean signature as $x^0\longrightarrow x^4=ix^0$ and as a  consequence $d^4x\longrightarrow d_E^4x=id^4x$, $\partial_0\longrightarrow \partial_4=-i\partial_0$ and $A_0\longrightarrow A_4=-iA_0$. We compute $F_{\mu\nu}F^{\mu\nu}\longrightarrow (F_{\mu\nu}^2)_E$ and $F_{\mu\nu}\tilde{F}^{\mu\nu}\longrightarrow i(F_{\mu\nu}\tilde{F}_{\mu\nu})_E$. We get then 

\begin{eqnarray}
Z_E=\int D A_{\mu}~\exp(-S_E).
\end{eqnarray}
\begin{eqnarray}
S_E&=&\frac{1}{2g^2}\int (d^4x)_E {\rm Tr}(F_{\mu\nu}^2)_E +\frac{i\theta}{16\pi^2}\int (d^4x)_E {\rm Tr} (F_{\mu\nu}\tilde{F}_{\mu\nu})_E.
\end{eqnarray}
We remark that the theta term is imaginary.  In the following we will drop the subscript $E$ for simplicity. Let us consider first the $\theta=0$ (trivial) sector. The pure Yang-Mills action is defined by
\begin{eqnarray}
S_{\rm YM}&=&\frac{1}{2g^2}\int d^4x {\rm Tr}F_{\mu\nu}^2.
\end{eqnarray}
The path integral is of the form

\begin{eqnarray}
\int D A_{\mu}~\exp(-\frac{1}{2g^2}\int d^4x {\rm Tr}F_{\mu\nu}^2).
\end{eqnarray}
First we find the equations of motion. We have 
\begin{eqnarray}
\delta S_{\rm YM} &=&\frac{1}{g^2}\int d^4x ~{\rm Tr}F_{\mu\nu}\delta F_{\mu\nu}\nonumber\\
&=&\frac{2}{g^2}\int d^4x ~{\rm Tr}F_{\mu\nu} D_{\mu}\delta A_{\nu}\nonumber\\
&=& -\frac{2}{g^2}\int d^4x ~{\rm Tr}D_{\mu}F_{\mu\nu}.\delta A_{\nu}+\frac{2}{g^2}\int d^4x ~{\rm Tr}D_{\mu}(F_{\mu\nu}\delta A_{\nu})\nonumber\\
&=& -\frac{2}{g^2}\int d^4x ~{\rm Tr}D_{\mu}F_{\mu\nu}.\delta A_{\nu}+\frac{2}{g^2}\int d^4x ~{\rm Tr}\partial_{\mu}(F_{\mu\nu}\delta A_{\nu}).
\end{eqnarray}
The equations of motion for variations of the gauge field which vanish at infinity are therefore given by
\begin{eqnarray}
D_{\mu}F_{\mu\nu}=0.
\end{eqnarray}
Equivalently 
\begin{eqnarray}
\partial_{\mu}F_{\mu\nu}-i[A_{\mu},F_{\mu\nu}]=0.
\end{eqnarray}
We can  reduce to zero dimension  by assuming that the configurations $A_a$ are constant configurations, i.e.  are $x-$independent. We employ the notation $A_a=X_a$. We obtain immediately the action and the equations of motion  
\begin{eqnarray}
S_{\rm YM}&=&-\frac{V_{R^4}}{2g^2} {\rm Tr}[X_{\mu},X_{\nu}]^2.\label{ym1}
\end{eqnarray}
\begin{eqnarray}
[X_{\mu},[X_{\mu},X_{\nu}]]=0.
\end{eqnarray}
\subsection{Chern-Simons Action: Myers Term}
Next we consider the general sector $\theta\neq 0$. First we show that the second term in the action $S_E$ does not affect the equations of motion. In other words, the theta term is only a surface term. We define
 \begin{eqnarray}
{\cal L}_{\theta}=\frac{1}{16\pi^2} {\rm Tr} F_{\mu\nu}\tilde{F}_{\mu\nu}.
\end{eqnarray}
We compute the variation 
 \begin{eqnarray}
\delta {\cal L}_{\theta}&=&\frac{1}{16\pi^2} \epsilon_{\mu\nu\alpha\beta}{\rm Tr} F_{\mu\nu}\delta {F}_{\alpha\beta}\nonumber\\
&=&\frac{1}{8\pi^2} \epsilon_{\mu\nu\alpha\beta}{\rm Tr} F_{\mu\nu}D_{\alpha}\delta {A}_{\beta}.
\end{eqnarray}
We use the Jacobi identity 
 \begin{eqnarray}
\epsilon_{\mu\nu\alpha\beta}D_{\alpha}F_{\mu\nu}&=&\epsilon_{\mu\nu\alpha\beta}(\partial_{\alpha}F_{\mu\nu}-i[A_{\alpha},F_{\mu\nu}])\nonumber\\
&=&-\epsilon_{\mu\nu\alpha\beta}[A_{\alpha},[A_{\mu},A_{\nu}]]\nonumber\\
&=&0.
\end{eqnarray}
Thus
 \begin{eqnarray}
\delta {\cal L}_{\theta}
&=&\frac{1}{8\pi^2} \epsilon_{\mu\nu\alpha\beta}{\rm Tr}D_{\alpha}(F_{\mu\nu}\delta {A}_{\beta})\nonumber\\
&=&\frac{1}{8\pi^2} \epsilon_{\mu\nu\alpha\beta}{\rm Tr}\bigg(\partial_{\alpha}(F_{\mu\nu}\delta {A}_{\beta})-i[A_{\alpha},F_{\mu\nu}\delta {A}_{\beta}]\bigg)\nonumber\\
&=&\partial_{\alpha}\delta {\cal K}_{\alpha}.
\end{eqnarray}
 \begin{eqnarray}
\delta {\cal K}_{\alpha}=\frac{1}{8\pi^2}\epsilon_{\alpha\mu\nu\beta}{\rm Tr} F_{\mu\nu}\delta A_{\beta}.
\end{eqnarray}
This shows explicitly that the theta term will not contribute to the equations of motion for variations of the gauge field which vanish at infinity.

 In order to find the current ${\cal K}_{\alpha}$ itself we adopt the method of \cite{Polyakov:1987ez}. We consider a one-parameter family of gauge fields $A_{\mu}(x,\tau)=\tau A_{\mu}(x)$ with $0\leq \tau\leq 1$. By using the above result we have immediately 
\begin{eqnarray}
\frac{\partial}{\partial \tau} {\cal K}_{\alpha}&=&\frac{1}{8\pi^2}\epsilon_{\alpha\mu\nu\beta}{\rm Tr} F_{\mu\nu}(x,\tau)\frac{\partial}{\partial \tau} A_{\beta}\nonumber\\
&=&\frac{1}{8\pi^2}\epsilon_{\alpha\mu\nu\beta}{\rm Tr} \bigg(\tau \partial_{\mu}A_{\nu}-\tau\partial_{\nu}A_{\mu}-i\tau^2[A_{\mu},A_{\nu}]\bigg).A_{\beta}(x).
\end{eqnarray}
By integrating both sides with respect to $ \tau$ between $\tau=0$ and $\tau=1$ and setting ${\cal K}_{\alpha}(x,1)={\cal K}_{\alpha}(x)$ and  ${\cal K}_{\alpha}(x,0)=0$ we get
\begin{eqnarray}
{\cal K}_{\alpha}
&=&\frac{1}{8\pi^2}\epsilon_{\alpha\mu\nu\beta}{\rm Tr} \bigg(\frac{1}{2}\partial_{\mu}A_{\nu}-\frac{1}{2}\partial_{\nu}A_{\mu}-\frac{i}{3}[A_{\mu},A_{\nu}]\bigg).A_{\beta}(x).
\end{eqnarray}
The theta term is proportional to an integer $k$ (known variously as the Pontryagin class, the winding number, the instanton number and the topological charge) defined by
\begin{eqnarray}
k&=&\int d^4x {\cal L}_{\theta}\nonumber\\
&=&\int d^4x \partial_{\alpha} {\cal K}_{\alpha}.
\end{eqnarray}
Now we imagine that the four-dimensional Euclidean spacetime is bounded by a large three-sphere $S^3$ in the same way that we can imagine that the plane is bounded by a large $S^1$, viz
\begin{eqnarray}
\partial { R}^4=S^3_{\infty}.
\end{eqnarray}
Then
\begin{eqnarray}
k
&=&\int_{\partial{ R}^4=S_{\infty}^3} d^3\sigma_{\alpha} {\cal K}_{\alpha}\nonumber\\
&=&\frac{1}{16\pi^2} \epsilon_{\alpha\mu\nu\beta}\int_{\partial{ R}^4=S_{\infty}^3} d^3\sigma_{\alpha} {\rm Tr}\bigg[F_{\mu\nu}A_{\beta}+i\frac{2}{3}A_{\mu}A_{\nu}A_{\beta}\bigg].\label{winding}
\end{eqnarray}
The Chern-Simons action is defined by
\begin{eqnarray}
S_{\rm CS}=i\theta k.
\end{eqnarray}
A Yang-Mills instanton is a solution of the equations of motion which has finite action. In order to have a finite action the field strength $F_{\mu\nu}$ must approach $0$ at infinity at least as $1/x^2$, viz\footnote{The requirement of finite action can be neatly satisfied if we compactify ${ R}^4$ by adding one point at $\infty$ to obtain the four-sphere $S^4$.}

\begin{eqnarray}
F_{\mu\nu}^I(x)=o(1/x^2)~,~x\longrightarrow \infty.
\end{eqnarray}
We can immediately  deduce that the gauge field must approach a pure gauge at infinity, viz
\begin{eqnarray}
A_{\mu}^I(x)=ig^{-1}\partial_{\mu}g+o(1/x)~,~x\longrightarrow \infty.
\end{eqnarray}
This can be checked by simple substitution in $F_{\mu\nu}=\partial_{\mu}A_{\nu}-\partial_{\nu}A_{\mu}-i[A_{\mu},A_{\nu}]$. 
Now a gauge configuration $A_{\mu}^I(x)$ at infinity (on the sphere $S^3_{\infty}$) defines a group element $g$ which satisfies (from the above asymptotic behavior) the equation $\partial_{\mu}g^{-1}=iA_{\mu}^Ig^{-1}$ or equivalently 
\begin{eqnarray}
\frac{d}{ds}g^{-1}(x(s),x_0)=i\frac{dx^{\mu}}{ds}A_{\mu}^I(x(s))g^{-1}(x(s),x_0).
\end{eqnarray}
The solution is given by the path-ordered Wilson line
\begin{eqnarray}
g^{-1}(x,x_0)={\cal P}\exp\bigg(i\int_0^1 ds\frac{dy^{\mu}}{ds}A_{\mu}^I(y(s))\bigg).
\end{eqnarray}
The path is labeled by the parameter $s$ which runs from $s=0$ ($y=x_0$) to $s=1$ ($y=x$) and the path-ordering operator ${\cal P}$ is defined such that terms with higher values of $s$ are always put on the left in every order in the Taylor expansion of the exponential .

In the above formula for $g^{-1}$ the points $x$ and $x_0$ are both at infinity, i.e. on the sphere $S^3_{\infty}$. In other words gauge configurations with finite action   (the instanton configurations  $A_{\mu}^I(x)$) define a map from  $S^3_{\infty}$ into $G$, viz
\begin{eqnarray}
g^{-1}:S^3_{\infty}\longrightarrow G.
\end{eqnarray}
These maps are classified by homotopy theory. 

As an example we take the group $G=SU(2)$. The group $SU(2)$ is topologically a three-sphere since any element $g\in SU(2)$ can be expanded (in the fundamental representation) as $g=n_4+i\vec{n}\vec{\tau}$ and as a consequence the unitarity condition $g^+g=1$ becomes $n_4^2+\vec{n}^2=1$. In this case we have therefore maps from the three-sphere to the three-sphere, viz

\begin{eqnarray}
g^{-1}:S^3_{\infty}\longrightarrow SU(2)=S^3.
\end{eqnarray}
These maps are characterized precisely by the integer $k$ introduced above. 
This number measures how many times the second $S^3$ (group) is wrapped  (covered) by the first sphere $S_{\infty}^3$ (space). In fact this is the underlying reason why $k$ must be quantized. In other words $k$ is an element of the third homotopy group $\pi_3(S^3)$, viz \footnote{In general $\pi_n(S^n)=Z$. It is obvious that $\pi_1(S^1)=\pi_2(S^2)=Z$.}
  \begin{eqnarray}
k\in \pi_3(SU(2))=\pi_3(S^3)=Z.
\end{eqnarray}
For general $SU(N)$ we consider instanton configurations obtained by embedding the $SU(2)$ instanton configurations into $SU(N)$ matrices as
\begin{eqnarray}
A_{\mu}^{SU(N)}= \left( \begin{array}{cc}
0 &  0 \\
0 & A_{\mu}^{SU(2)}
\end{array} \right). 
\end{eqnarray}
We can obviously use any spin $j$ representation of $SU(2)$  provided it fits inside the $N\times N$ matrices of $SU(N)$. The case $N=2j+1$ is equivalent to choosing the generators of $SU(2)$ in the spin $j$ representation as the first $3$ generators of $SU(N)$ and hence $A_{\mu}^{SU(N) a}$, $a=1,2,3$ are given by the $SU(2)$ instanton configurations whereas the other components  $A_{\mu}^{SU(N) a}$, $a=4,...,N^2-1$ are zero identically. The explicit constructions of all these instanton solutions will not be given here.

The story of instanton calculus is beautiful but long and complicated and we can only here refer the reader to the vast literature on the subject. See for example the pedagogical lectures \cite{Vandoren:2008xg}.

We go back to the main issue for us which is the zero  dimensional reduction of the Chern-Simons term. By using the fact that on $S_{\infty}^3$ we have $F_{\mu\nu}=0$ we can rewrite  (\ref{winding}) as
\begin{eqnarray}
k
&=&\frac{i}{24\pi^2} \epsilon_{\alpha\mu\nu\beta}\int_{\partial{ R}^4=S_{\infty}^3} d^3\sigma_{\alpha} {\rm Tr}A_{\mu}A_{\nu}A_{\beta}.
\end{eqnarray}
By using also the fact that $A_{\mu}=A_{\mu}^I=ig^{-1}\partial_{\mu}g=iX_{\mu}$ on $S_{\infty}^3$ we have
\begin{eqnarray}
k
&=&\frac{1}{24\pi^2} \epsilon_{\alpha\mu\nu\beta}\int_{\partial{ R}^4=S_{\infty}^3} d^3\sigma_{\alpha} {\rm Tr}X_{\mu}X_{\nu}X_{\beta}.
\end{eqnarray}
By introducing now a local parametrization $\xi_a=\xi_a(x)$ of the $G$ group elements we can rewrite $k$ as (with $X_{a}=g^{-1}\partial_{a}g$)
 \begin{eqnarray}
k
&=&\frac{1}{24\pi^2} \epsilon_{\alpha \mu\nu\beta}\int_{\partial{ R}^4=S_{\infty}^3} d^3\sigma_{\alpha} \frac{\partial \xi_a}{\partial x_{\mu}}\frac{\partial \xi_b}{\partial x_{\nu}}\frac{\partial \xi_c}{\partial x_{\beta}}{\rm Tr}X_{a}X_{b}X_{c}.\nonumber\\
\end{eqnarray}
Next we use
\begin{eqnarray}
d^3\sigma_{\alpha}=\frac{1}{6}\epsilon_{\alpha\mu\nu\beta}dx_{\mu}\wedge dx_{\nu}\wedge dx_{\beta}.
\end{eqnarray}
\begin{eqnarray}
\epsilon_{\alpha\mu\nu\beta}\epsilon_{\alpha\mu^{'}\nu^{'}\beta^{'}}=\delta_{[\mu\nu\beta]}^{\mu^{'}\nu^{'}\beta^{'}}=\delta^{\mu^{'}}_{\mu}(\delta^{\nu^{'}}_{\nu}\delta^{\beta^{'}}_{\beta}-\delta^{\nu^{'}}_{\beta}\delta^{\beta^{'}}_{\nu})+\delta^{\mu^{'}}_{\nu}(\delta^{\nu^{'}}_{\beta}\delta^{\beta^{'}}_{\mu}-\delta^{\nu^{'}}_{\mu}\delta^{\beta^{'}}_{\beta})+\delta^{\mu^{'}}_{\beta}(\delta^{\nu^{'}}_{\mu}\delta^{\beta^{'}}_{\nu}-\delta^{\nu^{'}}_{\nu}\delta^{\beta^{'}}_{\mu}).
\end{eqnarray}
We get
\begin{eqnarray}
k
&=&\frac{1}{24\pi^2} \frac{1}{6}\delta_{[\mu\nu\beta]}^{\mu^{'}\nu^{'}\beta^{'}} \int_{\partial{ R}^4=S_{\infty}^3}dx_{\mu^{'}}\wedge dx_{\nu^{'}}\wedge dx_{\beta^{'}}  \frac{\partial \xi_a}{\partial x_{\mu}}\frac{\partial \xi_b}{\partial x_{\nu}}\frac{\partial \xi_c}{\partial x_{\beta}}{\rm Tr}X_{a}X_{b}X_{c}\nonumber\\
&=&\frac{1}{24\pi^2} \int_{\partial{ R}^4=S_{\infty}^3}d\xi_{a}\wedge d\xi_{b}\wedge d\xi_{c}  {\rm Tr}X_{a}X_{b}X_{c}\nonumber\\
&=&\frac{1}{24\pi^2} \int_{\partial{ R}^4=S_{\infty}^3}d^3\xi \epsilon_{abc} {\rm Tr}X_{a}X_{b}X_{c}.
\end{eqnarray}
The trace ${\rm Tr}$ is generically $(2j+1)-$dimensional, and not $N-$dimensional, corresponding to the spin $j$ representation of $SU(2)$. The Chern-Simons action becomes
\begin{eqnarray}
S_{\rm CS}
&=&\frac{i\theta}{24\pi^2} \int_{\partial{ R}^4=S_{\infty}^3}d^3\xi \epsilon_{abc} {\rm Tr}X_{a}X_{b}X_{c}.
\end{eqnarray}
As before we can reduce to zero dimension  by assuming that the configurations $X_a$ are constant. We obtain immediately 
\begin{eqnarray}
S_{\rm CS}
&=&\frac{i\theta V_{S^3}}{24\pi^2}  \epsilon_{abc} {\rm Tr}X_{a}X_{b}X_{c}.\label{cs1}
\end{eqnarray}
By putting (\ref{ym1}) and (\ref{cs1}) we obtain the matrix action
\begin{eqnarray}
S_{E}
&=&-\frac{V_{R^4}}{2g^2} {\rm Tr}[X_{\mu},X_{\nu}]^2+\frac{i\theta V_{S^3}}{24\pi^2}  \epsilon_{abc} {\rm Tr}X_{a}X_{b}X_{c}.
\end{eqnarray}
We choose to perform the scaling
\begin{eqnarray}
X_{\mu}\longrightarrow \bigg(\frac{Ng^2}{2V_{R^4}}\bigg)^{1/4} X_{\mu}.
\end{eqnarray}
The action becomes
\begin{eqnarray}
S_{E}
&=&-\frac{N}{4} {\rm Tr}[X_{\mu},X_{\nu}]^2+i\frac{2N\alpha}{3}  \epsilon_{abc} {\rm Tr}X_{a}X_{b}X_{c}.
\end{eqnarray}
The new coupling constant $\alpha$ is given by
\begin{eqnarray}
\alpha=\frac{1}{16\pi^2}\frac{\theta  V_{S^3}}{N}  \bigg(\frac{Ng^2}{2V_{R^4}}\bigg)^{3/4}.
\end{eqnarray}
\section{Metropolis Accept/Reject Step}
In the remainder we only consider the basic Yang-Mills matrix action to be of interest. This is given by
\begin{eqnarray}
S_{\rm YM}[X]&=&-\frac{N}{4}Tr[X_{\mu},X_{\nu}]^2\nonumber\\
&=&-N\sum_{\mu=1}^d\sum_{\nu=\mu+1}^d(X_{\mu}X_{\nu}X_{\mu}X_{\nu}-X_{\mu}^2X_{\nu}^2).
\end{eqnarray}
The path integral or partition function of this model is given by
\begin{eqnarray}
Z=\int \prod_{\mu}dX_{\mu}\exp(-S_{\rm YM}).
\end{eqnarray}
The meaning of the meausre is obvious since $X_{\mu}$ are $N\times N$ matrices. The corresponding probability distribution for the matrix configurations $X_{\mu}$ is given by
\begin{eqnarray}
P(X)=\frac{1}{Z}\exp(-S_{\rm YM}[X]).
\end{eqnarray}
We want to sample this probability distribution in Monte Carlo using the Metropolis algorithm. Towards this end, we need to compute the variation of the action under the following arbitrary change 
\begin{eqnarray}
X_{\lambda}\longrightarrow X_{\lambda}^{'}=X_{\lambda}+\Delta X_{\lambda},
\end{eqnarray}
where
\begin{eqnarray}
(\Delta X_{\lambda})_{nm}=d\delta_{ni}\delta_{mj}+d^*\delta_{nj}\delta_{mi}.
\end{eqnarray}
The corresponding variation of the action is
\begin{eqnarray}
\Delta S_{\rm YM}&=&\Delta S_1+\Delta S_2.
\end{eqnarray}
The two pieces $\Delta S_1$ and $\Delta S_2$ are given respectively by
\begin{eqnarray}
\Delta S_1&=&-N \sum_{\sigma}Tr[X_{\sigma},[X_{\lambda},X_{\sigma}]]\Delta X_{\lambda}\nonumber\\
&=&-N d\sum_{\sigma}[X_{\sigma},[X_{\lambda},X_{\sigma}]]_{ji}-N d^*\sum_{\sigma}[X_{\sigma},[X_{\lambda},X_{\sigma}]]_{ij}.
\end{eqnarray}
\begin{eqnarray}
\Delta S_2&=&-\frac{N}{2}\sum_{\sigma\neq\lambda}[\Delta X_{\lambda},X_{\sigma}]^2\nonumber\\
&=&-\frac{N}{2} d\sum_{\sigma\ne\lambda}[X_{\sigma},[\Delta X_{\lambda},X_{\sigma}]]_{ji}-\frac{N}{2} d^*\sum_{\sigma\ne \lambda}[X_{\sigma},[\Delta X_{\lambda},X_{\sigma}]]_{ij}\nonumber\\
&=&-N\sum_{\sigma\ne\lambda}\bigg[d^2(X_{\sigma})_{ji}(X_{\sigma})_{ji}+(d^*)^2(X_{\sigma})_{ij}(X_{\sigma})_{ij}+2dd^*(X_{\sigma})_{ii}(X_{\sigma})_{jj}-dd^*\big((X_{\sigma}^2)_{ii}+(X_{\sigma}^2)_{jj}\big)\nonumber\\
&-&\frac{1}{2}(d^2+(d^*)^2)\big((X_{\sigma}^2)_{ii}+(X_{\sigma}^2)_{jj}\big)\delta_{ij}\bigg].
\end{eqnarray}
The Metropolis accept/reject step is based on the probability distribution
\begin{eqnarray}
P[X]={\rm min}(1,\exp(-\Delta S_{\rm YM}).
\end{eqnarray}
It is not difficult to show that this probability distribution satisfies detailed balance, and as a consequence, this algorithm is exact, i.e. free from systematic errors.
\section{Statistical Errors}
We use the Jacknife method to estimate statistical errors. Given a set of $T=2^P$ ( with $P$ some integer ) data points $f(i)$ we proceed by removing $z$ elements from the set in such a way that we end up with $n=T/z$ sets ( or bins). The minimum number of data points we can remove is $z=1$ and the maximum number  is $z=T-1$. The average of the elements of the $i$th bin is
\begin{eqnarray}
< y(j)>_i=\frac{1}{T-z}\bigg(\sum_{j=1}^{T}f(j)-\sum_{j=1}^{z}f((i-1)z+j )\bigg)~,~i=1,n.
\end{eqnarray}
For a fixed partition given by $z$ the corresponding error is  computed as follows
\begin{eqnarray}
e(z)=\sqrt{\frac{n-1}{n}\sum_{i=1}^{n}(<y(j)>_i-<f>)^2}~,~<f>=\frac{1}{T}\sum_{j=1}^Tf(j).
\end{eqnarray}
We start with $z=1$ and we compute the error $e(1)$  then we go to $z=2$ and compute the error $e(2)$. The true error is the largest value. Then we go to $z=3$, compute  $e(3)$, compare it with the previous error and again retain the largest value  and so on until we reach $z=T-1$.
\section{Auto-Correlation Time}
In any given ergodic process we obtain a sequence (Markov chain) of field/matrix configurations $ \phi_1$, $\phi_2$,....,$\phi_T$. We will assume that $\phi_i$ are thermalized configurations. Let $f$ some (primary) observable with values $f_i\equiv f(\phi_i)$ in the configurations $\phi_i$ respectively. The average value $<f>$ of $f$ and the statistical error $\delta f$ are given by the usual formulas
\begin{eqnarray}
<f>=\frac{1}{T}\sum_{i=1}^Tf_i.
\end{eqnarray}
\begin{eqnarray}
\delta f=\frac{\sigma}{\sqrt{T}}.
\end{eqnarray}
The standard deviation (the variance) is given by 
\begin{eqnarray}
\sigma^2=<f^2>-<f>^2.
\end{eqnarray}
The above theoretical estimate of the error is valid provided the thermalized configurations  $ \phi_1$, $\phi_2$,....,$\phi_T$ are statistically uncorrelated, i.e. independent. In real simulations, this is certainly not the case. In general, two consecutive configurations will be dependent, and the average number of configurations which separate two really uncorrelated configurations is called the auto-correlation time. The correct estimation of the error must depend on the auto-correlation time.

We define the auto-correlation function $\Gamma_j$ and the normalized auto-correlation function $\rho_j$ for the observable $f$ by
\begin{eqnarray}
\Gamma_j=\frac{1}{T-j}\sum_{i=1}^{T-j}(f_i-<f>)(f_{i+j}-<f>).
\end{eqnarray}
\begin{eqnarray}
\rho_j=\frac{\Gamma_j}{\Gamma_0}.
\end{eqnarray}
These function vanish if there is no auto-correlation. Obviously $\Gamma_0$ is the variance $\sigma^2$, viz $\Gamma_0=\sigma^2$. In the generic case, where the auto-correlation function is not zero, the statistical error in the average $<f>$ will be given by
\begin{eqnarray}
\delta f=\frac{\sigma}{\sqrt{T}}\sqrt{2\tau_{\rm int}}.
\end{eqnarray}
The so-called integrated auto-correlation time $\tau_{\rm int}$ is given in terms of the normalized  auto-correlation function $\rho_j$  by
\begin{eqnarray}
\tau_{\rm int}=\frac{1}{2}+\sum_{j=1}^{\infty}\rho_j.
\end{eqnarray}
The  auto-correlation function $\Gamma_j$, for large $j$,  can not be precisely determined, and hence, one must truncate the sum over $j$ in $\tau_{\rm int}$ at some cut-off $M$, in order to not increase the error $\delta \tau_{\rm int}$ in  $\tau_{\rm int}$ by simply summing up noise. The integrated auto-correlation time $\tau_{\rm int}$ should then be defined by
\begin{eqnarray}
\tau_{\rm int}=\frac{1}{2}+\sum_{j=1}^{M}\rho_j.
\end{eqnarray}
The value $M$ is chosen as the first integer between $1$ and $T$ such that
\begin{eqnarray}
M\geq 4\tau_{\rm int}+1.
\end{eqnarray}
The error  $\delta \tau_{\rm int}$ in  $\tau_{\rm int}$ is given by
\begin{eqnarray}
\delta \tau_{\rm int}=\sqrt{\frac{4M+2}{T}}\tau_{\rm int}.
\end{eqnarray}
This formalism can be generalized to secondary observables $F$ which are functions of $n$ primary observables $f^{\alpha}$, viz $F=F(f^1,f^2,...,f^n)$. See for example \cite{schaefer}. 

In general two among the three parameters of the molecular dynamics (the time step $dt$, the number of iterations $n$ and the time interval $T=ndt$) should be optimized in such a way that the acceptance rate is fixed, for example, between $70$ and $90$ per cent. We  fix $n$ and optimize $dt$ along the line discussed  in previous chapters. We make, for every $N$,  a  reasonable  guess for the value of the number of iterations $n$, based on trial and error, and then work with that value throughout. For example, for $N$ between $N=4$ and $N=8$, we found the value $n=10$, to be sufficiently reasonable. 

\section{Code and Sample Calculation}
Typically, we run $T_{\rm ther}+T_{\rm meas}$ Monte Carlo steps where thermalization is supposed to occur within the first $T_{\rm ther}$ steps, which are then discarded, while measurements are performed on a sample consisting of the subsequent $T_{\rm meas}$ configurations. We choose, for $N=4-8$, $T_{\rm ther}=2^{11}$ and $T_{\rm meas}=2^{11}$.  The interval from which we draw the variations $d$ and $d^*$ is updated after each Metropolis step by requiring that the acceptance rate is fixed  between $25$ and $30$ per cent. We generate our random numbers using the algorithm ran2. We do not discuss auto-correlations while error bars are estimated using the jackknife method as discussed above. A FORTRAN code along these lines is included in the last chapter for illustrative purposes. This seems to go as fast as $N^4$.

\begin{figure}[htbp]
\begin{center}
\includegraphics[width=9.0cm,angle=-0]{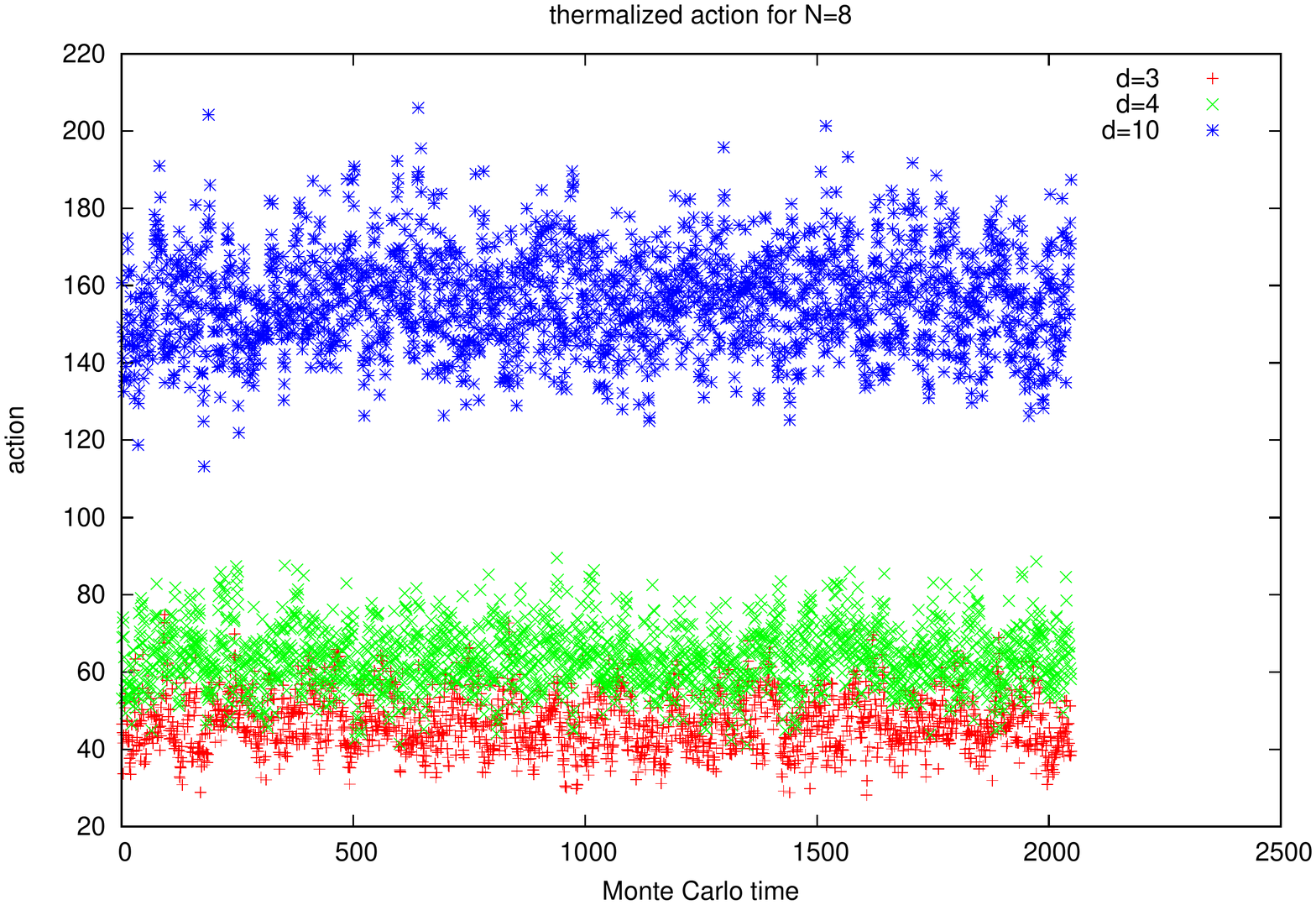}
\includegraphics[width=9.0cm,angle=-0]{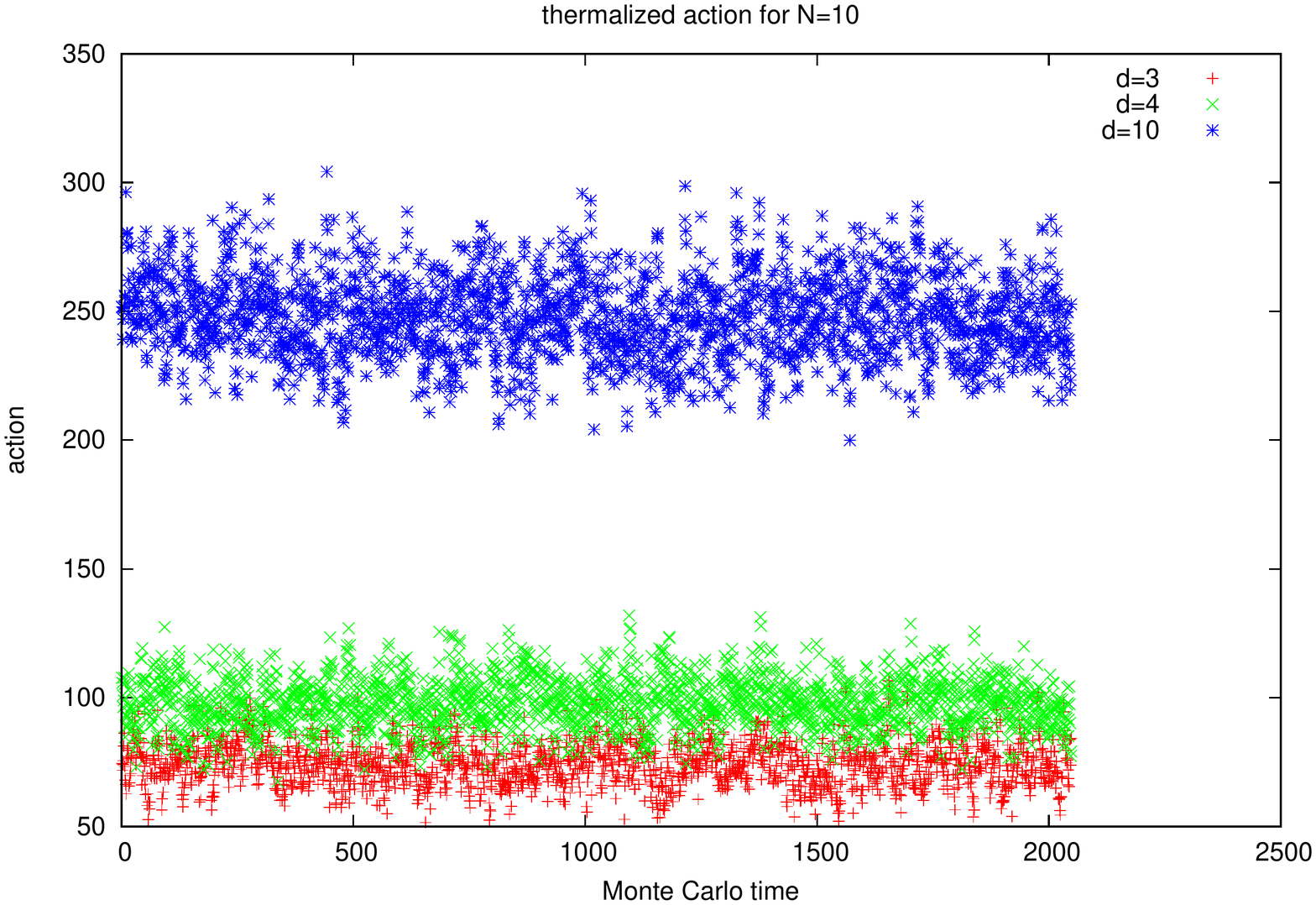}
\includegraphics[width=9.0cm,angle=-0]{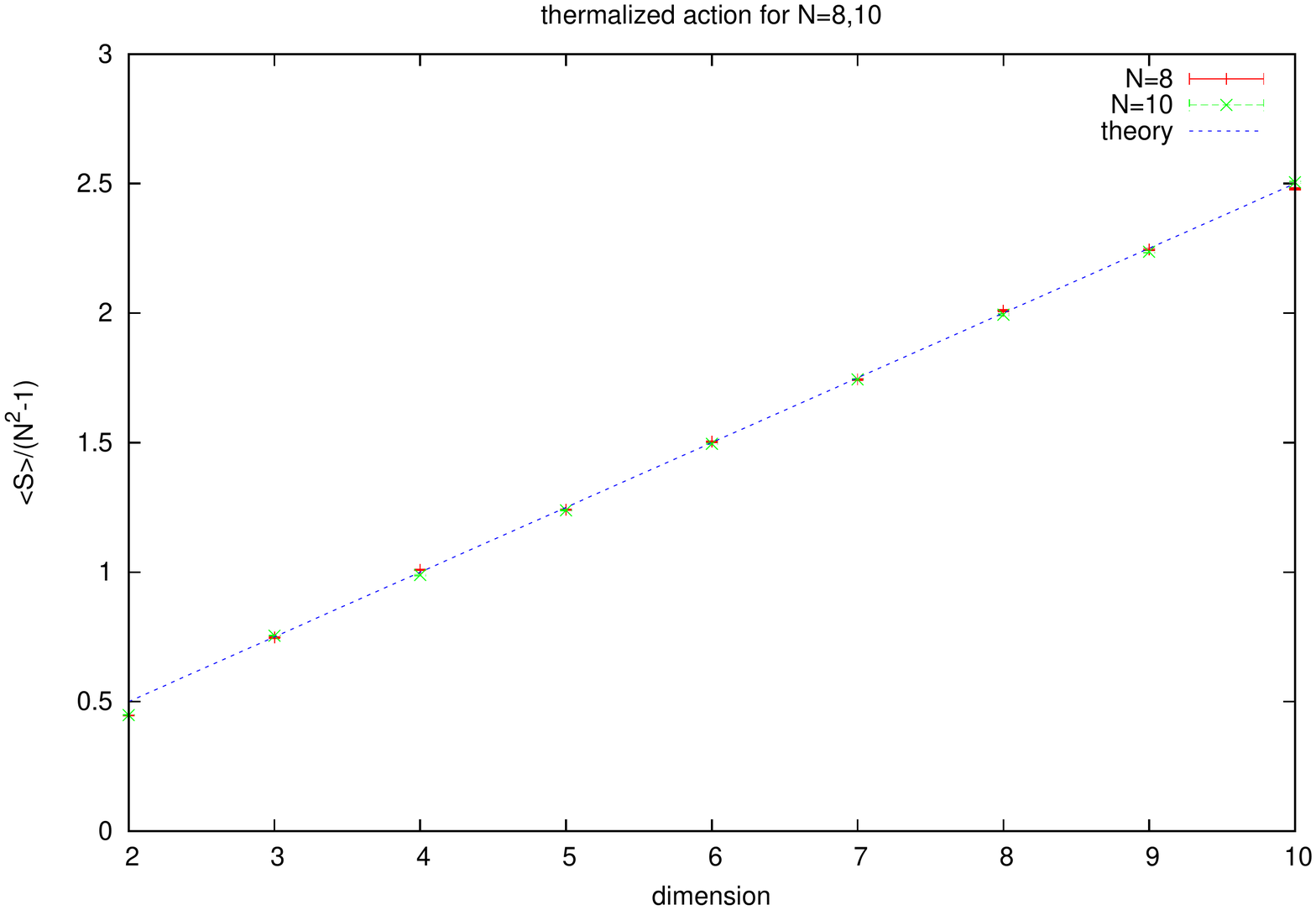}
\end{center}
\caption{}\label{testM}
\end{figure}

Some thermalized results for $N=8,10$, for dimensions between $d=2$ and $d=10$, are shown on figure (\ref{testM}). The observed linear fit for the average action is in excellent agreement with the exact analytic result

\begin{eqnarray}
\frac{<S>}{N^2-1}=\frac{d}{4}.
\end{eqnarray}
This identity follows from the invariance of the path integral under the translations $X_{\mu}\longrightarrow X_{\mu}+\epsilon X_{\mu}$. 

\chapter{Hybrid Monte Carlo Algorithm for Yang-Mills Matrix Models}
\section{The Yang-Mills Matrix Action} 
The hybrid Monte Carlo algorithm is a combination of the molecular dynamics method and the Metropolis algorithm. In this section we will follow \cite{Montvay:1994cy,Rothe:2005nw} and \cite{Ambjorn:2000bf1,Ambjorn:2000dx1, Anagnostopoulos:2005cy1}.

We are still interested in the Euclidean Yang-Mills matrix model
\begin{eqnarray}
S_{\rm YM}&=&-\frac{N\gamma}{4}\sum_{\mu,\nu=1}^dTr[X_{\mu},X_{\nu}]^2+V(X).
\end{eqnarray}
$\gamma$ is some parameter, and $V$ is some $U(N)-$invariant potential in the $d$ matrices $X_{\mu}$. In this chapter we will take a potential consisting of a harmonic oscillator term and a Chern-Simons term in the three directions $X_1$, $X_2$ and $X_3$ given by
\begin{eqnarray}
V=\frac{1}{2}m^2 Tr X_{\mu}^2+\frac{2N i\alpha}{3}\epsilon_{abc}Tr X_aX_bX_c.
\end{eqnarray}
The path integral we wish to sample in Monte Carlo simulation is

\begin{eqnarray}
Z_{\rm YM}=\int \prod_{\mu=1}^d dX_{\mu}~ \exp(-S_{\rm YM}[X]).
\end{eqnarray}
Firstly, we will think of the gauge configurations $X_{\mu}$ as evolving in some  fictitious time-like parameter $t$, viz
\begin{eqnarray}
X_{\mu}\equiv X_{\mu}(t).
\end{eqnarray}
The above path integral is then equivalent to the Hamiltonian dynamical system
\begin{eqnarray}
Z_{\rm YM}=\int \prod_{\mu}dP_{\mu} \prod_{\mu}dX_{\mu}~ \exp(-\frac{1}{2}\sum_{\mu=1}^dTr P_{\mu}^2-S_{\rm YM}[X]).
\end{eqnarray}
In other words, we have introduced $d$ Hermitian matrices $P_{\mu}$ which are obviously $N\times N$, and which are conjugate to $X_{\mu}$. The Hamiltonian is clearly given by 
\begin{eqnarray}
H=\frac{1}{2}\sum_{\mu=1}^dTr P_{\mu}^2+S_{\rm YM}[X].
\end{eqnarray}
In summary, we think of the matrices $X_{\mu}$ as  fields in one dimension with corresponding
conjugate momenta $P_{\mu}$. The Hamiltonian equations of motion read
 \begin{eqnarray}
\frac{\partial H}{\partial (P_{\mu})_{ij}}=(\dot{X}_{\mu})_{ij}~,~\frac{\partial H}{\partial (X_{\mu})_{ij}}=-(\dot{P}_{\mu})_{ij}.
\end{eqnarray}
We have then the equations of motion
 \begin{eqnarray}
(P_{\mu})_{ji}=(\dot{X}_{\mu})_{ij}.
\end{eqnarray}
 \begin{eqnarray}
\frac{\partial S_{\rm YM}}{\partial (X_{\mu})_{ij}}=-N\gamma\sum_{\nu=1}^d[X_{\nu},[X_{\mu},X_{\nu}]]_{ji}+\frac{\partial V}{\partial (X_{\mu})_{ij}}=-(\dot{P}_{\mu})_{ij}.
\end{eqnarray}
We will define 
 \begin{eqnarray}
(V_{\mu})_{ij}(t)&=&\frac{\partial S_{\rm YM}}{\partial (X_{\mu})_{ij}(t)}\nonumber\\
&=&-N\gamma\sum_{\nu=1}^d[X_{\nu},[X_{\mu},X_{\nu}]]_{ji}+\frac{\partial V}{\partial (X_{\mu})_{ij}}\nonumber\\
&=&-N\gamma\bigg(2X_{\nu}X_{\mu}X_{\nu}-X_{\nu}^2X_{\mu}-X_{\mu}X_{\nu}^2\bigg)_{ji}+m^2(X_{\mu})_{ji}\nonumber\\
&+&2i\alpha N[X_2,X_3]_{ji}\delta_{\mu 1}+2i\alpha N[X_3,X_1]_{ji}\delta_{\mu 2}+2i\alpha N[X_1,X_2]_{ji}\delta_{\mu 3}.
\end{eqnarray}
\section{The Leap Frog Algorithm} The first task we must face up with is to solve the above differential equations.

The numerical solution of these differential equations is formulated as follows. We consider Taylor expansions of $(X_{\mu})_{ij}(t+\delta t)$ and $(P_{\mu})_{ij}(t+\delta t)$ up to order $\delta t^2$ given by
 \begin{eqnarray}
(X_{\mu})_{ij}(t+\delta t)=(X_{\mu})_{ij}(t)+\delta t (\dot{X}_{\mu})_{ij}(t)+\frac{\delta t^2}{2}(\ddot{X}_{\mu})_{ij}(t)+...
\end{eqnarray}
 \begin{eqnarray}
(P_{\mu})_{ij}(t+\delta t)=(P_{\mu})_{ij}(t)+\delta t (\dot{P}_{\mu})_{ij}(t)+\frac{\delta t^2}{2}(\ddot{P}_{\mu})_{ij}(t)+...
\end{eqnarray}
We calculate that

\begin{eqnarray}
(\ddot{X}_{\mu})_{ij}=(\dot{P}_{\mu})_{ji}&=&-\frac{\partial S_{\rm YM}}{\partial (X_{\mu})_{ji}}\nonumber\\
&=&N\sum_{\nu=1}^d[X_{\nu},[X_{\mu},X_{\nu}]]_{ij}-\frac{\partial V}{\partial (X_{\mu})_{ji}}.
\end{eqnarray}
 \begin{eqnarray}
(\ddot{P}_{\mu})_{ij}&=&-\sum_{kl,\nu}\frac{\partial^2 S_{\rm YM}}{\partial (X_{\nu})_{kl}\partial (X_{\mu})_{ij}}(\dot{X}_{\nu})_{kl}\nonumber\\
&=&N\sum_{\nu=1}^d\bigg([P_{\nu}^T,[X_{\mu},X_{\nu}]]+[X_{\nu},[P_{\mu}^T,X_{\nu}]]+[X_{\nu},[X_{\mu},P_{\nu}^T]]\bigg)_{ji}-\sum_{kl,\nu}\frac{\partial^2 V}{\partial (X_{\nu})_{kl}\partial (X_{\mu})_{ij}}(\dot{X}_{\nu})_{kl}.\nonumber\\
\end{eqnarray}
For generic non-local potentials $V$ the second equation will be approximated by

 \begin{eqnarray}
(\ddot{P}_{\mu})_{ij}&=&\frac{(\dot{P}_{\mu})_{ij}(t+\delta t)-(\dot{P}_{\mu})_{ij}(t)}{\delta t}\nonumber\\
&=&-\frac{1}{\delta t}\bigg(\frac{\partial S_{\rm YM}}{\partial (X_{\mu})_{ij}(t+\delta t)}-\frac{\partial S_{\rm YM}}{\partial (X_{\mu})_{ij}(t)}\bigg).
\end{eqnarray}
 Taylor expansions of $(X_{\mu})_{ij}(t+\delta t)$ and $(P_{\mu})_{ij}(t+\delta t)$ become
\begin{eqnarray}
(X_{\mu})_{ij}(t+\delta t)=(X_{\mu})_{ij}(t)+\delta t (P_{\mu})_{ji}(t)-\frac{\delta t^2}{2}\frac{\partial S_{\rm YM}}{\partial (X_{\mu})_{ji}(t)}+...
\end{eqnarray}
 \begin{eqnarray}
(P_{\mu})_{ij}(t+\delta t)=(P_{\mu})_{ij}(t)-\frac{\delta t}{2} \frac{\partial S_{\rm YM}}{\partial (X_{\mu})_{ij}(t)}-\frac{\delta t}{2}\frac{\partial S_{\rm YM}}{\partial (X_{\mu})_{ij}(t+\delta t)}+...
\end{eqnarray}
We write these two equations as the three equations
\begin{eqnarray}
(P_{\mu})_{ij}(t+\frac{\delta t}{2})=(P_{\mu})_{ij}(t)-\frac{\delta t}{2}\frac{\partial S_{\rm YM}}{\partial (X_{\mu})_{ij}(t)}.
\end{eqnarray}
\begin{eqnarray}
(X_{\mu})_{ij}(t+\delta t)=(X_{\mu})_{ij}(t)+\delta t (P_{\mu})_{ji}(t+\frac{\delta t}{2}).
\end{eqnarray}
 \begin{eqnarray}
(P_{\mu})_{ij}(t+\delta t)=(P_{\mu})_{ij}(t+\frac{\delta t}{2})-\frac{\delta t}{2}\frac{\partial S_{\rm YM}}{\partial (X_{\mu})_{ij}(t+\delta t)}.
\end{eqnarray}
By construction $ (X_{\mu})_{ij}(t+\delta t)$ and $(P_{\mu})_{ij}(t+\delta t)$ solve Hamilton equations. 

What we have done here is to integrate Hamilton equations of motion according to the so-called leap-frog algorithm. The main technical point to note is that the coordinates $ (X_{\mu})_{ij}$ at time $t+\delta t$ are computed in terms of the coordinates  $ (X_{\mu})_{ij}$ at time $t$ and the conjugate momenta $(P_{\mu})_{ij}$ not at time $t$ but at time $t+\delta t/2$. The conjugate momenta $(P_{\mu})_{ij}$ at time $t+\delta t$ are then computed using the new coordinates $ (X_{\mu})_{ij}$ at time $t+\delta t$ and the conjugate momenta $(P_{\mu})_{ij}$ at time $t+\delta t/2$. The conjugate momenta $(P_{\mu})_{ij}$ at time $t+\delta t/2$ are computed first in terms of the  coordinates $ (X_{\mu})_{ij}$ and the conjugate momenta $(P_{\mu})_{ij}$ at time $t$.

We consider a lattice of points $t =n {\delta}t$, $n=0,1,2,...,\nu-1,\nu$ where $(X_{\mu})_{ij}(t)=(X_{\mu})_{ij}(n)$ and $(P_{\mu})_{ij}(t)=(P_{\mu})_{ij}(n)$. The point $n=0$ corresponds to  the initial configuration $(X_{\mu})_{ij}(0)=(X_{\mu})_{ij}$ whereas $n=\nu$ corresponds to  the final configuration $(X_{\mu})_{ij}(T)=(X_{\mu})_{ij}^{'}$ where $T=\nu \delta t$. The momenta $(P_{\mu})_{ij}(t)$ at the  middle points $n+1/2$, $n=0,...,\nu-1$ will be denoted by $(P_{\mu})_{ij}(n+1/2)$. The above equations take then the form 

\begin{eqnarray}
(P_{\mu})_{ij}(n+\frac{1}{2})=(P_{\mu})_{ij}(n)-\frac{\delta t}{2}(V_{\mu})_{ij}(n).
\end{eqnarray}

\begin{eqnarray}
(X_{\mu})_{ij}(n+1)=(X_{\mu})_{ij}(n)+\delta t (P_{\mu})_{ji}(n+\frac{1}{2}).
\end{eqnarray}
 \begin{eqnarray}
(P_{\mu})_{ij}(n+1)=(P_{\mu})_{ij}(n+\frac{1}{2})-\frac{\delta t}{2}(V_{\mu})_{ij}(n+1).
\end{eqnarray}
This algorithm applied to the solution of the equations of motion is essentially the molecular dynamics method.
\section{Metropolis Algorithm}
Along any classical trajectory we know that:
\begin{itemize}
\item{$1)$} The Hamiltonian is invariant.
\item{$2)$} The motion is reversible in phase space.
\item{$3)$} The phase space volume is preserved defined by the condition
\begin{eqnarray}
\frac{{\partial}(X(\tau),P(\tau))}{{\partial}(X(0),P(0))}=1.
\end{eqnarray}
\end{itemize} 
In other words detailed balance holds along a classical trajectory . The leap-frog method used to solve the above differential equations maintains only the last two properties. The violation of the first property introduces systematic errors and as a consequence detailed balance is violated. It is a well established fact that  introducing a Metropolis accept/reject step at the end of each classical trajectory will eliminate the systematic error completely. The algorithm becomes therefore exact and it is known-together with the initial generation  of the $P$'s according to the Gaussian distribution-as the hybrid Monte Carlo algorithm. The hybrid algorithm is the  hybrid Monte Carlo algorithm in which the Metropolis accept/reject step is omitted. 

The difference between the hybrid algorithm and the ordinary molecular dynamics algorithm is that in the hybrid algorithm we refresh the momenta $(P_{\mu})_{ij}(t)$ at the beginning  of each molecular dynamics trajectory in such a way that they are chosen from a Gaussian ensemble. In this way we avoid the ergodicity problem. 

The hybrid Monte Carlo algorithm can be summarized as follows:  
\begin{itemize}
\item{$1)$} Choose an initial configuration $X_{\mu}=X_{\mu}(0)$.
\item{$2)$}Choose $P_{\mu}=P_{\mu}(0)$ according to the Gaussian probability distribution $\exp(-\frac{1}{2}TrP_{\mu}^2)$.
\item{$3)$}Find the configuration $(X_{\mu}^{'},P_{\mu}^{'})$ by solving the above differential equations of motion, i.e. $(X_{\mu}^{'},P_{\mu}^{'})=(X_{\mu}(T),P_{\mu}(T))$.
\item{$4)$}Accept the configuration  $(X_{\mu}^{'},P_{\mu}^{'})$ with a probability ${\rm min}(1,e^{-\Delta H[X,P]})$ where $\Delta H$ is the change in the Hamiltonian..
\item{$5)$} Go back to step $2$ and repeat.
\end{itemize}
Steps $2-4$ consists one sweep or one unit of Hybrid Monte Carlo time. The Metropolis accept/reject step guarantees detailed balance of this algorithm and absence of systematic errors which are caused by the non-invariance of the Hamiltonian due to the discretization.  

\section{Gaussian Distribution}
We have
\begin{eqnarray}
\int dP_{\mu}~e^{-\frac{1}{2}Tr P_{\mu}^2}=\int d(P_{\mu})_{ii}e^{-\frac{1}{2}\sum_{\mu}\sum_i(P_{\mu})^2_{ii}}\int d(P_{\mu})_{ij}d(P_{\mu})_{ij}^*~e^{-\sum_{\mu}\sum_{i}\sum_{j=i+1}(P_{\mu})_{ij}(P_{\mu})_{ij}^*}.
\end{eqnarray}
We are therefore interested in the probability distribution 
\begin{eqnarray}
\int dx~e^{-\frac{1}{2}a x^2},
\end{eqnarray}
where $a=1/2$ for diagonal and $a=1$ for off-diagonal. By squaring and including normalization we have 
\begin{eqnarray}
\frac{a}{\pi}\int dx dy~e^{-\frac{1}{2}a (x^2+y^2)}=\int_0^1 dt_1\int_0^1 dt_2.
\end{eqnarray}
\begin{eqnarray}
t_1=\frac{\phi}{2\pi}~,~t_2=e^{-ar^2}.
\end{eqnarray}
We generate therefore two uniform random numbers $t_1$ and $t_2$ and write down for diagonal elements $(P_{\mu})_{ii}$ the following equations 
\begin{eqnarray}
&&\phi=2\pi t_1\nonumber\\
&& r=\sqrt{-2\ln (1-t_2)}\nonumber\\
&& (P_{\mu})_{ii}=r\cos\phi.
\end{eqnarray}
For off-diagonal elements $P_{ij}$ we write the following equations 
\begin{eqnarray}
&&\phi=2\pi t_1\nonumber\\
&& r=\sqrt{-\ln (1-t_2)}\nonumber\\
&& (P_{\mu})_{ij}=r\cos\phi +i r\sin\phi\nonumber\\
&& (P_{\mu})_{ji}= (P_{\mu})_{ij}^*.
\end{eqnarray}

\section{Physical Tests} The following tests can be conducted to verify the reliability of the written  code based on the above algorithm:
\begin{itemize}
\item{}{\bf Test $1$:}For $\gamma=\alpha=0$ the problem reduces to a harmonic oscillator problem. Indeed the system in this case is equivalent to $N^2d$ independent harmonic oscillators with frequency and period given by
 \begin{eqnarray}
\omega=m~,~T=\frac{2\pi}{m}.
\end{eqnarray}
The Hamiltonian is conserved with error seen to be periodic with period
\begin{eqnarray}
T_H=\frac{T}{2}=\frac{\pi}{m}.
\end{eqnarray}
\item{}{\bf Test $2$:}In the harmonic oscillator problem we know that the $X$'s are distributed according to the Gaussian distribution 
\begin{eqnarray}
\int dX_{\mu}~e^{-\frac{m^2}{2}Tr X_{\mu}^2}.
\end{eqnarray}
The Metropolis must generate this distribution.
\item{}{\bf Test $3$:}On general ground we must have
\begin{eqnarray}
<e^{-\Delta H}>&=&\frac{1}{Z}\int dP dX~e^{-H[X,P]}~e^{-\Delta H}\nonumber\\
&=&\frac{1}{Z}\int dP dX~e^{-H[X^{'},P^{'}]}\nonumber\\
&=&\frac{1}{Z}\int dP^{'} dX^{'}~e^{-H[X^{'},P^{'}]}\nonumber\\
&=&1.
\end{eqnarray}
\item{}{\bf Test $4$:}On general ground we must also have the Schwinger-Dyson identity (exact result) given by
\begin{eqnarray}
4\gamma <{\rm YM}>+3\alpha <{\rm CS}>+2m^2<{\rm HO}>=d(N^2-1).\label{SDB}
\end{eqnarray}
\begin{eqnarray}
{\rm YM}=-\frac{N}{4}\sum_{\mu,\nu=1}^dTr[X_{\mu},X_{\nu}]^2.
\end{eqnarray}
\begin{eqnarray}
{\rm CS}=\frac{2N i}{3}\epsilon_{abc}Tr X_aX_bX_c.
\end{eqnarray}
\begin{eqnarray}
{\rm HO}=\frac{1}{2} Tr X_{\mu}^2.
\end{eqnarray}
\item{}{\bf Test $5$:} We compute $<S_{\rm YM}>$ and $C_{\rm v}=<S_{\rm YM}^2>-<S_{\rm YM}>^2$ for $\gamma=1$ and $m=0$. There must be an emergent geometry phase transition in $\alpha$ for $d=3$ and $d=4$.
\item{}{\bf Test $6$:} We compute the eigenvalues distributions of the $X$'s in $d=3$ and $d=4$ for $\gamma=1$ and $\alpha=m=0$. 

\item{}{\bf Test $7$:} The Polyakove line is defined by
\begin{eqnarray}
P(k)=\frac{1}{N} Tr e^{ikX_1}.
\end{eqnarray}
We compute $<P(k)>$ as a function of $k$ for $m=\alpha=0$.
\end{itemize}
\section{Emergent Geometry: An Exotic Phase Transition}
As a concrete example we consider the Bosonic $d=3$ Yang-Mills matrix model with only a Chern-Simons term, i.e. $\gamma=1$, $\alpha\ne 0$ and $m=0$. This model depends on a single (scaled) parameter 
\begin{eqnarray}
\tilde{\alpha}=\alpha\sqrt{N}.
\end{eqnarray}
The order parameter in this problem is given by the observable ${\rm radius}$ defined by
\begin{eqnarray}
{\rm radius}=Tr X_a^2.
\end{eqnarray}
The radius of the sphere is related to this observable by
\begin{eqnarray}
r=\frac{\tilde{\alpha}^2 c_2}{{\rm radius}}~,~c_2=\frac{N^2-1}{4}.
\end{eqnarray}
A more powerful set of order parameters is given by the eigenvalues distributions of the matrices $X_3$, $i[X_1,X_2]$, and $X_a^2$.  Other useful observables are
\begin{eqnarray}
S_3={\rm YM}+{\rm CS}~,~{\rm YM}=-\frac{N}{4}[X_{\mu},X_{\nu}]^2~,~{\rm CS}=\frac{2iN\alpha}{3}\epsilon_{abc} Tr X_aX_bX_c.
\end{eqnarray}
The specific heat is
\begin{eqnarray}
C_v=<S_3^2>-<S_3>^2.
\end{eqnarray}
An exact Schwinger-Dyson identity is given by 
\begin{eqnarray}
{\rm identity}=4<{\rm YM}>+3<{\rm CS}>\equiv d N^2.
\end{eqnarray}
For this so-called ARS model it is important that we remove the trace part of the matrices $X_a$ after each molecular dynamics step because this mode can never be thermalized. In other words, we should consider in this case the path integral (partition function) given by
\begin{eqnarray}
Z=\int dX_a~\exp(-S_3)\delta(Tr X_a).
\end{eqnarray}
The corresponding hybrid Monte Carlo code is included in the last chapter. We skip here any further technical details and report only few physical results.

\begin{figure}[htbp]
\begin{center}
\includegraphics[width=10.0cm,angle=-0]{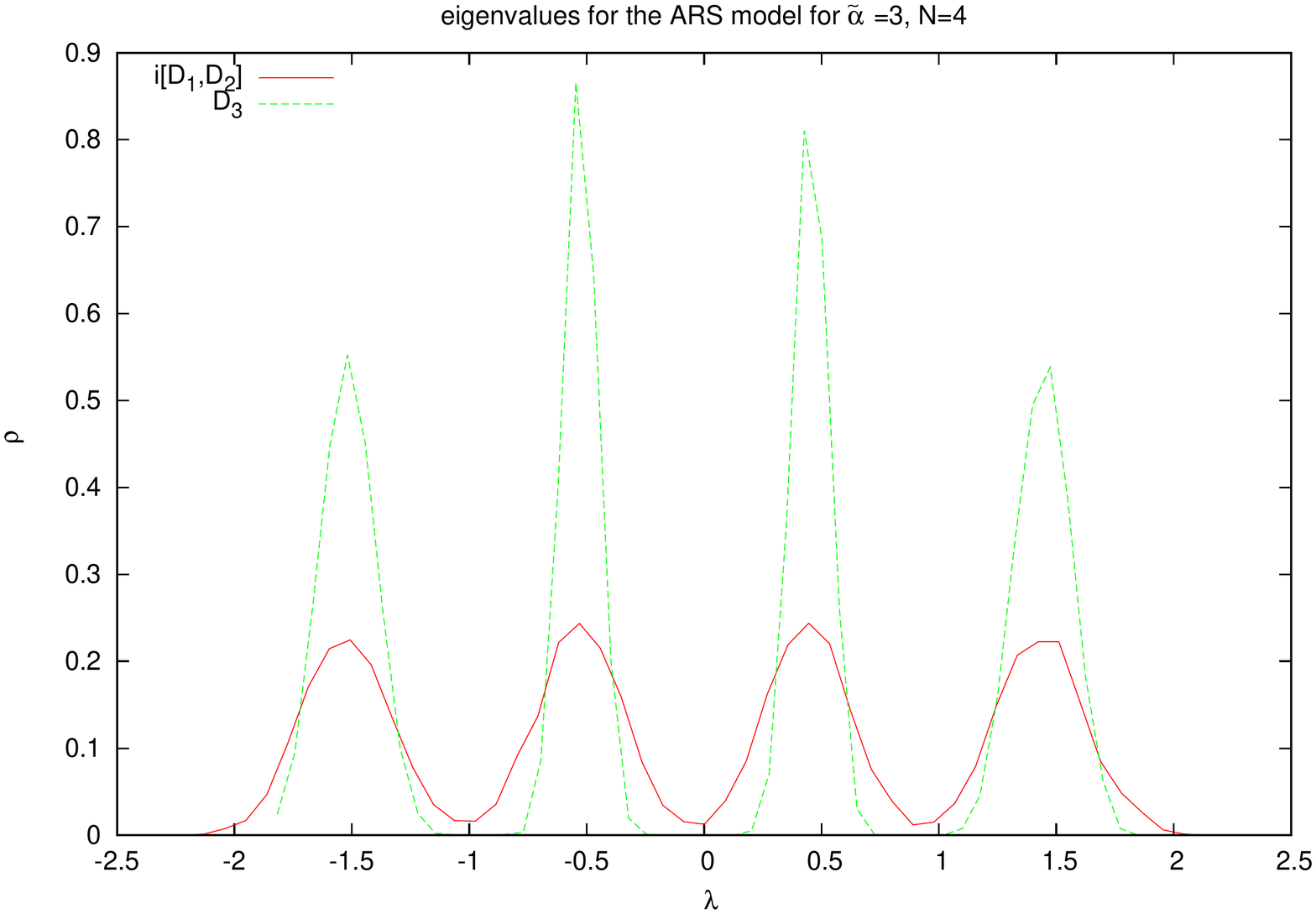}
\end{center}
\caption{}\label{Emerg1}
\end{figure}

\begin{figure}[htbp]
\begin{center}
\includegraphics[width=10.0cm,angle=-0]{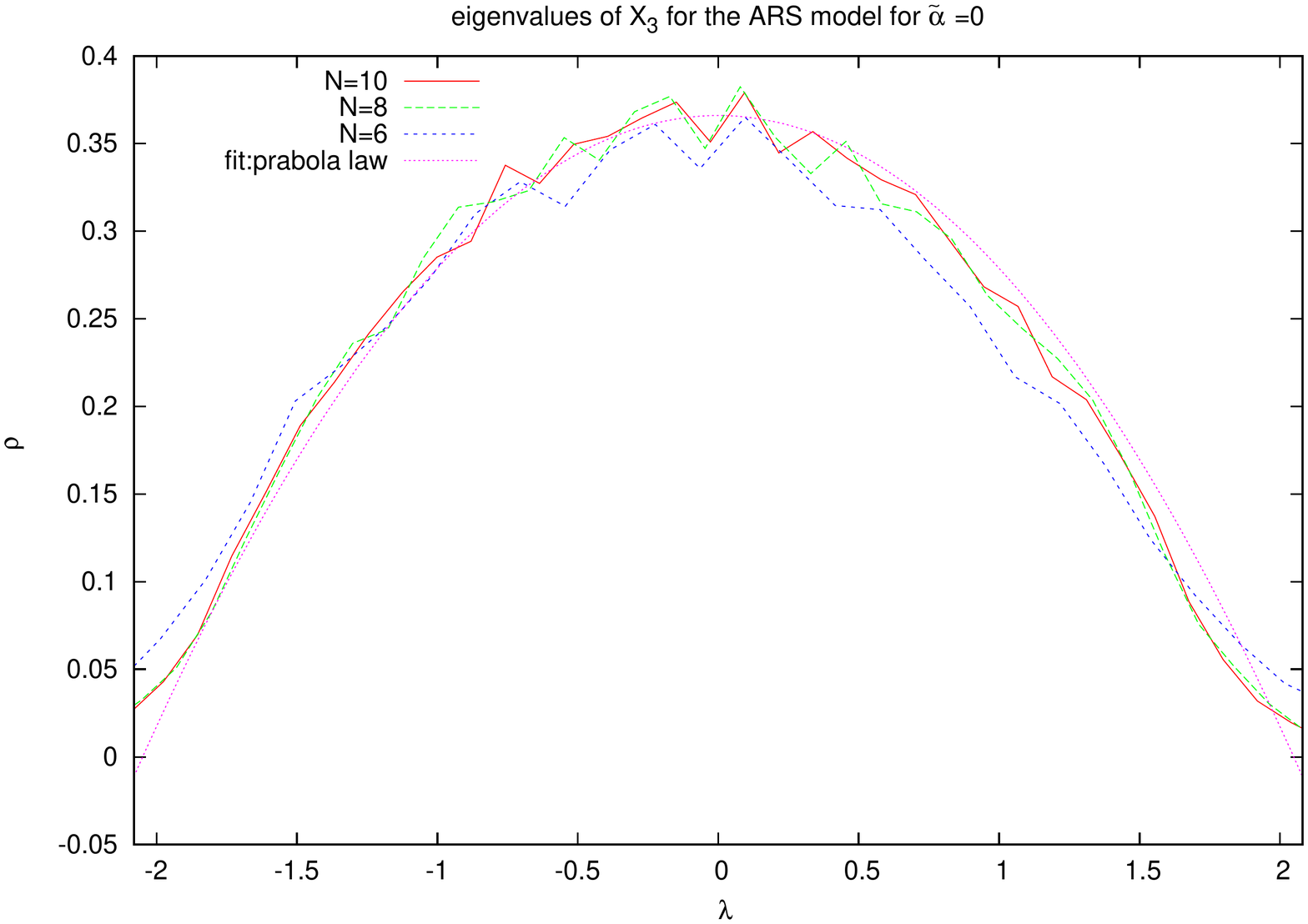}
\end{center}
\caption{}\label{Emerg2}
\end{figure}

\begin{figure}[htbp]
\begin{center}
\includegraphics[width=10.0cm,angle=-0]{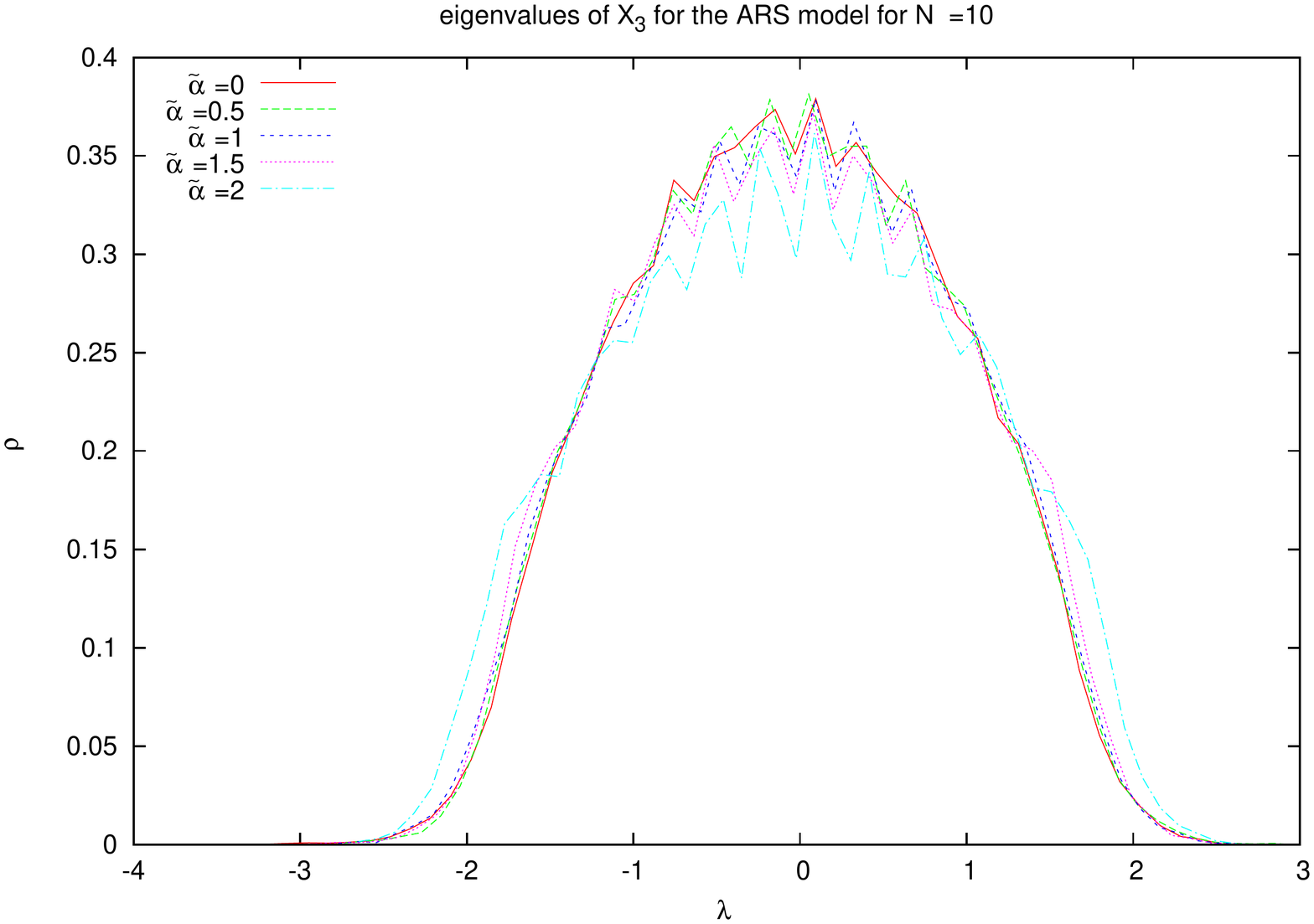}
\end{center}
\caption{}\label{Emerg3}
\end{figure}

\begin{figure}[htbp]
\begin{center}
\includegraphics[width=8.0cm,angle=-0]{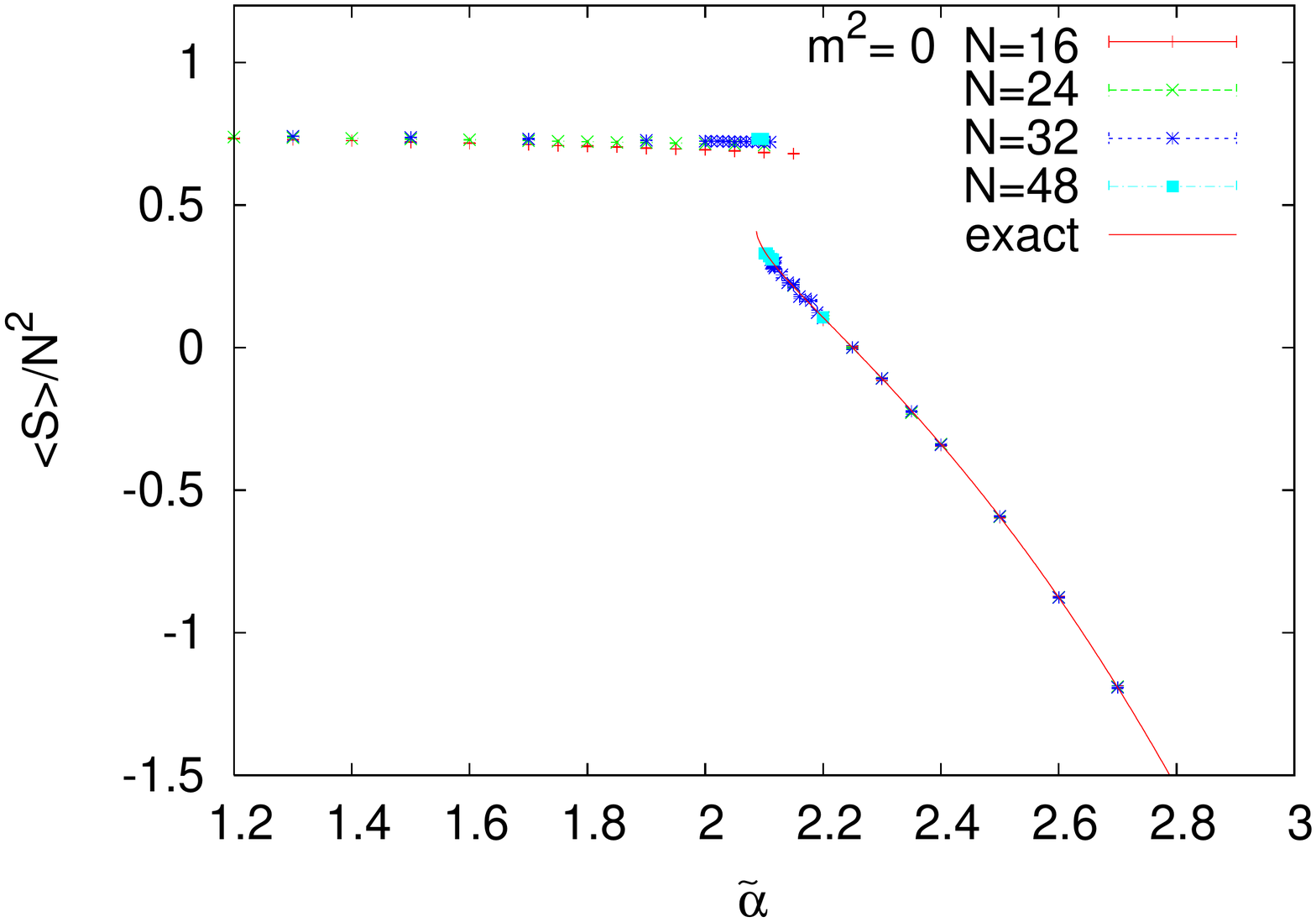}
\includegraphics[width=8.0cm,angle=-0]{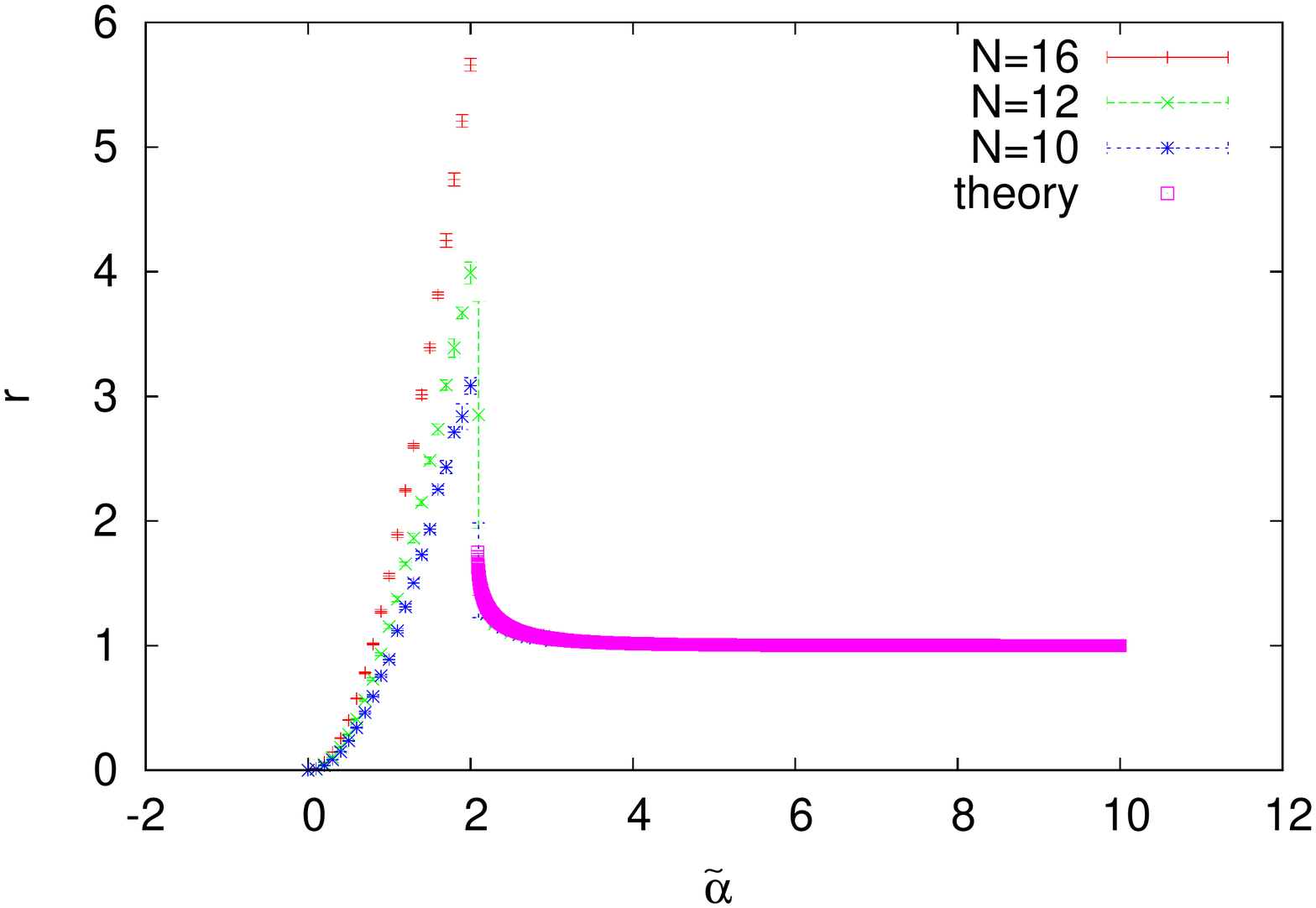}
\includegraphics[width=8.0cm,angle=-0]{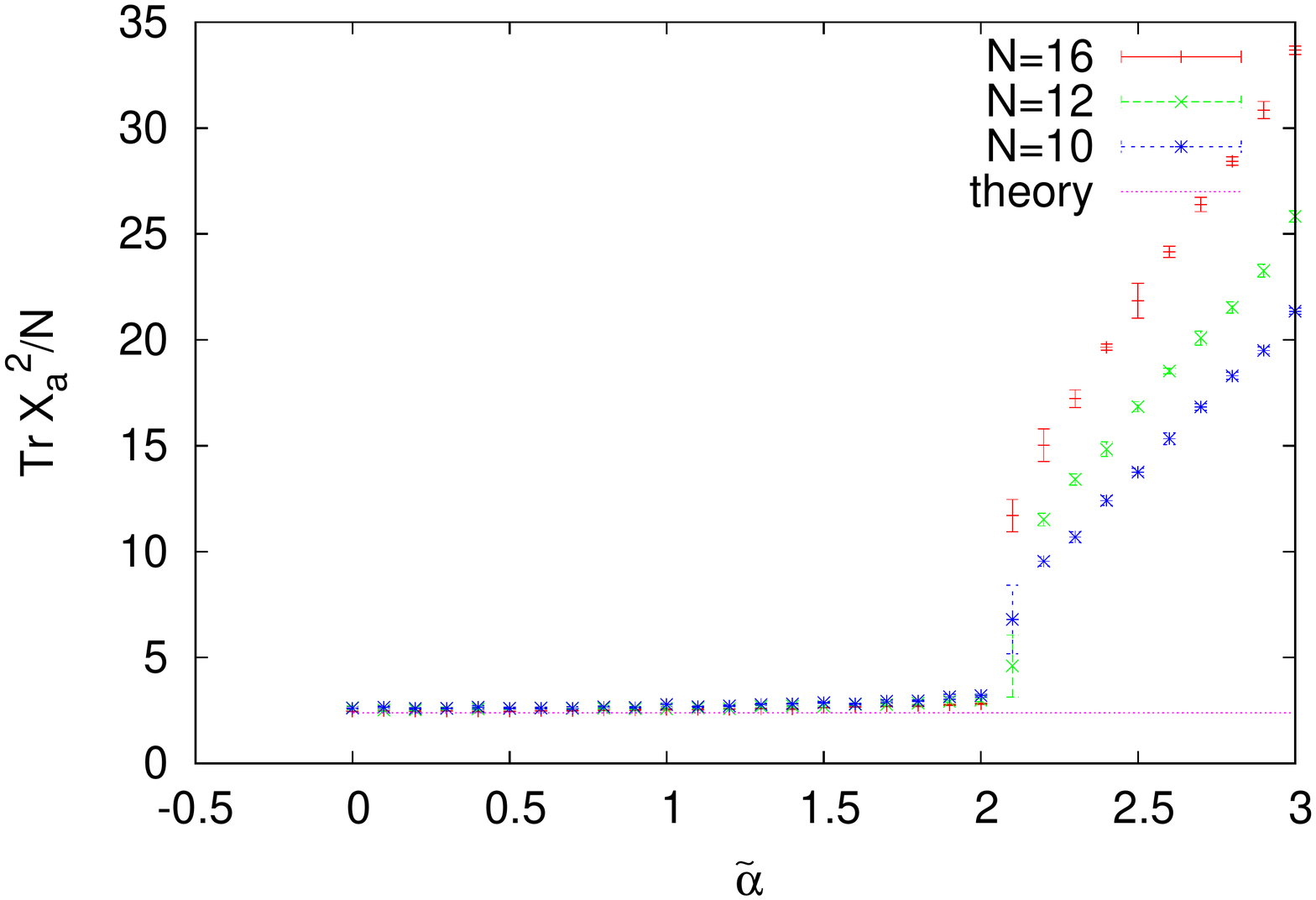}
\includegraphics[width=8.0cm,angle=-0]{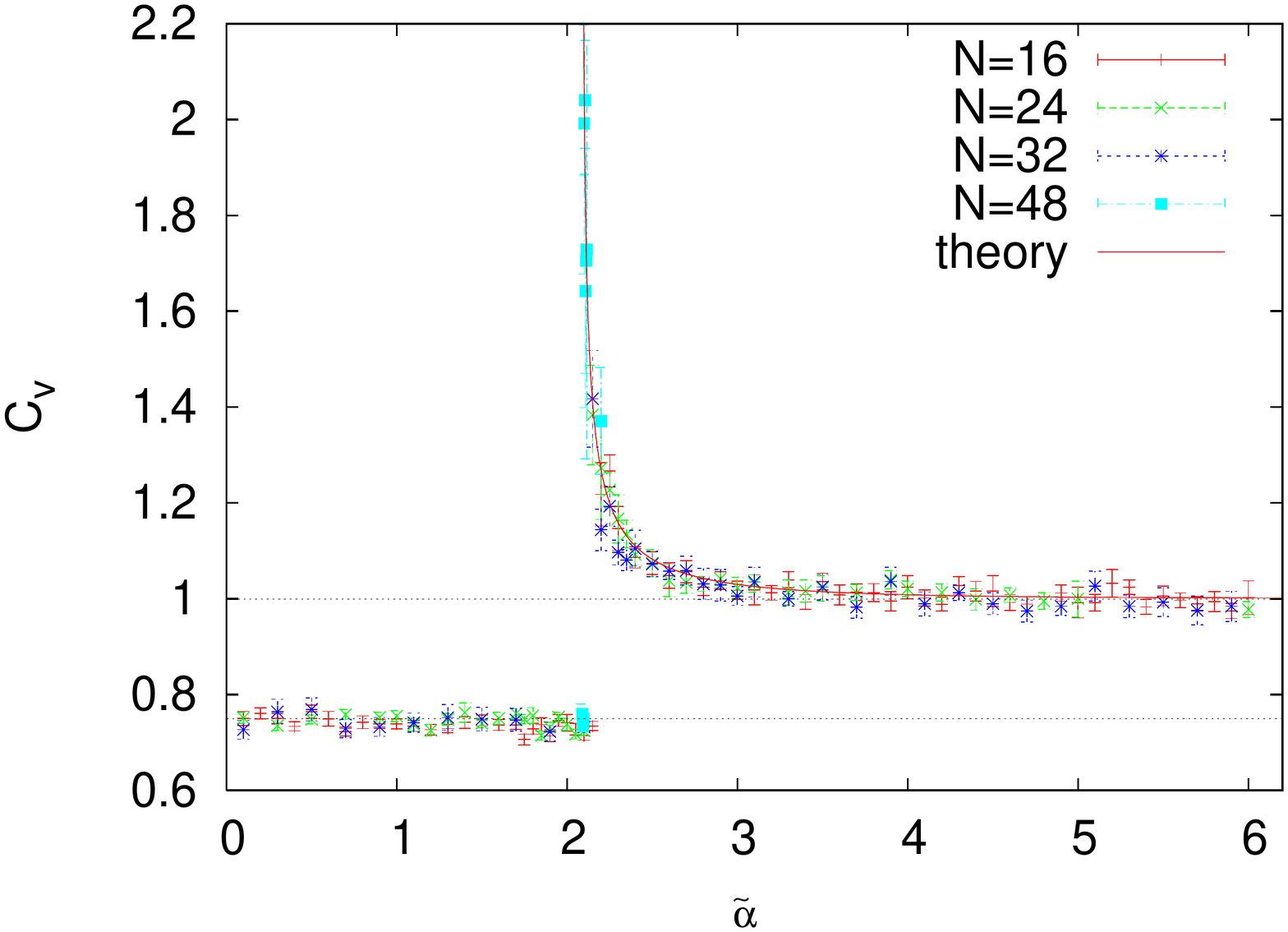}
\caption{}\label{figpublished}
\end{center}
\end{figure}

The ARS model is characterized by two phases: the fuzzy sphere phase and the Yang-Mills phase. Some of the fundamental results are:
\begin{enumerate}
\item{}{\bf The Fuzzy Sphere Phase:} 
\begin{itemize}
\item This appears for large values of $\tilde{\alpha}$. It corresponds to the class of solutions of the equations of motion given by
\begin{eqnarray}
[X_a,X_b]=i\alpha\phi\epsilon_{abc}X_c~,~\phi=1.
\end{eqnarray}
The global minimum is given by the largest irreducible representation of $SU(2)$ which fits in $N\times N$ matrices. This corresponds to the spin $l=(N-1)/2$ irreducible representation, viz
  \begin{eqnarray}
X_a=\phi\alpha L_a.
\end{eqnarray}
\begin{eqnarray}
[L_a,L_b]=i\epsilon_{abc}L_c~,~c_2=\sum_aL_a^2=l(l+1).{\bf 1}_N=\frac{N^2-1}{4}.{\bf 1}_N.
\end{eqnarray}
The values of the various observables in these configurations are
\begin{eqnarray}
S_3=\phi^3\tilde{\alpha}^4c_2(\frac{\phi}{2}-\frac{2}{3})~,~{\rm YM}=\frac{\phi^4\tilde{\alpha}^4c_2}{2}~,~{\rm CS}=-\frac{2\phi^3\tilde{\alpha}^4 c_2}{3}~,~{\rm radius}=\phi^2\tilde{\alpha}^2c_2.
\end{eqnarray}
\item{}The eigenvalues of $D_3=X_3/\alpha$ and $i[D_1,D_2]=i[X_1,X_2]/\alpha^2$ are given by
\begin{eqnarray}
\lambda_i=-\frac{N-1}{2},...,+\frac{N-1}{2}.
\end{eqnarray}
The spectrum of $[D_1,D_2]$ is a better measurement of the geometry since all fluctuations around $L_3$ are more suppressed. Some illustrative data for $\tilde{\alpha}=3$ and $N=4$ is shown on figure (\ref{Emerg1}).

\end{itemize}
\item{}{\bf The Yang-Mills (Matrix) Phase:} 
\begin{itemize}
\item
This appears for small values of $\tilde{\alpha}$. It corresponds to the class of solutions of the equations of motion given by
\begin{eqnarray}
[X_a,X_b]=0.
\end{eqnarray}
This is the phase of almost commuting matrices. It is characterized by the eigenvalues distribution 
\begin{eqnarray}
\rho(\lambda)=\frac{3}{4R^3}(R^2-\lambda^2).\label{for0}
\end{eqnarray}
It is believed that $R=2$. We compute
 \begin{eqnarray}
<{\rm radius}>&=&3 <Tr X_3^2>\nonumber\\
&=&3N\int_{-R}^{R}d\lambda \rho(\lambda)\lambda^2\nonumber\\
&=&\frac{3}{5}R^2 N.\label{for1}
\end{eqnarray}
\item The above eigenvalues distribution can be derived by assuming that the joint eigenvalues distribution of the the three commuting matrices  $X_1$, $X_2$ and $X_3$ is uniform inside a solid ball of radius $R$. This can be actually proven by quantizing the system in the Yang-Mills phase around commuting matrices \cite{Filev:2014jxa}.
\item{}The value of the radius $R$ is determined numerically as follows:
\begin{itemize}
\item{}The first measurement $R_1$ is obtained by comparing the numerical result for $<{\rm radius}>$, for the biggest value of $N$, with the formula (\ref{for1}).
\item{}We use $R_1$ to restrict the range of the eigenvalues of $X_3$.
\item{}We fit the numerical result for the density of  eigenvalues of $X_3$, for the biggest value of $N$, to the parabola (\ref{for0}) in order to get a second  measurement $R_2$.
\item{}We may take the average of $R_1$ and $R_2$.
\end{itemize} 
Example: For $\alpha=0$, we find the values $R_1=2.34(N=6)$, $R_1=2.15(N=8)$, $R_1=2.08(N=10)$, and $R_2=2.05\pm 0.01(N=10)$.  Sample data for $\tilde{\alpha}=0$ with $N=6,8$ and $10$ is shown on figure (\ref{Emerg2}).  
\item{}It is found that the eigenvalues distribution, in the Yang-Mills phase, is independent of $\tilde{\alpha}$. Sample data for $\tilde{\alpha}=0-2$ and $N=10$ is shown on figure (\ref{Emerg3}).  
\end{itemize}
\item {\bf Critical Fluctuations:}  The transition between the two phases occur at $\tilde{\alpha}=2.1$. The specific heat diverges at this point from the Yang-Mills side while it remains constant from the fuzzy sphere side.  This indicates a second order behaviour with critical fluctuations only from one side of the transition. The Yang-Mills and Chern-Simons actions, and as a consequence the total action, as well as the radii ${\rm radius}$ and $r$ suffer a discontinuity at this point reminiscent of a first order behavior. The different phases of the model are characterized by 
\begin{center}
\begin{tabular}{|c|c|}
\hline
fuzzy sphere ($\tilde{\alpha}>\tilde{\alpha}_*$ )& matrix phase ($\tilde{\alpha}<<\tilde{\alpha}_*$)\\
$r=1$ & $
r=0$\\
$C_v=1$  & $C_v=0.75$  \\
\hline
\end{tabular}
\end{center}
The Monte Carlo results of \cite{DelgadilloBlando:2007vx}, derived using the Metropolis algorithm of the previous chapter and shown on figure (\ref{figpublished}), should be easily obtainable using the attached hybrid Monte Carlo code.
\end{enumerate}

\chapter{Hybrid Monte Carlo Algorithm for Noncommutative Phi-Four}
\section{The Matrix Scalar Action} 
The hybrid Monte Carlo algorithm is a combination of the molecular dynamics method and the Metropolis algorithm. In this section we will apply this algorithm to matrix $\Phi^4$ on the fuzzy sphere. This problem was studied using other techniques in \cite{GarciaFlores:2009hf0,GarciaFlores:2005xc0,Martin:2004un0,Panero:2006bx0}. We will follow here \cite{Ambjorn:2000bf0,Ambjorn:2000dx0}.

We are  interested in the Euclidean matrix model
\begin{eqnarray}
S&=&{\rm Tr}\big(- a[L_a,{\Phi}]^2+b{\Phi}^2+c{\Phi}^4\big).\label{sf}
\end{eqnarray}
The  scaled (collapsed) parameters are given by 
\begin{eqnarray}
\tilde{b}=\frac{b}{aN^{\frac{3}{2}}}~,~\tilde{c}=\frac{c}{a^2N^2}.
\end{eqnarray}
The path integral we wish to sample in Monte Carlo simulation is

\begin{eqnarray}
Z=\int d\Phi~ \exp(-S[\Phi]).
\end{eqnarray}
As before, we will first think of the configurations $\Phi$ as evolving in some  fictitious time-like parameter $t$, viz
\begin{eqnarray}
\Phi\equiv \Phi(t).
\end{eqnarray}
The above path integral is then equivalent to the Hamiltonian dynamical system
\begin{eqnarray}
Z=\int dP d\Phi~ \exp(-\frac{1}{2}Tr P^2-S[\Phi]).
\end{eqnarray}
In other words, we have introduced a Hermitian  $N\times N$ matrix $P$ which is conjugate to $\Phi$. The Hamiltonian is clearly given by 
\begin{eqnarray}
H=\frac{1}{2}Tr P^2+S[\Phi].
\end{eqnarray}
In summary, we think of the matrix $\Phi$ as  a field in one dimension with corresponding
conjugate momentum $P$. The Hamiltonian equations of motion read
 \begin{eqnarray}
\frac{\partial H}{\partial P_{ij}}=(\dot{\Phi})_{ij}=P_{ji}~,~\frac{\partial H}{\partial \Phi_{ij}}=-(\dot{P})_{ij}=\frac{\partial S}{\partial \Phi_{ij}}.
\end{eqnarray}
We will define the scalar force by
 \begin{eqnarray}
V_{ij}(t)&=&\frac{\partial S}{\partial \Phi_{ij}(t)}\nonumber\\
&=&a\bigg(-4L_{a}\Phi L_{a}+2L_{a}^2\Phi+2\Phi L_{a}^2\bigg)_{ji}+2b \Phi_{ji}+4c(\Phi^3)_{ji}.
\end{eqnarray}
\section{The Leap Frog Algorithm} 

The numerical solution of the above differential equations can be given by the leap frog equations

\begin{eqnarray}
(P)_{ij}(t+\frac{\delta t}{2})=(P)_{ij}(t)-\frac{\delta t}{2}V_{ij}(t).
\end{eqnarray}

\begin{eqnarray}
\Phi_{ij}(t+\delta t)=\Phi_{ij}(t)+\delta t P_{ji}(t+\frac{\delta t}{2}).
\end{eqnarray}
 \begin{eqnarray}
P_{ij}(t+\delta t)=P_{ij}(t+\frac{\delta t}{2})-\frac{\delta t}{2}V_{ij}(t+\delta t).
\end{eqnarray}
Let us recall that $t =n {\delta}t$, $n=0,1,2,...,\nu-1,\nu$ where the point $n=0$ corresponds to  the initial configuration $\Phi_{ij}(0)$ whereas $n=\nu$ corresponds to  the final configuration $\Phi_{ij}(T)$ where $T=\nu \delta t$.

\section{Hybrid Monte Carlo Algorithm}

The hybrid Monte Carlo algorithm can be summarized as follows:  
\begin{itemize}
\item{$1)$} Choose $P(0)$ such that $P(0)$ is distributed according to the Gaussian probability distribution $\exp(-\frac{1}{2}TrP^2)$.
\item{$2)$}Find the configuration $(\Phi(T),P(T))$ by solving the above differential equations of motion.
\item{$3)$}Accept the configuration  $(\Phi(T),P(T))$  with a probability 
\begin{eqnarray}
{\rm min}(1,e^{-\Delta H[\Phi,P]}),
\end{eqnarray}
where $\Delta H$ is the corresponding change in the Hamiltonian when we go from $(\Phi(0),P(0))$ to $(\Phi(T),P(T))$.
\item{$4)$} Repeat.
\end{itemize}

\section{Optimization} 
\subsection{Partial Optimization}
We start with some general comment which is not necessarily a part of the optimization process. The scalar field $\Phi$ is a hermitian matrix, i.e. the diagonal elements are real, while the off diagonal elements are complex conjugate of each other. We find it crucial that we implement, explicitly in the code, the reality of the diagonal elements by subtracting from $\Phi_{ii}$ the imaginary part (error) which in each molecular dynamics iteration is small but can accumulate. The implementation of the other condition is straightforward. 

\begin{figure}[htbp]
\begin{center}
\includegraphics[width=9.0cm,angle=-0]{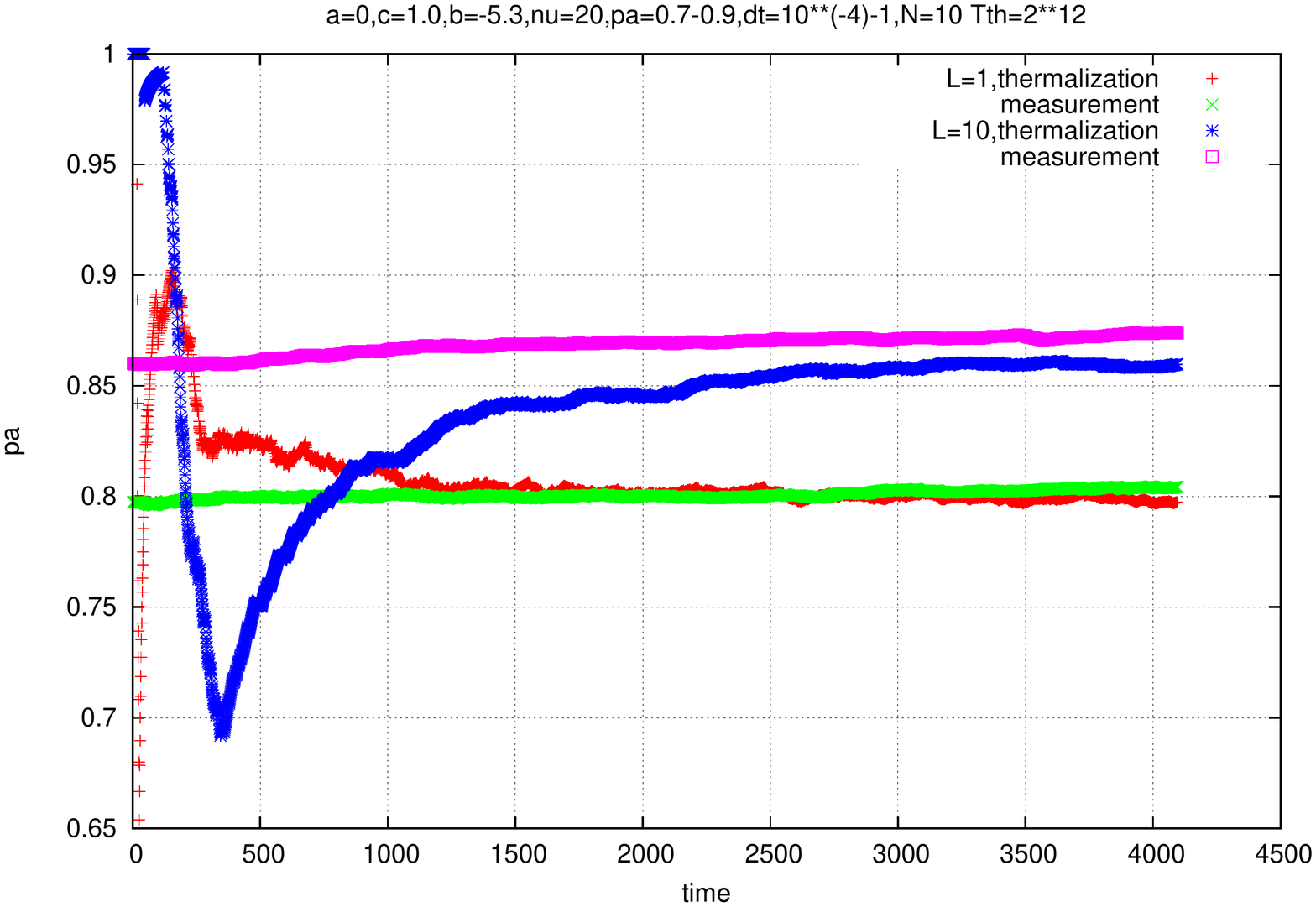}
\includegraphics[width=9.0cm,angle=-0]{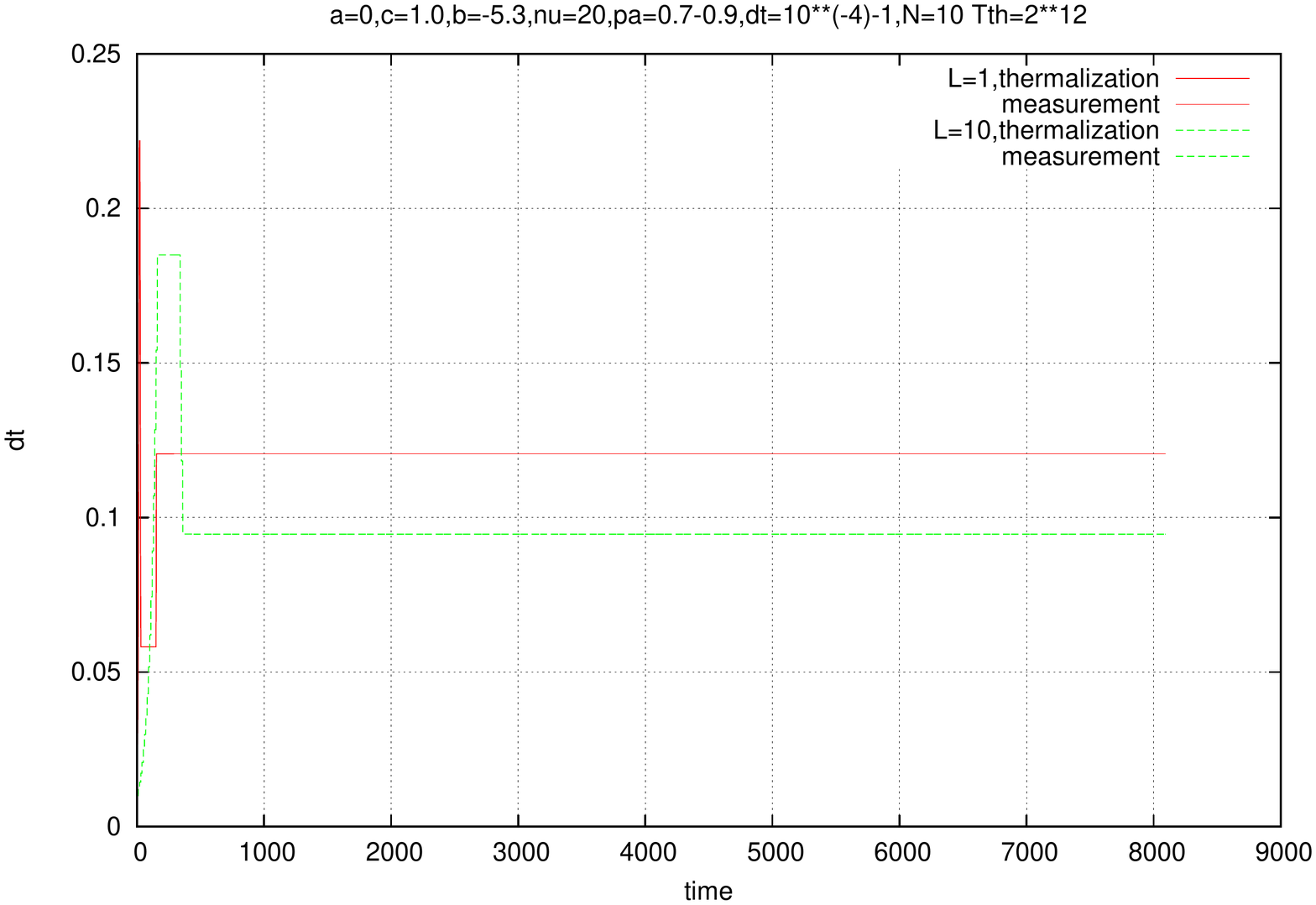}
\includegraphics[width=9.0cm,angle=-0]{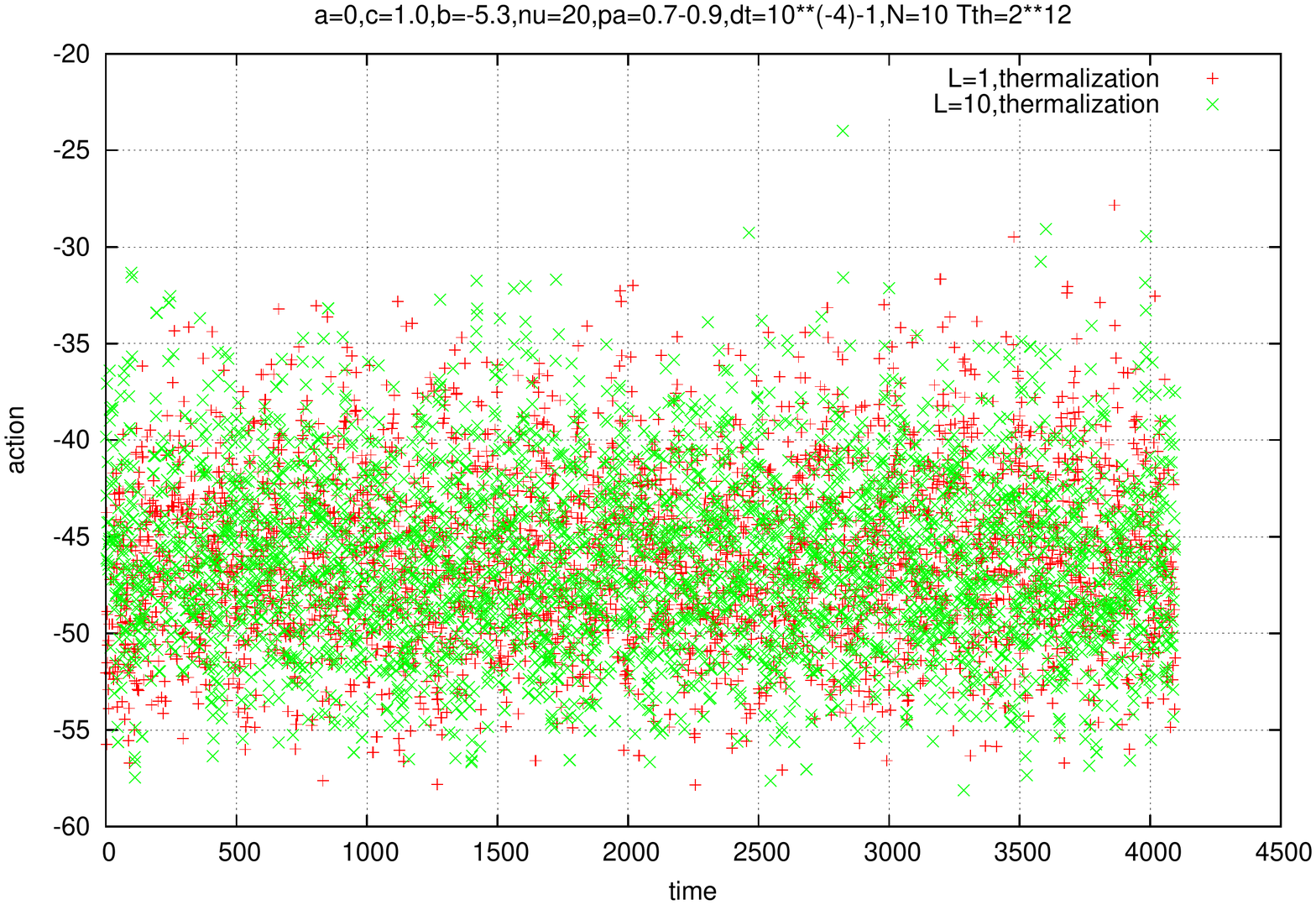}
\caption{}\label{fig_opt}
\end{center}
\end{figure}

 In actual simulations we can fix $\nu$, for example we take $\nu=20$, and adjust the step size $\delta t$, in some interval $[\delta t_{\rm min},\delta t_{\rm max}]$, in such a way that the acceptance rate ${\rm pa}$ is held fixed between some target acceptance rates say ${\rm pa}_{\rm low}=70$ and ${\rm pa}_{\rm high}=90$ per cents. If the acceptance rate becomes larger than the target acceptance rate  ${\rm pa}_{\rm high}$, then we increase the step size $\delta t$ by a factor ${\rm inc}=1.2$ if the outcome is within the interval  $[\delta t_{\rm min},\delta t_{\rm max}]$. Similarly, if the acceptance rate becomes smaller than the target acceptance rate  ${\rm pa}_{\rm low}$, we decrease the step size by a factor ${\rm dec}=0.8$ if the outcome is within the interval  $[\delta t_{\rm min},\delta t_{\rm max}]$. The adjusting of $\delta t$ can be done at each Monte Carlo step, but it can also be performed only each $L$ simulations. We take $L=1$. A sample pseudo code is attached below. A sample of the results is shown in figure (\ref{fig_opt}).

\begin{verbatim}
      pa=(Accept)/(Rejec+Accept)        
      cou=mod(tmc,L)        
        if (cou.eq.0)then
           if (pa.ge.target_pa_high) then 
              dtnew=dt*inc
                if (dtnew.le.dt_max)then
                   dt=dtnew
                else
                dt=dt_max
                endif
           endif
           if (pa.le.target_pa_low) then
              dtnew=dt*dec
                if (dtnew.ge.dt_min)then
                   dt=dtnew
                else
                   dt=dt_min
                endif
           endif                  
        endif
\end{verbatim}
\subsection{Full Optimization}
A more thourough optimization of the algorithm can also be done as follows  \cite{Ambjorn:2000bf,Ambjorn:2000dx,Anagnostopoulos:2005cy}. We take $\delta \tau$ small so that the acceptance rate ${\rm pa}$ is kept sufficiently large. Then we fix $\nu$ and look for the value of $\delta\times \tau$ where the speed of motion in the phase space defined by $\delta \tau \times {\rm pa}$ is maximum. Then we fix $\delta \tau$ at its optimal value and look for the value of $\nu$ where the autocorrelation time $T_{\rm au}$ is minimum. The number of iterations $\nu$ must also be kept relatively small so that the systematic error (which is of order $\nu \times\delta{\tau}^2$ for every hybrid Monte Carlo unit of time) is kept small. Clearly a small value of $\nu$ is better for the effeciency of the algorithm. 
\section{The Non-Uniform Order: Another Exotic Phase}
\subsection{Phase Structure}
 The theory (\ref{sf}) is a three-parameter model with the following three known phases: 
\begin{itemize}
\item The usual $2$nd order Ising phase transition between disordered $<\Phi>=0$ and uniform ordered $<\Phi>\sim {\bf 1}$ phases. This appears for small values of $c$. This is the only transition observed in commutative phi-four.
\item  A matrix transition between disordered $<\Phi>=0$ and non-uniform ordered $<\Phi>\sim \gamma$ phases with $\gamma^2={\bf 1}$. This transition coincides, for very large values of $c$,  with the $3$rd order transition of the real quartic matrix model, i.e. the model with $a=0$, which occurs at $b=-2\sqrt{Nc}$. See next chapter.
\item  A transition between uniform ordered  $<\Phi>\sim {\bf 1}$ and non-uniform ordered $<\Phi>\sim \gamma$ phases.  The non-uniform phase, in which translational/rotational invariance is spontaneously broken, is absent in the commutative theory. The non-uniform phase is essentially  the stripe phase observed originally on  Moyal-Weyl spaces in \cite{Gubser:2000cd1,Ambjorn:2002nj1}. 
\end{itemize}
The above three phases are already present in the pure potential model $V={\rm Tr}({b}\Phi^2+{c}\Phi^4)$. The ground state configurations  are given by the matrices
\begin{eqnarray}
\Phi_0=0.
\end{eqnarray}
\begin{eqnarray}
\Phi_{\gamma}=\sqrt{-\frac{b}{2c}}U\gamma U^+~,~{\gamma}^2={\bf 1}_N~,~UU^+=U^+U={\bf 1}_N.
\end{eqnarray}
We compute $V[\Phi_0]=0$ and $V[\Phi_{\gamma}]=-b^2/4c$. The first configuration corresponds to the disordered phase characterized by $<\Phi>=0$. The second solution makes sense only for $b<0$, and it corresponds to the ordered phase characterized by $<\Phi>\ne 0$. As mentioned above, there is a non-perturbative transition between the two phases which occurs quantum mechanically, not at $b=0$, but at $b=b_*=-2\sqrt{Nc}$, which is known as the one-cut to two-cut transition. The idempotent $\gamma$ can always be chosen such that $\gamma=\gamma_k={\rm diag}({\bf 1}_{k},-{\bf 1}_{N-k})$. The orbit of $\gamma_k$ is the Grassmannian manifold $U(N)/(U(k)\times U(N-k))$ which is $d_k-$dimensional where  $d_k=2kN-2k^2$. It is not difficult to show that this dimension is maximum at $k=N/2$, assuming that $N$ is even, and hence from entropy argument, the most important two-cut solution is the so-called stripe configuration given by $\gamma={\rm diag}({\bf 1}_{{N}/{2}},-{\bf 1}_{{N}/{2}})$. 

In this real quartic matrix model, we have therefore three possible phases characterized by the following order parameters:
\begin{eqnarray}
&&<\Phi>=0~~{\rm disordered}~{\rm phase}.
\end{eqnarray}
\begin{eqnarray}
&&<\Phi>=\pm\sqrt{-\frac{b}{2c}}{\bf 1}_N~~{\rm Ising}~({\rm uniform})~{\rm phase}.
\end{eqnarray}
\begin{eqnarray} 
&&<\Phi>=\pm\sqrt{-\frac{b}{2c}}\gamma~~{\rm matrix}~({\rm nonuniform}~{\rm or}~{\rm stripe})~{\rm phase}.
\end{eqnarray}
However, as one can explicitly check by calculating the free energies of the respective phases, the uniform ordered phase is not  stable in the real quartic matrix model  $V={\rm Tr}({b}\Phi^2+{c}\Phi^4)$.

The above picture is expected to hold for noncommutative/fuzzy phi-four theory in any dimension, and the three phases are all stable and are expected to meet at a triple point. This structure was confirmed in two dimensions by means of Monte Carlo simulations on the fuzzy sphere in  \cite{GarciaFlores:2009hf0,GarciaFlores:2005xc0}. 
\subsection{Sample Simulations}
We run simulations for every $N$ by running $T_{\rm th}$ thermalization steps, and then measuring observables in a sample containing $T_{\rm mc}$ thermalized configurations $\Phi$, where each two successive configurations are separated by $T_{\rm co}$ Monte Carlo steps in order  to reduce auto-correlation effects. Most of the detail of the simulations have already been explained. We only mention again that we estimate error bars using the jackknife method and use the random number generator ran2. A sample code is attached in the last chapter. 

\begin{figure}[htbp]
\begin{center}
\includegraphics[width=9.0cm,angle=-0]{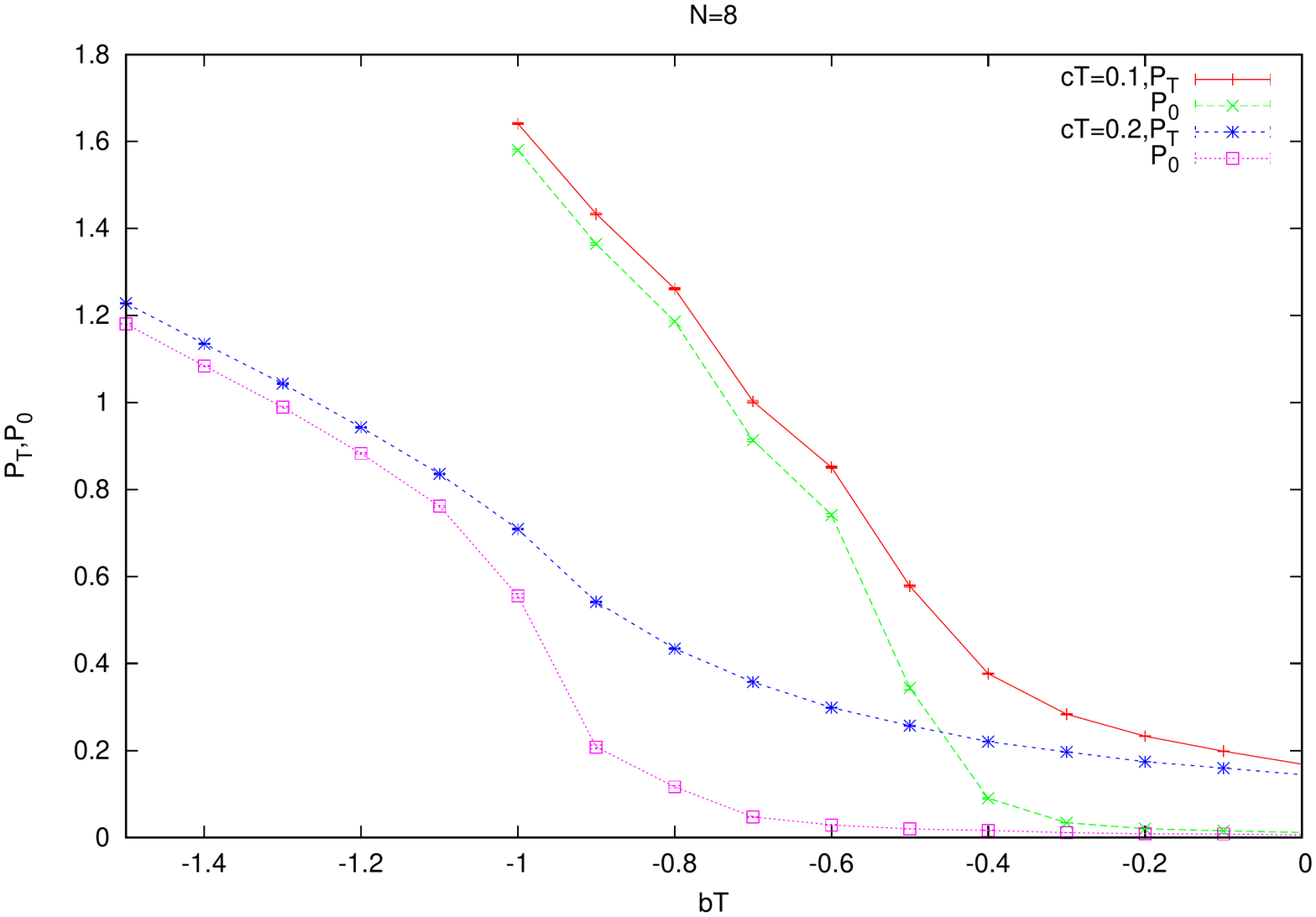}
\includegraphics[width=9.0cm,angle=-0]{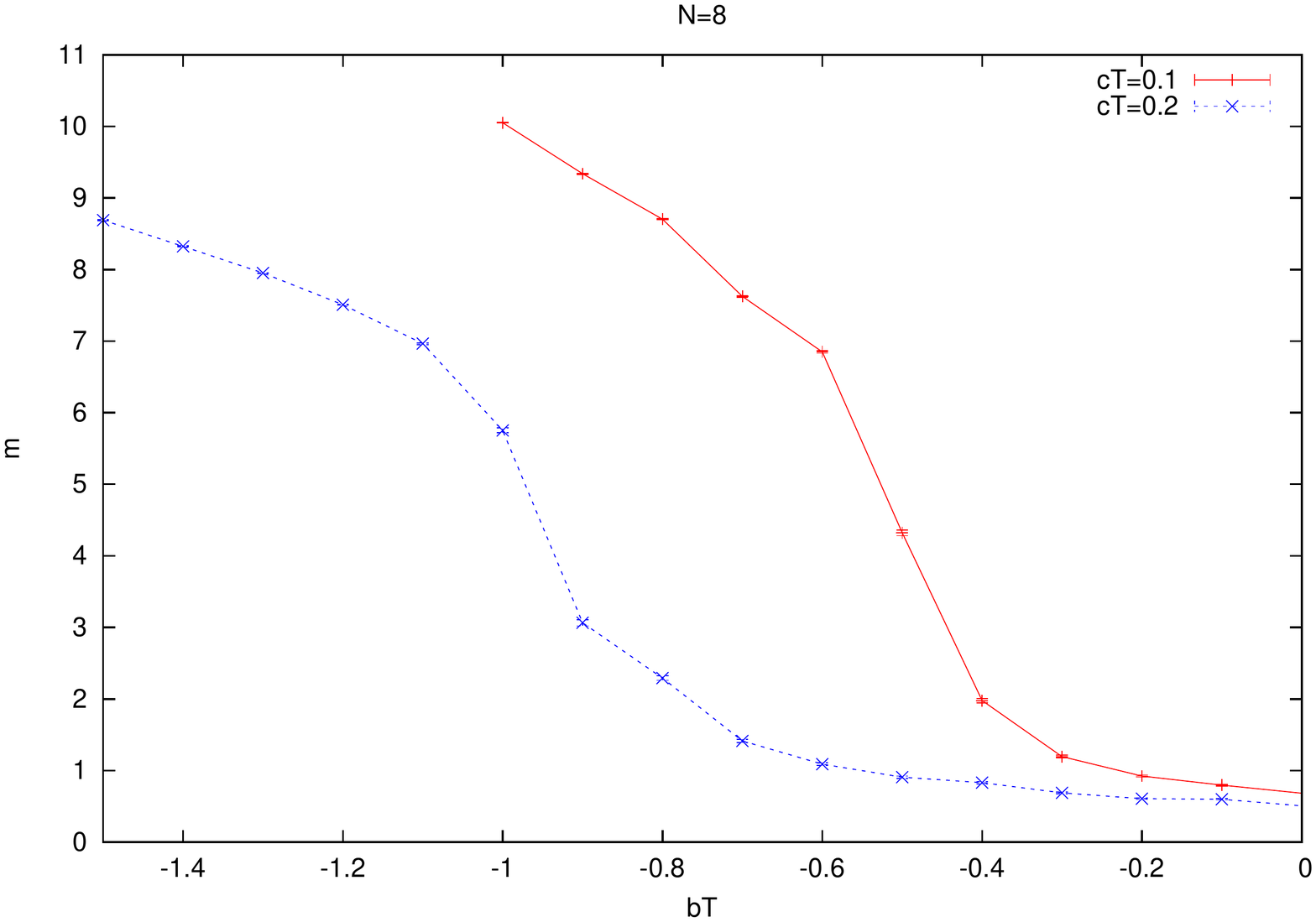}
\includegraphics[width=9.0cm,angle=-0]{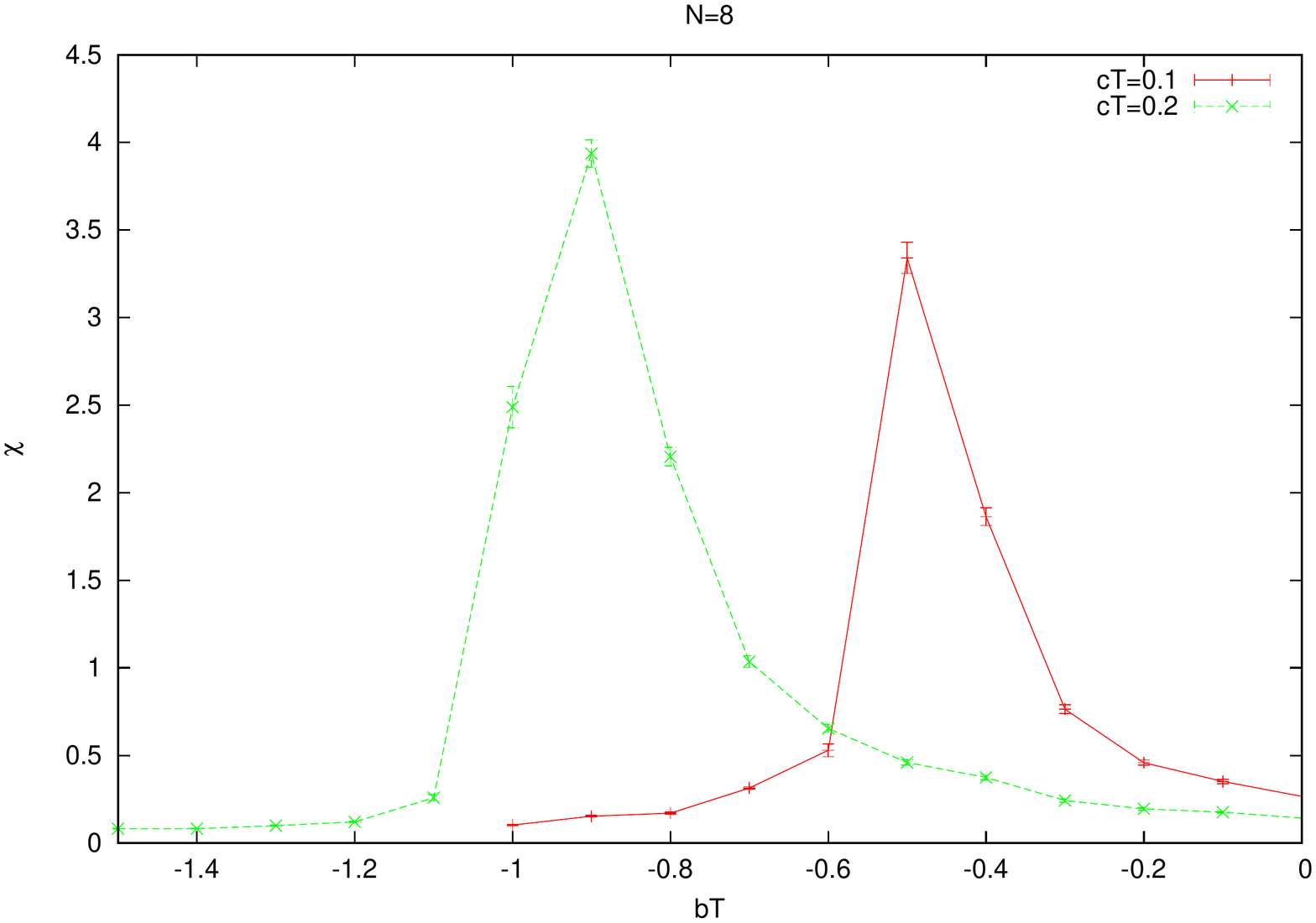}
\end{center}
\caption{}\label{testf4}
\end{figure}

\begin{figure}[htbp]
\begin{center}
\includegraphics[width=7.0cm,angle=-0]{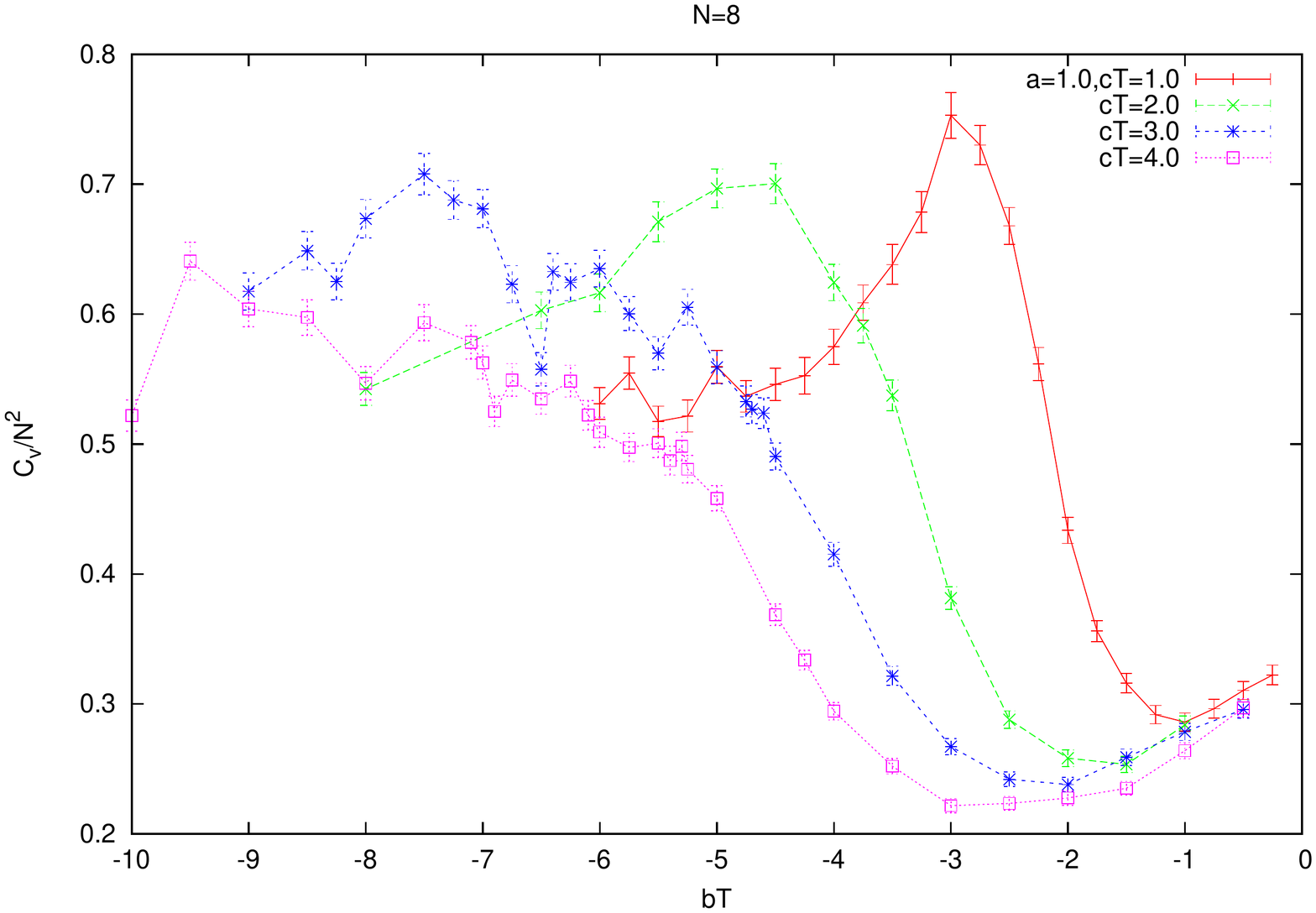}
\includegraphics[width=7.0cm,angle=-0]{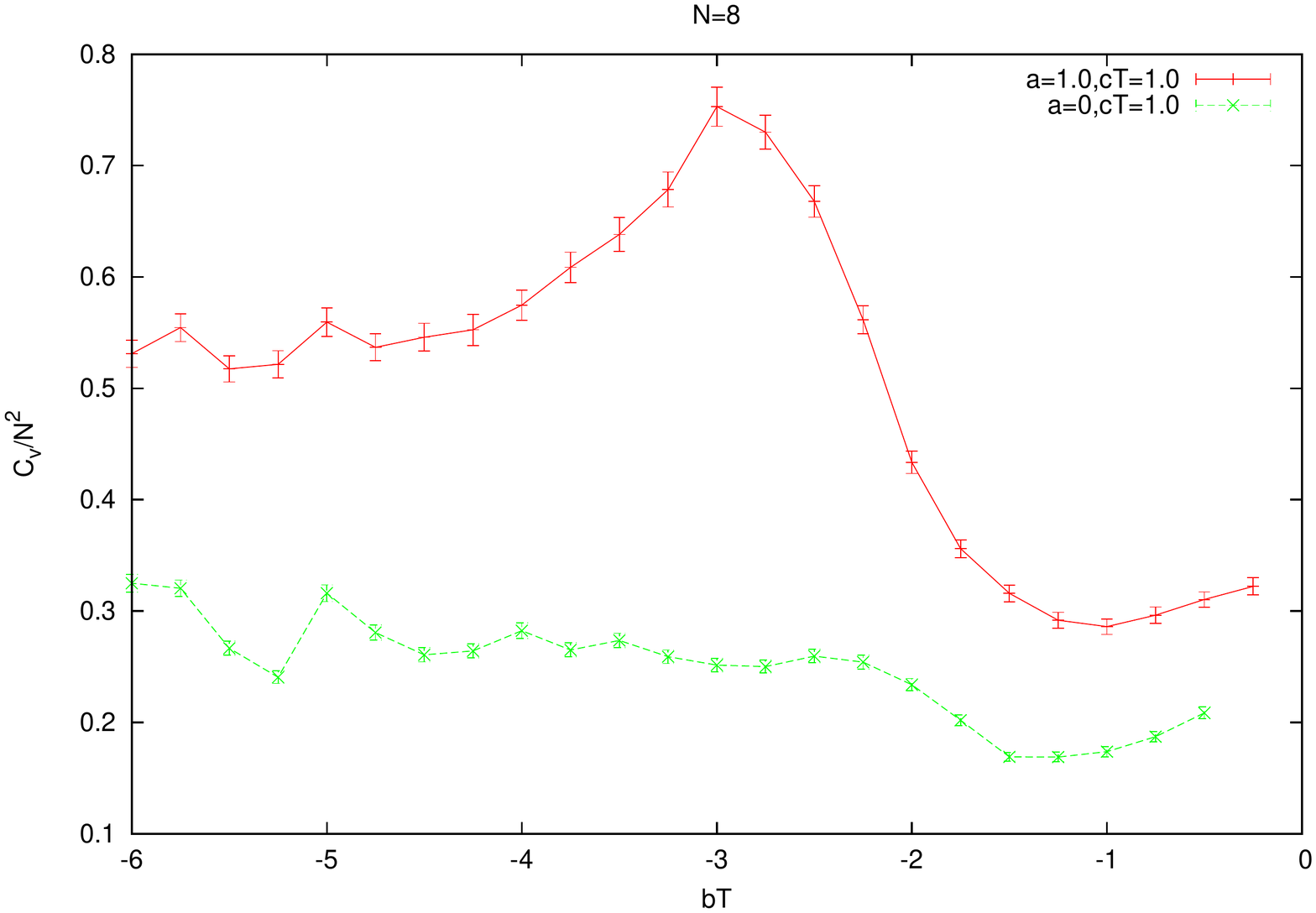}
\includegraphics[width=7.0cm,angle=-0]{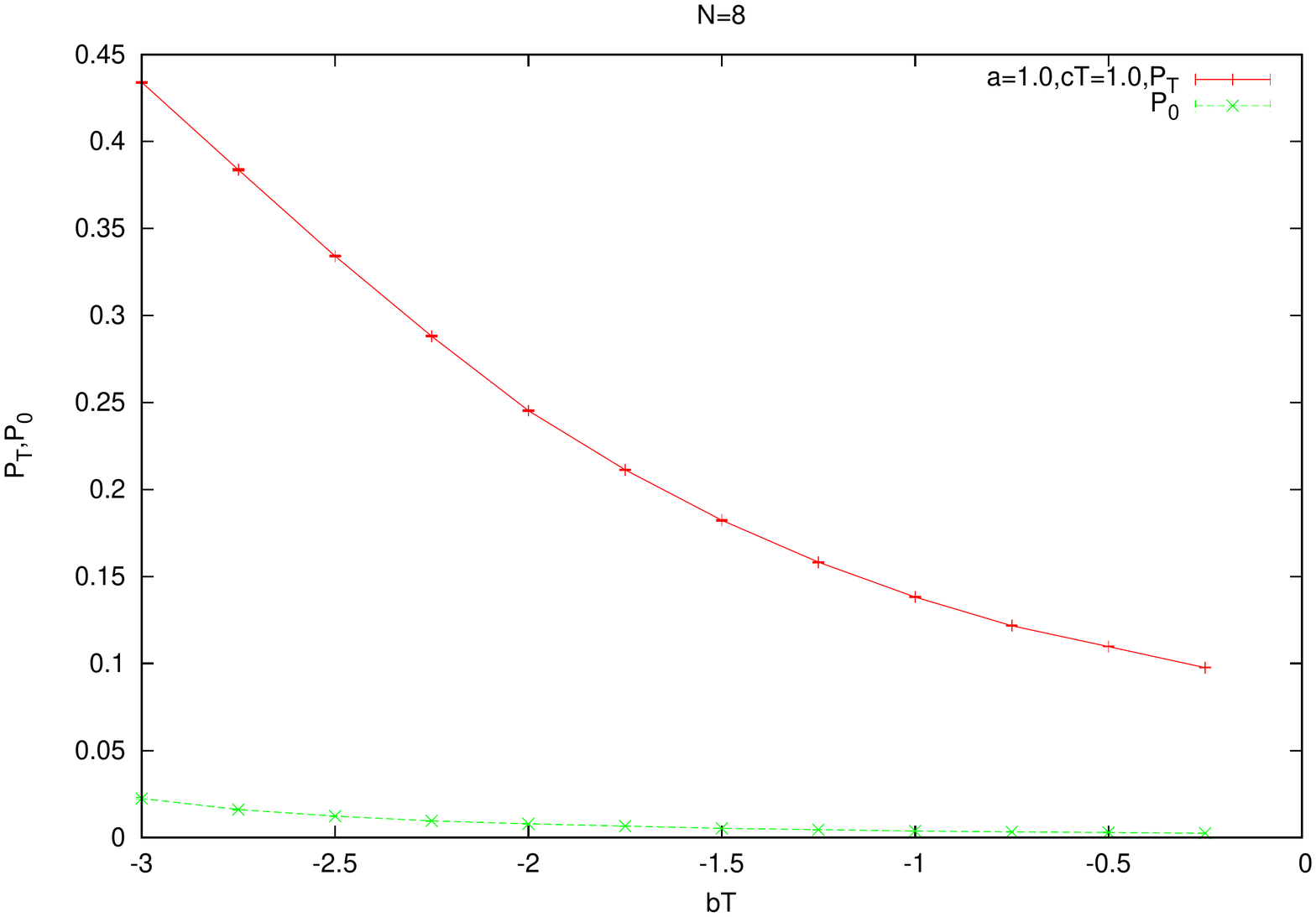}
\includegraphics[width=7.0cm,angle=-0]{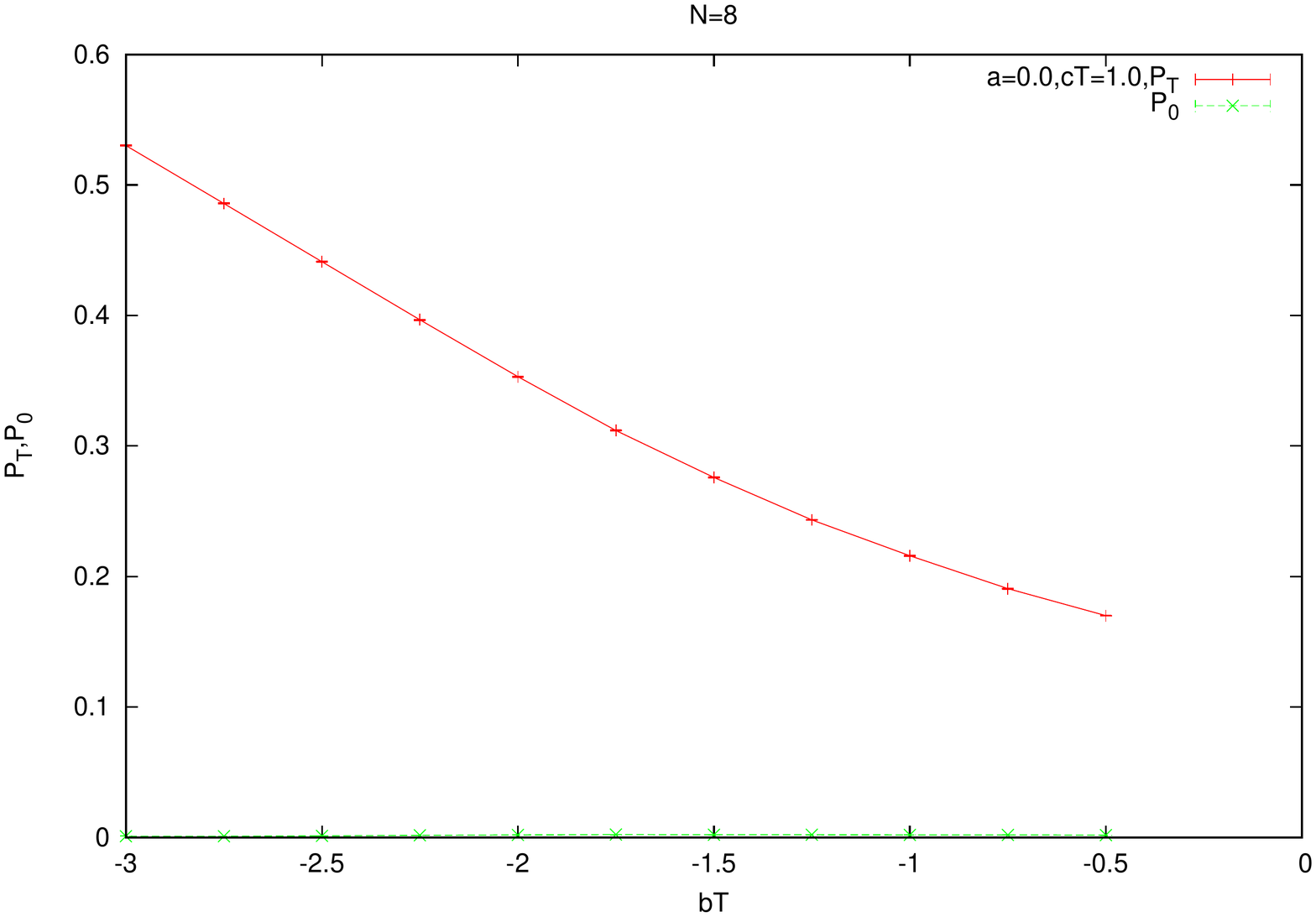}
\includegraphics[width=7.0cm,angle=-0]{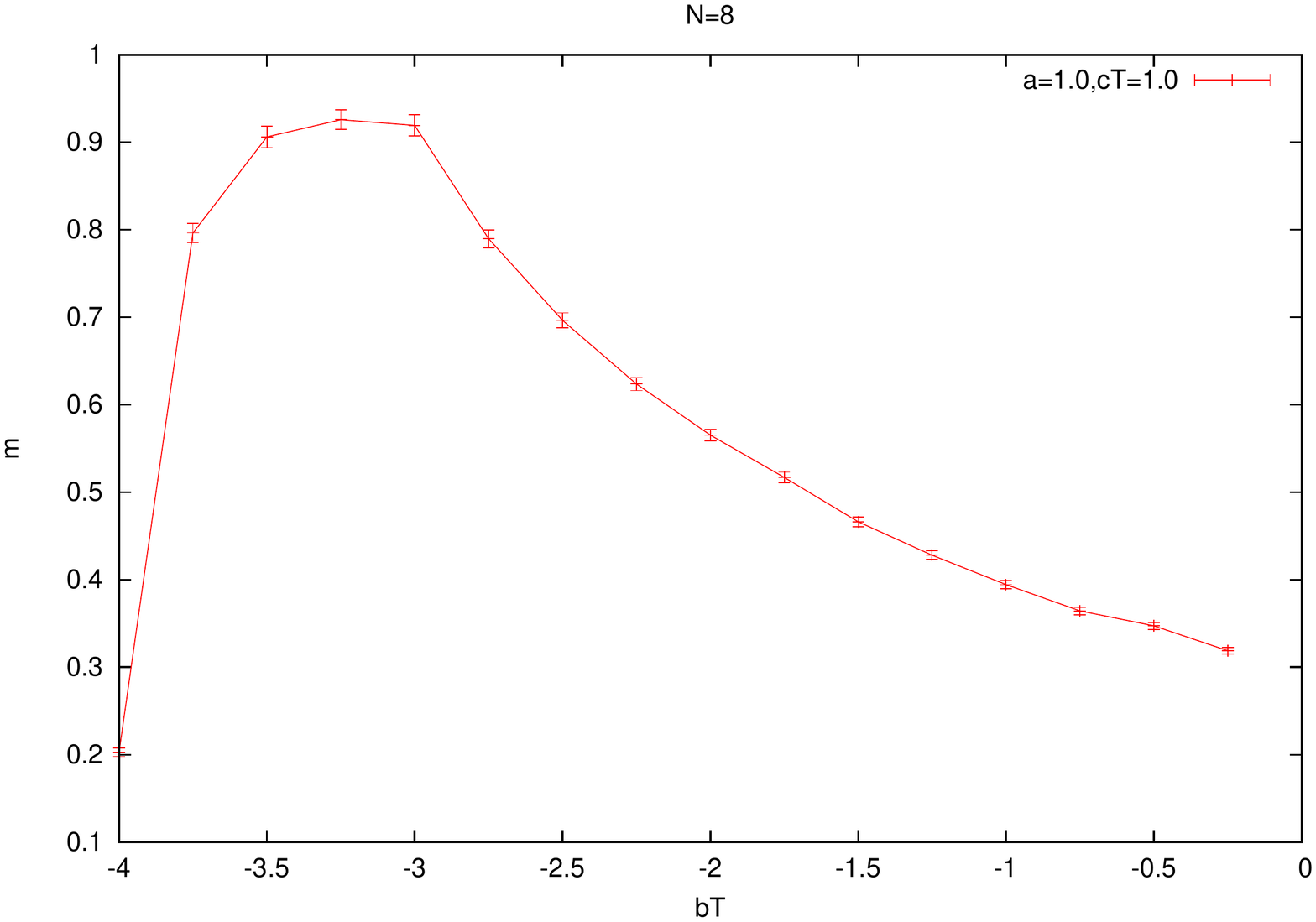}
\includegraphics[width=7.0cm,angle=-0]{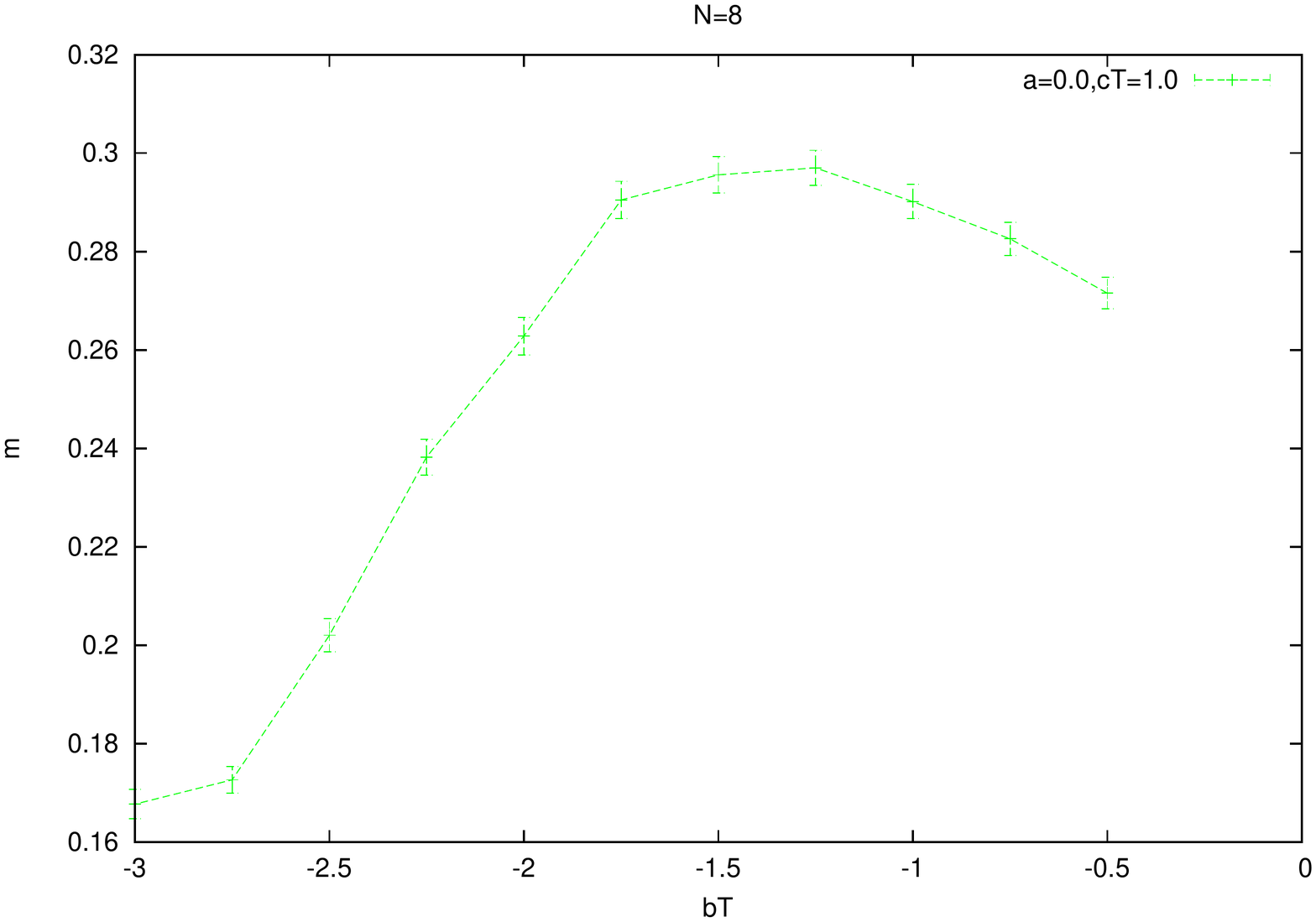}
\includegraphics[width=7.0cm,angle=-0]{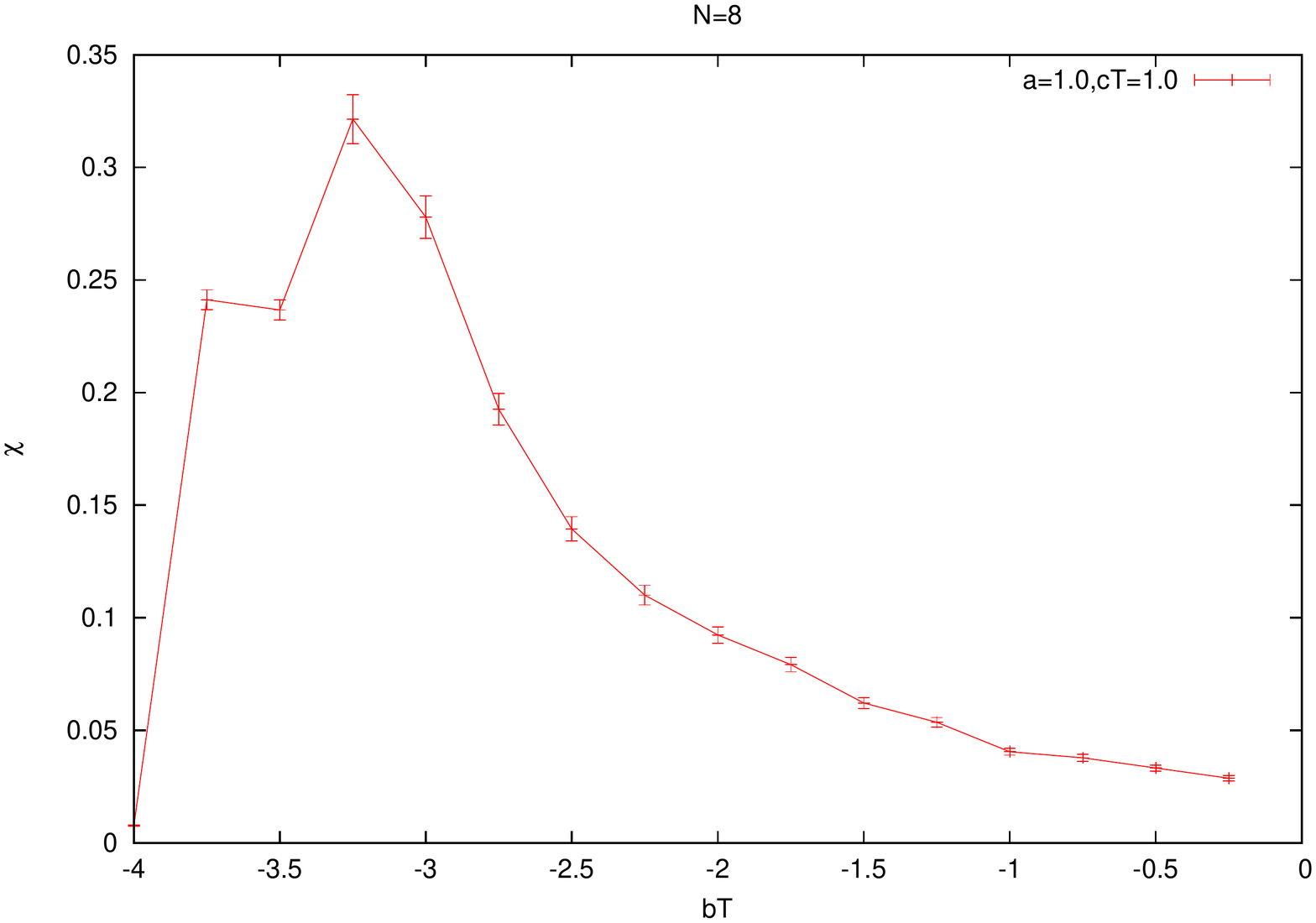}
\includegraphics[width=7.0cm,angle=-0]{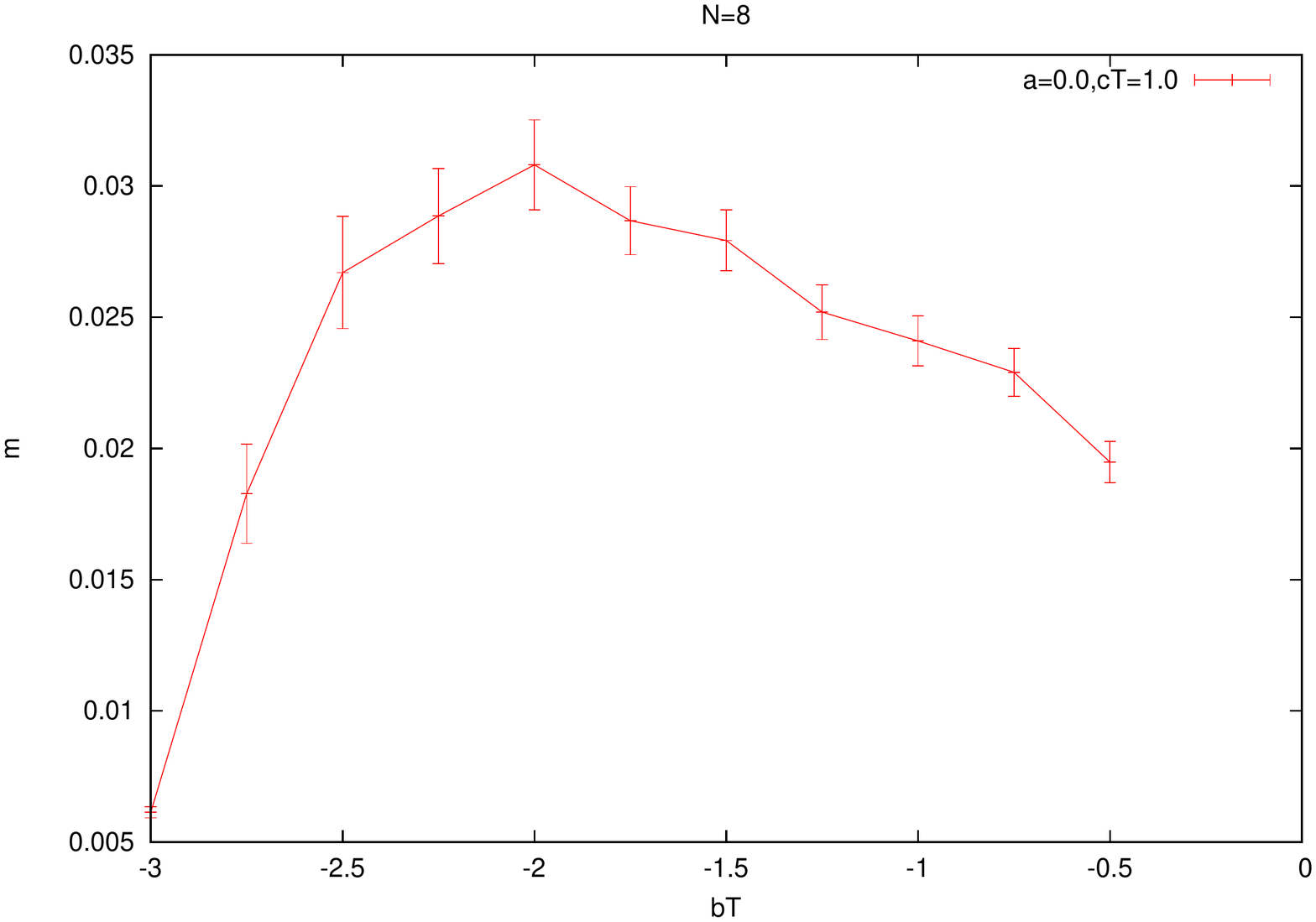}
\end{center}
\caption{}\label{testf5}
\end{figure}

\begin{figure}[htbp]
\begin{center}
\includegraphics[width=9.0cm,angle=-0]{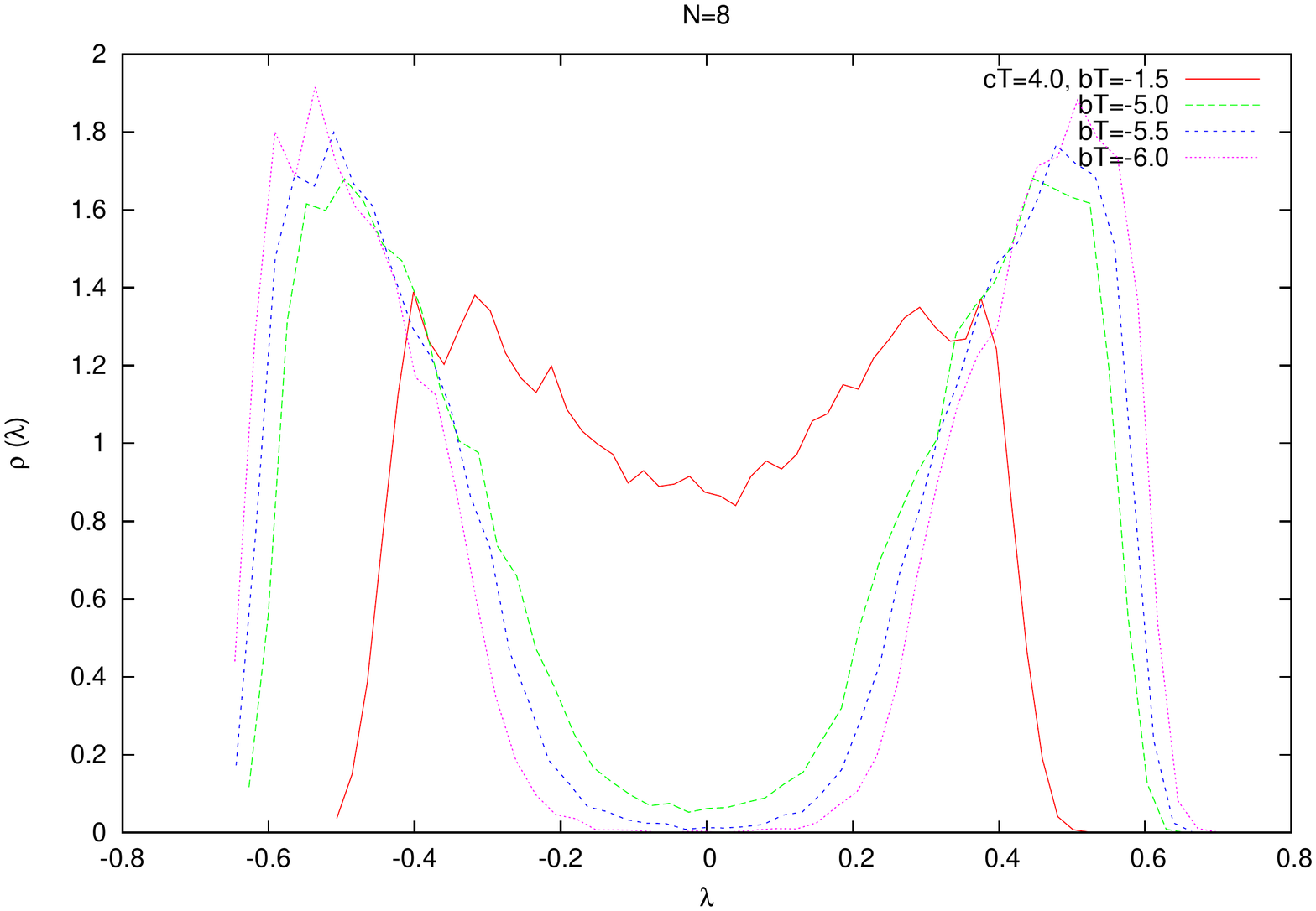}
\includegraphics[width=9.0cm,angle=-0]{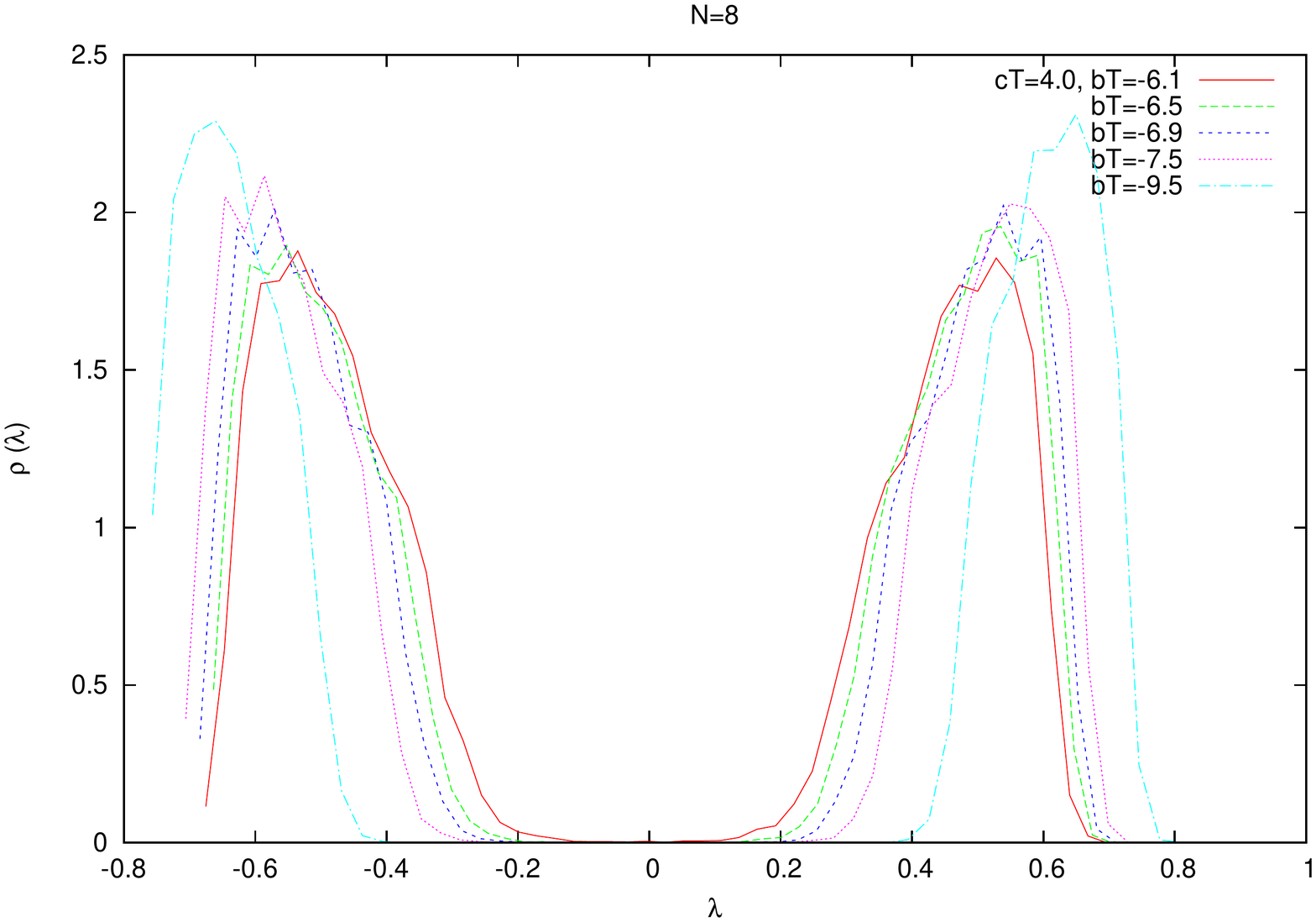}
\includegraphics[width=9.0cm,angle=-0]{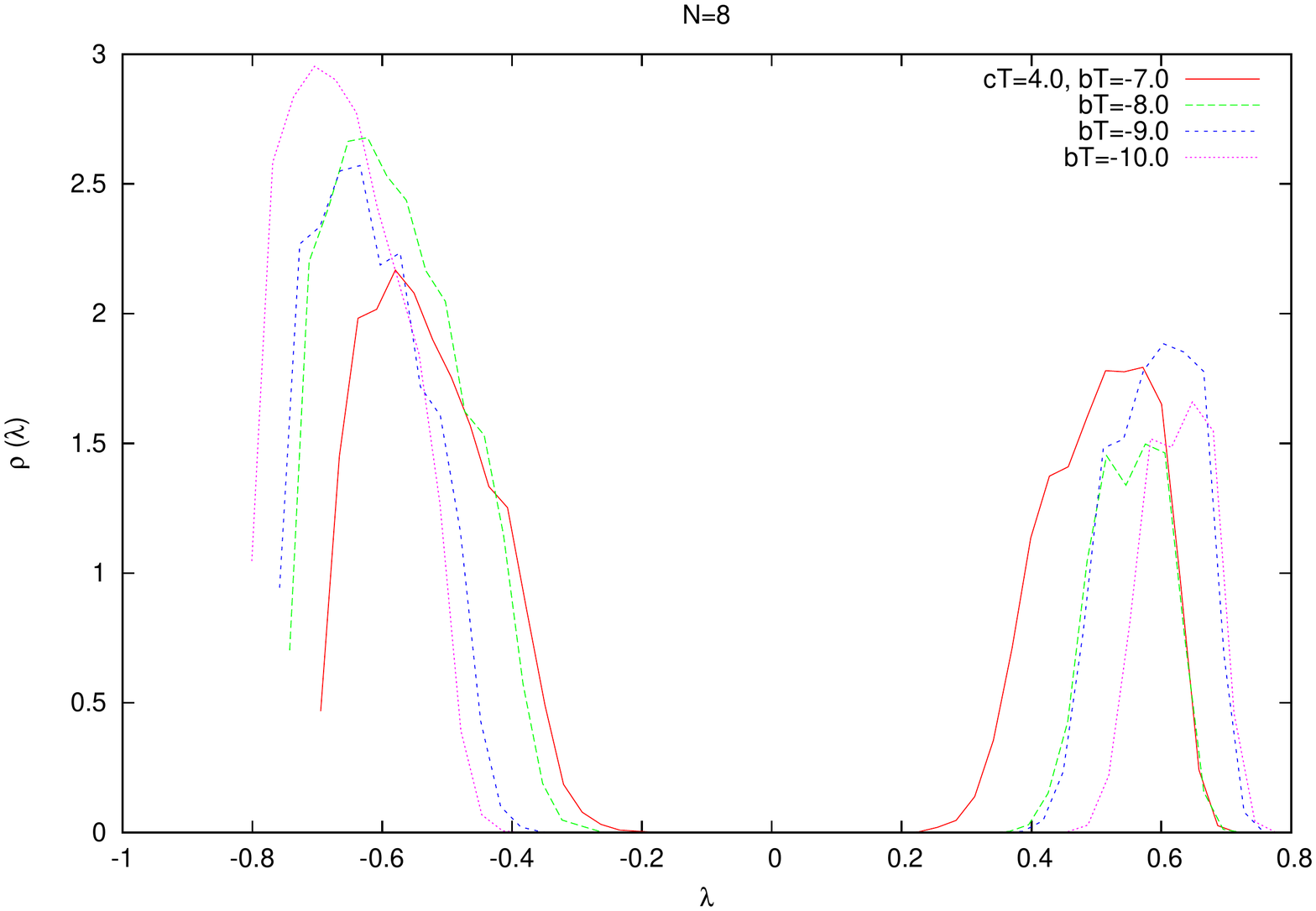}
\end{center}
\caption{}\label{testf6}
\end{figure}

We measure the action $<S>$, the specific heat $C_v$, the magnetization $m$ and the associated susceptibility $\chi$, the total power $P_T$, and the power in the zero modes $P_0$ defined respectively by 
\begin{eqnarray}
C_v=<S^2>-<S>^2.
\end{eqnarray}
\begin{eqnarray}
m=<|Tr\Phi|>.
\end{eqnarray}
\begin{eqnarray}
\chi=<|Tr\Phi|^2>-<|Tr\Phi|>^2.
\end{eqnarray}
\begin{eqnarray}
P_T=\frac{1}{N}Tr \Phi^2.
\end{eqnarray}
\begin{eqnarray}
P_0=\frac{1}{N^2}(Tr \Phi)^2.
\end{eqnarray}
We will also compute the eigenvalues of the matrix $\Phi$ by calling the library LAPACK and then construct appropriate histograms using known techniques. 
\paragraph{Ising:}
The Ising transition appears for small values of $\tilde{c}$ and is the easiest one to observe in Monte Carlo simulations.  We choose, for $N=8$, the Monte Carlo times  $T_{\rm th}=2^{11}$, $T_{\rm mc}=2^{11}$ and $T_{\rm co}=2^{0}$, i.e. we ignore to take into account auto-correlations for simplicity. The data for  $\tilde{c}=0.1,0.2$ is shown on figure (\ref{testf4}). The transition, marked by the peak of the susceptibility, occurs, for $\tilde{c}=0.1$, $0.2$, $0.3$ and $0.4$,  at $\tilde{b}=-0.5$, $-0.9$, $-1.4$ and $-1.75$ respectively. The corresponding linear fit which goes through the origin is  given by
\begin{eqnarray}
\tilde{c}=-0.22 \tilde{b}_*.\label{fit}
\end{eqnarray}
\paragraph{Matrix:}
The disorder-to-non-uniform phase transition appears for large values of $\tilde{c}$ and is quite difficult to observe in Monte Carlo simulations due to the fact that configurations, which have slightly different numbers of pluses and minuses, strongly competes for finite $N$, with the physically relevant stripe configuration with an equal numbers of pluses and minuses.  In principle then we should run the simulation until a symmetric eigenvalues distribution is reached which can be very difficult to achieve in practice. We choose, for $N=8$, the Monte Carlo times  $T_{\rm th}=2^{11}$, $T_{\rm mc}=2^{12}$ and $T_{\rm co}=2^{4}$. The data for  the specific heat for $\tilde{c}=1-4$ is shown on figure (\ref{testf5}). We also plot the data for the pure quartic matrix model for $\tilde{c}=1$ for comparison. The transition for smaller value of $\tilde{c}$ is marked, as before, by the peak in specific heat. However, this method becomes unreliable for larger values of $\tilde{c}$ since the peak disappears. Fortunately, the transition is always marked by the point where the eigenvalues distribution splits at $\lambda=0$.   The corresponding eigenvalues distributions are shown on (\ref{testf6}).  We include symmetric and slightly non-symmetric distributions since both were taken into account in the data of the specific heat. The non-symmetric distributions cause typically large fluctuations of the magnetization and peaks in the susceptibility which are very  undesirable finite size effects. But, on the other hand, as we increase the value of $|\tilde{b}|$ we are  approaching the non-symmetric uniform phase and thus the appearance of these  non-symmetric distributions is very natural. This makes the determinantion of the transition point very hard from the behavior of these observables. 

\begin{figure}[htbp]
\begin{center}
\includegraphics[width=8.0cm,angle=-0]{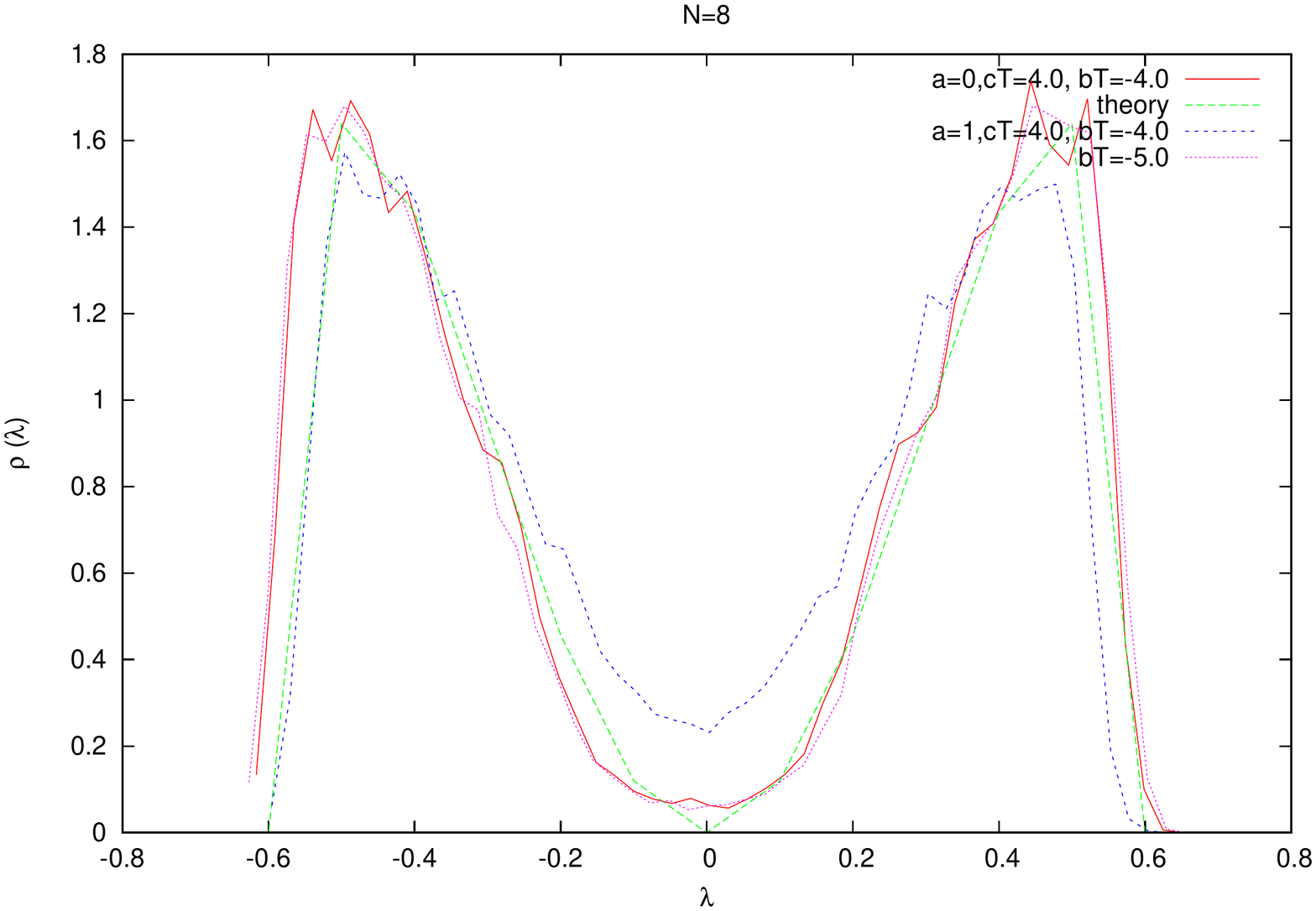}
\includegraphics[width=8.0cm,angle=-0]{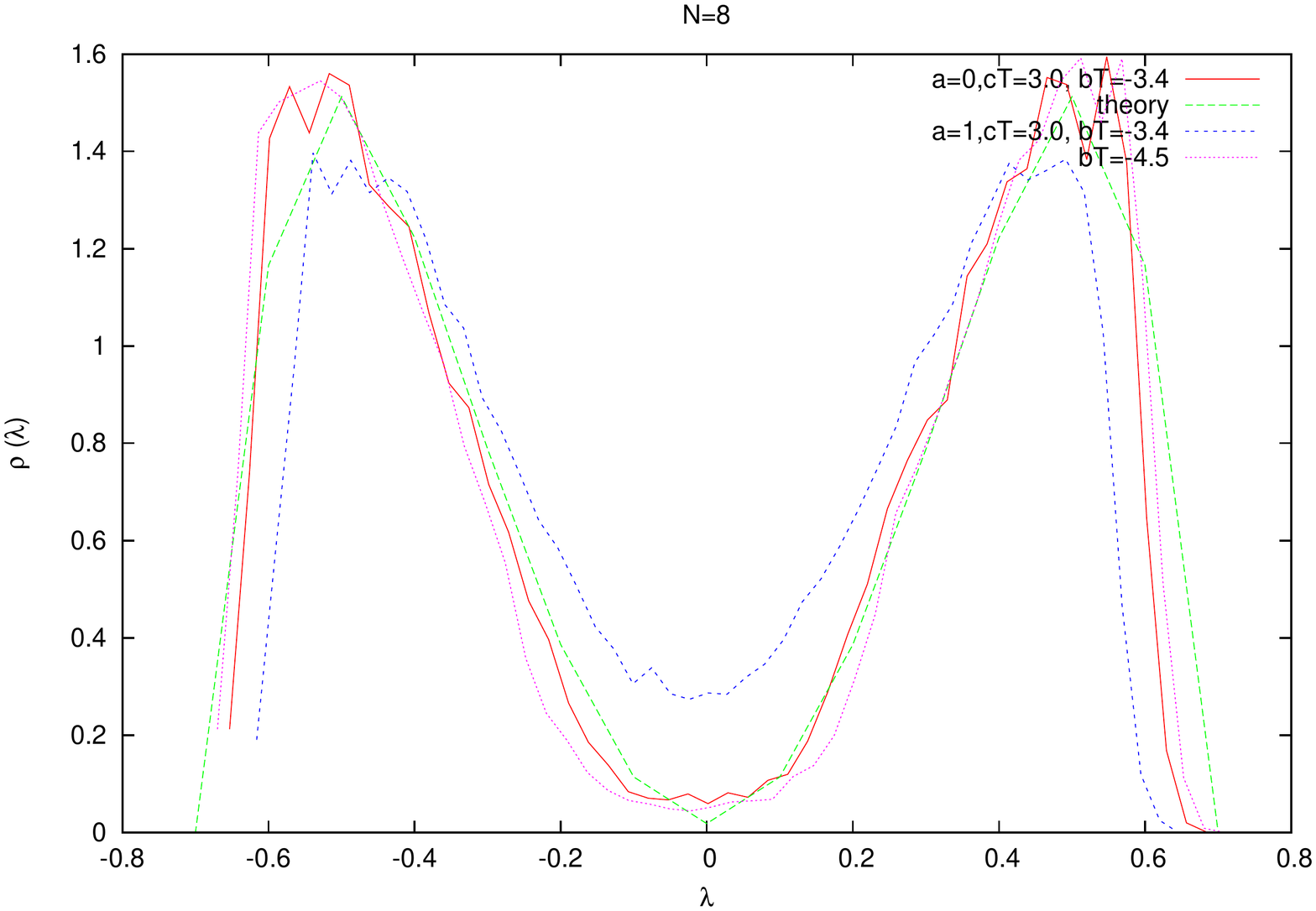}
\includegraphics[width=8.0cm,angle=-0]{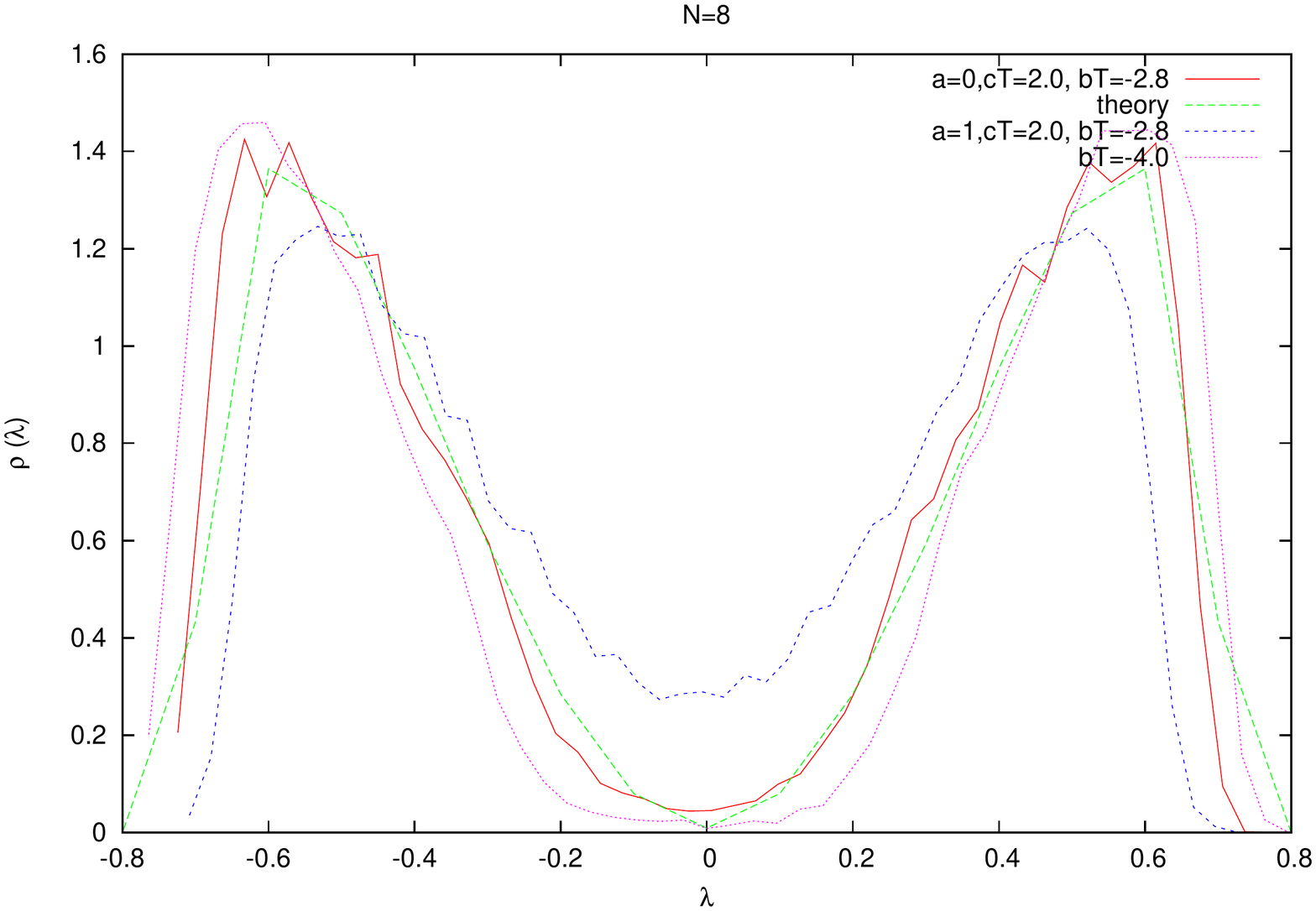}
\includegraphics[width=8.0cm,angle=-0]{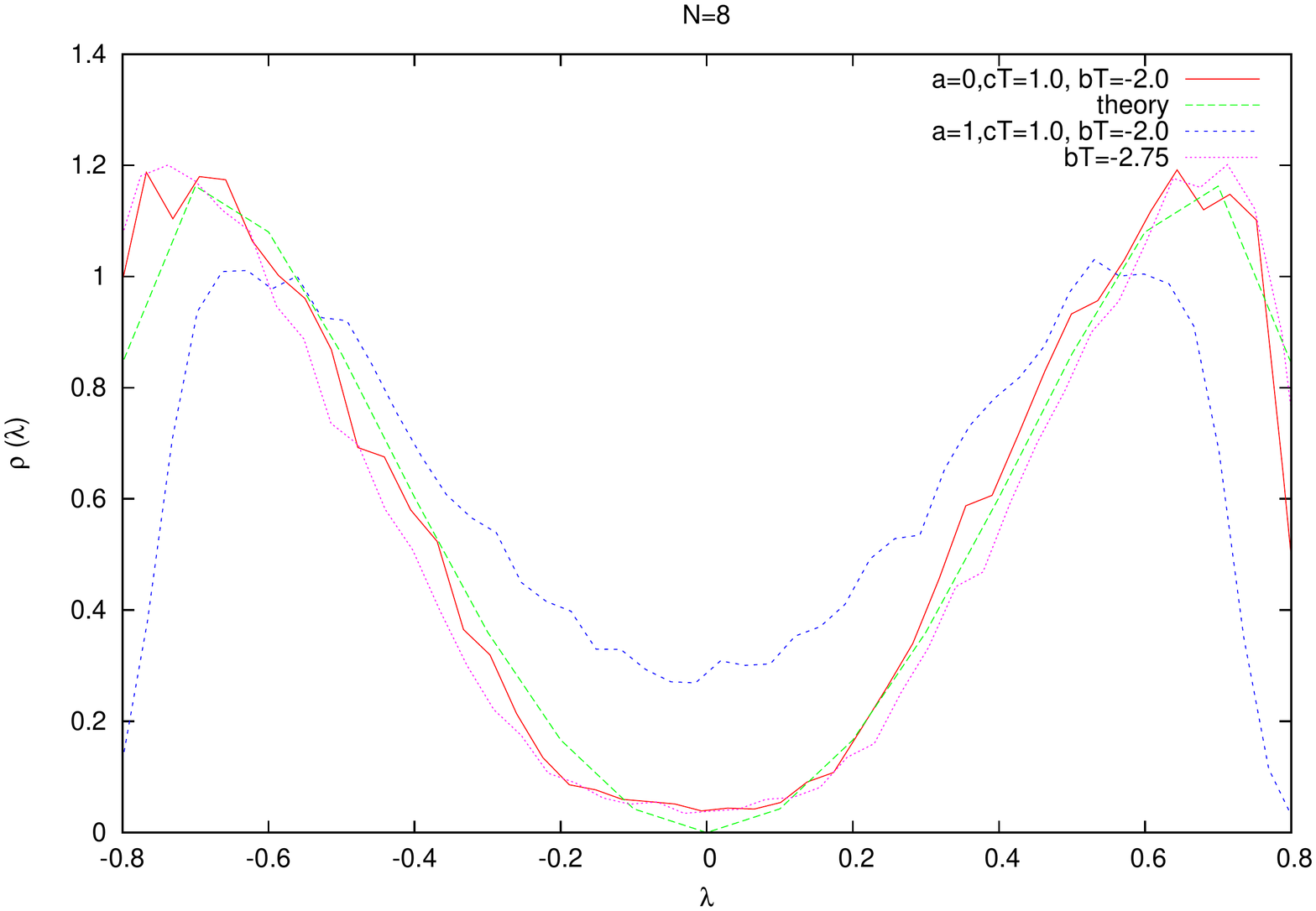}
\includegraphics[width=8.0cm,angle=-0]{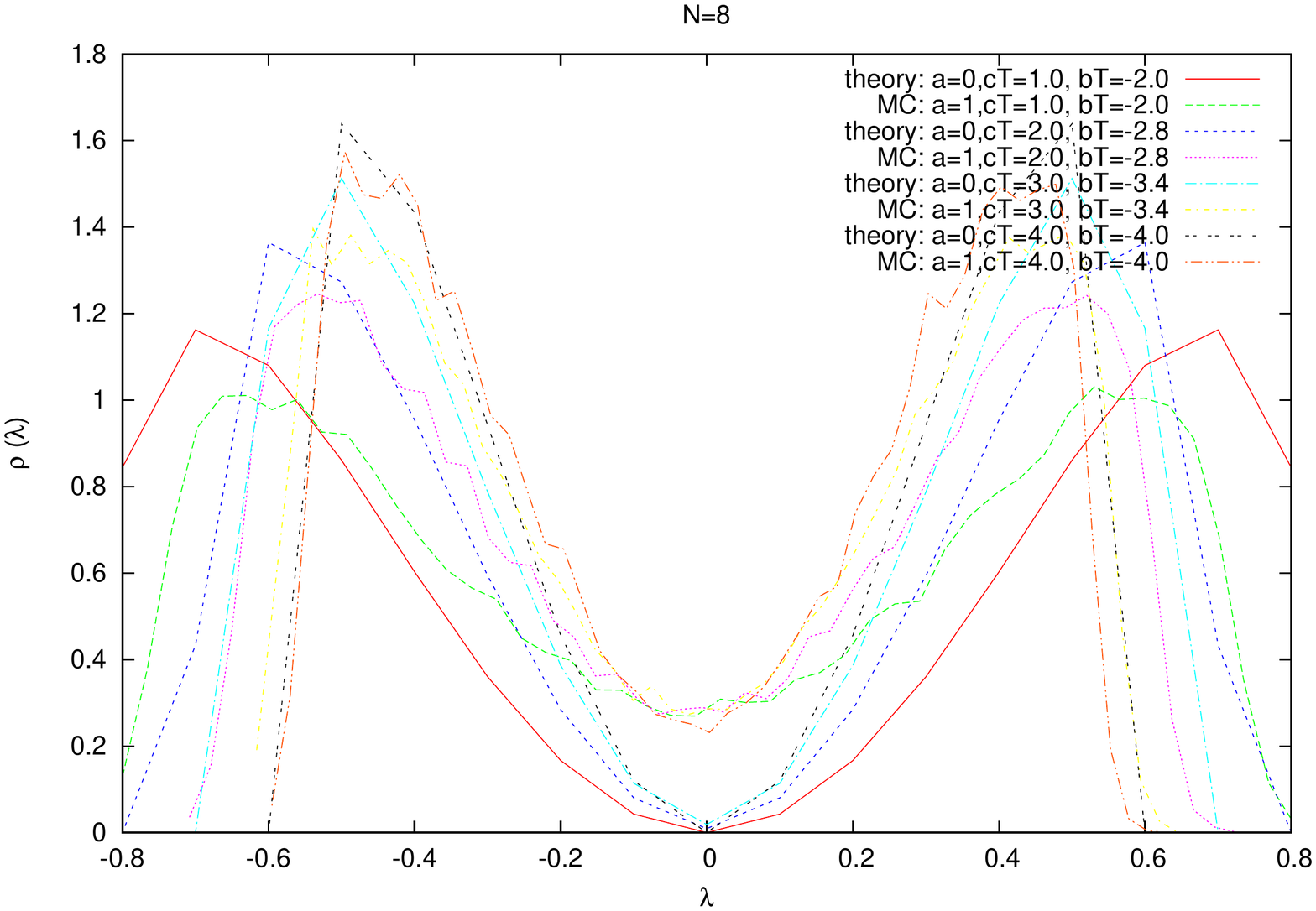}
\end{center}
\caption{}\label{testf7}
\end{figure}

\begin{figure}[htbp]
\begin{center}
\includegraphics[width=9.0cm,angle=-0]{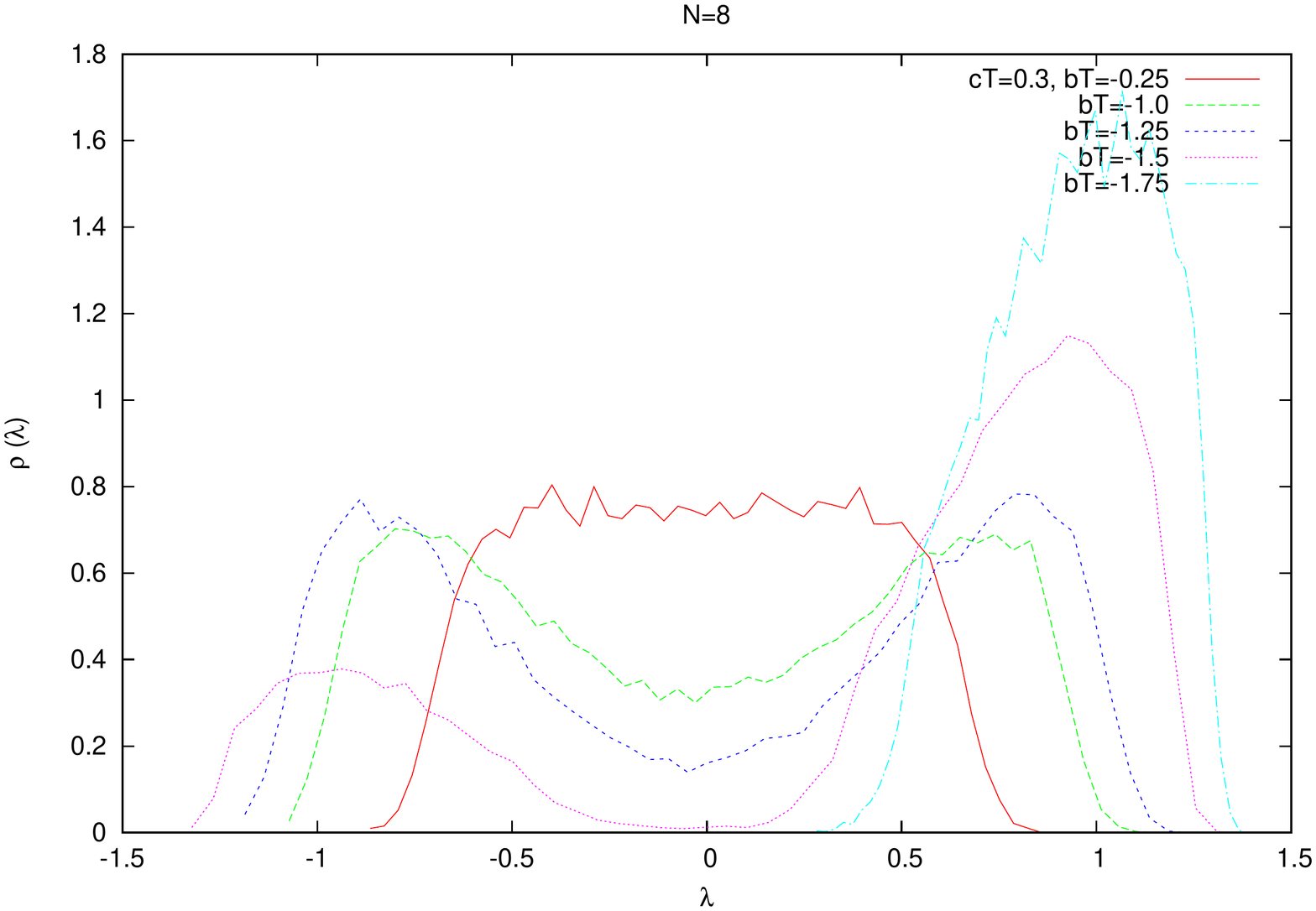}
\includegraphics[width=9.0cm,angle=-0]{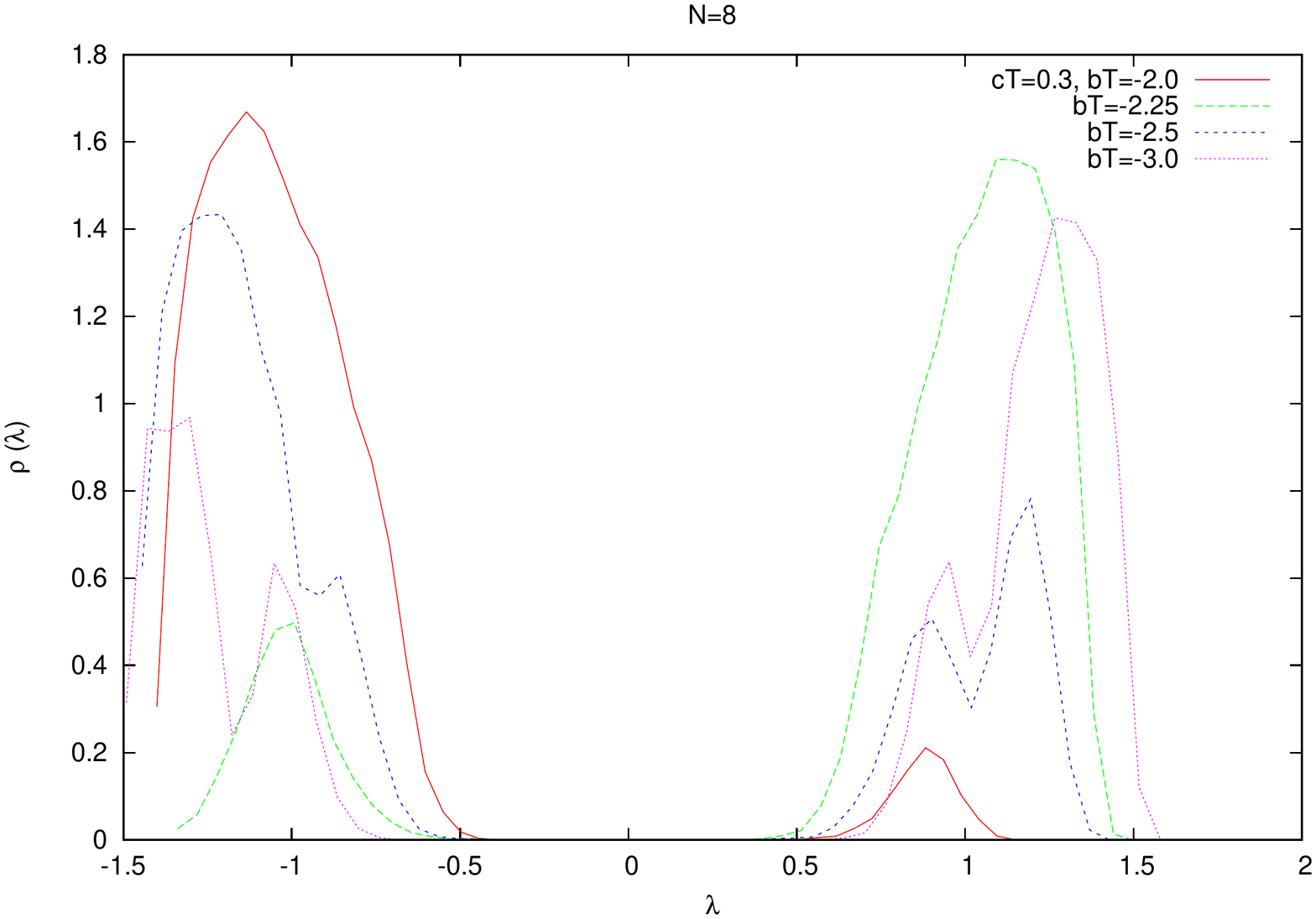}
\includegraphics[width=9.0cm,angle=-0]{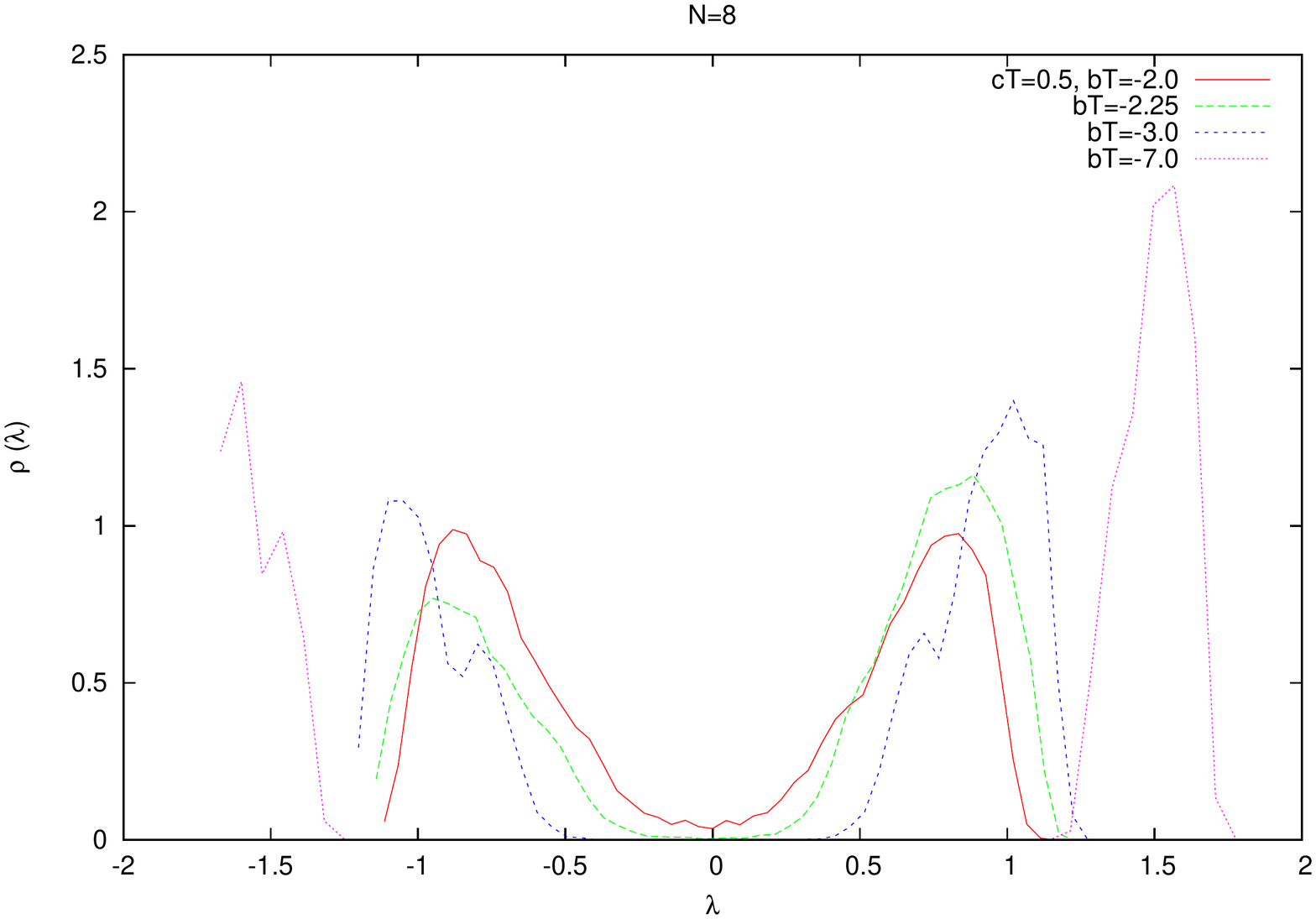}
\end{center}
\caption{}\label{testf9}
\end{figure}

\begin{figure}[htbp]
\begin{center}
\includegraphics[width=10.0cm,angle=-0]{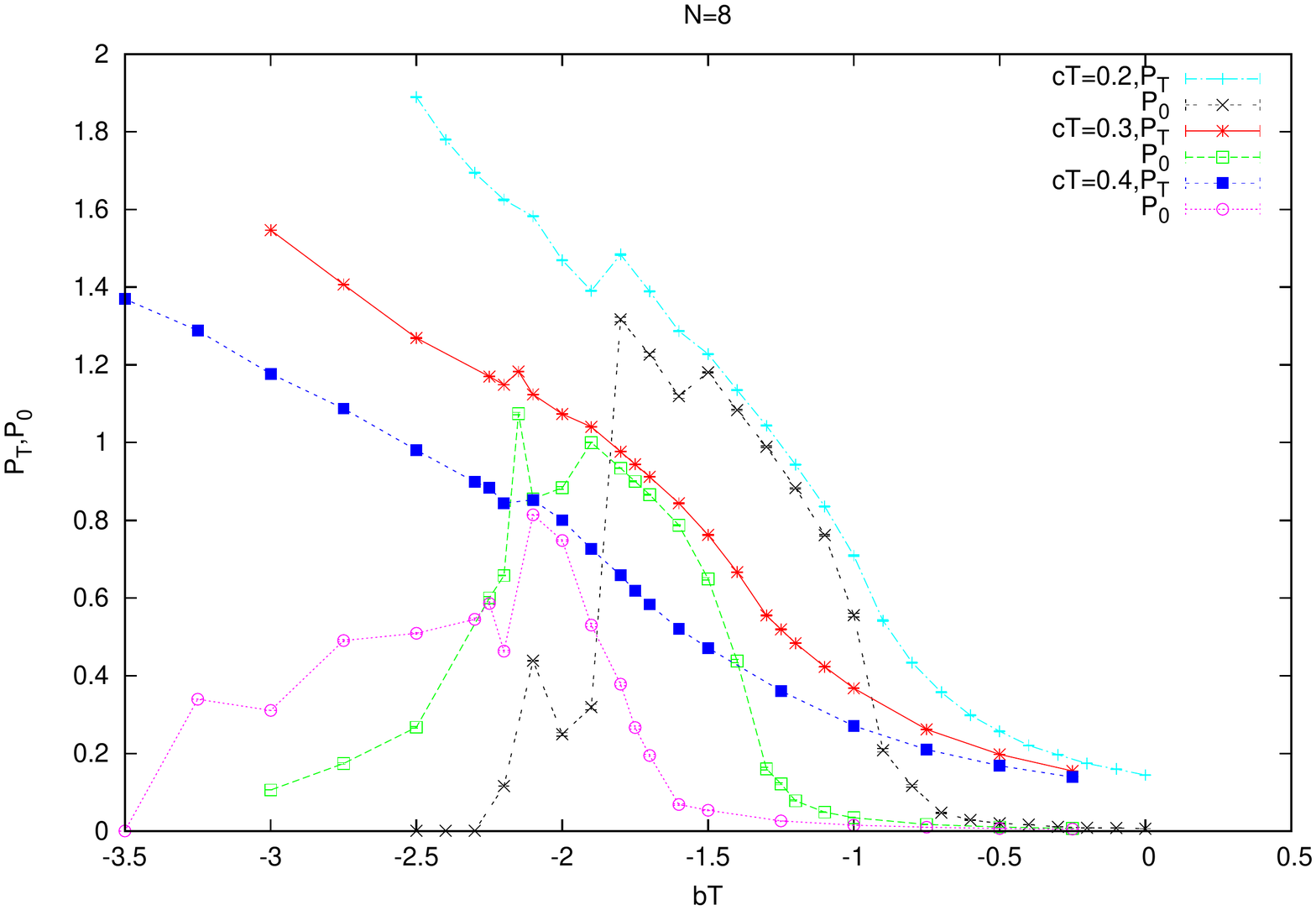}
\includegraphics[width=10.0cm,angle=-0]{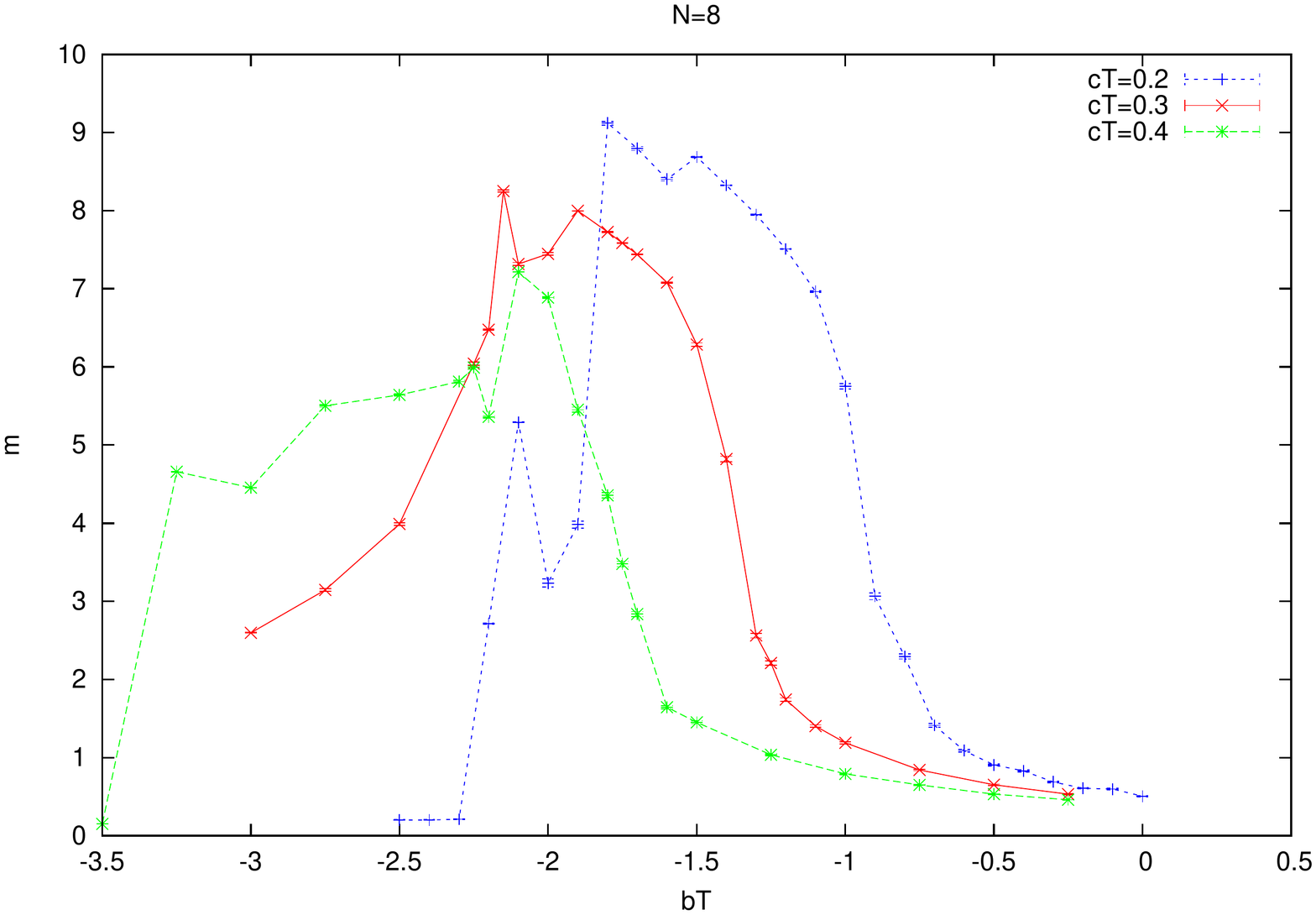}
\end{center}
\caption{}\label{testf10}
\end{figure}

We have determined instead the transition point by simulating, for a given $\tilde{c}$, the pure matrix model with $a=0$, in which we know that the transition occurs at $\tilde{b}_*=-2\sqrt{\tilde{c}}$, and then searching in the full model with $a=1$ for the value of $\tilde{b}$ with an eigenvalues distribution similar to the eigenvalues distribution found for  $a=0$ and  $\tilde{b}_*=-2\sqrt{\tilde{c}}$. This exercise is repeated for $\tilde{c}=4,3$, $2$ and $1$ and we found the transition points given  respectively by $\tilde{b}_*=-5$, $-4.5$, $-4$, and $-2.75$. See graphs on figure (\ref{testf7}). The corresponding  linear fit is given by
\begin{eqnarray}
\tilde{c}=-1.3 \tilde{b}_*-2.77.\label{fit1}
\end{eqnarray}
Two more observations concerning this transition are in order:
\begin{itemize}
\item The eigenvalues distribution for the pure matrix model with $a=0$ is such that it depends only on a single parameter given by $g=4Nc/b^2$. See next chapter for more detail. From the Monte Carlo data the same statement seems to hold in the full model with $a=1$ along the disorder-to-non-uniform boundary. See last graph on figure  (\ref{testf7}).
\item The  disorder-to-non-uniform transition line seems to be better approximated by a shift of the result  $\tilde{b}_*=-2\sqrt{\tilde{c}}$ by a single unit in the $-\tilde{b}$ direction. This is roughly in accord with the analytic result for the critical point found in \cite{Ydri:2014uaa0} for the multitrace approximation (see next chapter) which is given, for $a=1$, by
\begin{eqnarray}
\tilde{b}_*=-\frac{\sqrt{N}}{2}-2\sqrt{\tilde{c}}+\frac{N}{6\sqrt{\tilde{c}}}.
\end{eqnarray}
\end{itemize}
\paragraph{Stripe:}
The uniform-to-non-uniform phase transition is even more difficult to observe in Monte Carlo simulations but it is expected, according to\cite{GarciaFlores:2009hf0,GarciaFlores:2005xc0}, to only be a continuation of the disorder-to-uniform transition line (\ref{fit}).   The intersection point between the above two fits (\ref{fit}) and (\ref{fit1}) is therefore an estimation of the triple point. This is given by 
\begin{eqnarray}
(\tilde{c},\tilde{b})=(0.56,-2.57).
\end{eqnarray}
However, this is not really what we observe using our code here. The  uniform-to-non-uniform phase transition is only observed for small values of $\tilde{c}$ from the uniform phase to the non-uniform phase as we increase $-\tilde{b}$. The transition for these small values of $\tilde{c}$, such as $\tilde{c}=0.1,0.2, 0.3, 0.4$, corresponds to a second peak in the susceptibility and the specific heat. It corresponds to a transition from a one-cut eigenvalues distribution symmetric around $0$ to a one-cut eigenvalues distribution symmetric around a non-zero value. The eigenvalues distributions for $\tilde{c}=0.3$ are shown on the first two graphs of figure (\ref{testf9}). In this case we have found it much easier to determine the transition points from the behavior of the magnetization and the powers. In particular, we have determined the transition point from the broad maximum of the magnetization which corresponds to the discontinuity of the power in the zero modes. The magnetization and the powers, for  $\tilde{c}=0.1,0.2, 0.3, 0.4$, are shown on figure (\ref{testf10}). The transition points were found to be $-1.5$, $-1.7$, $-2$ and $-2.1$ respectively. 

The uniform phase becomes narrower as we approach the value $\tilde{c}=0.5$. The specific heat and the susceptibility have a peak around $\tilde{b}=-2.25$ which is consistent with the Ising transition but the powers and the magnetization show the behavior of the disorder-to-non-uniform-order transition. The eigenvalues distribution is also consistent with the disorder-to-non-uniform-order transition. See last graph of  figure (\ref{testf9}). The value  $\tilde{c}=0.5$ is roughly the location of the triple point.

\begin{figure}[htbp]
\begin{center}
\includegraphics[width=10.0cm,angle=-0]{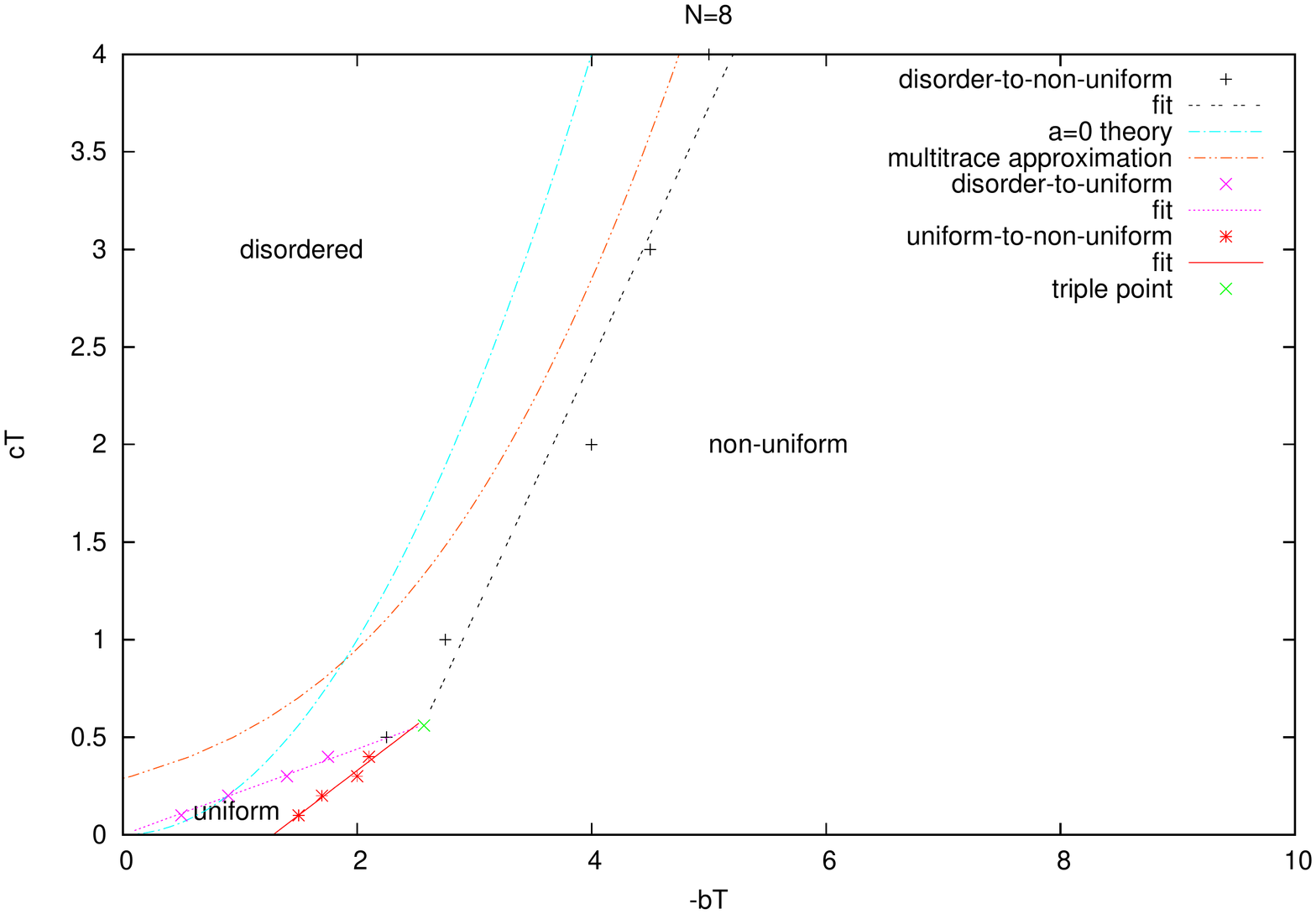}
\end{center}
\caption{}\label{testf8}
\end{figure}

The phase diagram is shown on figure (\ref{testf8}).


\chapter{Lattice HMC Simulations of $\Phi_2^4$: A Lattice Example}
References for this chapter include the elegant quantum field theory textbook \cite{Smit:2002ugp4} and the original articles \cite{Loinaz:1997az,Schaich:2009jk,Chang:1976ek}.

\section{Model and Phase Structure}
The Euclidean $\phi^4$ action with $O(N)$ symmetry is given by 
\begin{eqnarray}
S[\phi]=\int d^dx \bigg(\frac{1}{2}(\partial_{\mu}\phi^i)^2+\frac{1}{2}m^2\phi^i\phi^i+\frac{\lambda}{4}(\phi^i\phi^i)^2\bigg).\label{contaction}
\end{eqnarray}
We will employ lattice regularization in which $x=an$, $\int d^dx =a^d\sum_n$, $\phi^i(x)=\phi_n^i$ and $\partial_{\mu}\phi^i=(\phi_{n+\hat{\mu}}^i-\phi_n^i)/a$. The lattice action reads 
\begin{eqnarray}
S[\phi]
&=&\sum_n\bigg(-2\kappa \sum_{\mu}\Phi_n^i\Phi_{n+\hat{\mu}}^i+{\Phi}_n^i\Phi_n^i+g(\Phi_n^i\Phi_n^i-1)^2 \bigg).
\end{eqnarray}
The mass parameter $m^2$ is replaced by the so-called hopping parameter $\kappa$ and the coupling constant $\lambda$ is replaced by the coupling constant $g$ where
\begin{eqnarray}
m^2a^2=\frac{1-2g}{\kappa}-2d~,~\frac{\lambda}{a^{d-4}}=\frac{g}{\kappa^2}.
\end{eqnarray}
The fields $\phi_n^i$ and $\Phi_n^i$ are related by
\begin{eqnarray}
\phi_n^i=\sqrt{\frac{2\kappa}{a^{d-2}}}\Phi_n^i.
\end{eqnarray}
The partition function is given by
\begin{eqnarray}
Z&=&\int \prod_{n,i} d\Phi_n^{i}~e^{-S[\phi]}\nonumber\\
&=&\int d\mu(\Phi)~e^{2\kappa\sum_n\sum_{\mu}\Phi_n^i\Phi_{n+\hat{\mu}}^i}.
\end{eqnarray}
The measure $d\mu(\phi)$ is given by
\begin{eqnarray}
d\mu(\Phi)&=&\prod_{n,i} d\Phi_n^i ~e^{-\sum_n\big({\Phi}_n^i\Phi_n^i+g(\Phi_n^i\Phi_n^i-1)^2 \big)}\nonumber\\
&=&\prod_n \bigg(d^N\vec{\Phi}_n ~e^{-\vec{\Phi}_n^2-g(\vec\Phi_n^2-1)^2 }\bigg)\nonumber\\
&\equiv &\prod_n d\mu(\Phi_n).
\end{eqnarray}
This is a generalized Ising model. Indeed in the limit $g\longrightarrow \infty $ the dominant configurations are such that $\Phi_1^2+...+\Phi_N^2=1$, i.e.  points on the sphere $S^{N-1}$. Hence

\begin{eqnarray}
\frac{\int d\mu(\Phi_n) f(\vec{\Phi}_n)}{\int d\mu(\Phi_n)}=\frac{\int d\Omega_{N-1} f(\vec{\Phi}_n)}{\int d\Omega_{N-1}}~,~g\longrightarrow\infty.
\end{eqnarray}
For $N=1$ we obtain 
\begin{eqnarray}
\frac{\int d\mu(\Phi_n) f(\vec{\Phi}_n)}{\int d\mu(\Phi_n)}=\frac{1}{2}(f(+1)+f(-1))~,~g\longrightarrow\infty .
\end{eqnarray}
Thus the limit  $g\longrightarrow \infty $ of the $O(1)$ model is precisely the Ising model in $d$ dimensions. The limit $g\longrightarrow \infty $ of the $O(3)$ model corresponds to the Heisenberg model in $d$ dimensions. The $O(N)$ models on the lattice are thus intimately related to spin models.

There are two phases in this model. A disordered (paramagnetic) phase characterized by $<\Phi_n^i>=0$  and an ordered (ferromagnetic) phase characterized by $<\Phi_n^i>=v_i\neq 0$. This can be seen in various ways. The easiest way is to look for the minima of the classical potential 
\begin{eqnarray}
V[\phi]=-\int d^dx \bigg(\frac{1}{2}m^2\phi^i\phi^i+\frac{\lambda}{4}(\phi^i\phi^i)^2\bigg).
\end{eqnarray}
The equation of motion reads
\begin{eqnarray}
[m^2+\frac{\lambda}{2}\phi^j\phi^j]\phi^i=0.
\end{eqnarray}
For $m^2>0$ there is a unique solution  $\phi^i=0$ whereas for $m^2<0$ there is a second  solution given by $\phi^j\phi^j=-2m^2/\lambda$.

A more precise calculation is as follows. Let us compute the expectation value $<\Phi_n^i>$ on the lattice which is defined by
\begin{eqnarray}
<\phi_n^i>
&=&\frac{\int d\mu(\Phi)~\Phi_n^i e^{2\kappa\sum_n\sum_{\mu}\Phi_n^i\Phi_{n+\hat{\mu}}^i}}{\int d\mu(\Phi)~e^{2\kappa\sum_n\sum_{\mu}\Phi_n^i\Phi_{n+\hat{\mu}}^i}}\nonumber\\
&=&\frac{\int d\mu(\Phi)~\Phi_n^i e^{\kappa \sum_n\Phi_n^i \sum_{\mu}(\Phi_{n+\hat{\mu}}^i+\Phi_{n-\hat{\mu}}^i)}}{\int d\mu(\Phi)~e^{\kappa \sum_n\Phi_n^i \sum_n\sum_{\mu}(\Phi_{n+\hat{\mu}}^i+\Phi_{n-\hat{\mu}}^i)}}.\label{mean}
\end{eqnarray}
Now we approximate the spins $\Phi_n^i$ at the $2d$ nearest neighbors of each spin $\Phi_n^i$ by the average $v^i=<\Phi_n^i>$, viz
\begin{eqnarray}
\frac{\sum_{\mu}(\Phi_{n+\hat{\mu}}^i+\Phi_{n-\hat{\mu}}^i)}{2d}=v^i.
\end{eqnarray}
This is a crude form of the mean field approximation. Equation  (\ref{mean}) becomes
\begin{eqnarray}
v^i
&=&\frac{\int d\mu(\Phi)~\Phi_n^i e^{4\kappa d \sum_n \Phi_n^i v^i}}{\int d\mu(\Phi)~e^{4\kappa d \sum_n\Phi_n^i v^i}}\nonumber\\
&=&\frac{\int d\mu(\Phi_n)~\Phi_n^i e^{4\kappa d  \Phi_n^i v^i}}{\int d\mu(\Phi_n^i)~e^{4\kappa d \Phi_n^i v^i}}.
\end{eqnarray}
The extra factor of $2$ in the exponents comes from the fact that the coupling between any two nearest neighbor spins on the lattice occurs twice. We write the above equation as
\begin{eqnarray}
v^i=\frac{\partial}{\partial J^i}\ln Z[J]|_{J^i=4\kappa dv^i}.
\end{eqnarray}
\begin{eqnarray}
Z[J]
&=&\int d\mu(\Phi_n)~e^{\Phi_n^i J^i}\nonumber\\
&=&\int d^N\Phi_n^i~e^{-\Phi_n^i\Phi_n^i-g(\Phi_n^i\Phi_n^i-1)^2+\Phi_n^i J^i}.
\end{eqnarray}
\paragraph{The limit $g\longrightarrow 0$:} In this case we have
\begin{eqnarray}
Z[J]
&=&\int d^N\Phi_n^i~e^{-\Phi_n^i\Phi_n^i+\Phi_n^i J^i}=Z[0]~e^{\frac{J^i J^i}{4}}.
\end{eqnarray}
In other words
\begin{eqnarray}
v^i=2\kappa_c d v^i\Rightarrow \kappa_c=\frac{1}{2d}.
\end{eqnarray}
\paragraph{The limit $g\longrightarrow \infty$:} In this case we have
\begin{eqnarray}
Z[J]
&= &{\cal N}\int d^N\Phi_n^i~\delta(\Phi_n^i\Phi_n^i-1)~e^{\Phi_n^i J^i}\nonumber\\
&= &{\cal N}\int d^N\Phi_n^i~\delta(\Phi_n^i\Phi_n^i-1)~\bigg[1+\Phi_n^iJ^i+\frac{1}{2}\Phi_n^i\Phi_n^jJ^iJ^j+...\bigg].
\end{eqnarray}
By using rotational invariance in $N$ dimensions we obtain
\begin{eqnarray}
\int d^N\Phi_n^i~\delta(\Phi_n^i\Phi_n^i-1)~\Phi_n^i=0.
\end{eqnarray}
\begin{eqnarray}
\int d^N\Phi_n^i~\delta(\Phi_n^i\Phi_n^i-1)~\Phi_n^i\Phi_n^j=\frac{\delta^{ij}}{N}\int d^N\Phi_n^i~\delta(\Phi_n^i\Phi_n^i-1)~\Phi_n^k\Phi_n^k=\frac{\delta^{ij}}{N}\frac{Z[0]}{{\cal N}}.
\end{eqnarray}
Hence
\begin{eqnarray}
Z[J]&=&Z[0]\bigg[1+\frac{J^iJ^i}{2N}+...\bigg].
\end{eqnarray}
Thus
\begin{eqnarray}
v^i=\frac{J^i}{N}=\frac{4\kappa_c d v^i}{N}\Rightarrow \kappa_c=\frac{N}{4d}.
\end{eqnarray}
\paragraph{The limit of The Ising Model:} In this case we have
\begin{eqnarray}
N=1~,~g\longrightarrow \infty.
\end{eqnarray}
We compute then
\begin{eqnarray}
Z[J]
&= &{\cal N}\int d\Phi_n~\delta(\Phi_n^2-1)~e^{\Phi_n J}\nonumber\\
&= &Z[0]\cosh J.
\end{eqnarray}
Thus
\begin{eqnarray}
v=\tanh 4\kappa d v.
\end{eqnarray}
A graphical sketch of the solutions of this equation will show that for $\kappa<\kappa_c$ there is only one intersection point at $v=0$ whereas for $\kappa>\kappa_c$ there are two intersection points away from the zero, i.e. $v\neq 0$.  Clearly for $\kappa$ near $\kappa_c$ the solution $v$ is near $0$ and thus we can expand the above equation as
\begin{eqnarray}
v=4\kappa d v-\frac{1}{3}(4\kappa d)^3v^2+....
\end{eqnarray}
The solution is
\begin{eqnarray}
\frac{1}{3}(4d)^2\kappa^3 v^2=\kappa-\kappa_c.
\end{eqnarray}
Thus only for $\kappa>\kappa_c$ there is a non zero solution. 

In summary we have the two phases
\begin{eqnarray}
\kappa>\kappa_c~:~{\rm broken, ordered, ferromagnetic}
\end{eqnarray}
\begin{eqnarray}
\kappa<\kappa_c~:~{\rm symmetric, disordered, paramagnetic}.
\end{eqnarray}
The critical line $\kappa_c=\kappa_c(g)$ interpolates in the $\kappa-g$ plane between the two lines  given by
\begin{eqnarray}
\kappa_c=\frac{N}{4d}~,~g\longrightarrow\infty.
\end{eqnarray}
\begin{eqnarray}
\kappa_c=\frac{1}{2d}~,~g\longrightarrow 0.
\end{eqnarray}
For $d=4$ the critical value at $g=0$ is $\kappa_c=1/8$ for all $N$. This critical value  can be derived in a different way as follows. We know that the renormalized mass at one-loop order in the continuum $\phi^4$ with $O(N)$ symmetry is given by the equation 
\begin{eqnarray}
m_R^2&=&m^2+(N+2)\lambda I(m^2,\Lambda)\nonumber\\
&=&m^2+\frac{(N+2)\lambda}{16\pi^2}\Lambda^2+\frac{(N+2)\lambda}{16\pi^2}m^2\ln \frac{m^2}{\Lambda^2}+\frac{(N+2)\lambda}{16\pi^2}m^2{\bf C}+{\rm finite}~{\rm terms}.\nonumber\\
\end{eqnarray}
This equation reads in terms of dimensionless quantities as follows
\begin{eqnarray}
a^2m_R^2
&=&am^2+\frac{(N+2)\lambda}{16\pi^2}+\frac{(N+2)\lambda}{16\pi^2}a^2m^2\ln a^2m^2+\frac{(N+2)\lambda}{16\pi^2}a^2m^2{\bf C}+a^2\times {\rm finite}~{\rm terms}.\nonumber\\
\end{eqnarray}
The lattice space $a$ is formally identified with the inverse cut off $1/\Lambda$, viz
 \begin{eqnarray}
a=\frac{1}{\Lambda}.
\end{eqnarray}
Thus we obtain in the continuum limit $a\longrightarrow 0$ the result
\begin{eqnarray}
a^2m^2\longrightarrow -\frac{(N+2)\lambda}{16\pi^2}+\frac{(N+2)\lambda}{16\pi^2}a^2m^2\ln a^2m^2+\frac{(N+2)\lambda}{16\pi^2}a^2m^2{\bf C}+a^2\times {\rm finite}~{\rm terms}.\nonumber\\
\end{eqnarray}
In other words (with $r_0=(N+2)/8\pi^2$)
\begin{eqnarray}
a^2m^2\longrightarrow a^2m_c^2=-\frac{r_0}{2}\lambda+O(\lambda^2).
\end{eqnarray}
This is the critical line for small values of the coupling constant as we will now show. Expressing this equation in terms of $\kappa$ and $g$ we obtain
\begin{eqnarray}
\frac{1-2g}{\kappa}-8\longrightarrow -\frac{r_0}{2}\frac{g}{\kappa^2}+O(\lambda^2).
\end{eqnarray}
This can be brought to the form
\begin{eqnarray}
\bigg[\kappa-\frac{1}{16}(1-2g)\bigg]^2\longrightarrow \frac{1}{256}\bigg[1+16r_0g-4g\bigg]+O(g^2/\kappa^2).
\end{eqnarray}
We get the result
\begin{eqnarray}
\kappa \longrightarrow \kappa_c=\frac{1}{8}+(\frac{r_0}{2}-\frac{1}{4})g+O(g^2).
\end{eqnarray}
This result is of fundamental importance. The continuum limit $a\longrightarrow 0$ corresponds precisely to the limit in which the mass approaches its critical value. This happens for every value of the coupling constant and hence the continuum limit  $a\longrightarrow 0$ is the limit in which we approach the critical line. The continuum limit is therefore a second order phase transition. 
\section{The HM Algorithm}
We start by considering the Hamiltonian 
\begin{eqnarray}
H[\phi,P]
&=&\frac{1}{2}\sum_nP_n^iP_n^i+\sum_n\bigg(-2\kappa \sum_{\mu}\Phi_n^i\Phi_{n+\hat{\mu}}^i+{\Phi}_n^i\Phi_n^i+g(\Phi_n^i\Phi_n^i-1)^2 \bigg).
\end{eqnarray}
The Hamilton equations of motion are
\begin{eqnarray}
&&\frac{\partial H}{\partial P_n^i}=\dot{\Phi}_n^i=P_n^i\nonumber\\
&&\frac{\partial H}{\partial \Phi_n^i}=-\dot{P}_n^i=V_n^i.
\end{eqnarray}
The force is given by
\begin{eqnarray}
V_n^i&=&\frac{\partial S}{\partial \Phi_n^i}\nonumber\\
&=&-2\kappa\sum_{\mu}(\Phi_{n+\hat{\mu}}^i+\Phi_{n-\hat{\mu}}^i)+2\Phi_n^i+4g\Phi_n^i(\Phi_n^j\Phi_n^j-1).
\end{eqnarray}
The leap frog, or Stormer-Verlet, algorithm, which maintains the symmetry under time reversible
and the conservation of the phase space volume of the above Hamilton equations, is then given by the equations

\begin{eqnarray}
P_{n}^i(t+\frac{\delta t}{2})=(P)_{n}^i(t)-\frac{\delta t}{2}V_{n}^i(t).
\end{eqnarray}

\begin{eqnarray}
\Phi_{n}^i(t+\delta t)=\Phi_{n}^i(t)+\delta t P_{n}^i(t+\frac{\delta t}{2}).
\end{eqnarray}
 \begin{eqnarray}
P_{n}^i(t+\delta t)=P_{n}^i(t+\frac{\delta t}{2})-\frac{\delta t}{2}V_{n}^i(t+\delta t).
\end{eqnarray}
We recall that $t =n {\delta}t$, $n=0,1,2,...,\nu-1,\nu$ where the point $n=0$ corresponds to  the initial configuration $\Phi_{n}^i(0)$ whereas $n=\nu$ corresponds to  the final configuration $\Phi_{n}^i(T)$ where $T=\nu \delta t$. This algorithm does not conserve the Hamiltonian due to the systematic error associated with the discretization,  which goes as $O(\delta t^2)$, but as can be shown the addition of a Metropolis accept-reject step will nevertheless lead to an exact algorithm.

The hybrid Monte Carlo algorithm in this case can be summarized as follows:  
\begin{itemize}
\item{$1)$} Choose $P(0)$ such that $P(0)$ is distributed according to the Gaussian probability distribution $\exp(-\frac{1}{2}\sum_n P_n^iP_n^i)$. In particular we choose $P_n^i$ such that
\begin{eqnarray}
P_{n}^i=\sqrt{-2\ln(1-x_1)}\cos 2\pi(1-x_2),
\end{eqnarray}
where $x_1$ and $x_2$ are two random numbers uniformly distributed in the interval $[0,1]$. This step is crucial if we want to avoid ergodic problems.
\item{$2)$}Find the configuration $(\Phi(T),P(T))$ by solving the above differential equations of motion.
\item{$3)$}Accept the configuration  $(\Phi(T),P(T))$  with a probability 
\begin{eqnarray}
{\rm min}(1,e^{-\Delta H[ \Phi,P]}),
\end{eqnarray}
where $\Delta H$ is the corresponding change in the Hamiltonian when we go from $(\Phi(0),P(0))$ to $(\Phi(T),P(T))$.
\item{$4)$} Repeat.
\end{itemize}
\section{Renormalization and Continuum Limit}
The continuum and lattice actions for $\Phi^4$ theory in two dimensions with $N=1$ are given, with some slight change of notation, by

\begin{eqnarray}
S[\phi]=\int d^2x \bigg(\frac{1}{2}(\partial_{\mu}\phi)^2+\frac{1}{2}\mu_0^2\phi^2+\frac{\lambda}{4}\phi^4\bigg).
\end{eqnarray}
\begin{eqnarray}
S[\phi]
&=&\sum_n\bigg(-2\kappa \sum_{\mu}\Phi_n\Phi_{n+\hat{\mu}}+{\Phi}_n^2+g(\Phi_n^2-1)^2 \bigg).
\end{eqnarray}
\begin{eqnarray}
\mu_0^2=m^2.
\end{eqnarray}
\begin{eqnarray}
\mu_{0l}^2\equiv \mu_0^2a^2=\frac{1-2g}{\kappa}-4~,~\lambda_l\equiv \lambda a^2=\frac{g}{\kappa^2}.
\end{eqnarray}
In the simulations we will start by fixing the lattice quartic coupling $\lambda_l$ and the lattice mass parameter $\mu_{0l}^2$ which then allows us to fix $\kappa$ and $g$ as
\begin{eqnarray}
\kappa=\frac{\sqrt{8\lambda_l+(\mu_{0l}^2+4)^2}-(\mu_{0l}^2+4)}{4\lambda_l}.
\end{eqnarray}
\begin{eqnarray}
g=\kappa^2\lambda_l.
\end{eqnarray}
The phase diagram will be drawn originally in the $\mu_{0l}^2-\lambda_l$ plane. This is the lattice phase diagram. This should be extrapolated to the infinite volume limit $L=Na\longrightarrow \infty$. 

The Euclidean quantum field theory phase diagram should be drawn in terms of the renormalized parameters and is obtained from the lattice phase diagram by taking the limit $a\longrightarrow 0$. In two dimensions the $\Phi^4$ theory requires only mass renormalization while the quartic coupling constant is finite. Indeed, the bare mass $\mu_0^2$ diverges logarithmically when we remove the cutoff, i.e. in the limit $\Lambda\longrightarrow \infty$ where $\Lambda=1/a$ while $\lambda$ is independent of $a$. As a consequence, the lattice parameters will go to zero in the continuum limit $a\longrightarrow 0$.

We know that mass renormalization is due to the tadpole diagram which is the only divergent Feynman diagram in the theory and takes the form of a simple reparametrization given by
\begin{eqnarray}
\mu_0^2=\mu^2-\delta\mu^2,\label{ren}
\end{eqnarray}
where $\mu^2$ is the renormalized mass parameter and $\delta\mu^2$ is the counter term which is fixed via an appropriate renormalization condition. The unltraviolet divergence $\ln\Lambda$  of $\mu_0^2$ is contained in $\delta\mu^2$ while the renormalization condition will split the finite part of $\mu_0^2$ between $\mu^2$ and $\delta\mu^2$. The choice of the renormalization condition can be quite arbitrary. A convenient choice suitable for Monte Carlo measurements and which distinguishes between the two phases of the theory is given by the usual normal ordering prescription \cite{Loinaz:1997az} .

Quantization at one-loop gives explicitly the $2-$point function
\begin{eqnarray}
\Gamma^{(2)}(p)=p^2+\mu_0^2+3\lambda\int\frac{d^2k}{(2\pi)^2}\frac{1}{k^2+\mu_0^2}.
\end{eqnarray}
A self-consistent Hartree treatment gives then the result
\begin{eqnarray}
\Gamma^{(2)}(p)&=&p^2+\mu_0^2+3\lambda\int\frac{d^2k}{(2\pi)^2}\frac{1}{\Gamma^{(2)}(k)}\nonumber\\
&=&p^2+\mu^2+3\lambda\int\frac{d^2k}{(2\pi)^2}\frac{1}{\Gamma^{(2)}(k)}-\delta\mu^2\nonumber\\
&=&p^2+\mu^2+3\lambda\int\frac{d^2k}{(2\pi)^2}\frac{1}{k^2+\mu^2}-\delta\mu^2+{\rm two-loop}\nonumber\\
\end{eqnarray}
This should certainly work in the symmetric phase where $\mu^2>0$. We can also write this as
\begin{eqnarray}
\Gamma^{(2)}(p)
&=&p^2+\mu^2+\Sigma(p)~,~\Sigma(p)=3\lambda A_{\mu^2}-\delta\mu^2+{\rm two-loop}.
\end{eqnarray}
$A_{\mu^2}$ is precisely the value of the tadpole diagram given by
\begin{eqnarray}
A_{\mu^2}=\int\frac{d^2k}{(2\pi)^2}\frac{1}{k^2+\mu^2}.
\end{eqnarray}
The renormalization condition which is equivalent to normal ordering the interaction in the interaction picture in the symmetric phase is equivalent to the choice
 \begin{eqnarray}
\delta\mu^2=3\lambda A_{\mu^2}.\label{ren1}
\end{eqnarray}
A dimensionless coupling constant can the be defined by
\begin{eqnarray}
f=\frac{\lambda}{\mu^2}.
\end{eqnarray}
The action becomes
\begin{eqnarray}
S[\phi]=\int d^2x \bigg(\frac{1}{2}(\partial_{\mu}\phi)^2+\frac{1}{2}\mu^2(1-3fA_{\mu^2})\phi^2+\frac{f\mu^2}{4}\phi^4\bigg).
\end{eqnarray}
For sufficiently small $f$ the exact effective potential is well approximated by the classical potential with a single minimum at $\phi_{\rm cl}=0$. For larger $f$, the coefficient of the mass term in the above action can become negative and as a consequence a transition to the broken symmetry phase is possible, although in this regime the effective potential is no longer well approximated by the classical potential.  Indeed, a transition to the broken symmetry phase was shown to be present in \cite{Chang:1976ek}, where a duality between the strong coupling regime of the above action and a weakly coupled theory normal ordered with respect to the broken phase was explicitly constructed.

The sites on the lattice are located at $x_{\mu}=n_{\mu}a$ where $n_{\mu}=0,...,N-1$ with $L=Na$. The plane waves on a finite volume lattice with periodic boundary conditions are $\exp(ipx)$ with $p_{\mu}=m_{\mu}2\pi/L$ where $m_{\mu}=-N/2+1,-N/2+2,...,N/2$ for $N$ even. This means that the zero of the $x-$space is located at the edge of the box while the zero of the $p-$space is located in the middle of the box. We have therefore the normalization conditions $\sum_x\exp(-i(p-p^{'})x)=\delta_{p,p^{'}}$ and $\sum_p\exp(-i(x-x^{'})p)=\delta_{x,x^{'}}$ where, for example, $\sum_p=\sum_m/L^2$. In the infinite volume limit defined by $L=Na\longrightarrow \infty$ with $a$ fixed we have $\sum_p\longrightarrow \int_{-\pi/a}^{\pi/a}d^2p/(2\pi)^2$. It is not difficult to show that on the lattice the propagator $1/(p^2+\mu^2)$ becomes $a^2/(4\sum_{\mu}\sin^2a{p}_{\mu}/2+\mu_l^2)$ \cite{Smit:2002ugp4}. Thus on a finite volume lattice with periodic boundary conditions the Feynman diagram $A_{\mu^2}$ takes the form
\begin{eqnarray}
A_{\mu^2}&=&\sum_{p_1,p_2}^{}\frac{a^2}{4\sin^2a{p}_{1}/2+4\sin^2a{p}_{2}/2+\mu_l^2}\nonumber\\
&=&\frac{1}{N^2}\sum_{m_1=1}^N\sum_{m_2=1}^N\frac{1}{4\sin^2 {\pi m_1}/{N}+4\sin^2{\pi m_2}/{N}+\mu_l^2}.
\end{eqnarray}
In the last line we have shifted the integers $m_1$ and $m_2$ by $N/2$.
Hence on a finite volume lattice with periodic boundary conditions equation (\ref{ren}), together with equation (\ref{ren1}), becomes
\begin{eqnarray}
F(\mu_l^2)=\mu_l^2-3\lambda_l A_{\mu_l^2}-\mu_{0l}^2=0.\label{reno}
\end{eqnarray}
Given the critical value of $\mu_{0l}^2$ for every value of $\lambda_l$ we need then to determine the corresponding critical value of $\mu_{l}^2$. This can be done numerically using the Newton-Raphson algorithm. The continuum limit $a\longrightarrow 0$ is then given by extrapolating the results into the origin, i.e. taking $\lambda_l=a^2\lambda\longrightarrow 0$, $\mu_l^2=a^2\mu^2\longrightarrow 0$ in order to determine the critical value
\begin{eqnarray}
f_c={\rm lim}_{\lambda_l,\mu_l^2\longrightarrow 0}\frac{\lambda_l}{\mu_{lc}^2}.
\end{eqnarray}
\section{HMC Simulation Calculation of The Critical Line}

We measure as observables the average value of the action, the specific heat, the magnetization, the susceptibility and the Binder cumulant defined respectively by
\begin{eqnarray}
<S>.
\end{eqnarray}
\begin{eqnarray}
C_v=<S^2>-<S>^2.
\end{eqnarray}
\begin{eqnarray}
M=\frac{1}{N^2}<m>~,~m=|\sum_n\phi_n|.
\end{eqnarray}
\begin{eqnarray}
\chi=<m^2>-<m>^2.
\end{eqnarray}
\begin{eqnarray}
U=1-\frac{<m^4>}{3<m^2>^2}.
\end{eqnarray}
We note  the use of the absolute value in the definition of the magnetization since the usual definition $M=<\sum_n\phi_n>/N^2$ is automatically zero on the lattice because of the symmetry $\phi\longrightarrow -\phi$. The specific heat diverges at the critical point logarithmically as the lattice size is sent to infinity. The susceptibility shows also a peak at the critical point whereas the Binder cumulant exhibits a fixed point for all values of $N$. 

We run simulations with $T_{\rm th}+T_{\rm mc}\times T_{\rm co}$ steps with $T_{\rm th}=2^{13 }$ thermalization steps and $T_{\rm mc}=2^{14}$ measurement steps. Every two successive measurements are separated by $T_{\rm co}=2^3$ steps to reduce auto-correlations. We use ran2 as our random numbers generator and the Jackknife method to estimate error bars. The hybrid Monte Carlo code used in these simulations can be found in the last chapter. 

\begin{figure}[htbp]
\begin{center}
\includegraphics[width=8.0cm,angle=-0]{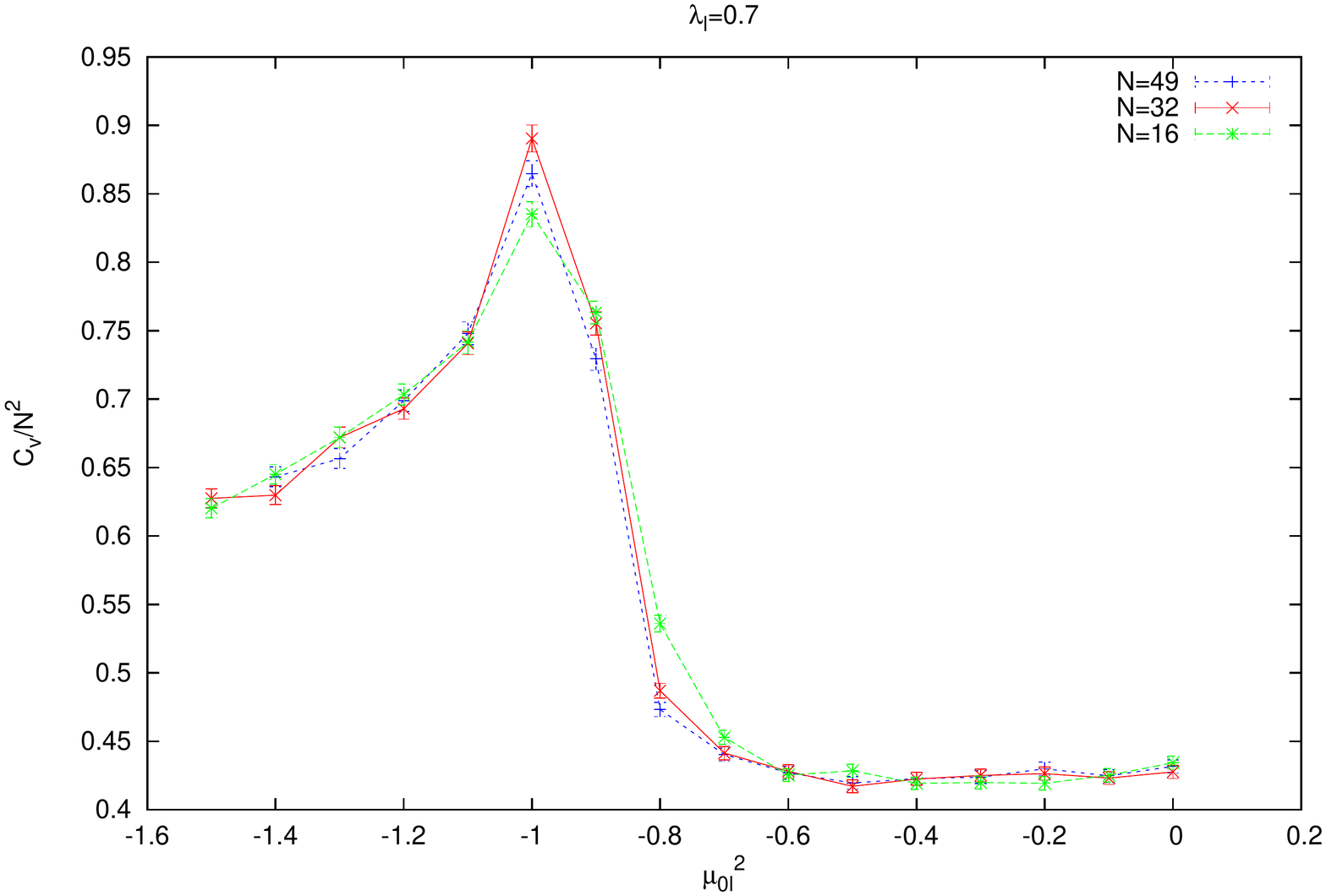}
\includegraphics[width=8.0cm,angle=-0]{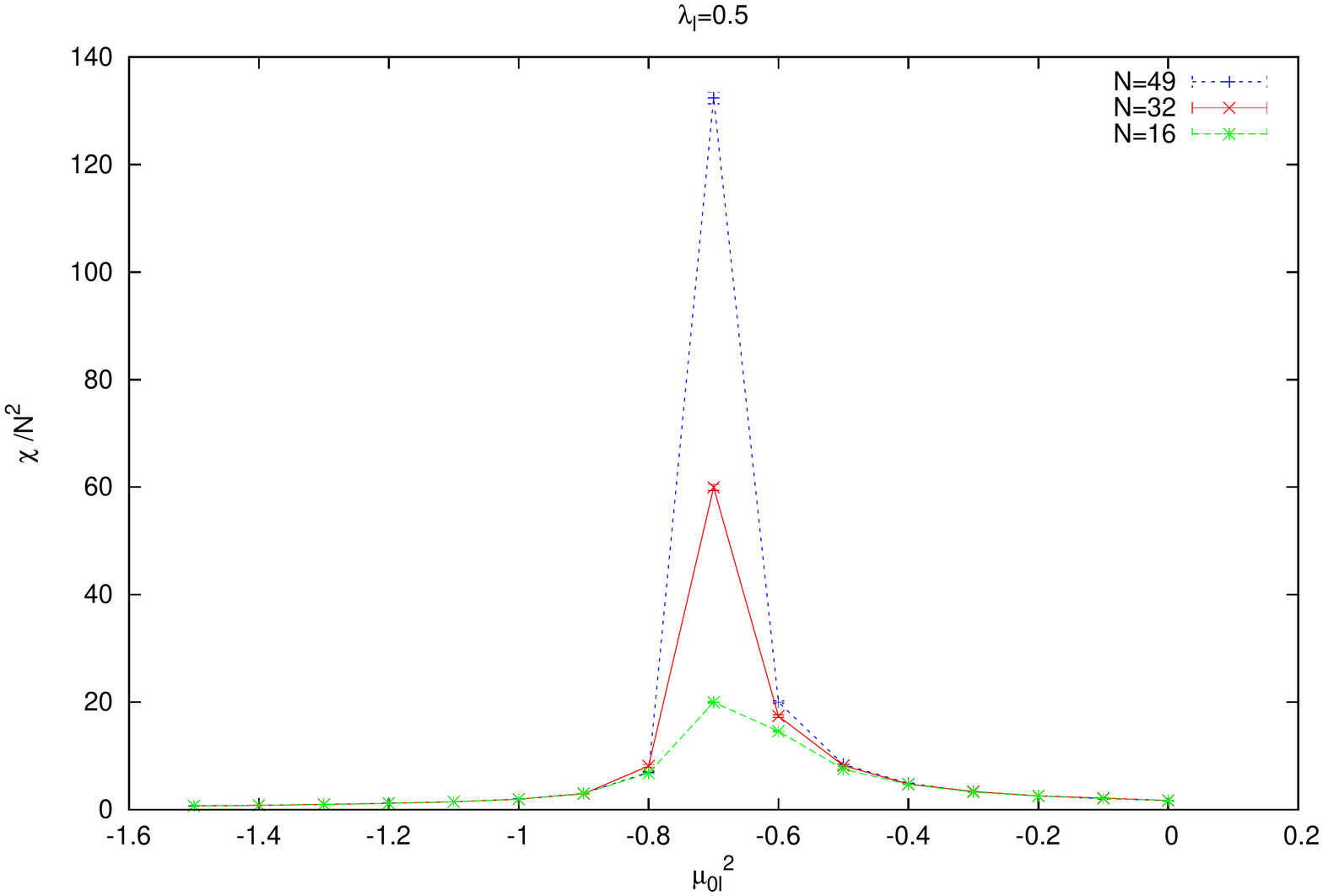}
\includegraphics[width=8.0cm,angle=-0]{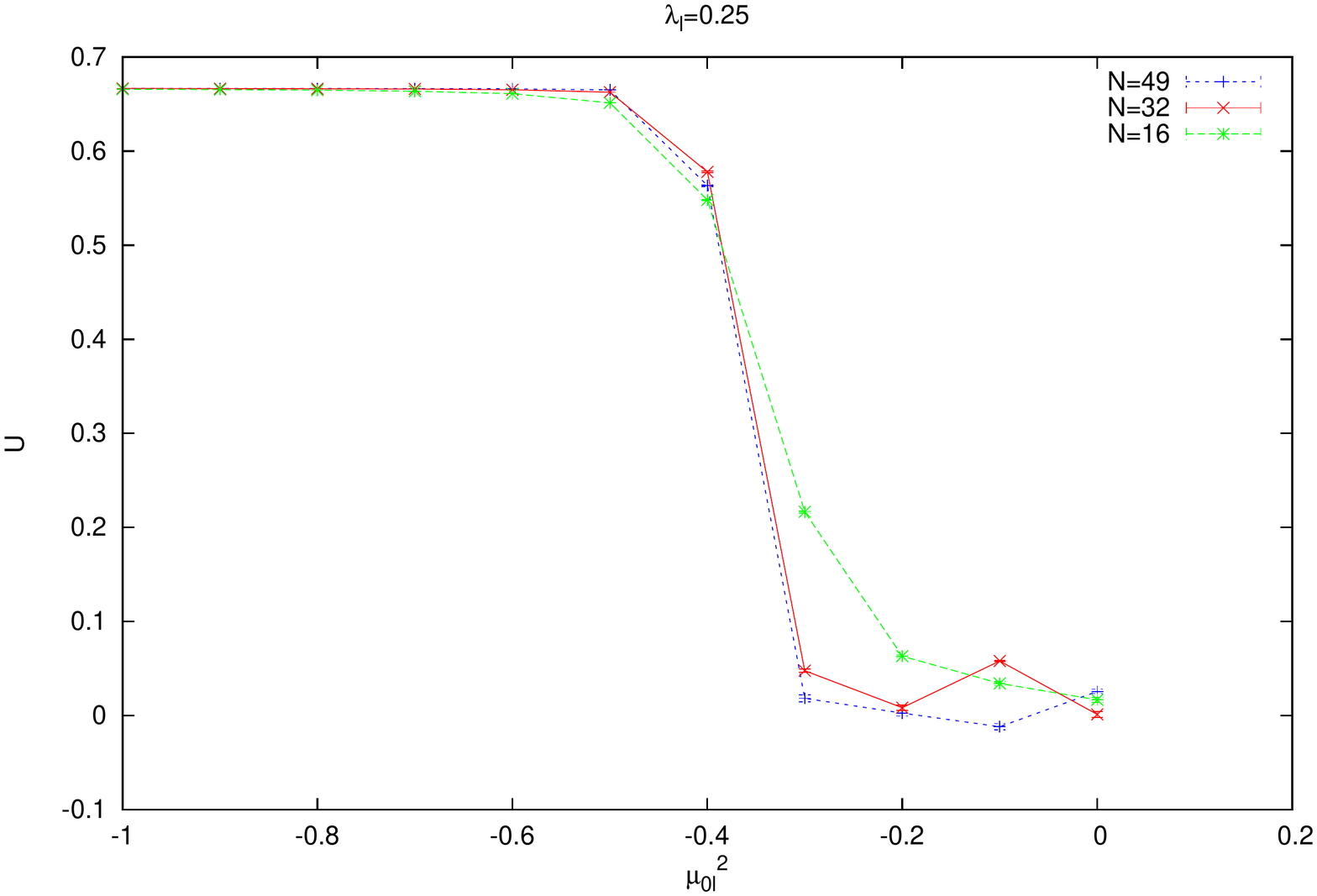}
\includegraphics[width=8.0cm,angle=-0]{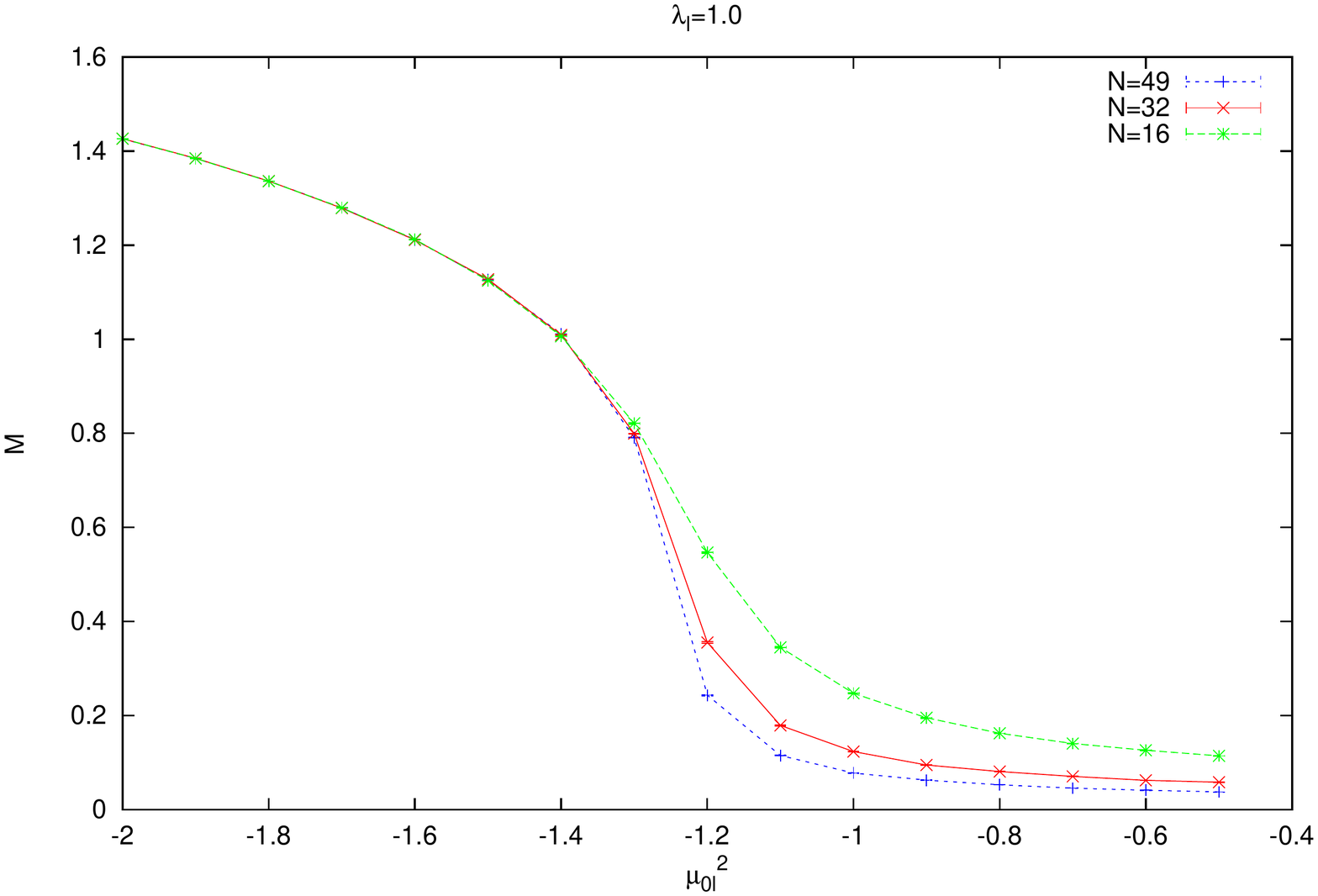}
\end{center}
\caption{}\label{latticephi4}
\end{figure}

\begin{figure}[htbp]
\begin{center}
\includegraphics[width=10.0cm,angle=-0]{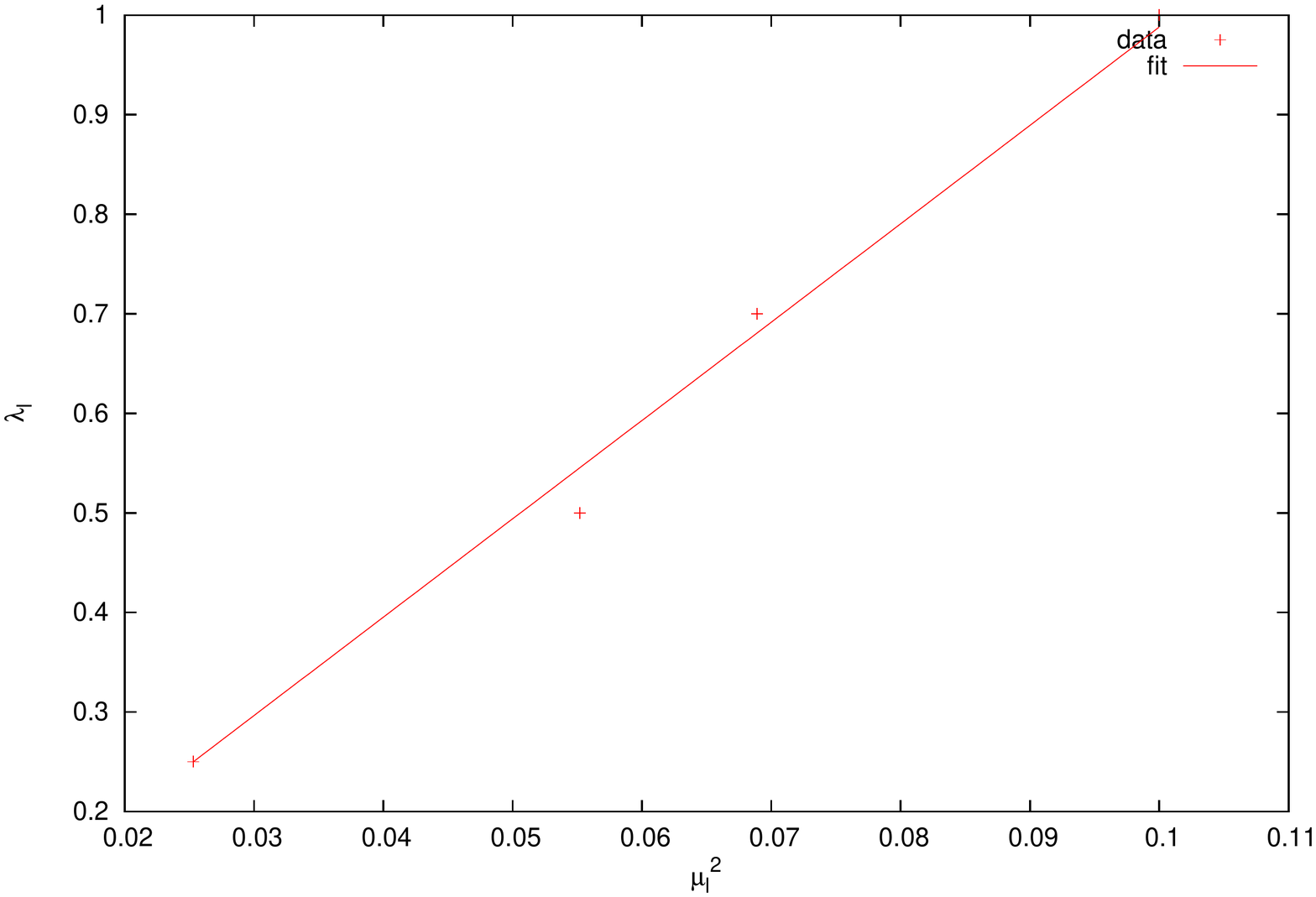}
\end{center}
\caption{}\label{latticephi4_e}
\end{figure}

We have considered lattices with $N=16,32$ and $49$ and values of the quartic coupling given by $\lambda_l=1, 0.7, 0.5, 0.25$. Some results are shown on figure (\ref{latticephi4}). The critical value $\mu_{0l*}^2$ for each value of $\lambda_l$ is found from averaging the values at which the peaks in the specific heat and the susceptibility occur. The results are shown on  the second column of table (\ref{tablephi4}). The final step is take the continuum limit $a\longrightarrow 0$ in order to find the critical value $\mu_{l*}^2$ by solving the renormalization condition (\ref{reno}) using the Newton-Raphson method. This is an iterative method based on a single iteration given by $\mu_{l*}^2=\mu_{l*}^2-F/F^{'}$. The corresponding results are shown  on  the third column of table (\ref{tablephi4}). The critical line is shown on figure (\ref{latticephi4_e}) with a linear fit going through the origin given by
\begin{eqnarray}
\lambda_l=(9.88\pm 0.22)\mu_{l*}^2.
\end{eqnarray}
This should be compared with the much more precise result $\lambda_l=10.8\mu_{l*}^2$ published in \cite{Schaich:2009jk}. The above result is sufficient for our purposes here.
\begin{table}[h]
\centering
\begin{tabular}{|l|c|c| }
\hline
$\lambda_l$ & $\mu_{0l*}^2$ &  $\mu_{l*}^2$    \\
\hline 
$1.0$ &  $-1.25\pm 0.05$ & $1.00\times 10^{-2}$  \\ 
$0.7$ &  $-0.95\pm 0.05$ &  $6.89\times 10^{-2}$ \\ 
$0.5$ &  $-0.7\pm 0.00$ & $5.52\times 10^{-2}$  \\
$0.25$ &  $-0.4\pm 0.00$ & $2.53\times 10^{-2}$  \\ 
\hline 
\end{tabular}
\caption{ }\label{tablephi4}
\end{table}


\chapter{(Multi-Trace) Quartic Matrix Models}
\section{The Pure Real Quartic Matrix Model}
This is a very well known, and a very well studied, model which depends on a single hermitian matrix $M$. This is given by
\begin{eqnarray}
V&=&BTr M^2+CTr M^4\nonumber\\
&=&\frac{N}{g}(-Tr M^2+\frac{1}{4} Tr M^4).\label{quarticM}
\end{eqnarray}
The model depends actually on a single coupling $g$ such that
\begin{eqnarray}
B=-\frac{N}{g}~,~C=\frac{N}{4g}.
\end{eqnarray}
There are two stable phases in this model:
\paragraph{Disordered phase (one-cut) for $g\geq g_c$:} This is characterized by the eigenvalues distribution of the matrix $M$ given by

\begin{eqnarray}
\rho(\lambda)&=&\frac{1}{N\pi}(2C\lambda^2+B+C\delta^2)\sqrt{\delta^2-\lambda^2}\nonumber\\
&=&\frac{1}{g\pi}(\frac{1}{2}\lambda^2-1+r^2)\sqrt{4r^2-\lambda^2}.\label{pred1}
\end{eqnarray}
This is a single cut solution with the cut defined by
\begin{eqnarray}
-2r\leq \lambda\leq 2r.
\end{eqnarray}
\begin{eqnarray}
r=\frac{1}{2}\delta.
\end{eqnarray}
\begin{eqnarray}
\delta^2&=&\frac{1}{3C}(-B+\sqrt{B^2+12 NC})\nonumber\\
&=&\frac{1}{3}(1+\sqrt{1+3g}).
\end{eqnarray}
\paragraph{Non-uniform ordered phase (two-cut) for $g\le g_c$:}This is characterized by the eigenvalues distribution of the matrix $M$ given by

\begin{eqnarray}
\rho(\lambda)&=&\frac{2C|\lambda|}{N\pi}\sqrt{(\lambda^2-\delta_1^2)(\delta_2^2-\lambda^2)}\nonumber\\
&=&\frac{|\lambda|}{2g\pi}\sqrt{(\lambda^2-r_{-}^2)(r_{+}^2-\lambda^2)}.\label{pred2}
\end{eqnarray}
Here there are two cuts defined by
\begin{eqnarray}
r_{-}\leq |\lambda|\leq r_{+}.
\end{eqnarray}
\begin{eqnarray}
r_{-}=\delta_1~,~r_{+}=\delta_2.
\end{eqnarray}
\begin{eqnarray}
r_{\mp}^2&=&\frac{1}{2C}(-B\mp 2\sqrt{NC})\nonumber\\
&=&2(1\mp \sqrt{g}).
\end{eqnarray}
A third order transition between the above two phases occurs at the critical point 
\begin{eqnarray}
g_c=1\leftrightarrow B_c^2=4NC \leftrightarrow B_c=-2\sqrt{NC}.
\end{eqnarray}
There is a third phase in this model: the so-called Ising or uniform ordered phase, which despite the fact that it is not stable, plays an important role in generalizations of this model, such as the one discussed in the next section, towards noncommutative $\Phi^4$.
\section{The Multi-Trace Matrix Model}
Our primary interest here is the theory of noncommutative $\Phi ^4$ on the fuzzy sphere given by the action
\begin{eqnarray}
S=\frac{4\pi R^2}{N+1} Tr\bigg(\frac{1}{2R^2}{\Phi}\Delta{\Phi}+\frac{1}{2}m^2{\Phi}^2+\frac{\lambda}{4!}{\Phi}^4\bigg).
\end{eqnarray}
The Laplacian is $\Delta=[L_a,[L_a,...]]$. Equivalently with the substitution ${\Phi}={\cal M}/\sqrt{2\pi\theta}$, where ${\cal M}=\sum_{i,j=1}^NM_{ij}|i><j|$, this action  reads 
\begin{eqnarray}
S=Tr\bigg(a{\cal M}\Delta {\cal M}+b{\cal M}^2+c{\cal M}^4\bigg).\label{ac}
\end{eqnarray}
The parameters are\footnote{The noncommutativity parameter on the fuzzy sphere is related to the radius of the sphere by $\theta=2R^2/\sqrt{N^2-1}$.}
\begin{eqnarray}
a=\frac{1}{2R^2}~,~b=\frac{1}{2}m^2~,~c=\frac{\lambda}{4!}\frac{1}{2\pi\theta}.
\end{eqnarray}
In terms of the matrix $M$ the action reads
\begin{eqnarray}
S[M]&=&r^2K[M]+Tr\big[b M^2+c M^4\big].
\end{eqnarray}
The kinetic matrix is given by
\begin{eqnarray}
K[M]&=&Tr\bigg[-\Gamma^+M\Gamma M-\frac{1}{N+1}\Gamma_3M\Gamma_3M+EM^2\bigg].
\end{eqnarray}
The matrices $\Gamma$, $\Gamma_3$ and $E$ are given by

\begin{eqnarray}
(\Gamma_3)_{lm}= l{\delta}_{lm}~,~(\Gamma)_{lm}=\sqrt{(m-1)(1- \frac{m}{N+1})}{\delta}_{lm-1}~,~(E)_{lm}=(l-\frac{1}{2}){\delta}_{lm}.
\end{eqnarray}
The relationship between the parameters $a$ and $r^2$ is given by  
\begin{eqnarray}
r^2=2aN
\end{eqnarray}
We start from the path integral  

\begin{eqnarray}
Z&=&\int d M ~\exp\big(-S[M]\big)\nonumber\\
&=&\int d \Lambda~\Delta^2(\Lambda) ~\exp\bigg(-Tr\big(b{\Lambda}^2+c{\Lambda}^4\big)\bigg)\int dU~\exp\bigg(-r^2K[U\Lambda U^{-1}]\bigg).
\end{eqnarray}
The second line involves the diagonalization of the matrix $M$ (more on this below). The calculation of the integral over $U\in U(N)$ is a very long calculation done in \cite{Ydri:2014uaa,O'Connor:2007ea}. The end result is a multi-trace effective potential given by (assuming the symmetry $M\longrightarrow -M$)
\begin{eqnarray}
S_{\rm eff}&=&\sum_{i}(b\lambda_i^2+c\lambda_i^4)-\frac{1}{2}\sum_{i\neq j}\ln(\lambda_i-\lambda_j)^2\nonumber\\
&+&\bigg[\frac{r^2}{8}v_{2,1}\sum_{i\ne j}(\lambda_i-\lambda_j)^2+\frac{r^4}{48}v_{4,1}\sum_{i\ne j}(\lambda_i-\lambda_j)^4-\frac{r^4}{24N^2}v_{2,2}\big[\sum_{i\ne j}(\lambda_i-\lambda_j)^2\big]^2+...\bigg].\label{Seff}\nonumber\\
\end{eqnarray}
The coefficients $v$ will be given below. If we do not assume the symmetry $M\longrightarrow -M$ then obviously there will be extra terms with more interesting consequences for the phase structure as we will discuss briefly below.

This problem (\ref{Seff}) is a generalization of the quartic  Hermitian matrix potential model. Indeed, this effective potential corresponds to the matrix model given by

\begin{eqnarray}
V&=&\bigg({b}+\frac{aN^2v_{2,1}}{2}\bigg) Tr M^2+\big({c}+\frac{a^2N^3v_{4,1}}{6}\big) Tr M^4-\frac{2\eta a^2N^2}{3}\bigg[ Tr M^2\bigg]^2.\label{mmm}
\end{eqnarray}
This can also be solved exactly as shown in  \cite{Ydri:2014uaa}. The strength of the multi-trace term $\eta$ is given by
\begin{eqnarray}
\eta=v_{2,2}-\frac{3}{4}v_{4,1}.
\end{eqnarray}
The coefficients $v_{2,1}$, $v_{4,1}$ and $v_{2,2}$ are given by the following two competing calculations of \cite{Ydri:2014uaa} and \cite{O'Connor:2007ea} given respectively by
\begin{eqnarray}
v_{2,1}=1~,~v_{4,1}=0~,~v_{2,2}=\frac{1}{8}.
\end{eqnarray}
\begin{eqnarray}
v_{2,1}=-1~,~v_{4,1}=\frac{3}{2}~,~v_{2,2}=0.
\end{eqnarray}
This discrepancy is discussed in \cite{Ydri:2014uaa}.

\section{Model and Algorithm}

We thus start from the potential and the partition function

\begin{eqnarray}
V
&=&Tr\bigg(B M^2+C M^4\bigg)+D\bigg(Tr M^2\bigg)^2.
\end{eqnarray}
We may include the odd terms found in \cite{Ydri:2014uaa} without any real extra effort. We will not do this here for simplicity, but we will include them for completeness in the attached code. The partition function (path integral) is given by

\begin{eqnarray}
Z=\int d M ~\exp\big(-V\big).
\end{eqnarray}
The relationship between the two sets of parameters $\{a,b,c\}$ and $\{B,C,D\}$ is given by
\begin{eqnarray}
B={b}+\frac{aN^2v_{2,1}}{2}~,~C={c}+\frac{a^2N^3v_{4,1}}{6}~,~D=-\frac{2\eta a^2N^2}{3}.
\end{eqnarray}
The collpased parameters are
\begin{eqnarray}
\tilde{B}=\frac{B}{N^{\frac{3}{2}}}=\tilde{b}+\frac{\tilde{a}v_{2,1}}{2}~,~\tilde{C}=\frac{C}{N^2}=\tilde{c}+\frac{\tilde{a}^2v_{4,1}}{6}~,~D=-\frac{2\eta \tilde{a}^2N}{3}.
\end{eqnarray}
Only two of these three parameters are independent. For consistency of the large $N$ limit, we must choose $\tilde{a}$ to be any fixed number. We then choose for simplicity $\tilde{a}=1$ or equivalently $D=-2\eta N/3$\footnote{The authors of \cite{GarciaFlores:2009hf} chose instead $a=1$.}.

We can now diagonalize the scalar matrix $M$ as
\begin{eqnarray}
M=U\Lambda U^{-1}.
\end{eqnarray}
We compute 
\begin{eqnarray}
\delta M=U\bigg(\delta\Lambda +[U^{-1}\delta U,\Lambda]\bigg)U^{-1}.
\end{eqnarray}
Thus (with $U^{-1}\delta U=i\delta V$ being an element of the Lie algebra of SU(N))
\begin{eqnarray}
Tr (\delta M)^2&=&Tr (\delta\Lambda)^2+Tr[U^{-1}\delta U,\Lambda]^2\nonumber\\
&=&\sum_i(\delta\lambda_i)^2+\sum_{i\neq j}(\lambda_i-\lambda_j)^2\delta V_{ij}\delta V_{ij}^*.
\end{eqnarray}
We count $N^2$ real degrees of freedom as there should be. The measure is therefore given by
\begin{eqnarray}
dM&=&\prod_id\lambda_i\prod_{i\neq j}dV_{ij}dV_{ij}^*\sqrt{{\rm det}({\rm metric})}\nonumber\\
&=&\prod_id\lambda_i\prod_{i\neq j}dV_{ij}dV_{ij}^*\sqrt{\prod_{i\neq j}(\lambda_i-\lambda_j)^2}.
\end{eqnarray}
We write this as
\begin{eqnarray}
  d  M= d\Lambda dU \Delta^2(\Lambda).
\end{eqnarray}
The $dU$ is the usual Haar measure over the group SU(N) which is normalized such that $\int dU=1$, whereas the Jacobian $\Delta^2(\Lambda)$ is precisely the so-called Vandermonde determinant defined by
\begin{eqnarray}
\Delta^2(\Lambda)= \prod_{i>j}(\lambda_i-\lambda_j)^2.
\end{eqnarray}
The partition function becomes 
\begin{eqnarray}
Z=\int d \Lambda~\Delta^2(\Lambda) ~\exp\bigg(-Tr\big(B{\Lambda}^2+C{\Lambda}^4\big)-D\bigg(Tr \Lambda^2\bigg)^2\bigg).
\end{eqnarray}
We are therefore dealing with an effective potential given by
\begin{eqnarray}
V_{\rm eff}=B\sum_{i=1}\lambda_i^2+C\sum_{i=1}\lambda_i^4+D\bigg(\sum_{i=1}\lambda_i^2\bigg)^2-\frac{1}{2}\sum_{i\neq j}\ln (\lambda_i-\lambda_j)^2.
\end{eqnarray}
We will use the Metropolis  algorithm to study this model. Under the change $\lambda_i\longrightarrow \lambda_i+h$ of the eigenvalue $\lambda_i$ the above effective potential changes as $V_{\rm eff}\longrightarrow V_{\rm eff}+\Delta V_{i,h}$ where 
\begin{eqnarray}
\Delta V_{i,h}=B\Delta S_2+C\Delta S_4+D (2S_2\Delta S_2+\Delta S_2^2)+\Delta S_{\rm Vand}.
\end{eqnarray}
The monomials $S_n$ are defined by $S_n=\sum_i\lambda_i^n$ while the variations $\Delta S_n$ and $\Delta S_{\rm Vand}$ are given by
\begin{eqnarray}
\Delta S_2=h^2+2h\lambda_i.
\end{eqnarray}
\begin{eqnarray}
\Delta S_4=6h^2\lambda_i^2+4h\lambda_i^3+4h^3\lambda_i+h^4.
\end{eqnarray}

\begin{eqnarray}
\Delta S_{\rm Vand}=-2\sum_{j \ne i}\ln|1+\frac{h}{\lambda_i-\lambda_j}|.
\end{eqnarray}
\section{The Disorder-to-Non-Uniform-Order Transition}
The pure quartic matrix model (\ref{quarticM}) is characterized by a third-order phase transition between a disordered  phase characterized by $<M>=0$ and a non-uniform ordered phase characterized by $<M>=-B\gamma/2C$ where $\gamma$ is an $N-$dimensional idempotent, viz $\gamma^2=1$. This transition is also termed one-cut-to-two-cut transition. Thus the eigenvalues distribution of the scalar field $M$ will go from a one-cut solution centered around $0$ in the disordered phase to a two-cut solution with two peaks symmetric around $0$ in the uniform ordered phase. The transition should occur around $g=g_c=1$. This transition is critical since the two different eigenvalues distributions in the two phases become identical at the transition point. 

\begin{figure}[htbp]
\begin{center}
\includegraphics[width=8.0cm,angle=-0]{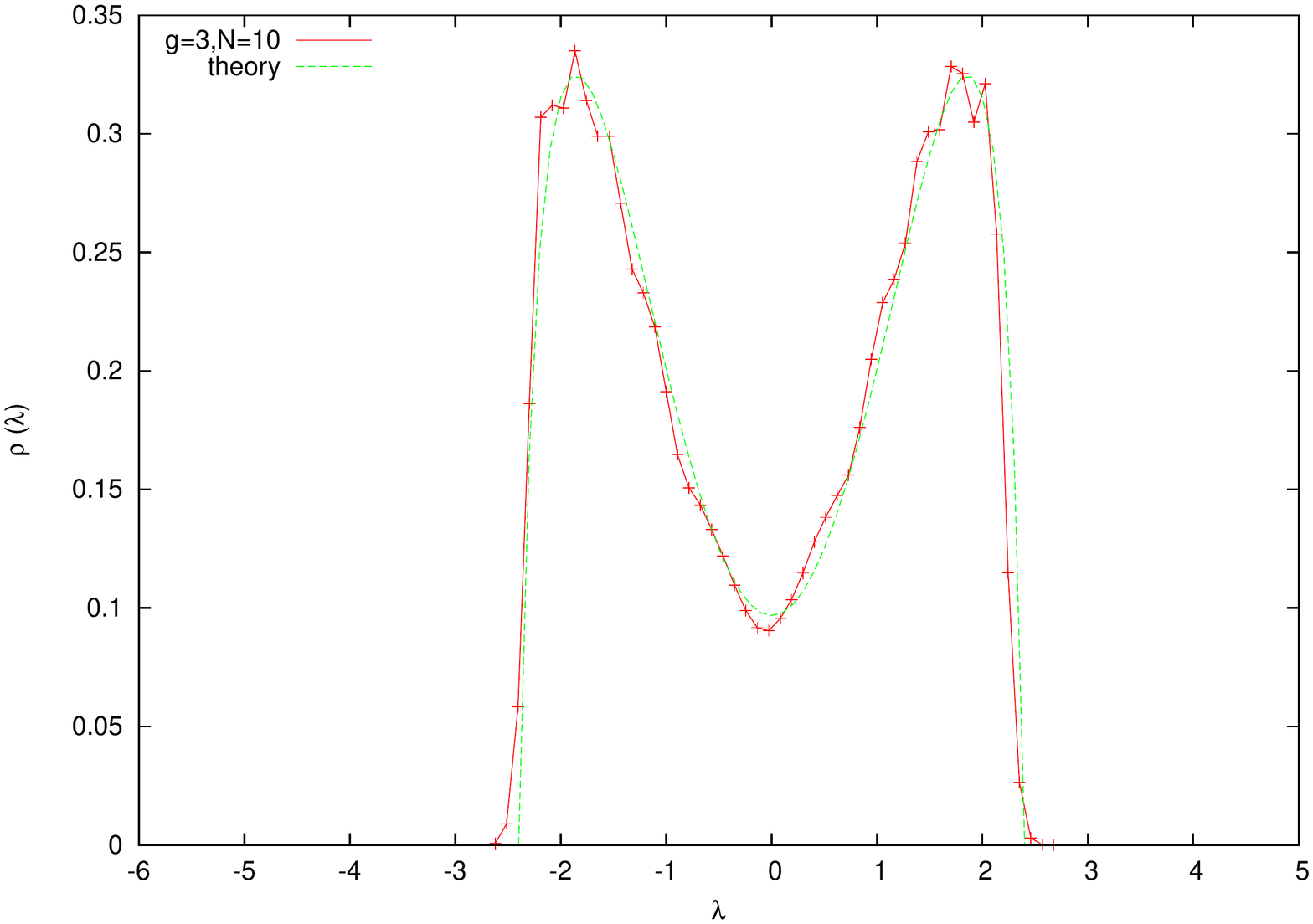}
\includegraphics[width=8.0cm,angle=-0]{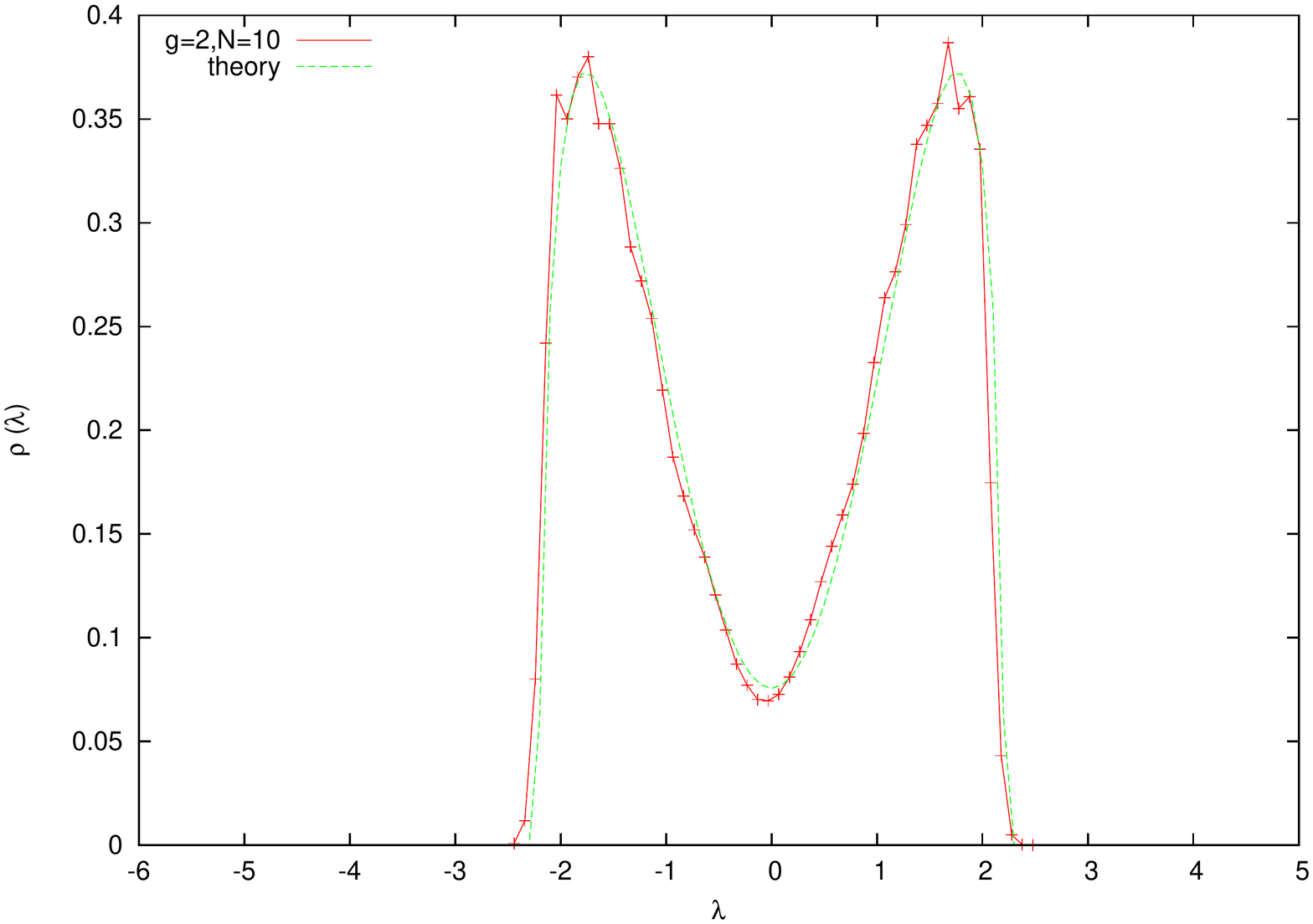}
\includegraphics[width=8.0cm,angle=-0]{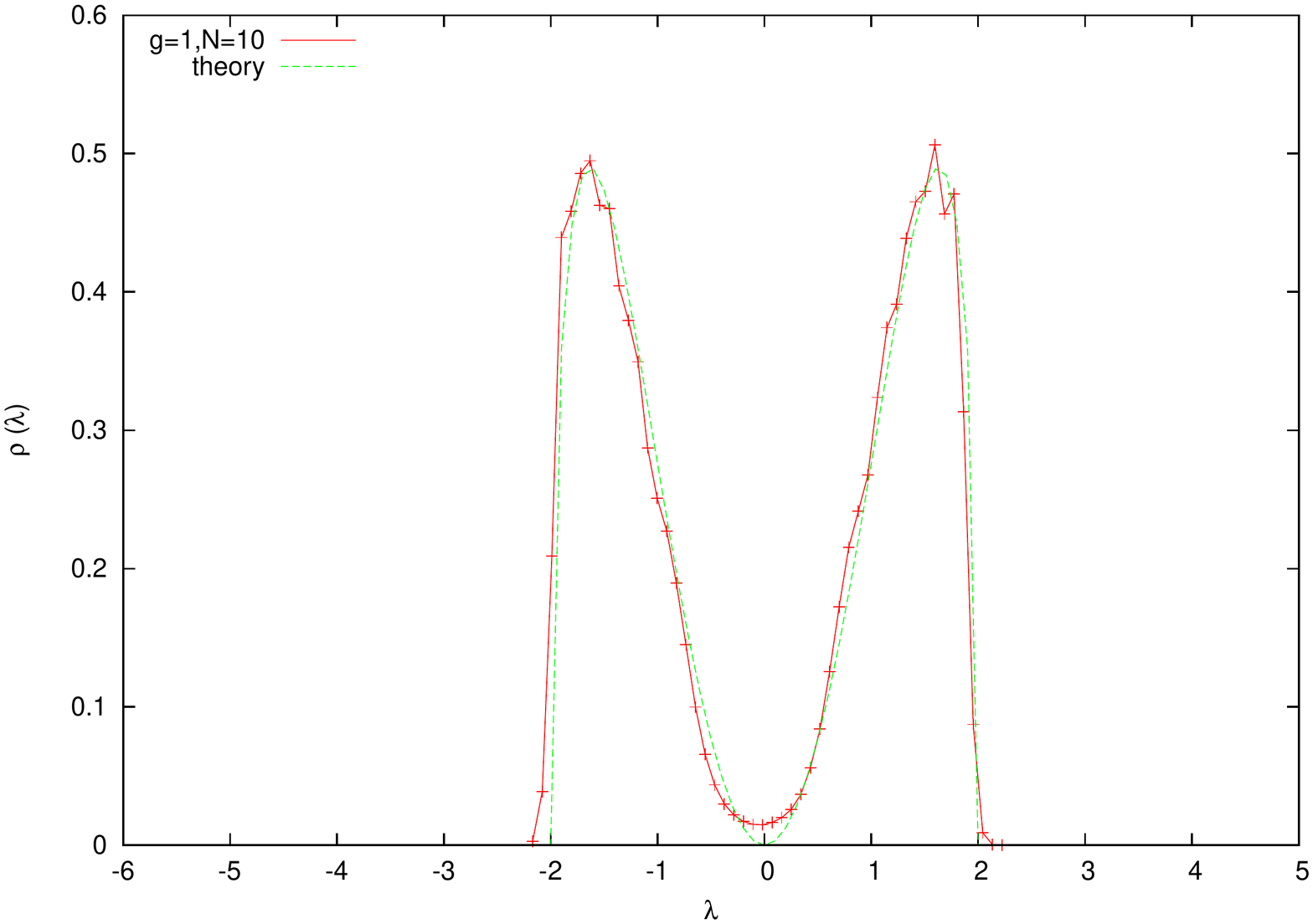}
\includegraphics[width=8.0cm,angle=-0]{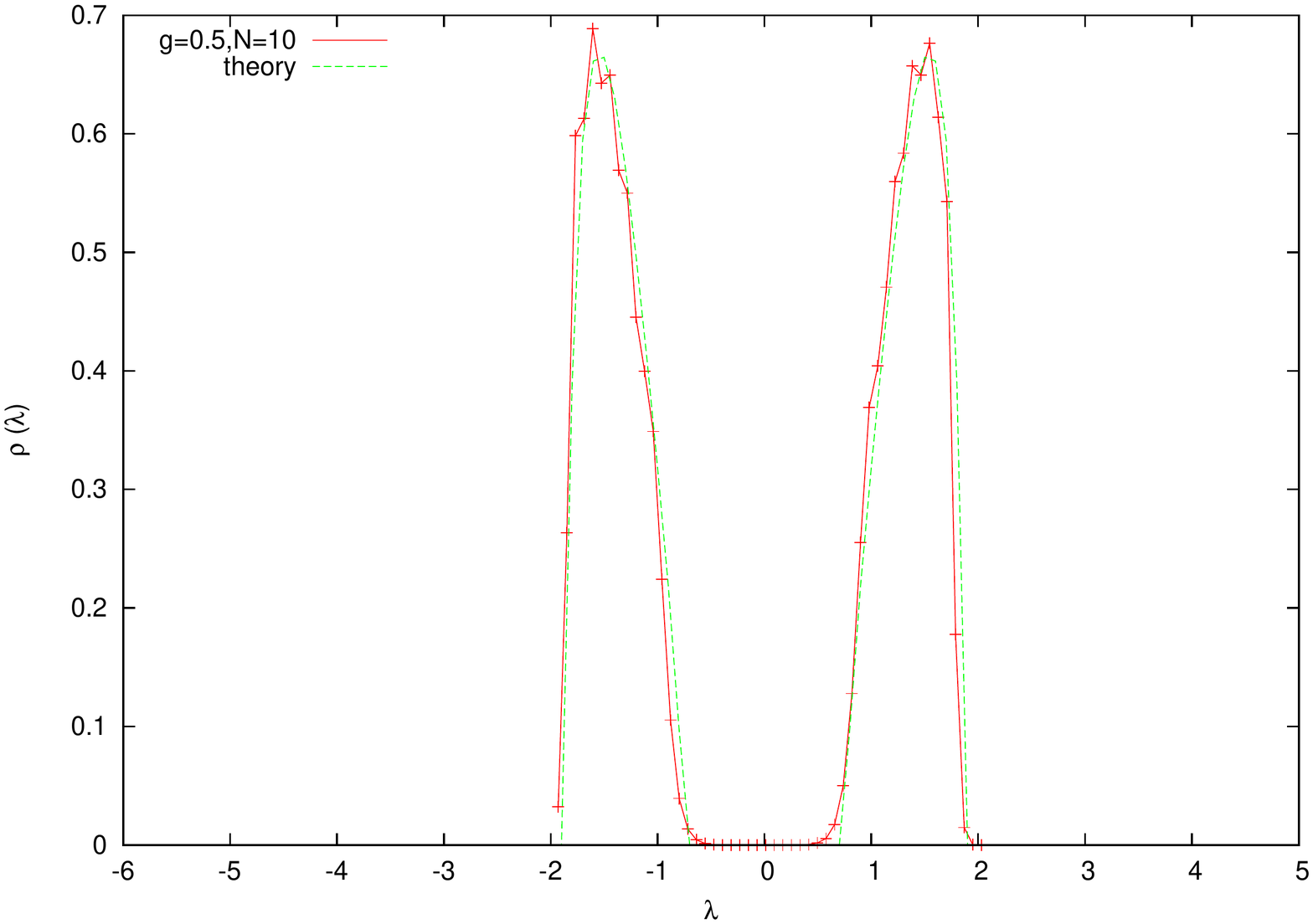}
\includegraphics[width=8.0cm,angle=-0]{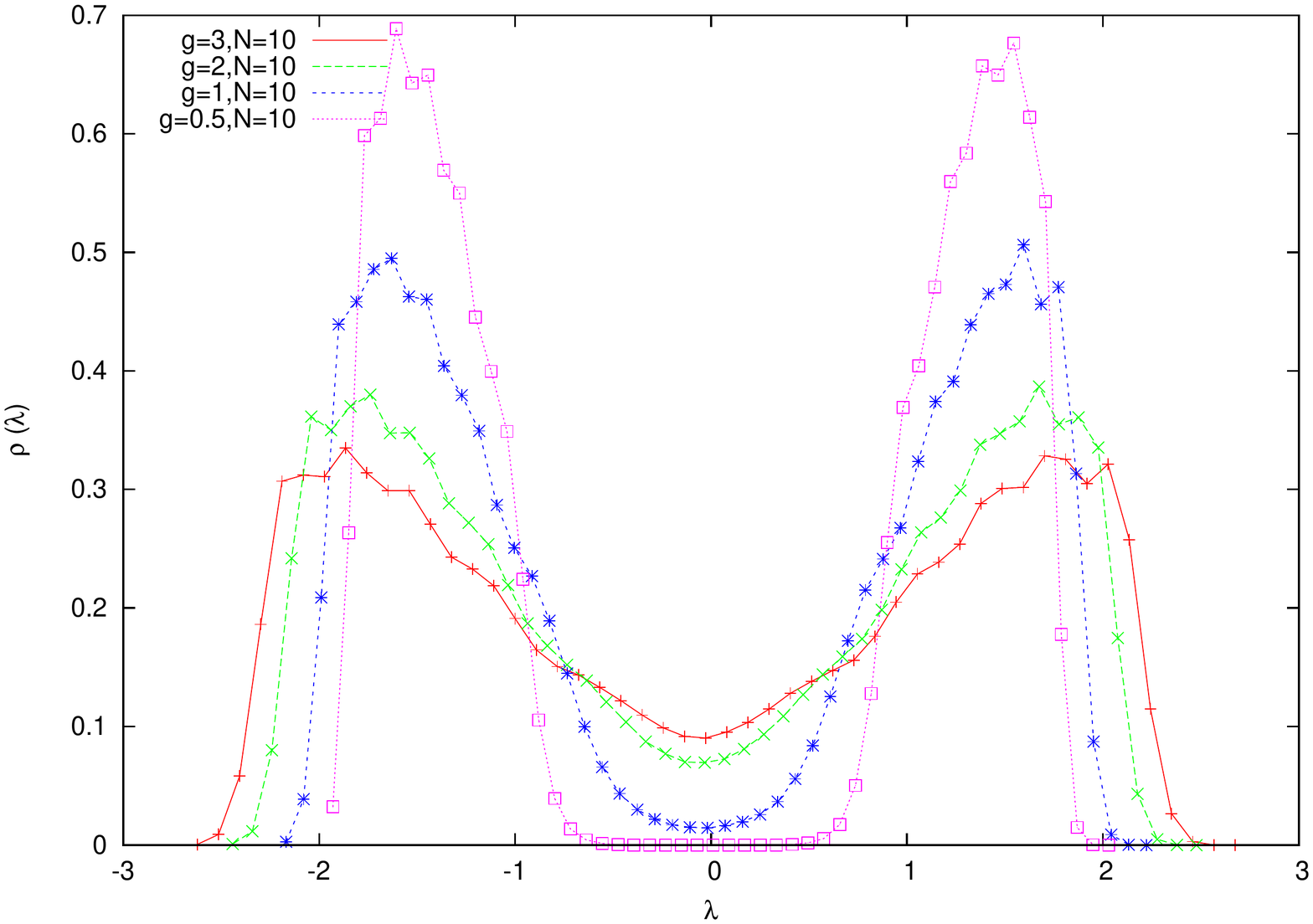}
\end{center}
\caption{}\label{testMT1}
\end{figure}

Monte Carlo tests of the above effects, and other physics, can be done using the code found in the last chapter. An illustration with $2^{20}$ thermalized configurations, where each two successive configurations are separated by $2^5$ Monte Carlo steps to reduce auto-correlation effects, and with $N=10$ and $g=2,1.5,1,0.5$, is shown on figure (\ref{testMT1}). The pure quartic matrix model is obtained from the multitrace matrix model by setting the kinetic parameter $\tilde{a}$ zero. We observe an excellent with the theoretical predictions (\ref{pred1}) and (\ref{pred2}).

\begin{figure}[htbp]
\begin{center}
\includegraphics[width=10.0cm,angle=-0]{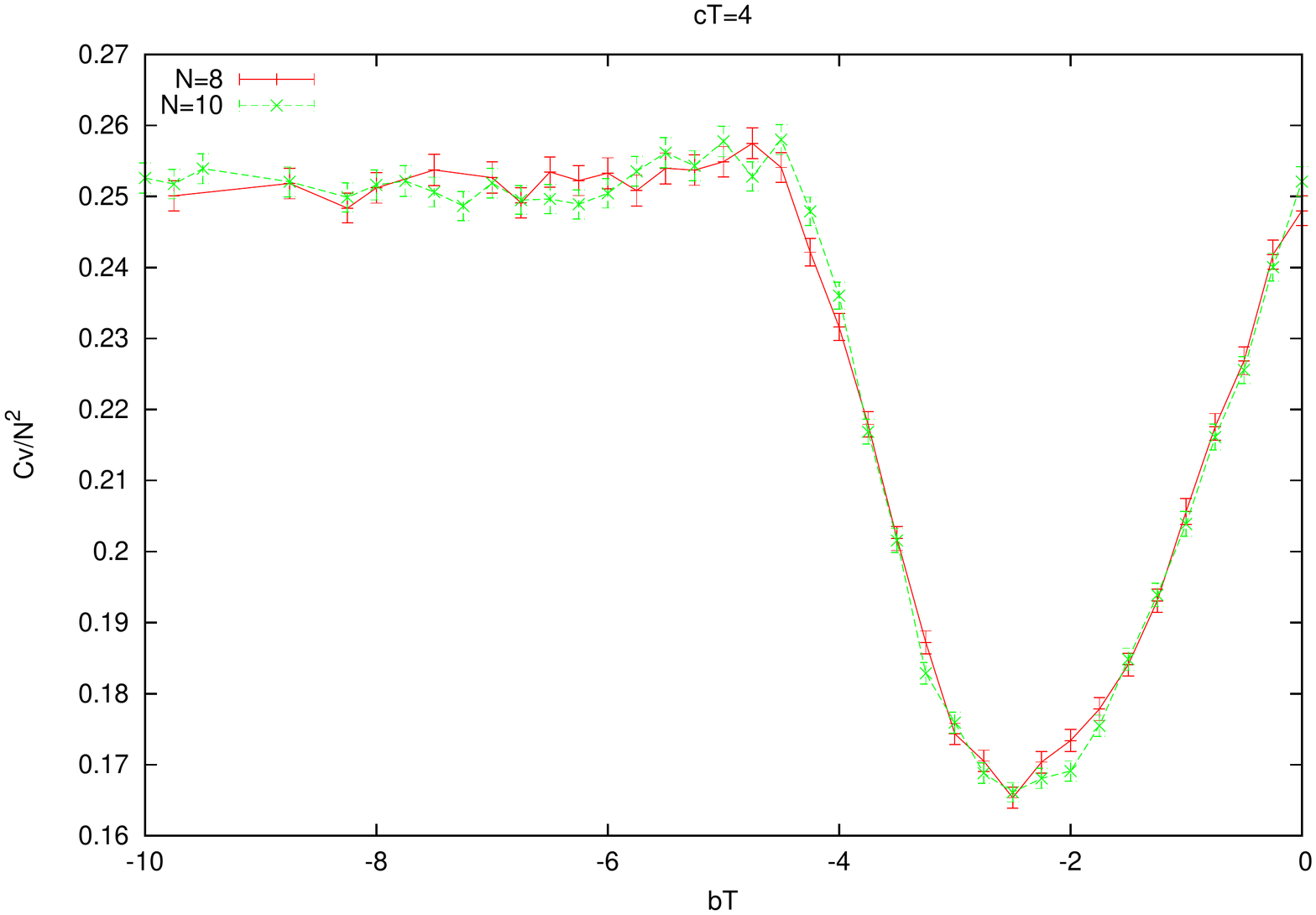}
\end{center}
\caption{}\label{testMT2}
\end{figure}

The above transition is third-order, as we said, since the first derivative of the specific heat has a finite discontinuity at $\bar{r}=B/|B_c|=-1$ as is obvious from the exact analytic result
\begin{eqnarray}
\frac{C_v}{N^2}=\frac{1}{4}~,~\bar{r}<-1.
\end{eqnarray}
\begin{eqnarray}
\frac{C_v}{N^2}=\frac{1}{4}+\frac{2\bar{r}^4}{27}-\frac{\bar{r}}{27}(2\bar{r}^2-3)\sqrt{\bar{r}^2+3}~,~\bar{r}>-1.
\end{eqnarray}
This behavior is also confirmed in Monte Carlo simulation as shown for $\tilde{c}=4$ and $N=8$ and $N=10$ on  figure (\ref{testMT2}).

The above one-cut-to-two-cut transition persists largely unchanged in the quartic multitrace matrix model (\ref{mmm}). On the other hand, and similarly to the above pure quartic matrix model, the Ising phase is not stable in this case and as a consequence the transition between non-uniform order and uniform-order is not observed in Monte Carlo simulations. The situation is drastically different if odd multitrace terms are included.

\section{Other Suitable Algorithms}
\subsection{Over-Relaxation Algorithm}
In the case of scalar $\Phi^4$ matrix models two more algorithms are available to us. The first is the over-relaxation algorithm which is very useful in the case of noncommutative $\Phi ^4$ on the fuzzy sphere given by the action
\begin{eqnarray}
S=\frac{4\pi R^2}{N+1} Tr\bigg(\frac{1}{2R^2}{\Phi}\Delta{\Phi}+\frac{1}{2}m^2{\Phi}^2+\frac{\lambda}{4!}{\Phi}^4\bigg).
\end{eqnarray}
We define
\begin{eqnarray}
S_2=\frac{4\pi R^2}{N+1} Tr\bigg(\frac{1}{2R^2}{\Phi}\Delta{\Phi}+\frac{1}{2}m^2{\Phi}^2\bigg)~,~S_4=\frac{4\pi R^2}{N+1} Tr\bigg(\frac{\lambda}{4!}{\Phi}^4\bigg).
\end{eqnarray}
Let $\Phi_0$ be some initial configuration obtained at the end of some ergodic procedure such as the Metropolis algorithm or the hybrid Monte Carlo algorithm. Let $\Phi_*$ be some new completely random configuration and thus completely  independent configuration from $\Phi_0$. If $S_*=S[\Phi_*]<S_0=S[\Phi_0]$ then $\Phi_*$ will be accepted as the new configuration. We want to devise an algorithm in which the system is forced to accept the new configuration $\Phi_*$ even if $S_*\geq S_0$. This is equivalent to heating up the system again and then letting it cool down slowly. Towards this end, we scale the configuration $\Phi_*$ as
 \begin{eqnarray}
\Phi_1=\alpha\Phi_*.
\end{eqnarray}
The scale $\alpha$ is chosen such that
\begin{eqnarray}
S_1=S[\Phi_1]=S_0.
\end{eqnarray}
Equivalently 
\begin{eqnarray}
S_{4*}\alpha^4+S_{2*}\alpha^2-S_0=0.
\end{eqnarray}
The solution is given by
\begin{eqnarray}
{\rm if}~S_0>0~:~\alpha^2=\frac{\sqrt{S_{2*}^2+4S_0S_{4*}}-S_{2*}}{2S_{4*}}.
\end{eqnarray}
\begin{eqnarray}
{\rm if}~S_0<0~{\rm and}~\{S_{2*}<-\sqrt{-4S_0S_{4*}}<0\}:~\alpha^2=\frac{\pm \sqrt{S_{2*}^2+4S_0S_{4*}}-S_{2*}}{2S_{4*}}.
\end{eqnarray}
If the conditions in the above two equations are not met then we should redefine the matrix $\Phi_*$ iterativley  as
\begin{eqnarray}
\Phi_*\longrightarrow \frac{\Phi_*+\Phi_0}{2}.
\end{eqnarray}
Then repeat. This iterative procedure will obviously create unwanted autocorrelations due to the fact that $\Phi_*$ becomes closer in each iteration to $\Phi_0$. However, the process will terminate in a finite number of steps and the obtained final configuration $\Phi_1$ has a greater probability in falling in a different orbit than the original $\Phi_0$.

The claim of \cite{Panero:2006bx} is that this algorithm solves the ergodic problem observed in Monte Carlo simulations of  noncommutative $\Phi ^4$ on the fuzzy sphere.

\subsection{Heat-Bath Algorithm}
The second algorithm is the heat-bath algorithm which works very nicely for the unbounded $\Phi^4$ potential
\begin{eqnarray}
V
&=&\frac{N}{g}(Tr M^2-\frac{1}{4} Tr M^4).
\end{eqnarray}
Remark the minus sign in front of the quartic term. Although this potential is unbounded from below it has a well defined large $N$ limit due to the metastability of the origin. The path integral is given by
 \begin{eqnarray}
Z&=&\int dM \exp(-\frac{N}{g}Tr  M^2)\exp(\frac{N}{4g}Tr  M^4)\nonumber\\
&=&\int dM dQ \exp(-\frac{N}{g}Tr  M^2- Tr Q^2+\sqrt{\frac{N}{g}}Tr QM^2).
\end{eqnarray}
The matrices $M$ and $Q$ are fully Gaussian. Let us then consider a Gaussian distribution 
 \begin{eqnarray}
\sqrt{\frac{a}{\pi}}\int dx \exp(-a x^2).
\end{eqnarray}
The Gaussian random number $x$ must be chosen, in any Monte Carlo routine, as 
 \begin{eqnarray}
&&R=\sqrt{-\frac{1}{a}\ln(1-r_1)}\nonumber\\
&&\phi=2\pi r_2\nonumber\\
&&x=R\cos\phi.
\end{eqnarray}
The $r_1$ and $r_2$ are two uniform random numbers between $0$ and $1$.

The part of the above path integral which depends on $Q$ is Gaussian given by
 \begin{eqnarray}
\int  dQ \exp(-Tr (Q-\frac{1}{2}\sqrt{\frac{N}{g}}M^2)^2).
\end{eqnarray}
The diagonal element $Q_{ii}$ comes with a factor $a=1$ while the off diagonal elements comes with a factor $a=2$. Thus we choose
\begin{eqnarray}
Q_{ii}=z_{ii}|_{a=1}+\frac{1}{2}\sqrt{\frac{N}{g}}(M^2)_{ii}~,~Q_{ij}=\frac{x_{ij}+iy_{ij}}{\sqrt{2}}|_{a=1}+\frac{1}{2}\sqrt{\frac{N}{g}}(M^2)_{ij}.
\end{eqnarray}
The $x$, $y$ and $z$ are  Gaussian random numbers with $a=1$.

The part of the path integral which depends on the diagonal element $M_{ii}$ is given by
\begin{eqnarray}
\int   \prod_{i} dM_{ii} \exp\sum_{i}\bigg(-\frac{N}{g}(1-\sqrt{\frac{g}{N}}Q_{ii})(M_{ii})^2+\frac{1}{2}\sqrt{\frac{N}{g}}\sum_{j\neq i}(Q_{ij}M_{ji}+Q_{ji}M_{ij})M_{ii}\bigg)&=&\nonumber\\
\int  \prod_{i} dM_{ii} \exp\sum_{i}\bigg(-l_i(M_{ii}-\frac{h_i}{2l_i})^2+...\bigg).
\end{eqnarray}
\begin{eqnarray}
l_i=\frac{N}{g}(1-\sqrt{\frac{g}{N}}Q_{ii})~,~h_i=\frac{1}{2}\sqrt{\frac{N}{g}}\sum_{j\neq i}(Q_{ij}M_{ji}+Q_{ji}M_{ij}).
\end{eqnarray}
Thus the diagonal elements $M_{ii}$ are Gaussian numbers which come with factors $a=l_i$. Thus we choose
\begin{eqnarray}
M_{ii}=\frac{x_{ii}}{\sqrt{l_i}}|_{a=1}+\frac{h_i}{2l_i}.
\end{eqnarray}
Finally, the part of the path integral which depends on the off diagonal element $M_{ij}$ is given by
\begin{eqnarray}
\int  \prod_{i\neq j}dM_{ij}dM_{ij}^* \exp\sum_{i\ne j}\bigg(-l_{ij}M_{ij}^*M_{ij}+h_{ij}M_{ij}^*+h_{ij}^*M_{ij}\bigg)&=&\nonumber\\
\int   \prod_{i\neq j}dM_{ij}dM_{ij}^* \exp\sum_{i\ne j}\bigg(-l_{ij}|M_{ij}-\frac{h_{ij}}{l_{ij}}|^2+...\bigg).
\end{eqnarray}
\begin{eqnarray}
l_{ij}=\frac{N}{g}\bigg(1-\frac{1}{2}\sqrt{\frac{g}{N}}(Q_{ii}+Q_{jj})\bigg)~,~h_{ij}=\frac{1}{4}\sqrt{\frac{N}{g}}\bigg(\sum_{k\neq i}Q_{ik}M_{kj}+\sum_{k\neq j}Q_{kj}M_{ik}\bigg).
\end{eqnarray}
Hence the off diagonal elements $M_{ij}$ are Gaussian numbers which come with factors $a=l_{ij}$. Thus we choose
\begin{eqnarray}
M_{ij}=\frac{x_{ij}+iy_{ij}}{\sqrt{l_{ij}}}|_{a=1}+\frac{h_{ij}}{l_{ij}}.
\end{eqnarray}
This algorithms can also be applied quite effectively to simple Yang-Mills matrix models as done for example in \cite{Hotta:1998en,Azuma:2004zq}.

\chapter{The Remez Algorithm and The Conjugate Gradient Method}

\section{Minimax Approximations}
The rational hybrid Monte Carlo algorithm (RHMC) uses in an essential way a rational approximation to the fermionic determinant. Thus in this section we will first review the issue of rational and polynomial approximations of functions. We will follow \cite{Kennedy:2004tj,Press:1992zz}.

\subsection{Minimax Polynomial Approximation and Chebyshev Polynomials}
\paragraph{Chebyshev norm:}We start by introducing the Chebyshev norm (also called uniform, infinity, supremum norm) of a continuous function $f$ over the unit interval $[0,1]$ by the relation
\begin{eqnarray}
||f||_{\infty}&=&{\rm lim}_{n\longrightarrow\infty} ||f||_n\nonumber\\
&=&{\rm lim}_{n\longrightarrow\infty}\bigg(\int_0^1 dx |f(x)|^n\bigg)^{1/n}\nonumber\\
&=&{\rm max}_x |f(x)|.
\end{eqnarray}
\paragraph{Minimax approximation:} A minimax polynomial (or rational) approximation of $f$ is a polynomial (or rational) function $p$ which minimizes the  Chebyshev norm of $p-f$, viz
\begin{eqnarray}
||p-f||_{\infty}
&=&{\rm min}_p{\rm max}_x |p(x)-f(x)|.
\end{eqnarray}
\paragraph{Weierstrass theorem:} The fundamental theorem of approximation theorem is Weierstrass’ theorem. This can be stated as follows. For every continuous function $f(x)$ over a closed interval $[a,b]$, and for every specified tolerance $\epsilon >0$, there exists a polynomial $p_n(x)$ of some degree $n$ such that for all $x\in[a,b]$, we have $||f(x)-p_n(x)||_{\infty}<\epsilon$. Thus any continuous function can be arbitrarily well approximated by a polynomial. This means in particular that the space of polynomials is dense in the space of continuous functions with respect to the topology induced by the Chebyshev norm.

\paragraph{Chebyshev theorem (minimax polynomial approximation):}

We consider a function $f$ defined on the unit interval. For any given degree $n$, there exists always a unique polynomial $p_n$ of degree $n$ which minimizes the error function

\begin{eqnarray}
||e||_{\infty}={\rm max}_{0\leq x\leq 1}|e(x)|={\rm max}_{0\leq x\leq 1} |p_n(x)-f(x)|,
\end{eqnarray}
iff the error function $e(x)$ takes its maximum absolute value at at least $n+2$ points on the unit interval, which may include the end points, and furthermore the sign of the error alternate between the successive extrema.

We can go from the function $f(x)$ defined in the interval $[-1,+1]$ to a function $f(y)$ defined in a generic interval $[a,b]$ by considering the transformation $x\longrightarrow y$ given by
\begin{eqnarray}
x=\frac{y-\frac{1}{2}(b+a)}{\frac{1}{2}(b-a)}.
\end{eqnarray}
A simple proof of this theorem can be found in \cite{Kennedy:2004tj}. This goes as follows:
\begin{itemize}
\item{\bf Chebyshev's criterion is necessary: If the error has fewer than $n+2$ alternating extrema then the approximation can be improved.} Let $p(x)$ be a polynomial for which the error $e(x)=p(x)-f(x)$ has fewer than $n+2$ alternating extrema. The next largest extremum of the error, corresponding to a local extremum,  is therefore smaller by some non zero  gap $\Delta$. Between any two successive alternating extrema the error obviously will pass by zero at some point $z_i$. If we assume that we have $d+1$ alternating extrema, then we will $d$ zeros $z_i$. We can trivially construct the polynomial 
\begin{eqnarray}
u(x)=A\prod_i(x-z_i).
\end{eqnarray}
We choose $A$ such that the sign of $u(x)$ is opposite to the sign of $e(x)$ and its magnitude $\Delta^{'}$ is less than $\Delta$, viz
\begin{eqnarray}
u(x_i)e(x_i)<0~,~\Delta^{'}={\rm max}_{0\leq x\leq 1} |u(x)|<\Delta.
\end{eqnarray}
We consider now the polynomial $p^{'}(x)=p(x)+u(x)$ with corresponding error function $e^{'}(x)=e(x)+u(x)$. The first condition $u(x_i)e(x_i)<0$ yields directly to the conclusion that the error  $e^{'}(x)$ is less than  $e(x)$ in the domain of the alternating extrema, whereas it is the condition $\Delta^{'}<\Delta$ that yields to the conclusion that $e^{'}(x)$ is less than  $e(x)$ in the domain of the next largest extremum. Thus $e^{'}(x)<e(x)$ throughout and hence  $p^{'}(x)$ is a better polynomial approximation.
\item{\bf Chebyshev's criterion is sufficient: If the error is extremal at exactly $n+2$ alternating points then the approximation is optimal.} Let us assume that there is another polynomial $p^{'}(x)$ which provides a better approximation. This means that the uniform norm $||e^{'}||_{\infty}={\rm max}_{0\leq x\leq 1}|e^{'}(x)|={\rm max}_{0\leq x\leq 1} |p^{'}(x)-f(x)|$ is less than $||e^{}||_{\infty}={\rm max}_{0\leq x\leq 1}|e^{}(x)|={\rm max}_{0\leq x\leq 1} |p^{}(x)-f(x)|$. Equivalently we must have at the $n+2$  extrema of $e(x_i)$ the inequalities
\begin{eqnarray}
|e^{'}(x_i)|<|e^{}(x_i)|.
\end{eqnarray}
By the requirement of continuity there must therefore exist $n+1$ points $z_i$ between the extrema at which we have
\begin{eqnarray}
e^{'}(z_i)=e^{}(z_i).
\end{eqnarray}
This leads immediately to
\begin{eqnarray}
p^{'}(z_i)=p^{}(z_i).
\end{eqnarray}
In other words, the polynomial $p^{'}(x)-p^{}(x)$ has $n+1$ zeros, but since this polynomial is of degree $n$, it must vanish identically. Hence $p^{'}(x)=p^{}(x)$.
\end{itemize}
\paragraph{Chebyshev polynomials:} The Chebyshev polynomial of degree $n$ is defined by 

\begin{eqnarray}
T_n(\cos\theta)=\cos n\theta\leftrightarrow T_n(x)=\cos(n \cos^{-1} x).
\end{eqnarray}
We have the explicit expressions 
\begin{eqnarray}
&&T_0=1~,~T_1=x~,~T_2=2x^2-1~,~...
\end{eqnarray}
From the results $T_{n\pm 1}=\cos n\theta \cos\theta\mp\sin n\theta\sin\theta$ we deduce the recursion relation
\begin{eqnarray}
T_{n+1}=2x T_n-T_{n-1}.
\end{eqnarray}
These polynomials are orthogonal in the interval  $[-1,1]$ with a weight $1/(1-x^2)^{1/2}$, viz
\begin{eqnarray}
\int_{-1}^{+1}\frac{dx}{\sqrt{1-x^2}}T_i(x)T_j(x)=\frac{\pi}{2}\delta_{ij}.
\end{eqnarray}
\begin{eqnarray}
\int_{-1}^{+1}\frac{dx}{\sqrt{1-x^2}}T_0(x)T_0(x)=\pi.
\end{eqnarray}
The zeros of the polynomial $T_n(x)$ are given by
\begin{eqnarray}
T_n(\cos\theta)=0\Rightarrow \cos n\theta=0\Rightarrow n\theta=(2k-1)\frac{\pi}{2}\Rightarrow x=\cos \frac{(2k-1)\pi}{2n}~,~k=1,2,...,n.
\end{eqnarray}
 Since the angle $\theta$ is in the interval between $0$ and $\pi$.  There are therefore $n$ zeros.
 
The derivative of $T_n$ is given by
\begin{eqnarray}
\frac{d}{dx}T_n&=&-n\frac{d}{dx}\cos^{-1}x.\sin (n\cos^{-1}x)\nonumber\\
&=&\frac{n}{\sqrt{1-x^2}} \sin (n\cos^{-1}x).
\end{eqnarray}
The extrema of the polynomial $T_n(x)$ are given by
\begin{eqnarray}
\frac{d}{dx}T_n=0\Rightarrow \sin (n\theta)=0\Rightarrow n\theta=k\pi\Rightarrow x=\cos \frac{k\pi}{n}~,~k=0,2,...,n.
\end{eqnarray}
There are $n+1$ extrema. The maxima satisfy $T_n(x)=1$ while the minima satisfy $T_n(x)=-1$.

The  Chebyshev polynomials satisfy also the following discrete orthogonality relation:
\begin{eqnarray}
\sum_{k=1}^m T_i(x_k)T_j(x_k)=\frac{m}{2}\delta_{ij}.
\end{eqnarray}
\begin{eqnarray}
\sum_{k=1}^m T_0(x_k)T_0(x_k)=m.
\end{eqnarray}
In the above two equations $i,j <m$ and $x_k$, $k=1,...,m$, are the $m$ zeros of the  Chebyshev polynomial $T_m(x)$. 

Since $T_n (x)$ has $n + 1$
extrema which alternate in value between $-1$
and $+1$ for $-1 \leq x \leq 1$, and since the leading coefficient of $T_n(x)$ is $2 ^{n-1}$; the polynomial  $p_n (x)=
x^n - 2^{1-n} T_n (x)$ is the best polynomial approximation of degree $n - 1$ with uniform weight to
the function $x^ n$ over the interval $[-1, 1]$. This is because by construction the
error $e_n (x)= p_n (x)- x^n = 2^{1-n} T_ n (x)$ satisfies
Chebyshev’s criterion. The magnitude of the error is just  $||e_n||_{\infty}= 2^{1-n} = 2e^{-n \ln 2}$, i.e. the error decreases exponentially with $n$.

\paragraph{Chebyshev approximation:} Let $f(x)$ be an arbitrary function in the interval $[-1,+1]$. The Chebyshev approximation of this function can be constructed as follows. Let $N$ be some large degree and $x_k$, $k=1,...,N$, be the zeros of the Chebyshev polynomial $T_N(x)$. The function $f(x)$ can be approximated by the polynomial of order $N$ defined by 
\begin{eqnarray}
f_N(x)=\sum_{k=1}^Nc_kT_{k-1}(x)-\frac{1}{2}c_1.
\end{eqnarray}
The coefficients $c_k$ are given by
\begin{eqnarray}
c_j=\frac{2}{N}\sum_{k=1}^Nf(x_k)T_{j-1}(x_k).
\end{eqnarray}
This approximation is exact for $x$ equal to all of the $N$ zeros of $T_N(x)$. Indeed, we can show
\begin{eqnarray}
\sum_{k=1}^N T_{l-1}(x_k)f_N(x_k)&=&\sum_{k=1}^Nc_k\sum_{k=1}^N T_{l-1}(x_k)T_{k-1}(x_k)-\frac{1}{2}c_1\sum_{k=1}^N T_{l-1}(x_k)\nonumber\\
&=&\frac{N}{2}c_l~,~l=1,...,N.
\end{eqnarray}
In other words,
\begin{eqnarray}
f_N(x_k)&=&f(x_k).
\end{eqnarray}
For very large $N$, the polynomial $f_N$ becomes very close to the function $f$. The polynomial $f_N$ can be "gracefully", by using the words of \cite{Press:1992zz}, truncated to a lower degree $m<<N$ by considering 
\begin{eqnarray}
f_{m}(x)=\sum_{k=1}^mc_kT_{k-1}(x)-\frac{1}{2}c_1.
\end{eqnarray}
The error for rapidly decreasing $c_k$, which is given by the difference between $f_N$ and $f_{m}$,  is dominated by $c_{m+1}T_{m}$ which has $m+1$ equal extrema distributed smoothly and uniformly in the interval $[-1,+1]$. Since the $T$'s are bounded between $-1$ and $+1$ the total error is the sum of the neglected $c_k$, $k=m+1,...,N$. The Chebyshev approximation  $f_{m}(x)$ is very close to the minimax polynomial which has the smallest maximum deviation from the function $f(x)$. Although the calculation of the  Chebyshev polynomial  $f_{m}(x)$ is very easy,  finding the actual minimax polynomial is very difficult in practice.

\paragraph{Economization of power series:} This will be explained by means of a specific example. We consider the function $f(x)=\sin x$. A quintic polynomial approximation of this function is given by the Taylor expansion 
\begin{eqnarray}
\sin x=x-\frac{x^3}{6}+\frac{x^5}{120}.
\end{eqnarray}
The domain of definition of $\sin x$ can be taken to be the interval $[-\pi,\pi]$. By making the replacement $x\longrightarrow x/\pi$ we convert the domain of definition $[-\pi,\pi]$ into the domain $[-1,1]$, viz
\begin{eqnarray}
\sin x=\pi x-\frac{\pi^3x^3}{6}+\frac{\pi^5x^5}{120}.
\end{eqnarray}
The error in the above quintic approximation is estimated by the first neglected term evaluated at the end points $x=\pm 1$, viz
\begin{eqnarray}
\frac{\pi^7x^7}{ 7!}|_{x=\pi}=0.6.
\end{eqnarray}
The error in the $7$th degree polynomial approximation can be found in the same way. We get in this case  ${\pi^9 x^9}/{ 9!}|_{x=\pi}=0.08$.

The monomials $x^k$ can be given in terms of Chebyshev polynomials by the formulas 
\begin{eqnarray}
x^k=\frac{1}{2^{k-1}}\bigg[T_k(x)+\frac{k!}{1!(k-1)!}T_{k-2}(x)+\frac{k!}{2!(k-2)!}T_{k-4}(x)+...+\frac{k!}{\frac{k-1}{2}!(k-\frac{k-1}{2})!}T_{1}(x)\bigg]~,~k~{\rm odd}.
\end{eqnarray}
\begin{eqnarray}
x^k=\frac{1}{2^{k-1}}\bigg[T_k(x)+\frac{k!}{1!(k-1)!}T_{k-2}(x)+\frac{k!}{2!(k-2)!}T_{k-4}(x)+...+\frac{k!}{\frac{k}{2}!(k-\frac{k}{2})!}T_{0}(x)\bigg]~,~k~{\rm even}.
\end{eqnarray}
For example
\begin{eqnarray}
x=T_1(x).
\end{eqnarray}
\begin{eqnarray}
x^3=\frac{1}{4}[T_3(x)+3T_1(x)].
\end{eqnarray}
\begin{eqnarray}
x^5=\frac{1}{16}[T_5(x)+5T_3(x)+10T_1(x)].
\end{eqnarray}
By substitution we get the result
\begin{eqnarray}
\sin x&=&\pi x-\frac{\pi^3x^3}{6}+\frac{\pi^5 x^5}{120}\nonumber\\
&=&\frac{\pi(192-24\pi^2+\pi^3)}{192}T_1-\frac{\pi^3(16-\pi^2)}{384}T_3+\frac{\pi^5}{1920}T_5.
\end{eqnarray}
Since $|T_n|\leq 1$, the last term is of the order of $0.16$. This is smaller than the error found in the quintic approximation above. By truncating this term we obtain a cubic approximation of the sine function given by
\begin{eqnarray}
\sin x=\frac{\pi(192-24\pi^2+\pi^3)}{192}T_1-\frac{\pi^3(16-\pi^2)}{384}T_3
\end{eqnarray}
By substituting the  Chebyshev polynomials by their expressions in terms of the $x^k$, and then changing back to the interval $[-\pi,+\pi]$, we obtain  the cubic polynomial 
\begin{eqnarray}
\sin x&=&\frac{383}{384}x-\frac{5x^3}{32}.
\end{eqnarray}
By construction this cubic approximation is better than the above considered quintic approximation.

\subsection{Minimax Rational Approximation and Remez Algorithm}
\paragraph{Chebyshev theorem revisited:}

Chebyshev theorem can be extended to the case of minimax rational approximation of functions as follows. Again we consider a function $f$ defined on the unit interval. For any given degree $(n,d)$, there exists always a unique rational function $r_{n,d}$ of degree $(n,d)$ which minimizes the error function given by

\begin{eqnarray}
||e||_{\infty}={\rm max}_{0\leq x\leq 1}|e(x)|={\rm max}_{0\leq x\leq 1} |r_{n,d}(x)-f(x)|,
\end{eqnarray}
iff the error function $e(x)$ takes its maximum absolute value at at least $n+d+2$ points on the unit interval, which may include the end points, and furthermore the sign of the error alternate between the successive extrema.  

A simple proof of this theorem can be found in \cite{Kennedy:2004tj}. As it can be shown rational approximations are far more superior to polynomial ones since, for some functions and some intervals, we can achieve substantially higher accuracy with the same number of coefficients. However, it should also be appreciated that constructing the rational approximation is much more difficult than the polynomial one.

We will further explain this very important theorem following the discussion of \cite{Press:1992zz}. The rational function $r_{n,d}$ is the ratio of two polynomials $p_n$ and $q_d$ of degrees $n$ and $d$ respectively, viz
\begin{eqnarray}
r_{n,d}(x)=\frac{p_n(x)}{q_d(x)}.\label{rat}
\end{eqnarray}
The polynomials $p_n$ and $q_d$ can be written as
\begin{eqnarray}
p_n(x)=\alpha_0+\alpha_1 x+...+\alpha_n x^n~,~q_d(x)=1+\beta_1 x+...+\beta_d x^d.
\end{eqnarray}
We will assume that $r_{n,d}$ is non degenerate, i.e. it has no common polynomial factors in numerator and denominator. The error function $e(x)$ is the deviation of $r_{n,d}$ from $f(x)$ with a maximum absolute value $e$, viz
\begin{eqnarray}
e(x)=r_{n,d}(x)-f(x)~,~e={\rm max}_{0\leq x\leq 1} |e(x)|.
\end{eqnarray}
Equation (\ref{rat}) can be rewritten as
\begin{eqnarray}
\alpha_0+\alpha_1 x+...+\alpha_n x^n=(f(x)+e(x))\bigg(1+\beta_1 x+...+\beta_d x^d\bigg).
\end{eqnarray}
There are $n+d+1$ unknowns $\alpha_i$ and $\beta_i$ plus one which is the error function $e(x)$.  We can choose the rational approximation $r_{n,x}(x)$ to be exactly equal to the function $f(x)$ at $n+d+1$ points $x_i$ in the interval $[-1,1]$,viz
 \begin{eqnarray}
f(x_i)=r_{n,d}(x_i)~,~e(x_i)=0.
\end{eqnarray}
As a consequence the  $n+d+1$ unknowns $\alpha_i$ and $\beta_i$ will be given by the $n+d+1$ linear equations 
\begin{eqnarray}
\alpha_0+\alpha_1 x_i+...+\alpha_n x_i^n=f(x_i)\bigg(1+\beta_1 x_i+...+\beta_d x_i^d\bigg).\label{rat1}
\end{eqnarray}
This can be solved any standard method such as LU decomposition.

 The points $x_i$ which are chosen in the interval  $[-1,1]$ will generically be such that there exists an extremum of the error function $e(x)$ in each subinterval $[x_i,x_{i+1}]$ plus two more extrema at the endpoints $\pm -1$ for a total of $n+d+1$ extrema. In general, the magnitudes of $r(x)$ at the extrema are not the same.
 
Alternatively, we can choose the rational approximation $r_{n,x}(x)$, at $n+d+1$ points $x_i$, to be equal to $f(x)+y_i$ with some fixed values $y_i$ of the error function $e(x)$. Equation (\ref{rat1}) becomes
\begin{eqnarray}
\alpha_0+\alpha_1 x_i+...+\alpha_n x_i^n=(f(x_i)+y_i)\bigg(1+\beta_1 x_i+...+\beta_d x_i^d\bigg).\label{rat2}
\end{eqnarray}
If we choose the $x_i$ to be the extrema of the error function $e(x)$ then the $y_i$ will be exactly $\pm e$ where $e$ is the maximal value of $|e(x)|$. We get then $n+d+2$ (not $n+d+1$) equations  for the unknowns  $\alpha_i$, $\beta_i$ and $e$ given by
 \begin{eqnarray}
\alpha_0+\alpha_1 x_i+...+\alpha_n x_i^n=(f(x_i)\pm e)\bigg(1+\beta_1 x_i+...+\beta_d x_i^d\bigg).\label{rat3}
\end{eqnarray}
The $\pm$ signs are due to the fact that successive extrema are alternating between $-e$ and $+e$. Although, this is not exactly a linear system since $e$ enters non linearly, it can still be solved using for example methods such as Newton-Raphson.
\paragraph{Remez algorithm:}
A practical constructive approach to the minimax rational approximation of functions is given by Remez (or Remes) algorithm. This is a very difficult algorithm to get to work completely and properly and some people such as the authors \cite{Press:1992zz} dislike it.

The Remez algorithm involves two nested iterations; the first on $e$ and  the second on the $x_i$'s. Explicitly, it goes through the following steps:
\begin{itemize}
\item We choose or guess $n+d+2$ initial values of the points $x_i$ in the interval $[0,1]$. The goal is to make these points converge to the  alternating extrema discussed above. 
\item {\bf The first iteration:} We keep the $x_i$'s fixed  and find the best rational approximation which goes through the points $(x_i, f(x_i)+(-1)^i\Delta)$. Towards this end, we need to solve the $n+d+2$ equations 
 \begin{eqnarray}
\alpha_0+\alpha_1 x_i+...+\alpha_n x_i^n=(f(x_i)+(-1)^i\Delta))\bigg(1+\beta_1 x_i+...+\beta_d x_i^d\bigg).\label{rat4}
\end{eqnarray}
The unknowns are  $\alpha_i$, $\beta_i$ and $\Delta$. We write this equation as
 \begin{eqnarray}
Mv=0.
\end{eqnarray}
The $(n+d+2)-$dimensional vector $v$ is formed from the coefficients $\alpha_i$, $i=0,...,n$ and $\beta_j$, $j=0,...,d$ with $\beta_0=1$. This linear system has a non trivial solution iff ${\rm det}M=0$. This condition is a  polynomial in $\Delta$. The real roots of this polynomial are the allowed values of $\Delta$ and each one of them will correspond to a solution $\alpha_i$ and $\beta_j$. Each solution $(\alpha_i,\beta_j)$  corresponds to a certain rational approximation $r_{n,d}(x)$. We pick the solution which minimizes the error function.

\item {\bf The second iteration:} We keep $e$ or $\Delta$ fixed  and choose a new set of points $x_i$'s which is the best alternating set for $e(x)$. This is done as follows. We choose an arbitrary partition $\{I_i\}$ of the interval $[0,1]$ where $I_i$ is such that $x_i\in I_i$. Then we choose a new set of points $x_i^{'}$ such that
 \begin{eqnarray}
x_i^{'}\in I_i~,~(-1)^ie(x_i^{'})={\rm max}_{x\in I_i}(-1)^ie(x_i^{}).
\end{eqnarray}
\end{itemize}
Several drawbacks of this algorithm are noted in \cite{Kennedy:2004tj,Press:1992zz}. Among these, we mention here the slow rate of convergence and the necessity of multiple precision arithmetic. 

\paragraph{Zolotarev’s Theorem:} The case of rational approximations of the sign function, the square root and the inverse square root are known analytically in the sense that the coefficients of the optimal and unique Chebyshev rational approximations are known exactly. This result is due to Zolotarev.

\paragraph{The Numerical Recipes algorithm:} 

A much simpler but very sloppy approximation, which is claimed in \cite{Press:1992zz} to be "within a fraction of a least
significant bit of the minimax one", and in which we try to bring the error not to zero as in the minimax case but to $\pm$ some consistent value, can be constructed as follows:
\begin{itemize}
\item We start from $n+d+1$ values of $x_i$, or even a larger number of $x_i$, which are spaced approximately like the zeros of a higher order Chebyshev polynomials.
\item We solve for  $\alpha_i$ and $\beta_j$ the linear system:
\begin{eqnarray}
\alpha_0+\alpha_1 x_i+...+\alpha_n x_i^n=f(x_i)\bigg(1+\beta_1 x_i+...+\beta_d x_i^d\bigg).
\end{eqnarray}
In the case that the number of $x_i$'s is larger than $n+d+1$ we can use the singular value decomposition method to solve this system. The solution will provide our starting rational approximation $r_{n,d}(x)$. Compute $e(x_i)$ and $e$.
\item We solve  for  $\alpha_i$ and $\beta_j$ the linear system: 
 \begin{eqnarray}
\alpha_0+\alpha_1 x_i+...+\alpha_n x_i^n=(f(x_i)\pm e)\bigg(1+\beta_1 x_i+...+\beta_d x_i^d\bigg).
\end{eqnarray}
The $\pm$ is chosen to be the sign of the observed error function $e(x_i)$ at each point $x_i$.
\item We repeat the second step several times.
\end{itemize}
\subsection{The Code "AlgRemez"}
This code can be found in \cite{algremez0}.
\section{Conjugate Gradient Method}
\subsection{Construction}
Our presentation of the conjugate gradient method in this section will follow the pedagogical note \cite{cg}. See also \cite{cg1,cg2}.

\paragraph{The basic problem:} We consider a symmetric and positive definite $n\times n$ matrix $A$ and an $n-$dimensional vector $\vec{v}$. The basic problem here is to solve for the $n-$dimensional vector $\vec{x}$ which satisfies the equation 
\begin{eqnarray}
A\vec{x}=\vec{v}.
\end{eqnarray}
We will find the solution by means of the conjugate gradient method which is an iterative algorithm suited for large sparse matrices $A$.
\paragraph{Principles of the method:} The above problem is equivalent to finding the minimum $\vec{x}$ of the function $\Phi(\vec{x})$ defined by
\begin{eqnarray}
\Phi(\vec{x})=\frac{1}{2}\vec{x}A\vec{x}-\vec{x}\vec{v}.
\end{eqnarray}
The gradient of $\Phi$ is given by
\begin{eqnarray}
\vec{\nabla}\Phi(\vec{x})=A\vec{x}-\vec{v}.
\end{eqnarray}
This vanishes at the minimum. If not zero, it gives precisely the direction of steepest ascent  of the surface $\Phi$. The residual of the above set of equations is defined by
\begin{eqnarray}
\vec{r}=-\vec{\nabla}\Phi(\vec{x})=\vec{v}-A\vec{x}.
\end{eqnarray}
We will denote the $n$ linearly independent vectors in the vector space to which $\vec{x}$ belongs by $\vec{p}^{(i)}$, $i=1,...,n$. They form a basis in this vector space. The vector $\vec{x}$ can be expanded as
\begin{eqnarray}
\vec{x}=\sum_{i=1}^ns_i\vec{p}^{(i)}=P\vec{s}.
\end{eqnarray}
$P$ is the $n\times n$ matrix of the linearly independent vectors $\vec{p}^{(i)}$, i.e. $P_{ij}=p_i^{(j)}$, and $\vec{s}$ is the vector of the coefficients $s_i$. Typically, we will start from a reference vector $\vec{x}_0$. Thus we write
 \begin{eqnarray}
\vec{x}=\vec{x}_0+P\vec{s}.
\end{eqnarray}
The vectors  $\vec{p}^{(i)}$ are $A-$conjugate to each other iff
\begin{eqnarray}
\vec{p}^{(i)}A\vec{p}^{(j)}=0~,~i\neq j.
\end{eqnarray}
Thus we can write
\begin{eqnarray}
P^{T}AP=D.
\end{eqnarray}
$D$ is a diagonal matrix with elements given by
 \begin{eqnarray}
d_i=\vec{p}^{(i)}A\vec{p}^{(i)}.
\end{eqnarray}
The gradient of $\Phi$ takes the form
 \begin{eqnarray}
\vec{\nabla}\Phi=AP\vec{s}-\vec{r}_0~,~\vec{r}_0=\vec{v}-A\vec{x}_0.
\end{eqnarray}
Next, multiplication with the transpose $P^{T}$ yields
\begin{eqnarray}
P^T\vec{\nabla}\Phi&=&P^TAP\vec{s}-P^T\vec{r}_0\nonumber\\
&=&D\vec{s}-P^T\vec{r}_0.
\end{eqnarray}
The solution to $\vec{\nabla}\Phi=0$ is then
\begin{eqnarray}
D\vec{s}-P^T\vec{r}_0=0\Rightarrow s_i=\frac{\vec{p}^{(i)}\vec{r}_0}{\vec{p}^{(i)}A\vec{p}^{(i)}}.
\end{eqnarray}
The solution $s_i$ found by globally minimizing $\Phi$, also locally minimizes $\Phi$ along the direction $\vec{p}^{(i)}$. Thus starting from a vector $\vec{x}_0$ we obtain the solution 
\begin{eqnarray}
\vec{x}_1=\vec{x}_0+s_1\vec{p}^{(1)}~,~s_1=\frac{\vec{p}^{(1)}\vec{r}_0}{\vec{p}^{(1)}A\vec{p}^{(1)}}~,~\vec{r}_0=\vec{v}-A\vec{x}_0.
\end{eqnarray}
This is the local minimum of $\Phi$ along a line from $\vec{x}_0$ in the direction $\vec{p}^{(1)}$. Indeed, we can check that
 \begin{eqnarray}
\vec{p}^{(1)}\vec{\nabla}\Phi=0\Rightarrow s_1=\frac{\vec{p}^{(1)}\vec{r}_0}{\vec{p}^{(1)}A\vec{p}^{(1)}}.
\end{eqnarray}
The vector $\vec{r}_0$ is the first residual at the point $\vec{x}_0$ given by
\begin{eqnarray}
\vec{\nabla}\Phi|_{\vec{x}_0}=-\vec{r}_0.
\end{eqnarray}
 Next, starting from the vector $\vec{x}_1$ we obtain the solution 
\begin{eqnarray}
\vec{x}_2=\vec{x}_1+s_2\vec{p}^{(2)}~,~s_2=\frac{\vec{p}^{(2)}\vec{r}_1}{\vec{p}^{(2)}A\vec{p}^{(2)}}~,~\vec{r}_1=\vec{v}-A\vec{x}_1.
\end{eqnarray}
This is the local minimum of $\Phi$ along a line from $\vec{x}_1$ in the direction $\vec{p}^{(2)}$. The vector $\vec{r}_1$ is the new residual at the point $\vec{x}_1$, viz
\begin{eqnarray}
\vec{\nabla}\Phi|_{\vec{x}_1}=-\vec{r}_1.
\end{eqnarray}
 In general  starting from the vector $\vec{x}_i$ we obtain the solution
\begin{eqnarray}
\vec{x}_{i+1}=\vec{x}_i+s_{i+1}\vec{p}^{(i+1)}~,~s_{i+1}=\frac{\vec{p}^{(i+1)}\vec{r}_i}{\vec{p}^{(i+1)}A\vec{p}^{(i+1)}}~,~\vec{r}_i=\vec{v}-A\vec{x}_i.
\end{eqnarray}
This is the local minimum of $\Phi$ along a line from $\vec{x}_i$ in the direction $\vec{p}^{(i+1)}$. The vector $\vec{r}_i$ is the residual at the point $\vec{x}_{i}$, viz
\begin{eqnarray}
\vec{\nabla}\Phi|_{\vec{x}_i}=-\vec{r}_i.
\end{eqnarray}
The residual vectors provide the directions of steepest descent  of the function  $\Phi$ at each iteration step.  Thus if we know the conjugate vectors $\vec{p}^{(i)}$ we can compute the coefficients $s_i$ and write down the solution $\vec{x}$. Typically, a good approximation of the true minimum of $\Phi$ may be obtained only after a small subset of the conjugate vectors are visited.

\paragraph{Choosing the conjugate vectors:} The next step is to choose a set of conjugate vectors. An obvious candidate is the set of eigenvectors of the symmetric matrix $A$. However, in practice this choice is made as follows. Given that we have reached the iteration step $i$, i.e. we have reached the vector  $\vec{x}_{i}$ which minimizes  $\Phi$ in the direction $\vec{p}^{(i)}$, the search direction  $\vec{p}^{(i+1)}$ will be naturally chosen in the direction of steepest descent  of the function  $\Phi$ at the point $\vec{x}_{i}$, which since $A$ is positive definite is given by the direction of the residual $\vec{r}_{i}$, but conjugate to the previous search direction  $\vec{p}^{(i)}$. We start then from the ansatz 

\begin{eqnarray}
\vec{p}^{(i+1)}=\vec{r}_{i}-\lambda \vec{p}^{(i)}.
\end{eqnarray}
This must be $A-$conjugate to  $\vec{p}^{(i)}$, viz
\begin{eqnarray}
\vec{p}^{(i)}A\vec{p}^{(i+1)}=0.
\end{eqnarray}
This yields the value
\begin{eqnarray}
\lambda=\frac{\vec{p}^{(i)}A\vec{r}_{i}}{\vec{p}^{(i)}A\vec{p}^{(i)}}.
\end{eqnarray}
The gradient $\vec{\nabla}\Phi$ at the point $\vec{x}_i$ is orthogonal to all previous search directions  $\vec{p}^{(j)}$, $j< i$. Indeed, we compute
\begin{eqnarray}
\vec{p}^{(j)}\vec{\nabla}\Phi|_{\vec{x}_i}&=&\vec{p}^{(j)}\big(A\vec{x}_i-\vec{v}\big)\nonumber\\
&=&\vec{p}^{(j)}\big(A\vec{x}_0+\sum_{k=1}^i s_kA\vec{p}^{(k)}-\vec{v}\big)\nonumber\\
&=&\vec{p}^{(j)}\big(\sum_{k=1}^i s_kA\vec{p}^{(k)}-\vec{r}_0\big)\nonumber\\
&=&\sum_{k=1}^i s_k\vec{p}^{(j)}A\vec{p}^{(k)}-\vec{p}^{(j)}\vec{r}_0\nonumber\\
&=&s_j\vec{p}^{(j)}A\vec{p}^{(j)}-\vec{p}^{(j)}\vec{r}_0\nonumber\\
&=&0.
\end{eqnarray}
This formula works also for $j=i$. The gradients  $\vec{\nabla}\Phi|_{\vec{x}_i}$  is also orthogonal to all previous gradients  $\vec{\nabla}\Phi|_{\vec{x}_j}$, $j< i$. Indeed, we have
\begin{eqnarray}
\vec{\nabla}\Phi|_{\vec{x}_j}\vec{\nabla}\Phi|_{\vec{x}_i}&=&-\vec{r}_j\vec{\nabla}\Phi|_{\vec{x}_i}\nonumber\\
&=&-(\lambda \vec{p}^{(j)}+\vec{p}^{(j+1)})\vec{\nabla}\Phi|_{\vec{x}_i}\nonumber\\
&=&0.
\end{eqnarray}
The first search direction can be chosen arbitrarily. We can for example choose $\vec{p}^{(1)}=\vec{r}_0=-\vec{\nabla}\Phi|_{\vec{x}_0}$. The next search direction $\vec{p}^{(2)}$ is by construction $A-$conjugate to $\vec{p}^{(1)}$. At the third iteration step we obtain $\vec{p}^{(3)}$ which is $A-$conjugate to $\vec{p}^{(2)}$. The remaining question is whether $\vec{p}^{(3)}$ is $A-$conjugate to $\vec{p}^{(1)}$  or not. In general we would like to show that the search direction  $\vec{p}^{(i)}$ generated at the $i$th iteration step, which is $A-$conjugate to  $\vec{p}^{(i-1)}$, is also $A-$conjugate to all previously generated search directions  $\vec{p}^{(j)}$, $j< i-1$. Thus we need to show that
\begin{eqnarray}
\vec{p}^{(j)}A\vec{p}^{(i)}&=&0~,~j<i-1.
\end{eqnarray}
We compute
\begin{eqnarray}
\vec{p}^{(j)}A\vec{p}^{(i)}&=&\vec{p}^{(j)}A(\vec{r}_{i-1}-\lambda \vec{p}^{(i-1)})\nonumber\\
&=&\vec{p}^{(j)}A\vec{r}_{i-1}-\lambda \vec{p}^{(j)}A\vec{p}^{(i-1)}\nonumber\\
&=&\frac{1}{s_j}(\vec{x}_j-\vec{x}_{j-1})A\vec{r}_{i-1}-\lambda \vec{p}^{(j)}A\vec{p}^{(i-1)}\nonumber\\
&=&\frac{1}{s_j}(-\vec{r}_j+\vec{r}_{j-1})\vec{r}_{i-1}-\lambda \vec{p}^{(j)}A\vec{p}^{(i-1)}\nonumber\\
&=&-\lambda \vec{p}^{(j)}A\vec{p}^{(i-1)}\nonumber\\
&=&0.
\end{eqnarray}
\paragraph{Summary:} Let us now summarize the main ingredients of the above algorithm. We have the following steps:
\begin{itemize}
\item[$1)$] We choose a reference vector $\vec{x}_0$. We calculate the initial residual $\vec{r}_0=\vec{v}-A\vec{x}_0$.
\item[$2)$] We choose the first search direction  as $\vec{p}^{(1)}=\vec{r}_0$.
\item[$3)$] The first iteration towards the solution is
\begin{eqnarray}
\vec{x}_1=\vec{x}_0+s_1\vec{p}^{(1)}~,~s_1=\frac{\vec{p}^{(1)}\vec{r}_0}{\vec{p}^{(1)}A\vec{p}^{(1)}}.
\end{eqnarray}

\item[$4)$] The above three steps are iterated as follows:
\begin{eqnarray}
\vec{r}_i=\vec{v}-A\vec{x}_i.\label{31}
\end{eqnarray}
\begin{eqnarray}
\vec{p}^{(i+1)}=\vec{r}_{i}-\lambda \vec{p}^{(i)}~,~\lambda=\frac{\vec{p}^{(i)}A\vec{r}_{i}}{\vec{p}^{(i)}A\vec{p}^{(i)}}.
\end{eqnarray}
\begin{eqnarray}
s_{i+1}=\frac{\vec{p}^{(i+1)}\vec{r}_i}{\vec{p}^{(i+1)}A\vec{p}^{(i+1)}}.
\end{eqnarray}
\begin{eqnarray}
\vec{x}_{i+1}=\vec{x}_i+s_{i+1}\vec{p}^{(i+1)}.\label{34}
\end{eqnarray}
By using equations (\ref{31}) and (\ref{34}) we can show that equation (\ref{31}) can be replaced by the equation
\begin{eqnarray}
\vec{r}_i=\vec{r}_{i-1}-s_iA\vec{p}^{(i)}
\end{eqnarray}
Also we can derive the more efficient formulas
\begin{eqnarray}
s_{i+1}=\frac{\vec{r}_{i}\vec{r}_i}{\vec{p}^{(i+1)}A\vec{p}^{(i+1)}}~,~\lambda=-\frac{\vec{r}_{i}\vec{r}_{i}}{\vec{r}_{i-1}\vec{r}_{i-1}}.
\end{eqnarray}
\item[$5)$] The above procedure continues as long as $|\vec{r}|\ge \epsilon$ where $\epsilon$ is some tolerance, otherwise stop. 
\end{itemize}
\subsection{The Conjugate Gradient Method as a Krylov Space Solver}
We start this section by introducing some slight change of notation. By making the replacements $\vec{p}^{(i+1)}\longrightarrow \vec{p}_{i}$, $s_{i+1}\longrightarrow -\beta_i$, $\lambda\longrightarrow -\alpha_i$ the conjugate gradient algorithm will read
\begin{eqnarray}
\vec{x}_{i+1}=\vec{x}_i-\beta_{i}\vec{p}_{i}~,~\beta_i=-\frac{\vec{r}_{i}\vec{r}_i}{\vec{p}_{i}A\vec{p}_{i}}.
\end{eqnarray}
\begin{eqnarray}
\vec{r}_{i+1}=\vec{r}_i+\beta_{i}A\vec{p}_{i}.
\end{eqnarray}
\begin{eqnarray}
\vec{p}_{i+1}=\vec{r}_{i+1}+\alpha_{i+1}\vec{p}_{i}~,~\alpha_{i+1}=\frac{\vec{r}_{i+1}\vec{r}_{i+1}}{\vec{r}_{i}\vec{r}_{i}}.
\end{eqnarray}
We start iterating from
\begin{eqnarray}
\vec{x}_{0}=0~,~\vec{r}_{0}=\vec{v}-A\vec{x}_{0}=\vec{v}~,~\vec{p}_{0}=\vec{r}_{0}=\vec{v}.
\end{eqnarray}
Remark now the following. We have
\begin{eqnarray}
\vec{r}_{0}=\vec{v}-A\vec{x}_{0}\in {\rm span}\{\vec{r}_{0}\}.
\end{eqnarray}
\begin{eqnarray}
\vec{r}_{1}=\vec{r}_0+\beta_{0}A\vec{r}_{0}\in {\rm span}\{\vec{r}_{0},A\vec{r}_{0}\}.
\end{eqnarray}
\begin{eqnarray}
\vec{r}_{2}=\vec{r}_0+\beta_{0}A\vec{r}_{0}+\beta_1A(\vec{r}_0+\beta_{0}A\vec{r}_{0})+\alpha_1\beta_{1}A\vec{r}_{0}\in {\rm span}\{\vec{r}_{0},A\vec{r}_{0},A^2\vec{r}_{0}\}.
\end{eqnarray}
In general we will have
\begin{eqnarray}
\vec{r}_{n}=P_n(A)\vec{r}_0\in {\rm span}\{\vec{r}_{0},A\vec{r}_{0},A^2\vec{r}_{0},...,A^n\vec{r}_{0}\}.
\end{eqnarray}
The $P_n(A)$ is a polynomial of degree $n$ which obviously satisfy $P_n(0)=1$. It is called the residual polynomial. On the other hand, the space ${\rm span}\{\vec{r}_0, A\vec{r}_0,..., A^n\vec{r}_{0}\}$ is called a Krylov subspace. Since the residues $\vec{r}_{n}$ are orthogonal the polynomials  $P_n(A)$ are also orthogonal.

Similarly, we observe that
\begin{eqnarray}
\vec{p}_{0}=\vec{r}_{0}\in {\rm span}\{\vec{r}_{0}\}.
\end{eqnarray}
\begin{eqnarray}
\vec{p}_{1}=\vec{r}_1+\alpha_{1}\vec{r}_{0}\in {\rm span}\{\vec{r}_{0},A\vec{r}_{0}\}.
\end{eqnarray}
\begin{eqnarray}
\vec{p}_{2}=\vec{r}_2+\alpha_{2}\vec{r}_{1}+\alpha_1\alpha_2\vec{r}_0\in {\rm span}\{\vec{r}_{0},A\vec{r}_{0},A^2\vec{r}_{0}\}.
\end{eqnarray}
Thus in general
\begin{eqnarray}
\vec{p}_{n}\in {\rm span}\{\vec{r}_{0},A\vec{r}_{0},A^2\vec{r}_{0},...,A^n\vec{r}_{0}\}.
\end{eqnarray}
Also
\begin{eqnarray}
\vec{x}_{n}&=&\vec{x}_{0}-\sum_{i=0}^{n-1}\beta_i\vec{p}_{i}.
\end{eqnarray}
Thus
\begin{eqnarray}
\vec{x}_{n}-\vec{x}_{0}=Q_{n-1}(A)\vec{r}_{0}\in {\rm span}\{\vec{r}_{0},A\vec{r}_{0},A^2\vec{r}_{0},...,A^{n-1}\vec{r}_{0}\}.
\end{eqnarray}
The $Q_{n-1}(A)$ is a polynomial of exact degree $n-1$. Hence both the conjugate gradient directions $\vec{p}_{n}$ and the solutions $\vec{x}_{n}-\vec{x}_{0}$ belong to various Krylov subspaces.

The conjugate gradient method is an example belonging to a large class of Krylov subspace methods. It is due to Hestenes and Stiefel \cite{HS} and it is the method of choice for solving linear systems that are symmetric positive definite or Hermitian positive definite. We conclude this section by the following two definitions.

\paragraph{Definition $1$:} Given a non-singular matrix $A\in {\bf C}^{n\times n}$ and a non-zero vector $r\in {\bf C}^n$, the $n$th
Krylov (sub)space ${\cal K}_n(A, r)$ generated by $A$ from $r$ is
\begin{eqnarray}
{\cal K}_n(A, r)={\rm  span} (r, Ar, . . . , A^{n-1} r).
\end{eqnarray}
\paragraph{Definition $2$:} A standard Krylov space method for solving a linear
system $Ax = b$ is an iterative method  which starts from some initial guess $x_0$ with residual $r_0=b-Ax_0$ and then generates better approximations $x_n$ to the exact solution $x_*$ as follows 
\begin{eqnarray}
{x}_{n}-{x}_{0}=Q_{n-1}(A){r}_{0}\in {\cal K}_n(A,r_0)={\rm span}\{{r}_{0},A{r}_{0},A^2{r}_{0},...,A^{n-1}{r}_{0}\}.
\end{eqnarray}
The residuals $r_n$ of the above so-called Krylov space solver will satisfy
\begin{eqnarray}
{r}_{n}=P_n(A){r}_0\in {\cal K}_{n+1}(A,r_0)={\rm span}\{{r}_{0},A{r}_{0},A^2{r}_{0},...,A^n{r}_{0}\}.
\end{eqnarray}
It is not difficult to show that
\begin{eqnarray}
P_n(A)=1-AQ_{n-1}(A).\label{ide}
\end{eqnarray}
\subsection{The Multi-Mass Conjugate Gradient Method}
The goal now is to solve a multi-mass linear system of the form
\begin{eqnarray}
(A+\sigma)\vec{x}=\vec{v}.
\end{eqnarray}
By a direct application of the conjugate gradient method we get the solution
\begin{eqnarray}
\vec{x}_{i+1}^{\sigma}=\vec{x}_i^{\sigma}-\beta_{i}^{\sigma}\vec{p}_{i}^{\sigma}~,~\beta_i^{\sigma}=-\frac{\vec{r}_{i}^{\sigma}\vec{r}_i^{\sigma}}{\vec{p}_{i}^{\sigma}(A+\sigma)\vec{p}_{i}^{\sigma}}.
\end{eqnarray}
\begin{eqnarray}
\vec{r}_{i+1}^{\sigma}=\vec{r}_i^{\sigma}+\beta_{i}^{\sigma}(A+\sigma)\vec{p}_{i}^{\sigma}.\label{mul0}
\end{eqnarray}
\begin{eqnarray}
\vec{p}_{i+1}^{\sigma}=\vec{r}_{i+1}^{\sigma}+\alpha_{i+1}^{\sigma}\vec{p}_{i}^{\sigma}~,~\alpha_{i+1}^{\sigma}=\frac{\vec{r}_{i+1}^{\sigma}\vec{r}_{i+1}^{\sigma}}{\vec{r}_{i}^{\sigma}\vec{r}_{i}^{\sigma}}.\label{mul1}
\end{eqnarray}
\begin{eqnarray}
\vec{x}_{0}^{\sigma}=0~,~\vec{r}_{0}^{\sigma}=\vec{v}^{\sigma}-(A+\sigma)\vec{x}_{0}^{\sigma}=\vec{v}~,~\vec{p}_{0}^{\sigma}=\vec{r}_{0}^{\sigma}=\vec{v}.
\end{eqnarray}
There is clearly a loop over $\sigma$ which could be very expensive in practice. Fortunately we can solve, by following \cite{Jegerlehner:1996pm}, the above multi-mass linear system using only a single set of vector-matrix operations as follows. First we note that
\begin{eqnarray}
\vec{r}_{i+1}^{\sigma}=\vec{r}_i^{\sigma}+\beta_{i}^{\sigma}(A+\sigma)\vec{p}_{i}^{\sigma}=P_{i+1}^{\sigma}(A+\sigma)\vec{r}_0^{\sigma}\in {\cal K}_{i+2}(A+\sigma,\vec{r}_0).
\end{eqnarray}
As discussed before the polynomials $P_{i+1}^{\sigma}$ are orthogonal in $A+\sigma$. This follows from the fact that $\vec{r}_{i+1}^{\sigma}\perp \vec{r}_{i}^{\sigma}$ and as a consequence 
\begin{eqnarray}
P_{i+1}^{\sigma}(A+\sigma)\vec{r}_0^{\sigma}\perp {\cal K}_{i+1}(A+\sigma,\vec{r}_0).
\end{eqnarray}
However, we have the obvious and fundamental fact that
\begin{eqnarray}
{\cal K}_{i+1}(A+\sigma,\vec{r}_0)={\cal K}_{i+1}(A,\vec{r}_0).
\end{eqnarray}
In other words, the polynomials $P_{i+1}^{\sigma}$ are orthogonal in $A$ as well. We must therefore have
 \begin{eqnarray}
P_{i+1}^{\sigma}(A+\sigma)=\zeta_{i+1}^{\sigma} P_{i+1}(A).\label{shifted}
\end{eqnarray}
The polynomials $P_{i+1}^{\sigma}$  are thus of a shifted structure. By the identity (\ref{ide}) it follows that the polynomials $Q_{i}^{\sigma}$ are not of a shifted structure. This single observation will allow us to reduce the problem to a single set of vector-matrix operations. 

By multiplying equation (\ref{mul1}) by $\beta_{i+1}^{\sigma}(A+\sigma)$ and using equation (\ref{mul0}) we get
\begin{eqnarray}
\beta_{i+1}^{\sigma}(A+\sigma)\vec{p}_{i+1}^{\sigma}=\beta_{i+1}^{\sigma}(A+\sigma)\vec{r}_{i+1}^{\sigma}+\frac{\beta_{i+1}^{\sigma}\alpha_{i+1}^{\sigma}}{\beta_i^{\sigma}}(\vec{r}_{i+1}^{\sigma}-\vec{r}_{i}^{\sigma}).
\end{eqnarray}
By substitution in  equation (\ref{mul0}) we get the $3-$term recurrence given by
\begin{eqnarray}
\vec{r}_{i+2}^{\sigma}=(1+\frac{\beta_{i+1}^{\sigma}\alpha_{i+1}^{\sigma}}{\beta_i^{\sigma}})\vec{r}_{i+1}^{\sigma}+\beta_{i+1}^{\sigma}(A+\sigma)\vec{r}_{i+1}^{\sigma}-\frac{\beta_{i+1}^{\sigma}\alpha_{i+1}^{\sigma}}{\beta_i^{\sigma}}\vec{r}_{i}^{\sigma}.
\end{eqnarray}
By using (\ref{shifted}) we obtain
\begin{eqnarray}
\zeta_{i+2}^{\sigma}\vec{r}_{i+2}=(1+\frac{\beta_{i+1}^{\sigma}\alpha_{i+1}^{\sigma}}{\beta_i^{\sigma}})\zeta_{i+1}^{\sigma}\vec{r}_{i+1}+\beta_{i+1}^{\sigma}(A+\sigma)\zeta_{i+1}^{\sigma}\vec{r}_{i+1}-\frac{\beta_{i+1}^{\sigma}\alpha_{i+1}^{\sigma}}{\beta_i^{\sigma}}\zeta_i^{\sigma}\vec{r}_{i}.\label{recu}
\end{eqnarray}
However, the no-sigma recurrence reads
\begin{eqnarray}
\vec{r}_{i+2}=(1+\frac{\beta_{i+1}\alpha_{i+1}}{\beta_i})\vec{r}_{i+1}+\beta_{i+1}A\vec{r}_{i+1}-\frac{\beta_{i+1}\alpha_{i+1}}{\beta_i}\vec{r}_{i}.
\end{eqnarray}
By comparing the $A\vec{r}_{i+1}$ terms we obtain 
\begin{eqnarray}
\beta_n^{\sigma}=\beta_n\frac{\zeta_{n+1}^{\sigma}}{\zeta_n^{\sigma}}.
\end{eqnarray}
By comparing the $\vec{r}_{i}$ terms and also using the above result we obtain
\begin{eqnarray}
\alpha_n^{\sigma}=\alpha_n\frac{\zeta_{n}^{\sigma}\beta_{n-1}^{\sigma}}{\zeta_{n-1}^{\sigma}\beta_{n-1}}.
\end{eqnarray}
By comparing the $\vec{r}_{i+1}$ terms and also using the above two results we find after some calculation  
\begin{eqnarray}
\zeta_{n+1}^{\sigma}=\frac{\zeta_{n}^{\sigma}\zeta_{n-1}^{\sigma}\beta_{n-1}}{\alpha_n\beta_n(\zeta_{n-1}^{\sigma}-\zeta_{n}^{\sigma})+\zeta_{n-1}^{\sigma}\beta_{n-1}(1-\sigma\beta_n)}.
\end{eqnarray}
Let us conclude by summarizing the main ingredients of this algorithm. These are:
\begin{enumerate}
\item We start from 
\begin{eqnarray}
\vec{x}=\vec{x}_{0}^{\sigma}=0~,~\vec{r}_0=\vec{r}_{0}^{\sigma}=\vec{v}~,~\vec{p}=\vec{p}_{0}^{\sigma}=\vec{v}.\label{cg1}
\end{eqnarray}
By setting $i=-1$ in (\ref{recu}) we see that we must also start from
\begin{eqnarray}
\alpha_0=\alpha_0^{\sigma}=0~,~\beta_{-1}=\beta_{-1}^{\sigma}=1~,~\zeta_{0}^{\sigma}=\zeta_{-1}^{\sigma}=1.\label{cg2}
\end{eqnarray}
\item We solve the no-sigma problem (we start from $n=0$):
\begin{eqnarray}
&&\beta_n=-\frac{\vec{r}_{n}\vec{r}_n}{\vec{p}_{n}A\vec{p}_{n}}\nonumber\\
&&\vec{x}_{n+1}=\vec{x}_n-\beta_{n}\vec{p}_{n}.\label{cg3}
\end{eqnarray}
\begin{eqnarray}
\vec{r}_{n+1}=\vec{r}_n+\beta_{n}A\vec{p}_{n}.\label{cg4}
\end{eqnarray}
\begin{eqnarray}
&&\alpha_{n+1}=\frac{\vec{r}_{n+1}\vec{r}_{n+1}}{\vec{r}_{n}\vec{r}_{n}}\nonumber\\
&&\vec{p}_{n+1}=\vec{r}_{n+1}+\alpha_{n+1}\vec{p}_{n}.\label{cg5}
\end{eqnarray}
\item We generate solutions of the sigma problems by the relations (we start from $n=0$):
\begin{eqnarray}
\zeta_{n+1}^{\sigma}=\frac{\zeta_{n}^{\sigma}\zeta_{n-1}^{\sigma}\beta_{n-1}}{\alpha_n\beta_n(\zeta_{n-1}^{\sigma}-\zeta_{n}^{\sigma})+\zeta_{n-1}^{\sigma}\beta_{n-1}(1-\sigma\beta_n)}.\label{cg7}
\end{eqnarray}
\begin{eqnarray}
\beta_n^{\sigma}=\beta_n\frac{\zeta_{n+1}^{\sigma}}{\zeta_n^{\sigma}}.\label{cg6}
\end{eqnarray}
\begin{eqnarray}
\vec{x}_{n+1}^{\sigma}=\vec{x}_n^{\sigma}-\beta_{n}^{\sigma}\vec{p}_{n}^{\sigma}.\label{cg9}
\end{eqnarray}
\begin{eqnarray}
\vec{r}_{n+1}^{\sigma}=\zeta_{n+1}^{\sigma}\vec{r}_{n+1}.\label{cg10}
\end{eqnarray}
\begin{eqnarray}
\alpha_{n+1}^{\sigma}=\alpha_{n+1}\frac{\zeta_{n+1}^{\sigma}\beta_{n}^{\sigma}}{\zeta_{n}^{\sigma}\beta_{n}}.\label{cg8}
\end{eqnarray}

\begin{eqnarray}
\vec{p}_{n+1}^{\sigma}=\vec{r}_{n+1}^{\sigma}+\alpha_{n+1}^{\sigma}\vec{p}_{n}^{\sigma}.\label{cg11}
\end{eqnarray}

Remark how the residues are generated directly from the residues of the no-sigma problem.
\item The above procedure continues as long as $|\vec{r}|\ge \epsilon$ where $\epsilon$ is some tolerance, otherwise stop. Thus
\begin{eqnarray}
|\vec{r}|\ge \epsilon~,~{\rm continue}.\label{cg12}
\end{eqnarray}
\end{enumerate}
We finally note that in the case of a hermitian matrix, i.e. $A^+=A$, we must replace in the above formulas the transpose by hermitian conjugation. For example, we replace $\vec{p}_n^TA\vec{p}_n$ by $\vec{p}_n^+A\vec{p}$. The rest remains unchanged.

\chapter{Monte Carlo Simulation of Fermion Determinants}
 As it is well known, simulation of fermion determinants and Pfaffians  is crucial to lattice QCD, but as it trurns out, it is also crucial to all supersymmetric matrix models and quantum mechanical matrix models encountered or needed in matrix field theory, matrix/fuzzy geometry and matrix formulation of noncommutative geometry, supersymmetry and strings. As done before in  this part of the book, the theoretical background will be kept to a minimum, otherwise we will stray too far afield,  and we will mostly focus on practical problems.  The main reference for this chapter is \cite{Ambjorn:2000bf,Ambjorn:2000dx}. See also \cite{Anagnostopoulos:2005cy,Anagnostopoulos:2013xga}. For some subtle details of the rational hybrid Monte Carlo algorithm see \cite{Kennedy:1998cu,Clark:2003na,Clark:2005sq,Clark:2006wq}.
\section{The Dirac Operator}

The basic problem we want to solve in this section is to simulate the partition function of ${\cal N}=1$ supersymmetric Yang-Mills matrix model in $d=4$ dimensions given by

\begin{eqnarray}
Z_{\rm YM}=\int \prod_{\mu=1}^4X_{\mu}~d\bar{\theta}d\theta~\exp\bigg(\bar{\theta}\big(i[X_4,..]+\sigma_a[X_a,..]+\xi\big)\theta\bigg) \exp(-S_{\rm BYM}[X]).
\end{eqnarray}

\begin{eqnarray}
S_{\rm BYM}&=&-\frac{N\gamma}{4}\sum_{\mu,\nu=1}^4Tr[X_{\mu},X_{\nu}]^2.
\end{eqnarray}
The parameter $\gamma$ will be set to one and we may add to the bosonic Yang-Mills action a Chern-Simons term and a harmonic oscillator term with parameters $\alpha$ and $m^2$ respectively. The spinors $\bar{\theta}$ and $\theta$ are two independent complex two-component Weyl spinors. They contain the same number of degrees of Freedom as the four-component real Majorana spinors in four dimensions. The scalar curvature or fermion mass parameter is given by $\xi$. The above theory is only supersymmetric for a restricted set of values of the parameters $\gamma$, $\alpha$,  $m^2$ and $\xi$. See \cite{Ydri:2012bq} and references therein for a discussion of this matter.

We have considered above the Dirac operator given by
\begin{eqnarray}
{\cal D}=iX_4-iX_4^R+{\sigma}_aX_a-{\sigma}_aX_a^R+ \xi.
\end{eqnarray}
The determinant of this Dirac operator is positive definite since the eigenvalues come in complex conjugate pairs \cite{Ambjorn:2000bf}. In $d=6$ and $d=10$ the determinant is, however, complex valued which presents a serious obstacle to numerical evaluation. In these three cases, i.e. for $d=4,6,10$, the supersymmetric path integral is well behaved. In $d=3$ the supersymmetric path integral is ill defined and only the bosonic "quenched" approximation makes sense. The source of the divergence lies in the so-called flat directions, i.e. the set of commuting matrices. See \cite{Austing:2001ib} and references therein.

It is possible to rewrite the Dirac action in the following form (with $X_{34}=X_3+iX_4$ and $X_{\pm}=X_1\pm iX_2$)
\begin{eqnarray}
Tr\bar{\theta}{\cal D}\theta&=&Tr\bigg[\bar{\theta}_1(X_{34} +\xi){\theta}_1+\bar{\theta}_1X_{-}{\theta}_2+\bar{\theta}_2X_{+}{\theta}_1+\bar{\theta}_2(-X_{34}^++\xi){\theta}_2\bigg]\nonumber\\
&-&Tr\bigg[X_{34}\bar{\theta}_1{\theta}_1+X_-\bar{\theta}_1{\theta}_2+X_+\bar{\theta}_2{\theta}_1-X_{34}^+\bar{\theta}_2{\theta}_2\bigg].
\end{eqnarray}
We expand the $N\times N$ matrices ${\theta}_1,{\theta}_2$ and $\bar{\theta}_1,\bar{\theta}_2$ as 
\begin{eqnarray}
{\theta}_{\alpha}=\sum_{A=1}^{N^2}{\theta}_{\alpha}^AT^A~,~\bar{\theta}_{\alpha}=\sum_{\bar{A}=1}^{N^2}\bar{\theta}_{\alpha}^{{A}}T^{{A}}.
\end{eqnarray}
The $N\times N$ matrices $T^A$ are defined by 
\begin{eqnarray}
(T^A)_{ij}={\delta}_{ii_A}{\delta}_{jj_A}~,~A=N(i_A-1)+j_A.
\end{eqnarray}
Then we find that
\begin{eqnarray}
Tr\bar{\theta}{\cal D}\theta = \bar{\chi}_1{\cal M}_{11}{\chi}_1+ \bar{\chi}_1{\cal M}_{12}{\chi}_2+ \bar{\chi}_2{\cal M}_{21}{\chi}_2+ \bar{\chi}_2{\cal M}_{22}{\chi}_2.
\end{eqnarray} 
The $N^2-$dimensional vectors ${\chi}_1$, ${\chi}_2$ and $\bar{\chi}_1$, $\bar{\chi}_2$ are defined by $({\chi}_{\alpha})_A={\theta}_{\alpha}^A$ and  $(\bar{\chi}_{\alpha})_A=\bar{\theta}_{\alpha}^{{A}}$. The matrices ${\cal M}_{\alpha\beta}^{{A}B}$ are $N^2\times N^2$ defined by
\begin{eqnarray}
&&({\cal M}_{11})^{{A}B}=Tr T^{{A}}(X_{34} +\xi)T^B-TrX_{34}T^{{A}}T^B.
\end{eqnarray}
\begin{eqnarray}
&&({\cal M}_{12})^{{A}B}=TrT^{{A}}X_{-}T^B-TrX_- T^{{A}}T^B.
\end{eqnarray}
\begin{eqnarray}
&&({\cal M}_{21})^{{A}B}=Tr T^{{A}}X_+T^B-TrX_+T^{{A}}T^B.
\end{eqnarray}
\begin{eqnarray}
&&({\cal M}_{22})^{{A}B}=Tr T^{{A}}(-X_{34}^+ +\xi)T^B+TrX_{34}^+T^{{A}}T^B.
\end{eqnarray}
We remark that
\begin{eqnarray}
&&Tr T^{{A}}XT^B-TrX T^{{A}}T^B=X_{j_Ai_B}{\delta}_{i_Aj_B}-X_{j_Bi_A}{\delta}_{j_Ai_B}.
\end{eqnarray}
\begin{eqnarray}
&&Tr(T^{{A}})^+T^B={\delta}_{i_Ai_B}{\delta}_{j_Aj_B}=\delta_{{A}B}~,~Tr T^{{A}}T^B={\delta}_{j_Ai_B}{\delta}_{j_Bi_A}=\delta_{\bar{A}B}.
\end{eqnarray}
In the above two equations $\bar{A}$ and $B$ are such that
\begin{eqnarray}
\bar{A}=N(j_A-1)+i_A~,~B=N(i_B-1)+j_B.
\end{eqnarray}
In summary, the Dirac operator in terms of the $2N^2-$dimensional vectors ${\chi}$ and $\bar{\chi}$ becomes
\begin{eqnarray}
Tr\bar{\theta}{\cal D}\theta = \bar{\chi}{\cal M}{\chi}.
\end{eqnarray} 
Next, we observe that the trace parts of the matrices $X_a$ drop from the partition function. Thus the measure should read $\int dX_a{\delta}(TrX_a)$ instead of simply $\int dX_a$.  Similarly, we observe that if we write  $\theta={\theta}_0+\eta {\bf 1}$, then the trace part $\eta$ will decouple from the rest since

\begin{eqnarray}
Tr\bar{\theta}\bigg(i[X_4,..]+{\sigma}_a[{X}_a,..]+\xi\bigg)\theta=Tr\bar{\theta}_0\bigg(i[X_4,..]+{\sigma}_a[{X}_a,..]+\xi\bigg){\theta}_0+\xi \bar{\eta}\eta.
\end{eqnarray}
Hence, the constant fermion modes ${\eta}_{\alpha}$ can also be integrated out from the partition function and thus we should consider the measure $\int d{\theta}d\bar{\theta}{\delta}(Tr{\theta}_{\alpha}){\delta}(Tr\bar {\theta}_{\alpha})$ instead of  $\int d{\theta}d\bar{\theta}$. These facts should be taken into account in the numerical study. We are thus led to consider the partition function
\begin{eqnarray}
&&Z_{\rm YM}=\int \prod_{\mu=1}^4dX_{\mu}~{\delta}(TrX_{\mu})~ \det {\cal D} ~\exp\big(-S_{\rm BYM}[X]\big).\label{PIbasic}
\end{eqnarray}
The determinant is given by 
\begin{eqnarray}
{\rm det}{\cal D}&=&\int d{\theta}d\bar {\theta}{\delta}(Tr{\theta}_{\alpha}){\delta}(Tr\bar{\theta}_{\alpha})\exp\big(Tr \bar{\theta}{\cal D}{\theta}\big)\nonumber\\&=&\int d{\chi}d\bar{\chi}{\delta}\bigg(\sum_{A=1}^{N^2}({\chi}_{\alpha})_A{\delta}_{i_Aj_A}\bigg){\delta}\bigg(\sum_{A=1}^{N^2}(\bar{\chi}_{\alpha})_A{\delta}_{i_Aj_A}\bigg)\exp\big(\bar{\chi}{{\cal M}}{\chi}\big)\nonumber\\
&=&\int d{\chi}^{'}d\bar{\chi}^{'}\exp\big( \bar{\chi}^{'}{{\cal M}^{'}}{\chi}^{'}\big).
\end{eqnarray}
The vectors ${\chi}_{\alpha}^{'}$, $\bar{\chi}_{\alpha}^{'}$ are $(N^2-1)-$dimensional. The matrix ${\cal M}^{'}$ is  $2(N^2-1)\times 2(N^2-1)$ dimensional, and it is given by 
\begin{eqnarray}
{\cal M}_{\alpha \beta}^{'A^{'}B^{'}}={\cal M}_{\alpha \beta}^{A^{'}B^{'}}-{\cal M}_{\alpha \beta}^{N^2B^{'}}{\delta}_{i_{A^{'}}j_{A^{'}}}-{\cal M}_{\alpha \beta}^{A^{'}N^2}{\delta}_{i_{B^{'}}j_{B^{'}}}+{\cal M}_{\alpha \beta}^{N^2N^2}{\delta}_{i_{A^{'}}j_{A^{'}}}{\delta}_{i_{B^{'}}j_{B^{'}}}.\label{21}
\end{eqnarray}
We remark that
\begin{eqnarray}
{\cal M}_{\alpha \beta}^{N^2N^2}=\xi {\delta}_{\alpha \beta}.
\end{eqnarray}
Thus we must have
\begin{eqnarray}
\det {\cal D}=\det {{\cal M}^{'}}.
\end{eqnarray}
The partition function thus reads 
\begin{eqnarray}
&&Z_{\rm YM}=\int \prod_{\mu=1}^4dX_{\mu}~{\delta}(TrX_{\mu}) ~\exp\big(-S_{\rm YM}[X]\big).
\end{eqnarray}
\begin{eqnarray}
S_{\rm YM}[X]=S_{\rm BYM}[X]+V[X]~,~V=-\ln\det{{\cal M}^{'}}.\label{susyym}
\end{eqnarray}
We will need
 \begin{eqnarray}
\frac{\partial S_{\rm BYM}}{\partial (X_{\mu})_{ij}(t)}&=&-N\gamma\sum_{\nu=1}^4[X_{\nu},[X_{\mu},X_{\nu}]]_{ji}\nonumber\\
&=&-N\gamma\bigg(2X_{\nu}X_{\mu}X_{\nu}-X_{\nu}^2X_{\mu}-X_{\mu}X_{\nu}^2\bigg)_{ji}.\label{Bforce}
\end{eqnarray}
The determinant is real positive definite since the eigenvalues are paired up. Thus, we can introduce the positive definite operator ${\Delta}$ by
\begin{eqnarray}
{\Delta}=({\cal M}^{'})^+{\cal M}^{'}.
\end{eqnarray}
The action $V$ can be rewritten as
\begin{eqnarray}
V=-\frac{1}{2}\ln\det{\Delta}.
\end{eqnarray}
The leap-frog algorithm for this problem is given by


\begin{eqnarray}
(P_{\mu})_{ij}(n+\frac{1}{2})=(P_{\mu})_{ij}(n)-\frac{\delta t}{2}\bigg[\frac{\partial S_{\rm BYM}}{\partial (X_{\mu})_{ij}}(n)+(V_{\mu})_{ij}(n)\bigg].\label{lf1}
\end{eqnarray}

\begin{eqnarray}
(X_{\mu})_{ij}(n+1)=(X_{\mu})_{ij}(n)+\delta t (P_{\mu})_{ji}(n+\frac{1}{2}).\label{lf2}
\end{eqnarray}
 \begin{eqnarray}
(P_{\mu})_{ij}(n+1)=(P_{\mu})_{ij}(n+\frac{1}{2})-\frac{\delta t}{2}\bigg[\frac{\partial S_{\rm BYM}}{\partial (X_{\mu})_{ij}}(n+1)+(V_{\mu})_{ij}(n+1)\bigg].\label{lf3}
\end{eqnarray}
The effect of the determinant is encoded in the matrix
 \begin{eqnarray}
(V_{\mu})_{ij}&=&\frac{\partial V}{\partial (X_{\mu})_{ij}}\nonumber\\
&=&-\frac{1}{2}Tr_{\rm ad}{\Delta}^{-1}\frac{\partial {\Delta}}{\partial  (X_{\mu})_{ij}}.\label{susyym1}
\end{eqnarray}
From (\ref{susyym}) and (\ref{susyym1}) we see that we must compute the inverse and the determinant of the Dirac operator at each hybrid Monte Carlo step. However, the Dirac operator is an ${\cal N}\times {\cal N}$ matrix where ${\cal N}=2N^2-2$.  This is proportional to the number of degrees of freedom. Since the computation of the determinant requires $O({\cal N}^3)$ operations at best, through Gaussian elimination, we see that the computational effort of the  above algorithm will be $O(N^6)$. Recall that the computational effort of the bosonic theory is $O(N^3)$\footnote{Compare also with field theory in which the number of degrees of freedom is proportional to the volume, the computational effort of the bosonic theory is $O(V)$ while that of the full theory, which includes a determinant,  is $O(V^2)$.} . 

\section{Pseudo-Fermions and Rational Approximations}
We introduce pseudo-fermions in the usual way as follows. The determinant can be rewritten in the form

\begin{eqnarray}
\det {\cal D}=\det {\cal M}^{'}&=&(\det {\Delta})^{\frac{1}{2}}\nonumber\\
&=&\int d\phi^+ d\phi~\exp(-  \phi^+\Delta^{-1/2}\phi).
\end{eqnarray}
Since ${\cal D}$, ${\cal M}{'}$ and $\Delta$ are ${\cal N}\times {\cal N}$ matrices organized as $2\times 2$ matrices, with components given by $\hat{\cal N}\times \hat{\cal N}$ matrices where $\hat{\cal N}={\cal N}/2$, the vectors $\phi^+$ and $\phi$ can be thought of as two-component spinors where each component is given by an $\hat{\cal N}-$dimensional vector. We will write
\begin{eqnarray}
\phi=\left(\begin{array}{c}
\phi_1\\
\phi_2
\end{array}\right)~,~\phi^+=\left(\begin{array}{c}
\phi_1^+~ \phi_2^+\\
\end{array}\right).
\end{eqnarray}
These are precisely the pseudo-fermions. They are complex-valued instead of Grassmann-valued degrees of freedom, and that is why they are pseudo-fermions, with a positive definite Laplacian and thus they can be sampled in Monte Carlo simulations in the usual way. 

Furthermore, we will use the so-called rational approximation, which is why the resulting hybrid Monte Carlo is termed rational, which allows us to write
\begin{eqnarray}
(\det {\Delta})^{\frac{1}{2}}
&=&\int d\phi^+ d\phi~\exp(-  \phi^+ r^2(\Delta) \phi).
\end{eqnarray}
The rational approximation $r(x)$ is given by

\begin{eqnarray}
x^{-1/4}\simeq r(x)=a_0+\sum_{\sigma=1}^M\frac{a_{\sigma}}{x+b_{\sigma}}.
\end{eqnarray}
The parameters $a_0$, $a_{\sigma}$, $b_{\sigma}$ and $M$ are real positive numbers which can be optimized for any strictly positive range such as $\epsilon \leq x\leq 1$. This point was discussed at great length previously.

Thus the pseudo-fermions are given by a heatbath, viz
\begin{eqnarray}
\phi=r^{-1}(\Delta)\xi,
\end{eqnarray}
where $\xi$ is given by the Gaussian noise $P(\xi)=\exp(- \xi^{+}\xi)$. We write
\begin{eqnarray}
\phi=\bigg(c_0+\sum_{\sigma=1}^M\frac{c_{\sigma}}{\Delta+d_{\sigma}}\bigg)\xi.\label{form1}
\end{eqnarray}
By using a different rational approximation $\bar{r}(x)$, in order to avoid double inversion (see below), we rewrite the original path integral in the form
\begin{eqnarray}
Z_{\rm YM}
&=&\int \prod_{\mu=1}^4dX_{\mu} \int d\phi^+ d\phi~{\delta}(TrX_{\mu})~\exp\big(-S_{\rm BYM}[X]\big)~\exp(-  \phi^+ \bar{r}(\Delta) \phi).
\end{eqnarray}
The new rational approximation is defined by
\begin{eqnarray}
x^{-1/2}\simeq \bar{r}(x)=a_0+\sum_{\sigma=1}^M\frac{a_{\sigma}}{x+b_{\sigma}}.
\end{eqnarray}
The full action becomes
\begin{eqnarray}
S_{\rm YM}&=&S_{\rm BYM}[X]+V[X].
\end{eqnarray}
The potential is given in this case by
\begin{eqnarray}
V&=&  \phi^+\bar{r}(\Delta)\phi\nonumber\\
&=&a_0 \phi^+\phi+\sum_{\sigma=1}^M a_{\sigma}   \phi^+(\Delta+b_{\sigma})^{-1}\phi\nonumber\\
&=&a_0  \phi^+\phi+\sum_{\sigma=1}^M a_{\sigma}   \phi^+G_{\sigma}=a_0 \phi^+_{\alpha}\phi_{\alpha}+\sum_{\sigma=1}^M a_{\sigma}   \phi^+_{\alpha}G_{\sigma\alpha}\nonumber\\
&=&a_0  \phi^+\phi+\sum_{\sigma=1}^M a_{\sigma}   G_{\sigma}^+\phi=a_0  \phi^+_{\alpha}\phi_{\alpha}+\sum_{\sigma=1}^M a_{\sigma}   G^+_{\sigma\alpha}\phi_{\alpha}.
\end{eqnarray}
This can be rewritten compactly as
\begin{eqnarray}
V&=&W_{\alpha}\phi_{\alpha}~,~W_{\alpha}=a_0(\phi^{*}_{\alpha})_A+\sum_{\sigma=1}^Ma_{\sigma} (G^*_{\sigma\alpha})_A.
\end{eqnarray}
The vectors (pseudo-fermions) $G_{\sigma}$ are defined by
\begin{eqnarray}
G_{\sigma}=(\Delta+b_{\sigma})^{-1}\phi.\label{form31}
\end{eqnarray}
We introduce a fictitious time parameter $t$ and a Hamiltonian $H$ given by
\begin{eqnarray}
H&=&\frac{1}{2}TrP_{\mu}^2+ Q^+Q+S_{\rm YM}\nonumber\\
&=&\frac{1}{2}TrP_{\mu}^2+ Q^+_{\alpha}Q_{\alpha}+S_{\rm YM}.
\end{eqnarray}
The equation of motion associated with the matrix $\phi$ is given by
\begin{eqnarray}
-(\dot{Q}_{\alpha})_{A}&=&\frac{\partial H}{\partial (\phi_{\alpha})_{A}}\nonumber\\
&=&\frac{\partial V}{\partial (\phi_{\alpha})_{A}}\nonumber\\
&=&a_0(\phi^{*}_{\alpha})_A+\sum_{\sigma=1}^Ma_{\sigma} (G^*_{\sigma\alpha})_A\nonumber\\
&\equiv &(W_{\alpha})_{A}.\label{form32}
\end{eqnarray}
\begin{eqnarray}
(\dot{\phi}_{\alpha})_{A}&=&\frac{\partial H}{\partial (Q_{\alpha})_{A}}\nonumber\\
&\equiv &(Q^*_{\alpha})_{A}.
\end{eqnarray}
This last equation is equivalent to
\begin{eqnarray}
(\dot{\phi}_{\alpha}^*)_{A}
&\equiv &(Q_{\alpha})_{A}.
\end{eqnarray}
The leap-frog algorithm for this part of the problem is given by
\begin{eqnarray}
(Q_{\alpha})_{A}(n+\frac{1}{2})=(Q_{\alpha})_{A}(n)-\frac{\delta t}{2}(W_{\alpha})_{A}(n).\label{form21}
\end{eqnarray}

\begin{eqnarray}
(\phi_{\alpha})_{A}(n+1)=(\phi_{\alpha})_{A}(n)+\delta t (Q_{\alpha}^*)_{A}(n+\frac{1}{2}).\label{form22}
\end{eqnarray}
 \begin{eqnarray}
(Q_{\alpha})_{A}(n+1)=(Q_{\alpha})_{A}(n+\frac{1}{2})-\frac{\delta t}{2}(W_{\alpha})_{A}(n+1).\label{form23}
\end{eqnarray}
The first set of equations of motion associated with the matrices $X_{\mu}$ are given by
\begin{eqnarray}
-(\dot{P}_{\mu})_{ij}&=&\frac{\partial H}{\partial (X_{\mu})_{ij}}\nonumber\\
&=&\frac{\partial S_{\rm BYM}}{\partial (X_{\mu})_{ij}}+\frac{\partial V}{\partial (X_{\mu})_{ij}}\nonumber\\
&=&\frac{\partial S_{\rm BYM}}{\partial (X_{\mu})_{ij}}-\sum_{\sigma=1}^M  a_{\sigma} G_{\sigma\alpha}^+\frac{\partial \Delta_{\alpha\beta}}{\partial (X_{\mu})_{ij}} G_{\sigma\beta}.
\end{eqnarray}
The effect of the determinant is now encoded in the matrix (the force)
 \begin{eqnarray}
(V_{\mu})_{ij}=-\sum_{\sigma=1}^M  a_{\sigma} G_{\sigma\alpha}^+\frac{\partial \Delta_{\alpha\beta}}{\partial (X_{\mu})_{ij}} G_{\sigma\beta}.\label{Fforce}
\end{eqnarray}
The second set of equations associated with the matrices $X_{\mu}$ are given by
\begin{eqnarray}
(\dot{X}_{\mu})_{ij}&=&\frac{\partial H}{\partial (P_{\mu})_{ij}}\nonumber\\
&=&(P_{\mu})_{ji}.
\end{eqnarray}
The leap-frog algorithm for this part of the problem is given by the equations (\ref{lf1}), (\ref{lf2}) and (\ref{lf3}) with the appropriate re-interpretation of the meaning of $(V_{\mu})_{ij}$.
\section{More on The Conjugate-Gradient}
\subsection{Multiplication by ${\cal M}^{'}$ and $({\cal M}^{'})^+$}
Typically we will need to find $x^{'}$, given  $v$, which solves the linear system
\begin{eqnarray}
(\Delta +b)x^{'}=v.
\end{eqnarray}
We will use the conjugate gradient method to do this. The product $\Delta x^{'}$ involves the products ${\cal M}^{'}x^{'}$ and $({\cal M}^{'})^+y^{'}$, viz
\begin{eqnarray}
y^{'}&=&{\cal M}^{'}x^{'}~\leftrightarrow (y^{'}_{\alpha})_{A^{'}}={\cal M}_{\alpha\beta}^{'A^{'}B^{'}}(x^{'}_{\beta})_{B^{'}}.\label{54m}
\end{eqnarray}
\begin{eqnarray}
z^{'}&=&({\cal M}^{'})^+y^{'}~\leftrightarrow (z^{'}_{\alpha})_{A^{'}}=({\cal M}_{\beta\alpha}^{'*})^{B^{'}A^{'}}(y^{'}_{\beta})_{B^{'}}.\label{56m}
\end{eqnarray}
\paragraph{Multiplication by ${\cal M}^{'}$:} By using (\ref{21}) we have
\begin{eqnarray}
(y^{'}_{\alpha})_{A^{'}}&=&{\cal M}_{\alpha \beta}^{'A^{'}B^{'}}(x^{'}_{\beta})_{B^{'}}\nonumber\\
&=&{\cal M}_{\alpha \beta}^{A^{'}B^{'}}(x^{'}_{\beta})_{B^{'}}-{\cal M}_{\alpha \beta}^{N^2B^{'}}{\delta}_{i_{A^{'}}j_{A^{'}}}(x^{'}_{\beta})_{B^{'}}-{\cal M}_{\alpha \beta}^{A^{'}N^2}{\delta}_{i_{B^{'}}j_{B^{'}}}(x^{'}_{\beta})_{B^{'}}+{\cal M}_{\alpha \beta}^{N^2N^2}{\delta}_{i_{A^{'}}j_{A^{'}}}{\delta}_{i_{B^{'}}j_{B^{'}}}(x^{'}_{\beta})_{B^{'}}.\label{55}\nonumber\\
\end{eqnarray}
Recall that the primed indices run from $1$ to $N^2-1$ while unprimed indices run from  $1$ to $N^2$. We introduce then
\begin{eqnarray}
(y_{\alpha})_{A^{}}&=&{\cal M}_{\alpha \beta}^{A^{}B^{}}(x_{\beta})_{B^{}}\nonumber\\
&=&{\cal M}_{\alpha \beta}^{A^{}B^{'}}(x_{\beta})_{B^{'}}+{\cal M}_{\alpha \beta}^{A^{}N^{2}}(x_{\beta})_{N^{2}}.
\end{eqnarray}
We define
\begin{eqnarray}
(x_{\beta})_{B^{'}}=(x^{'}_{\beta})_{B^{'}}~,~(x_{\beta})_{N^{2}}=-(x^{'}_{\beta})_{B^{'}} {\delta}_{i_{B^{'}}j_{B^{'}}}.\label{59m}
 \end{eqnarray}
Thus
\begin{eqnarray}
(y_{\alpha})_{A^{}}
&=&{\cal M}_{\alpha \beta}^{A^{}B^{'}}(x^{'}_{\beta})_{B^{'}}-{\cal M}_{\alpha \beta}^{A^{}N^{2}}(x^{'}_{\beta})_{B^{'}} {\delta}_{i_{B^{'}}j_{B^{'}}}.
\end{eqnarray}
The next definition is obviously then
\begin{eqnarray}
(y^{'}_{\alpha})_{A^{'}}=(y^{}_{\alpha})_{A^{'}}-(y^{}_{\alpha})_{N^{2}}{\delta}_{i_{A^{'}}j_{A^{'}}}.\label{61m}
\end{eqnarray}
This leads immediately to
\begin{eqnarray}
(y^{'}_{\alpha})_{A^{'}}={\cal M}_{\alpha \beta}^{A^{'}B^{'}}(x^{'}_{\beta})_{B^{'}}-{\cal M}_{\alpha \beta}^{A^{'}N^{2}}(x^{'}_{\beta})_{B^{'}} {\delta}_{i_{B^{'}}j_{B^{'}}}-{\cal M}_{\alpha \beta}^{N^{2}B^{'}}(x^{'}_{\beta})_{B^{'}}+{\cal M}_{\alpha \beta}^{N^{2}N^{2}}(x^{'}_{\beta})_{B^{'}} {\delta}_{i_{B^{'}}j_{B^{'}}}.
\end{eqnarray}
This is precisely (\ref{55}).

Next we introduce the $N\times N$ matrices $\hat{x}_{\alpha}$, $\hat{y}_{\alpha}$ associated with the vectors $x_{\alpha}$ and $y_{\alpha}$ by the relations
\begin{eqnarray}
\hat{x}_{\alpha}=\sum_{A=1}^{N^2}(x_{\alpha})_AT^A~,~\hat{y}_{\alpha}=\sum_{A=1}^{N^2}(y_{\alpha})_AT^A.
\end{eqnarray}
Thus
\begin{eqnarray}
(x_{\alpha})_{\bar{A}}=Tr \hat{x}_{\alpha}T^{{A}}=(\hat{x}_{\alpha})_{j_Ai_A}~,~(y_{\alpha})_{\bar{A}}=Tr \hat{y}_{\alpha}T^{{A}}=(\hat{y}_{\alpha})_{j_Ai_A}.
\end{eqnarray}
And
\begin{eqnarray}
(x_{\alpha})_{{A}}=Tr \hat{x}_{\alpha}(T^{{A}})^{+}=(\hat{x}_{\alpha})_{i_Aj_A}~,~(y_{\alpha})_{{A}}=Tr \hat{y}_{\alpha}(T^{{A}})^{+}=(\hat{y}_{\alpha})_{i_Aj_A}.\label{65m}
\end{eqnarray}
We verify that
\begin{eqnarray}
{\cal M}_{\alpha\beta}^{{A}B}(x_{\beta})_B=Tr T^{{A}}({\cal D}\hat{x})_{\alpha}.
\end{eqnarray}
By comparing with
\begin{eqnarray}
(y_{\alpha})_A=Tr T^A(\hat{y})_{\alpha},
\end{eqnarray}
we get
\begin{eqnarray}
\hat{y}^T={\cal D}\hat{x}.
\end{eqnarray}
We recall the Dirac operator 
\begin{eqnarray}
{\cal D}=\left(\begin{array}{cc}
 X_{34}-X_{34}^{R}+\xi & X_{-}-X_-^{R}\\
X_{+}-X_+^{R} &  - X_{34}^++(X_{34}^{R})^++\xi
\end{array}\right).
\end{eqnarray}
Thus $\hat{y}^T={\cal D}\hat{x}$ is equivalent to
\begin{eqnarray}
(\hat{y}_1)_{ij}=({\cal D}_{1\alpha}\hat{x}_{\alpha})_{ji}=[X_{34},\hat{x}_1]_{ji}+[X_-,\hat{x}_2]_{ji}+\xi (\hat{x}_1)_{ji}.\label{70m}
\end{eqnarray}
\begin{eqnarray}
(\hat{y}_2)_{ij}=({\cal D}_{2\alpha}\hat{x}_{\alpha})_{ji}=-[X_{34}^+,\hat{x}_2]_{ji}+[X_+,\hat{x}_1]_{ji}+\xi (\hat{x}_2)_{ji}.\label{71m}
\end{eqnarray}
For completeness we remark  
\begin{eqnarray}
(y_{\alpha})_A^{*}{\cal M}_{\alpha\beta}^{AB}(x_{\beta})_B=Tr \hat{y}^*_{\alpha}({\cal D}\hat{x})_{\alpha}.
\end{eqnarray}
\paragraph{Multiplication by $({\cal M}^{'})^+$:}  As before the calculation of 
\begin{eqnarray}
(z^{'}_{\alpha})_{A^{'}}=({\cal M}_{\beta\alpha}^{'*})^{B^{'}A^{'}}(y^{'}_{\beta})_{B^{'}}
\end{eqnarray}
can be reduced to the calculation of
\begin{eqnarray}
(z^{}_{\alpha})_{A^{}}=({\cal M}_{\beta\alpha}^{*})^{B^{}A^{}}(y^{}_{\beta})_{B^{}},
\end{eqnarray}
with the definitions
\begin{eqnarray}
(y_{\beta})_{B^{'}}=(y^{'}_{\beta})_{B^{'}}~,~(y_{\beta})_{N^{2}}=-(y^{'}_{\beta})_{B^{'}} {\delta}_{i_{B^{'}}j_{B^{'}}}.
 \end{eqnarray}
\begin{eqnarray}
(z^{'}_{\alpha})_{A^{'}}=(z^{}_{\alpha})_{A^{'}}-(z^{}_{\alpha})_{N^{2}}{\delta}_{i_{A^{'}}j_{A^{'}}}.
\end{eqnarray}
The next step is to note that
\begin{eqnarray}
{\cal M}_{\beta\alpha}^{*B{A}}(y_{\beta})_B=Tr T^{{A}}({\cal D}^+\hat{y})_{\alpha}.
\end{eqnarray}
The hermitian conjugate of the Dirac operator is defined by the relation
\begin{eqnarray}
{\cal D}^+=-\left(\begin{array}{cc}
 X_{34}^*-(X_{34}^{R})^*+\xi & X_{+}^*-(X_+^{R})^*\\
X_{-}^*-(X_-^{R})^* &  - X_{34}^T+(X_{34}^{R})^T+\xi
\end{array}\right).
\end{eqnarray}
Hence
\begin{eqnarray}
\hat{z}^T={\cal D}^+\hat{y}.
\end{eqnarray}
Equivalently 
\begin{eqnarray}
(\hat{z}_1)_{ij}=({\cal D}^+_{1\alpha}\hat{y}_{\alpha})_{ji}=-[X_{34}^*,\hat{y}_1]_{ji}-[X_+^*,\hat{y}_2]_{ji}+\xi (\hat{y}_1)_{ji}.\label{80m}
\end{eqnarray}
\begin{eqnarray}
(\hat{z}_2)_{ij}=({\cal D}^+_{2\alpha}\hat{y}_{\alpha})_{ji}=[X_{34}^T,\hat{y}_2]_{ji}-[X_-^*,\hat{y}_1]_{ji}+\xi (\hat{y}_2)_{ji}.\label{81m}
\end{eqnarray}
\subsection{The Fermionic Force}
Also we will need to compute explicitly in the molecular dynamics part the fermionic force (with $ ({\cal M}^{'+})_{\alpha\beta}= ({\cal M}^{'}_{\beta\alpha})^{+}$)
 \begin{eqnarray}
(V_{\mu})_{ij}&=&-\sum_{\sigma=1}^M  a_{\sigma} G_{\sigma\alpha}^+\frac{\partial \Delta_{\alpha\beta}}{\partial (X_{\mu})_{ij}} G_{\sigma\beta}\nonumber\\
&=&-\sum_{\sigma=1}^M  a_{\sigma} G_{\sigma\alpha}^+\frac{\partial ({\cal M}^{'}_{\beta\alpha})^{+}}{\partial (X_{\mu})_{ij}} F_{\sigma\beta}-\sum_{\sigma=1}^M  a_{\sigma} F_{\sigma\beta}^+\frac{\partial {\cal M}^{'}_{\beta\alpha}}{\partial (X_{\mu})_{ij}} G_{\sigma\alpha}\nonumber\\
&=&-\sum_{\sigma=1}^Ma_{\sigma}\bigg(  F_{\sigma\beta}^+\frac{\partial {\cal M}^{'}_{\beta\alpha}}{\partial (X_{\mu})^{*}_{ij}} G_{\sigma\alpha}\bigg)^{*}-\sum_{\sigma=1}^M  a_{\sigma} F_{\sigma\beta}^+\frac{\partial {\cal M}^{'}_{\beta\alpha}}{\partial (X_{\mu})_{ij}} G_{\sigma\alpha}.
\end{eqnarray}
The vectors $F_{\sigma\alpha}$ and  $F^{+}_{\sigma\alpha}$ are defined by
\begin{eqnarray}
F_{\sigma\alpha}={\cal M}^{'}_{\alpha\beta}G_{\sigma\beta}~,~F_{\sigma\alpha}^{+}=G_{\sigma\beta}^{+}({\cal M}^{'}_{\alpha\beta})^{+}.
\end{eqnarray}
We can expand the bosonic matrices $X_{\mu}$ similarly to the fermionic matrices as
\begin{eqnarray}
X_{\mu}=\sum_{A=1}^{N^2}X_{\mu}^AT^A.
\end{eqnarray}
Equivalently
\begin{eqnarray}
(X_{\mu})_{i_Aj_A}=X_{\mu}^A~,~A=N(i_A-1)+j_A.
\end{eqnarray}
Reality of the bosonic matrices gives
\begin{eqnarray}
(X_{\mu})_{i_Aj_A}^{*}=X_{\mu}^{\bar{A}}=(X_{\mu}^A)^{*}~,~\bar{A}=N(j_A-1)+i_A.
\end{eqnarray}
Hence we have
 \begin{eqnarray}
V_{\mu}^A&\equiv &(V_{\mu})_{i_Aj_A}\nonumber\\
&=&-\sum_{\sigma=1}^Ma_{\sigma}\bigg(  F_{\sigma\beta}^+\frac{\partial {\cal M}^{'}_{\beta\alpha}}{\partial X_{\mu}^{\bar{A}}} G_{\sigma\alpha}\bigg)^{*}-\sum_{\sigma=1}^M  a_{\sigma} F_{\sigma\beta}^+\frac{\partial {\cal M}^{'}_{\beta\alpha}}{\partial X_{\mu}^A } G_{\sigma\alpha}\nonumber\\
&=&-\sum_{\sigma=1}^Ma_{\sigma}\big({\cal T}_{\sigma\mu}^{\bar{A}}\big)^*-\sum_{\sigma=1}^Ma_{\sigma}{\cal T}_{\sigma\mu}^A.\label{Fforce1}
\end{eqnarray}
The definition of ${\cal T}_{\sigma\mu}^A$ is obviously given by
\begin{eqnarray}
{\cal T}_{\sigma\mu}^A=F_{\sigma\beta}^+\frac{\partial {\cal M}^{'}_{\beta\alpha}}{\partial X_{\mu}^A } G_{\sigma\alpha}.
\end{eqnarray}
For simplicity we may denote the derivations with respect to $X_{\mu}^A $ and $X_{\mu}^{\bar{A}} $ by $\partial$ and $\bar{\partial}$ respectively. As before we introduce the vectors in the full Hilbert space:
\begin{eqnarray}
(\tilde{G}_{\sigma\alpha})_{B^{'}}=(G_{\sigma\alpha})_{B^{'}}~,~(\tilde{G}_{\sigma\alpha})_{N^{2}}=-(G_{\sigma\alpha})_{B^{'}} {\delta}_{i_{B^{'}}j_{B^{'}}}.
 \end{eqnarray}
\begin{eqnarray}
(\tilde{F}_{\sigma\alpha})_{B^{'}}=(F_{\sigma\alpha})_{B^{'}}~,~(\tilde{F}_{\sigma\alpha})_{N^{2}}=-(F_{\sigma\alpha})_{B^{'}} {\delta}_{i_{B^{'}}j_{B^{'}}}.
 \end{eqnarray}
 A straightforward calculation gives
 \begin{eqnarray}
(F_{\sigma\beta}^*)_{A^{'}}({\cal M}^{'}_{\beta\alpha})^{A^{'}B^{'}} (G_{\sigma\alpha})_{B^{'}}=(\tilde{F}_{\sigma\beta}^*)_{A^{}} ({\cal M}^{}_{\beta\alpha})^{A^{}B^{}} (\tilde{G}_{\sigma\alpha})_{B^{}}.
\end{eqnarray}

 \begin{eqnarray}
(F_{\sigma\beta}^*)_{A^{'}}\partial ({\cal M}^{'}_{\beta\alpha})^{A^{'}B^{'}} (G_{\sigma\alpha})_{B^{'}}=(\tilde{F}_{\sigma\beta}^*)_{A^{}} \partial ({\cal M}^{}_{\beta\alpha})^{A^{}B^{}} (\tilde{G}_{\sigma\alpha})_{B^{}}.
\end{eqnarray}
Thus
\begin{eqnarray}
{\cal T}_{\sigma\mu}^A=\tilde{F}_{\sigma\beta}^+\frac{\partial {\cal M}^{}_{\beta\alpha}}{\partial X_{\mu}^A } \tilde{G}_{\sigma\alpha}.
\end{eqnarray}
Explicitly we have
\begin{eqnarray}
{\cal T}_{\sigma\mu}^A=(\tilde{F}_{\sigma\beta}^*)_C\frac{\partial {\cal M}^{CD}_{\beta\alpha}}{\partial X_{\mu}^A } (\tilde{G}_{\sigma\alpha})_D.
\end{eqnarray}
We use the result
\begin{eqnarray}
\frac{\partial {\cal M}^{CD}_{\beta\alpha}}{\partial X_{\mu}^A }=Tr \frac{\partial {M}^{}_{\beta\alpha}}{\partial X_{\mu}^A }[T^D,T^C],
\end{eqnarray}
where
\begin{eqnarray}
M_{11}=X_{34}~,~M_{12}=X_{-}~,~M_{21}=X_+~,~M_{22}=-X_{34}^+.
\end{eqnarray}
We also introduce the matrices $\hat{F}$ and $\hat{G}$  given by
\begin{eqnarray}
\hat{F}_{\alpha}=\sum_{A=1}^{N^2}(\tilde{F}_{\alpha})_AT^A~,~\hat{G}_{\alpha}=\sum_{A=1}^{N^2}(\tilde{G}_{\alpha})_AT^A.
\end{eqnarray}
The reverse of these equations is
\begin{eqnarray}
(\tilde{F}_{\alpha})_A=Tr \hat{F}_{\alpha} (T^A)^+~,~(\tilde{G}_{\alpha})_A=Tr \hat{G}_{\alpha} (T^A)^+.
\end{eqnarray}
We use also the identity 
\begin{eqnarray}
\sum_A (T^A)_{ij}(T^A)^+_{kl}~=\delta_{il}\delta_{jk}.
\end{eqnarray}
A direct calculation yields then the fundamental results
\begin{eqnarray}
{\cal T}_{\sigma\mu}^A=Tr \frac{\partial {M}^{}_{\beta\alpha}}{\partial X_{\mu}^A }[\hat{G}_{\sigma\alpha},\hat{F}_{\sigma\beta}^*]~,~{\cal T}_{\sigma\mu}^{\bar{A}}=Tr \frac{\partial {M}^{}_{\beta\alpha}}{\partial X_{\mu}^{\bar{A}} }[\hat{G}_{\sigma\alpha},\hat{F}_{\sigma\beta}^*].
\end{eqnarray}
Explicitly we have
\begin{eqnarray}
{\cal T}_{\sigma 1}^A=[\hat{G}_{\sigma 1},\hat{F}_{\sigma 2}^*]_{j_Ai_A}+[\hat{G}_{\sigma 2},\hat{F}_{\sigma 1}^*]_{j_Ai_A}~,~{\cal T}_{\sigma 1}^{\bar{A}}=[\hat{G}_{\sigma 1},\hat{F}_{\sigma 2}^*]_{i_Aj_A}+[\hat{G}_{\sigma 2},\hat{F}_{\sigma 1}^*]_{i_Aj_A}.
\end{eqnarray}
\begin{eqnarray}
{\cal T}_{\sigma 2}^A=-i [\hat{G}_{\sigma 1},\hat{F}_{\sigma 2}^*]_{j_Ai_A}+i [\hat{G}_{\sigma 2},\hat{F}_{\sigma 1}^*]_{j_Ai_A}~,~{\cal T}_{\sigma 2}^{\bar{A}}=-i [\hat{G}_{\sigma 1},\hat{F}_{\sigma 2}^*]_{i_Aj_A}+i [\hat{G}_{\sigma 2},\hat{F}_{\sigma 1}^*]_{i_Aj_A}.
\end{eqnarray}
\begin{eqnarray}
{\cal T}_{\sigma 3}^A=[\hat{G}_{\sigma 1},\hat{F}_{\sigma 1}^*]_{j_Ai_A}-[\hat{G}_{\sigma 2},\hat{F}_{\sigma 2}^*]_{j_Ai_A}~,~{\cal T}_{\sigma 3}^{\bar{A}}=[\hat{G}_{\sigma 1},\hat{F}_{\sigma 1}^*]_{i_Aj_A}-[\hat{G}_{\sigma 2},\hat{F}_{\sigma 2}^*]_{i_Aj_A}.
\end{eqnarray}
\begin{eqnarray}
{\cal T}_{\sigma 4}^A=i [\hat{G}_{\sigma 1},\hat{F}_{\sigma 1}^*]_{j_Ai_A}+i[\hat{G}_{\sigma 2},\hat{F}_{\sigma 2}^*]_{j_Ai_A}~,~{\cal T}_{\sigma 4}^{\bar{A}}=i [\hat{G}_{\sigma 1},\hat{F}_{\sigma 1}^*]_{i_Aj_A}+i[\hat{G}_{\sigma 2},\hat{F}_{\sigma 2}^*]_{i_Aj_A}.
\end{eqnarray}


\section{The Rational Hybrid Monte Carlo Algorithm}
\subsection{Statement}
In summary the rational hybrid Monte Carlo algorithm in the present setting consists of the following steps:
\begin{enumerate}
\item {\bf Initialization of $X$}: Start $X$ (the fundamental field in the problem) from  a random configuration.
\item {\bf Initialization of Other Fields}:
\begin{itemize}
\item Start $P$ (the conjugate field to $X$) from a Gaussian distribution according to the probability $\exp(-Tr P_{\mu}^2/2)$. Both $X_{\mu}$ and $P_{\mu}$ are hermitian $N\times N$ matrices.
\item Start $\xi$ from a Gaussian distribution according to the probability $\exp(-\xi^+\xi)$.
\item Calculate $\phi$ (the pseudo-fermion) using the formula (\ref{form1}). This is done using the conjugate gradient method (see below). The coefficients $c$ and $d$ are computed using the Remez algorithm from the rational approximation of $x^{1/4}$. 
\item Start $Q$ (the conjugate field to $\phi$) from a Gaussian distribution according to the probability $\exp(-Q^+Q)$. The spinors $Q_{\alpha}$ and $\phi_{\alpha}$, as well as $\xi_{\alpha}$, are $(N^2-1)-$dimensional complex vectors.
\end{itemize}
\item {\bf Molecular Dynamics}: This consists of two parts:
\begin{itemize}
\item {\bf Pseudo-Fermion}: We evolve the pseudo-fermion $\phi$ and its conjugate field $Q$ using the Hamilton equations (\ref{form21}), (\ref{form22}) and (\ref{form23}). This is done using the conjugate gradient method which, given the input $\phi$,  computes as output the spinors $G_{\sigma}$ given by equation (\ref{form31}) and the spinor $W$ given by equation (\ref{form32}). On the other hand, in the initialization step above we call the conjugate gradient method with input $\xi$ to obtain the output $\phi=W^*$.  Here and below, the coefficients $a$ and $b$ are computed using the Remez algorithm from the rational approximation of $x^{-1/2}$. 
\item {\bf Gauge Field}: We evolve $X_{\mu}$ and $P_{\mu}$ using the Hamilton equations (\ref{lf1}), (\ref{lf2}) and (\ref{lf3}). This requires the calculation of the boson contribution to the force given by equation (\ref{Bforce}) and the fermion contribution given by equation (\ref{Fforce}). The numerical evaluation of the fermion force is quite involved and uses the formula (\ref{Fforce1}). This requires, among other things, the calculation of the spinors $G_{\sigma}$ and $F_{\sigma}={\cal M}^{'}G_{\sigma}$ using the conjugate gradient.
\end{itemize}
\item {\bf Metropolis Step}: After obtaining the solution $(X(T),P(T),\phi(T),Q(T))$ of the molecular dynamics evolution starting from the initial configuration $(X(0),P(0),\phi(0),Q(0))$ we compute the resulting variation $\Delta H$ in the Hamiltonian. The new configuration is accepted with probability 
\begin{eqnarray}
{\rm probability}={\rm min}(1,\exp(-\Delta H)).
\end{eqnarray}  
\item {\bf Iteration:} Repeat starting from $2$.
\item {\bf Other Essential Ingredients}: The two other essential ingredients of this algorithm are:
\begin{enumerate}
\item {\bf Conjugate Gradient}: This plays a fundamental role in this algorithm. The multimass Krylov space solver employed here is based on the fundamental equations (\ref{cg1})-(\ref{cg12}). This allows us to compute the  $G_{\sigma}$ for all $\sigma$ given by equation (\ref{form31}) at once. The multiplication by $\Delta$ is done in two steps: first we multiply by ${\cal M}^{'}$ then we multiply by $({\cal M}^{'})^+$. This is done explicitly by reducing (\ref{54m}) to (\ref{70m})+(\ref{71m}) and reducing (\ref{56m}) to   (\ref{80m})+(\ref{81m}). Here, we obviously need to convert between a given traceless vector and its associated matrix and vice versa. The relevant equations are (\ref{59m}), (\ref{61m}) and (\ref{65m}).
\item {\bf Remez Algorithm}: This is discussed at length in the previous chapter. We only need to re-iterate here that the real coefficients $c$, $d$, for the rational approximation of $x^{1/4}$,  and $a$ and $b$, for the rational approximation of $x^{-1/2}$, as well as the integer $M$ are obtained using the Remez algorithm of  \cite{algremez}. The integer $M$ is supposed to be determined separately for each function by requiring some level of accuracy whereas the range over which the functions are approximated by their rational approximations should be determined on a trial and error basis by inspecting the spectrum of the Dirac operator.
\end{enumerate}
\end{enumerate}
\subsection{Preliminary Tests}
\begin{figure}[htbp]
\begin{center}
\includegraphics[width=8.0cm,angle=-0]{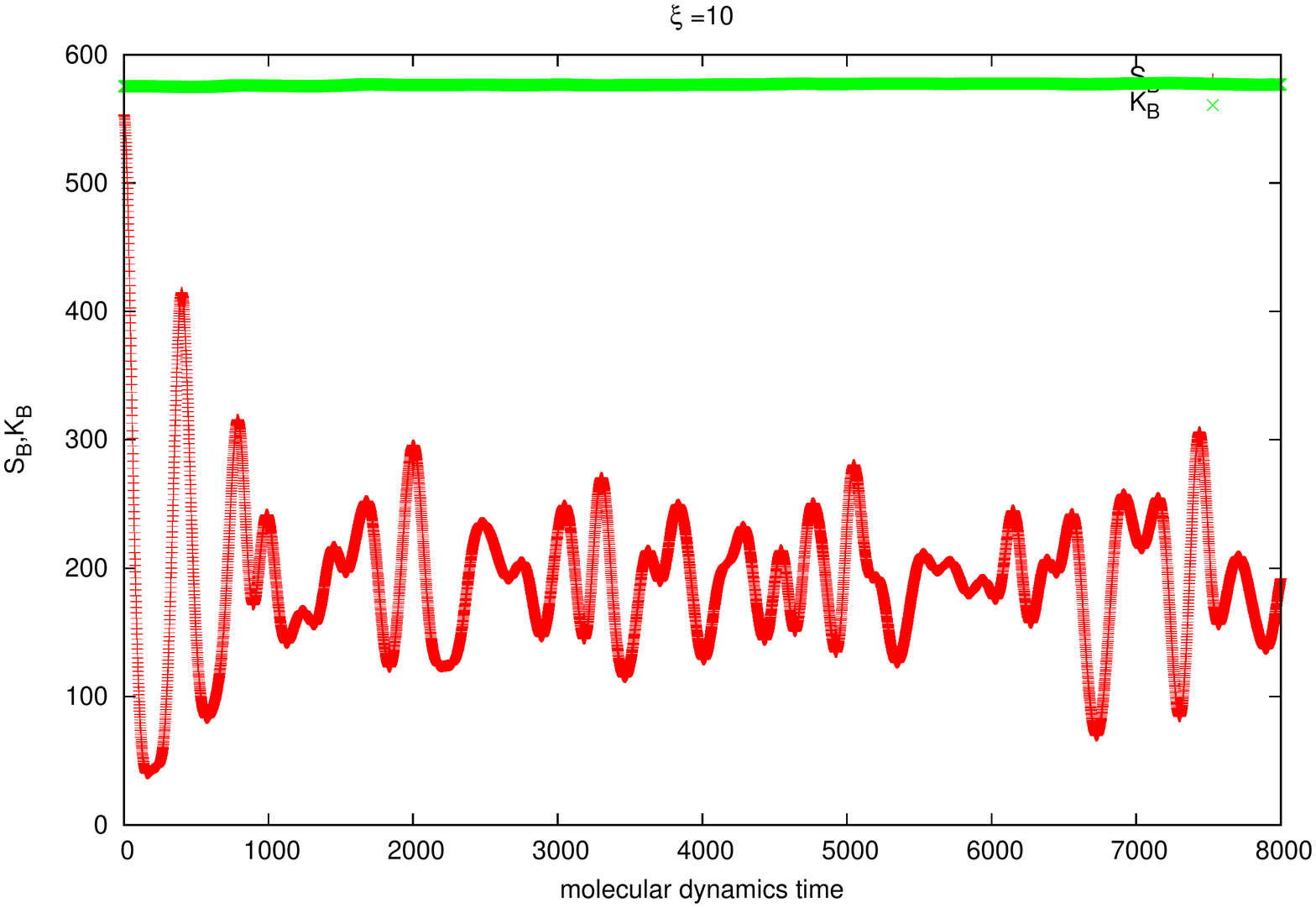}
\includegraphics[width=8.0cm,angle=-0]{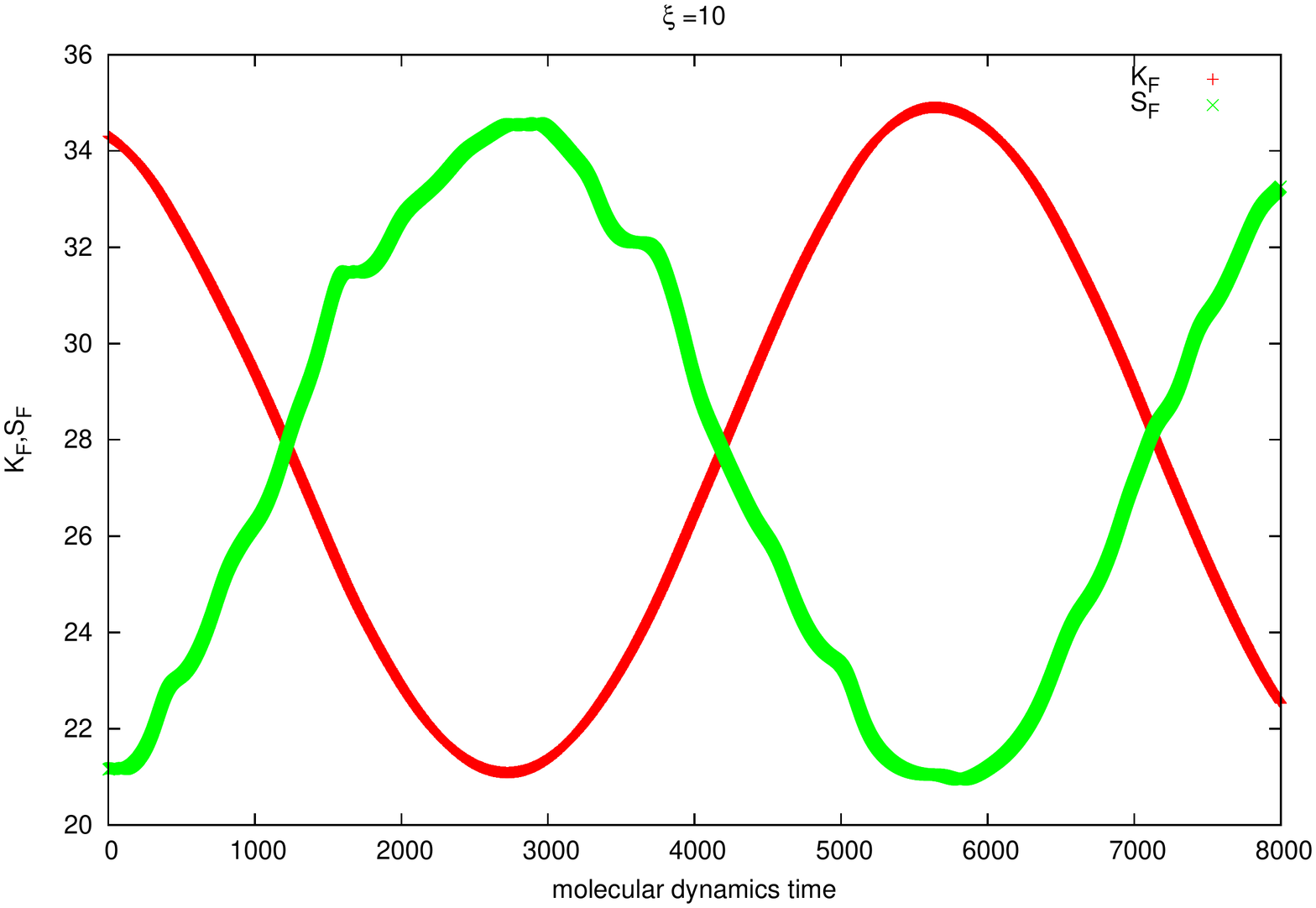}
\includegraphics[width=8.0cm,angle=-0]{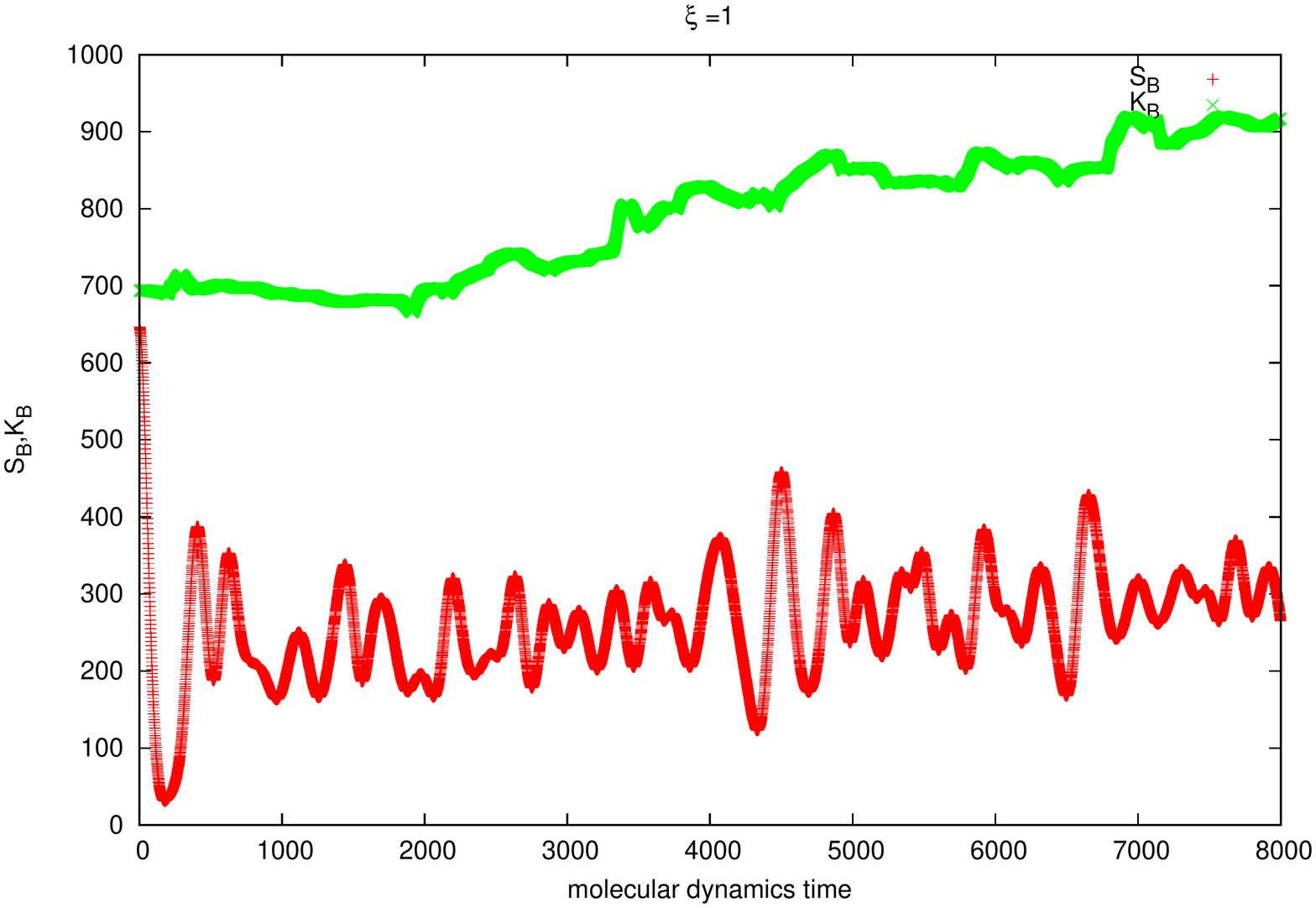}
\includegraphics[width=8.0cm,angle=-0]{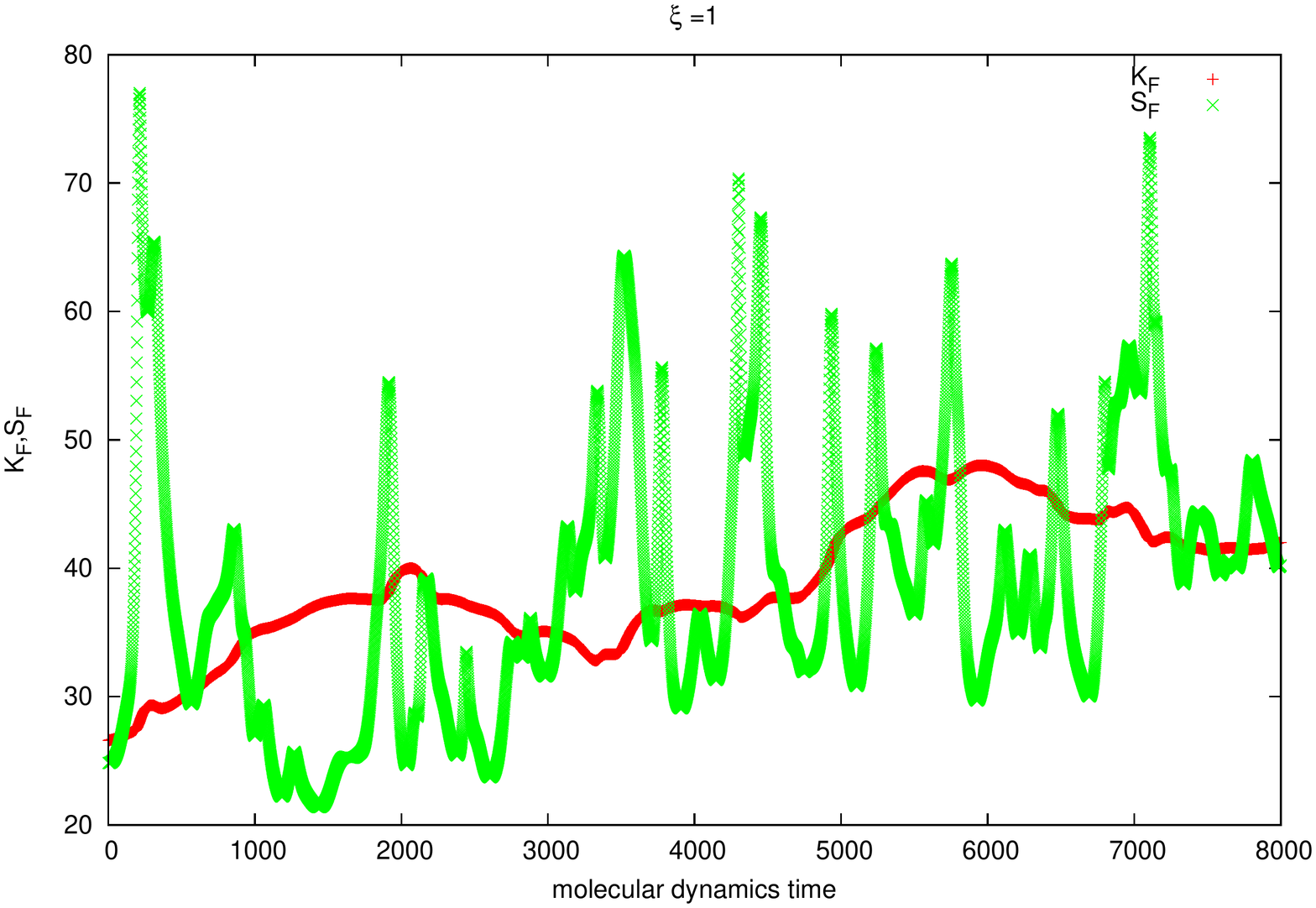}
\includegraphics[width=8.0cm,angle=-0]{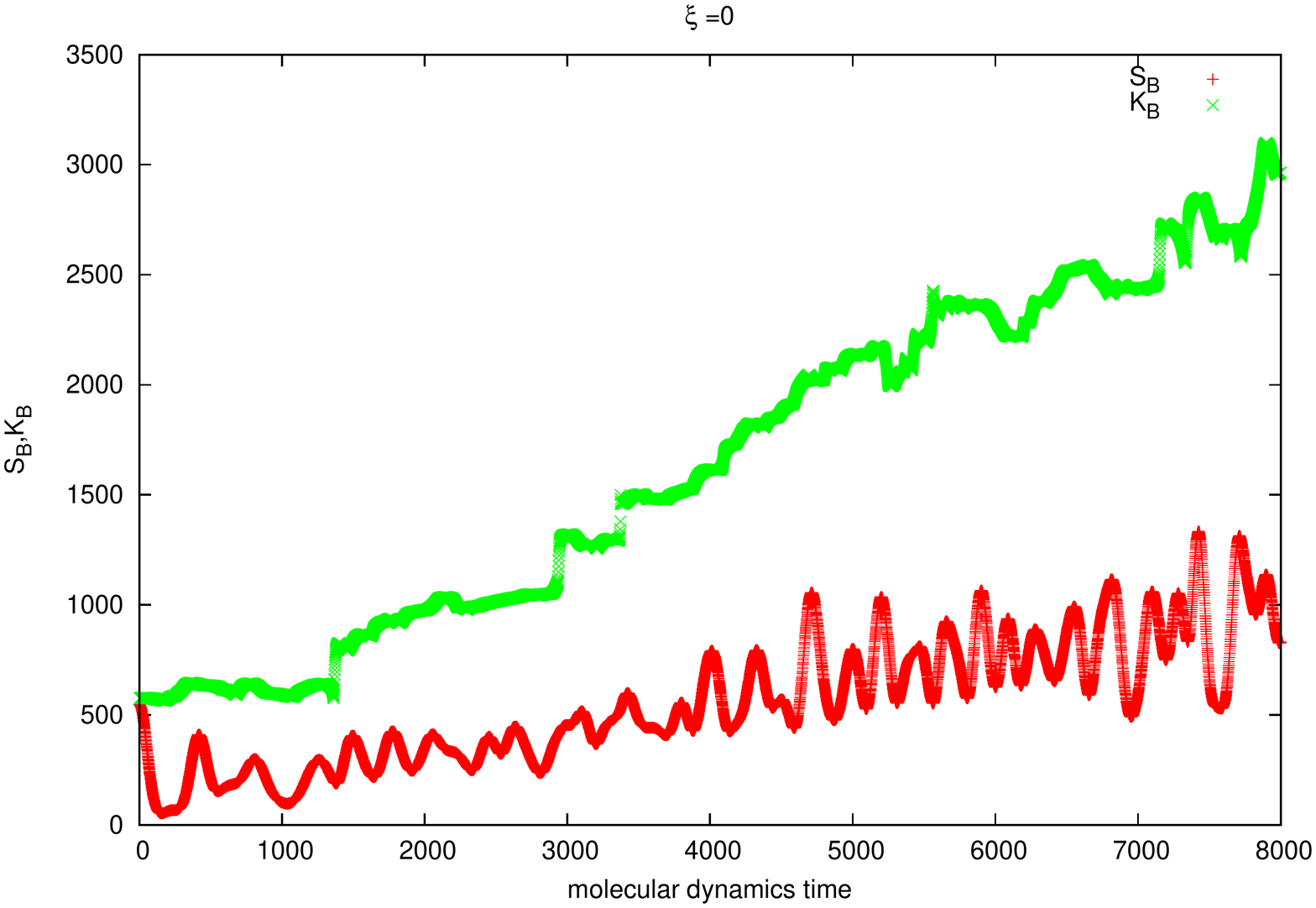}
\includegraphics[width=8.0cm,angle=-0]{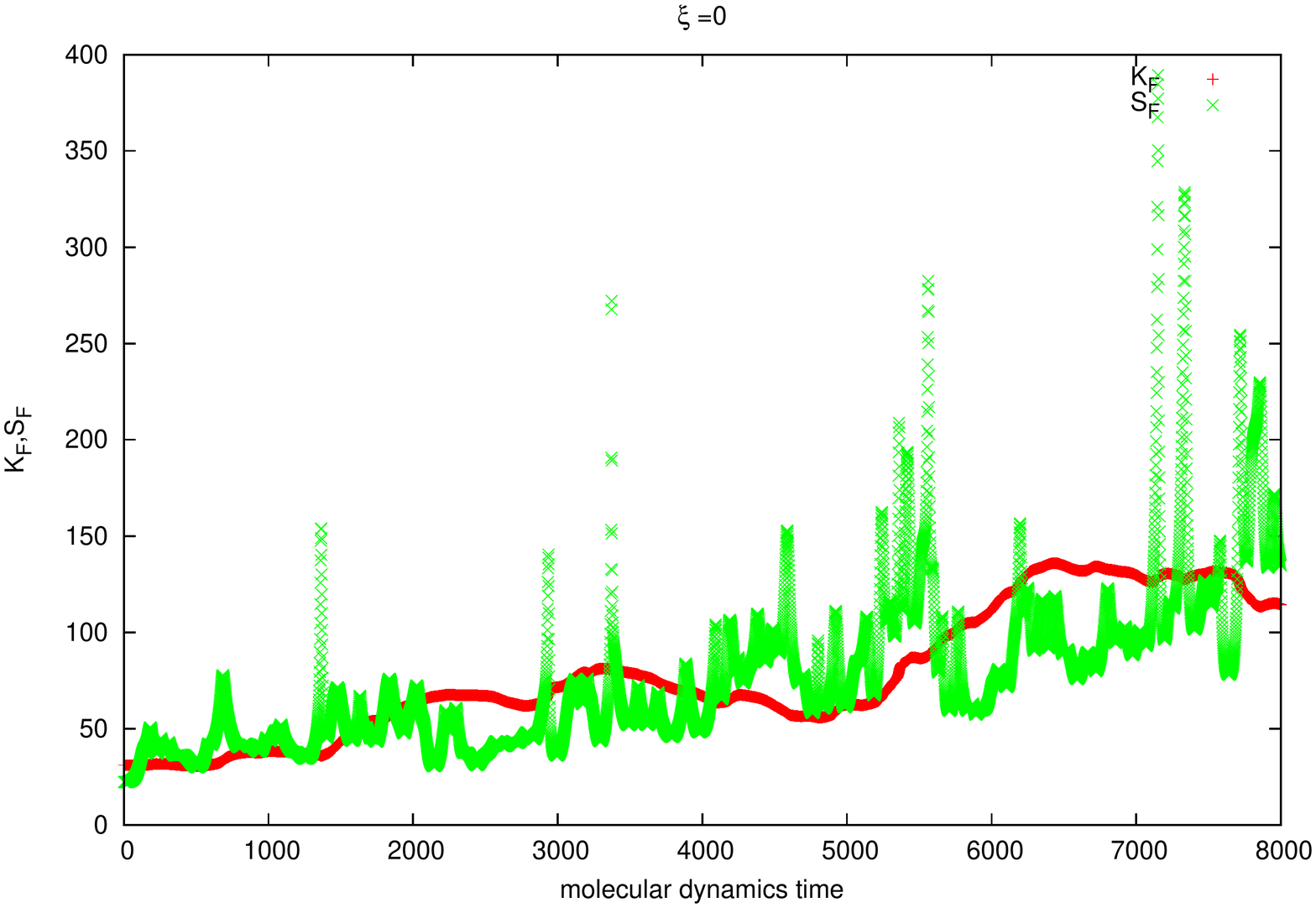}
\end{center}
\caption{}\label{testMD1}
\end{figure}

\begin{figure}[htbp]
\begin{center}
\includegraphics[width=8.0cm,angle=-0]{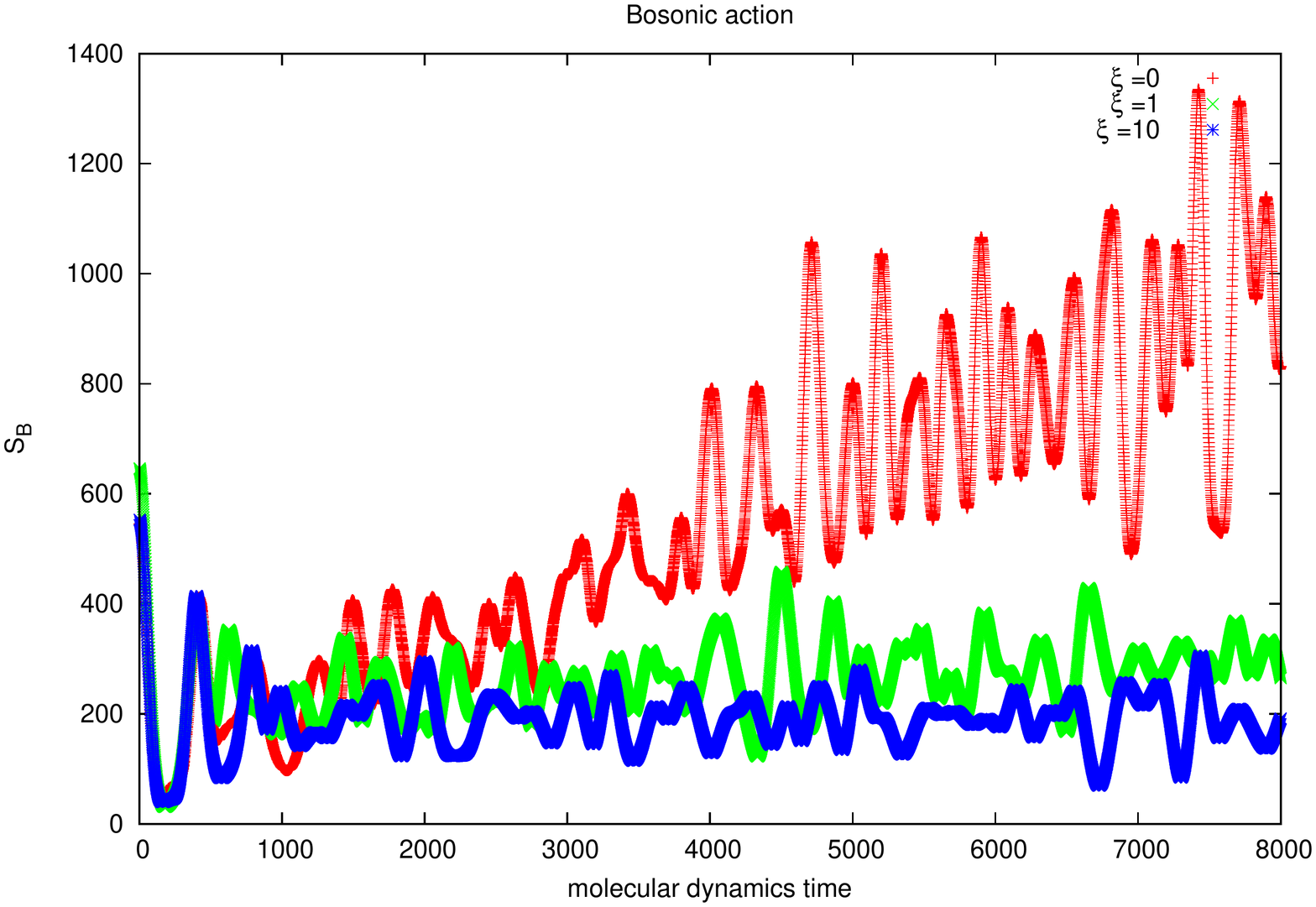}
\includegraphics[width=8.0cm,angle=-0]{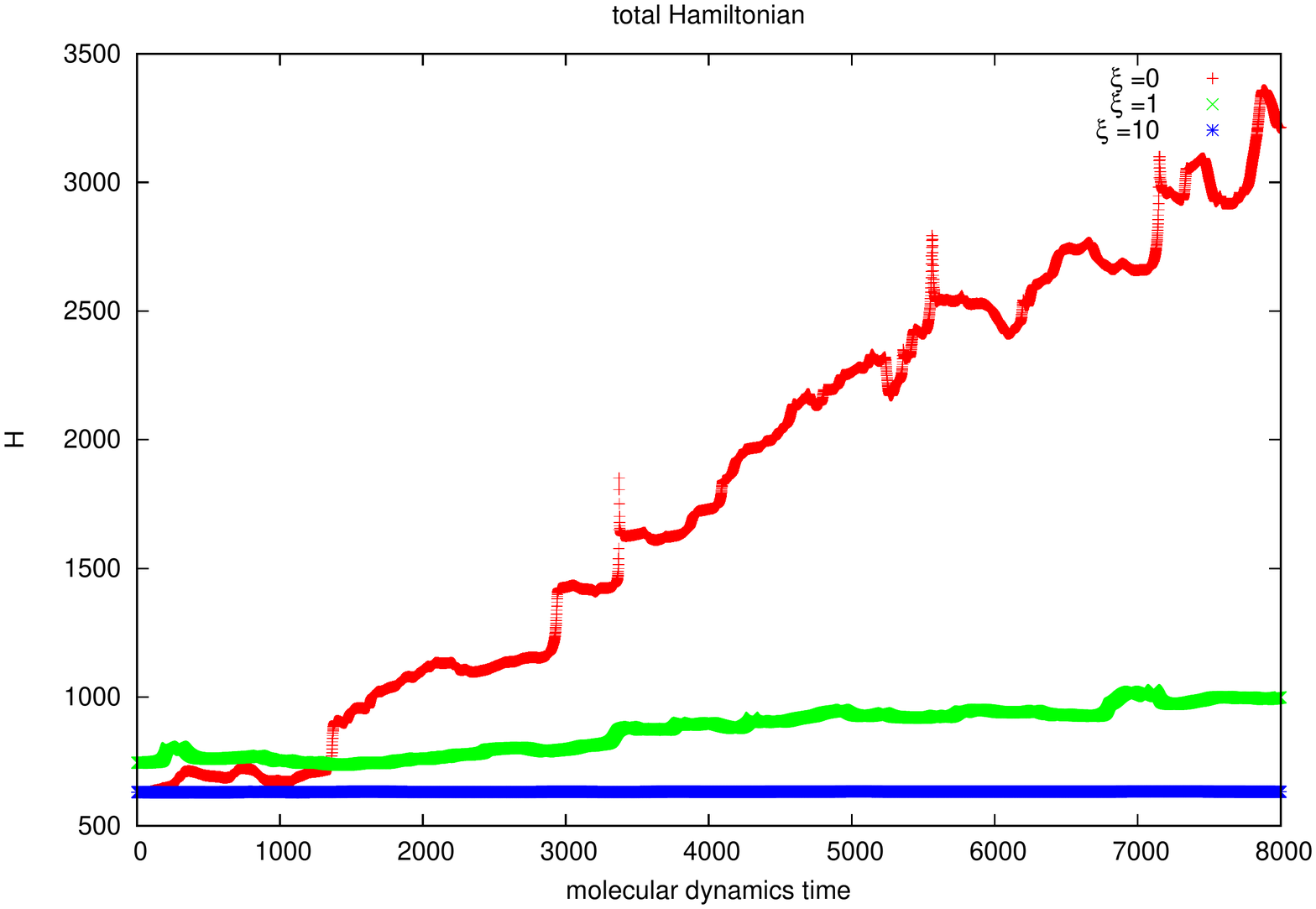}
\includegraphics[width=8.0cm,angle=-0]{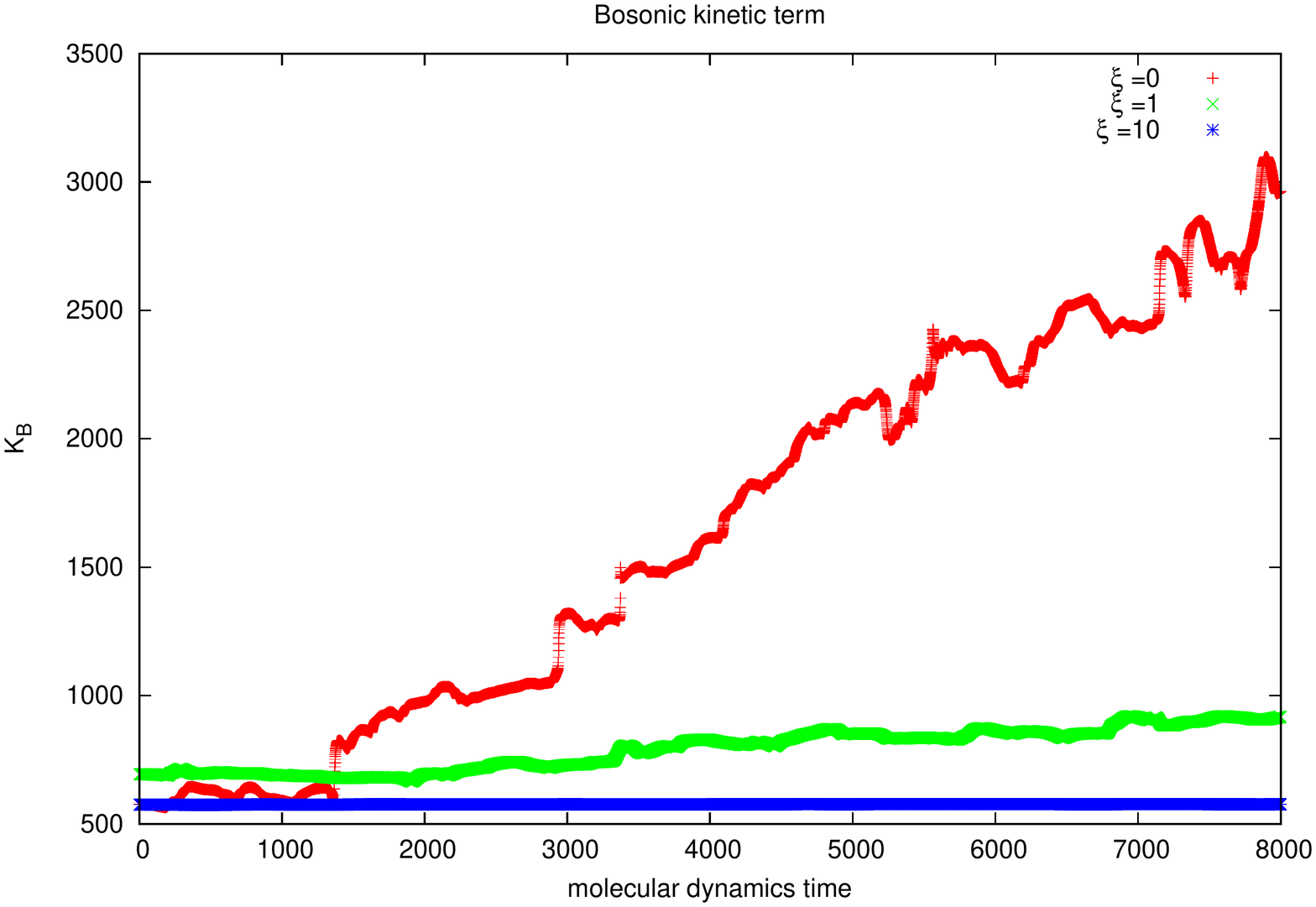}
\includegraphics[width=8.0cm,angle=-0]{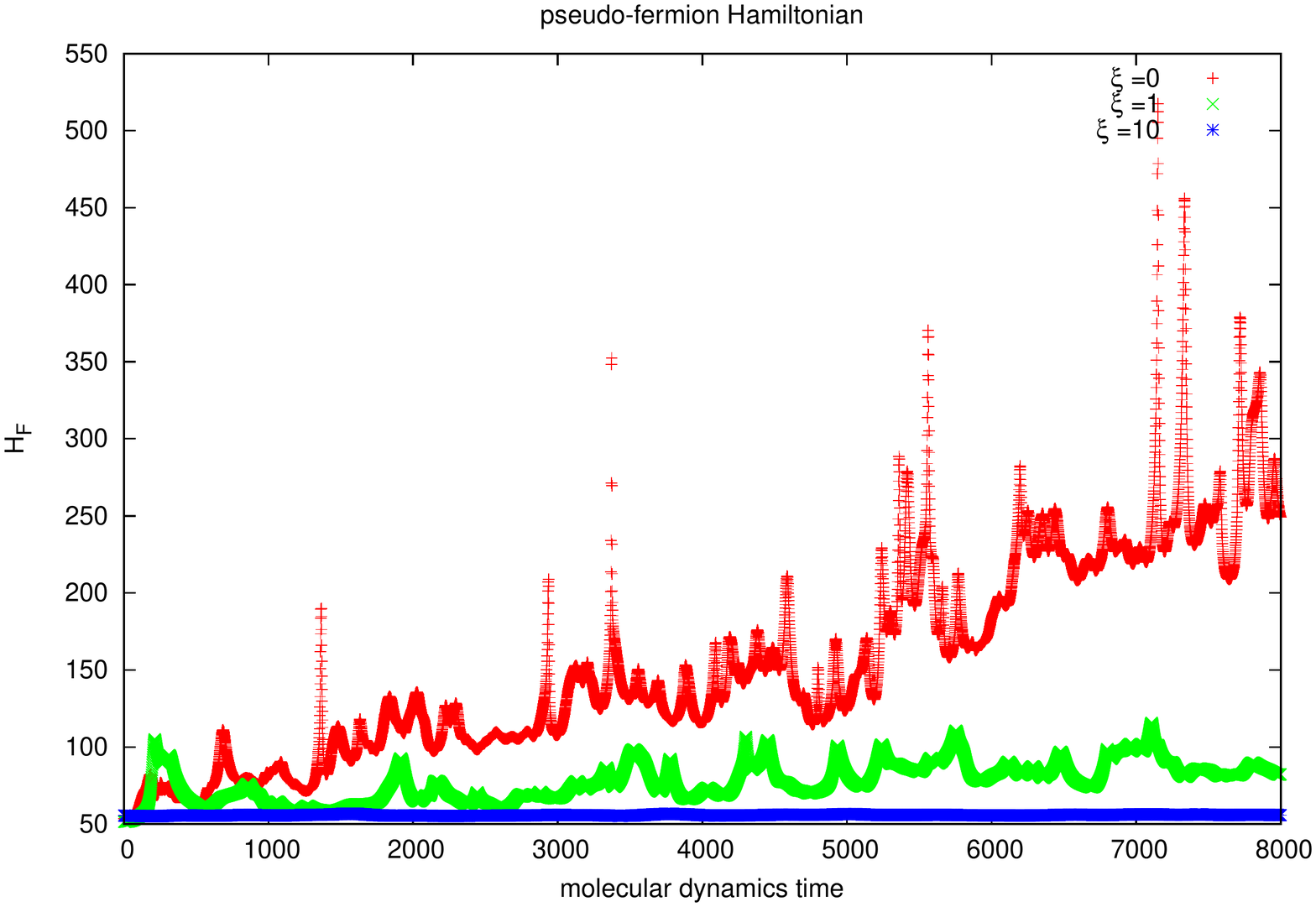}
\end{center}
\caption{}\label{testMD2}
\end{figure}

\begin{figure}[htbp]
\begin{center}
\includegraphics[width=8.0cm,angle=-0]{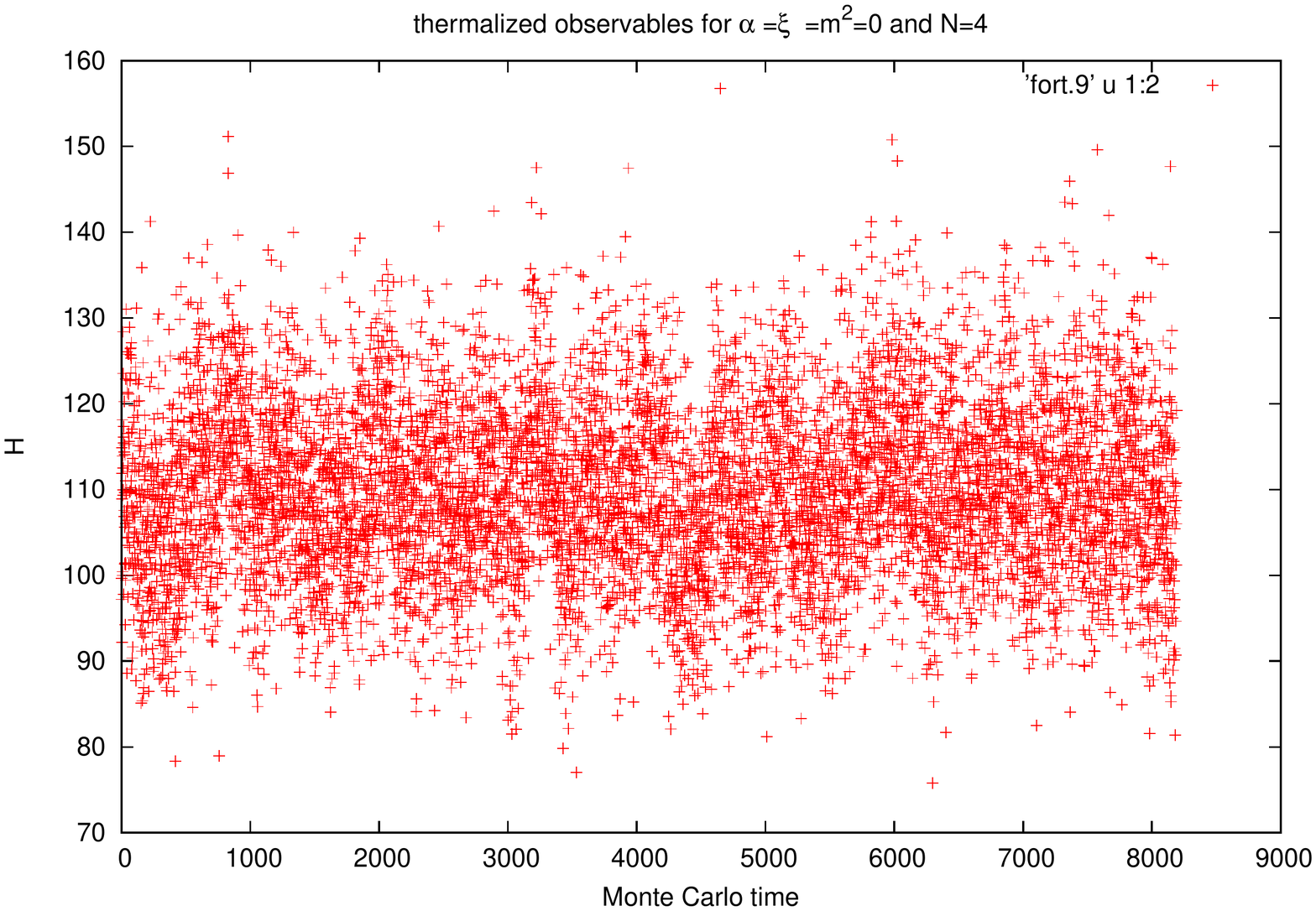}
\includegraphics[width=8.0cm,angle=-0]{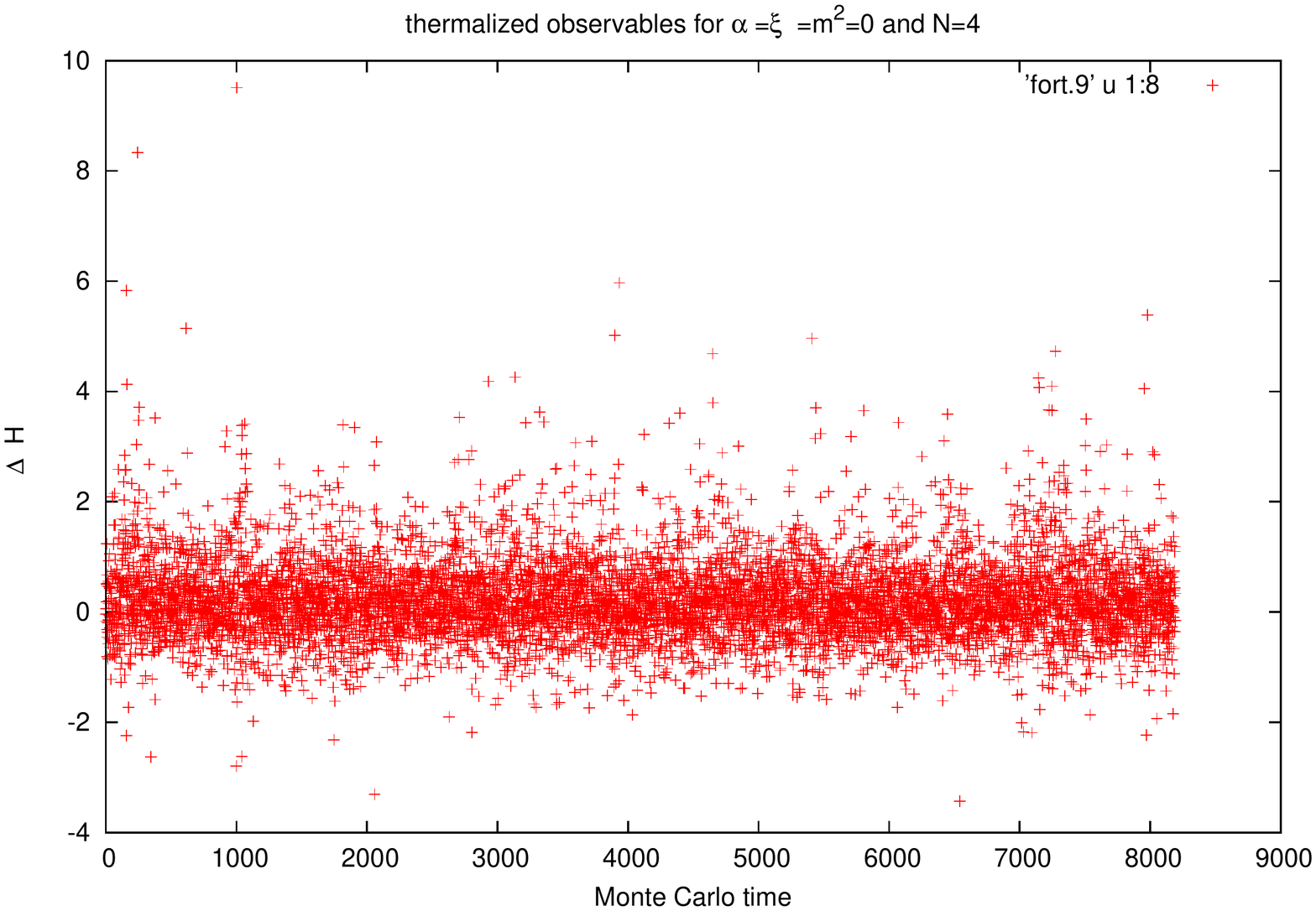}
\includegraphics[width=8.0cm,angle=-0]{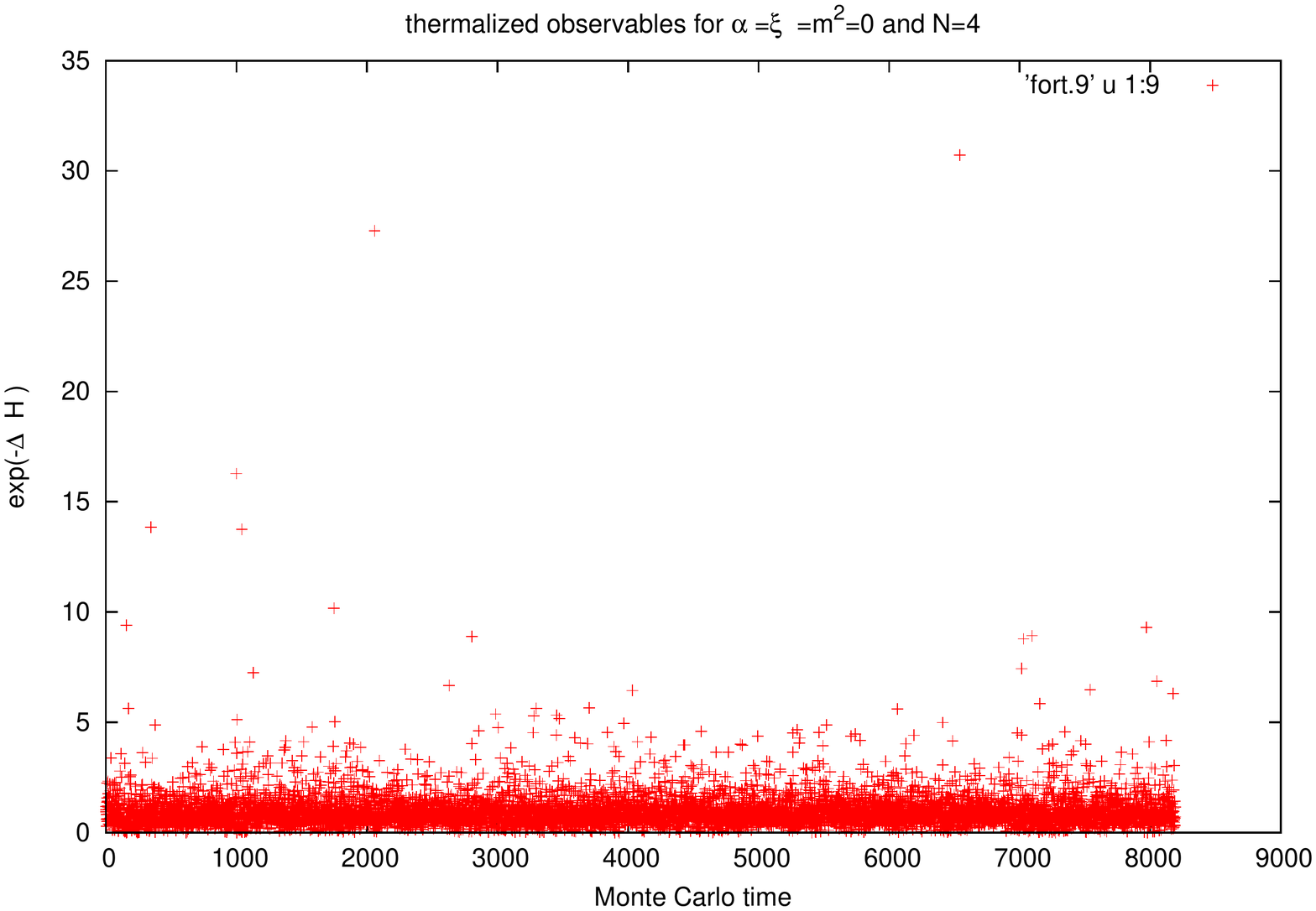}
\includegraphics[width=8.0cm,angle=-0]{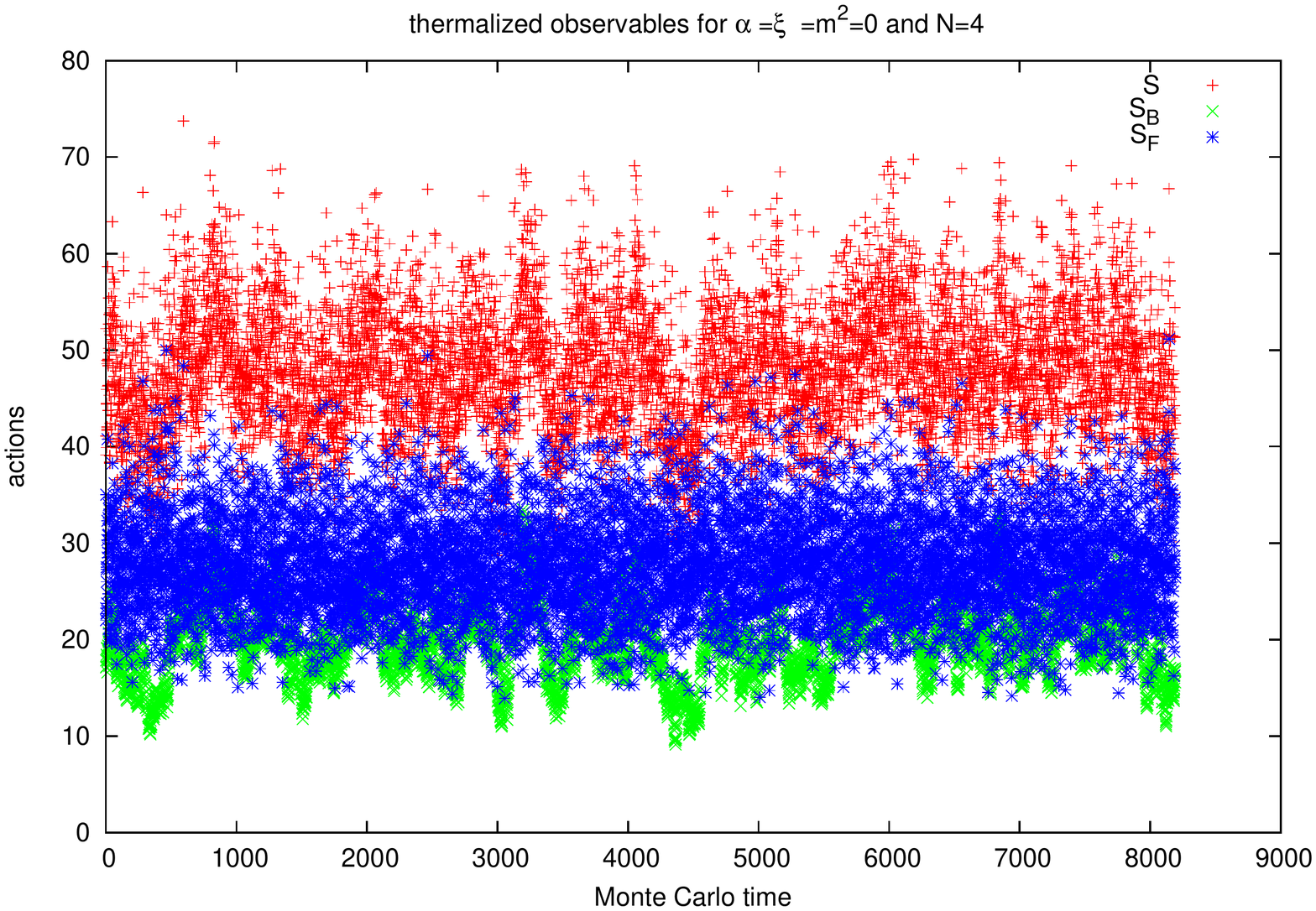}
\includegraphics[width=8.0cm,angle=-0]{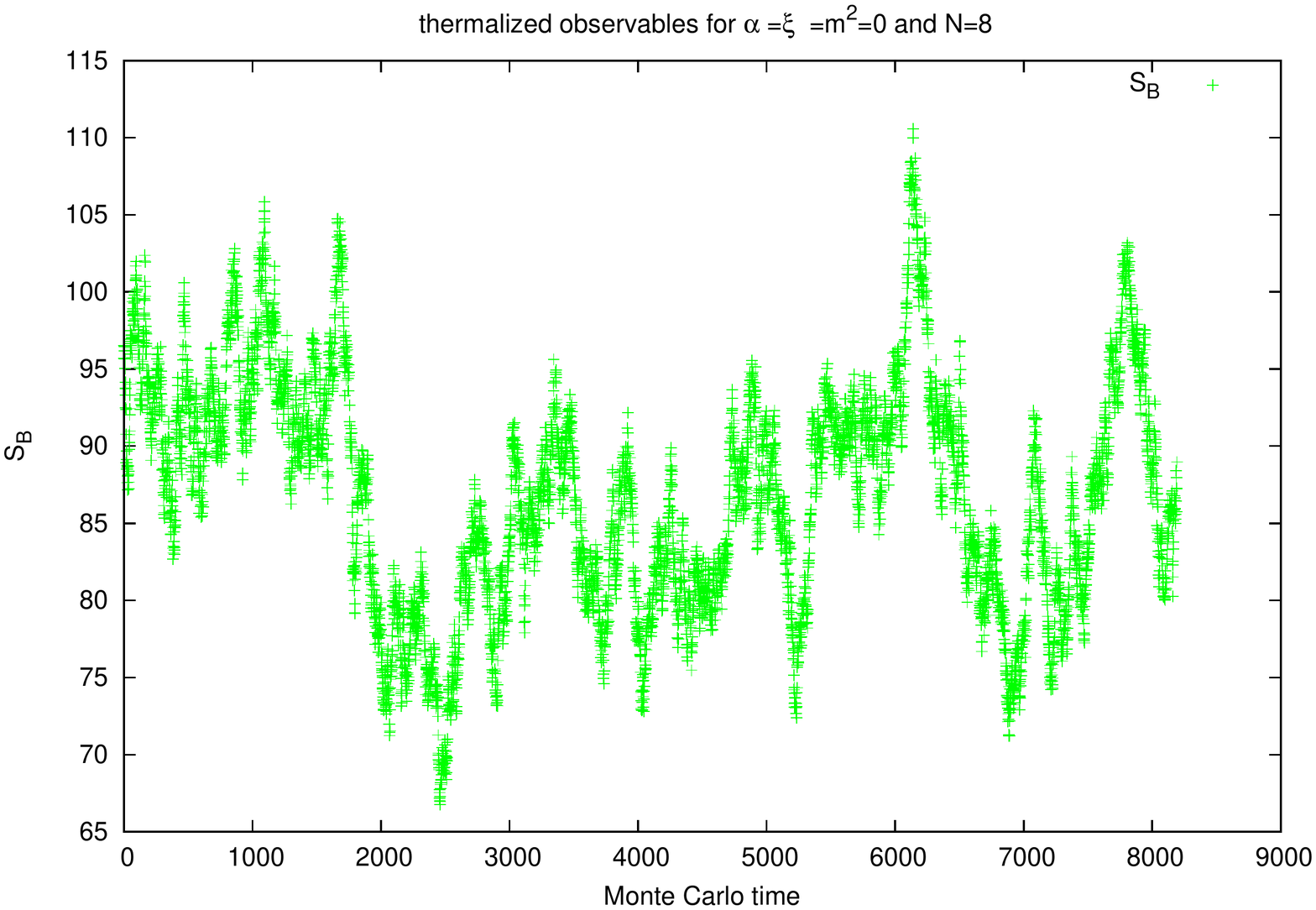}
\includegraphics[width=8.0cm,angle=-0]{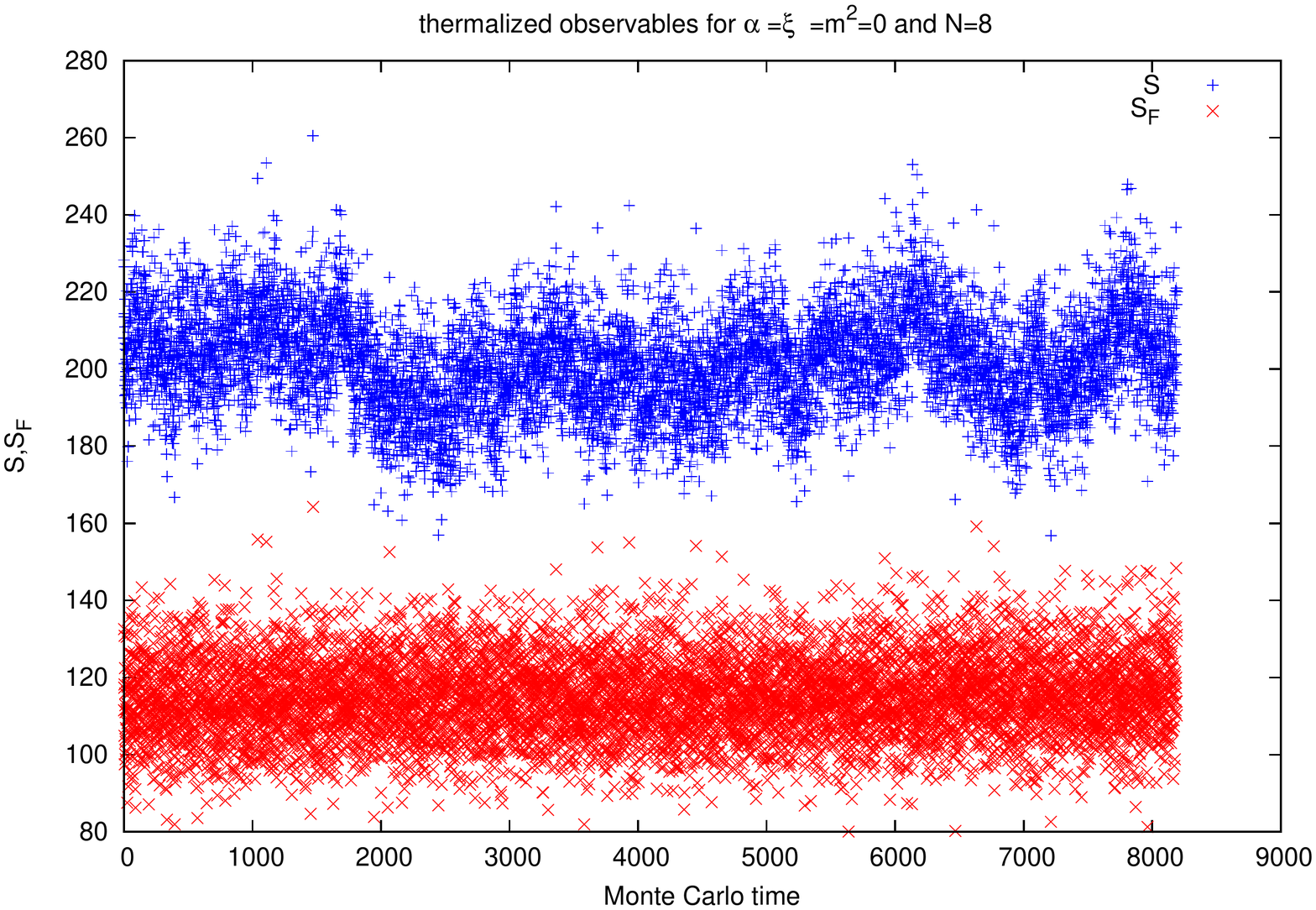}
\end{center}
\caption{}\label{testHM1}
\end{figure}

\begin{figure}[htbp]
\begin{center}
\includegraphics[width=8.0cm,angle=-0]{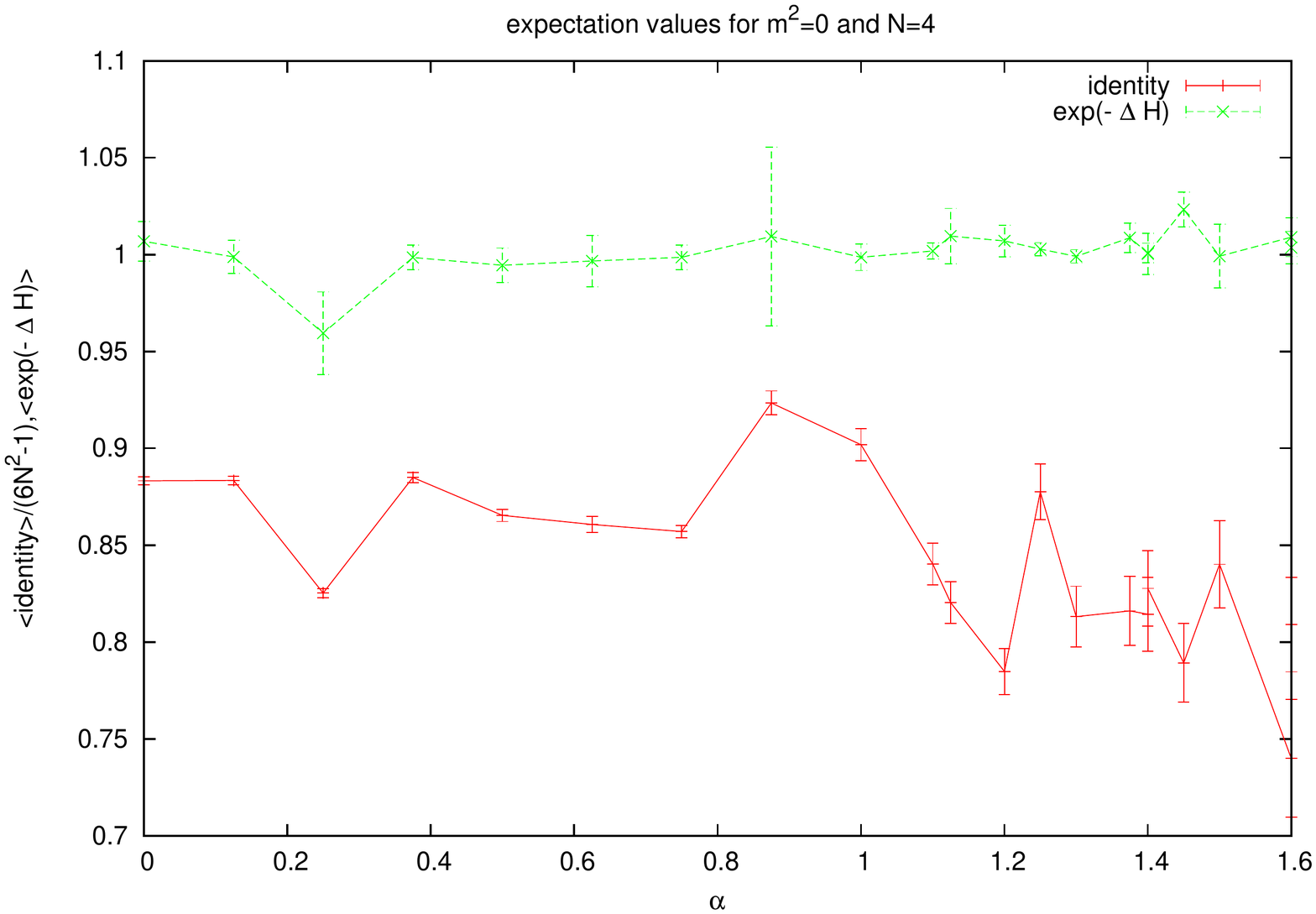}
\includegraphics[width=8.0cm,angle=-0]{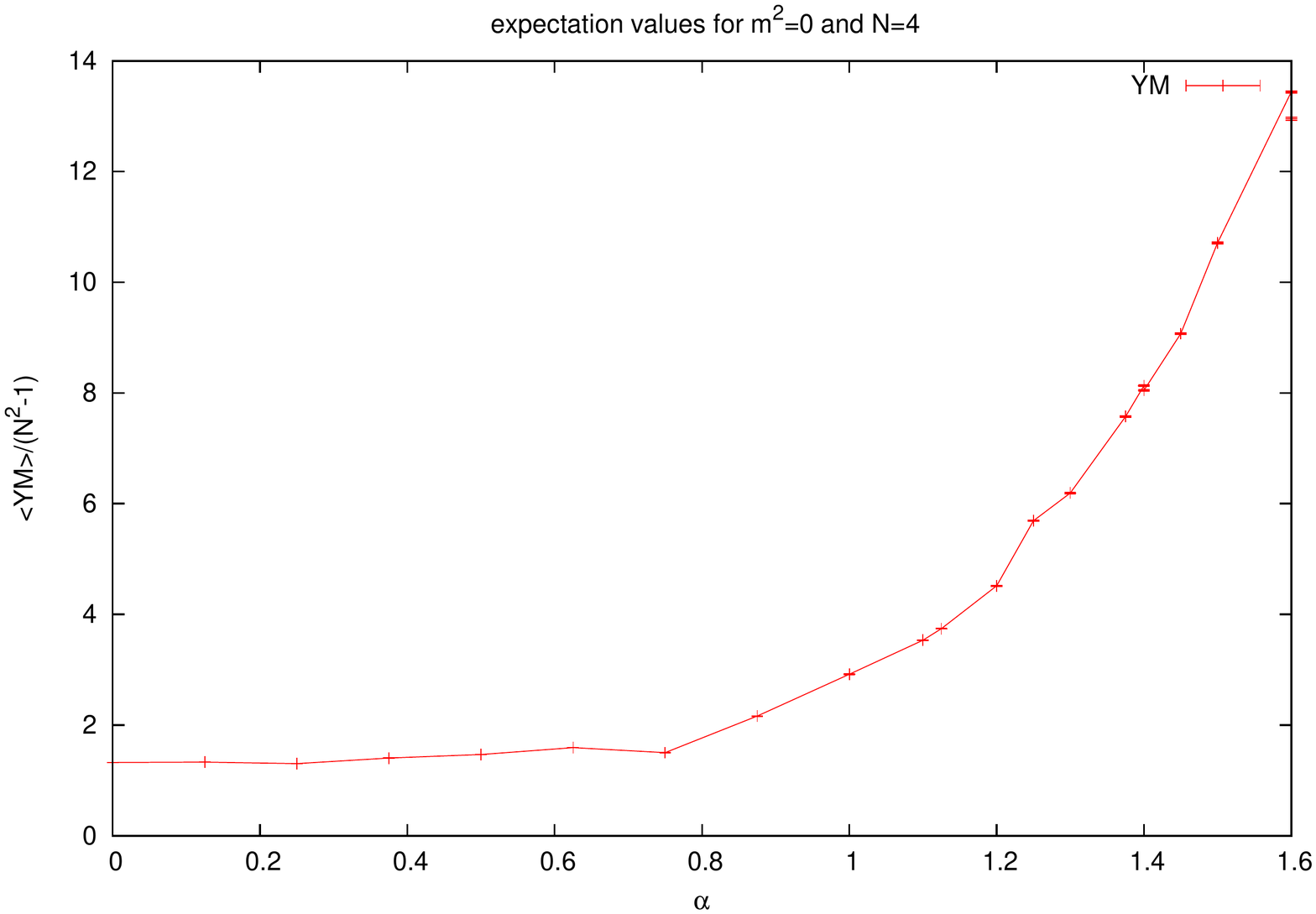}
\includegraphics[width=8.0cm,angle=-0]{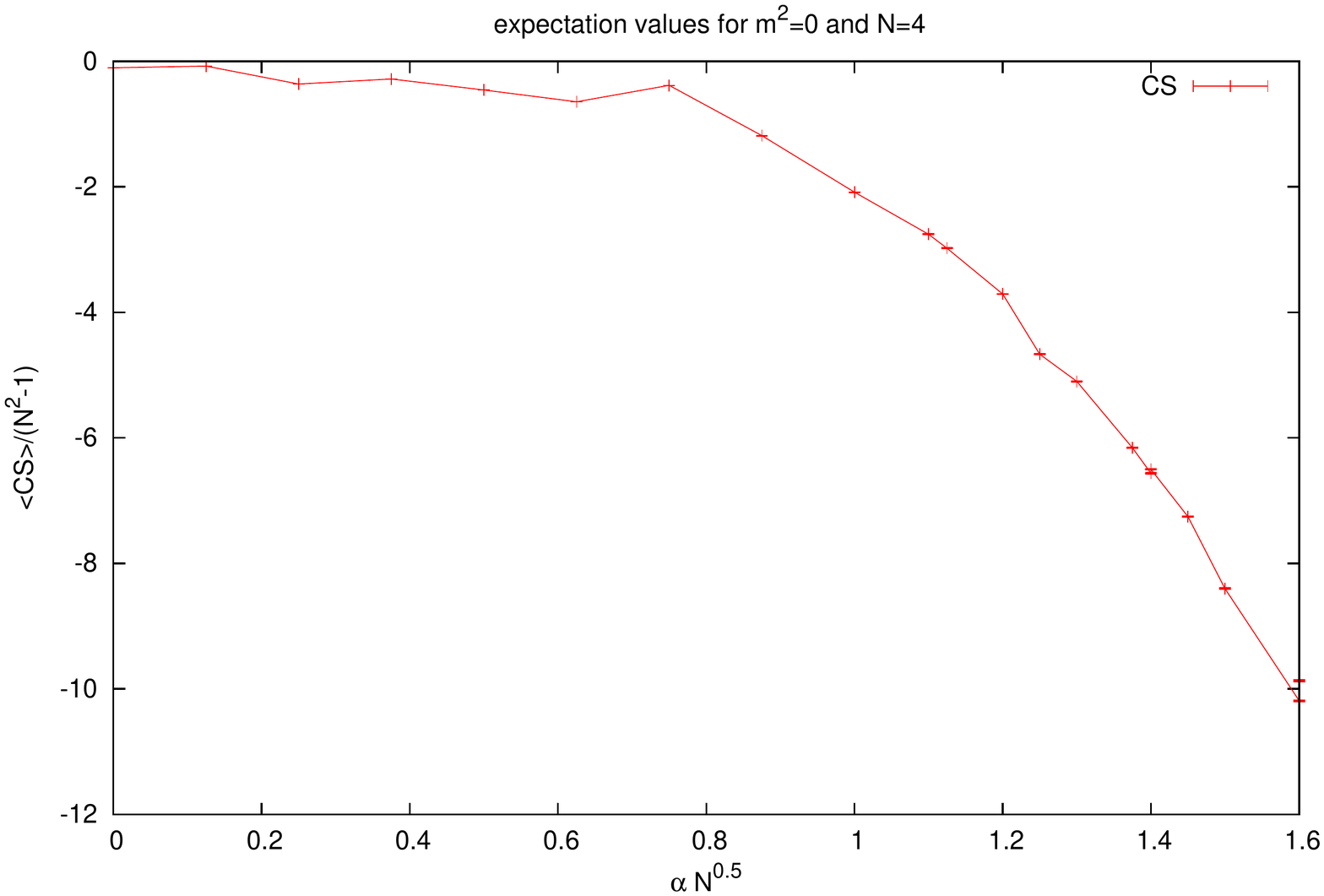}
\includegraphics[width=8.0cm,angle=-0]{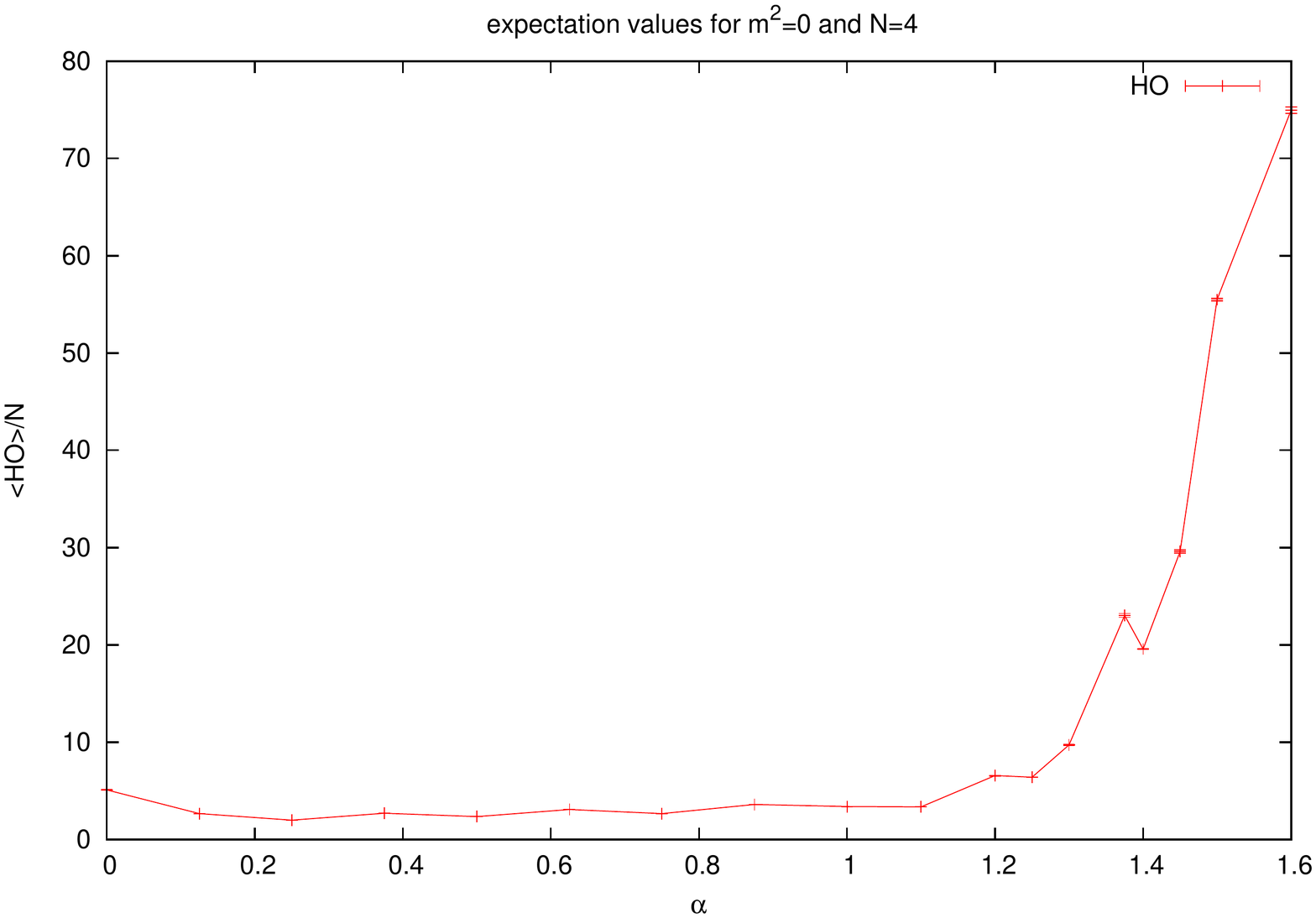}
\includegraphics[width=8.0cm,angle=-0]{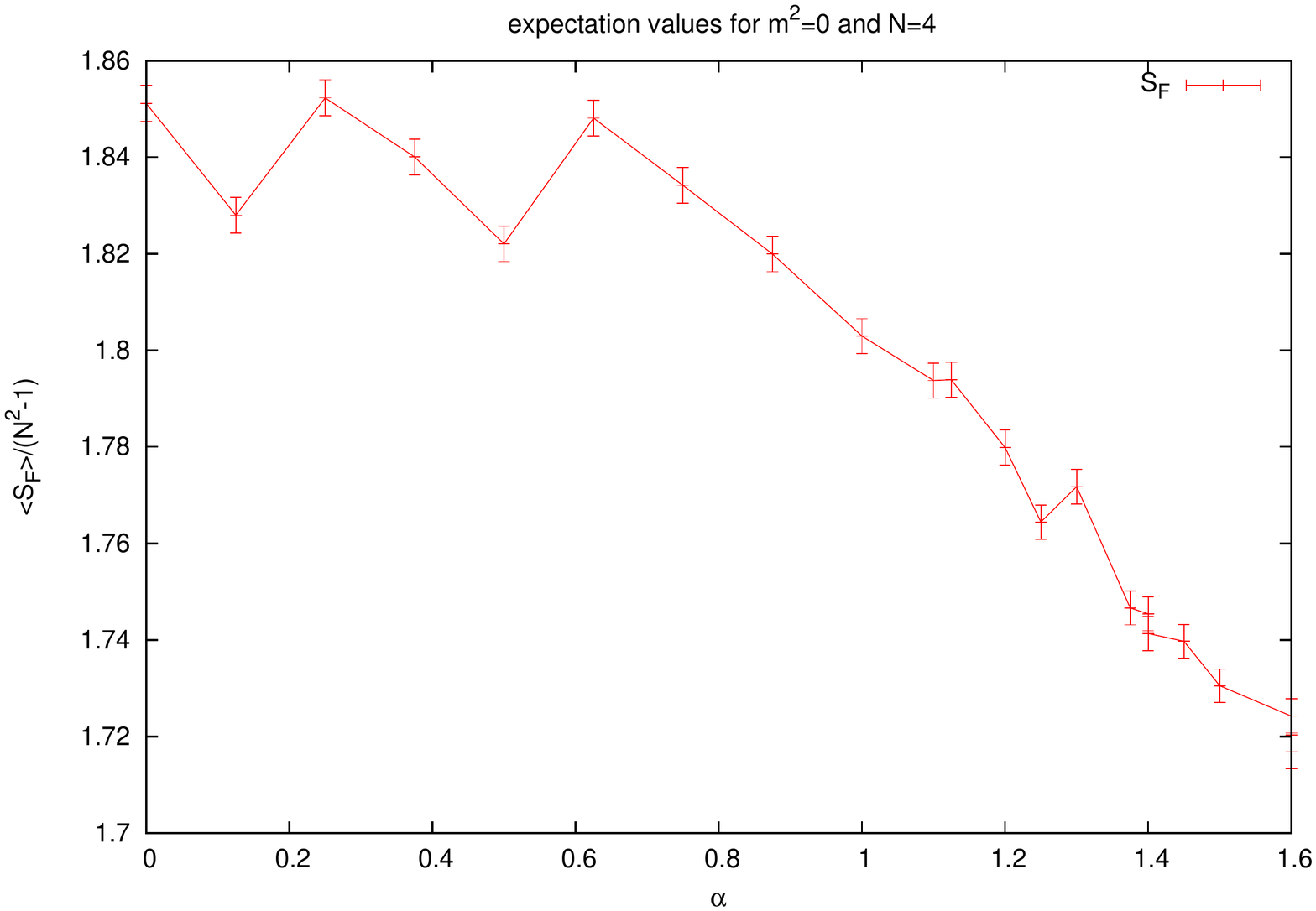}
\end{center}
\caption{}\label{testHM2}
\end{figure}

\begin{enumerate}
\item {\bf The rational approximations}: The first thing we need to do is to fix the parameters $a$, $b$, $c$ and $d$ of the rational approximations by invoking the Remez algorithm. For a tolerance equal $10^{-4}$ and over the interval $[0.0004,1]$ with precision $40$ we have found that the required degrees of the rational approximations, for $x^{-1/2}$ and $x^{1/4}$, are $M=6$ and $M_0=5$ respectively; $M$ is the minimum value for which the uniform norm $|r-f|_{\infty}={\bf max}|r-f|$ is smaller than the chosen tolerance. We can plot these rational approximations versus the actual functions to see whether or not these approximations are sufficiently good over the fixed range.

\item {\bf The conjugate gradient}: The conjugate gradient is a core part in this algorithm and it must be checked thoroughly. A straightforward check is to verify that $(\Delta +b_{\sigma})G_{\sigma}=\phi$ for all values of $\sigma$.  We must be careful that the matrix-vector multiplication $\Delta.G_{\sigma}$ does not vanish. Thus the no-sigma problem should be defined, not with zero mass  $b_{\sigma}=0$, but with the smallest possible value of the mass $b_{\sigma}$ which presumably corresponds to the least convergent linear system. In the results included below we fix the tolerance of the conjugate gradient at $10^{-5}$.

\item {\bf The decoupled theory}: This is the theory in which the gauge field $(X_{\mu})_{ij}$ and the pseudo-fermion field $\phi_{\alpha}^A$ are completely decoupled from each other. This is then equivalent to the bosonic theory. This is expected to be obtained for sufficiently large values of the fermion mass $\xi$. In this theory the fermion field behaves exactly as a harmonic oscillator. The decoupled theory can also be obtained, both in the molecular dynamics part and the hybrid Monte Carlo part which includes in addition the metropolis step, by setting
\begin{eqnarray}
c_0=\frac{1}{\sqrt{a_0}}~,~a_i=c_i=0.
\end{eqnarray}
In this case the pseudo-fermions decouple from the gauge fields and behave as  harmonic oscillators with  period $T=2\pi$. The corresponding action should then be periodic with period $T=\pi$.
\item {\bf The molecular dynamics}: We can run the molecular dynamics on its own to verify the prediction of the decoupled theory. In general, it is also useful to monitor the classical dynamics for its own interest and monitor in particular the systematic error due to the non-conservation of the Hamiltonian.

In the molecular dynamics we need to fix the time step $dt$ and the number of iterations $n$.  Thus we run the molecular dynamics for a time interval $T=n.dt$. We choose $dt=10^{-3}$ and $n=2^{13}-2^{14}$. Some results with $N=4$ are included in figures (\ref{testMD1}) and (\ref{testMD2}). We remark that the drift in the Hamiltonian becomes pronounced as $\xi\longrightarrow 0$. This systematic error will be canceled by the Metropolis step (see below).
 
We can use the molecular dynamics to obtain an estimation of the range of the rational approximations needed as follows. Starting from $\xi=0$, we increase the value of $\xi$ until the behavior of the theory becomes that of the decoupled (bosonic) theory. The value of $\xi$ at which this happens will be taken as an estimation of the range. In the above example (figures  (\ref{testMD1}) and (\ref{testMD2})) we observe that the pseudo-fermion sector becomes essentially a harmonic oscillator around the value $\xi=10$. Thus a reasonable range should be taken between $0$ and $10$.

\item {\bf The metropolis step}: 
In general two among the three parameters of the molecular dynamics (the time step $dt$, the number of iterations $n$ and the time interval $T=ndt$) should be optimized in such a way that the acceptance rate is fixed, for example, between $70$ and $90$ per cent. We  fix $n$ and optimize $dt$ along the line discussed  in previous chapters. We make, for every $N$,  a  reasonable  guess for the value of the number of iterations $n$, based on trial and error, and then work with that value throughout. For example, for $N$ between $N=4$ and $N=8$, we found the value $n=10$, to be sufficiently reasonable. 

Typically, we run $T_{\rm ther}+T_{\rm meas}$ Monte Carlo steps where thermalization is supposed to occur within the first $T_{\rm ther}$ steps which are discarded while measurements are performed on a sample consisting of the subsequent $T_{\rm meas}$ configurations. We choose, for $N=4-8$, $T_{\rm ther}=2^{11}$ and $T_{\rm meas}=2^{13}$. We do not discuss in the following auto-correlation issues while error bars are computed using the jackknife method. As always, we generate our random numbers using the algorithm ran2. Some thermalized results for $N=4,8$ and $\alpha=m^2=\xi=0$ are shown on figure (\ref{testHM1}).

There are two powerful tests (exact analytic results) which can be used to calibrate the simulations. We must have the identities:
\begin{itemize}
\item We must have on general grounds the identity:
\begin{eqnarray}
<\exp(-\Delta H)>=1.
\end{eqnarray}
\item We must also have the Schwinger-Dyson identity: 
\begin{eqnarray}
<4\gamma {\rm YM}>+<3\alpha {\rm CS}>+<2m^2{\rm HO}>+<\xi {\rm COND}>=(d+2)(N^2-1).
\end{eqnarray}
We have included for completeness the effects of a Chern-Simons term and a harmonic oscillator term in the bosonic action. This identity is a generalization of (\ref{SDB}) where the definition of the condensation ${\rm COND}$ can be found in \cite{Ydri:2012bq}. This identity follows from the invariance of the path integral (\ref{PIbasic}) under the translations $X_{\mu}\longrightarrow X_{\mu}+\epsilon X_{\mu}$. For the flat space supersymmetric model for which $\xi=0$ the above Schwinger-Dyson identity reduces to
 \begin{eqnarray}
<4\gamma {\rm YM}>+<3\alpha {\rm CS}>+<2m^2{\rm HO}>=(d+2)(N^2-1).
\end{eqnarray}
\end{itemize}
As an illustration some expectation values as functions of $\alpha$ for $N=4$ and  $m^2=\xi=0$ are shown on figure (\ref{testHM2}).
\item {\bf Emergent geometry:} We observe from the graph of $Tr X_{\mu}^2$ that something possibly interesting happens around $\alpha\sim 1.2$. In fact, this is the very dramatic phenomena of emergent geometry which is known to occur in these models when there is a non-zero mass term (here the Chern-Simons term)  included. This can be studied in great detail using as order parameters the eigenvalues distributions of $X_4$ and $X_a$. In the matrix or Yang-Mills phase (small values of $\alpha$) the matrices $X_{\mu}$ are nearly commuting with eigenvalues distributed uniformly inside a solid ball with a parabolic eigenvalues distributions, or a generalization thereof, whereas in the fuzzy sphere phase (large values of $\alpha$) the matrix $X_4$ decouples from $X_a$ and remains distributed as in the matrix phase, while the matrices $X_a$ will be dominated by fluctuations around the $SU(2)$ generators in the spin $(N-1)/2$ irreducible representation. 
\item {\bf Code:} The attached code can be used to study the above emergent geometry effect, and many other issues, in great detail. On an intel dual core E$4600$ processor ($2.40$GHz) running Ubuntu 14.04 LTS this codes goes as $N^5$.
\end{enumerate}
\section{Other Related Topics}
Many other important topics, requiring techniques similar to the ones discussed in this chapter, and which have been studied extensively by the Japan group, includes:
\begin{enumerate}
\item {\bf IKKT models:} The extension of the problem to higher dimensions; for example $d=6$; but in particular $d=10$ which is the famous IKKT model which provides a non-perturbative definition of string theory, is the first obvious generalization. However, the determinant in these cases is complex-valued which makes its numerical evaluation very involved. 

\item {\bf Cosmological Yang-Mills matrix models:} In recent years a generalization from Euclidean Yang-Mills matrix models to Minkowski signature was carried out with dramatic, interesting and novel consequences for cosmological models. The  problem with the complex-valued Pfaffians and determinants is completely resolved in these cases. 

\item {\bf Quantum mechanical Yang-Mills matrix models:} The extension of Yang-Mills matrix models to quantum mechanical Yang-Mills matrix models, such as the BFSS and BMN models which also provide non-perturbative definitions of string theory and M-theory, involves the introduction of time. This new continuous variable requires obviously a lattice regularization.  There is so much physics here relevant to the dynamics of black holes, gauge-gravity duality, strongly coupled gauge theory and many other fundamental problems.
\item {\bf The noncommutative torus:} The noncommutative torus provides another, seemingly different, non-perturbative regularization of noncommutative field theory besides fuzzy spaces. The phenomena of emergent geometry is also observed here, as well as the phenomena of stripe phases, and furthermore, we can add fermions and supersymmetry in an obvious way. The connection to commutative theory and the commutative limit is more transparent in this case which is an advantage.
\item {\bf Supersymmetry:} A non-perturbative definition of supersymmetry which allows Monte Carlo treatment is readily available from the above discussed, and much more, matrix models. These non-lattice simulations seem very promising to strongly coupled gauge theories. 

\end{enumerate}

\chapter{$U(1)$ Gauge Theory on the Lattice: Another Lattice Example}

In this chapter we will follow the excellent pedagogical textbook \cite{Gattringer:2010zz} especially on practical detail regarding the implementation of the Metropolis and other algorithms to lattice gauge theories. The classic textbooks \cite{Creutz:1984mgf,Smit:2002ugf,Rothe:2005nwf,Montvay:1994cyf} were also very useful.

\section{Continuum Considerations}

A field theory is a dynamical system with $N$ degrees of freedom where $N\longrightarrow \infty$. The classical description is given in terms of the Lagrangian  and the action while the quantum description is given in terms of the Feynman path integral and the correlation functions. In a scalar field theory the basic field has spin $j=0$ with respect to Lorentz transformations. Scalar field theories are relevant to critical phenomena. In gauge theories the basic fields have spin $j=1$ (gauge vector fields) and spin $j=1/2$ (fermions) and they are relevant to particle physics. The requirement of renormalizability restricts severely the set of quantum field theories to only few possible models. Quantum electrodynamics or QED is a  renormalizable field theory given by the action
\begin{eqnarray}
S_{\rm QED}
&=&\int d^4x\bigg[-\frac{1}{4}F_{\mu\nu}F^{\mu\nu}+\bar{\psi}(i\gamma^{\mu}\partial_{\mu}-M)\psi-e\bar{\psi}\gamma_{\mu}\psi A^{\mu}\bigg].\label{QED}
\end{eqnarray}
 The $\gamma^{\mu}$ are the famous $4\times 4$ Dirac gamma matrices which appear in any theory containing a spin $1/2$ field. They satisfy $\{{\gamma}^{\mu},{\gamma}^{\nu}\}=2\eta^{\mu\nu}$ where $\eta^{\mu\nu}={\rm diag}(1,-1,-1,-1)$. The electromagnetic field is given by the $U(1)$ gauge vector field $A^{\mu}$ with field strength $F_{\mu\nu}=\partial_{\mu}A_{\nu}-\partial_{\nu}A_{\mu}$ while the fermion (electron) field is given by the spinor field $\psi$ with mass $M$. The spinor $\psi $ is a $4-$component field and $\bar{\psi}={\psi}^{+}{\gamma}^{0}$. The interaction term is proportional to the electric charge $e$ given by the last term $-e\bar{\psi}\gamma_{\mu}\psi A^{\mu}$. The Euler-Lagrange classical equations of motion derived from the above action are precisely  the Maxwell equations $\partial_{\mu}F^{\mu\nu}~=~j^{\nu}$ with $j^{\mu}=e\overline\psi\gamma^{\mu}\psi$ and the Dirac equation $(i\gamma^{\mu}\partial_{\mu}-m-e\gamma_{\mu} A^{\mu})\psi=0$. The above theory is also invariant under the following $U(1)$ gauge transformations 

\begin{eqnarray}
A_{\mu}~\longrightarrow ~A_{\mu} + \partial_{\mu}\Lambda~~,~~\psi~\longrightarrow~\exp(-ie\Lambda)\psi~~,~~\bar{\psi}\longrightarrow \bar{\psi} \exp(ie\Lambda).
\end{eqnarray}
The Feynman path integral is
\begin{eqnarray}
Z
&=&\int {\cal D}A^{\mu}{\cal D}\bar{\psi}{\cal D}\psi\exp(iS_{\rm QED}).
\end{eqnarray}
Before we can study this theory numerically using the Monte Carlo method we need to:
\begin{enumerate}
\item Rotate to Euclidean signature in order to convert the theory into a statistical field theory.
\item Regularize the UV behavior of the theory by putting it on a lattice.
\end{enumerate}
As a consequence we obtain an ordinary statistical system accessible to ordinary sampling techniques such as the Metropolis algorithm.

We start by discussing a little further the above action. The free fermion action in Minkowski spacetime is given by
\begin{eqnarray}
S_F=\int d^4x\bar{\psi}(x)(i{\gamma}^{\mu}{\partial}_{\mu}-M)\psi(x).
\end{eqnarray}
This action is invariant under the global $U(1)$ transformation $\psi(x)\longrightarrow G\psi(x)$ and $\bar{\psi}(x)\longrightarrow \bar{\psi}(x)G^{-1}$ where $G=\exp(-i\Lambda)$. The symmetry $U(1)$ can be made local (i.e. $G$ becomes a function of $x$) by replacing the ordinary derivative ${\partial}_{\mu}$ with the covariant derivative $D_{\mu}={\partial}_{\mu}+ie A_{\mu}$ where the $U(1)$ gauge field $A_{\mu}$ is the electromagnetic $4-$vector potential. The action becomes
\begin{eqnarray}
S_F=\int d^4x\bar{\psi}(x)(i{\gamma}^{\mu}D_{\mu}-M)\psi(x).
\end{eqnarray}
This action is invariant under 
\begin{eqnarray}
&&{\psi}\longrightarrow G(x){\psi}~,~\bar{{\psi}}\longrightarrow \bar{{\psi}}G^{-1}(x),
\end{eqnarray}
provided we also transform the covariant derivative and the gauge field as follows
\begin{eqnarray}
D_{\mu}\longrightarrow G D_{\mu}G^{-1}~\Longleftrightarrow ~A_{\mu}\longrightarrow G(x)A_{\mu}G^{-1}(x)-\frac{i}{e}G(x){\partial}_{\mu}G^{-1}(x).
\end{eqnarray}
Since $A_{\mu}$ and $G(x)=\exp(-i\Lambda(x))$ commute the transformation law of the gauge field reduces to $A_{\mu}\longrightarrow A_{\mu}+{\partial}_{\mu}\Lambda/e$. The dynamics of the gauge field $A_{\mu}$ is given by the Maxwell action
 \begin{eqnarray}
S_G=-\frac{1}{4}\int d^4x F_{\mu\nu}F^{\mu\nu}~,~F_{\mu\nu}={\partial}_{\mu}A_{\nu}-{\partial}_{\nu}A_{\mu}.
\end{eqnarray}
This action is also invariant under the local $U(1)$ gauge symmetry $A_{\mu}\longrightarrow A_{\mu}+{\partial}_{\mu}\Lambda/e$. The total action is then
 \begin{eqnarray}
S_{\rm QED}=-\frac{1}{4}\int d^4x F_{\mu\nu}F^{\mu\nu}+\int d^4x\bar{\psi}(x)(i{\gamma}^{\mu}{D}_{\mu}-M)\psi(x).
\end{eqnarray}
This is precisely (\ref{QED}).

The Euclidean action $S_F^{\rm eucl}$ is obtained by i) making the replacement $x_0\longrightarrow -ix_4$ wherever $x_0$ appears explicitly, ii) substituting ${\psi}^{E} (x)=\psi(\vec{x},x_4)$ for $\psi(x)=\psi(\vec{x},t)$, iii) making the replacements $A^0\longrightarrow iA_4$ and  $D^0\longrightarrow iD_4$ and iv) multiplying the obtained expression by $-i$. Since in Euclidean space the Lorentz group is replaced by the $4-$dimensional rotation group we introduce new $\gamma-$matrices ${\gamma}_{\mu}^{E}$ as follows ${\gamma}_4^{E}={\gamma}^{0}$,${\gamma}_i^{E}=-i{\gamma}^{i}$. They satisfy $\{{\gamma}_{\mu}^{E},{\gamma}_{\nu}^{E}\}=2{\delta}_{\mu\nu}$. The fermion Euclidean action is then
\begin{eqnarray}
S_F^{\rm Eucl}=\int d^4x\bar{\psi}^{E}(x)({\gamma}_{\mu}^{E}{D}_{\mu}+M){\psi}^{E}(x).
\end{eqnarray}
Similarly the Euclidean action $S_G^{\rm eucl}$ is obtained by i) making the replacement $x_0\longrightarrow -ix_4$ wherever $x_0$ appears explicitly, ii) making the replacement $A^0\longrightarrow iA_4$ and iii) multiplying the obtained expression by $-i$. We can check that $F_{\mu\nu}F^{\mu\nu}$, $\mu,\nu=0,1,2,3$ will be replaced with $F_{\mu\nu}^2$, $\mu=1,2,3,4$. The gauge Euclidean  action is then
\begin{eqnarray}
S_G^{\rm Eucl}=\frac{1}{4}\int d^4x F_{\mu\nu}^2.
\end{eqnarray}
The full Euclidean action is
\begin{eqnarray}
S_{QED}^{\rm Eucl}=\frac{1}{4}\int d^4x F_{\mu\nu}^2+\int d^4x\bar{\psi}^{E}(x)({\gamma}_{\mu}^{E}{D}_{\mu}+M){\psi}^{E}(x).
\end{eqnarray}
 We will drop the labels ${\rm Eucl}$ in the following.
\section{Lattice Regularization}
\subsection{Lattice Fermions and Gauge Fields}
\paragraph{Free Fermions on the Lattice:}

The continuum free fermion action in Euclidean $4$d spacetime is
\begin{eqnarray}
S_F=\int d^4x\bar{\psi}^{E}(x)({\gamma}_{\mu}^{E}{\partial}_{\mu}+M){\psi}^{E}(x).\label{startingfermionicaction}
\end{eqnarray}
This has the symmetry ${\psi}{\longrightarrow}e^{i{\theta}}{\psi}$ and the symmetry 
${\psi}{\longrightarrow}e^{i{\theta}{\gamma}_5}{\psi}$  when $M=0$. The associated conserved currents are known to be given by
$J_{\mu}=\bar{\psi}{\gamma}_{\mu}{\psi}$ and $J_{\mu}^{5}=\bar{\psi}{\gamma}_{\mu}{\gamma}_5{\psi}$ where
${\gamma}_5={\gamma}_1{\gamma}_2{\gamma}_3{\gamma}_4$. It is also a known result that in the quantum theory one
can not maintain the conservation of both of these currents simultaneously in the presence of gauge fields.

A regularization which maintains exact chiral invariance of the above action can be achieved by
replacing the Euclidean four dimensional spacetime by a four dimensional  hypercubic lattice of $N^4$ sites.  Every point on the lattice is specified by $4$ integers  which we denote collectively by $n=(n_1,n_2,n_3,n_4)$ where $n_4$ denotes Euclidean time. Clearly each component of the $4-$vector $n$ is an integer in the range $-{N}/{2}{\leq}n_{\mu}{\leq}{N}/{2}$ with $N$ even. The lattice is assumed to be periodic. Thus $x_{\mu}=an_{\mu}$ where $a$ is the lattice spacing and $L=aN$ is the linear size of the lattice.  Now to each site $x=an$ we associate a spinor variable
${\psi}(n)=\psi(x)$ and the derivative ${\partial}_{\mu}{\psi}(x)$ is replaced by
\begin{equation}
{\partial}_{\mu}{\psi}(x){\longrightarrow}\frac{1}{a}\hat{\partial}_{\mu}{\psi}(n)=\frac{1}{2a}\Big[{\psi}(n+\hat{\mu})-{\psi}(n-\hat{\mu})\Big].
\end{equation}
The vector $\hat{\mu}$ is the unit vector in the $\mu-$direction. With this prescription the action (\ref{startingfermionicaction}) becomes (with $\hat{M}=aM$ and $\hat{\psi}=a^{{3}{/2}}\psi$)

\begin{eqnarray}
S_F&=&\sum_{n}\sum_m\sum_{\alpha}\sum_{\beta}\bar{\hat{\psi}}_{\alpha}(n)K_{\alpha \beta}(n,m)\hat{\psi}_{\beta}(m)\nonumber\\
K_{\alpha \beta}(n,m)&=&\frac{1}{2}\sum_{\mu}({\gamma}_{\mu})_{\alpha\beta}\bigg({\delta}_{m,n+\hat{\mu}}-{\delta}_{m,n-\hat{\mu}}\bigg)+\hat{M}{\delta}_{\alpha\beta}{\delta}_{m,n}.\label{discretetheory}
\end{eqnarray}

\paragraph{$U(1)$ Lattice Gauge Fields:}
The free fermion action on the lattice is therefore given by
\begin{eqnarray}
S_F&=&\hat{M} \sum_{n}\sum_{\alpha}\bar{\hat{\psi}}_{\alpha}(n)\hat{\psi}_{\alpha}(n)\nonumber\\
&-&\frac{1}{2}\sum_{n}\sum_{\alpha}\sum_{\beta}\sum_{\mu}\bigg[({\gamma}_{\mu})_{\alpha\beta}\bar{\hat{\psi}}_{\alpha}(n+\hat{\mu})\hat{\psi}_{\beta}(n)-({\gamma}_{\mu})_{\alpha\beta}\bar{\hat{\psi}}_{\alpha}(n)\hat{\psi}_{\beta}(n+\hat{\mu})\bigg].\nonumber\\
\end{eqnarray}
This action has the following global $U(1)$ symmetry
\begin{eqnarray}
&&\hat{\psi}_{\alpha}(n)\longrightarrow G\hat{\psi}_{\alpha}(n)~,~\bar{\hat{\psi}}_{\alpha}(n)\longrightarrow \bar{\hat{\psi}}_{\alpha}(n)G^{-1}.
\end{eqnarray}
The phase $G=\exp(-i\Lambda)$ is an element of $U(1)$. By requiring the theory to be invariant under local $U(1)$ symmetry, i.e. allowing $G$ to depend on the lattice site we arrive at a gauge invariant fermion action on the lattice. The problem lies in how we can make the bilinear fermionic terms (the second and third terms) in the above action gauge invariant.

We go back to the continuum formulation and see how this problem is solved. In the continuum the fermionic bilinear $\bar{\psi}(x)\psi(y)$ transforms under a local $U(1)$ transformation as follows
\begin{eqnarray}
\bar{\psi}(x)\psi(y)\longrightarrow \bar{\psi}(x)G^{-1}(x)G(y)\psi(y).
\end{eqnarray}
This bilinear can  be made gauge covariant by inserting the Schwinger line integral 
\begin{eqnarray}
U(x,y)=e^{ie\int_x^y dz_{\mu}A_{\mu}(z)},
\end{eqnarray}
which transforms as 
\begin{eqnarray}
U(x,y)\longrightarrow G(x)U(x,y)G^{-1}(y).
\end{eqnarray}
Therefore the fermionic bilinear 
\begin{eqnarray}
\bar{\psi}(x)U(x,y)\psi(y)=\bar{\psi}(x)e^{ie\int_x^y dz_{\mu}A_{\mu}(z)}\psi(y)
\end{eqnarray}
is  U$(1)$ gauge invariant. For $y=x+\epsilon$ we have
\begin{eqnarray}
U(x,x+\epsilon)=e^{ie{\epsilon}_{\mu}A_{\mu}(x)}.
\end{eqnarray}
We conclude that in order to get local $U(1)$ gauge invariance we replace  the second and third bilinear fermionic terms in the above action as follows
\begin{eqnarray}
&&\bar{\hat{\psi}}(n)(r-{\gamma}_{\mu})\hat{\psi}(n+\hat{\mu})\longrightarrow \bar{\hat{\psi}}(n)(r-{\gamma}_{\mu})U_{n,n+\hat{\mu}}\hat{\psi}(n+\hat{\mu})\nonumber\\
&&\bar{\hat{\psi}}(n+\hat{\mu})(r-{\gamma}_{\mu})\hat{\psi}(n)\longrightarrow \bar{\hat{\psi}}(n+\hat{\mu})(r-{\gamma}_{\mu})U_{n+\hat{\mu},n}\hat{\psi}(n).
\end{eqnarray}
We obtain then the action
\begin{eqnarray}
S_F&=&\hat{M}\sum_{n}\sum_{\alpha}\bar{\hat{\psi}}_{\alpha}(n)\hat{\psi}_{\alpha}(n)\nonumber\\
&-&\frac{1}{2}\sum_{n}\sum_{\alpha}\sum_{\beta}\sum_{\mu}\bigg[({\gamma}_{\mu})_{\alpha\beta}\bar{\hat{\psi}}_{\alpha}(n+\hat{\mu})U_{n+\hat{\mu},n}\hat{\psi}_{\beta}(n)-({\gamma}_{\mu})_{\alpha\beta}\bar{\hat{\psi}}_{\alpha}(n)U_{n,n+\hat{\mu}}\hat{\psi}_{\beta}(n+\hat{\mu})\bigg].\nonumber\\
\end{eqnarray}
The $U(1)$ element $U_{n,n+\hat{\mu}}$ lives on the lattice link connecting the two points $n$ and $n+\hat{\mu}$. This link variable is therefore a directed quantity given explicitly by
\begin{eqnarray}
U_{n,n+\hat{\mu}}=e^{i{\phi}_{\mu}(n)}\equiv U_{\mu}(n)~,~U_{n+\hat{\mu},n}=U^+_{n,n+\hat{\mu}}=e^{-i{\phi}_{\mu}(n)}\equiv U_{\mu}^+(n).
\end{eqnarray}
The second equality is much clearer in the continuum formulation but on the lattice it is needed for the reality of the action. The phase ${\phi}_{\mu}(n)$ belongs to the compact interval $ [0,2\pi]$. Alternatively we can work with $A_{\mu}(n)$ defined through
\begin{eqnarray}
{\phi}_{\mu}(n)=eaA_{\mu}(n).
\end{eqnarray}
Let us now consider the  product of link variables around the smallest possible closed loop on the lattice, i.e. a plaquette. For a plaquette in the $\mu-\nu$ plane we have 
\begin{eqnarray}
U_P\equiv U_{\mu\nu}(n)=U_{\mu}(n)U_{\nu}(n+\hat{\mu})U_{\mu}^+(n+\hat{\nu})U_{\nu}^+(n).
\end{eqnarray}
The links are path-ordered. We can immediately compute
\begin{eqnarray}
U_P\equiv U_{\mu\nu}(n)=e^{iea^2F_{\mu\nu}(n)}~,~F_{\mu\nu}=\frac{1}{a}\bigg[A_{\nu}(n+\hat{\mu})-A_{\nu}(n)-A_{\mu}(n+\hat{\nu})+A_{\mu}(n)\bigg].
\end{eqnarray}
In other words in the continuum limit $a\longrightarrow 0$ we have
\begin{eqnarray}
\frac{1}{e^2}\sum_n\sum_{\mu <\nu}\bigg[1-\frac{1}{2}\big(U_{\mu\nu}(n)+U_{\mu\nu}^+(n)\big)\bigg]=\frac{a^4}{4}\sum_n\sum_{\mu,\nu}F_{\mu\nu}^2.
\end{eqnarray}
The $U(1)$ gauge action on the lattice is therefore
\begin{eqnarray}
S_G=\frac{1}{e^2}\sum_{P}\bigg[1-\frac{1}{2}\big(U_p+U_p^+\big)\bigg].
\end{eqnarray}

\subsection{Quenched Approximation}
The QED partition function on a lattice $\Lambda$ is given by

\begin{eqnarray}
Z=\int {\cal D}U~ {\cal D}\bar{\hat{\psi}} {\cal D}\hat{\psi}~ e^{-S_G[U]-S_F[U , \bar{\hat{\psi}} , \hat{\psi}]}.
\end{eqnarray}
The measures are defined by

\begin{eqnarray}
 {\cal D}U=\prod_{n\in\Lambda}\prod_{\mu=1}^4dU_{\mu}(n)~,~{\cal D} \bar{\hat{\psi}} = \prod_{n\in\Lambda} d\bar{\hat{\psi}}(n)~,~
 {\cal D} \hat{\psi} = \prod_{n\in\Lambda} d\hat{\psi}(n).
\end{eqnarray}
The plaquette and the link variable are given by 
\begin{eqnarray}
U_{\mu\nu}(n)=U_{\mu}(n)U_{\nu}(n+\hat{\mu})U_{\mu}^+(n+\hat{\nu})U_{\nu}^+(n)~,~ U_{\mu}(n)=e^{i{\phi}_{\mu}(n)}.
\end{eqnarray}
The action of  a $U(1)$ gauge theory on a lattice  is given by (with $\beta=1/e^2$)
\begin{eqnarray}
S_G[U]&=&\beta \sum_{n\in\Lambda}\sum_{\mu<\nu}\bigg[1-\frac{1}{2}\big(U_{\mu\nu}(n)+U_{\mu\nu}^+(n)\big)\bigg]=\beta \sum_{n\in\Lambda}\sum_{\mu<\nu}{\rm Re}\bigg[1-U_{\mu\nu}(n)\bigg].
\end{eqnarray}
The action of fermions coupled to a $U(1)$ gauge field on a lattice is given by
\begin{eqnarray}
S_F[U , \bar{\hat{\psi}} , \hat{\psi}] = \sum_{\alpha} \sum_{\beta} \sum_{n} \sum_{m} \bar{\hat{\psi_{\alpha}}}(n)  {\cal D}_{\alpha\beta}(U)_{n,m} \hat{\psi_{\beta}}(m).
\end{eqnarray}
Where
\begin{eqnarray}
{\cal D}_{\alpha\beta}(U)_{n,m} = \hat{M} \delta_{\alpha\beta} \delta_{n,m} - \frac{1}{2}(\gamma_{\mu})_{\alpha\beta}~\delta_{n,m+\hat{\mu}}~U_{n+\hat{\mu},n}+\frac{1}{2}(\gamma_{\mu})_{\alpha\beta}~\delta_{m,n+\hat{\mu}}~U_{n,n+\hat{\mu}}.
\end{eqnarray}
Using the result

\begin{eqnarray}
\int {\cal D}\bar{\hat{\psi}} {\cal D}\hat{\psi}~e^{- \sum_{\alpha} \sum_{\beta} \sum_{n} \sum_{m} \bar{\hat{\psi_{\alpha}}}(n)  {\cal D}_{\alpha\beta}(U)_{n,m} \hat{\psi_{\beta}}(m)} ~= ~{\rm det} {\cal D}_{\alpha \beta} (U)_{n,m} .
\end{eqnarray}
The partition  function becomes

\begin{eqnarray}
Z=\int {\cal D} U~ {\rm det} {\cal D}_{\alpha\beta} (U)_{n,m}~ e^{-S_G[U]} .
\end{eqnarray}
At this stage we will make the approximation that we can set the determinal equal $1$, i.e. the QED partition function will be approximated by
\begin{eqnarray}
Z=\int {\cal D} U~ e^{-S_G[U]}
\end{eqnarray}
This is called the quenched approximation.
\subsection{Wilson Loop, Creutz Ratio and Other Observables}
The first observable we would like to measure is the expectation value of the action which after dropping the constant term is given by
\begin{eqnarray}
<S_G[U]>&=&-\beta \sum_{n\in\Lambda}\sum_{\mu<\nu} <{\rm Re}~U_{\mu\nu}(n)>.
\end{eqnarray}
The specific heat is the corresponding second moment, viz
\begin{eqnarray}
C_v&=&<S_G[U]^2>-<S_G[U]>^2.
\end{eqnarray}
We will also measure the expectation value of the so-called Wilson loop which has a length $I$ in one of the spatial direction (say $1$) and a width $J$ in the temporal direction $4$. This rectangular loop $C$ is defined by
\begin{eqnarray}
W_C[U]=S(n,n+I\hat{1})T(n+I\hat{1},n+I\hat{1}+J\hat{4})S^+(n+J\hat{4},n+I\hat{1}+J\hat{4})T^+(n,n+J\hat{4}).
\end{eqnarray}
The Wilson lines are
\begin{eqnarray}
S(n,n+I\hat{1})=\prod_{i=0}^{I-1}U_1(n+i\hat{1})~,~S(n+J\hat{4},n+I\hat{1}+J\hat{4})=\prod_{i=0}^{I-1}U_1(n+i\hat{1}+J\hat{4}).
\end{eqnarray}
The temporal transporters are
\begin{eqnarray}
T(n+I\hat{1},n+I\hat{1}+J\hat{4})=\prod_{j=0}^{J-1}U_4(n+I\hat{1}+j\hat{4})~,~T(n,n+J\hat{4})=\prod_{j=0}^{J-1}U_4(n+j\hat{4}).
\end{eqnarray}
The expectation value of $W_C[U]$ will be denoted by
\begin{eqnarray}
W[I,J]=\frac{\int {\cal D}U~W_C[U]~e^{-S_G[U]}}{\int {\cal D}U~e^{-S_G[U]}}.
\end{eqnarray}
 By using the fact that under  $\phi_{\mu}(n)\longrightarrow -\phi_{\mu}(n)$, the partition function is invariant while the Wilson loop changes its orientation, i.e. $W_C[U]\longrightarrow W_C[U]^+$, we obtain
\begin{eqnarray}
W[I,J]=<{\rm Re}~W_C[U]>.\label{expWilson}
\end{eqnarray}
It is almost obvious that in the continuum limit
 \begin{eqnarray}
W[I,J]\longrightarrow W[R,T]=<\exp(ie\oint_C dx_{\mu}A_{\mu})>.
\end{eqnarray}
The loop $C$ is now a rectangular contour with spatial length $R=Ia$ and timelike  length $T=Ja$. This represents the probability amplitude for the process of creating an infinitely heavy, i.e. static, quark-antiquark \footnote{For $U(1)$ we should really speak of an electron-positron pair.} pair at time $t=0$ which are separated by a distance $R$, then allowing them to evolve in time and then eventually annihilate after a long time $T$.

The precise meaning of the expectation value (\ref{expWilson}) is as follows
\begin{eqnarray}
< {\cal O} >=\frac{1}{L}\sum_{i=1}^L\bigg(\frac{1}{N^3N_T}\sum_n{\rm Re}~W_C[U_i]\bigg).
\end{eqnarray}
In other words we also take the average over the lattice which is necessary in order to reduce noise in the measurment of the Creutz ratio (see below).

The above Wilson loop is the order parameter of the pure $U(1)$ gauge theory. For large time $T$ we expect the behavior 
\begin{eqnarray}
W[R,T\longrightarrow\infty]\longrightarrow e^{-V(R)T}=e^{-aV(R)J},
\end{eqnarray}
where $V(R)$ is the static quark-antiquark potential. For strong coupling (small $\beta$) we can show that the potential is linear, viz
\begin{eqnarray}
V(R)=\sigma R.
\end{eqnarray}
The constant $\sigma$ is called the string tension from the fact that the force between the quark and the antiquark can be modeled by the force in a string attached to the quark and antiquark. For a linear potential the Wilson loop follows an area law $W[R,T]=\exp(-\sigma A)$ with $A=a^2IJ$. This behavior is typical in a confining phase which occurs at high temperature.

For small coupling (large $\beta$,low temperature) the lattice $U(1)$ gauge field becomes weakly coupled and as a consequence we expect the Coulomb potential to dominate the static quark-antiquark potential, viz
\begin{eqnarray}
V(R)=\frac{Z}{R}.
\end{eqnarray}
Hence for large $R$ the quark and antiquark become effectively free and their energy is simply the sum of their self-energies. The Wilson loop in this case follows a perimeter law $W[R,T]=\exp(-2\epsilon T)$.

In summary for a rectangular $R\times T$ Wilson loop with perimeter $P=2(R+T)$ and area $A=RT$ we expect the behavior 
\begin{eqnarray}
W[R,T]=e^{-\sigma A}~,~{\rm confinement}~{\rm phase}.
\end{eqnarray}
\begin{eqnarray}
W[R,T]=e^{-\epsilon P}~,~{\rm coulomb}~{\rm phase}.
\end{eqnarray}
In general the Wilson loop will behave as
\begin{eqnarray}
W[R,T]=e^{-B-\sigma A-\epsilon P}.
\end{eqnarray}
The perimeter piece actually dominates for any fixed size loop. To measure the string tension we must therefore eliminate the perimeter behavior which can be achieved using the so-called Creutz ratio defined by
\begin{eqnarray}
\chi(I,J)=-\ln\frac{W[I,J]W[I-1,J-1]}{W[I,J-1]W[I-1,J]}.
\end{eqnarray}
For large loops clearly
\begin{eqnarray}
\chi(I,J)=a^2\sigma.
\end{eqnarray}
This should holds especially in the confinement phase whereas in the Coulomb phase we should expect $\chi(I,J)\sim 0$.

 The $1\times 1$ Wilson loop $W(1,1)$ is special since it is related to the average action per  plaquette. We have
\begin{eqnarray}
W[1,1]=<{\rm Re}~U_1(n)U_4(n+\hat{1})U_4^+(n)U_1^+(n+\hat{4})>.
\end{eqnarray}
Next we compute straightforwardly 
\begin{eqnarray}
-\frac{\partial \ln Z}{\partial \beta}=\sum_n\sum_{\mu<\nu}<[1-{\rm Re}~U_{\mu\nu}(n)]>.
\end{eqnarray}
Clearly all the planes $\mu\nu$ are equivalent and thus we should have
\begin{eqnarray}
-\frac{\partial \ln Z}{\partial \beta}&=&6\sum_n<[1-{\rm Re}~U_{14}(n)]>\nonumber\\
&=&6\sum_n<[1-{\rm Re}~U_1(n)U_4(n+\hat{1})U_4^+(n)U_1^+(n+\hat{4})]>.
\end{eqnarray}
Remark that there are $N^3N_T$ lattice sites. Each site corresponds to $4$ plaquettes in every plane $\mu\nu$ and thus it corresponds to $4\times 6$ plaquettes in all. Each plaquette in a plane $\mu\nu$ corresponds to $4$ sites and thus to avoid overcounting we must divide by $4$. In summary we have $4\times 6\times N^3\times N_T/4$ plaquettes in total. Six is therefore the ratio of the number of plaquettes to the number of sites. 

We have then
\begin{eqnarray}
-\frac{1}{6N^3N_T}\frac{\partial \ln Z}{\partial \beta}
&=&1-\frac{1}{N^3N_T}\sum_n<{\rm Re}~U_1(n)U_4(n+\hat{1})U_4^+(n)U_1^+(n+\hat{4})>.
\end{eqnarray}
We can now observe that all lattice sites $n$ are the same under the expectation value, namely
\begin{eqnarray}
-\frac{1}{6N^3N_T}\frac{\partial \ln Z}{\partial \beta}
&=&1-<{\rm Re}~U_1(n)U_4(n+\hat{1})U_4^+(n)U_1^+(n+\hat{4})>.
\end{eqnarray}
This is the average action per  plaquette (the internal energy) denoted by
\begin{eqnarray}
P=-\frac{1}{6N^3N_T}\frac{\partial \ln Z}{\partial \beta}
&=&1-W[1,1].\label{P}
\end{eqnarray}

\section{Monte Carlo Simulation of Pure $U(1)$ Gauge Theory}
\subsection{The Metropolis Algorithm}
The action of pure U$(1)$ gauge theory, the corresponding partition function and the measure of interest are given  on a lattice $\Lambda$ respectively by (with $\beta=1/e^2$)
\begin{eqnarray}
S_G[U]
&=&\beta \sum_{n\in\Lambda}\sum_{\mu<\nu}{\rm Re}\bigg[1-U_{\mu\nu}(n)\bigg].
\end{eqnarray}
\begin{eqnarray}
Z=\int {\cal D}U~e^{-S_G[U]}.
\end{eqnarray}
\begin{eqnarray}
{\cal D}U=\prod_{n\in\Lambda}\prod_{\mu=1}^4dU_{\mu}(n).
\end{eqnarray}
The vacuum expectation value of any observable ${\cal O}={\cal O}(U)$ is given by
\begin{eqnarray}
< {\cal O} >=\frac{1}{Z}\int {\cal D}U~{\cal O}~e^{-S_G[U]}.\label{measurment}
\end{eqnarray}
For $U(1)$ gauge theory we can write
\begin{eqnarray}
U_{\mu}(n)=e^{i\phi_{\mu}(n)}.
\end{eqnarray}
Hence
\begin{eqnarray}
{\cal D}U=\prod_{n\in\Lambda}\prod_{\mu=1}^4d\phi_{\mu}(n).
\end{eqnarray}
We will use the Metropolis algorithm to solve this problem. This  goes as follows. Starting from a given gauge field configuration, we choose a lattice point $n$ and a direction $\mu$, and change the link variable there, which is $U_{\mu}(n)$, to $U_{\mu}(n)^{'}$. This link is shared by $6$ plaquettes. The corresponding variation of the action is
 \begin{eqnarray}
\Delta S_G[U_{\mu}(n))]&=&S_G[U^{'}]-S_G[U].
\end{eqnarray}
The gauge field configurations $U$ and $U^{'}$ differ only by the value of the link variable $U_{\mu}(n)$. We need to isolate the contribution of $U_{\mu}(n)$ to the action $S_G$. Note the fact that $U_{\mu\nu}^+=U_{\nu\mu}$. 
We write
\begin{eqnarray}
S_G[U]&=&\beta \sum_{n\in\Lambda}\sum_{\mu<\nu}1-\frac{\beta}{2}\sum_{n\in\Lambda}\sum_{\mu<\nu}\big(U_{\mu\nu}(n)+U_{\mu\nu}^+(n)\big).
\end{eqnarray}
The second term is
\begin{eqnarray}
-\frac{\beta}{2}\sum_{n\in\Lambda}\sum_{\mu<\nu}U_{\mu\nu}(n)
&=&-\frac{\beta}{2}\sum_{n\in\Lambda}\sum_{\mu<\nu}U_{\mu}(n)U_{\nu}(n+\hat{\mu})U_{\mu}^+(n+\hat{\nu})U_{\nu}^+(n).
\end{eqnarray}
In the $\mu-\nu$ plane, the link variable $U_{\mu}(n)$ appears twice corresponding to the two lattice points $n$ and $n-\hat{\nu}$. For every $\mu$ there are three relevant planes. The six relevant terms are therefore given by
\begin{eqnarray}
-\frac{\beta}{2}\sum_{n\in\Lambda}\sum_{\mu<\nu}U_{\mu\nu}(n)
\longrightarrow &-&\frac{\beta}{2}\sum_{\nu\neq \mu}\bigg(U_{\mu}(n)U_{\nu}(n+\hat{\mu})U_{\mu}^+(n+\hat{\nu})U_{\nu}^+(n)\nonumber\\
&+&U_{\mu}^+(n)U_{\nu}^+(n-\hat{\nu})U_{\mu}(n-\hat{\nu})U_{\nu}(n-\hat{\nu}+\hat{\mu})\bigg)+...
\end{eqnarray}
By adding the complex conjugate terms we obtain
 \begin{eqnarray}
-\frac{\beta}{2}\sum_{n\in\Lambda}\sum_{\mu<\nu}(U_{\mu\nu}(n)+U_{\mu\nu}^+(n))
\longrightarrow &-&\frac{\beta}{2}\bigg(U_{\mu}(n){\cal A}_{\mu}(n)+U_{\mu}^+(n){\cal A}_{\mu}^+(n)\bigg)+...
\end{eqnarray}
The ${\cal A}_{\mu}(n)$ is the sum over the six so-called staples which are  the products over the other three link variables which together with  $U_{\mu}(n)$  make up the six plaquettes which share  $U_{\mu}(n)$. Explicitly we have
\begin{eqnarray}
{\cal A}_{\mu}(n)=\sum_{\nu\neq \mu}\bigg(U_{\nu}(n+\hat{\mu})U_{\mu}^+(n+\hat{\nu})U_{\nu}^+(n)+U_{\nu}^+(n+\hat{\mu}-\hat{\nu})U_{\mu}^+(n-\hat{\nu})U_{\nu}(n-\hat{\nu})\bigg).
\end{eqnarray}
We have then the result
\begin{eqnarray}
-\frac{\beta}{2}\sum_{n\in\Lambda}\sum_{\mu<\nu}(U_{\mu\nu}(n)+U_{\mu\nu}^+(n))
\longrightarrow &-&\beta{\rm Re}(U_{\mu}(n){\cal A}_{\mu}(n))+...
\end{eqnarray}
We compute then
 \begin{eqnarray}
\Delta S_G[U_{\mu}(n))]&=&S_G[U^{'}]-S_G[U]\nonumber\\
&=&-\beta(U_{\mu}(n)^{'}-U_{\mu}(n)){\cal A}_{\mu}(n).
\end{eqnarray}
Having computed the variation $\Delta S_G[U_{\mu}(n))]$, next we inspect its sign. If this variation is negative then the proposed change  $U_{\mu}(n)\longrightarrow U_{\mu}(n)^{'}$ will be accepted (classical mechanics). If the variation is positive, we compute the Boltzmann probability
 \begin{eqnarray}
\exp(-\Delta S_G[U_{\mu}(n))])
&=&\exp(\beta(U_{\mu}(n)^{'}-U_{\mu}(n)){\cal A}_{\mu}(n)).
\end{eqnarray}
The proposed change  $U_{\mu}(n)\longrightarrow U_{\mu}(n)^{'}$ will be accepted according to this probability (quantum mechanics). In practice we will pick a uniform random number $r$ between $0$ and $1$ and compare it with $\exp(-\Delta S_G[U_{\mu}(n))])$. If $\exp(-\Delta S_G[U_{\mu}(n))])<r$ we accept this change otherwise we reject it.

We go through the above steps for every link in the lattice which constitutes one Monte Carlo step. Typically equilibration (thermalization) is reached after a large number of Monte Carlo steps at which point we can start taking measurements based on the formula (\ref{measurment}) written as
\begin{eqnarray}
< {\cal O} >=\frac{1}{L}\sum_{i=1}^L{\cal O}_i~,~{\cal O}_i={\cal O}(U_i).
\end{eqnarray}
The $L$ configurations $U_i=\{U_{\mu}(n)\}_i$ are $L$ thermalized gauge field configurations distributed according to $\exp(-S_G[U])$. 

The error bars in the  different measurements will be estimated using the jackknife method. We can also compute auto-correlation time and take it into account by separating the measured gauge field configurations $U_i$ by at least one unit of auto-correlation time.
 
Let us also comment on how we choose the proposed configurations $U_{\mu}(n)^{'}$. The custom is to take $U_{\mu}(n)^{'}=XU_{\mu}(n)$ where $X$ is an element in the gauge group (which is here $U(1)$) near the identity. In order to maintain a symmetric selection probability, $X$ should be drawn randomly from a set of $U(1)$ elements  which contains also $X^{-1}$. For $U(1)$ gauge group we have $X=\exp(i\phi)$ where $\phi\in[0,2\pi]$. In principle the acceptance rate can be maintained around at least $0.5$ by tuning appropriately the angle $\phi$. Reunitarization of  $U_{\mu}(n)^{'}$ may also be applied to reduce rounding errors. 

The final technical remark is with regard to boundary conditions. In order to reduce edge effects we usually adopt periodic boundary conditions, i.e. 
\begin{eqnarray}
&&U_{\mu}(N,n_2,n_3,n_4) = U_{\mu}(0,n_2,n_3,n_4)  ,  U_{\mu}(n_1,N,n_3,n_4) = U_{\mu}(n_1,0,n_3,n_4),\nonumber\\
&&U_{\mu}(n_1,n_2,N,n_4) = U_{\mu}(n_1,n_2,n,0,n_4)   ,  U_{\mu}(n_1,n_2,n_3,N_T) = U_{\mu}(n_1,n_2,n_3,0).
\end{eqnarray}
This means in particular that the lattice is actually a four dimensional torus. In the actual code this is implemented by replacing $i\pm1$ by ${\rm ip}(i)$ and ${\rm im}(i)$, ${\rm ipT}(i)$ and ${\rm imT}(i)$ respectively which are defined by

\begin{verbatim}
do i=1,N
   ip(i)=i+1
   im(i)=i-1
enddo
   ip(N)=1
   im(1)=N
do i=1,NT
   ipT(i)=i+1
   imT(i)=i-1
enddo
   ipT(NT)=1
   imT(1)=NT
\end{verbatim}
A code written along the above lines is attached in the last chapter.     

\subsection{Some Numerical Results}
\begin{figure}[htbp]
\begin{center}
\includegraphics[width=10cm,angle=-0]{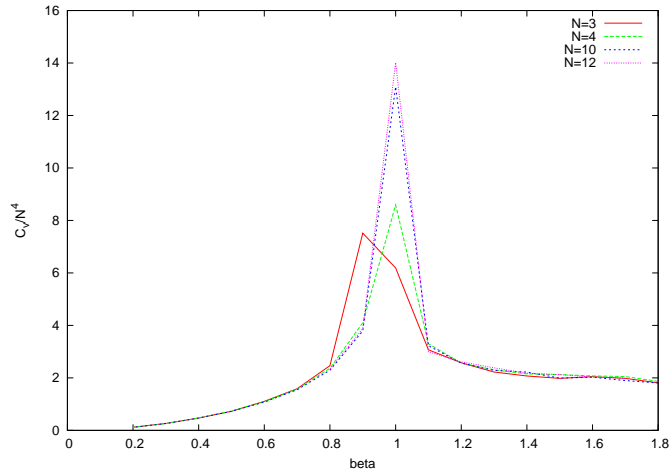}
\caption{The specific heat on a $3^4$, $4^4$, $10^4$ and $12^4$ lattices.}\label{latticeU1cv}
\end{center}
\end{figure}

\begin{figure}[htbp]
\begin{center}
\includegraphics[width=10cm,angle=-0]{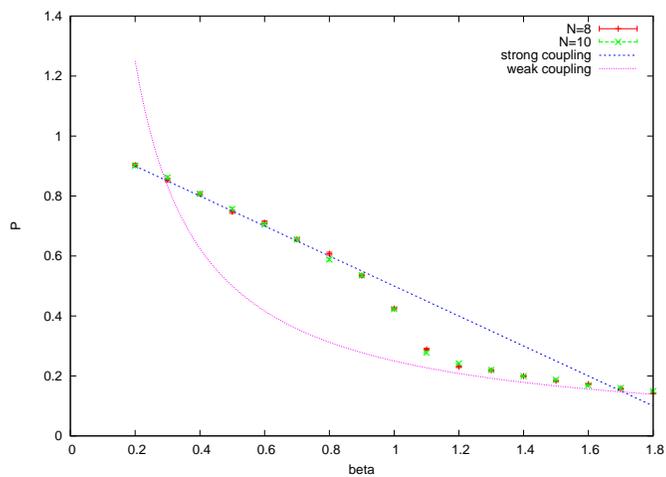}
\caption{The action per plaquette on a $8^4$ and $10^4$ lattices.}\label{latticeU1P}
\end{center}
\end{figure}

\begin{figure}[htbp]
\begin{center}
\includegraphics[width=10cm,angle=-0]{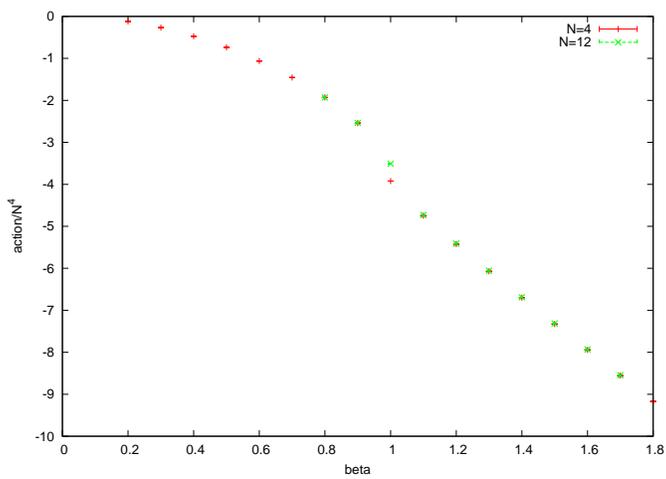}
\caption{The action on a $4^4$ and $12^4$ lattices.}\label{latticeU1S}
\end{center}
\end{figure}

\begin{figure}[htbp]
\begin{center}
\includegraphics[width=10cm,angle=-0]{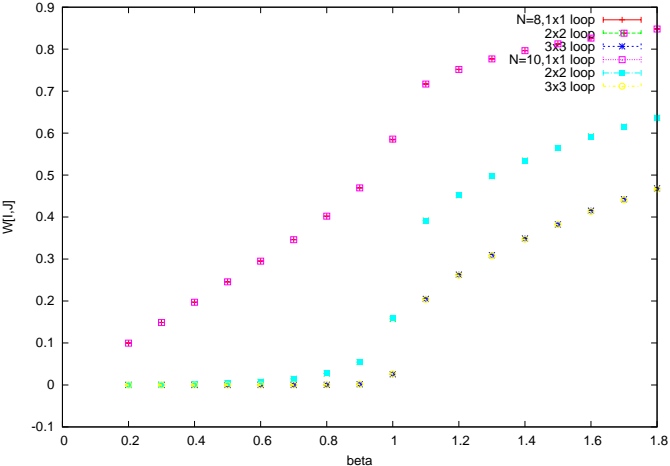}
\caption{The Wilson loop as a function of the inverse coupling strength $\beta$.}\label{latticeU1W}
\end{center}
\end{figure}

\begin{figure}[htbp]
\begin{center}
\includegraphics[width=10cm,angle=-0]{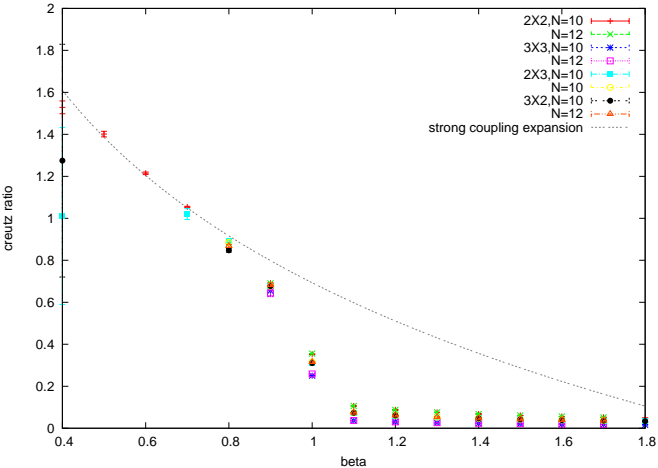}
\caption{String tension from Creutz ratio as a function of $\beta$ on a $12^4$ lattice.}\label{latticeU1C}
\end{center}
\end{figure}

\begin{enumerate}
\item We run simulations for $N = 3,4,8,10,12$ with the coupling constant in the range $\beta=2,...,12$. We use typically $2^{14}$ thermalization steps and $2^{14}$ measurements steps.

\item We measure the specific heat (figure (\ref{latticeU1cv})). We observe a peak in the specific heat at around $\beta=1$. The peak grows with N which signals a critical behavior typical of 2nd order transition.
\\
\item The simplest order parameter is the action per plaquette $P$, defined in equation (\ref{P}), which is shown on figure (\ref{latticeU1P}). We observe good agreement between the high-temperature and low-temperature expansions of $P$ from one hand and the corresponding observed behavior in the strong coupling and weak coupling regions respectively from the other hand. We note that  the high-temperature and low-temperature expansions of the pure $U(1)$ gauge field are given by
\begin{eqnarray}
P=1-\frac{\beta}{2}+O(\beta^3)~,~{\rm high}~T.
\end{eqnarray}
\begin{eqnarray}
P=1-\frac{1}{4\beta}+O(1/\beta^2)~,~{\rm low}~T.
\end{eqnarray}
We do not observe a clear-cut discontinuity in $P$ which is, in any case, consistent with the conclusion that this phase is second order. We note that for higher $U(N)$ the transition is first order \cite{Creutz:1984mgf}. 

A related object to $P$ is the total action shown on figure (\ref{latticeU1S}).

\item A more powerful order parameters are the Wilson loops which are shown on figure  (\ref{latticeU1W}). We observe that the Wilson loop in the strong coupling region averages to zero very quickly as we increase the size of the loop. This may be explained by an area law behavior. In the weak coupling region, the evolution as a function of the area is much more slower. The demarcation  between the two phases becomes very sharp (possibly a jump) for large loops at $\beta=1$. 

\item Calculating the expectation value of the Wilson loop and then extracting the string tension is very difficult since the perimeter law is dominant more often. The Creutz ratios  (figure  (\ref{latticeU1C})) allow us to derive the string tension in a direct way without measuring the Wilson loop. The string tension is the coefficient of the linearly rising part
of the potential for large (infinite) separations of a quark-antiquark pair in the absence of pair production processes.
In this way, we hope to measure the physical string tension in a narrow
range of the coupling constant. 

We observe that the string tension in the weak coupling regime is effectively independent of the coupling constant and it is essentially zero. In the strong coupling regime we reproduce the strong coupling behavior 
\begin{eqnarray}
\sigma=-\ln\frac{\beta}{2}.
\end{eqnarray}

\end{enumerate}

\subsection{Coulomb and Confinement Phases}
The physics of the compact $U(1)$ theory is clearly different in the weak- and strong-coupling regions. This can be understood from the fact that there is a phase transition as a function of the bare coupling constant. The compact $U(1)$ theory at weak coupling is not confining and contains no glueballs but simply the photons of the free Maxwell theory. One speaks of a Coulomb phase at weak coupling and a confining phase at strong coupling.
In the Coulomb phase photons are massless and the static potential has the standard Coulomb form
\begin{eqnarray}
V=-\frac{e^2}{4\pi r} + {\rm constant},
\end{eqnarray}
whereas in the confinement phase photons become massive and the potential is linearly confining at large distances
\begin{eqnarray}
V=\sigma r.
\end{eqnarray}
There is a phase transition at a critical coupling $\beta \approx 1$ at which the string tension $\sigma(\beta)$ vanishes in the Coulomb phase. In the confinement phase topological configurations are important such as monopoles and glueballs.

The strong-coupling expansion is an expansion in powers of $1/g^2$. It has the advantage over the weak-coupling expansion that it has a non-zero radius of convergence. A lot of effort has been put into using it as a method of computation similar to the high-temperature or the hopping parameter expansion for scalar field theories.
One has to be able to tune on the values of the coupling constant where the theory exhibits continuum behavior. This turns out to be difficult for gauge theories.
However, a very important aspect of the strong-coupling expansion is that it gives insight into the qualitative behavior of the theory such as confinement and the particle spectrum. 

The strong-coupling expansion of  compact $U(1)$ theory shows explicitly that the theory is confining, i.e. the potential is linear with a string tension given by (with $a_1=\beta/2$)
\begin{eqnarray}
\sigma &=&-\ln a_{1} - 2(d-2)a_{1}^{4} +....  
\end{eqnarray}


\chapter{Codes}
\newpage
\addcontentsline{toc}{section}{$9.1$~metropolis-ym.f}  

\includepdf[pages=-,]{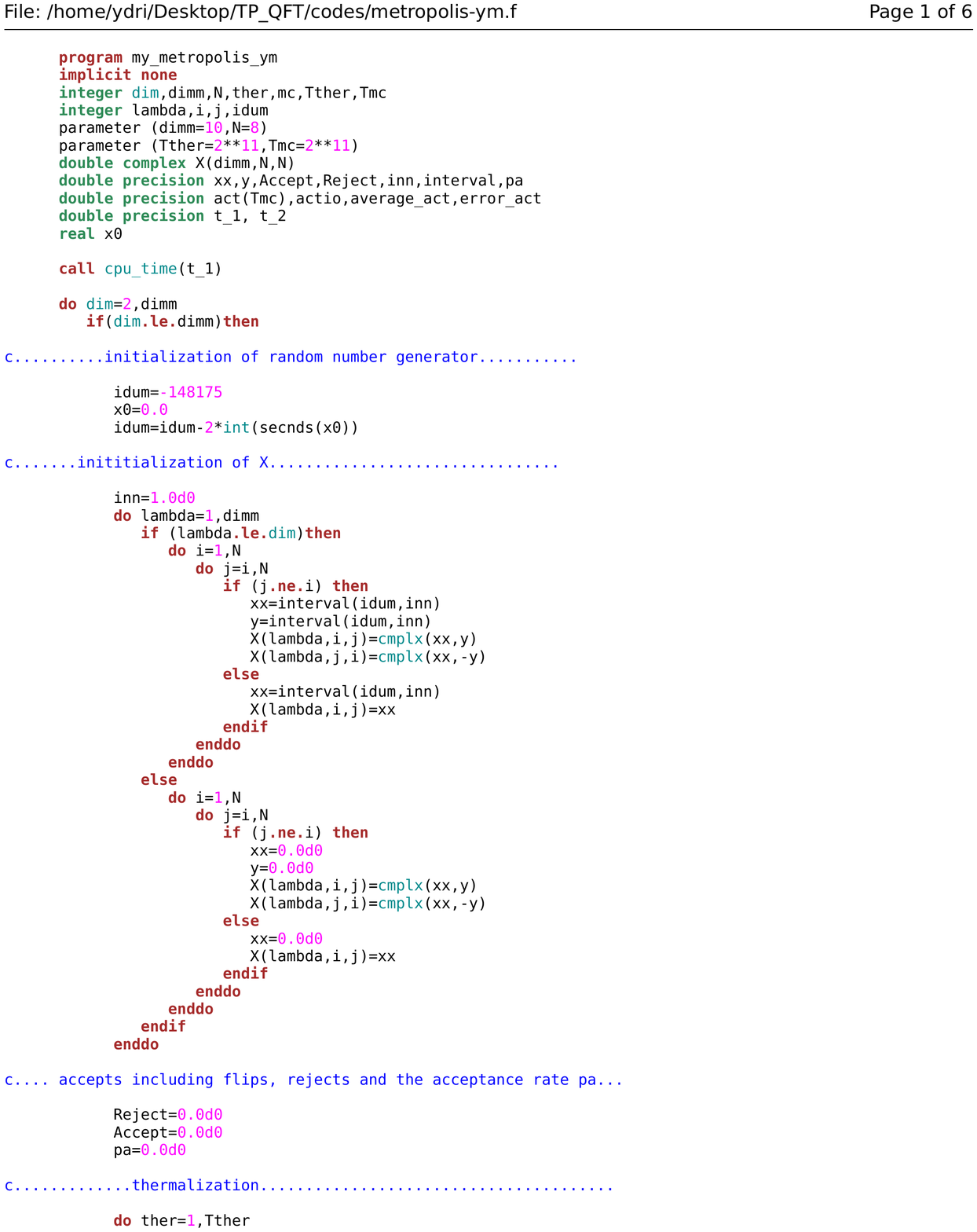}

\addcontentsline{toc}{section}{$9.2$~hybrid-ym.f}  

\includepdf[pages=-,]{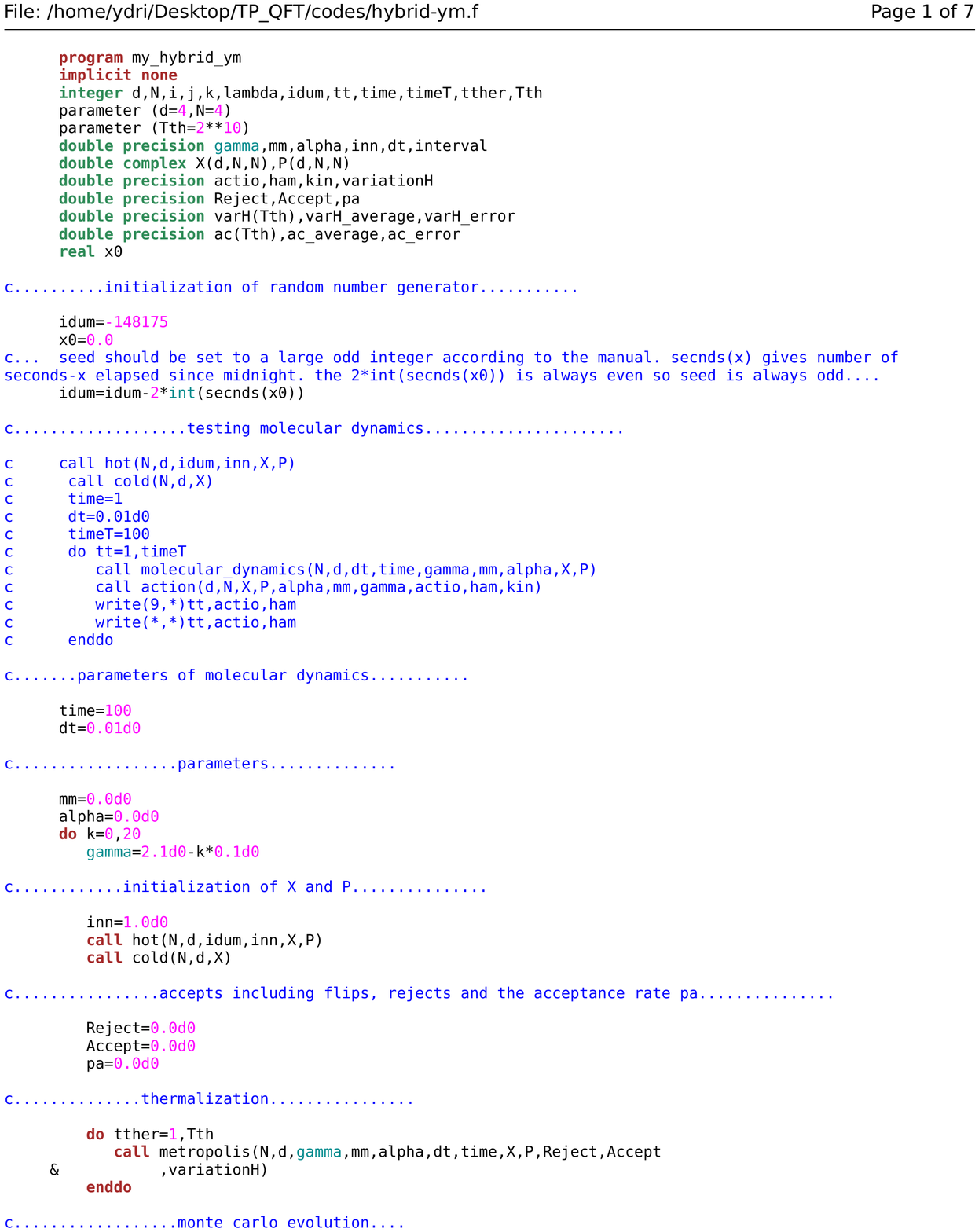}

\addcontentsline{toc}{section}{$9.3$~hybrid-scalar-fuzzy.f} 

\includepdf[pages=-,]{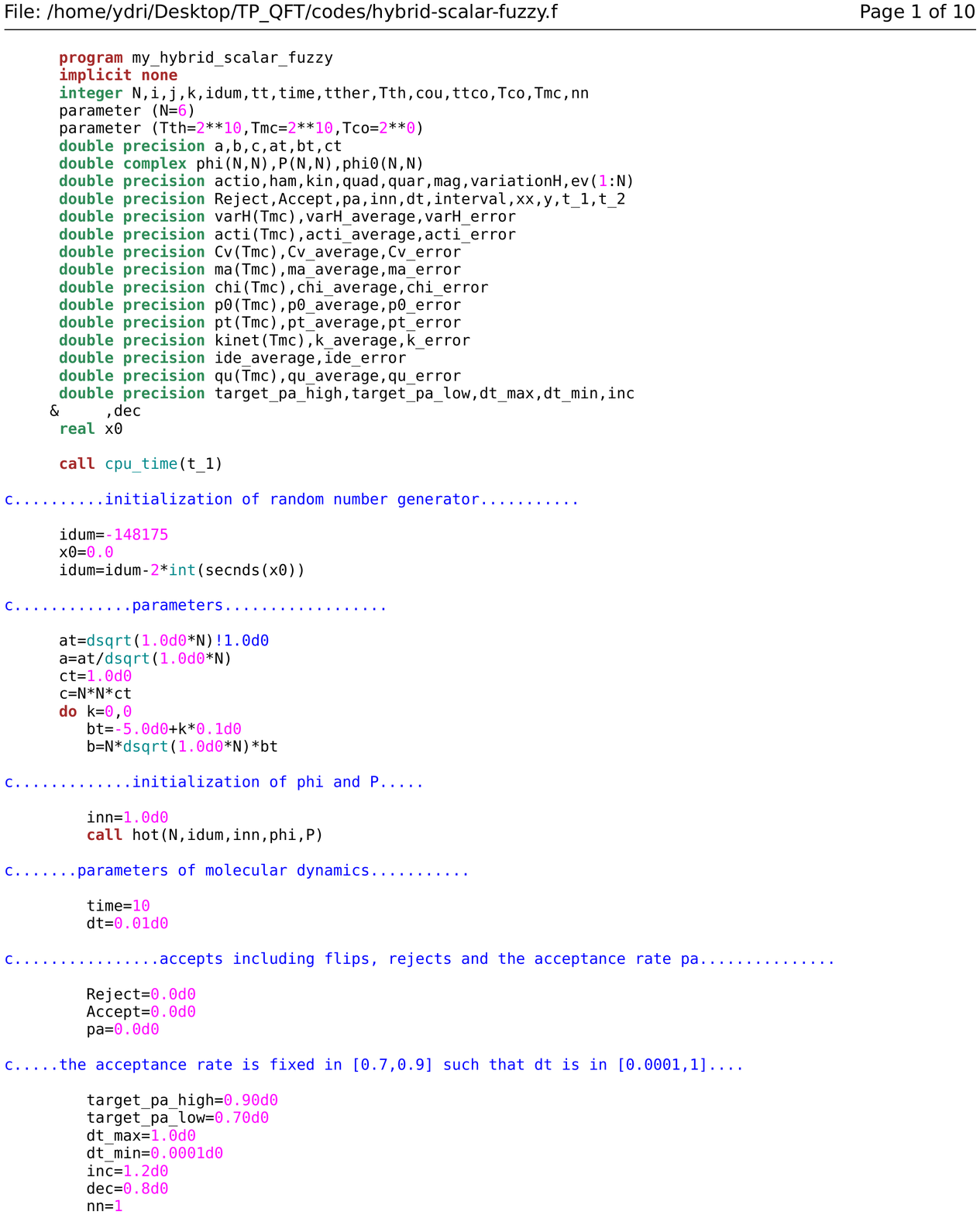}

\addcontentsline{toc}{section}{$9.4$~phi-four-on-lattice.f} 

\includepdf[pages=-,]{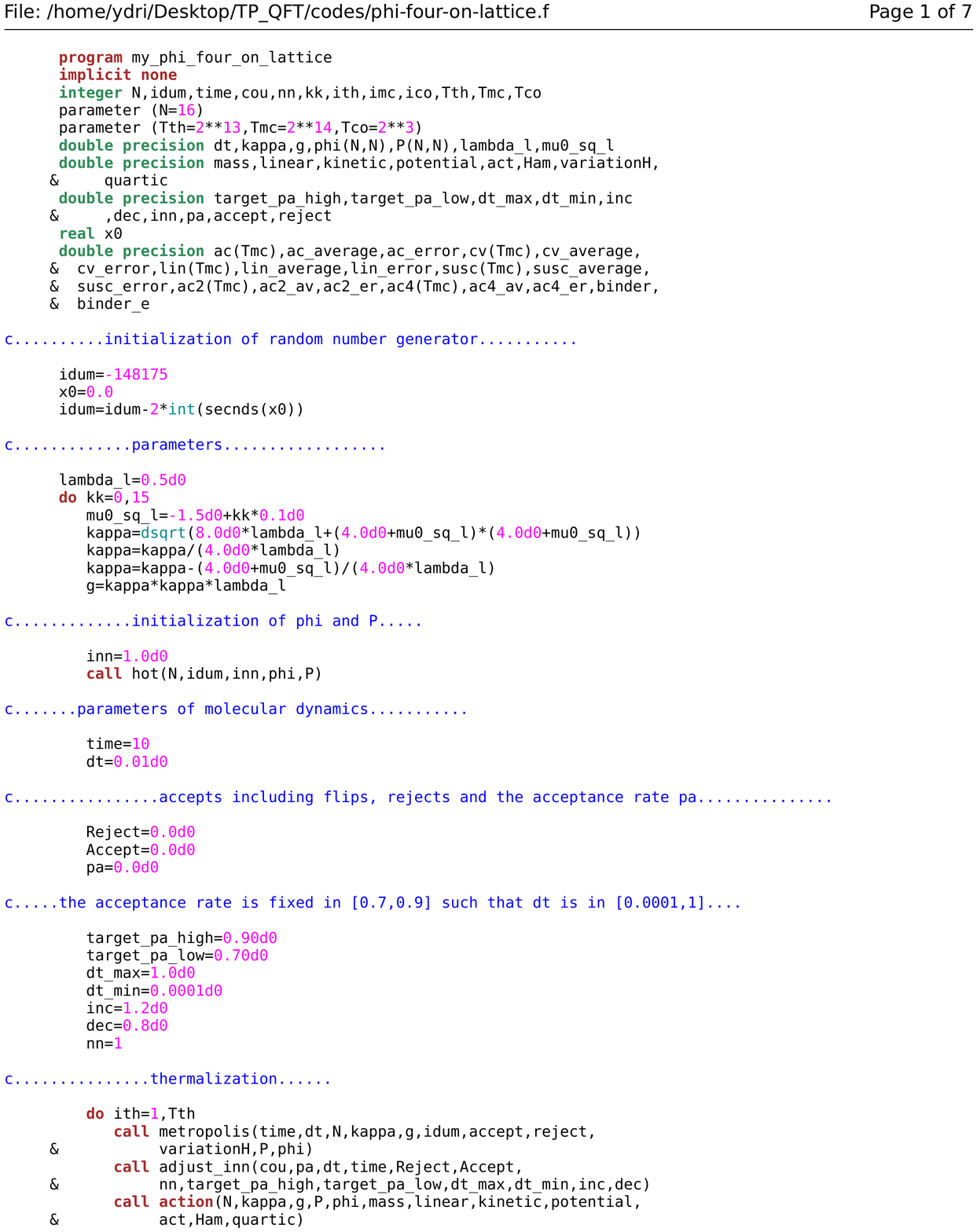}

\addcontentsline{toc}{section}{$9.5$~metropolis-scalar-multitrace.f} 

\includepdf[pages=-,]{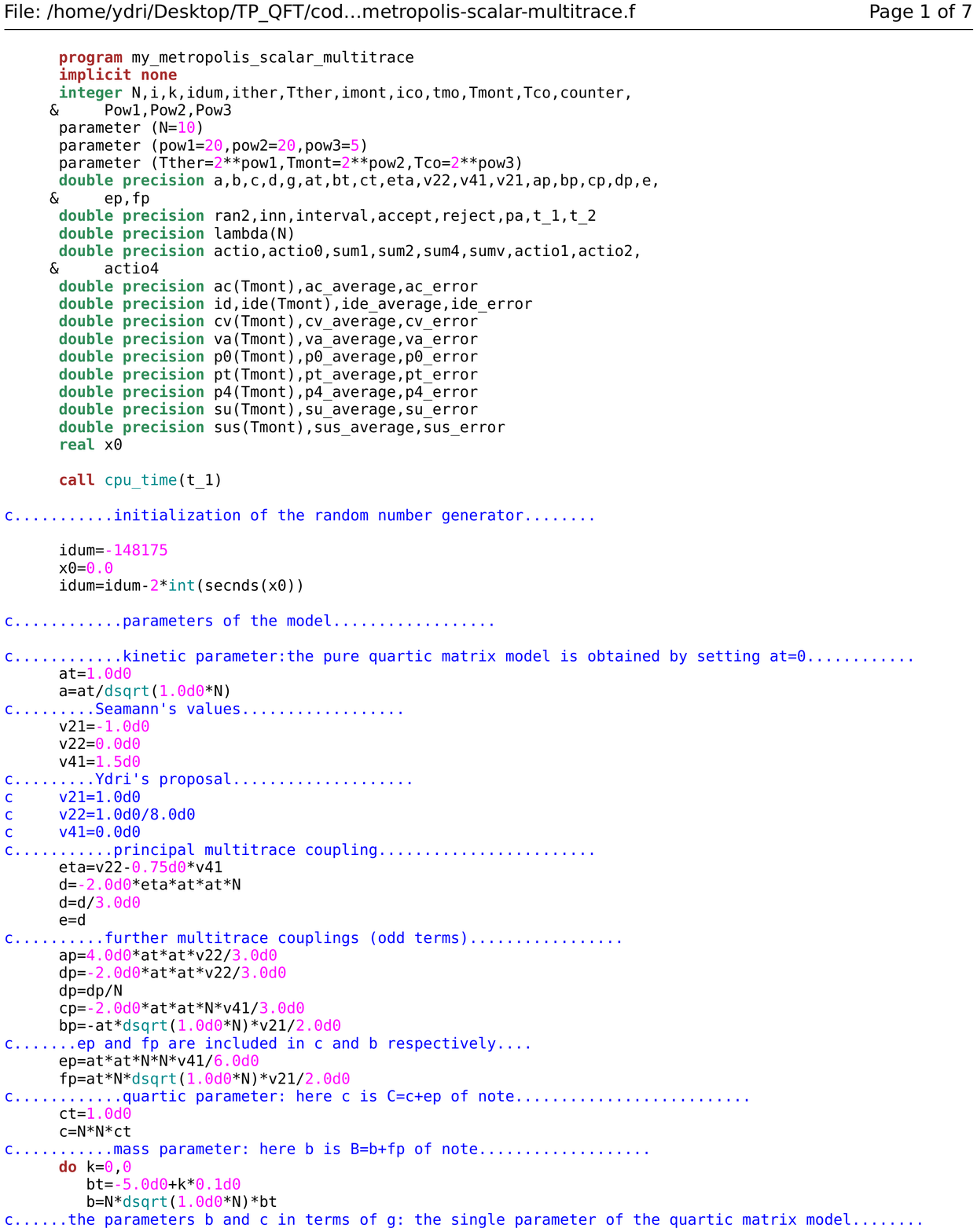}

\addcontentsline{toc}{section}{$9.6$~remez.f} 

\includepdf[pages=-,]{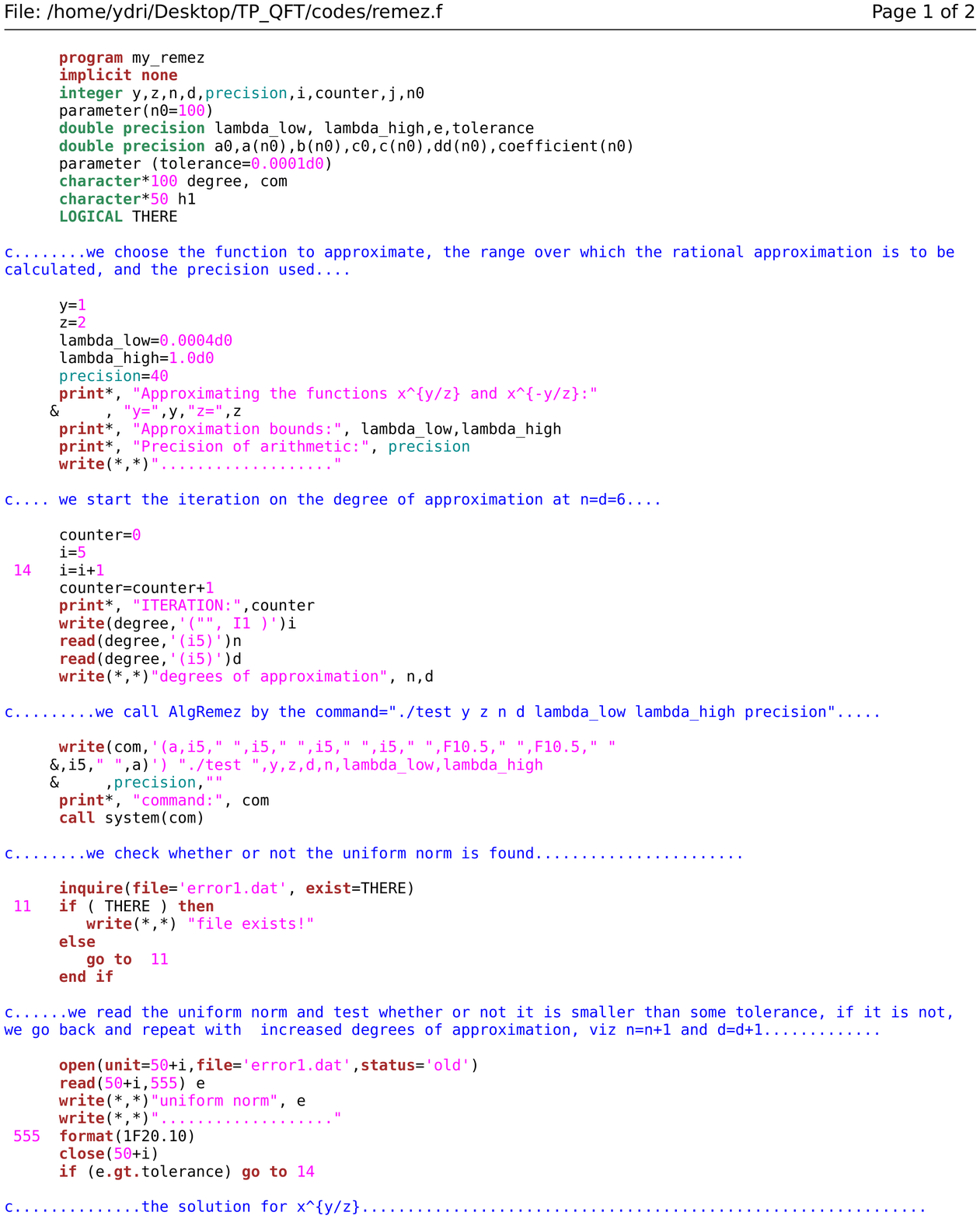}

\addcontentsline{toc}{section}{$9.7$~conjugate-gradient.f} 

\includepdf[pages=-,]{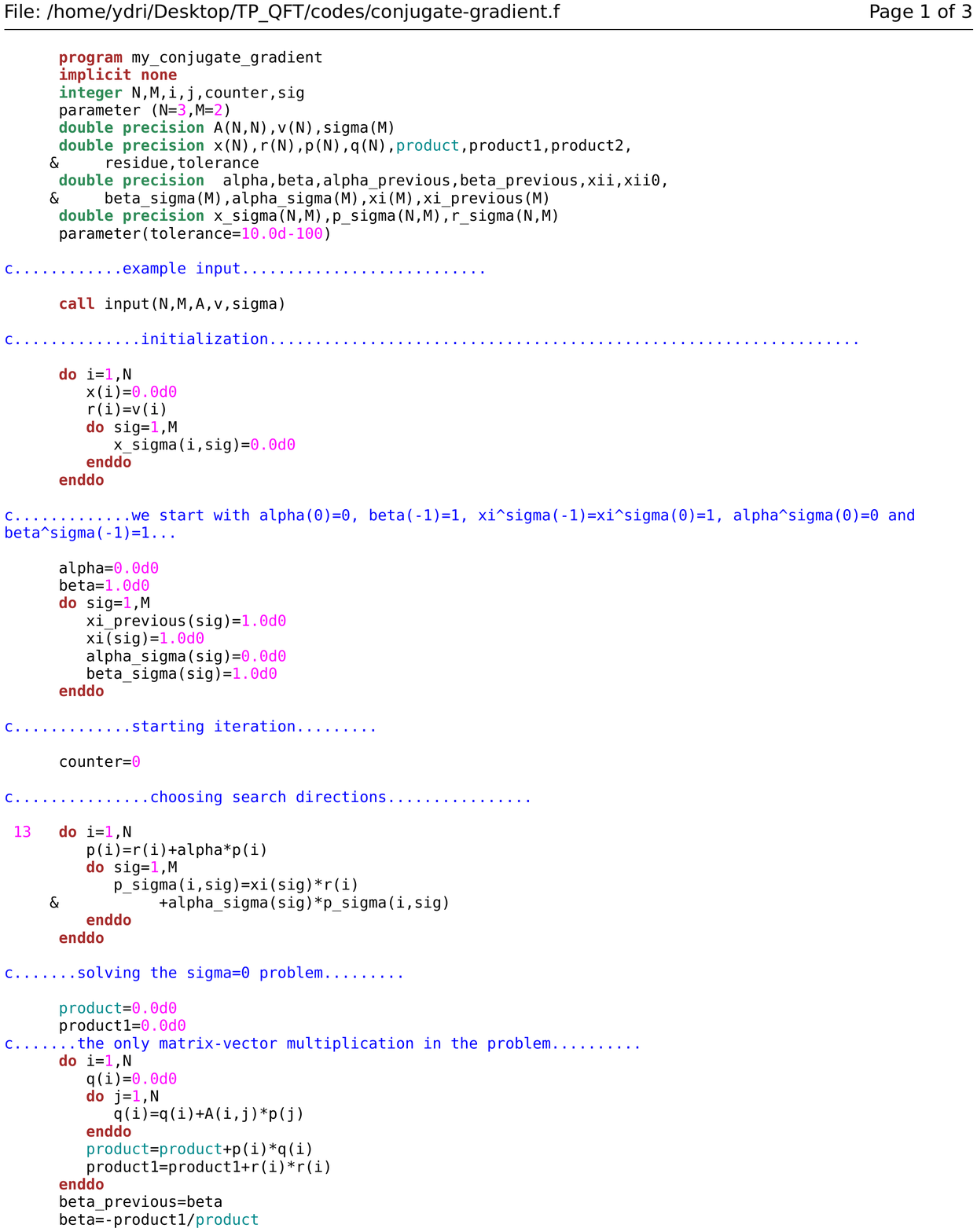}

\addcontentsline{toc}{section}{$9.8$~hybrid-supersymmetric-ym.f} 

\includepdf[pages=-,]{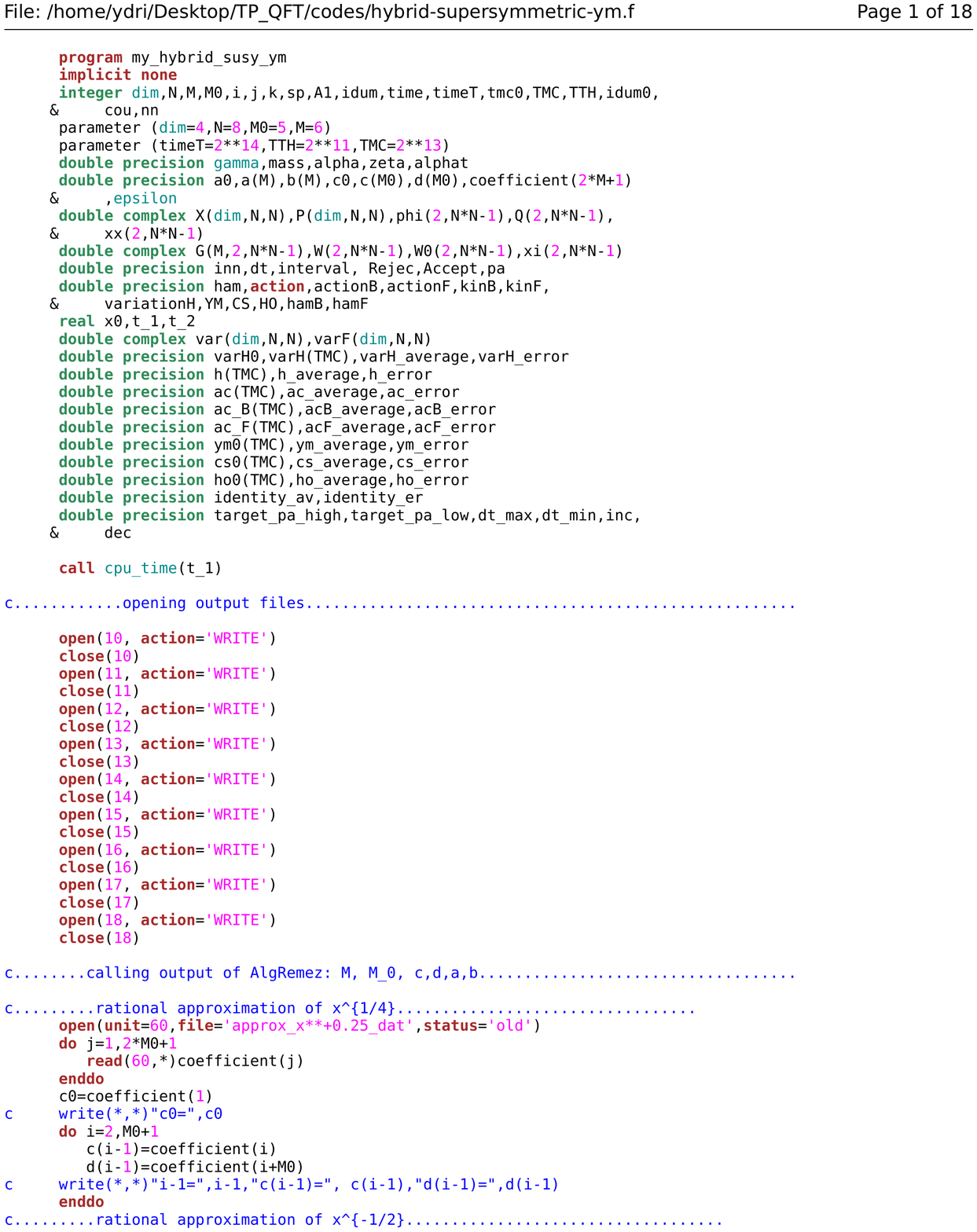}

\addcontentsline{toc}{section}{$9.9$~u-one-on-the-lattice.f} 

\includepdf[pages=-,]{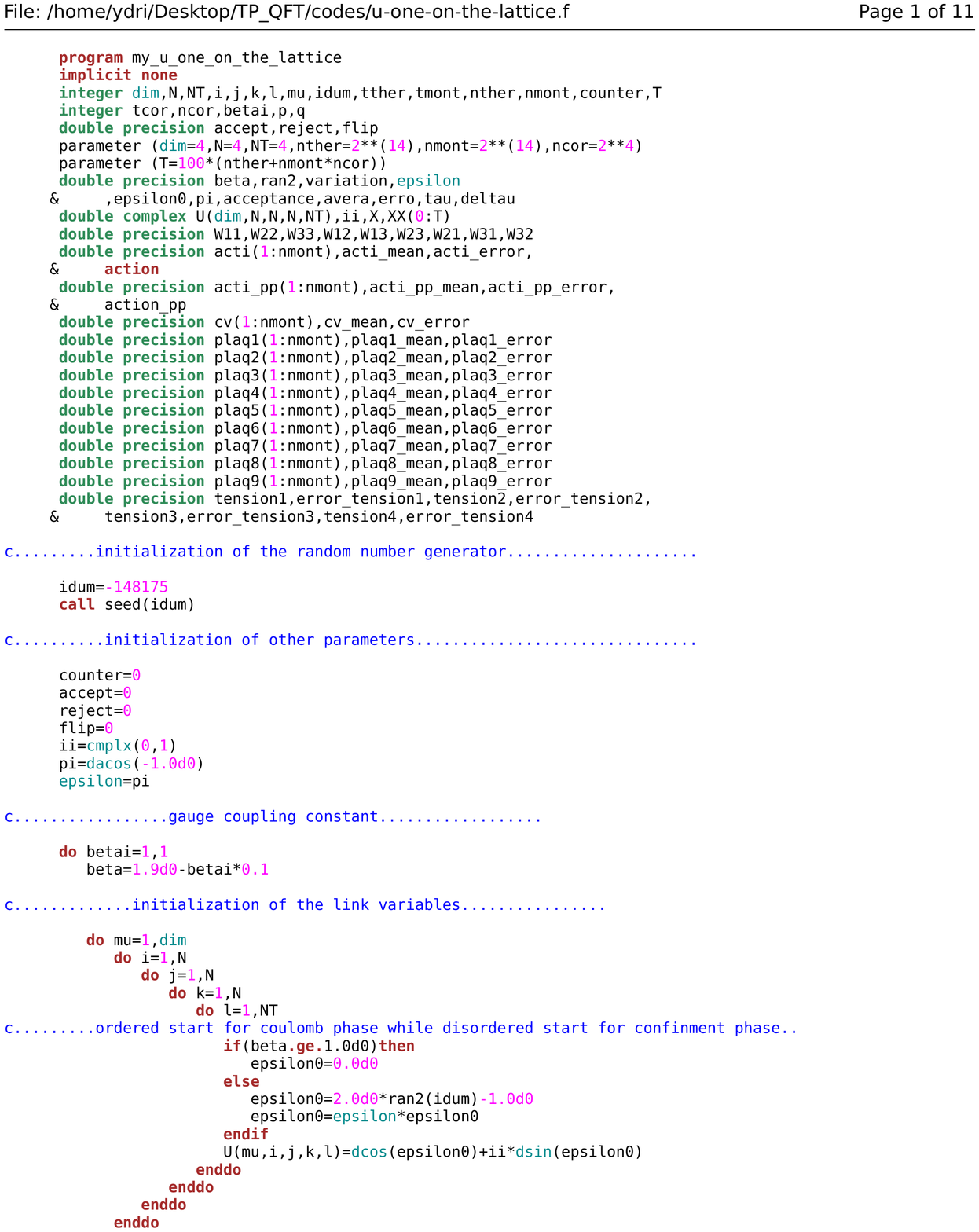}

\begin{appendices}

\chapter{Floating Point Representation, Machine Precision and Errors }

\paragraph{Floating Point Representation:}
Any real number $x$ can be put in the following binary form
\begin{eqnarray}
x=\pm m \times 2^{e-{\rm bias}}~,~1{\leq}m<2~,~m=b_0.b_1b_2b_3...
\end{eqnarray}
We consider a $32-$bit computer. Since $1{\leq}m<2$ we must have $b_0=1$. This binary expansion is called {\it normalized}. For single precision floating-point numbers (singles or floats) we use a $32-$bit word with one bit for the {\it sign}, $8$ bits for the {\it exponent} $e$ and $23$ bits for the {\it significand} $m$. Since only $8$ bits are used to store the exponent we must have $e$ in the range $0{\leq}e{\leq}255$. The bias is chosen ${\rm bias}=127$ so that the actual exponent is in the range $-127{\leq}e-{\rm bias}{\leq}128$. This way we can have very small numbers while  the stored exponent is always positive. Since the first  bit of the significand is $1$ the stored bits of the significand are only $b_1b_2...b_{23}$. If $b_{24},b_{25},..$ are not all zero the floating point representation is not exact. Strictly speaking a floating point number is a number for which $b_{24}=b_{25}=..0$. The floating point representation of a non-zero real number is unique because of the condition $1{\leq}m<2$. In 
summary the above real number is represented on the computer by
\begin{eqnarray}
x_{\rm normal~float}=(-1)^{s} 1.f\times 2^{e-127}~,~0<e<255.
\end{eqnarray}
These are normal numbers.  The terminology floating point is now clear. The binary point can be moved (floated) to any position in the bitstring by choosing the appropriate exponent.

The smallest normalized number is $2^{-126}$. The subnormal numbers are represented by
\begin{eqnarray}
x_{\rm subnormal~float}=(-1)^{s} 0.f\times 2^{-126}.
\end{eqnarray}
These are not normalized numbers. In fact the space between $0$ and the smallest  positive normalized  number is filled by the subnormal numbers.

Explicitly
\begin{center}
\begin{tabular}{||p{2.5cm}|p{2.5cm}|p{2.5cm}|p{2.5cm}||}
\hline
 & s & e &f \\ \hline
 Bit Position  & 31 & 30-23 & 22-0 \\ \hline
\end{tabular}
\end{center}
Because only a finite number of bits is used the set of {\it machine numbers} (the numbers that the computer can store exactly or approximately) is much smaller than the set of real numbers. There is a maximum and a minimum.  Exceeding the maximum  we get the error condition known as overflow. Falling below the minimum we get the error condition known as underflow. 

The largest number corresponds to the normal floating number with $s=0$, $e=254$ and $1.f=1.111..1$ (with $23$ $1$s after the binary point). We compute $1.f=1+0.5+0.25+0.125+...=2$. Hence $x_{\rm normal~float~max}=2\times 2^{127}\simeq 3.4\times 10^{38}$. The smallest number corresponds to the subnormal floating number with $s=0$ and  $0.f=0.00...1=2^{-23}$. Hence $x_{\rm subnormal~float~min}=2^{-149}\simeq 1.4\times 10^{-45}$. We get for single precision floats the range 
\begin{eqnarray}
1.4\times 10^{-45}{\leq}~{\rm single~precision}~{\leq}3.4\times 10^{38}.
\end{eqnarray}
We remark that
\begin{eqnarray}
2^{-23}\simeq 10^{-6.9}.
\end{eqnarray}
Thus single precision numbers have $6-7$ decimal places of significance.

There are special cases. The zero can not be normalized. It is represented by two floats $\pm 0$. Also $\pm \infty$ are special numbers. Finally {\rm NaN} (not a number) is also a special case. Explicitly we have
\begin{eqnarray}
\pm 0=(-1)^{s} 0.0...0 \times 2^{-126}.
\end{eqnarray}

\begin{eqnarray}
\pm \infty=(-1)^{s} 1.0...0 \times 2^{127}.
\end{eqnarray}

\begin{eqnarray}
{\rm NaN}=(-1)^{s} 1.f \times 2^{127}~,~f\neq 0.
\end{eqnarray}
The double precision floating point numbers (doubles) occupy $64$ bits. The first bit is for the sign, $11$ bits for the exponent and $52$ bits for the significand. They are stored as two $32-$bist words. Explicitly
\begin{center}
\begin{tabular}{||p{2.5cm}|p{2.5cm}|p{2.5cm}|p{2.5cm}|p{2.5cm}||}
\hline
 & s & e &f & f\\ \hline
 Bit Position  & 63 & 62-52 & 51-32& 31-0 \\ \hline
\end{tabular}
\end{center}
In this case the bias is ${\it bias}=1023$. They correspond approximately to $16$ decimal places of precision. They are in the range

\begin{eqnarray}
4.9\times 10^{-324}{\leq}~{\rm double~precision}~{\leq}1.8\times 10^{308}.
\end{eqnarray}
The above description corresponds to the IEEE $754$ standard adopted in $1987$ by the Institute of Electrical and Electronics Engineers (IEEE) and American National Standards Institute (ANSI).

\paragraph{Machine Precision and Roundoff Errors:}The gap $\epsilon$ between the number $1$ and the next largest number is called the machine precision. For single precision we get $\epsilon= 2^{-23}$. For double precision we get  $\epsilon= 2^{-52}$. 

Alternatively the machine precision ${\epsilon}_m$ is the largest positive number which if added to the number stored as $1$ will not change this stored $1$, viz
\begin{eqnarray}
1_c+{\epsilon}_m=1_c.
\end{eqnarray}
Clearly ${\epsilon}_m< \epsilon$. The number $x_c$ is the computer representation of of the number $x$. The relative error ${\epsilon}_x$ in $x_c$ is therefore such that
\begin{eqnarray}
|{\epsilon}_x|=|\frac{x_c-x}{x}|{\leq}{\epsilon}_m.
\end{eqnarray}
All single precision numbers contain an error in their $6$th decimal place and all double precision numbers contain an error in their $15$th decimal place.

An operation on the computer will therefore only approximate the analytic answer since numbers are stored approximately. For example the difference $a=b-c$ is on the computer $a_c=b_c-c_c$. We compute
\begin{eqnarray}
\frac{a_c}{a}=1+{\epsilon}_b\frac{b}{a}-{\epsilon}_c\frac{c}{a}.
\end{eqnarray}
In particular the subtraction of two very large nearly equal numbers $b$ and $c$ may lead to a very large error in the answer $a_c$. Indeed we get the error
\begin{eqnarray}
{\epsilon}_a\simeq \frac{b}{a}({\epsilon}_b-{\epsilon}_c).
\end{eqnarray}
In other words the large number $b/a$ can magnify the error considerably. This is called subtractive cancellation.

Let us next consider the operation of multiplication of two numbers $b$ and $c$ to produce a number $a$, viz $a=b\times c$. This operation is represented on the computer by $a_c=b_c\times c_c$. We get the error
\begin{eqnarray}
{\epsilon}_a={\epsilon}_b+{\epsilon}_c.
\end{eqnarray}
Let us now consider an operation involving a large number $N$ of steps. The question we want to ask is  how does the roundoff error  accumulate. 

The main observation is that roundoff errors grow {\it slowly} and {\it randomly} with  $N$. They diverge as $N$ gets very large. By assuming that the roundoff errors in the individual steps of the operation are not correlated we can view the accumulation of error as a random walk problem with step size equal to the machine precison ${\epsilon}_m$. We know from the study of the random walk problem  in statistical mechanics that the total roundoff error will  be proportional to $\sqrt{N}$, namely
\begin{eqnarray}
{\epsilon}_{\rm ro}=\sqrt{N}{\epsilon}_m.
\end{eqnarray}
This is the most conservative estimation of the roundoff errors. The roundoff errors are analogous to the uncertainty in the measurement of a physical quantity.

\paragraph{Systematic (Algorithmic) Errors:}
This type of errors arise from the use of approximate numerical solutions. In general the algorithmic (systematic) error is inversely proportional to some power of the number of steps $N$, i.e.
\begin{eqnarray}
{\epsilon}_{\rm sys}=\frac{\alpha}{N^{\beta}}.
\end{eqnarray}
The total error is obtained by adding the roundoff error, viz
\begin{eqnarray}
{\epsilon}_{\rm tot}={\epsilon}_{\rm sys}+{\epsilon}_{\rm ro}=\frac{\alpha}{N^{\beta}}+\sqrt{N}{\epsilon}_m.
\end{eqnarray}
There is a competition between the two types of errors. For small $N$ it is the systematic error which dominates while for large $N$ the roundoff error dominates.  This is very interesting because it means that by trying to decrease the systematic error (by increasing $N$) we will increase the roundoff error. The best algorithm is the algorithm which gives an acceptable approximation in a small number of steps so that there will be no time for roundoff errors to grow large.

As an example let us consider the case $\beta=2$ and $\alpha=1$. The total error is
\begin{eqnarray}
{\epsilon}_{\rm tot}=\frac{1}{N^{2}}+\sqrt{N}{\epsilon}_m.
\end{eqnarray}
This error is minimum when
\begin{eqnarray}
\frac{d{\epsilon}_{\rm tot}}{dN}=0.
\end{eqnarray}
For single precision calculation (${\epsilon}_m=10^{-7}$) we get $N=1099$. Hence ${\epsilon}_{\rm tot}=4\times 10^{-6}$. Most of the error is roundoff. In order to decrease the roundoff error and hence the total error in this example we need to decrease the number of steps. Furthermore in order for the systematic error to not increase when we decrease the number of steps we must find another algorithm which converges faster with $N$. For an algorithm with $\alpha=2$ and $\beta=4$ the total error is 
\begin{eqnarray}
{\epsilon}_{\rm tot}=\frac{2}{N^{4}}+\sqrt{N}{\epsilon}_m.
\end{eqnarray}
This error is minimum now at $N=67$ for which ${\epsilon}_{\rm tot}=9\times 10^{-7}$. We have only $1/16$ as many steps with an error smaller by a factor of $4$.

\chapter{Executive Arabic Summary of Part I}



\section*{Supplementary material}

# Appendix B

# Executive Arabic Summary of Part I

# اعمال تطبيقية في الفيزياء العددية

## باديس يدري

معهد الفيزياء، جامعة باجي مختار، عنابة، الجزائر

جانفي 2015

# الفهرس

## مقدمة

الفيزياء العددية هي احد فروع العلوم العددية التي تعرف ايضا باسم الحاسوبية العلمية و التي ظهرت و تبلورت خلال $30 - 40$ سنة الاخيرة مع التقدم الهائل الذي حصل في التكنولوجيا الرقمية خاصة في الولايات المتحدة الامريكية.

يمكن اعتبار الفيزياء العددية قسم من اقسام الفيزياء النظرية او يمكن اعتبارها جسر يربط بين الفيزياء النظرية و الفيزياء التجريبية و هناك حتي من يعتبرها تخصص قائم لخدمة الفيزياء التجريبية لا حسب. تقليديا هناك مقاربتان متكاملتان، علي الاقل منذ عصر نيوتن، للفيزياء. فهناك من ناحية المقاربة النظرية و من ناحية اخري هناك المقاربة التجريبية. هناك الكثير من العاملين في هذا المجال و من غيرهم، من يعتبر انه توجد مقاربة ثالثة منفصلة و مختلفة للفيزياء هي المقاربة العددية. من وجهة النظر هذه فان الفيزياء العددية هي حقل منفصل بذاته غير مرتبط بالضرورة بالحقلين النظري و التجريبي. رغم هذا فان وجهة النظر التي سوف نتبناها في هذه المطوية هو الراي الاول الذي يعتبر ان الفيزياء العددية هو فرع من فروع الفيزياء النظرية.

في الفيزياء العددية يتم مزج عناصر من الفيزياء و خاصة الفيزياء النظرية و عناصر من الرياضيات التطبيقية مثل التحليل العددي مع عناصر من علوم الحاسوب مثل البرمجة من اجل هدف واحد هو حل مسالة فيزيائية معينة ليس لها حل كامل او حل معروف.

اهم استعمالات الكمبيوتر في الفيزياء هو اجراء المحاكيات (جمع محاكاة) العددية. المحاكيات العددية تلائم اكثر المسائل الفيزيائية التي تتحكم فيها معادلات رياضية غير خطية و التي لا تتوفر في معظمها علي حل تحليلي مضبوط. نقطة البدء لاي محاكاة عددية هو نموذج مثالي للجملة الفيزيائية قيد الدراسة و من الطبيعي اننا نريد ان نتأكد ما اذا كان تصرف هذا النموذج منسجم مع المشاهدة او لا في حالة توفر نتائج تجريبية للمقارنة اما في حالة عدم توفر اي نتائج تجريبية فان الهدف هو استشراف ما يمكن ان تعطيه التجربة اذا ما اجريت. الخطوة الاولي من اجل تحقيق هذا الهدف هو ايجاد خوارزمية رياضية من اجل انجاز هذا النموذج نظريا و ايضا علي الكمبيوتر. تنفيذ هذا الانجاز علي كمبيوتر هو ما نسميه بالمحاكاة العددية و هو يعتمد علي ترجمة الخوارزمية الرياضية الي شفرة مكتوبة باحدي لغات البرمجة يمكن للكمبيوتر ان يفهمها.

المحاكيات العددية هي اذن تجارب افتراضية. فمثلا يلعب النموذج الرياضي في المحاكاة العددية بالضبط دور العينة في التجربة المعملية اما الخوارزمية او الشفرة التي تستعملها المحاكاة العددية فهي تقوم بدور جهاز القياس في التجربة المعملية. قبل البدء في استعمال المحاكاة العددية في الدراسة الفيزيائية فانه علينا اختبار او معايرة الشفرة تماما كما اننا نقوم بمعايرة جهاز القياس في التجربة المعملية قبل البدء في اجراء اي قياس. القياس الذي نقوم به في التجربة المعملية يقابله الحساب الذي تجريه المحاكاة العددية و نختتم كلا العمليتين بنفس الأمر و هو تحليل المعطيات.

من الواضح جدا و من الطبيعي ان اهم وسائل الفيزياء العددية هي لغات البرمجة. في معظم المحاكيات العددية التي نجدها في الاعمال البحثية الفيزيائية تكتب الشفرات في احدي اللغات المجمعة مثل الفرترون (Fortran) او لغة سي ($C$). في هذه المحاكيات يمكن ايضا عند الحاجة مناداة مكتبات الروتينات العددية مثل لاباك (Lapack) و غيرها. استعمال البرمجيات العددية الجاهزة مثل ماتلاب (Matlab) و ماثيماتيكا (Mathematica) في هذه المحاكيات العددية، خاصة التي تعتمد علي طريقة المونتي كارلو (Monte Carlo)، غير عملي بالمرة لانه يؤدي الي زمن

سير طويل جدا للشفرة علي الكمبيوتر و هذا راجع بالخصوص الي كون البرمجيات الجاهزة هي لغات مترجمة و ليست لغات مجمعة. ليس هناك ادني شك في ان البرمجيات الجاهزة مفيدة للغاية في الحسابات العددية التي لا تعتمد علي التكرار لكنها غير ملائمة تماما في المحاكيات العددية التي تعتمد بالاساس علي تكرار نفس الخطوة عدد هائل من المرات. في هذه المطوية سوف نتبع بالضبط هذا الطريق اي سوف نكتب جميع شفراتنا في لغة مجمعة و نتجنب استخدام البرمجيات الجاهزة. سوف نستخدم بالخصوص الفرترون 77 او 90 علي نظام التشغيل لينيكس (Linux) توزيع يوبنتو (Ubuntu) .

هذه المطوية تحتوي علي مجموع الاعمال التطبيقية المرفقة بمحاضرات الفيزياء العددية التي القاها المؤلف باديس ايدري في معهد الفيزياء منذ العام 2009 علي طلبة الليسانس و الماستر في اطار مقاييس التحليل العددي (ليسانس فيزياء)، الفيزياء العددية (ماستر فيزياء نظرية) و الاعلام الالي (باقي تخصصات الماستر). يمكن الحصول علي مطوية ـمحاضرات في الفيزياء العدديةـ عن طريق الاتصال بالمؤلف باديس ايدري عبر البريد الالكتروني badis.ydri@univ − annaba.org او تصفح الموقع الرسمي خاصته علي http : //homepages.dias.ie/ydri/.

في الختام يتقدم المؤلف باديس ايدري اصالة عن نفسه و عن باقي فريق العمل بالشكر الجزيل للمدير السابق لمعهد الفيزياء الاستاذ مصطفي بن شهاب و كذلك للمدير الحالي الاستاذ علاوة شيباني للمساعدات الجليلة و التسهيلات الكثيرة التي قدماها منذ البداية من اجل ادخال هذه المحاضرات و التطبيقات في البرامج الرسمية للفيزياء.

باديس يدري
سرايدي، عنابة، الجزائر
الاثنين 8 جويلية 2013

# خوارزمية اولر ـ مقاومة الهواء

يقود رياضي دراجة هوائية علي طريق مستقيمة و مسطحة بسرعة $v$ . القوة التي يطبقها
الرياضي علي الدراجة تكافئ استطاعة ثابتة $P$ تساوي 200 واط يوفرها الرياضي لفترة زمنية
تقدر بساعة واحدة. قوة مقاومة الهواء (التي تعرف ايضا بقوة الجر الهوائي) تكون معاكسة
للحركة و متناسبة طردا مع مربع السرعة معطاة بالعلاقة

$$F_{\text{drag}} = -C\rho A v^2.$$

في هذه المعادلة $\rho$ هي كثافة الهواء، $C$ هو معامل الجر و $A$ هي مساحة المقطع العرضي
لجملة الرياضي زائد الدراجة. قانون نيوتن الثاني يأخذ الشكل التالي

$$\frac{dv}{dt} = \frac{P}{mv} - \frac{C\rho A v^2}{m}.$$

المطلوب هو حساب السرعة $v$ كدالة في الزمن. المقاربة العددية لهذه المسألة تعتمد علي
خوارزمية اولر. نقطع المجال الزمني $T$ الي $N$ مجال زمني صغير

$$\Delta t = \frac{T}{N}$$

اي

$$t = i\Delta t \ , \ i = 0, ..., N.$$

نعرف

$$\hat{v}(i) = v(t - \Delta t).$$

الحل المعطي بتقريب اولر يأخذ الشكل

$$\hat{v}(i + 1) = \hat{v}(i) + \Delta t \left( \frac{P}{m\hat{v}(i)} - \frac{C\rho A \hat{v}^2(i)}{m} \right) . \ i = 1, ..., N + 1$$

اللحظات الزمنية المرافقة تعطي ب

$$\hat{t}(i + 1) = i\Delta t \ , \ i = 1, ..., N + 1.$$

(1) احسب السرعة كدالة في الزمن في حالة وجود مقاومة الهواء وفي حالة عدم وجود
مقاومة الهواء. ماذا تلاحظ. في هذا السؤال نأخذ ثابت الجر $C$ يساوي $0.5$. نعطي ايضا
القيم

$$m = 70\text{kg} \ , \ A = 0.33 m^2 \ , \ \rho = 1.2\text{kg}/m^3 \ , \ \Delta t = 0.1 s \ , \ T = 200 s.$$

السرعة الابتدائية تعطي ب

$$\hat{v}(1) = 4m/s \ , \ \hat{t}(1) = 0.$$

(2) ماذا تلاحظ في حالة تغيير ثابت الجر و/او الاستطاعة. ماذا تلاحظ عندما يتم تصغير
الخطوة الزمنية.

# حركة القذائف تحت تأثير مقاومة الهواء

نعتبر حركة قذيفة تحت تأثير قوة مقاومة الهواء التي تعمل عكس اتجاه الحركة و تكون متناسبة مع مربع السرعة. نرمز لثابت التناسب ب $B$. قانون نيوتن الثاني يؤدي الي معادلات الحركة التالية

$$\frac{dx}{dt} = v_x \ , \ m\frac{dv_x}{dt} = -Bvv_x.$$

$$\frac{dy}{dt} = v_y \ , \ m\frac{dv_y}{dt} = -mg - Bvv_y.$$

حل هذه المعادلات التفاضلية المعطي بخوارزمية اولر بأخذ الشكل التالي

$$v_x(i+1) = v_x(i) - \Delta t \frac{Bv(i)v_x(i)}{m}.$$

$$v_y(i+1) = v_y(i) - \Delta t g - \Delta t \frac{Bv(i)v_y(i)}{m}.$$

$$v(i+1) = \sqrt{v_x^2(i+1) + v_y^2(i+1)}.$$

$$x(i+1) = x(i) + \Delta t \ v_x(i).$$

$$y(i+1) = y(i) + \Delta t \ v_y(i).$$

القيم الابتدائية للموضع و السرعة توافق القيمة 1 ل $i$ و $i$ يأخذ القيم من 1 الي $N$.

(1) اكتب شفرة فورترون منجز فيها الحل المعطي بخوارزمية اولر لمسألة.

(2) نأخذ القيم التالية

$$\frac{B}{m} = 0.00004 m^{-1} \ , \ g = 9.8 m/s^2.$$

$$v(1) = 700 m/s \ , \ \theta = 30 \ \text{degree}.$$

$$v_x(1) = v(1)\cos\theta \ , \ v_y(1) = v(1)\sin\theta.$$

$$N = 10^5 \ , \ \Delta t = 0.01 s.$$

احسب المسار بدون و ب مقاومة الهواء. ماذا تلاحظ.

(3) باستخدام التصريح الشرطي $if$ يمكن تعيين مدي القذيفة. هذا التصريح يضاف داخل حلقة $do$ كالتالي

$$if(y(i+1).le.0)exit.$$

عين مدي القذيفة في حالة وجود و في حالة عدم وجود مقاومة للهواء.

(4) في حالة عدم وجود مقاومة للهواء نعرف ان المدي يأخذ اعظم قيمة له لما تكون الزاوية الابتدائية تساوي 45 درجة. تحقق من هذا الامر عدديا باختبار عدة قيم للزاوية الابتدائية. يمكن ايضا اضافة حلقة $do$ في الزاوية ثم دراسة المدي كدالة في الزاوية و البحث عن قيمته العظمي.

(5) في حالة وجود مقاومة الهواء احسب الزاوية التي يكون فيها المدي اعظمي.

# الهزاز التوافقي- خوارزميات اولر- كرومر و فيرلات

نعتبر هزاز توافقي بسيط عبارة عن كتلة $m$ مربوطة بخيط طوله $l$ معلق في مرتكز ثابت تحت تأثير الثقالة $g$. نفترض ان الحركة خطية اي ان الزاوية التي يصنعها النواس مع المحور الشاقولي تبقى دائما صغيره. معادلة الاهتزاز المشتقة من قانون نيوتن الثاني تأخذ الشكل

$$\frac{d^2\theta}{dt^2} + \frac{g}{l}\theta = 0.$$

هذه المعادلة التفاضلية من الرتبه الثانيه يمكن تعويضها بمعادلتين تفاضليتين من الرتبة الاولى كالتالي

$$\frac{d\theta}{dt} = \Omega \ , \ \frac{d\Omega}{dt} = -\frac{g}{l}\theta.$$

الحل العددي الاول الذي سنعتبره هنا هو الحل المعطي بخوارزمية اولر

$$\Omega_{i+1} = \Omega_i - \frac{g}{l}\theta_i \ \Delta t.$$

$$\theta_{i+1} = \theta_i + \Omega_i \ \Delta t.$$

الحل العددي الاخر الذي سنعتبره هنا هو الحل المعطي بخوارزمية اولر- كرومر. هذا الحل يعطي بالمعادلات التالية

$$\Omega_{i+1} = \Omega_i - \frac{g}{l}\theta_i \ \Delta t.$$

$$\theta_{i+1} = \theta_i + \Omega_{i+1} \ \Delta t.$$

نعتبر هنا ايضا الحل المعطي بخوارزمية فيرلات الذي يأخذ الشكل

$$\theta_{i+1} = 2\theta_i - \theta_{i-1} - \frac{g}{l}\theta_i(\Delta t)^2.$$

(1) اكتب شفرة فورترون منجز فيها الحلول المعطاة بخوارزميات اولر و اولر- كرومر لمسألة الهزاز التوافقي.

(2) احسب الزاوية، السرعة الزاوية و الطاقة كدوال في الزمن. طاقة الهزاز تعطي ب

$$E = \frac{1}{2}\Omega^2 + \frac{1}{2}\frac{g}{l}\theta^2.$$

نأخذ القيم العددية

$$g = 9.8m/s^2 \ , l = 1m \ .$$

نأخذ عدد الخطوات و الخطوة الزمنية

$$N = 10000 \ , \ \Delta t = 0.05s.$$

نأخذ الزاوية و السرعة الزاوية الابتدائيتان

$$\theta_1 = 0.1\text{radian} \ , \ \Omega_1 = 0.$$

باستعمال التصريح الشرطي $if$ يمكن تحديد زمن الحركة بخمسة اضعاف الدور كالتالي

$$if(t(i+1).ge.5 * \text{period}) \ \text{exit}.$$

(3) قارن بين قيمة الطاقة المحسوبة باولر و قيمة الطاقة المحسوبة باولر- كرومر. ماذا تلاحظ و ماذا تستنتج.

(4) اعد الحساب باستخدام خوارزمية فيرلات. لنلاحظ ان هذه الطريقة لا يمكنها الانطلاق فقط من القيم الابتدائية $\theta_1$ و $\Omega_1$. يجب ايضا اعطاء الزاوية $\theta_2$ التي يمكن حسابها باستعمال طريقة اولر اي

$$\theta_2 = \theta_1 + \Omega_1 \, \Delta t.$$

لنلاحظ ايضا ان خوارزمية فيرلات لا تحتاج الي حساب السرعة الزاوية. لكن من اجل حساب الطاقة نحتاج الي معرفة السرعة الزاوية التي نحسبها باستعمال العبارة

$$\Omega_i = \frac{\theta_{i+1} - \theta_{i-1}}{2\Delta t}.$$

## التكاملات العددية

نعتبر التكاملات في بعد واحد من الشكل

$$I = \int_a^b dx f(x).$$

نعتبر الحالة العامة التي لا يمكن فيها اجراء التكامل تحليليا و يبقي الحل العددي هو الخيار الوحيد . الخوارزميات التي سنستعملها هي التقريب بالمستطيلات و بأشباه المنحرف و التقريب بقطوع المكافئ . في كل هذه الطرق نقسم مجال التعريف الي $N$ مجال طوله $\Delta x$ كالتالي

$$x_i = x_0 + i\Delta x \ , \ i = 0, ..., N \ , \ \Delta x = \frac{b-a}{N} \ , \ x_0 = a, x_N = b.$$

التقريب بالمستطيلات يعطي ب

$$F_N = \Delta x \sum_{i=0}^{N-1} f(x_i).$$

التقريب باشباه المنحرف يعطي ب

$$T_N = \Delta x \left[ \frac{1}{2} f(x_0) + \sum_{i=1}^{N-1} f(x_i) + \frac{1}{2} f(x_N) \right].$$

التقريب بقطوع المكافئ (قاعدة سيمبسون) يعطي ب ( هنا $N$ يجب ان يكون زوجي)

$$S_N = \frac{\Delta x}{3} \left[ f(x_0) + 4 \sum_{i=0}^{\frac{N-2}{2}} f(x_{2i+1}) + 2 \sum_{i=0}^{\frac{N-2}{2}} f(x_{2i}) + f(x_N) \right].$$

الخطأ في هذه التقريبات الثلاث متناسب مع $1/N$، $1/N^2$ و $1/N^4$ علي التوالي .

(1) نأخذ التكامل

$$I = \int_0^1 f(x) dx \ ; \ f(x) = 2x + 3x^2 + 4x^3.$$

احسب قيمة هذا التكامل باستعمال طريقة المستطيلات. قارن مع قيمة التكامل التحليلية. ملاحظة: شفر الدالة باستعمال subroutine او function .

(2) غير عدد المجالات $N$ . احسب الخطأ المرتكب بدلالة $N$. قارن مع النظري.

(3) اعد السؤالين السابقين باستعمال طريقة اشباه المنحرف و قاعدة سيمبسون.

(4) خذ الان التكاملات التالية

$$I = \int_0^{\frac{\pi}{2}} \cos x dx \ , \ \ I = \int_1^e \frac{1}{x} dx \ , \ I = \int_{-1}^{+1} \lim_{\epsilon \longrightarrow 0} \left( \frac{1}{\pi} \frac{\epsilon}{x^2 + \epsilon^2} \right) dx.$$

# خوارزمية نيوتن – رافسون

جسيم ذو كتلة $m$ يتحرك في بئر كمون ارتفاعه $V$ و طوله $2a$ يمتد من $-a$ الي $+a$. نهتم بحالات الجملة ذات الطاقات الأصغر من ارتفاع البئر اي الحالات المرتبطة. حالة الجملة قد تكون زوجية او فردية. الطاقات المسموح بها المرفقة بالدوال الموجية الزوجية تعطي بحلول المعادلة المتسامية

$$\alpha \tan \alpha a = \beta.$$

$$\alpha = \sqrt{\frac{2mE}{\hbar^2}} \ , \ \beta = \sqrt{\frac{2m(V-E)}{\hbar^2}}.$$

في حالة بئر الكمون اللانهائي نجد الحلول

$$E_n = \frac{(n+\frac{1}{2})^2 \pi^2 \hbar^2}{2ma^2} \ , \ n = 0, 1....$$

نختار (مع اهمال كتابة الوحدات)

$$\hbar = 1 \ , \ a = 1 \ , \ 2m = 1.$$

من اجل ايجاد الطاقات $E$ نستعمل خوارزمية نيوتن – رافسون التي تسمح لنا بايجاد جذور اي معادلة $f(x) = 0$ كالتالي. انطلاقا من تخمين معين $x_0$ فاننا نقرب حل المعادلة $f(x) = 0$ بنقطة تقاطع مماس الدالة $f(x)$ في النقطة $x_0$ مع محور السينات. نسمي هذا التقريب الاول $x_1$ و هو يعطي بالمعادلة

$$x_1 = x_0 - \frac{f(x_0)}{f'(x_0)}.$$

انطلاقا من $x_1$ نقوم بنفس الخطوة من اجل ايجاد التقريب الثاني $x_2$ ثم نستخدم $x_2$ من اجل ايجاد التقريب الثالث $x_3$ و هكذا. التقريب $x_{i+1}$ يعطي بدلالة التقريب $x_i$ بالعلاقة

$$x_{i+1} = x_i - \frac{f(x_i)}{f'(x_i)}.$$

(1)  من اجل $V = 10$ بين عدد الحلول $E$ باستعمال الطريقة البيانية عبر دراسة الدالتين

$$f(\alpha) = \tan \alpha a \ , \ g(\alpha) = \frac{\beta}{\alpha} = \sqrt{\frac{V}{\alpha^2} - 1}.$$

(2)  جد باستعمال طريقة نيوتن–رافسون الحلين بدقة اقل او تساوي من $10^{-8}$. من اجل ايجاد الحل الاول نأخذ التخمين الاول لطريقة نيوتن–رافسون في نقطة التباعد الاول لدالة الظل اي $\alpha = \pi/a$. من اجل ايجاد الحل الثاني نأخذ التخمين الاول في نقطة التباعد الثانية اي $\alpha = 2\pi/a$.

(3)  اعد السؤال من اجل $V = 20$.

(4)  حدد الحلول الاربعة من اجل $V = 100$. استعن بالطريقة البيانية من اجل تحديد التخمين الاول كل مرة.

(5)  اعد الاسئلة السابقة باستعمال طريقة التنصيف.

# خوارزمية رونج ـ كوتا ـ المجموعة الشمسية

نعتبر مجموعة شمسية مشكلة من كوكب واحد يتحرك حول الشمس. نفترض ان كتلة الشمس ثقيلة جدا بالمقارنة مع كتلة الكوكب بحيث يمكن اعتبارها ساكنة في مركز النظام. قانون نيوتن الثاني يعطي معادلات الحركة التالية

$$v_x = \frac{dx}{dt} \ , \ \frac{dv_x}{dt} = -\frac{GM_s}{r^3}x \ , \ v_y = \frac{dy}{dt} \ , \ \frac{dv_y}{dt} = -\frac{GM_s}{r^3}y.$$

$$r = \sqrt{x^2 + y^2}.$$

نستخدم الوحدات الفلكية حيث

$$GM_s = 4\pi^2 AU^3/yr^2.$$

حل معادلات الحركة الانفة الذكر المعطي بخوارزمية رونج ـ كوتا يأخذ الشكل

$$k_1 = \Delta t \, v_x(i) \ , \ p_1 = \Delta t \, v_y(i).$$

$$r(i) = \sqrt{x(i)^2 + y(i)^2}.$$

$$k_3 = -\frac{GM_s}{r(i)^3}x(i)\Delta t \ , \ p_3 = -\frac{GM_s}{r(i)^3}y(i)\Delta t.$$

$$k_2 = (v_x(i) + \frac{1}{2}k_3)\Delta t \ , \ p_2 = (v_y(i) + \frac{1}{2}p_3)\Delta t.$$

$$R(i) = \sqrt{(x(i) + \frac{1}{2}k_1)^2 + (y(i) + \frac{1}{2}p_1)^2}.$$

$$k_4 = -\frac{GM_s}{R(i)^3}(x(i) + \frac{1}{2}k_1)\Delta t \ , \ p_4 = -\frac{GM_s}{R(i)^3}(y(i) + \frac{1}{2}p_1)\Delta t.$$

$$x(i+1) = x(i) + k_2.$$

$$v_x(i+1) = v_x(i) + k_4.$$

$$y(i+1) = y(i) + p_2.$$

$$v_y(i+1) = v_y(i) + p_4.$$

في المعادلات السابقة $i$ يأخذ القيم من 1 الي $N$ . القيم الابتدائية للموضع و السرعة توافق القيمة 1 ل $i$.

(1) اكتب شفرة فورترون منجز فيها الحل المعطي بخوارزمية رونج ـ كوتا لمسألة النظام الشمسي.

(2) احسب المسار و السرعة و كذلك الطاقة كدوال في الزمن. ماذا تلاحظ بالنسبة للطاقة. استخدم الوحدات الفلكية. للتذكير فان طاقة الكوكب في وحدة الكتلة تعطي ب

$$E = \frac{1}{2}v^2 - \frac{GM_s}{r}.$$

(3) حسب قانون كبلر الاول فان جميع المدارات هي قطوع ناقصة مع وجود الشمس في احد المحرقين. في ما يلي سنعتبر فقط الكواكب التي نعلم من المشاهدة ان مداراتها دائرية الي حد كبير. هذه الكواكب هي الزهرة و الارض و المريخ و المشتري و ساتورن. انصاف الاقطار تعطي في وحدة الوحدات الفلكية ب

$$a_{\text{venus}} = 0.72 \ , \ a_{\text{earth}} = 1 \ , \ a_{\text{mars}} = 1.52 \ , \ a_{\text{jupiter}} = 5.2 \ , \ a_{\text{saturn}} = 9.54.$$

تحقق من قانون كبلر الاول من اجل كل هذه الكواكب.

من اجل الاجابة علي السؤالين السابقين 2 و 3 نأخذ الشروط الابتدائية

$$x(1) = a \ , \ y(1) = 0 \ , \ v_x(1) = 0 \ , \ v_y(1) = v.$$

القيمة التي تأخذها السرعة الابتدائية مهمة جدا من اجل الحصول علي المسار الصحيح و يتم تعيينها مثلا من افتراض ان المسار هو فعلا دائري و بالتالي فان قوة الجذب الثقالي تكون متوازنة مع قوة الطرد المركزي. نحصل علي

$$v = \sqrt{\frac{GM_s}{a}}.$$

ايضانأخذ الخطوة الزمنية و عدد التكرارات كالاتي

$$\Delta t = 0.01 yr \ , \ N = 10^3 - 10^4.$$

(4) حسب قانون كبلر الثالث فان مربع الدور يكون متناسب طردا مع مكعب نصف القطر. من اجل المدارات الدائرية فان ثابت التناسب يساوي بالضبط واحد. تحقق من هذا الامر من اجل كل الكواكب المذكورة اعلاه. يتم قياس الدور مثلا بمراقبة متي يرجع الكوكب الي ابعد نقطة له عن الشمس.

(5) بتغيير السرعة الابتدائية بطريقة مناسبة فانه يمكن الحصول علي مدار قطع ناقص. جرب هذا الامر.

(6) القانونان الاساسيان اللذان يحكمان حركة النظام الشمسي المبسط الذي اعتبرناه في هذا التمرين هو قانون نيوتن للجذب الثقالي بين الكتل من جهة و قانون نيوتن الثاني من جهةاخري.

قانون الجذب الثقالي ينص في اهم بنوده في ان القوة بين الشمس و الكوكب مركزية موجهة من الكوكب نحو الشمس و متناسبة عكسا مع مربع المسافة. نفترض في الاتي ان قوة الجذب الثقالي هي متناسبة عكسا مع اس اخر للمسافة مختلف عن اثنين. غير الشفرة من اجل اخذ هذا التصرف الجديد للقوة بعين الاعتبار. احسب المدارات من اجل اساسات بين ثلاثة و واحد. ماذا تلاحظ و ماذا تستنتج.

# مسألة دوران الحضيض الشمسي لكوكب عطارد

حسب قانون كبلر الاول فان مدارات جميع الكواكب السيارة و من ضمنها عطارد تعطي بقطوع ناقصة مع وجود الشمس في احد المحرقين . هذا القانون يمكن اشتقاقه من تطبيق قوانين نيوتن علي تفاعل الكواكب مع الشمس مع افتراض انه يمكننا اهمال تفاعل الكواكب نفسها فيما بينها . تأثير الكواكب علي بعضها البعض يؤدي الي ظاهرة دوران محاور القطوع الناقصة حول الشمس و بالتالي دوران الحضيض الشمسي حول الشمس الذي هو اقرب نقطة في مسار الكوكب من الشمس . هذا الدوران للحضيض الشمسي حول الشمس يحدث لجميع الكواكب لكن مشاهدته صعبة للغاية بسبب كون اغلب المدارات هي دائرية الي حد كبير . فقط بلوتو و عطارد لها مدارات قطوع ناقصة ذات لاتراكزية كبيرة. لكن بالنسبة لبلوتو فان سرعته المدارية المنخفضة لا تسمح بمشاهدة دوران حضيضه الشمسي. يبقي عطارد الذي يمكن قياس دوران حضيضه حول الشمس بدقة معتبرة. قام الفلكيون بقياس سرعة الدوران التالية

$$566 \ \text{arcsecond/century}.$$

اي ان حضيض عطارد يصنع دورة كاملة حول الشمس كل 240000 سنة. من جهة اخري فانه بتطبيق قوانين نيوتن علي تفاعل كوكب عطارد مع الشمس مع اخذ بعين الاعتبار تأثير باقي الكواكب علي عطارد نحسب سرعة الدوران

$$523 \ \text{arcsecond/century}.$$

الفرق هو

$$43 \ \text{arcsecond/century}.$$

هذه الكمية لا يمكن تفسيرها الا من خلال النسبية العامة اي من خلال فهمنا للثقالة علي انها قوة يتوسطها انحناء الفضاء−زمن. القوة الناجمة عن انحناء الفضاء−زمن بسبب كتلة الشمس والتي يستشعرها عطارد اكثر من غيره من الكواكب يمكن تقريبها ب

$$F = \frac{GM_s M_m}{r^2}\left(1 + \frac{\alpha}{r^2}\right) \ , \ \alpha = 1.1.10^{-8} AU^2.$$

الهدف هو التحقق عدديا من ان هذه القوة تؤدي فعلا الي كمية دوران للحضيض الشمسي لعطارد تساوي 43 قوس ثانية في القرن.

(1) عدل شفرة الفورترون التي استخدمناها في التطبيق السابق من اجل الاخذ بعين الاعتبار القوة المذكورة اعلاه .

أختيار الشروط الابتدائية مهم للغاية . الشرط الابتدائي الاول هو موضع عطارد . نختار

$$x_0 = (1+e)a \ , \ y_0 = 0.$$

اي اننا نختار الكوكب في اللحظة الابتدائية في ابعد نقطة له عن الشمس. نصف القطر الكبير $a$ لعطارد هو $0.39$ وحدة فلكية و لاتراكزية عطارد $e$ تساوي $0.206$. المسافة $ea$ هي بعد الشمس التي توجد في احد المحرقين عن مركز القطع الناقص . الشرط الابتدائي الثاني هو سرعة عطارد في اللحظة الابتدائية التي تعطي ب

$$v_{x0} = 0 \ , \ v_{y0} = \sqrt{\frac{GM_s}{a}\frac{1-e}{1+e}}.$$

هذه السرعة يمكن حسابها من تطبيق قانوني انحفاظ العزم الحركي و انحفاظ الطاقة بين النقطة الابتدائية اعلاه و النقطة $(x = 0, y = b)$ حيث $b$ هو نصف القطر الصغير لعطارد اي

$$b = a\sqrt{1 - e^2}.$$

(2) لان قيمه $\alpha$ التي تعطيها النسبية العامة صغيره جدا فان كمية دوران الحضيض الشمسي لعطارد ضئيلة يصعب ملاحظتها في اي محاكاه عددية ذات وقت محدود . نختار قيمة اكبر بكثير ل $\alpha$ مثلا

$$\alpha = 0.0008 AU^2.$$

نختار ايضا

$$N = 20000 \; , \;\; dt = 0.0001.$$

احسب المدار من اجل هذه القيم . احسب الزاوية $\theta$ التي يصنعها الشعاع الذي يربط عطارد و الشمس مع المحور الافقي بدلالة الزمن . احسب ايضا المسافة بين الشمس و عطارد و مشتقتها بالنسبة للزمن اي

$$\frac{dr}{dt} = \frac{xv_x + yv_y}{r}.$$

هذه المشتقة تغير اشارتها كلما بلغ عطارد ابعد نقطة له عن الشمس او بلغ اقرب نقطة له (اي الحضيض الشمسي) عن الشمس . استخدم هذه الملاحظة من اجل رسم الزاوية $\theta_p$ لما يكون عطارد في ابعد نقطة له عن الشمس بدلالة الزمن . ماذا تلاحظ .

عين الميل $d\theta_p/dt$ الذي هو بالضبط كمية دوران الحضيض الشمسي لعطارد حول الشمس من اجل قيمة $\alpha$ المختارة اعلاه .

(3) اعد السؤال السابق من اجل قيم اخري ل $\alpha$ . نقترح

$$\alpha = 0.001, 0.002, 0.004.$$

في كل مرة احسب $d\theta_p/dt$ . ارسم $d\theta_p/dt$ بدلالة $\alpha$ . ماذا تلاحظ . اوجد الميل . استنتج كمية دوران الحضيض الشمسي لعطارد من اجل القيمة

$$\alpha = 1.1.10^{-8} AU^2.$$

(4) باستخدام معطيات السؤال السابق اوجد الدالة

$$\frac{d\theta_p}{dt} = f(\alpha).$$

باستعمال طريقة المربعات الاصغرية .

# النواس الفوضوي 1 : تأثير الفراشة

نواس عبارة عن كتلة $m$ مربوطة بخيط طوله $l$ معلق في مرتكز ثابت تحت تأثير الثقالة $g$ . حركة اي نواس هي حركة غير خطية عموما لان الزاوية التي يصنعها النواس مع المحور الشاقولي ليست بالضرورة صغيرة و يمكن ان تبلغ القيمة العظمى $\pi$ او القيمة الصغرى $-\pi$ و بالتالي فان النواس يمكن ان يدور دورة كاملة تساوي 360 درجة حول نقطة ارتكازه . نأخذ بعين الاعتبار تأثير قوة مقاومة الهواء علي الكتلة $m$ و نفترض انها تعطي بقانون ستوكس الذي ينص علي ان مقاومة الهواء تكون معاكسه للحركة و متناسبة خطيا مع السرعة $d\theta/dt$ مع ثابت تناسب يساوي $mlq$ :

$$F_{\text{drag}} = -mlq\frac{d\theta}{dt}.$$

الاحتكاك مع الهواء يؤدي الي تخامد حركة النواس و توقفه عن الحركةبعد استهلاك النواس لكامل طاقته الابتدائية . من اجل الحفاظ علي حركة النواس ضد مقاومة الهواء من الضروري اضافة قوة تحريك خارجية التي نفترض انها قوة دورية في الزمن ذات تواتر $\nu_D$ و سعة ثابتة $mlF_D$ :

$$F_{\text{drive}} = mlF_D \sin 2\pi\nu_D t.$$

معادلة الاهتزاز المشتقة من قانون نيوتن الثاني تأخذ الشكل

$$\frac{d^2\theta}{dt^2} = -\frac{g}{l}\sin\theta - q\frac{d\theta}{dt} + F_D \sin 2\pi\nu_D t.$$

نأخذ دائما التواتر الزاوي $\sqrt{g/l}$ المرفق بالاهتزازات البسيطة للنواس يساوي واحد اي $l = g$ . الحل العددي الذي سنعتبره هنا هو الحل المعطي بخوارزمية اولر – كرومر :

$$\Omega_{i+1} = \Omega_i + \left( -\frac{g}{l}\sin\theta_i - q\Omega_i + F_D \sin 2\pi\nu_D t_i \right)\Delta t \ , \ \theta_{i+1} = \theta_i + \Omega_{i+1}\,\Delta t.$$

هذه الجملة الديناميكية تعرف باسم النواس الفوضوي (chaotic pendulum) و من اهم ما تتميز به الحساسية المفرطة للشروط الابتدائية . هذه الخاصية تعرف ايضا باسم تأثير الفراشة (butterfly effect) .

يمكن للهزازالفوضوي ان يتصرف بطريقتين مختلفتين . في المنطقة الخطية للهزاز الفوضوي الحركة دورية ذات دور يساوي دور قوة التحريك الخارجية اذا اهملنا الحركة الابتدائية العابرة . في المنطقة الفوضوية الحركة غير دورية لا تكرر نفسها ابدا و بالاضافة الي ذلك فان اي خطأ مهما كان متناه في الصغر في تحديد الشروط الابتدائية يؤدي الي حركة مختلفة بالكامل .

(1) اكتب شفرة منجز فيها الحل المعطي بخوارزمية اولر – كرومر لمسألة الهزاز الفوضوي . لنلاحظ ان الزاوية $\theta$ يمكن دائما اخذها محصورة في المجال $[-\pi,\pi]$ و في الحالة التي تكون فيها خارج هذا المجال نقوم باضافة $\pm 2\pi$ من اجل اعادة حصرها في المجال و هذا كالتالي

$$\text{if}(\theta_i.\text{lt.} \mp \pi)\ \theta_i = \theta_i \pm 2\pi.$$

<div dir="rtl">

(2) **نأخذ القيم و الشروط الابتدائية**

</div>

$$dt = 0.04s \ , \ 2\pi\nu_D = \frac{2}{3}s^{-1} \ , \ q = \frac{1}{2}s^{-1} \ , \ N = 1000 - 2000.$$

$$\theta_1 = 0.2 \text{ radian} \ , \ \Omega_1 = 0 \text{ radian}/s.$$

$$F_D = 0 \text{ radian}/s^2 \ , \ F_D = 0.1 \text{ radian}/s^2 \ , \ F_D = 1.2 \text{ radian}/s^2.$$

<div dir="rtl">

**ارسم الزاوية $\theta$ بدلالة الزمن. ماذا تلاحظ بالنسبة للقيمة الاولي لقوة التحريك الخارجية، ماهو تواتر الاهتزاز. ماذا تلاحظ بالنسبة للقيمة الثانية لقوة التحريك الخارجية، ماهو تواتر الاهتزاز من اجل الازمنة الصغري و ماهو تواتر الاهتزاز من اجل باقي الازمنة. ماذا تلاحظ بالنسبة للقيمة الثالثة . هل الحركة دورية .**

</div>

# النواس الفوضوي 2 : مقاطع بوانكري

الحركة في المنطقة الفوضوية هي حركة حتمية لان تصرف الهزاز في جميع الازمنة اللاحقة يحسب من حل معادلة الحركة اعلاه مع اعطاء شروط ابتدائية ملائمة لكن لا يمكن التنبؤ بها. لكن هذا لا يعني ان الهزاز الفوضوي هو جملة عشوائية و هي خاصية يمكن رؤيتها بوضوح في مقاطع بوانكري.
عوض رسم المدار من اجل كل الازمنة يمكن ان نرسم فقط النقاط $(\theta, \Omega)$ في فضاء الطور من اجل الازمنة التي تحقق الشرط $\nu_D t = n$ . مجموعة النقاط التي نحصل عليها بهذه الطريقة تسمى مقطع بوانكري.
في المنطقة الخطية للهزاز الفوضوي تتكون حركة الهزاز من حركة ابتدائية عابرة في الازمنة الصغرى و حركة دورية في باقي الازمنة . الجزء الدوري لا يتعلق بالشروط الابتدائية و لذلك يسمى مقطع المدار في فضاء الطور بالجاذب الدوري للهزاز الفوضوي . يتكون مقطع بوانكري من نقطة واحدة لان هذا المقطع هو جاذب لانه لا يتعلق بالشروط الابتدائية . من الواضح ان هذا المقطع هو جاذب لانه لا يتعلق بالشروط الابتدائية .
مقطع بوانكري في المنطقة الفوضوية هو ايضا جاذب في فضاء الطور اي مدار لا يتعلق بالشروط الابتدائية يسمى بالجاذب الغريب مما يؤكد حقيقة ان الهزاز الفوضوي رغم انه جملة حتمية لا يمكن التنبؤ بتصرفها في المنطقة الفوضوية الا انه ليس بجملة عشوائية .

(1) نعتبر الان هزازان فوضويان $A$ و $B$ متماثلان في كل شئ لكن شروطهما الابتدائية مختلفة اختلافا طفيفا . مثلا نأخذ

$$\theta_1^A = 0.2 \text{ radian} , \ \theta_1^B = 0.201 \text{ radian}.$$

يقاس الاختلاف بين الحركتين $A$ و $B$ بالفرق بين الزاويتين $\theta_A$ و $\theta_B$:

$$\Delta \theta_i = \theta_i^A - \theta_i^B.$$

احسب $\ln \Delta \theta$ بدلالة الزمن من اجل

$$F_D = 0.1 \text{ radian}/s^2 , \ F_D = 1.2 \text{ radian}/s^2.$$

ماذا تلاحظ . هل الحركتان $A$ و $B$ متماثلتان . ماذا يحدث في الازمنة الكبرى . هل حركة الهزاز الفوضوي هي من النوع الذي يمكن التنبؤ به . بالنسبة للقيمة الثانية استعمل

$$N = 10000 , \ dt = 0.01s.$$

(2) احسب السرعة الزاوية $\Omega$ بدلالة الزاوية $\theta$ من اجل

$$F_D = 0.5 \text{ radian}/s^2 , \ F_D = 1.2 \text{ radian}/s^2.$$

ماهو المدار في فضاء الطور من اجل الازمنة الصغرى و ماذا يمثل. كيف يصبح المدار في الازمنة الكبرى . قارن بين الهزازين $A$ و $B$ . هل يتعلق المدار في الازمنة الكبرى بالشروط الابتدائية .

(3) **للحصول علي مقطع بوانكري عدديا نرسم النقاط $(\theta, \Omega)$ في الازمنة التي تنعدم فيها الدالة $\sin \pi \nu_D t$ اي في الازمنة التي تغير فيها هذه الدالةاشارتها:**

$$\text{if}(\sin \pi \nu_D t_i \sin \pi \nu_D t_{i+1}.\text{lt}.0)\text{then}$$

$$\text{write}(*, *)t_i, \theta_i, \Omega_i.$$

**تحقق من ان مقطع بوانكري في المنطقة الخطية هو معطي بنقطة و حيدة في فضاء الطور . خذ مثلا**

$$F_D = 0.5 \text{ radian}/s^2.$$

**و استعمل**

$$N = 10^4 - 10^7 \ , \ dt = 0.001s.$$

**تحقق من ان مقطع بوانكري في المنطقة الفوضوية هو ايضا جاذب . خذ مثلا**

$$F_D = 1.2 \text{ radian}/s^2.$$

**و استعمل**

$$N = 10^5 \ , \ dt = 0.04s.$$

**قارن بين مقطع بوانكري للهزاز $A$ و مقطع بوانكري للهزاز $B$ . ماذا تلاحظ و ماذا تستنتج .**

# النواس الفوضوي 3 : ظاهرة تضاعف الدور

من اهم الخصائص الفوضوية التي يتميز بها النواس الفوضوي هو ظاهرة تضاعف الدور. المدارات الدورية التي لها نفس دور قوة التحريك الخارجية تسمي الحركة ذات الدور واحد (period-1 motion) . لكن توجد ايضا مدارات ذات دور يساوي ضعف دور القوة الخارجية و مدارات ذات دور يساوي اربعة اضعاف دور القوة الخارجية و بصفة عامة مدارات ذات دور يساوي $2^N$ ضعف دور القوة الخارجية . المدارات التي دورها يساوي $2^N$ ضعف دور قوة التحريك الخارجية تسمي الحركة ذات الدور $N$ (period-$N$ motion) . في عالم الاهتزازات و الامواج المدارات التي نحصل عليها في العادة هي مدارات دورية ذات ادوار تساوي دور قوة التحريك الخارجية تقسيم $2^N$ و هي ظاهرة تعرف باسم المزج (mixing). اذن ظاهرة تضاعف الدور التي تشاهد في النواس الفوضوي هي ظاهرة جديدة تنتمي الي عالم الفوضي . التحول الي الفوضي يحدث بالضبط لما $N \longrightarrow \infty$ .

من اجل الحركة ذات الدور $N$ نتوقع ان توجد $N$ قيمة مختلفة للزاوية $\theta$ من اجل كل قيمة ل $F_D$ . الدالة $\theta$ بدلالة $F_D$ تسمي مخطط انشطار (bifurcation) و هو منحني ذو بنية منكسرة (fractal) في المنطقة الفوضوية. من هذا المخطط يمكن حساب متي يحدث بالضبط التحول نحو الفوضي.

(1) نأخذ القيم و الشروط الابتدائية

$$l = g \ , \ 2\pi\nu_D = \frac{2}{3}s^{-1} \ , \ q = \frac{1}{2}s^{-1} \ , \ N = 3000 - 100000 \ , \ dt = 0.01s.$$

$$\theta_1 = 0.2 \text{ radian} \ , \ \Omega_1 = 0 \text{ radian}/s.$$

عين دور الحركة من اجل القيم

$$F_D = 1.35 \text{ radian}/s^2 \ , \ F_D = 1.44 \text{ radian}/s^2 \ , \ F_D = 1.465 \text{ radian}/s^2.$$

ماذا يحدث للدور عندما نزيد في قيمة $F_D$. هل القيمتان الثانيتان ل $F_D$ تقعان في المنطقة الخطية ام في المنطقة الفوضوية للهزاز الفوضوي.

(2) احسب الزاوية $\theta$ بدلالة $F_D$ من اجل الازمنة التي تحقق الشرط $2\pi\nu_D t = 2n\pi$. نأخذ $F_D$ في المجال

$$F_D = (1.34 + 0.005k) \text{ radian}/s^2 \ , \ k = 1, ..., 30.$$

عين مجال قوة التحريك الخارجية الذي تكون فيه المدارات تنتمي الي الحركات ذات الدور واحد، اثنان و اربعة.

في هذا السؤال من المهم جدا ازالة الحركة الابتدائية العابرة قبل البدء في قياس مخطط الانشطار. يمكن انجاز هذا الامر كالتالي. نقوم بحساب الحركة لمدة $2N$ خطوة ثم نأخذ بعين الاعتبار فقط ال $N$ خطوة الاخيرة عند حساب مقطع بوانكري من اجل كل قيمة ل $F_D$.

# النواس الفوضوي 4: مخططات الانشطار و الانكسار التلقائي للتناظر

الهزاز الفوضوي يعطي بمعادلة الحركة

$$\frac{d^2\theta}{dt^2} = -\sin\theta - \frac{1}{Q}\frac{d\theta}{dt} + F_D \cos 2\pi\nu_D t.$$

نأخذ عبر كل هذه المحاكاة القيم التالية

$$F_D = 1.5 \text{ radian}/s^2 \ , \ 2\pi\nu_D = \frac{2}{3}s^{-1}.$$

من اجل تحري دقة عددية اعلي نستخدم هذه المرة خوارزمية رونج – كوتا:

$$k_1 = \Delta t \ \Omega(i).$$

$$k_3 = \Delta t \left[ -\sin\theta(i) - \frac{1}{Q}\Omega(i) + F_D \cos 2\pi\nu_D \Delta t(i-1) \right].$$

$$k_2 = \Delta t \left( \Omega(i) + \frac{1}{2}k_3 \right).$$

$$k_4 = \Delta t \left[ -\sin\left(\theta(i) + \frac{1}{2}k_1\right) - \frac{1}{Q}\left(\Omega(i) + \frac{1}{2}k_3\right) + F_D \cos 2\pi\nu_D \Delta t(i - \frac{1}{2}). \right].$$

$$\theta(i+1) = \theta(i) + k_2.$$

$$\Omega(i+1) = \Omega(i) + k_4.$$

$$t(i+1) = \Delta t \ i.$$

في المنطقة الخطية المدارات هي قطوع ناقصة تتميز بالتناظر

$$\theta \longrightarrow -\theta.$$

هذه المدارات بالاضافة الي كونها دورية ذات دور $T_D$ يساوي دور القوة الخارجية فهي تتميز بتناظر كامل بين اليمين و اليسار و بالتالي فان الوقت الذي يصرفه النواس في حركته الي يمين محوره الشاقولي يساوي الوقت الذي يصرفه في حركته الي يسار محوره الشاقولي. من المثير للاهتمام وجود حلول اخري لمعادلات حركة الهزاز الفوضوي دورية ذات دور يساوي $T_D$ لكنها لا تتميز بالتناظر $-\theta \longrightarrow \theta$. في هذه الحلول نجد ان الهزاز يصرف معظم وقته اما في المنطقة $0 > \theta$ او في المنطقة $0 < \theta$. يمكن و صف هذه الحلول غير المتناظرة بمخطط انشطار

$$\Omega = \Omega(Q).$$

من اجل كل قيمة لمعامل الجودة $Q$ فاننا نحسب مقطع بوانكري اي قيم $\theta$ و $\Omega$ في اللحظات $t = nT_D$. نلاحظ ان مقطع بوانكري ينشطر من اجل قيمة معينة $Q_*$ لـ $Q$. تحت هذه القيمة نحصل علي خط واحد لان الحركة ذات دور $T_D$ و فوق $Q_*$ نحصل علي خطين رغم ان دورالحركة ما زال يساوي $T_D$. الخطان يقابلان الحلان اللذان يصرف فيهما الهزاز اغلب وقته في المنطقة اليمني ($0 < \theta$) او المنطقة اليسري ($0 < \theta$). الوصول الي احد الحلين انطلاقا من

الحل المتناظر يتعلق بالشروط الابتدائية و يكون عبر زيادة قيمة $Q$ تدريجيا. هذا مثال لظاهرة الانكسار التلقائي للتناظر.

كما رأينا في المحاكاة السابقة يمكن ايضا وصف ظاهرة تضاعف الدور بمخطط انشطار. هذه الظاهرة هي ايضا مثال لظاهرة الانكسار التلقائي للتناظر. في هذه الحالة فان التناظر الذي ينكسر هو

$$t \longrightarrow t + T_D.$$

فقط الحركات التي دورها يساوي $T_D$ تتميز بهذا التناظر. لنلاحظ ان الحركات ذات الدور $\mathcal{N}$ اي المدارات التي لها دور يساوي $2^{\mathcal{N}} T_D$ لا تتميز ايضا بالتناظر $\theta \longrightarrow -\theta$.

لتكن $Q_{\mathcal{N}}$ قيمة $Q$ التي يحدث فيها الانشطار رقم $\mathcal{N}$. اي ان $Q_{\mathcal{N}}$ هي القيمة التي يتحول عندها المدار من مدار ذو دور يساوي $2^{\mathcal{N}-1} T_D$ الي مدار ذو دور يساوي $2^{\mathcal{N}} T_D$. نسبة فاينباوم تعرف كالتالي

$$F_{\mathcal{N}} = \frac{Q_{\mathcal{N}-1} - Q_{\mathcal{N}-2}}{Q_{\mathcal{N}} - Q_{\mathcal{N}-1}}.$$

لما نقترب من المنطقة الفوضوية اي لما $\mathcal{N} \longrightarrow \infty$ فان $F_{\mathcal{N}}$ يقترب بسرعة من القيمة الثابتة

$$F = 4.669.$$

هذه النتيجة عامة لا تختص بالهزاز الفوضوي دون غيره من الجمل الفوضوية. في اي جملة ديناميكية يمكنها ان تتحول الي الفوضي عبر سلسلة غير منتهية من الانشطارات المرفقة بتضاعف للدور فان ثابت فاينباوم يقترب من نفس القيمة 4.669 لما $\mathcal{N} \longrightarrow \infty$.

(1) اعد كتابة الشفرة باستخدام رونج – كوتا.

(2) نأخذ مجموعتين مختلفتين من الشروط الابتدائية

$$\theta = 0.0 \text{ radian} , \ \Omega = 0.0 \text{ radian}/s.$$

$$\theta = 0.0 \text{ radian} , \ \Omega = -3.0 \text{ radian}/s .$$

ادرس طبيعة المدار من اجل القيم

$$Q = 0.5s \ , \ Q = 1.24s \ , \ Q = 1.3s.$$

ماذا تلاحظ.

احسب مقطع بوانكري من اجل قيم $Q$ في المجال

$$[1.2, 1.3].$$

ارسم مخطط الانشطار $\Omega = \Omega(Q)$. ماهي القيمة $Q_*$ التي ينكسر فيها التناظر $\theta \longrightarrow -\theta$ تلقائيا.

(3) احسب المدار و مقطع بوانكري من اجل

$$Q = 1.36s.$$

ماهو دور الحركة. هل المدار متناظر تحت تأثير $t \longrightarrow t + T_D$. هل المدار متناظر تحت تأثير $\theta \longrightarrow -\theta$. ارسم مخطط الانشطار $\Omega = \Omega(Q)$ من اجل مجموعتين مختلفتين من الشروط الابتدائية.

ماهي القيمة $Q_1$ التي يتضاعف فيها الدور اي القيمة التي ينكسر فيها التناظر $t \longrightarrow t + T_D$ .

(4) في هذا السؤال و الذي يليه نستخدم الشروط الابتدائية

$$\theta = 0.0 \text{ radian} , \ \Omega = 0.0 \text{ radian}/s.$$

احسب المدار و مقطع بوانكري و ارسم مخطط الانشطار $\Omega = \Omega(Q)$ من اجل قيم $Q$ في المجال

$$[1.34, 1.38].$$

عين من مخطط الانشطار القيم $Q_{\mathcal{N}}$ من اجل $\mathcal{N} = 1, 2, 3, 4, 5$. احسب ثابت فاينباوم و نقطة التراكم $Q_\infty$ التي يحدث عندها التحول نحو الفوضي.

(5) حتي نفهم التحول نحو الفوضي بطريقة افضل نعتبر هزازان فوضويان مختلفان اختلافا طفيفا. مثلا نأخذ

$$\Delta\theta = 10^{-6} \text{ radian} , \ \Delta\Omega = 10^{-6} \text{ radian}/s.$$

احسب المدار و مقطع بوانكري و عين الدور و كذلك احسب $\ln |\Delta\Omega|$ من اجل قيم $Q$ التالية

$$Q = 1.372s , \ 1.375s , \ 1.3757s , \ 1.376s.$$

ماذا تلاحظ لما نقترب من منطقة الفوضي.

# الديناميك الجزيئي $1$: توزيع ماكسويل

نعتبر حركة $N$ ذرة ارغون في بعدين داخل علبة مساحتها $L^2$. طاقة التفاعل بين اي ذرتين مفصولتين بمسافة $r$ تعطي بكمون لينارد-جونز $u$ المعرف ب

$$u = 4\epsilon \left[ \left( \frac{\sigma}{r} \right)^{12} - \left( \frac{\sigma}{r} \right)^6 \right].$$

القوة التي تطبقها الذرة $k$ علي الذرة $i$ هي

$$f_{k,i} = \frac{24\epsilon}{r_{ki}} \left[ 2 \left( \frac{\sigma}{r_{ki}} \right)^{12} - \left( \frac{\sigma}{r_{ki}} \right)^6 \right].$$

معادلات حركة الذرة $i$ تعطي ب

$$\frac{d^2 x_i}{dt^2} = a_{x,i} = \frac{1}{m} \sum_{k \neq i} f_{k,i} \frac{x_i - x_k}{r_{ki}} \ , \ \frac{d^2 y_i}{dt^2} = a_{y,i} = \frac{1}{m} \sum_{k \neq i} f_{k,i} \frac{y_i - y_k}{r_{ki}}.$$

الخوارزمية العددية التي سنستعملها لحل هذه المعادلات التفاضلية هي خوارزمية قيرلات التي تعطي بالمعادلات

$$x_{i,n+1} = 2x_{i,n} - x_{i,n-1} + (\Delta t)^2 a_{x,i,n} \ , \ y_{i,n+1} = 2y_{i,n} - y_{i,n-1} + (\Delta t)^2 a_{y,i,n}.$$

سنحسب ايضا السرعات باستعمال المعادلات التالية

$$v_{x,i,n} = \frac{x_{i,n+1} - x_{i,n-1}}{2\Delta t} \ , \ v_{y,i,n} = \frac{y_{i,n+1} - y_{i,n-1}}{2\Delta t}.$$

من اجل التبسيط نستخدم الوحدات المختزلة $\sigma = \epsilon = m = 1$. ايضا من اجل التقليل من اثار الحواف نستعمل الشروط الحدية الدورية. اي نعتبر العلبة التي تحتوي علي الغاز علي انها تورص بدون حواف و بالتالي فانه عندما تصطدم ذرة ارغون بجدران العلبة في اي اتجاه فاننا نزيد او ننقص طول العلبة في ذلك الاتجاه كما يلي

if $(x_i > L)$ then $x_i = x_i - L$ , if $(x_i < 0)$ then $x_i = x_i + L$

if $(y_i > L)$ then $y_i = y_i - L$ , if $(y_i < 0)$ then $y_i = y_i + L$.

بسبب الشروط الحدية الدورية فان ألمسافة العظمي في الاتجاه $x$ بين اي ذرتين هو $L/2$ فقط و كذلك المسافة العظمي في الاتجاه $y$ بين اي ذرتين هو $L/2$. يتم تنفيذ هذا الامر كالتالي

if $(x_{ij} > L/2)$ then $x_{ij} = x_{ij} - L$ , if $(x_{ij} < -L/2)$ then $x_{ij} = x_{ij} + L$

if $(y_{ij} > L/2)$ then $y_{ij} = y_{ij} - L$ , if $(y_{ij} < -L/2)$ then $y_{ij} = y_{ij} + L$.

في هذه المسألة نأخذ $L$ فردي و $N$ مربع تام. الشبكة تتميز بطول الخطوة

$$a = \frac{L}{\sqrt{N}}.$$

اذن الشبكة تتشكل من $N$ خلية مساحة كل منها هي $a^2$. نختار $L$ و $N$ بحيث $a > 2\sigma$. نختار مواضع الذرات كالتالي. الذرة $j$ توضع في مركز الخلية ذات الاركان $k = \sqrt{N}(i-1)$، $(i,j)$، $(i+1,j)$، $(i,j+1)$ و $(i+1,j+1)$. نقوم بعد ذلك بادخال اضطراب عشوائي علي هذه الوضعيات الابتدائية عن طريق اضافة اعداد عشوائية في المجال $[-a/4, +a/4]$ الي احداثيات الذرات. نختار السرعات الابتدائية في اتجاهات عشوائية لكن بطويلة تساوي $v_0$ من اجل جميع الذرات.

(1) اكتب شفرة ديناميك جزيئي باتباع الخطوات اعلاه. خذ $L = 15$، $N = 25$، $\Delta t = 0.02$، Time $= 500$ و $v_0 = 1$. كاختبار اولي تحقق من ان الطاقة الكلية للجملة منحفظة. ارسم مسارات الجسيمات. ماذا تلاحظ.

(2) كاختبار ثاني نقترح قياس درجة الحرارة عن طريق ملاحظة كيفية اقتراب الغاز من التوازن. استعمل نظرية التقسيم المتساوي للطاقة التي تعطي ب

$$k_B T = \frac{m}{2N} \sum_{i=1}^{N} (v_{i,x}^2 + v_{i,y}^2).$$

ارسم $T$ كدالة في الزمن. خذ Time $= 1000 - 1500$. ماهي درجة حرارة الغاز عند التوازن.

(3) احسب توزيع سرعات ذرات الارغون عن طريق انشاء هيستوغرام السرعات. نأخذ القيمة Time $= 2000$. نعتبر سرعات كل الجسيمات في كل اللحظات. هناك $N$.Time قيمة للسرعة في هذه العينة. ننشئ هيستوغرام هذه العينة عن طريق:

○ ايجاد القيمة العظمى و القيمة الصغرى.
○ تقسيم المجال الي سلات.
○ تحديد عدد المرات التي تقع فيها قيمة معينة للسرعة داخل سلة معينة.
○ تنظيم التوزيع.

قارن مع توزيع ماكسويل

$$P_{\text{Maxwell}}(v) = C \frac{v^2}{k_B T} \, e^{-\frac{mv^2}{2k_B T}}.$$

استنتج درجة الحرارة من القيمة العظمي للتوزيع التي تعطي ب

$$k_B T = m v_{\text{peak}}^2.$$

قارن مع درجة الحرارة المحصل عليها من نظرية التقسيم المتساوي للطاقة. ماذا يحدث اذا زدنا السرعة الابتدائية.

# الديناميك الجزيئي 2: الانصهار

نريد في هذه المسألة دراسة الانصهار الذي هو التحول الطوري من الحالة الصلبة الي الحالة
السائلة. علينا اولا ان نحدد الشروط الصحيحة للحالة الصلبة. من الواضح ان درجة الحرارة
يجب ان تكون منخفضة بما فيه الكفاية و الكثافة مرتفعة بما فيه الكفاية حتي تكون الحالة
صلبة. من اجل خفض درجة الحرارة الي اقصي حد ممكن نبدأ من الحالة التي تكون فيها كثافة
الجسيمات في حالة سكون. من اجل الحصول علي تجاذب اعظمي بين الذرات نختار كثافة
مساوية لجسيم واحد في كل وحدة مساحة مختزلة. نختار بالخصوص $N = 16$ و $L = 4$.

(1)  بين انه باستعمال الشروط الابتدائية المذكورة اعلاه فاننا نحصل علي حالة صلبة بلورية
ذات شبكة مثلثية.

(2)  من اجل مشاهدة الانصهار يجب تسخين الجملة عن طريق زيادة الطاقة الحركية للذرات
يدويا. يمكننا تحقيق هذا الامر مثلا عن طريق تغيير مواضع الجسيمات كل 1000 خطوة
كالتالي

$$\text{hh} = \text{int}(n/1000)$$
$$\text{if } (\text{hh} * 1000.\text{eq}.n) \text{ then}$$
$$x(i, n) = x(i, n + 1) - R(x(i, n + 1) - x(i, n))$$
$$y(i, n) = y(i, n + 1) - R(y(i, n + 1) - y(i, n))$$
$$\text{endif}.$$

هذه العملية تؤدي الي ضرب السرعات بالقيمة $R$. نختار $R = 1.5$.

تحقق من اننا نحصل بالفعل علي الانصهار بهذه الطريقة. ماذا يحدث للطاقة و درجة
الحرارة.

# الاعداد العشوائية

**الجزء الاول**   نعتبر مولد اعداد شبه عشوائية يعتمد علي طريقة المتبقيات المتمثلة في العلاقة

$$r_{i+1} = \text{remainder}\left(\frac{ar_i + c}{M}\right).$$

الثوابت $a$، $c$ و $M$ هم الضارب، المضاف و الطويلة علي التوالي. العدد العشوائي الابتدائي $r_i$ يسمي البذرة. نعطي القيم

$$a = 899, c = 0, M = 32768, r_1 = 12 \ \text{"good"}$$
$$a = 57, c = 1, M = 256, r_1 = 10 \ , \ \text{"bad"}.$$

الدالة remainder تنفذ في الفورترن كالتالي

$$\text{remainder} \ \frac{a}{b} = \text{mod}(a, b).$$

(1) احسب سلسلة الاعداد العشوائية باستعمال القيم اعلاه. ارسم $r_i$ بدلالة $i$. انشئ مخطط التناثر ($x_i = r_{2i}, y_i = r_{2i+1}$).

(2) احسب متوسط الاعداد العشوائية. ماذا تلاحظ.

(3) ليكن $N$ عدد الاعداد العشوائية المولدة. احسب دوال الربط

$$\text{sum}_1(k) = \frac{1}{N-k} \sum_{i=1}^{N-k} x_i x_{i+k} \ , \ \text{sum}_2 = \frac{\text{sum}_1(k) - <x_i>^2}{\text{sum}_1(0) - <x_i>^2}.$$

ماهو تصرف هذه الدوال في $k$.

(4) احسب دور المولدات العشوائية اعلاه.

**الجزء الثاني**   نأخذ $N$ عدد عشوائي في المجال $[0, 1]$ الذي نقسمه الي $K$ مجال صغير او سلة طول كل واحدة هو $\delta = 1/K$. ليكن $N_i$ عدد الاعداد العشوائية التي تقع في السلة $i$. من اجل سلسلة اعداد عشوائية منتظمة عدد الاعداد العشوائية المتوقع في كل سلة هو $n_{\text{ideal}} = N/K$. تعرف احصائية $\chi^2-$ كالتالي

$$\chi^2 = \frac{1}{n_{\text{ideal}}} \sum_i (N_i - n_{\text{ideal}})^2.$$

(1) تحقق من النتيجة $n_{\text{ideal}} = N/K$ من اجل المولد rand الذي نجده في المكتبة المعيارية للفورترن. خذ القيم $K = 10$ و $N = 1000$. ارسم $N_i$ بدلالة الموضع $x_i$ للسلة $i$.

(2) عدد درجات الحرية هو $\nu = K - 1$. القيمة الاكثر احتمالا ل $\chi^2$ هي $\nu$. تحقق من هذه النتيجة من اجل عدد كلي من اختبارات السلة يساوي $L = 1000$ و $K = 11$. في كل مرة احسب عدد المرات $L_i$ من بين ال $L = 1000$ اختبار سلة التي نحصل فيها علي قيمة معينة ل $\chi^2$. ارسم $L_i$ بدلالة $\chi^2$. ماذا تلاحظ.

# المشاء العشوائي

**الجزء الاول**     نعتبر حركة مشاء عشوائي في بعد واحد. المشاء يمكنه الحركة الي اليمين خطوة تساوي $s_i = a$ باحتمال $p$ او الي اليسار خطوة تساوي $s_i = -a$ باحتمال $q = 1 - p$. بعد $N$ خطوة موضع المشاء يصبح $x_N = \sum_i s_i$. نأخذ القيم

$$p = q = \frac{1}{2} \ , \ a = 1.$$

**من اجل محاكاة حركة المشاء العشوائي نحتاج الي مولد للاعداد العشوائية. في هذه المسألة نستخدم المولد rand الذي نجده في المكتبة المعيارية للفورترن. نستدعي هذا المولد عن طريق اصدار الامر التالي**

       call srand(seed)

       rand()

**يمكن ان نستنسخ حركة المشاء العشوائي بالشفرة التالية**

       if (rand() < p) then

       $x_N = x_N + a$

       else

       $x_N = x_N - a$

       endif.

(1) احسب المواضع $x_i$ لثلاث مشاءات عشوائيات بدلالة رقم الخطوة $i$. نأخذ $i = 1, 100$. ارسم المسارات الثلاثة.

(2) نعتبر الان $K$ حركة مشاء عشوائي حيث $K = 500$. احسب المتوسطات

$$< x_N > = \frac{1}{K} \sum_{i=1}^{K} x_N^{(i)} \ , \ < x_N^2 > = \frac{1}{K} \sum_{i=1}^{K} (x_N^{(i)})^2.$$

**في المعادلات اعلاه $x_N^{(i)}$ هو موضع المشاء العشوائي $i$ بعد $N$ خطوة. ادرس تصرف هذه المتوسطات كدوال في $N$. قارن مع الحسابات النظرية.**

**الجزء الثاني**     نعتبر الان مشاء عشوائي في بعدين علي شبكة نقاط غير منتهية. انطلاقا من اي نقطة $(i, j)$ علي الشبكة يمكن للمشاء الوصول الي اي نقطة من نقاط الجوار الاقرب الاربعة $(i+1, j)$، $(i-1, j)$، $(i, j+1)$ و $(i, j-1)$ باحتمالات $p_x$، $q_x$، $p_y$ و $q_y$ علي التوالي حيث $p_x + q_x + p_y + q_y = 1$. من اجل التبسيط نفترض ان $p_x = q_x = p_y = q_y = 0.25$.

(1) احسب المتوسطات $< \vec{r}_N >$ و $< \vec{r}_N^2 >$ كدوال في عدد الخطوات $N$ من اجل $L = 500$ مشاء عشوائي. نعتبر القيم $N = 10, ..., 1000$.

# تقريبات النقطة الوسطي و مونتي كارلو

**الجزء الاول**    يعطي حجم كرة نصف قطرها $R$ في بعد $d$ بالعلاقة

$$
\begin{aligned}
V_d &= \int_{x_1^2+...+x_d^2 \leq R^2} dx_1...dx_d \\
&= 2 \int dx_1...dx_{d-1}\sqrt{R^2 - x_1^2 - ... - x_{d-1}^2} \\
&= \frac{R^d}{d} \frac{2\pi^{\frac{d}{2}}}{\Gamma(\frac{d}{2})}.
\end{aligned}
$$

(1) اكتب برنامج يحسب التكامل اعلاه في ثلاث ابعاد باستعمال طريقة النقطة الوسطي. نأخذ طول الخطوة $h = 2R/N$، نصف القطر $R = 1$ و عدد الخطوات في كل اتجاه يساوي $N = N_x = N_y = 2^p$ حيث $p = 1, 15$.

(2) بين ان الخطأ يتصرف مثل $1/N$. ارسم لوغاريتم القيمة المطلقة للخطأ المطلق بدلالة لوغاريتم $N$.

(3) جرب حساب التكامل في بعدين. استعمل فقط الربع الموجب للمستوي الحقيقي و خذ طول الخطوة $h = R/N$ حيث $R = 1$ و $N = 2^p$، $p = 1, 15$. نعلم من النظري ان الخطأ يجب ان يتصرف مثل $1/N^2$. ماهو الخطأ في هذه الحالة و لماذا الاختلاف. ملحوظة: المشتقة الثانية للدالة داخل التكامل غير معرفة عند $x = R$ مما يغير تصرف الخطأ من $1/N^2$ الي $1/N^{1.5}$.

**الجزء الثاني**    من اجل حساب حجم الكرة في اي بعد $d$ عدديا نستعمل علاقة التكرار

$$
V_d = \frac{V_{d-1}}{R^{d-1}} \int_{-R}^{+R} dx_d \, (R^2 - x_d^2)^{\frac{d-1}{2}}.
$$

(1) احسب الحجوم في الابعاد $d = 4, 5, 6, 7, 8, 9, 10, 11$. قارن بالنتيجة المضبوطة المعطاة اعلاه.

# الجزء الثالث

(1) استعمل طريقة المعاينة لمونتي كارلو المسماة طريقة الاصابة او الخطأ من اجل حساب التكاملات في الابعاد $d = 2, 3, 4$ و $d = 10$. هل استعمال طريقة مونتي كارلو هذه اسهل من استعمال طريقة النقطة الوسطي في اي بعد.

(2) استعمل طريقة القيمة الوسطي للعينة لمونتي كارلو من اجل حساب التكاملات في الابعاد $d = 2, 3, 4$ و $d = 10$. من اجل كل $d$ نجري $M$ قياس كل واحد مشكل من $N$ عينة. نعتبر $M = 1, 10, 100, 150$ و $N = 2^p$ حيث $p = 10, 19$. تحقق من ان الخطأ المضبوط يتصرف مثل $1/\sqrt{N}$.

ملحوظة: قارن الخطأ المضبوط الذي هو معروف في هذه الحالة مع الانحراف المعياري للمتوسط $\sigma_M$ و مع $\sigma/\sqrt{N}$ حيث $\sigma$ هو الانحراف المعياري في قياس واحد. هذه الكميات الثلاثة يجب ان تكون متساوية.

## الجزء الرابع

(1) يمكن ان تعطي قيمة $\pi$ بالتكامل

$$\pi = \int_{x^2+y^2 \leq R^2} dx \; dy.$$

استعمل طريقة المعاينة لمونتي كارلو (طريقة الاصابة او الخطأ) لحساب قيمة تقريبية ل $\pi$.

(2) التكامل اعلاه يمكن ايضا ان يكتب علي الشكل

$$\pi = 2 \int_{-1}^{+1} dx \; \sqrt{1-x^2}.$$

استعمل طريقة القيمة الوسطي للعينة لمونتي كارلو لحساب قيمة تقريبية ل $\pi$.

# توزيعات الاحتمال غير المنتظمة

**الجزء الاول**     توزيع غوس يعطي ب

$$P(x) = \frac{1}{\sqrt{2\pi\sigma^2}} \exp{-\frac{(x-\mu)^2}{2\sigma}}.$$

الوسيط $\mu$ هو المتوسط و $\sigma$ هو التفاوت اي الجذر التربيعي للانحراف المعياري. نختار $\mu = 0$
و $\sigma = 1$.

(1)  اكتب برنامج يحسب سلسلة من الاعداد العشوائية $x$ موزعة حسب $P(x)$ باستعمال طريقة
التحويل العكسي (خوارزمية بوكس و مولر) المعطاة بالمعادلات

$$x = r\cos\phi.$$

$$r^2 = -2\sigma^2\ln v \ , \ \phi = 2\pi w.$$

الاعداد $v$ و $w$ هي اعداد عشوائية منتظمة في المجال $[0,1]$.

(2)  ارسم هيستوغرام للاعداد العشوائية المحصل عليها في السؤال السابق باتباع الخطوات
التالية:

$a$-  عين مجال الاعداد العشوائية المحصل عليها في السؤال السابق.

$b$-  نقسم المجال الي $u$ سلة طول كل واحدة هو $h = \text{interval}/u$. نأخذ $u = 100$.

$c$-  نحدد موضع كل عدد عشوائي $x$ بين السلات. كل مرة نجد فيها عدد عشوائي في
سلة معينة نزيد واحد الي العداد المرفق بهذه السلة.

$d$-  نرسم نسبة الاعداد العشوائية في كل سلة بدلالة الموضع $x$. نسبة الاعداد العشوائية
في كل سلة تساوي عدد الاعداد العشوائية التي تقع في هذه السلة علي $hN$ حيث $N$
هو العدد الكلي للاعداد العشوائية. نأخذ $N = 10000$.

(3)  ارسم الهيستوغرام علي سلم لوغاريتمي اي ارسم $\log(\text{fraction})$ بدلالة $x^2$. اوجد الفت و
قارن مع النظرية.

## الجزء الثاني

(1)  طبق طريقة الرفض و القبول لمونتي كارلو علي المسألة اعلاه.

(2)  طبق طريقة فرنانداز و كريادو علي المسالة اعلاه. الخطوات هي كالتالي:

$a$-  نبدأ من $N$ نقطة $x_i$ حيث $x_i = \sigma$.

$b$ـ **نختار بشكل عشوائي زوج من النقاط** $(x_i, x_j)$ **من السلسلة و نقوم بالتغيير**

$$x_i \longrightarrow \frac{x_i + x_j}{\sqrt{2}}$$
$$x_j \longrightarrow -x_i + \sqrt{2}x_j.$$

$c$ـ **نكرر الخطوة الثانية حتي نصل الي التوازن. مثلا كرر الخطوة الثانية** $M$ **مرة حيث**
$M = 10, 100, ...$.

# خوارزمية ميتروبوليس و نموذج ايزينغ

**الجزء الاول**   نعتبر $N$ سبين علي شبكة مربعة حيث $L$ هو عدد مواقع الشبكة في كل اتجاه اي
ان $N = L^2$. كل سبين يمكنه ان يأخذ احدى القيمتين $s_i = +1$ (سبين علوي) او $s_i = -1$ (سبين
سفلي). كل سبين يتفاعل فقط مع جيرانه الاربعة الاقرب و ايضا مع حقل مغناطيسي خارجي
$H$. نموذج ايزينغ في بعدين يعطي بدالة الطاقة

$$E = -J \sum_{<ij>} s_i s_j - H \sum_i s_i.$$

السبين الموجود في نقطة تقاطع الخط $i$ و العمود $j$ يمثل بعنصر المصفوفة $\phi(i,j)$. الطاقة
يمكن ان تعطي اذن ب

$$E = -\frac{J}{2} \sum_{i,j=1}^{L} \phi(i,j)\Big(\phi(i+1,j) + \phi(i-1,j) + \phi(i,j+1) + \phi(i,j-1)\Big)$$
$$-H \sum_{i=1}^{} \phi(i,j).$$

نفرض الشروط الحدية الموافقة للتورص اي

$$\phi(0,j) = \phi(n,j) \ , \ \phi(n+1,j) = \phi(1,j) \ , \ \phi(i,0) = \phi(i,n) \ , \ \phi(i,n+1) = \phi(i,1).$$

نفترض ايضا ان الجملة في حالة توازن حراري مع خزان حراري ذو درجة حرارة $T$. التقلبات
الحرارية للجملة تحاكي بخوارزمية ميتروبوليس.

(1)  اكتب روتين جزئي يحسب الطاقة $E$ و المغنطة $M$ في التمثيلة $\phi$ لنموذج ايزينغ. المغنطة
هو وسيط ترتيب الجملة و هو معرف كالتالي

$$M = \sum_i s_i = \sum_{i,j=1}^{} \phi(i,j).$$

(2)  اكتب روتين جزئي ينفذ خوارزمية ميتروبوليس لهذه الجملة. الفرق في الطاقة الناجم
عن قلب السبين $\phi(i,j)$ يعطي ب

$$\Delta E = 2J\phi(i,j)\big(\phi(i+1,j) + \phi(i-1,j) + \phi(i,j+1) + \phi(i,j-1)\big) + 2H \sum_{i=1}^{} \phi(i,j).$$

(3)  نختار $L = 10$، $H = 0$، $J = 1$ و $\beta = 1/T$. نعتبر حالة الانطلاقة الباردة و كذلك حالة
الانطلاقة الساخنة المعرفتان علي التوالي ب

$$\phi(i,j) = +1 \ \ \forall \, i,j : \ \text{Cold Start.}$$

$$\phi(i,j) = \text{rand() } : \ \text{Hot Start.}$$

شغل خوارزمية ميتروبوليس من اجل زمن موازنة $\text{TTH} = 2^6$ و ادرس تاريخ الطاقة و
المغنطة من اجل درجات حرارة مختلفة. الطاقة و المغنطة تقتربان من القيم $E = 0$ و
$M = 0$ لما $T \longrightarrow \infty$ و من القيم $E = -2JN$ و $M = +1$ لما $T \longrightarrow 0$.

(4) ضف TTM $= 2^{10}$ خطوة مونتي كارلو و احسب متوسطات الطاقة و المغنطة.

(5) احسب السعة الحرارية و الحساسية المغناطيسية لهذه الجملة المعرفان ب

$$C_v = \frac{\partial}{\partial \beta} < E > = \frac{\beta}{T}(< E^2 > - < E >^2) \ , \ \chi = \frac{\partial}{\partial H} < M > = \beta(< M^2 > - < M >^2).$$

(6) احسب النقطة الحرجة و قارن بالنتيجة النظرية المضبوطة

$$k_B T_c = \frac{2J}{\ln(\sqrt{2}+1)}.$$

**الجزء الثاني**    ضف الي الشفرة روتين جزئي اخر ينفذ طريقة المطواة من اجل اي مجموعة من القياسات. احسب الاخطاء في الطاقة، المغنطة، السعة الحرارية و الحساسية المغناطيسية لنموذج ايزينغ باستعمال طريقة المطواة.

# التغير الطوري من الرتبة الثانية الفيرومغناطيسي

**الجزء الاول**    الاساس الحرج المرفق بالسعة الحرارية يعطي ب $\alpha = 0$، اي

$$\frac{C_v}{L^2} \sim (T_c - T)^{-\alpha} \, , \; \alpha = 0.$$

هذا يعني ان السعة الحرارية تتباعد لوغاريتميا عند $T = T_c$ وهذا يظهر علي شكل تزايد لقمة السعة الحرارية (القيمة الاعظمية) مع $L$ لوغاريتميا اي

$$\frac{C_v}{L^2}|_{\text{peak}} \sim \log L.$$

تحقق من هذا التصرف عدديا . استخدم شبكات بين $L = 10 - 30$ و $\text{TTH} = 2^{10}$، $\text{TMC} = 2^{13}$. درجات الحرارة تؤخذ في المجال

$$T = T_c - 10^{-2}\text{step} \, , \; \text{step} = -50, 50.$$

ارسم القيمة الاعظمية لـ $C_v/L^2$ بدلالة $\ln L$.

**الجزء الثاني**    المغنظة بجوار لكن تحت درجة الحرارة الحرجة تتصرف كالتالي

$$\frac{<M>}{L^2} \sim (T_c - T)^{-\beta} \, , \; \beta = \frac{1}{8}.$$

نقترح دراسة المغنطة بجوار $T_c$ من اجل تعيين قيمة $\beta$ عدديا. من اجل تحقيق هذا الهدف ارسم $|<M>|$ بدلالة $T_c - T$ حيث $T$ يؤخذ في المجال

$$T = T_c - 10^{-4}\text{step} \, , \; \text{step} = 0, 5000.$$

نعتبر شبكات كبيرة بين $L = 30 - 50$ مع $\text{TTH} = \text{TMC} = 2^{10}$. نذكر ان درجة الحرارة الحرجة في نموذج ايزينغ في بعدين تعطي ب

$$k_B T_c = \frac{2J}{\ln(\sqrt{2} + 1)}.$$

**الجزء الثالث**    الحساسية المغناطيسية بجوار درجة الحرارة الحرجة في نموذج ايزينغ في بعدين تتصرف كالتالي

$$\frac{\chi}{L^2} \sim |T - T_c|^{-\gamma} \, , \; \gamma = \frac{7}{4}.$$

عين $\gamma$ عدديا . استعمل $\text{TTH} = 2^{10}$، $\text{TMC} = 2^{13}$، $L = 50$ و خذ درجة الحرارة في المجالين

$$T = T_c - 5.10^{-4}\text{step} \, , \; \text{step} = 0, 100.$$

$$T = T_c - 0.05 - 4.5.10^{-3}\text{step} \, , \; \text{step} = 0, 100.$$

# دالة الربط (غرين) الثنائية

**في هذه المسألة نواصل دراسة التغير الطوري الفيرومغناطيسي. بالخصوص سوف نحسب في هذه المسالة دالة الربط (غرين) الثنائية المعرفة بالعبارة**

$$f(n) = <s_0 s_n>$$
$$= < \frac{1}{4L^2} \sum_{i,j} \phi(i,j) \Big( \phi(i+n,j) + \phi(i-n,j) + \phi(i,j+n) + \phi(i,j-n) \Big) > .$$

(1) **تحقق ان تصرف الدالة $f(n)$ عند $T = T_c$ يعطي ب**

$$f(n) \simeq \frac{1}{n^\eta} \ , \ \eta = \frac{1}{4}.$$

(2) **تحقق ان تصرف الدالة $f(n)$ من اجل $T$ اقل من $T_c$ يعطي ب**

$$f(n) = < M >^2 .$$

(3) **تحقق ان تصرف الدالة $f(n)$ من اجل $T$ اكبر من $T_c$ يعطي ب**

$$f(n) \simeq a \ \frac{1}{n^\eta} e^{-\frac{n}{\xi}} .$$

**في جميع الاسئلة اعلاه نأخذ شبكات فردية اي $L = 2LL + 1$ بين $LL = 20 - 50$. نعتبر ايضا القيم $TTH = 2^{10}$، $TTC = 2^{13}$.**

(4) **بالقرب من $T_c$ يتباعد طول الربط كالتالي**

$$\xi \simeq \frac{1}{|T - T_c|^\nu} \ , \ \nu = 1.$$

**في هذا السؤال نأخذ $LL = 20$. نعتبر ايضا القيم $TTH = 2^{10}$، $TTC = 2^{15}$ و درجات الحرارة**

$$T = T_c + 0.1.\text{step} \ , \ \text{step} = 0, 10.$$

**لاحظ ان جيران الدليل $i$ الذين يبعدون عنه مسافة $n$ يمكن ان يعطون بالشفرة التالية**

```
do i=1,L
do n=1,LL
if (i+n.le.L)then
ipn(i,n)=i+n
else
ipn(i,n)=(i+n)-L
endif
```

```
if ((i-n).ge.1)then
imn(i,n)=i-n
else
imn(i,n)=i-n+L
endif
enddo
enddo
```

# الهستريسيس و التغير الطوري من الرتبة الاولي

في هذه المسألة نعتبر تأثير حقل مغناطيسي علي فيزياء نموذج ايزينغ. سوف نلاحظ بالخصوص تغير طوري من الرتبة الاولي بالقرب من $H = 0$ و كذلك ظاهرة هستريسيس.

(1) نحسب المغنطة و الطاقة بدلالة $H$ من اجل درجات حرارة مختلفة. تجري الموازنة من اجل القيمة الاولي للحقل المغناطيسي و بعد حساب المغنطة المتوسطة نبدأ بتغيير الحقل المغناطيسي بشكل ادياباتيكي اي بشكل بطيء جدا عبر خطوات صغيرة حتي لا نخسر موازنة الجملة. نعتبر في هذا السؤال المجال $H = -5, 5$ مع خطوات تساوي $0.25$.

– عين من اجل $T < T_c$ (مثلا $T = 0.5$ و $T = 1.5$) موقع التغير الطوري من الرتبة الاولي من نقطة لا استمرارية (النقطة التي تقفز عندها) الطاقة و المغنطة. هذا التغير الطوري يحدث عند قيمة غير منعدمة للحقل المغناطيسي بسبب الهستريسيس. القفزة التي نلاحظها في الطاقة عند موقع التغير الطوري توافق قيمة غير منعدمة لكمية الحرارة الكامنة.

– بين ان المغنطة من اجل $T > T_c$ (مثلا $T = 3$ و $T = 5$) تصبح دالة سلسة (دالة مستمرة و قابلة للاشتقاق عدد كيفي من المرات) بالقرب من $H = 0$ و هذا يعني انه لا يوجد اي فرق بين الحالات الفيرومغناطيسية $M \geq 0$ و الحالات الفيرومغناطيسية $M \leq 0$.

(2) نعيد حساب المغنطة كدالة في $H$ من اجل المجال من $-5$ الي $5$ ذهابا و ايابا. يجب ان نلاحظ حلقة هستريسيس.

– تحقق من ان نافذة الهستريسيس تضيق بزيادة درجة الحرارة او بعد تراكم عدد اكبر من خطوات مونتي كارلو.

– ماذا يحدث عند زيادة حجم الشبكة.

تشير ظاهرة الهستريسيس الي ان تصرف الجملة يتعلق بحالتها الابتدائية و تاريخها او ان الجملة عالقة في حالات شبة مستقرة.



\end{appendices}
\end{document}